%% file: main.tex
\documentclass[11pt]{article}

\usepackage[numbers]{natbib}
\usepackage{amsmath}
\usepackage{graphicx}
\usepackage{amssymb}
\usepackage{mathrsfs}
\usepackage{amsthm}
\usepackage{xcolor}
\usepackage{array}
\usepackage{booktabs}
\usepackage{tabularx}
\usepackage{hyperref}
\usepackage{algorithm}
\usepackage{algpseudocode}
\usepackage{enumitem}\usepackage[margin=1in]{geometry}

\hypersetup{colorlinks=true,citecolor=blue,linkcolor=blue,hypertexnames=false}
\allowdisplaybreaks

\input{_0_vars}

\input{_1_maths}

\theoremstyle{plain}
\newtheorem{theorem}{Theorem}[section]
\newtheorem{lemma}[theorem]{Lemma}

\newtheorem{corollary}[theorem]{Corollary}
\theoremstyle{definition}

\begin{document}

\date{September 15, 2026}
\title{\paperTitle}
\author{\paperAuthor}
\maketitle

\begin{abstract}
\input{00_abstract}
\end{abstract} 
\newpage
\input{_2_body}

\bibliographystyle{alpha}
\bibliography{ref}

\appendix

\input{_3_app}

\end{document}

%% file: _0_vars.tex
\newcommand{\paperTitle}{Polynomial Time Algorithms for the Kadison-Singer Problem}
\newcommand{\paperAuthor}{Zhao Song\thanks{\texttt{magic.linuxkde@gmail.com}.} \and Song Yue}

%% file: _1_maths.tex
\DeclareMathOperator{\tr}{tr}
\newcommand{\E}{\mathbb E}
\newcommand{\R}{\mathbb R}
\providecommand{\C}{\mathbb C}
\renewcommand{\d}{\mathrm{d}}

%% file: 00_abstract.tex
Marcus, Spielman, and Srivastava~\cite{mss15} established the existence of Kadison--Singer partitions. We provide polynomial-time algorithms for the Kadison--Singer problem. For Hermitian matrices $A_1,\ldots,A_m\in\mathbb C^{n\times n}$ of rank at most one, we give two algorithms that find signs $\sigma\in\{\pm1\}^m$ satisfying $\|\sum_i \sigma_iA_i\|\le C\|\sum_i A_i^2\|^{1/2}$. The deterministic algorithm achieves $C=3.3443$ using $\widetilde O(mn^2+n^{4.75})$ arithmetic operations. The randomized algorithm achieves $C=4.8628$ using $\widetilde O(mn^2+n^{3.58})$ arithmetic operations in expectation. For vectors satisfying $\sum_i a_ia_i^*=I$ and $\|a_i\|^2\le\alpha$, the algorithms yield partitions $[m]=I_1\cup I_2$ satisfying $\|\sum_{i\in I_j}a_ia_i^*-I/2\|\le (C/2)\sqrt\alpha$ for $j=1,2$.

%% file: _2_body.tex
\input{01_intro}
\input{03_tech}

\input{20_better_constant}

\input{30_proof}

\input{40_bit_complexity}

%% file: 01_intro.tex
\section{Introduction}

Matrix discrepancy asks how small a signed sum of matrices can be in operator norm. For Hermitian matrices $A_1,\ldots,A_m\in\mathbb{C}^{n\times n}$ of rank at most one, the natural scale is $\|\sum_i A_i^2\|^{1/2}$. The rank-one matrix discrepancy problem seeks signs $\sigma_1,\ldots,\sigma_m\in\{\pm1\}$ satisfying
\begin{equation}\label{eq:intro:rank-one-discrepancy}
\|\sum_{i=1}^m \sigma_iA_i\|\le C\|\sum_{i=1}^m A_i^2\|^{1/2}
\end{equation}
for a universal constant $C$ independent of $m$ and $n$. Here $\|\cdot\|$ denotes the operator norm for matrices and the Euclidean norm for vectors. The operator norm requires the same signing to control every quadratic form $\sum_i \sigma_i u^*A_i u$ over unit vectors $u$.

A central special case is the partition of an isotropic family of vectors. Let $a_1,\ldots,a_m\in\mathbb{C}^n$ satisfy $\sum_i a_ia_i^*=I$ and $\|a_i\|^2\le\alpha$. The goal is to partition $[m]:=\{1,\ldots,m\}$ into two sets $I_1,I_2$ so that each matrix $\sum_{i\in I_j}a_ia_i^*$ approximates $I/2$. The algorithmic challenge is to find such a partition in polynomial time with a dimension-independent $O(\sqrt{\alpha})$ error. Weaver~\cite{w04} connected this finite-dimensional discrepancy problem to the Kadison--Singer problem~\cite{ks59} through his $\mathsf{KS}_2$ formulation. A signing satisfying Eq.~\eqref{eq:intro:rank-one-discrepancy} gives such a partition: for $A_i:=a_ia_i^*$, the identity $A_i^2=\|a_i\|^2A_i$ implies $\sum_i A_i^2\preceq\alpha I$. The two sign classes therefore satisfy
\begin{equation}\label{eq:intro:partition-from-signing}
\|\sum_{i\in I_j}a_ia_i^*-I/2\|\le (C/2)\sqrt{\alpha},\qquad j=1,2.
\end{equation}
The bound depends only on the largest squared vector norm and is uniform in the ambient dimension.

Marcus, Spielman, and Srivastava~\cite{mss15} resolved the Kadison--Singer problem by proving Weaver's conjecture. In the normalization above, their partition theorem gives
\[
\|\sum_{i\in I_j}a_ia_i^*\|\le (1/\sqrt{2}+\sqrt{\alpha})^2,\qquad j=1,2.
\]
Since the two parts sum to $I$, this also gives the two-sided estimate $\|\sum_{i\in I_j}a_ia_i^*-I/2\|\le\sqrt{2\alpha}+\alpha$. Kyng, Luh, and Song~\cite{kls20} subsequently proved the general rank-one discrepancy bound in Eq.~\eqref{eq:intro:rank-one-discrepancy} with $C=4$. Their theorem applies without an isotropy assumption and controls the discrepancy directly by the matrix variance. Xie, Xu, and Zhu~\cite{xxz22} improved this constant to $C=3$ under the same assumptions and matrix-variance normalization. Their proof uses interlacing polynomials. These results establish dimension-independent existence bounds.

Anari, Oveis Gharan, Saberi, and Srivastava~\cite{aoss18} gave an
algorithm for constructing Kadison--Singer partitions by approximating the
largest roots of interlacing polynomials. Its running time is
subexponential in the number of vectors for a fixed vector-norm bound.
Jourdan, Macgregor, and Sun~\cite{jms22} studied the algorithmic problem
$\mathsf{KS}_2(c)$ for real vectors, where the target partition error is
$c\sqrt{\alpha}$ for a fixed constant $c>0$. The algorithmic Kadison--Singer problem asks for a polynomial-time algorithm that finds signs yielding such a partition.

\subsection{Our Results}
We state our result as follows.

\begin{theorem}[Informal version of Theorem~\ref{rank_one_thm_main}, deterministic polynomial time algorithm]\label{thm_intro_constant}
For rational Hermitian matrices $A_1,\ldots,A_m\in\mathbb C^{n\times n}$
of rank at most one, a deterministic algorithm finds signs
$\sigma\in\{\pm1\}^m$ such that
\[
 \|\sum_i \sigma_iA_i\|\le3.3443\|\sum_iA_i^2\|^{1/2},
\]
using $\widetilde O(mn^2+n^{4.75})$ arithmetic operations.
If $A_i=a_ia_i^*$, $\sum_iA_i=I$, and $\|a_i\|^2\le\alpha$,
the two sign classes satisfy
\[
 \|\sum_{i\in I_j}a_ia_i^*-I/2\|\le 1.67215\sqrt\alpha,
 \qquad j=1,2.
\]
\end{theorem}

Our constant $3.3443$ is larger than the constant $3$ in the existence result of Xie, Xu, and Zhu~\cite{xxz22}. A natural direction for future work is to achieve $C=3$ with a polynomial-time algorithm.

Our second result uses randomization to obtain a faster construction
with a larger constant.

\begin{theorem}[Informal version of Theorem~\ref{soft_thm_algorithm}, randomized polynomial time algorithm]\label{thm_intro_randomized}
For rational Hermitian matrices $A_1,\ldots,A_m\in\mathbb C^{n\times n}$
of rank at most one, a zero-error randomized algorithm finds signs
$\sigma\in\{\pm1\}^m$ such that
\[
 \|\sum_i\sigma_iA_i\|\le4.8628\|\sum_iA_i^2\|^{1/2},
\]
using $\widetilde O(mn^2+n^{3.58})$ arithmetic operations in expectation.
If $A_i=a_ia_i^*$, $\sum_iA_i=I$, and
$\|a_i\|^2\le\alpha$, the two sign classes satisfy
\[
 \|\sum_{i\in I_j}a_ia_i^*-I/2\|\le2.4314\sqrt\alpha,
 \qquad j=1,2.
\]
\end{theorem}

When the rank-one matrices are given in factored form, the input consists of $O(mn)$ scalar entries. With our current techniques, it is difficult to exploit this representation without explicitly forming $n\times n$ matrices. A natural direction for future work is to use sketching techniques to achieve a running time of $O(mn+n^c)$ for a small constant $c$.

\paragraph{Organization.}
Section~\ref{sec_technical_overview} gives a technical overview of the proof.
Section~\ref{sec_rank_one_proof} establishes the main discrepancy bound.
Section~\ref{rank_one_sec_algorithm} gives the deterministic construction and the shared algorithmic tools.
Section~\ref{sec_fast_randomized} develops the randomized algorithm and
bounds its expected running time.
Section~\ref{sec_bit_complexity} gives the shared finite-precision
implementation and proves the bit-complexity bounds.

%% file: 03_tech.tex
\section{Technical overview}\label{sec_technical_overview}

Section~\ref{sec_tech_budgets} introduces the potential and explains how
the budgets determine the discrepancy constants.
Section~\ref{sec_tech_covariance} describes the covariance and
curvature estimates that give descent, and
Section~\ref{sec:tech:budget-exponents} explains the choice of exponents
and polynomial factors. Sections~\ref{sec:tech:trace-budget-properties}
and~\ref{sec:tech:fast-budget-properties} list the required properties
of the trace-corrected and fast budgets, respectively.
Section~\ref{sec:tech:weighted-descent} compares the two algorithms in a
common format, outlines their preprocessing and descent steps, and explains
the deterministic and randomized running-time bounds.

\subsection{The budgets and the discrepancy constants}\label{sec_tech_budgets}

After absorbing eigenvalue signs and applying a common scaling, the inputs
are PSD rank-one matrices $\bar A_1,\ldots,\bar A_m$ with $\sum_i\tr[\bar A_i]=1$.
Define $\zeta:=\|\sum_i \bar A_i^2\|$. A fractional signing $y\in[-1,1]^m$
determines $M(y):=\sum_i y_i\bar A_i$. We use a coupled semidefinite
potential based on Lehner's variational formula~\cite{l99}.
For a nonnegative budget $\beta$ and a parameter $\iota>0$, define
$\mathcal K_y(D):=\iota\sum_i\beta(y_i)\bar A_iD\bar A_i$. The matrix potential $\mathcal P(y)$
minimizes $b+\rho\tr[U+V]$ over $b\in\mathbb R$ and $U,V\succ0$
subject to
\begin{equation}\label{eq_tech_coupled_constraints}
 \begin{aligned}
 U^{-1}+M(y)+\mathcal K_y(V)&\preceq bI,\\
 V^{-1}-M(y)+\mathcal K_y(U)&\preceq bI.
 \end{aligned}
\end{equation}
The two constraints control the upper and lower ends of the spectrum of
$M(y)$. When a coordinate reaches an endpoint, its budget vanishes and
its contribution to the coupling is removed. The trace penalty gives
the regularity needed to differentiate $\mathcal P$; see
Lemma~\ref{rank_one_lem_potential}.

The square-root dependence on the matrix variance in
Eq.~\eqref{eq:intro:rank-one-discrepancy} suggests $1/2$ as a natural
starting exponent for the scalar budget. Appendix~\ref{sec_one_half}
analyzes the corresponding pure square-root budget. Our main theorems
further tune both the exponent and a polynomial factor:
\begin{align}
 \beta_{\rm sqrt}(t)&:=(1-t^2)^{1/2},
       &&\text{Theorem~\ref{one_half_rank_one_thm_main}},\label{eq_tech_budget_sqrt}\\*
 \beta_{\rm tr}(t)&:=(1-t^2)^{3/4}f_{\rm tr}(|t|),
       &&\text{Theorem~\ref{thm_intro_constant}},\label{eq_tech_budget_tr}\\*
 \beta_{\rm fast}(t)&:=(1-t^2)^{97/200}f(|t|),
       &&\text{Theorem~\ref{soft_thm_algorithm}}.\label{eq_tech_budget_fast}
\end{align}

\begin{figure}[!ht]
\centering
\includegraphics[width=\textwidth]{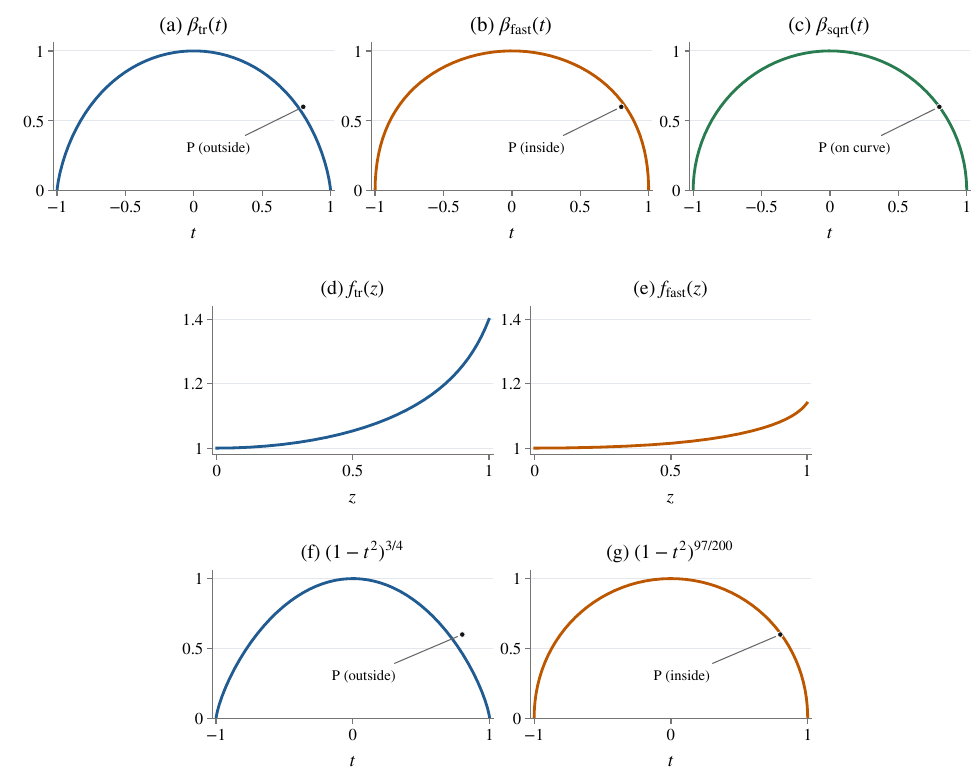}
\caption{Top row: the budgets $\beta_{\rm tr}(t)=(1-t^2)^{3/4}f_{\rm tr}(|t|)$,
$\beta_{\rm fast}(t)=(1-t^2)^{97/200}f_{\rm fast}(|t|)$, and
$\beta_{\rm sqrt}(t)=(1-t^2)^{1/2}$ on $[-1,1]$.
Middle row: the polynomial factors $f_{\rm tr}$ and $f_{\rm fast}$ on $[0,1]$.
Bottom row: the power functions $(1-t^2)^{3/4}$ and $(1-t^2)^{97/200}$ on $[-1,1]$.
Here $f_{\rm fast}=f$ in Eq.~\eqref{eq_tech_budget_fast}.
Each family of plots uses common axis limits.
Panels (a)--(c), (f), and (g) share the reference point $P=(0.8,0.6)$;
``inside'' and ``outside'' mean strictly below and strictly above the corresponding curve, respectively.}
\label{fig:budget:comparison}
\end{figure}

Here $f_{\rm tr}$ and $f$ are explicit rational polynomials of degree
$20$, specified in Section~\ref{sec_trace_budget} and
Section~\ref{sec_rank_one_proof}, respectively. Both have value $1$
and derivative $0$ at zero. The power determines how fast a budget
vanishes at an endpoint; the polynomial factor adjusts its curvature
in the interior.
All three of our budgets preserve
\[
 \beta(0)=1,\qquad \beta(\pm1)=0,\qquad
 1-t^2\le\beta(t)\le1.
\]
The trace-corrected budget also satisfies
$\beta_{\rm tr}(t)\ge(1-t^2)^{3/4}$, while
$\beta_{\rm fast}(t)\ge\sqrt{1-t^2}$.
Their covariance and curvature estimates are different, as described in
Section~\ref{sec_tech_covariance}, and have exact rational certificates.

\begin{figure}[!htb]
\centering
\setlength{\abovecaptionskip}{4pt}
\makebox[0pt][l]{\raisebox{6pt}[0pt][0pt]{\includegraphics[width=\linewidth,trim=0 4bp 0 234bp,clip]{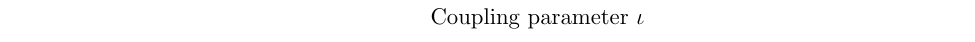}}}\raisebox{20bp}{\includegraphics[width=\linewidth,trim=0 24bp 0 0,clip]{figs/iota_feasibility.pdf}}
\par\vspace{-10pt}
\caption{Scalar feasibility for the deterministic algorithm in
Theorem~\ref{thm_intro_constant}. Define
$m_j^{\mathrm{tr}}(\iota):=\min_{(t,H,V)\in[0,1]^3}E_j(t,H,V;\iota)$,
where $(E_1,\ldots,E_6):=(F,B,\mathfrak f,\mathfrak u,\mathfrak z,\mathfrak p)$
are the six expressions in Eq.~\eqref{eq_trace_scalar_conditions}.
The curves estimate these minima numerically, with the budget and helper
polynomials fixed. The vertical axis uses a symmetric logarithmic scale.
The limiting crossing of $m_6^{\mathrm{tr}}$ is near $\iota=2.79592$.
The blue point marks $\iota_{\rm tr}=2.796$, for which
Lemma~\ref{lem_trace_surplus} proves strict positivity throughout the cube.}
\label{fig:iota:feasibility}
\end{figure}

For the trace-corrected budget, Figure~\ref{fig:iota:feasibility} shows
how the six scalar conditions in Section~\ref{sec_trace_budget} depend
on $\iota$ with the budget and helper polynomials held fixed.

\begin{figure}[!htb]
\centering
\setlength{\abovecaptionskip}{4pt}
\includegraphics[width=\linewidth,trim=0 4bp 0 0,clip]{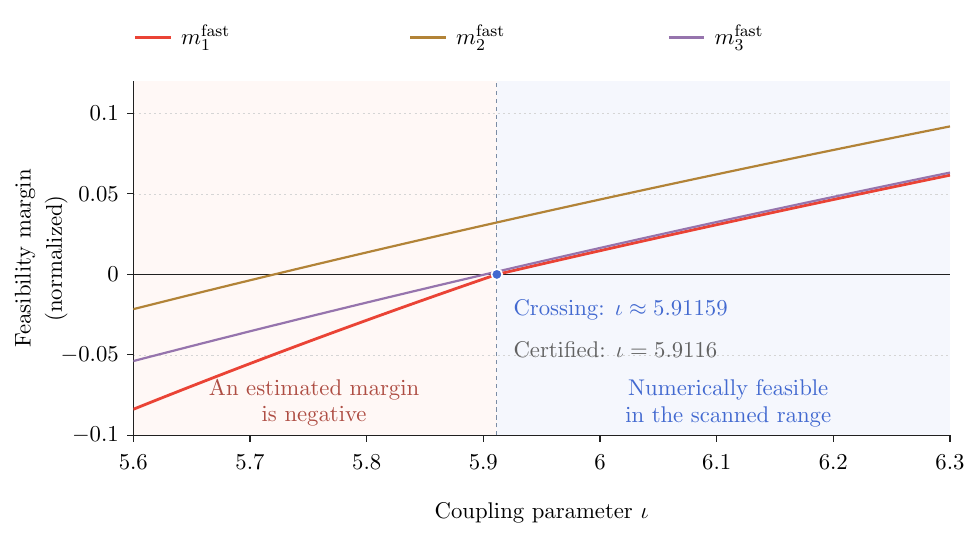}
\par\vspace{-10pt}
\caption{Scalar feasibility for the randomized algorithm in
Theorem~\ref{thm_intro_randomized}, with $\beta_{\rm fast}$, $\tau$,
$\kappa_{\rm fast}$, and $\vartheta_*$ fixed as in
Section~\ref{sec_rank_one_proof}.
The normalized margins $m_1^{\mathrm{fast}}$, $m_2^{\mathrm{fast}}$, and
$m_3^{\mathrm{fast}}$ measure joint curvature, budget drift, and the domain
bound, respectively. Define $m_1^{\mathrm{fast}}(\iota):=\min H_{\rm cert}/\mathcal A$
over the domain in Eq.~\eqref{rank_one_eq_joint_domain},
$m_2^{\mathrm{fast}}(\iota):=1-20(1+\tau/2)(1+\vartheta_*)/[9\iota(1-\vartheta_*^3)]$,
and $m_3^{\mathrm{fast}}(\iota):=1-(1+\vartheta_*)/(\iota\vartheta_*^2)$.
The curvature minimum is estimated numerically.
The movement condition $\iota>2$ holds throughout the plotted range.
The limiting crossing of $m_1^{\mathrm{fast}}$ is near $\iota=5.91159$.
The blue point marks $\iota_{\rm fast}=5.9116$, whose scalar conditions
are certified in Section~\ref{rank_one_sec_certificate}.}
\label{fig:iota:fast-feasibility}
\end{figure}

Figure~\ref{fig:iota:fast-feasibility} gives the corresponding scan for
the fast budget used by the randomized algorithm. The three margins
measure joint curvature, budget drift, and the domain bound used by
the scalar certificate.

To distinguish the scalar budget from its sum over coordinates, write
\[
 \mathcal B_{\rm tr}(y):=\sum_i\beta_{\rm tr}(y_i),\qquad
 \mathcal B_{\rm fast}(y):=\sum_i\beta_{\rm fast}(y_i).
\]
The matrix potential also depends on the chosen budget through $\mathcal K_y$.
For the fast budget, the descent analysis uses
$\mathcal F:=\mathcal P+\gamma\mathcal B_{\rm fast}$. The randomized
implementation smooths its matrix term as described below. For
Theorem~\ref{thm_intro_constant}, the budget $\beta_{\rm tr}$ enters $\mathcal P$
through the coupling, but the corrected drift gives strict descent of
$\mathcal P$ itself at light states; an auxiliary term $\gamma\mathcal B_{\rm tr}$
is not needed.

The effect of reducing $\iota$ is visible in the initial-value bound:
\[
 \|M(y)\|\le \mathcal P(y),\qquad
 \mathcal P(0)\le2\sqrt{\iota\zeta+2n\rho},\qquad
 (\mathcal P+\gamma\mathcal B)(0)\le2\sqrt{\iota\zeta+2n\rho}+\gamma m.
\]
Our covariance and curvature estimates permit
\[
 \iota_{\rm tr}:=\frac{699}{250},\qquad 2\sqrt{\iota_{\rm tr}}<3.3443,
 \qquad
 \iota_{\rm fast}:=\frac{14779}{2500},\qquad 2\sqrt{\iota_{\rm fast}}<4.8628.
\]
Small positive regularization and explicit error allowances give the
stated constants. Thus Theorem~\ref{thm_intro_constant} uses the
trace-corrected budget for the smaller constant, while
Theorem~\ref{soft_thm_algorithm} combines the other budget with
the faster randomized implementation.

\subsection{Sharper covariance and joint curvature}\label{sec_tech_covariance}

Our descent analysis combines the scalar budget with a covariance estimate
for the cost of matrix inversion. At the optimizer, let $W,Z\succ0$ be the dual
multipliers. For $\bar A_i=a_ia_i^*$, define
\[
 u_i:=a_i^*Ua_i,\quad v_i:=a_i^*Va_i,\quad
 w_i:=a_i^*Wa_i,\quad z_i:=a_i^*Za_i.
\]
Define $\beta_i:=\beta(y_i)$, $\beta_i':=\beta'(y_i)$, and
$\beta_i'':=\beta''(y_i)$. If $\iota\beta_i v_i\ge1-y_i$, coordinate $i$
can move to $1$ while preserving feasibility. The condition
$\iota\beta_i u_i\ge1+y_i$ similarly permits a move to $-1$. The remaining
case is a light state, where both inequalities are strict in the opposite
direction. There define
\[
 p_i:=1+\iota\beta_i'v_i,\qquad q_i:=1-\iota\beta_i'u_i,\qquad
 \Gamma_i:=\frac1{\iota\beta_i u_iv_i}
       >\frac{\iota\beta_i}{1-y_i^2}\ge \iota.
\]
The first two coefficients describe the response to a coordinate move;
$\Gamma_i$ measures the reciprocal of its local coupling strength.

Let $\mathcal Q(r)$ be the nonnegative quadratic cost of matrix inversion in the
first-order responses of $U,V$ to a direction $r$, with $b$ fixed.
Our weighted projection inequality,
Lemma~\ref{rank_one_lem_weighted}, retains the individual $\Gamma_i$.
For the faster algorithm, Lemma~\ref{rank_one_lem_covariance} gives a
centered direction with nonzero covariance
$K:=\E[rr^\top]$ satisfying
\begin{equation}\label{eq_tech_covariance_fast}
 \E[\mathcal Q(r)]\le\sum_i k_i
       (p_i^2u_iw_i+q_i^2v_iz_i)K_{ii},
\end{equation}
where
\[
 \vartheta_i:=\frac{2}{\sqrt{1+4\Gamma_i}-1},\qquad
 k_i:=\frac{1+\vartheta_i}{1-\vartheta_i^3}.
\]
As $\Gamma_i$ increases, $k_i$ decreases towards $1$. Keeping this
coordinate dependence preserves the stronger bound at weakly coupled
coordinates, in particular near an endpoint.

The scalar budget is matched to these same coefficients. With
$\beta:=\beta_{\rm fast}$ and $\iota:=\iota_{\rm fast}$,
Lemma~\ref{rank_one_lem_joint_curvature} proves
\begin{equation}\label{eq_tech_joint_curvature}
 \iota(-\beta_i''-\frac{\beta_i'^2}{\tau\beta_i})
 \ge(1+\kappa_{\rm fast})(2+\tau)k_ip_iq_i,
\end{equation}
for the fixed positive parameters $\tau$ and $\kappa_{\rm fast}$
specified in Section~\ref{sec_rank_one_proof}. The term $-\beta_i''$
is the concavity available for descent; the subtraction pays for the
mixed variation of the budget and the matrix optimizer. The right-hand
side is the cost after the covariance comparison. We bound
$k_ip_iq_i$ jointly with the endpoint position, and verify the
resulting scalar inequality by exact rational Bernstein certificates.
A coordinatewise drift then cancels the mismatch between the two
quadratic weights. The strict surplus in
Eq.~\eqref{eq_tech_joint_curvature} gives
Eq.~\eqref{eq_fast_strict_surplus}, which supports the faster algorithm.

Theorem~\ref{thm_intro_constant} uses a further refinement.
Lemma~\ref{lem_trace_covariance} retains a diagonal and both
off-diagonal local traces and controls the mixed response directly:
\[
 \E[\mathcal Q(r)]+\sum_i\Lambda_i\E[\mathsf Y_i\mathsf F_i]
 \le\sum_i\widetilde k_i
 (p_i^2u_iw_i+q_i^2v_iz_i)K_{ii}
       +v_{\rm corr}\cdot\nabla \mathcal P.
\]
Here $\mathsf Y_i,\mathsf F_i$ are paired response components, and
$\widetilde k_i,v_{\rm corr}$ are denoted in that lemma by $k_i,\nu$.
The exact mixed variation is the product on the left plus an explicit
gradient correction. The off-diagonal trace has a state-dependent
reference value whose deviation is also a gradient term. Retaining its
correlation with $\Lambda_i$ makes the direct comparison possible.
The other off-diagonal entry is also an exact gradient term. This lets
both row auxiliaries use both paired directions. The paired covariance
weight also retains an off-diagonal entry, controlled by a third helper
while preserving positive definiteness. All gradient corrections are
canceled by the drift.

The target coefficient is chosen as
\[
 \widetilde k_i:=\frac{\iota(-\beta_i'')}{2(1+\kappa_{\rm tr})p_iq_i},
 \qquad\kappa_{\rm tr}:=10^{-6}.
\]
Lemma~\ref{lem_trace_surplus} verifies that the covariance construction
works with this target at every light state. Thus the full budget
curvature remains available for descent. The covariance helpers
${j_{\rm 0}}_i:={j_{\rm 0}}(|y_i|,H_i)$ and $b_{{r_{\rm aux}},i}:=b_{r_{\rm aux}}(|y_i|,H_i)$, ${r_{\rm aux}}=0,1$, are fixed polynomials,
where $H_i:=\iota^2\beta_i^2u_iv_i/(1-y_i^2)$ is the normalized local product.
They are evaluated pointwise; no derivatives of them are needed. The
polynomially weighted $3/4$-power budget and
$\iota_{\rm tr}=699/250$ give strict descent of $\mathcal P$, as stated in
Eq.~\eqref{eq_trace_strict_descent}, and yield the constant $3.3443$.
The corrections also change the quantitative drift bounds:
Section~\ref{sec_trace_algorithm} combines the trace-corrected covariance
with entropy smoothing, deterministic grouped reduction, and block
response solves, giving dimension term $n^{4.742354}$. The comparison response holds both
matrix constraint expressions fixed, so smoothing leaves its second
variation unchanged. This is the source of the
constant--running-time tradeoff between our two theorems.

\subsection{Choice of the budget exponents}\label{sec:tech:budget-exponents}

The exponents $1/2$, $3/4$, and $97/200$ are paired with different
polynomial factors and covariance estimates. Their common purpose is
to permit a smaller coupling $\iota$, since the leading discrepancy bound
is $2\sqrt \iota$. The budget must supply enough concavity to pay for the
response of the matrix optimizer. The exponent and polynomial control
that concavity, while the covariance estimate controls its cost.

\paragraph{The tradeoff between center and endpoint curvature.}
Define $z:=|t|$. For $0<p<1$, let $\beta_p(t):=(1-t^2)^p f(|t|)$, where $f$ is a
polynomial with $f(0)=1$, $f'(0)=0$, and $f(1)>0$. Direct
differentiation gives
\begin{align}
 -\beta_p''(0)&=2p-f''(0),\label{eq:tech:exponent-center}\\
 \lim_{z\to1^-}\frac{\beta_p'(z)^2}{\beta_p(z)(-\beta_p''(z))}
 &=\frac{p}{1-p}.\label{eq:tech:exponent-endpoint}
\end{align}
Holding $f$ fixed, increasing $p$ increases the center curvature,
but also increases the endpoint mixed cost relative to concavity.
When the mixed term is bounded by Young's inequality with $\tau>0$,
the fraction of curvature remaining after this cost satisfies
\begin{equation}\label{eq:tech:exponent-remaining-curvature}
 \lim_{z\to1^-}
 \frac{-\beta_p''(z)-\beta_p'(z)^2/(\tau\beta_p(z))}{-\beta_p''(z)}
 =1-\frac{p}{\tau(1-p)}.
\end{equation}
Thus $\tau>p/(1-p)$ leaves a positive leading curvature margin,
whereas $\tau<p/(1-p)$ makes the useful curvature negative near an
endpoint. Lowering $p$ relaxes this endpoint constraint on $\tau$.
A smaller $\tau$ reduces the opposing factor $2+\tau$ in the fast joint
curvature inequality, although it also increases the mixed cost for a
fixed budget. The full inequality therefore determines the useful
balance. A regular polynomial factor adjusts the interior profile
while preserving the endpoint ratio in
Eq.~\eqref{eq:tech:exponent-endpoint}.

\paragraph{Why the trace-corrected budget uses $3/4$.}
A polynomial factor finite and positive at $1$ cannot preserve a lower
bound by a fixed positive multiple of $\sqrt{1-t^2}$ when $p>1/2$:
\begin{equation}\label{eq:tech:exponent-lower-bound}
 \frac{\beta_p(z)}{\sqrt{1-z^2}}=(1-z^2)^{p-1/2}f(z)
 \longrightarrow0\qquad(z\to1^-).
\end{equation}
The smaller-constant proof uses the weaker lower bound
$\beta_{\rm tr}(t)\ge(1-t^2)^{3/4}$ and accounts for it in its
quantitative estimates. At $p=3/4$, the endpoint mixed-cost ratio is
$3$, so a Young-inequality treatment would need $\tau>3$ for a
positive leading curvature margin. Our direct covariance comparison
includes the mixed variation itself, and both off-diagonal trace
identities are used. This stronger comparison permits the larger exponent.
The factor $f_{\rm tr}$ sets the center curvature to
$3/2-f_{\rm tr}''(0)=1.116899$ and adjusts the interior profile,
while the covariance helpers ${j_{\rm 0}}(|t|,H),b_0(|t|,H),b_1(|t|,H)$ also retain the local product.

The construction chooses the target covariance coefficient from the full
curvature and certifies its scalar conditions with coupling
$\iota_{\rm tr}=699/250$. After the stated numerical allowances, the
resulting discrepancy bound is $3.3443$.
This bound uses both the chosen budget and a covariance estimate that
retains the correlation between the mixed response and the local traces. The full-state certificate in Section~\ref{sec_trace_budget}
verifies this choice; it does not assert that the exponent is optimal.

\paragraph{Why the polynomial factor is useful.}
For the faster construction, a positive factor with $f'(z)\ge0$ has
logarithmic derivative
\begin{equation}\label{eq:tech:polynomial-log-derivative}
 \frac{\beta_p'(z)}{\beta_p(z)}
 =-\frac{2pz}{1-z^2}+\frac{f'(z)}{f(z)}.
\end{equation}
The term $f'/f$ offsets the negative logarithmic derivative of the pure
power, while $f''$ adjusts the curvature profile. We choose
$(p,f,\tau,\iota)$ jointly to satisfy the full joint curvature inequality
at every light configuration, together with the derivative and drift
conditions needed for the faster implementation. The degree-$20$ factor
provides this construction. The smaller-constant proof uses the stronger
direct mixed covariance and a separate degree-$20$ factor with
nonnegative derivative. Its certificate controls the full logarithmic
derivative and curvature, together with the covariance helpers ${j_{\rm 0}},b_0,b_1$.

\paragraph{Why the fast budget uses $97/200$.}
The fast analysis combines the covariance coefficient of
Lemma~\ref{rank_one_lem_covariance} with
$\tau_{\rm fast}:=\frac{4902302619}{5000000000}<1$.
For $p=97/200$, the endpoint ratio is $97/103<\tau_{\rm fast}$,
leaving a positive leading curvature margin. The degree-$20$
polynomial is matched to this exponent and to $\iota_{\rm fast}=14779/2500$;
Section~\ref{rank_one_sec_certificate} verifies the joint inequality
on its full domain. The derivative and zero-crossing properties needed
for the faster implementation are listed in
Section~\ref{sec:tech:fast-budget-properties}.
For comparison, $96/200$ and $98/200$ have endpoint ratios $12/13$
and $49/51$, respectively, both below $\tau_{\rm fast}$. Thus the
endpoint test permits both neighboring values; the full joint
certificate determines whether either works with specified polynomial
and coupling parameters. The value $97/200$ is the one certified here,
and changing it requires rechecking that certificate.

\subsection{Properties of the trace-corrected budget}\label{sec:tech:trace-budget-properties}

The budget $\beta_{\rm tr}(t):=(1-t^2)^{3/4}f_{\rm tr}(|t|)$ satisfies
the scalar conditions needed by the potential and direct mixed covariance.
Throughout this section, $z:=|t|$, $f:=f_{\rm tr}$, and
$\beta:=\beta_{\rm tr}$. Derivatives below are in $z$. Write
${r_{\rm aux}}:=1-z^2$, $p:=3/4$, $\mathfrak a:={r_{\rm aux}}f'-2pzf$, and
$\mathfrak d:=2p[1+(1-2p)z^2]f+4pz{r_{\rm aux}}f'-{r_{\rm aux}}^2f''$.

\begin{enumerate}
\item \textbf{Normalization and regularity at zero.}
The identities $\beta(0)=1$ and $\beta'(0)=0$ fix the initial potential
bound and make the even budget smooth on $(-1,1)$. In particular it has
the regularity required by Lemma~\ref{rank_one_lem_potential}.

\item \textbf{Positive, controlled growth.}
The budget satisfies $1-z^2\le\beta(z)\le1$ and
\begin{equation}\label{eq:tech:trace-log-growth}
 0\le-\frac{\beta'(z)}{\beta(z)}
 =-\frac{\mathfrak a(z)}{(1-z^2)f(z)}\le\frac{3}{4(1-z)}.
\end{equation}
The certificate gives $f\ge1$, $f'\ge0$, and $\mathfrak a\le0$.
At a light state with $y_i=z\ge0$, the logarithmic derivative bound gives
$p_i\ge1/4$ and $q_i\ge1$.
The roles exchange for negative coordinates. Also $\beta(\pm1)=0$,
so the coupling of a signed coordinate vanishes.

\item \textbf{Uniform concavity.}
Direct differentiation gives
\begin{equation}\label{eq:tech:trace-concavity}
 -\beta''(z)=\frac{\mathfrak d(z)}{(1-z^2)^{5/4}}>\frac12.
\end{equation}
This supplies a positive curvature scale for the descent estimates.

\item \textbf{Controlled derivatives through order three.}
At endpoint distance at least $\delta_{\rm end}$,
\begin{equation}\label{eq_trace_derivative_control}
 |\beta^{(j)}|\le C\delta_{\rm end}^{3/4-j},\qquad
 |\beta^{(j)}|/\beta\le C\delta_{\rm end}^{-j},\qquad 1\le j\le3.
\end{equation}
The fixed polynomial factor and its positive lower bound give these
estimates, which control the Taylor errors in the polynomial implementation.
This branch uses strict descent of $\mathcal P$ itself and does not require a
separate drift inequality for the scalar budget sum.

\item \textbf{Direct mixed covariance at every light configuration.}
Define $\iota:=699/250$ and $\kappa:=10^{-6}$. For every $z\in[0,1)$ and
every ${a_{\rm loc}},{b_{\rm loc}}>0$ satisfying $\iota\beta(z){a_{\rm loc}}<1+z$ and $\iota\beta(z){b_{\rm loc}}<1-z$, define
$p:=1+\iota\beta'(z){b_{\rm loc}}$, $q:=1-\iota\beta'(z){a_{\rm loc}}$, and
$\Gamma:=1/(\iota\beta(z){a_{\rm loc}}{b_{\rm loc}})$. The target coefficient is
\begin{equation}\label{eq:tech:trace-joint-curvature}
 k:=\frac{\iota(-\beta''(z))}{2(1+\kappa)pq}.
\end{equation}
Lemma~\ref{lem_trace_surplus} verifies the six scalar conditions that
make Lemma~\ref{lem_trace_covariance} hold with this target and
${j_{\rm 0}}:={j_{\rm 0}}(z,H)$ and $b_{r_{\rm aux}}:=b_{r_{\rm aux}}(z,H)$, ${r_{\rm aux}}=0,1$, where
$H:=\iota^2\beta(z)^2{a_{\rm loc}}{b_{\rm loc}}/(1-z^2)$. The proof
keeps the mixed coefficient and off-diagonal trace reference correlated
through the same state variables. Evenness covers negative coordinates
by exchanging ${a_{\rm loc}}$ and ${b_{\rm loc}}$. Once all gradient corrections are canceled,
the surplus $\kappa$ gives strict descent.
\end{enumerate}

Section~\ref{sec_trace_budget} specifies the ten integer coefficients
of $f_{\rm tr}$ and the helpers ${j_{\rm 0}},b_0,b_1$. Exact Bernstein checks establish
the budget inequalities, and an integer interval certificate with
$318239$ boxes proves the covariance inequalities on the full state domain.

\subsection{Properties of the fast budget}\label{sec:tech:fast-budget-properties}

Let $f_{\rm fast}$ denote the degree-$20$ polynomial denoted by $f$
in Section~\ref{sec_rank_one_proof}. Its role is to make
$\beta_{\rm fast}(t)=(1-t^2)^{97/200}f_{\rm fast}(|t|)$ satisfy both
the descent inequalities and the regularity estimates used by the faster
algorithm. Throughout this section, $z:=|t|$, $f:=f_{\rm fast}$,
$\beta:=\beta_{\rm fast}$, and $p:=97/200$. The following conditions
on a fixed rational polynomial $f$ supply the scalar properties used in
the proof.

\begin{enumerate}
\item \textbf{Normalization and regularity at zero.}
Require $f(0)=1$ and $f'(0)=0$. These give $\beta(0)=1$ and make
$\beta(t)=(1-t^2)^p f(|t|)$ a $C^2$ function on $(-1,1)$ with locally
Lipschitz second derivative. This is the regularity used in
Lemma~\ref{rank_one_lem_potential}. On each side of zero the budget is
smooth; derivatives of order three and higher at zero are interpreted
one-sided.

\item \textbf{Positive, controlled growth.}
Require $f(z)\ge1$ and $f'(z)\ge0$ on $[0,1]$, together with
$\beta'(z)\le0$ for $0\le z<1$. The derivative conditions are equivalent to
\begin{equation}\label{eq:tech:fast-log-growth}
 0\le\frac{f'(z)}{f(z)}\le\frac{2pz}{1-z^2}
 =\frac{97z}{100(1-z^2)}.
\end{equation}
Since $p<1/2$, these conditions imply
$\sqrt{1-z^2}\le(1-z^2)^p\le\beta(z)\le1$.
The factor $(1-t^2)^p$ gives $\beta(\pm1)=0$, while $f(1)$ remains
positive. These bounds control the coupling at light states and allow
the coupling of a signed coordinate to be removed.

\item \textbf{Uniform concavity.}
Require
\begin{equation}\label{eq:tech:fast-concavity}
 -\beta''(z)>\frac12,\qquad 0\le z<1.
\end{equation}
This gives a positive scale for the budget weights in
Lemma~\ref{fast_lem_dichotomy}. At zero it becomes
$2p-f''(0)>1/2$, or equivalently $f''(0)<47/100$.
The concavity condition applies to the full budget $\beta$.

\item \textbf{Drift compatibility.}
Require
\begin{equation}\label{eq:tech:fast-drift}
 (-\beta'(z))(1+z)\le\beta(z)(-\beta''(z)),\qquad 0\le z<1.
\end{equation}
This bounds the first-derivative contribution to the budget drift by
its negative second-derivative contribution. The drift analysis in
Section~\ref{sec_rank_one_proof} uses this inequality to prove strict
descent of the auxiliary budget sum $\mathcal B(y):=\sum_i\beta(y_i)$.
Together with matrix-potential descent, it yields descent of
$\mathcal F:=\mathcal P+\gamma\mathcal B$ for $\gamma>0$.

\item \textbf{Joint curvature at every light configuration.}
Define $\iota:=14779/2500$, $\tau:=\frac{4902302619}{5000000000}$, and
$\kappa:=10^{-8}$. For every $z\in[0,1)$ and every ${a_{\rm loc}},{b_{\rm loc}}>0$ satisfying
$\iota\beta(z){a_{\rm loc}}<1+z$ and $\iota\beta(z){b_{\rm loc}}<1-z$, define
$p:=1+\iota\beta'(z){b_{\rm loc}}$, $q:=1-\iota\beta'(z){a_{\rm loc}}$, and
$\Gamma:=1/(\iota\beta(z){a_{\rm loc}}{b_{\rm loc}})$. The growth conditions ensure
$p,q>0$ and $\Gamma>\iota>2$. Define
$\vartheta:=2/(\sqrt{1+4\Gamma}-1)$ and
$k:=(1+\vartheta)/(1-\vartheta^3)$, the coefficient from
Lemma~\ref{rank_one_lem_covariance}. Require
\begin{equation}\label{eq:tech:fast-joint-curvature}
 \iota(-\beta''(z)-\frac{\beta'(z)^2}{\tau\beta(z)})
 \ge(1+\kappa)(2+\tau)kp q.
\end{equation}
Evenness covers negative coordinates by exchanging ${a_{\rm loc}}$ and ${b_{\rm loc}}$.
The concavity remaining after the mixed-derivative cost must dominate
the entire covariance cost for every light configuration.
Lemma~\ref{rank_one_lem_joint_curvature} verifies this inequality.
Its strict surplus supports the weighted gradient/curvature alternative
in Lemma~\ref{fast_lem_dichotomy}. The same parameters ensure
$(1+\tau/2)k/\iota<9/20$, which supplies the strict margin in the budget
drift estimate.

\item \textbf{Controlled derivatives through order four.}
For $0<\delta_{\rm end}\le1/2$ and $|t|\le1-\delta_{\rm end}$, the fixed polynomial form
and positivity of $f$ give
\begin{align}
 |\beta^{(j)}(t)|&\le C\delta_{\rm end}^{p-j},\qquad 1\le j\le4,
 \label{eq:tech:fast-derivative-growth}\\
 \frac{|\beta^{(j)}(t)|}{\beta(t)}&\le C\delta_{\rm end}^{-j},\qquad 1\le j\le4.
 \label{eq:tech:fast-relative-derivatives}
\end{align}
Here $C$ depends only on the fixed polynomial, and derivatives at zero
are one-sided. With the algorithm's cutoff
$\delta_{\rm end}^{-1}=O(\log(2n))$, these bounds and uniform concavity imply
\begin{equation}\label{eq:tech:fast-weighted-derivatives}
 |\beta^{(j)}(t)|+\frac{|\beta^{(j)}(t)|^2}{\beta(t)}
 \le P_n(-\beta''(t)),\qquad 1\le j\le4,
\end{equation}
where $P_n$ is a fixed polynomial in $\log(2n)$.
These estimates control the directional responses in
Lemma~\ref{fast_lem_directional} and the smoothed local corrections
in Lemma~\ref{soft_lem_transport}.

\item \textbf{A nonpositive third-derivative jump at zero.}
Require $f'''(0)\le0$. Since $f'(0)=0$, the even budget satisfies
\begin{equation}\label{eq:tech:fast-zero-jump}
 \beta'''(0^+)-\beta'''(0^-)=2f'''(0)\le0.
\end{equation}
The chosen polynomial has cubic coefficient $-5706553/10^9$, so this
inequality is strict. Along a line in the coordinate space, each zero
crossing therefore contributes a nonpositive jump to the third
derivative of $\mathcal F$. The local estimates in
Section~\ref{soft_sec_geometry} use the appropriate one-sided derivatives
on each side of zero.
\end{enumerate}

Section~\ref{rank_one_sec_certificate} gives exact rational
Bernstein certificates for the scalar descent inequalities. Fixed
rational coefficients also permit evaluation of the polynomial and its
derivatives at the short precision used in
Section~\ref{sec_fast_precision}. The degree and coefficients provide
one construction satisfying these requirements; the proof uses the
stated properties of the resulting budget.

\subsection{Efficient implementations and running-time bounds}
\label{sec:tech:weighted-descent}
\label{sec:tech:simplified-algorithms}

Algorithms~\ref{alg:overview:deterministic} and~\ref{alg:overview:randomized}
summarize Algorithms~\ref{alg:trace:signing} and~\ref{alg:randomized:signing}.
The first gives constant $3.3443$ deterministically; the second gives
constant $4.8628$ with a smaller expected running time. Both use the
same input conversion and certified optimizer corrections. The randomized
algorithm holds a small background fixed and uses curved moves while
many coordinates remain, then switches to endpoint and gradient sweeps.
All randomized routines use finite caps, verified tests, and deterministic
fallbacks.

Both algorithms first round the input once to short PSD rank-one dyadic
matrices. The grouped-coordinate reduction in
Lemma~\ref{rank_one_lem_grouped_partial_signing} prepares their common
partial-signing framework. A balanced binary tree stores the matrix sums
of coordinate groups. Each group has one shared fractional coefficient,
and splitting it preserves the signed sum by copying this coefficient
to both children. Computing all group sums costs $\widetilde O(mn^2)$.

The shared local estimates use
\[
 \varpi_i:=(-\beta_i'')(\iota(v_iw_i+u_iz_i)+\gamma),\qquad
 G:=\operatorname{diag}(\varpi_i).
\]
Lemmas~\ref{soft_lem_ball} and~\ref{soft_lem_transport} give the smoothed
Newton neighborhood and optimizer correction.
Sections~\ref{sec_fast_reset}--\ref{sec_fast_arithmetic} give capped
initialization and operator actions. Section~\ref{sec_bit_complexity}
gives their finite-precision implementations.
The short input cache and strict discrepancy margins keep the original
binary input length additive. Section~\ref{rank_one_sec_algorithm}
accounts for input approximation, endpoint rounding, numerical
comparisons, and optimizer correction.

The running-time analysis depends on efficient matrix computations.
Related inverse-maintenance methods for semidefinite programming were
developed by Huang et al.~\cite{hjstz22}.
For rectangular matrix multiplication, we use the bounds of
Alman et al.~\cite[Table~1]{adwxxz25} and
Le Gall and Urrutia~\cite[Table~3]{lu18}.

\begin{algorithm}[!htb]
\caption{Simplified deterministic signing}\label{alg:overview:deterministic}
\begin{algorithmic}[1]
\Procedure{SimplifiedDeterministicSigning}{$A_1,\ldots,A_m$}
    \State $(\beta,\iota)\gets(\beta_{\rm tr},699/250)$.
    \State Preprocess \textbf{deterministically} and initialize the potential.
        \Comment{Lemma~\ref{trace_lem_deterministic_groups}}
    \While{active coordinates remain after phase updates and deferrals}
        \If{a regular endpoint move is certified}
            \State Fix its sign and continue the optimizer to the new face.
        \ElsIf{the large-gradient test succeeds}
            \State Perform a deterministic gradient sweep and correct the optimizer.
        \Else
            \State Take a certified step in a \textbf{deterministically selected direction}.
                \Comment{Lemma~\ref{trace_lem_block_responses}}
        \EndIf
    \EndWhile
    \State Round deferred coordinates \textbf{deterministically}. \Comment{Lemma~\ref{sr_lem_round}}
    \State Recover the original-input signing $\sigma$.
    \State \Return $\sigma$.
\EndProcedure
\end{algorithmic}
\end{algorithm}

\begin{algorithm}[!htb]
\caption{Simplified zero-error randomized signing}\label{alg:overview:randomized}
\begin{algorithmic}[1]
\Procedure{SimplifiedRandomizedSigning}{$A_1,\ldots,A_m$}
    \State $(\beta,\iota)\gets(\beta_{\rm fast},14779/2500)$.
    \State Preprocess by certified orientations and a smaller sketch; initialize the potential.
        \Comment{Lemma~\ref{bs_lem_grouped}}
    \While{the resident count exceeds $K_c$}
        \State Hold near-endpoint coordinates fixed until a bulk deletion is due.
            \Comment{Lemma~\ref{br_lem_flushes}}
        \State Mask the pending and non-light coordinates, retaining their coupling.
        \State Take a certified drift-adjusted curved move using capped trials and fallback.
            \Comment{Lemmas~\ref{md_lem_step} and~\ref{br_lem_sum}}
    \EndWhile
    \State Finish using endpoint continuation, gradient sweeps, and checked spread moves.
        \Comment{Section~\ref{am_sec_improvement}}
    \State Round deferred coordinates by verified sampling with fallback.
        \Comment{Lemma~\ref{oe_lem_round}}
    \State Recover the original-input signing $\sigma$.
    \State \Return $\sigma$.
\EndProcedure
\end{algorithmic}
\end{algorithm}

\paragraph{Deterministic algorithm.}
Algorithm~\ref{alg:overview:deterministic} completes a fractional signing
while controlling the norm of its matrix sum. After input conversion
and variance scaling, the matrices $C_i$ satisfy $\sum_iC_i^2\preceq I$.
The deterministic reduction in Lemma~\ref{trace_lem_deterministic_groups}
fixes all but at most $2n^2$ coefficients, with a prescribed residual,
using $\widetilde O(mn^2+n^{2\omega_0})$ arithmetic operations.
Padding the matrices by zeros increases their order by only a constant
factor and restores the active-count condition. The subsequent state
consists of an active set $I$, fractional coefficients $y_i$, and a fixed
contribution $F$, giving the modeled sum $M(y):=F+\sum_{i\in I}y_iC_i$.
The initial residual is charged once when recovering the final signing.

The algorithm uses the trace-corrected budget $\beta_{\rm tr}$ and
coupling $\iota=699/250$. It maintains the primal matrices $U,V$ and
dual matrices $W,Z$ defining the smoothed potential $\mathcal P_{\rm s}$.
Lemmas~\ref{soft_lem_ball} and~\ref{soft_lem_transport} give the
Newton-neighborhood and optimizer-correction estimates.
Entropy smoothing replaces the largest-eigenvalue objective by a
log-trace-exponential, giving differentiable optimizer equations while
changing the potential by at most its reserved smoothing allowance.
Eq.~\eqref{soft_eq_sandwich} gives
$\mathcal P_{\rm s}(y)\ge\|M(y)\|$, so controlling this potential
controls discrepancy. In a phase starting with $K$ active coordinates,
the descent potential is
\[
 \mathcal F_{\rm s}(y):=\mathcal P_{\rm s}(y)
             +\lambda_K\sum_{i\in I}\beta_{\rm tr}(y_i).
\]
The small positive reservoir supplies curvature in the coordinate
directions. Its coefficient is fixed during the phase and renewed when
the active count halves; the total potential added by these renewals
fits a fixed allowance. Each phase also uses Lemma~\ref{trace_lem_small_rounding},
with the phase cutoff in Section~\ref{sec_trace_algorithm}, to round
sufficiently small-trace inputs and remove their coupling contributions
in a batch.

The local loop first tests whether a coordinate can be fixed at $1$ or
$-1$. Endpoint scores computed from the current optimizer certify that
such a move respects the potential allowance. When a test succeeds,
the algorithm records the sign and continues the optimizer to the
smaller active problem using the trace-budget application of
Lemma~\ref{am_lem_endpoint}. Otherwise it tests the weighted gradient.
A large gradient gives first-order descent. A gradient sweep then uses
scalar coordinate models evaluated at one fixed optimizer to make
several updates, followed by an optimizer correction. The bounds in
Lemmas~\ref{am_lem_frozen} and~\ref{am_lem_sweep} specify how far the
sweep can move before those models must be refreshed.

If both tests fail, the trace-corrected covariance estimate guarantees
negative curvature. A polynomial spectral filter isolates negative-curvature directions,
and the conditional-expectation construction in
Lemma~\ref{trace_lem_block_responses} fixes auxiliary signs one at a
time to construct a direction $r$.
It controls both the largest coordinate of $r$ and the four relative
matrix responses: the changes of $U,V,W,Z$ measured against their
current values. These bounds permit a finite step while keeping the
matrices positive and the coordinates inside the cube.
The direction is oriented so its first-order contribution is
nonpositive. The finite-step bound in Eq.~\eqref{eq_trace_matrix_step_bound},
with the step size in Eq.~\eqref{eq_trace_matrix_steps}, gives a definite
potential decrease. The new optimizer is obtained by the initialization
homotopy in Section~\ref{soft_sec_init}, and the decrease is checked.

Coordinates within the endpoint cutoff are handled separately from
regular endpoint moves. They are frozen at their current fractional
values, whose contributions are retained in $F$, and their coupling
terms are removed together. Once no active coordinates remain,
deterministic matrix-exponential rounding assigns all deferred signs
jointly, with the discrepancy allowance of Lemma~\ref{round_lem_round}
and the faster implementation of Lemma~\ref{sr_lem_round}. This permits an
endpoint cutoff whose reciprocal is logarithmic in $n$, keeping the
budget derivatives controlled throughout the local loop.

The operation count combines these mechanisms. During a phase with
$K/2<k\le K$, there are
$\widetilde O(\min\{n,n^2/K\})$ curvature moves.
Computing all response columns at one optimizer together gives total
curvature-search work over the
phases of $\widetilde O(n^{2\omega_0})$. Gradient sweeps reuse an
optimizer across updates, and endpoint continuations are charged to
their potential decreases. Adding initialization, preprocessing, and
final rounding gives $\widetilde O(mn^2+n^{4.742354})$ arithmetic
operations. The discrepancy allowances preserve the constant $3.3443$
in Theorem~\ref{rank_one_thm_main}; Section~\ref{sec_bit_complexity}
gives the corresponding bit bound with additive $L_{\rm in}$.

\paragraph{Randomized algorithm.}
Algorithm~\ref{alg:overview:randomized} uses random sketches and sampled
directions to reduce the work of constructing a signing. After the common
input conversion and variance scaling, Lemma~\ref{bs_lem_grouped}
leaves at most $2n^2$ fractional original coefficients using
$\widetilde O(mn^2)$ arithmetic operations in expectation. Its sketches
are formed through actions of the original rank-one matrices. Residual
and norm tests certify the reduced instance before it is used.
Constant-factor padding restores the active-count condition, and the
single preprocessing residual is reserved in the final discrepancy
allowance.

The local algorithm uses the fast budget $\beta_{\rm fast}$, coupling
$\iota=14779/2500$, and the smoothed potential with its scalar reservoir.
It maintains the fractional coefficients and a corrected primal-dual
optimizer. A coordinate is \emph{resident} while its matrix and coupling
term remain in this model. On reaching the endpoint cutoff, its
coefficient is held fixed and marked as pending. Entering this pending
set changes neither the potential nor the optimizer. The algorithm also
holds fixed the coordinates whose endpoint scores do not certify the
lightness conditions needed for a curved move. Lemma~\ref{br_lem_mask}
bounds this combined mask, leaving most resident coordinates movable.

At a state with many resident coordinates, the algorithm modifies the
Hessian by subtracting a diagonal term involving the gradient and the
coordinate drift. Lemma~\ref{md_lem_inertia} gives many negative
eigenvalues for this modified operator. A polynomial filter applied to
a random sign vector produces a candidate direction. Random sketches
accelerate the inverse actions needed to evaluate this filter.
Lemma~\ref{md_lem_direction} checks the candidate's negative curvature,
coordinate spread, and relative primal and dual matrix responses
against the true model. These checks ensure that an accepted direction
has both sufficient descent and controlled changes in the optimizer.
Each trial has an absolute conditional success probability.
Unsuccessful trials are discarded; a finite trial cap is followed by
a certified deterministic move.

The drift correction is realized by a quadratic path. For each movable
coordinate, Eq.~\eqref{md_eq_path} prescribes
\[
 y_i(s):=y_i+s r_i-s^2d_i r_i^2,
\]
where $r$ is the accepted direction and $d_i$ is the drift coefficient
computed at the base optimizer. Masked coordinates stay fixed.
The second derivative along this path equals the quadratic form of
the modified Hessian. The sign of $s$ is chosen to make the initial
linear contribution nearly nonpositive. A first-order optimizer
prediction is followed by a root correction and a true potential
comparison. Lemma~\ref{md_lem_step} gives decrease
$\widetilde\Omega(\sqrt{k}/n^{3/2})$ when the resident count is $k$.

Pending coordinates are deleted only after they form a prescribed
inverse-polylogarithmic fraction of the resident set. Their combined
coupling is gradually attenuated, their fixed matrix contributions
are retained, and the optimizer is corrected for the resulting model.
Each deletion therefore reduces the resident count by that fraction.
Lemma~\ref{br_lem_flushes} charges the continuation stages to the
potential decrease of each batch, giving only
$\widetilde O(\sqrt n)$ stages in total. This batching shares the
optimizer-update work across many coordinates.

When $k$ falls to $K_c=nP_n$, where $P_n$ is a fixed polylogarithm,
the algorithm clears the pending batch and switches to the endpoint,
gradient-sweep, and verified spread-curvature moves of
Section~\ref{am_sec_improvement}. Finally, Lemma~\ref{oe_lem_round}
rounds the deferred coordinates. Matrices above its trace cutoff receive
their nearest signs; the remaining signs are sampled with the recorded
means. A norm certificate
accepts a signing only within the reserved error allowance, and capped
sampling has a deterministic fallback. Restoring the input eigenvalue
signs then produces the original-input signing.

The expected operation bound reflects both progress and evaluation.
Rational matrix-function approximations and shifted Lyapunov solves
supply the required actions. The sketch sizes and recursive solve depth
adapt to the resident count; Lemma~\ref{br_lem_sum} bounds a complete
trial, including response tests and optimizer correction. In a phase
$K/2<k\le K$, Lemma~\ref{br_lem_flushes} bounds successful curved moves
by $\widetilde O(n^{3/2}/\sqrt K)$. Combining these bounds and charging
the largest-count phases to $mn^2$ gives
$\widetilde O(mn^2+n^{3.575374})$ arithmetic operations in expectation;
Section~\ref{sec:randomized:operation-totals} includes the continuation,
finishing, and final-rounding work. The discrepancy allowances give
the constant $4.8628$ in Theorem~\ref{soft_thm_algorithm}.
Section~\ref{sec_bit_complexity} gives the corresponding bit bound
with additive $L_{\rm in}$.

%% file: 20_better_constant.tex
\section{A polynomial time algorithm with small constant}
\label{sec_rank_one_proof}

Section~\ref{sec:main:discrepancy:bound} states the main discrepancy bound.
Section~\ref{sec:constant:potential} introduces the normalization, budget, and matrix potential,
and Section~\ref{sec:constant:variations} derives the endpoint moves and variation formulas.
Sections~\ref{sec:constant:projection} and~\ref{sec:constant:covariance} prove a weighted projection
inequality and use it to construct the covariance.
Sections~\ref{sec:constant:descent} and~\ref{rank_one_sec_certificate} establish strict descent
and verify the required scalar inequalities.
Sections~\ref{sec_trace_correction} and~\ref{sec_trace_budget} correct the covariance using the
local trace and prove the improved constant with its exact certificate.
Section~\ref{rank_one_sec_algorithm} gives the algorithmic implementations.

\subsection{Main discrepancy bound}\label{sec:main:discrepancy:bound}

The formal version of Theorem~\ref{thm_intro_constant} combines the
trace-corrected discrepancy bound with an explicit arithmetic cost.
Theorem~\ref{bit_thm_totals} gives the corresponding bit-complexity bound
with the same discrepancy and partition guarantees.

\begin{theorem}[Formal version of Theorem~\ref{thm_intro_constant}, deterministic polynomial time algorithm]\label{rank_one_thm_main}
For rational Hermitian matrices $A_1,\ldots,A_m\in\C^{n\times n}$
of rank at most one and total binary length $L_{\rm in}$, a deterministic
algorithm finds signs $\sigma\in\{\pm1\}^m$ such that
\[
 \|\sum_i \sigma_iA_i\|\le3.3443\|\sum_iA_i^2\|^{1/2}.
\]
The number of arithmetic operations satisfies
\[
 T_{\rm arith}\le\widetilde O(mn^2+n^{4.742354}).
\]
If $A_i=a_ia_i^*$, $\sum_iA_i=I$, and $\|a_i\|^2\le\alpha$, the returned
signs determine a partition $[m]=I_1\cup I_2$ satisfying
\[
 \|\sum_{i\in I_j}a_ia_i^*-I/2\|\le1.67215\sqrt\alpha,
 \qquad j=1,2.
\]
The partition is computed within the same arithmetic bound.
Here $\widetilde O$ suppresses logarithmic factors in $m,n,L_{\rm in}$.
\end{theorem}
\begin{proof}
Use the budget $\beta_{\rm tr}$ and $\iota_{\rm tr}=699/250$ from
Section~\ref{sec_trace_budget}. For nonzero PSD inputs, write
$\zeta:=\|\sum_i\bar A_i^2\|$ and fix $\rho>0$. The continuous matrix
potential $\mathcal P$ attains a minimum on the compact cube. Among its minimizers,
choose one with the fewest active coordinates. An available endpoint move
does not increase $\mathcal P$ and removes a coordinate, contradicting this choice.
Otherwise all active coordinates are light. The active gradient vanishes
and the active Hessian is PSD, contradicting the strict generator bound
in Eq.~\eqref{eq_trace_strict_descent}. Thus some minimizer is a vertex,
and the scalar feasible point in Eq.~\eqref{rank_one_eq_initial} gives
\[
 \|\sum_i \sigma_i\bar A_i\|\le2\sqrt{\iota_{\rm tr}\zeta+2n\rho}.
\]
Let $\rho$ tend to zero and take a vertex that occurs infinitely often.
Since $4\iota_{\rm tr}<(3.3443)^2$, this proves the claimed bound.
For general Hermitian rank-one inputs, apply the result to $|A_i|$ and
absorb the eigenvalue signs; $|A_i|^2=A_i^2$ preserves the variance.
The deterministic construction and its arithmetic cost are proved in
Section~\ref{sec_trace_algorithm}. Finally $A_i^2\preceq\alpha A_i$ in the isotropic case,
and each partition error is half the signed sum.
\end{proof}

Algorithm~\ref{alg:trace:signing} records the deterministic construction for rational inputs.
For the active set $I$, it maintains the modeled sum
$M(y):=F+\sum_{i\in I}y_iC_i$ and the smoothed potential
$\mathcal F_{\rm s}:=\mathcal P_{\rm s}+\lambda_K\sum_{i\in I}\beta_i$
from Eq.~\eqref{eq_trace_matrix_reservoir}, with budget
$\beta:=\beta_{\rm tr}$ and coupling $\iota:=699/250$.
In preprocessing, replace the grouped reduction by
Lemma~\ref{trace_lem_deterministic_groups}, pad the retained matrices
to working order $N:=2^{\lceil\log_2(2n)\rceil}$, and use $N$ in
place of $n$ in all subsequent parameters and routines.
At the current optimizer, define $\beta_i:=\beta(y_i)$,
$u_i:=\tr[C_iU]$, and $v_i:=\tr[C_iV]$.
The scores $\iota\beta_i v_i-(1-y_i)$ and $\iota\beta_i u_i-(1+y_i)$
correspond to the endpoints $1$ and $-1$, respectively.
The parameters $\rho$, $\chi$, $\delta_{\rm end}$, $\delta$, and
$\lambda_K$, together with all precision and comparison margins,
are chosen as in Section~\ref{sec_trace_algorithm}.
Gradient sweeps use the original weights
$\kappa_i:=\iota(v_iw_i+u_iz_i)$ and two separated thresholds.
Curvature search uses $G_+$ and the matrix-response bounds in
Eq.~\eqref{eq_trace_matrix_direction}. Its oriented step has length
given by Eq.~\eqref{eq_trace_matrix_steps}; the new optimizer is
initialized by the explicit data homotopy. Compute all response columns
at one root together by Lemma~\ref{trace_lem_block_responses}.
Regular endpoints use the charged continuation of
Lemma~\ref{am_lem_endpoint}, as justified in Section~\ref{sec_trace_algorithm}.
The small-input cutoff and reservoir coefficient are renewed when
the active count halves. Near-endpoint slots are removed together.

\begin{algorithm}[!htb]
\caption{Deterministic rank-one signing with discrepancy constant $3.3443$}
\label{alg:trace:signing}
\begin{algorithmic}[1]
\Procedure{TraceSigning}{$A_1,\ldots,A_m$}
    \Comment{Theorem~\ref{rank_one_thm_main}}
    \State $(C,y,I,F,\mathcal R)\gets\Call{Preprocess}{A_1,\ldots,A_m;10^{-7},10^{-6}}$.
        \Comment{With Lemma~\ref{trace_lem_deterministic_groups}}
    \If{$I=\varnothing$}
        \State Recover and return the original-input signing using $\mathcal R$.
    \EndIf
    \State Pad the retained matrices to order $N$ and use $n\gets N$ below.
    \State Initialize $\mathcal D\gets\varnothing$, $K\gets2|I|$, $\lambda_K\gets0$, and $L\gets1+\lceil\log_2(2n^2)\rceil$.
    \State Build the common dyadic cache; initialize the smoothed optimizer.
        \Comment{Section~\ref{soft_sec_init}}
    \While{$I\ne\varnothing$}
        \If{$|I|\le K/2$}
            \State Set $K\gets|I|$ and select $S$ using cutoff $\chi/(4L\sqrt K)$.
            \State Round $S$ jointly and record its signs in $\mathcal R$.
            \State Preserve $M$, attenuate the slots of $S$ together, and remove them from $I$.
            \State Set the new reservoir coefficient $\lambda_K$.
                \Comment{Eq.~\eqref{eq_trace_matrix_reservoir}}
        \ElsIf{$S:=\{i\in I:1-|y_i|\le\delta_{\rm end}\}\ne\varnothing$}
            \State Append $(i,y_i)$ for $i\in S$ to $\mathcal D$; preserve $M$ and attenuate these slots together.
            \State Remove $S$ from $I$ and its reservoir contributions, correcting the optimizer.
        \Else
            \State Evaluate both endpoint scores for every $i\in I$ to error $\delta$.
            \If{an approximate score is at least $-\delta$}
                \State Record the corresponding endpoint sign $s$ at coordinate $i$.
                \State Set $F\gets F+sC_i$; remove $i$ and its coupling and reservoir terms.
                \State Continue the optimizer to the new face by the charged positive homotopy.
            \Else
                \State Compute the certified high gradient test using $\kappa_i$.
                \If{the high test succeeds}
                    \State Execute a frozen-root sweep and correct its optimizer.
                        \Comment{Section~\ref{sec_trace_algorithm}}
                \Else
                    \State Construct $r$ by the deterministic matrix-response selection.
                        \Comment{Eq.~\eqref{eq_trace_matrix_direction}}
                    \State Orient $r$ by the directional gradient and choose $t$ by Eq.~\eqref{eq_trace_matrix_steps}.
                    \State Set $y\gets y+tr$, round inward, initialize its optimizer, and certify the decrease.
                \EndIf
            \EndIf
        \EndIf
    \EndWhile
    \State Round $\mathcal D$ jointly using Lemma~\ref{sr_lem_round} and record its signs in $\mathcal R$.
    \State Recover and return the original-input signing using $\mathcal R$.
\EndProcedure
\end{algorithmic}
\end{algorithm}

\subsection{Normalization, budget, and potential}\label{sec:constant:potential}
Discard zero inputs. Define $L:=\sum_i|\tr[A_i]|$ and
$\bar A_i:=\operatorname{sign}(\tr[A_i])A_i/L$. Then $\bar A_i\succeq0$ has rank one,
$\sum_i\tr[\bar A_i]=1$, and, writing $\zeta:=\lVert \sum_i\bar A_i^2\rVert$,
\[
 (mn)^{-1}\le\zeta\le1,\qquad \lVert \bar A_i\rVert\le\sqrt\zeta.
\]
A signing for the $\bar A_i$ transfers to one for the $A_i$. For clarity we first
analyze nonzero PSD inputs with a variance upper bound $\bar\zeta$.
The identities also hold on a face with a fixed signed matrix contribution.
We use the variance upper bound $\bar\zeta:=\zeta$ for the existence proof.

Until Section~\ref{sec_trace_correction}, use the following budget,
denoted also by $\beta_{\rm fast}$ when both algorithms are discussed.
Let ${r_{\rm aux}}(z):=1-z^2$ and $f:=f_{\rm fast}$, where
$f_{\rm fast}$ is the explicit polynomial in Eq.~\eqref{eq:budget:fast:polynomial}.
Define
\[
 \beta(y):=(1-y^2)^p f(|y|),\qquad -1\le y\le1.
\]
Define
\[
 p:=\frac{97}{200},\qquad \iota:=\frac{14779}{2500}=5.9116,\qquad
 \tau:=\frac{4902302619}{5000000000},\qquad
 \vartheta_* :=\frac{101}{200}.
\]
The following properties are certified in
Section~\ref{rank_one_sec_certificate}:
\begin{equation}\label{rank_one_eq_budget_basic}
 \beta(0)=1,\quad \beta(\pm1)=0,\quad
 \sqrt{1-y^2}\le\beta(y)\le1,
\end{equation}
\begin{equation}\label{rank_one_eq_budget_signs}
 f'(z)\ge0,\qquad \beta'(z)\le0,\qquad -\beta''(z)>1/2,
 \qquad 0\le z<1,
\end{equation}
\begin{equation}\label{rank_one_eq_budget_drift}
 \frac{(-\beta'(z))(1+z)}{\beta(z)(-\beta''(z))}\le1.
\end{equation}
The even extension $\beta$ is $C^2$ and has locally Lipschitz second
derivative on $(-1,1)$. Its third derivative may jump at zero; no third
derivative at zero is needed.

For a fractional signing $y$, define $M(y):=\sum_i y_i\bar A_i$ and
$\mathcal K_y(D):=\iota\sum_i\beta(y_i)\bar A_iD\bar A_i$. For $\rho>0$ define
\begin{align}
 \mathcal P(y):=\min_{b,\ U,V\succ0}\ &b+\rho\tr[U+V]\label{rank_one_eq_potential}\\
 \text{subject to }&U^{-1}+M(y)+\mathcal K_y(V)\preceq bI,\notag\\
 &V^{-1}-M(y)+\mathcal K_y(U)\preceq bI.\notag
\end{align}
Our total potential is $\mathcal F:=\mathcal P+\gamma\mathcal B$, where
$\mathcal B(y):=\sum_i\beta(y_i)$ and $\gamma>0$.
For every feasible point the constraints imply $b>\|M(y)\|$.
At $y=0$, the scalar choice $U:=\ell I$, $V:=\ell I$, and $b:=\ell^{-1}+\iota\bar\zeta \ell$
is feasible. Minimizing its objective over $\ell>0$ gives
\begin{equation}\label{rank_one_eq_initial}
 \lVert M(y)\rVert\le \mathcal P(y),\qquad
 \mathcal F(0)\le2\sqrt{\iota\bar\zeta+2n\rho}+\gamma m.
\end{equation}

The trace penalty makes the optimizer unique. The PSD dual multipliers
$W,Z$ are positive definite and satisfy
\begin{gather}
 \tr[W+Z]=1,\qquad
 W=U\mathcal K_y(Z)U+\rho U^2,\qquad
 Z=V\mathcal K_y(W)V+\rho V^2,\label{rank_one_eq_kkt}\\
 U^{-1}+M+\mathcal K_y(V)=bI,\qquad
 V^{-1}-M+\mathcal K_y(U)=bI.\label{rank_one_eq_tight}
\end{gather}
The next lemma justifies differentiating the optimized potential twice
within each open face of the cube.

\begin{lemma}\label{rank_one_lem_potential}
On each open face of the cube, the program in
Eq.~\eqref{rank_one_eq_potential} has a unique optimizer. Its primal and
dual variables, and $\mathcal P$, are $C^2$ with locally Lipschitz second derivatives.
The optimizer satisfies Eqs.~\eqref{rank_one_eq_kkt} and
\eqref{rank_one_eq_tight}. The value $\mathcal P$ is continuous on the closed cube.
\end{lemma}
\begin{proof}
The map $\mathcal K_y$ is positive and self-adjoint. Matrix inversion is operator
convex, and $U:=I$ and $V:=I$ with large $b$ is strictly feasible. On any bounded
objective sublevel, the constraints bound $U^{-1},V^{-1}$ above, and the
trace penalty bounds $U,V$ above. The sublevel is therefore compact inside
the positive definite cone, so a minimizer exists. Convex duality gives
$W,Z\succeq0$. Differentiating the Lagrangian in $b,U,V$ gives
Eq.~\eqref{rank_one_eq_kkt}. In particular $W\succeq\rho U^2\succ0$
and $Z\succeq\rho V^2\succ0$, so complementary slackness gives
Eq.~\eqref{rank_one_eq_tight}.

For a nonzero Hermitian ${\mathsf A}$,
\[
 \frac{\d^2}{\d t^2}\tr[W(U+t{\mathsf A})^{-1}]|_{t=0}
 =2\tr[WU^{-1}{\mathsf A}U^{-1}{\mathsf A}U^{-1}]>0.
\]
Thus the Lagrangian is strictly convex in $(U,V)$; it has only one
minimizer, and the tight equations determine $b$.

For completeness, define the self-adjoint linearization
\[
 {\mathscr J}({\mathsf A},{\mathsf B}):=(U^{-1}{\mathsf A}U^{-1}-\mathcal K_y({\mathsf B}),
         V^{-1}{\mathsf B}V^{-1}-\mathcal K_y({\mathsf A})).
\]
The positive map $\mathcal N({\mathsf A},{\mathsf B}):=(U\mathcal K_y({\mathsf B})U,V\mathcal K_y({\mathsf A})V)$ satisfies
$\mathcal N(W,Z)=(W,Z)-\rho(U^2,V^2)\preceq r_0(W,Z)$ for some
$r_0<1$. Every Hermitian pair is bounded in absolute semidefinite order
by a multiple of $(W,Z)$, so the Neumann series for $I-\mathcal N$
converges. Hence ${\mathscr J}$ is invertible and ${\mathscr J}^{-1}$ preserves positivity.
This also proves uniqueness of the dual pair.

Let ${\mathscr E}$ be the positive definite Hessian of the Lagrangian in $(U,V)$.
A vector $({d_{\rm b}},{\mathsf A},{\mathsf B},W',Z')$ in the kernel of the full KKT Jacobian satisfies
\[
 {\mathscr J}({\mathsf A},{\mathsf B})=-{d_{\rm b}}(I,I),\qquad {\mathscr J}(W',Z')={\mathscr E}({\mathsf A},{\mathsf B}),\qquad\tr[W'+Z']=0.
\]
Pairing the middle equation with $({\mathsf A},{\mathsf B})$ shows
$\langle({\mathsf A},{\mathsf B}),{\mathscr E}({\mathsf A},{\mathsf B})\rangle=0$. Thus ${\mathsf A}={\mathsf B}=0$, then ${d_{\rm b}}=0$ and
$W'=Z'=0$. The implicit function theorem proves the asserted regularity;
the locally Lipschitz second derivatives follow from those of $\beta$.
Finally the sublevel bounds can be chosen uniformly in $y$ for fixed
$\rho>0$. A convergent sequence of optimizers gives lower
semicontinuity at every boundary point. Conversely, keeping $U,V$ fixed
and increasing $b$ by a vanishing amount restores feasibility after a
small change in $y$, proving upper semicontinuity.
\end{proof}

\subsection{Endpoint moves and variation formulas}\label{sec:constant:variations}
Fix a state and let its ${k_{\rm act}}$ active coordinates be indexed by $[{k_{\rm act}}]$. Write
$\bar A_i=a_ia_i^*$ and
\[
 u_i:=a_i^*Ua_i,\quad v_i:=a_i^*Va_i,\quad
 w_i:=a_i^*Wa_i,\quad z_i:=a_i^*Za_i.
\]
If $\iota\beta_i v_i\ge1-y_i$, rounding $y_i$ to $1$ preserves feasibility
with the same $(b,U,V)$ and decreases $\mathcal F$ by at least $\gamma\beta_i$.
Likewise $\iota\beta_i u_i\ge1+y_i$ permits rounding to $-1$.
Hence the remaining case is
\begin{equation}\label{rank_one_eq_light}
 \iota\beta_i u_i<1+y_i,\qquad \iota\beta_i v_i<1-y_i.
\end{equation}
Define
\[
 p_i:=1+\iota\beta'_iv_i,\quad q_i:=1-\iota\beta'_iu_i,\quad
 c_i:=\sqrt{\beta_ip_i/q_i},\quad
 d_i:=\sqrt{\beta_iq_i/p_i},\quad
 \tau_i:=\sqrt{p_iq_i/\beta_i}.
\]
For $y_i\ge0$, $f'\ge0$ gives
\[
 -\beta'_i/\beta_i\le \frac{y_i}{1-y_i^2},\qquad
 p_i\ge1/(1+y_i)\ge1/2,\quad q_i\ge1,\quad
 p_iq_i\le1-(1+y_i)\beta'_i/\beta_i.
\]
For negative $y_i$, exchange $p_i,q_i$ and $u_i,v_i$.
In particular all scales above are positive. Write
\[
 \widetilde u_i:=c_i u_i,\quad \widetilde v_i:=d_i v_i,
 \qquad {\widetilde w}_i:=c_i w_i,\quad {\widetilde z}_i:=d_i z_i.
\]
We will use $c_i d_i=\beta_i$, $\tau_ic_i=p_i$,
$\tau_i d_i=q_i$, and
\begin{equation}\label{rank_one_eq_product}
 \widetilde u_i\widetilde v_i=\beta_i u_i v_i
 <\frac{1-y_i^2}{\iota^2\beta_i}\le\frac1{\iota^2}.
\end{equation}

For a real direction $r$, hold $b$ fixed and differentiate the two tight
constraints. The resulting feasible local path has derivatives satisfying
\begin{align}
 U^{-1}\dot U U^{-1}&=\sum_ip_i r_i\bar A_i+\mathcal K_y(\dot V),\notag\\
 V^{-1}\dot V V^{-1}&=-\sum_iq_i r_i\bar A_i+\mathcal K_y(\dot U).
 \label{rank_one_eq_response}
\end{align}
Define the nonnegative quadratic form
\[
 \mathcal Q(r):=\tr[WU^{-1}\dot U U^{-1}\dot U U^{-1}]
       +\tr[ZV^{-1}\dot V V^{-1}\dot V V^{-1}].
\]
Differentiating the fixed-$b$ feasible path gives
\begin{equation}\label{rank_one_eq_gradient}
 \partial_i\mathcal P=p_iw_i-q_iz_i=\tau_i({\widetilde w}_i-{\widetilde z}_i).
\end{equation}
More explicitly, its second-variation calculation gives
\[
 r^\top\nabla^2\mathcal P r\le2\mathcal Q(r)+2\sqrt{\mathcal Q(r) \mathcal M(r)}
       -\iota\sum_i(-\beta''_i)(v_iw_i+u_iz_i)r_i^2,
\]
where $\mathcal M(r):=\iota\sum_i(\beta_i'^2/\beta_i)(v_iw_i+u_iz_i)r_i^2$.
Here is a direct derivation of the second-variation bound. Let
$\varphi_r(t):=b+\rho\tr[U(t)+V(t)]$ along the fixed-$b$ feasible path.
Then $\varphi_r(t)\ge \mathcal P(y+tr)$ with equality at zero. Also
${\mathscr J}(W,Z)=\rho(I,I)$, so differentiating the constraints once gives
$\varphi_r'(0)=\sum_i(p_iw_i-q_iz_i)r_i$. Differentiating twice gives
\[
 \varphi_r''(0)=2\mathcal Q(r)+2\iota\sum_i\beta'_ir_i
     (w_ia_i^*\dot Va_i+z_ia_i^*\dot Ua_i)
       +\iota\sum_i\beta''_i(v_iw_i+u_iz_i)r_i^2.
\]
By the dual equations,
\[
 \mathcal Q(r)\ge \iota\sum_i\beta_i
  (w_ia_i^*\dot VV^{-1}\dot Va_i+z_ia_i^*\dot UU^{-1}\dot Ua_i).
\]
The inequalities $|a_i^*\dot Ua_i|^2\le u_ia_i^*\dot UU^{-1}\dot Ua_i$
and its $V$ analogue, followed by scalar Cauchy--Schwarz, bound the
absolute mixed term by $2\sqrt{\mathcal Q(r)\mathcal M(r)}$. This proves the stated bound
because $r^\top\nabla^2\mathcal Pr\le \varphi_r''(0)$.

Apply $2\sqrt{EB}\le\tau E+B/\tau$. For any centered random $r$,
with $\pi_i:=\tau_i^2\E[r_i^2]$, this gives
\begin{equation}\label{rank_one_eq_curvature}
 \frac12\E[r^\top\nabla^2\mathcal Pr]
 \le(1+\tau/2)\E[\mathcal Q(r)]
 -\frac \iota2\sum_i(-\beta_i''-\frac{\beta_i'^2}{\tau\beta_i})
       (v_iw_i+u_iz_i)\E[r_i^2].
\end{equation}

\subsection{A weighted projection inequality}\label{sec:constant:projection}
The covariance must retain a different coupling coefficient at each
coordinate. The next lemma permits these coefficients without requiring
that the response projection commute with their weights.

\begin{lemma}\label{rank_one_lem_weighted}
Let $e_1,\ldots,e_{k_{\rm act}},f_1,\ldots,f_{k_{\rm act}}$ be an orthonormal basis of
$\R^{2{k_{\rm act}}}$. Let $\Pi_0$ project onto the span of the $e_i$, and let
$\Pi$ be a rank-${k_{\rm act}}$ orthogonal projection. Suppose
\[
 (I-\Pi_0)(I-T)\Pi=0,\qquad
 \langle e_i,Te_i\rangle+\langle f_i,Tf_i\rangle=0,\qquad
 \langle e_i,Te_i\rangle\le0,
\]
\[
 \langle e_i,{D_{\rm aux}}e_i\rangle=\langle f_i,{D_{\rm aux}}f_i\rangle=1,
 \qquad {D_{\rm aux}}\succeq T^\top\Gamma T,
\]
where $\Gamma e_i=\Gamma_i e_i$, $\Gamma f_i=\Gamma_i f_i$, and
$\Gamma_i>2$. Define $0<\vartheta_i<1$ and $k_i$ by
\[
 \vartheta_i:=\frac{2}{\sqrt{1+4\Gamma_i}-1},\qquad
 k_i:=\frac{1+\vartheta_i}{1-\vartheta_i^3}.
\]
There is a PSD matrix ${\mathsf A}$, positive definite on $\operatorname{range}\Pi$
and zero on its orthogonal complement, such that
\begin{equation}\label{rank_one_eq_weighted_comparison}
 \tr[{D_{\rm aux}}{\mathsf A}]\le\sum_ik_i
 \langle e_i,(I-T){\mathsf A}(I-T)^\top e_i\rangle.
\end{equation}
Moreover $\tr[(I-T){\mathsf A}(I-T)^\top]>0$.
\end{lemma}
\begin{proof}
For $0<\vartheta<1$ define
\[
 A:=1+\vartheta^2+\vartheta^3,\quad
 w_e:=\frac1{(1+\vartheta)A},\quad
 w_f:=\frac{\vartheta}{(1+\vartheta)(2+\vartheta)},\quad
 {h_{\rm aux}}:=\frac2A,\quad j:=\frac2{2+\vartheta}.
\]
With $\Gamma:=(1+\vartheta)/\vartheta^2$ and
$k:=(1+\vartheta)/(1-\vartheta^3)$, direct rational identities give
\begin{align}
 \frac {h_{\rm aux}}{w_e}-\frac{{h_{\rm aux}}^2}{4w_e^2}(1/k-1/\Gamma)&=1/w_e,
 \label{rank_one_eq_row_identity_e}\\
 \frac j{w_f}+\frac{j^2}{4\Gamma w_f^2}&=1/w_f,
 \label{rank_one_eq_row_identity_f}\\
 \frac{{h_{\rm aux}}^2}{4\Gamma w_e}+\frac{j^2}{4\Gamma w_f}&=1-w_e-w_f.
 \label{rank_one_eq_row_identity_sum}
\end{align}
Also
\[
 {h_{\rm aux}}-j=\frac{2(1-\vartheta)(1+\vartheta)^2}{(2+\vartheta)A}>0.
\]
Use these values at $\vartheta_i$ for each coordinate and define
\[
 {\mathsf D}:=\sum_i(w_i^e e_ie_i^\top+w_i^f f_if_i^\top),\qquad
 {\mathsf A}:={\mathsf E}({\mathsf E}^\top {\mathsf D}^{-1}{\mathsf E})^{-1}{\mathsf E}^\top,\qquad {\mathsf B}:={\mathsf D}-{\mathsf A},
\]
where ${\mathsf E}$ is an orthonormal basis matrix for $\operatorname{range}\Pi$.
Since ${\mathsf D}^{-1/2}{\mathsf A}{\mathsf D}^{-1/2}$ is an orthogonal projection of rank ${k_{\rm act}}$,
\[
 {\mathsf A},{\mathsf B}\succeq0,\quad {\mathsf A}{\mathsf D}^{-1}{\mathsf A}={\mathsf A},\quad {\mathsf A}{\mathsf D}^{-1}{\mathsf B}=0,
 \quad\tr[{\mathsf D}^{-1}{\mathsf A}]={k_{\rm act}}.
\]
The energy outside ${\mathsf A}$ satisfies
\[
 \tr[{D_{\rm aux}}{\mathsf A}]\le\tr[{D_{\rm aux}}{\mathsf D}]-\tr[\Gamma T{\mathsf B}T^\top],\qquad
 \tr[{D_{\rm aux}}{\mathsf D}]=\sum_i(w_i^e+w_i^f).
\]
We compare this gap and the movement by completing squares in each row.

For the $e_i$ row define $t_i:=e_i^\top T$ and $C_i:=\Gamma_i {\mathsf B}+k_i{\mathsf A}$.
This matrix is positive definite, with
\[
 C_i^{-1}=\Gamma_i^{-1}{\mathsf D}^{-1}
 +(k_i^{-1}-\Gamma_i^{-1}){\mathsf D}^{-1}{\mathsf A}{\mathsf D}^{-1}.
\]
Completing a square gives
\begin{align*}
 & ~ \Gamma_i t_i{\mathsf B}t_i^\top+k_i(e_i^\top-t_i){\mathsf A}(e_i-t_i^\top)+{h_{\rm aux}}_it_ie_i\\
 \ge & ~ k_i\langle e_i,{\mathsf A}e_i\rangle
 -(k_i{\mathsf A}e_i-{h_{\rm aux}}_ie_i/2)^\top C_i^{-1}(k_i{\mathsf A}e_i-{h_{\rm aux}}_ie_i/2)\\
 = & ~ \frac{\langle e_i,{\mathsf A}e_i\rangle}{w_i^e}
          -\frac{{h_{\rm aux}}_i^2}{4\Gamma_iw_i^e}.
\end{align*}
The last equality uses Eq.~\eqref{rank_one_eq_row_identity_e}.
For the $f_i$ row define $s_i:=f_i^\top T$.
The response condition gives $s_i{\mathsf A}=f_i^\top {\mathsf A}$, so
\[
 s_if_i=\frac{\langle f_i,{\mathsf A}f_i\rangle+s_i{\mathsf B}f_i}{w_i^f}.
\]
Completing a square with the PSD matrix ${\mathsf B}$, which need not be invertible,
and using Eq.~\eqref{rank_one_eq_row_identity_f}, we obtain
\begin{equation*}  \Gamma_i s_i{\mathsf B}s_i^\top+j_is_if_i  \ge\frac{j_i}{w_i^f}\langle f_i,{\mathsf A}f_i\rangle        -\frac{j_i^2}{4\Gamma_i(w_i^f)^2}\langle f_i,{\mathsf B}f_i\rangle  =\frac{\langle f_i,{\mathsf A}f_i\rangle}{w_i^f}        -\frac{j_i^2}{4\Gamma_iw_i^f}. \end{equation*}
Let
\[
 \mathfrak q:=\tr[\Gamma T{\mathsf B}T^\top]
       +\sum_ik_i\langle e_i,(I-T){\mathsf A}(I-T)^\top e_i\rangle.
\]
Summing the row inequalities and Eq.~\eqref{rank_one_eq_row_identity_sum}
yields
\[
 \mathfrak q+\sum_i({h_{\rm aux}}_i\langle e_i,Te_i\rangle+j_i\langle f_i,Tf_i\rangle)
 \ge\tr[{\mathsf D}^{-1}{\mathsf A}]-\sum_i(1-w_i^e-w_i^f)
 =\tr[{D_{\rm aux}}{\mathsf D}].
\]
The added sum equals
$\sum_i({h_{\rm aux}}_i-j_i)\langle e_i,Te_i\rangle\le0$.
Thus $\mathfrak q\ge\tr[{D_{\rm aux}}{\mathsf D}]$, proving Eq.~\eqref{rank_one_eq_weighted_comparison}.

Finally define $\Gamma_{\min}:=\min_i\Gamma_i>2$ and
$w_{\min}:=\lambda_{\min}({\mathsf D})>0$. Taking traces in the hypothesis gives
$\|T\|_F^2\le2{k_{\rm act}}/\Gamma_{\min}<{k_{\rm act}}$, and ${\mathsf A}\succeq w_{\min}\Pi$.
Consequently
\begin{equation}\label{rank_one_eq_projection_movement}
 \tr[(I-T){\mathsf A}(I-T)^\top]
 \ge w_{\min}{k_{\rm act}}(1-\sqrt{2/\Gamma_{\min}})^2>0.
\end{equation}
No dimension-independent lower bound on $w_{\min}$ is needed.
\end{proof}

\subsection{A covariance with coordinate-dependent coefficients}\label{sec:constant:covariance}
We apply the weighted projection inequality to the response system
to obtain a covariance for coordinate moves.

\begin{lemma}\label{rank_one_lem_covariance}
Under Eq.~\eqref{rank_one_eq_light}, define
\[
 \Gamma_i:=\frac1{\iota\widetilde u_i\widetilde v_i}
          =\frac1{\iota\beta_i u_i v_i},\qquad
 \vartheta_i:=\frac{2}{\sqrt{1+4\Gamma_i}-1},\qquad
 k_i:=\frac{1+\vartheta_i}{1-\vartheta_i^3}.
\]
There is a centered, finitely supported direction $r$ with nonzero
covariance such that, with $\pi_i:=\tau_i^2\E[r_i^2]$,
\begin{equation}\label{rank_one_eq_covariance}
 \E[\mathcal Q(r)]\le\sum_ik_i(\widetilde u_i {\widetilde w}_i+\widetilde v_i{\widetilde z}_i)\pi_i.
\end{equation}
\end{lemma}
\begin{proof}
Define symmetric PSD, entrywise nonnegative matrices
\[
 \mathscr{U}_{ij}:=\iota c_ic_j|a_i^*Ua_j|^2,\qquad
 \mathscr{V}_{ij}:=\iota d_id_j|a_i^*Va_j|^2.
\]
For $P,Q\in\R^{k_{\rm act}}$ let $M_P:=\sum_ic_iP_i\bar A_i$ and
$M_Q:=\sum_id_iQ_i\bar A_i$. The responses $\dot U:=UM_PU$,
$\dot V:=VM_QV$ satisfy Eq.~\eqref{rank_one_eq_response} precisely when
\begin{equation}\label{rank_one_eq_graph}
 J:=P-\mathscr{V}Q=\mathscr{U}P-Q,\qquad r_i:=J_i/\tau_i.
\end{equation}
The linearized constraint map is invertible, so these are the actual
responses. Define
\[
 E_P:=\operatorname{diag}(\widetilde u_i {\widetilde w}_i),\quad
 E_Q:=\operatorname{diag}(\widetilde v_i {\widetilde z}_i),\quad E:=\operatorname{diag}(E_P,E_Q).
\]
The energy $\tr[WM_PUM_P]+\tr[ZM_QVM_Q]$ is a PSD quadratic form
with matrix $M_{\rm raw}$ and diagonal $E$. Hence
$\widehat {D_{\rm aux}}:=E^{-1/2}M_{\rm raw}E^{-1/2}$ has unit diagonal and zero
cross-block entries. Define
\[
 \widehat T:=\begin{pmatrix}
 0&E_P^{1/2}\mathscr{V}E_Q^{-1/2}\\
 E_Q^{1/2}\mathscr{U}E_P^{-1/2}&0
 \end{pmatrix},\qquad
 \Gamma:=\operatorname{diag}(\Gamma_1,\ldots,\Gamma_{k_{\rm act}},\Gamma_1,\ldots,\Gamma_{k_{\rm act}}).
\]
The full coefficient at each coordinate gives
\begin{equation}\label{rank_one_eq_energy_dominance}
 \widehat {D_{\rm aux}}\succeq\widehat T^\top\Gamma\widehat T,
 \qquad \Gamma_i>\frac{\iota\beta_i}{1-y_i^2}\ge \iota>2.
\end{equation}
Indeed the dual equation and Cauchy--Schwarz imply, for every real $P$,
\begin{equation*}  \tr[WM_PUM_P]  \ge \iota\sum_i\beta_iz_i a_i^*UM_PUM_PUa_i  \ge\sum_i\frac{{\widetilde z}_i}{\iota\widetilde u_i}(\mathscr{U}P)_i^2  =\sum_i\Gamma_i\widetilde v_i{\widetilde z}_i(\mathscr{U}P)_i^2. \end{equation*}
Here $|a_i^*UM_PUa_i|^2\le u_ia_i^*UM_PUM_PUa_i$ and
$(\mathscr{U}P)_i=\iota c_i a_i^*UM_PUa_i$.
The analogous inequality for $Q$, followed by diagonal normalization,
proves the matrix inequality for all $(P,Q)$.
The light bounds give $\iota^2\beta_i^2u_iv_i<1-y_i^2$; also
$\beta_i\ge1-y_i^2$. These prove the remaining assertions.
The argument applies to complex Hermitian input matrices.

Let
\[
 \mathcal V_{\rm resp}:=\{(P,Q):(I-\mathscr{U})P+(I-\mathscr{V})Q=0\}.
\]
This space has dimension ${k_{\rm act}}$. In fact the strict dual equations give
$\mathscr{U}\widetilde z<\widetilde w$ and $\mathscr{V}\widetilde w<\widetilde z$ coordinatewise, so
$\rho(\mathscr{U}\mathscr{V})<1$. A nonzero left kernel vector of
$[I-\mathscr{U}\ I-\mathscr{V}]$ would be a common eigenvector with
eigenvalue one, a contradiction. Let $\Pi$ project onto $E^{1/2}\mathcal V_{\rm resp}$.
Writing $d_{P,i}:=(E_P)_{ii}$ and $d_{Q,i}:=(E_Q)_{ii}$, define
\[
 e_i:=\frac{\sqrt{d_{P,i}}e_{P,i}-\sqrt{d_{Q,i}}e_{Q,i}}
             {\sqrt{d_{P,i}+d_{Q,i}}},\qquad
 f_i:=\frac{\sqrt{d_{Q,i}}e_{P,i}+\sqrt{d_{P,i}}e_{Q,i}}
             {\sqrt{d_{P,i}+d_{Q,i}}}.
\]
Let $\Pi_0$ project onto the span of the $e_i$.
Eq.~\eqref{rank_one_eq_graph} gives $(I-\Pi_0)(I-\widehat T)\Pi=0$.
The local two-coordinate blocks of $\widehat {D_{\rm aux}}$ are identity matrices,
so its quadratic forms at $e_i,f_i$ equal one. Directly,
\[
 \langle e_i,\widehat Te_i\rangle
 =-\frac{d_{P,i}\mathscr{V}_{ii}+d_{Q,i}\mathscr{U}_{ii}}
            {d_{P,i}+d_{Q,i}}\le0,\qquad
 \langle f_i,\widehat Tf_i\rangle=-\langle e_i,\widehat Te_i\rangle.
\]
The matrix $\Gamma$ is scalar on each coordinate pair, so this change of
basis preserves its form. Apply Lemma~\ref{rank_one_lem_weighted}.
For independent uniform signs $\epsilon\in\{\pm1\}^{2{k_{\rm act}}}$ define
$(P,Q):=E^{-1/2}{\mathsf A}^{1/2}\epsilon$ and define $J,r$ by
Eq.~\eqref{rank_one_eq_graph}. This pair belongs to $\mathcal V_{\rm resp}$, and
\[
 \E[\mathcal Q(r)]=\tr[\widehat {D_{\rm aux}}{\mathsf A}],\qquad
 \langle e_i,(I-\widehat T){\mathsf A}(I-\widehat T)^\top e_i\rangle
 =(d_{P,i}+d_{Q,i})\E[J_i^2].
\]
Thus Eq.~\eqref{rank_one_eq_weighted_comparison} proves the desired
comparison, and Eq.~\eqref{rank_one_eq_projection_movement} implies
$\tr[\E[rr^\top]]>0$. The signs here supply only a finite-support
existence certificate.
\end{proof}

\subsection{Joint curvature and strict descent}\label{sec:constant:descent}
The coupling coefficient and endpoint position must be bounded together.
This replaces the use of a single lower bound for the scalar curvature
ratio.

\begin{lemma}\label{rank_one_lem_joint_curvature}
At every light state,
\begin{equation}\label{rank_one_eq_budget_curvature}
 \iota(-\beta_i''-\frac{\beta_i'^2}{\tau\beta_i})
 \ge(1+\kappa_{\rm fast})(2+\tau)k_ip_iq_i,\qquad \kappa_{\rm fast}:=10^{-8}.
\end{equation}
\end{lemma}
\begin{proof}
By symmetry suppose $y_i\ge0$ and define $z:=y_i$. Write
\[
 {q_{\rm log}}:=-\beta'/\beta\ge0,\qquad u:=\iota\beta_i u_i,\qquad v:=\iota\beta_i v_i.
\]
The light inequalities give $0<u<1+z$, $0<v<1-z$, and
\[
 p q=(1-{q_{\rm log}}v)(1+{q_{\rm log}}u),\qquad \Gamma=\frac{\iota\beta}{uv}.
\]
Here $1-{q_{\rm log}}v>0$. The function $k(\Gamma)$ decreases in $\Gamma$, since
$(1+\vartheta)/\vartheta^2$ decreases in $\vartheta$ whereas
$(1+\vartheta)/(1-\vartheta^3)$ increases. For fixed $v$, the product
$k(\Gamma)(1-{q_{\rm log}}v)(1+{q_{\rm log}}u)$ therefore increases with $u$.
Define $u:=1+z$ for an upper bound, define ${v_{\rm loc}}:=v/(1-z)$, and define
$\vartheta$ by
\[
 \vartheta:=\frac{2}{\sqrt{1+4\iota\beta/((1-z^2){v_{\rm loc}})}-1}.
\]
The case ${v_{\rm loc}}=0$ is its continuous limit $\vartheta=0$.
The constraint $0\le {v_{\rm loc}}\le1$ becomes
\begin{equation}\label{rank_one_eq_joint_domain}
 \iota f(z)\vartheta^2\le(1+\vartheta)(1-z^2)^{1-p}.
\end{equation}
All such parameters satisfy $0\le\vartheta<\vartheta_*$, because
$f\ge1$, $(1-z^2)^{1-p}\le1$, and
$\iota\vartheta_*^2>1+\vartheta_*$.
A cancellation retains the negative endpoint term:
\begin{equation}\label{rank_one_eq_joint_endpoint}
 k_ip_iq_i
 \le(1+(1+z){q_{\rm log}})
 \frac{1+\vartheta-\iota(-\beta')\vartheta^2/(1+z)}{1-\vartheta^3}.
\end{equation}
Indeed substitute ${v_{\rm loc}}=\iota\beta\vartheta^2/((1-z^2)(1+\vartheta))$
into $(1-{q_{\rm log}}v)k$.

Define
\[\begin{aligned}
 {r_{\rm aux}}&:=1-z^2,\quad q:={r_{\rm aux}}f'-2pzf,\quad v_0:={r_{\rm aux}}q'+2(1-p)zq,\\
 d_0&:={r_{\rm aux}}f-(1+z)q,\quad w_0:=-\tau fv_0-q^2.
\end{aligned}\]
Then $\beta'={r_{\rm aux}}^{p-1}q$ and $\beta''={r_{\rm aux}}^{p-2}v_0$.
After multiplying Eq.~\eqref{rank_one_eq_joint_endpoint} by the positive
denominators, Eq.~\eqref{rank_one_eq_budget_curvature} follows from
$\mathcal A\ge {r_{\rm aux}}^{1-p}{G_{\rm cert}}$, where
\begin{align}
 \mathcal A&:=\iota[(1+z)w_0(1-\vartheta^3)
                -(1+\kappa_{\rm fast})(2+\tau)\tau d_0q\vartheta^2],\notag\\
 {G_{\rm cert}}&:=(1+\kappa_{\rm fast})(2+\tau)\tau d_0(1+z)(1+\vartheta).
 \label{rank_one_eq_joint_polynomials}
\end{align}
The scalar certificate below proves $w_0,d_0>0$ and $q\le0$, so
$\mathcal A,{G_{\rm cert}}>0$ on
$[0,1]\times[0,\vartheta_*]$. It also proves the full-domain implication
\begin{equation}\label{rank_one_eq_joint_certificate}
 {{H_{\rm cert}}}:=\mathcal A-{r_{\rm aux}}^{1-p}{G_{\rm cert}}\ge0
 \quad\text{whenever}\quad
 \mathcal G:=\iota f\vartheta^2-{r_{\rm aux}}^{1-p}(1+\vartheta)\le0.
\end{equation}
The constraint is exactly Eq.~\eqref{rank_one_eq_joint_domain}.
All multipliers used above are positive for $z<1$, so this proves
Eq.~\eqref{rank_one_eq_budget_curvature}.
\end{proof}

Define $C_{2,i}:=(1+\tau/2)k_i$ and define
\[
 h_i:=-\frac{C_{2,i}}{\tau_i}(\widetilde u_i-\widetilde v_i)\pi_i,
 \qquad K:=\E[rr^\top].
\]
Define $\mathscr{L}g:=\nabla g\cdot h+
\frac12\tr[\nabla^2gK]$.
The cancellation is coordinatewise:
\[
 (\widetilde u_i {\widetilde w}_i+\widetilde v_i{\widetilde z}_i)
 -(\widetilde u_i{\widetilde z}_i+\widetilde v_i {\widetilde w}_i)
 =(\widetilde u_i-\widetilde v_i)({\widetilde w}_i-{\widetilde z}_i).
\]
Also
\[
 (\widetilde u_i{\widetilde z}_i+\widetilde v_i {\widetilde w}_i)\pi_i
 =p_iq_i(u_iz_i+v_iw_i)\E[r_i^2].
\]
Combining Eq.~\eqref{rank_one_eq_gradient},
Eq.~\eqref{rank_one_eq_curvature} and the two preceding lemmas gives
\begin{equation}\label{rank_one_eq_matrix_descent}
 \mathscr{L}\mathcal P\le\sum_i
 [C_{2,i}p_iq_i-
 \frac \iota2(-\beta_i''-\frac{\beta_i'^2}{\tau\beta_i})]
 (u_iz_i+v_iw_i)\E[r_i^2]\le0.
\end{equation}
No derivative of $k_i$ or $h_i$ is taken: the drift and covariance are
pointwise witnesses for this generator inequality.

To bound the budget drift suppose $y_i\ge0$. Since
\[
 \tau_i(\widetilde u_i-\widetilde v_i)
 =u_i-v_i+2\iota\beta'_iu_iv_i\le u_i,
\]
Eqs.~\eqref{rank_one_eq_light} and~\eqref{rank_one_eq_budget_drift} give
\[
 \beta'_ih_i\le\frac{C_{2,i}}\iota
 \frac{(-\beta'_i)(1+y_i)}{\beta_i}\E[r_i^2]
 \le\frac{C_{2,i}}\iota(-\beta''_i)\E[r_i^2].
\]
Negative coordinates follow by symmetry. The exact uniform bound is
\begin{equation}\label{rank_one_eq_parameter_condition}
 \frac{C_{2,i}}\iota\le
 \frac{(1+\tau/2)(1+\vartheta_*)}{\iota(1-\vartheta_*^3)}
 =\frac{4485593088319}{10300518152100}<\frac9{20}.
\end{equation}
Together with $-\beta''>1/2$, this proves
\begin{equation}\label{rank_one_eq_budget_descent}
 \mathscr{L}\mathcal B\le-\frac1{20}\sum_i(-\beta''_i)\E[r_i^2]
 \le-\frac1{40}\tr[K]<0.
\end{equation}
The covariance is nonzero by Eq.~\eqref{rank_one_eq_projection_movement}.
Thus the total potential has strict local descent whenever an endpoint
move is unavailable.

The same cancellation retains the strict surplus:
\begin{equation}\label{eq_fast_strict_surplus}
 \mathscr{L}\mathcal P\le-\kappa_{\rm fast}\sum_iC_{2,i}p_iq_i
                    (v_iw_i+u_iz_i)K_{ii}.
\end{equation}
This additional margin will be used in the faster algorithm. The earlier
budget estimate also gives
$\mathscr{L}\mathcal B\le-\frac1{20}\sum_i(-\beta_i'')K_{ii}$.
Minimizing $\mathcal F$ on the cube now proves the $4.8628$ discrepancy bound:
an endpoint move strictly decreases $\mathcal F$, and a light nonvertex minimum
contradicts its negative generator. Letting $\rho,\gamma$ tend to zero
in Eq.~\eqref{rank_one_eq_initial} gives $2\sqrt \iota<4.8628$.

\subsection{Exact scalar certificate}\label{rank_one_sec_certificate}
The scalar inequalities follow from finite rational coefficient tests.
With $r_{\rm aux},q,v_0,d_0,w_0$ defined above, these tests give
\begin{gather*}
 f-1\ge0,\quad f'\ge0,\quad -q\ge0,\quad -v_0-1/2>0,
 \quad d_0>0,\quad w_0>0,\\
 (-fv_0)^2-(1+z)^2q^2{r_{\rm aux}}>0
\end{gather*}
on $[0,1]$. Each inequality has a rational Bernstein certificate on
$[0,1/2]$ and $[1/2,1]$; the maximum degree is $84$.
The derivative identities give Eq.~\eqref{rank_one_eq_budget_signs}.
Since $p\le1/2$, we have ${r_{\rm aux}}^{1-p}\le\sqrt {r_{\rm aux}}$. Taking nonnegative
square roots in the last inequality therefore gives
\[
 \frac{(-\beta')(1+z)}{\beta(-\beta'')}
 ={r_{\rm aux}}^{1-p}\frac{-(1+z)q}{-fv_0}\le1.
\]
Also $f\ge1$, monotonicity of $\beta$, and $\beta(0)=1$ imply
$\sqrt {r_{\rm aux}}\le {r_{\rm aux}}^p\le\beta\le1$. The missing linear term in $f$ gives
a $C^2$ even extension with locally Lipschitz second derivative.

We describe the certificate for Eq.~\eqref{rank_one_eq_joint_certificate}
without replacing a fractional power by a high-degree polynomial.
Define ${p_{\rm cert}}:=1-p=103/200$. On each dyadic rectangle in
$[0,1]\times[0,\vartheta_*]$, let $z_0$ be the midpoint of its
$z$ interval and define $r_0:=1-z_0^2>0$. Define
$u_0:=k/2^{80}$, where $k$ is the least positive integer satisfying
$u_0^{200}\ge r_0^{103}$. This is an exact integer comparison.
Concavity of $t^{p_{\rm cert}}$ on $[0,1]$ gives the polynomial upper bound
\[
 {r_{\rm aux}}^{p_{\rm cert}}\le L(z):=u_0(1-{p_{\rm cert}}+{p_{\rm cert}}\frac{{r_{\rm aux}}(z)}{r_0}).
\]
Indeed the tangent with $r_0^{p_{\rm cert}}$ in place of $u_0$ is an upper bound,
and its multiplier $1-{p_{\rm cert}}+{p_{\rm cert}}{r_{\rm aux}}/r_0$ is positive.
Define two polynomials on this rectangle by
\[
 F_L:=\mathcal A-L{G_{\rm cert}},\qquad
 G_L:=\iota f\vartheta^2-L(1+\vartheta).
\]
Their common Bernstein bidegree is $(43,3)$. Each rectangle passes
one of three tests: all coefficients of $F_L$ are nonnegative; all
coefficients of $G_L$ are strictly positive; or all coefficients of
$F_L+\lambda G_L$ are nonnegative for some rational $\lambda\ge0$.
These tests suffice. Since ${G_{\rm cert}}>0$, the first gives
${{H_{\rm cert}}}\ge F_L\ge0$. The second excludes every feasible point,
because $\mathcal G\ge G_L$. For the third, feasibility gives
$G_L\le\mathcal G\le0$, and hence
${{H_{\rm cert}}}\ge F_L\ge-\lambda G_L\ge0$.

We give the coefficient rules used here and in
Appendix~\ref{one_half_sec_trace_budget}. For a polynomial $P$ on a box
$\prod_{j=1}^d[a_j,b_j]$, substitute $x_j=a_j+(b_j-a_j)u_j$ and expand
$P(x)=\sum_\nu A_\nu u^\nu$. If $n_j$ bounds its degree in $u_j$,
its tensor-product Bernstein coefficients are
\begin{equation}\label{eq:certificate:bernstein}
 B_\mu:=\sum_{\nu\le\mu}A_\nu
       \prod_{j=1}^d\frac{\binom{\mu_j}{\nu_j}}{\binom{n_j}{\nu_j}},
 \qquad 0\le\mu_j\le n_j.
\end{equation}
The corresponding basis functions are nonnegative and sum to one.
Thus $P$ lies between the smallest and largest coefficients on the box.
For midpoint subdivision in one coordinate, apply
\begin{equation}\label{eq:certificate:subdivision}
 c_i^{(0)}:=B_i,\qquad
 c_i^{(r+1)}:=\frac{c_i^{(r)}+c_{i+1}^{(r)}}2
 \quad(0\le i<n-r).
\end{equation}
The left and right coefficient lists are $(c_0^{(r)})_{r=0}^n$ and
$(c_i^{(n-i)})_{i=0}^n$, respectively. Apply this operation to every
coefficient line parallel to the selected coordinate. All operations
are rational.

For the joint inequality, use degree $(43,3)$ and scale $\vartheta$
by $\vartheta_*$. Initialize coefficient arrays for the six polynomials
\[
 \mathcal A,\quad G_{\rm cert},\quad r_{\rm aux}G_{\rm cert},\quad
 \iota f\vartheta^2,\quad 1+\vartheta,\quad r_{\rm aux}(1+\vartheta).
\]
Keep a common positive scale through subdivision. This preserves the
relative coefficients when forming $F_L$ and $G_L$.
If their coefficient pairs are $(a_\mu,b_\mu)$, define
\begin{equation}\label{eq:certificate:multiplier}
 \lambda_-:=\max(\{0\}\cup\{-a_\mu/b_\mu:b_\mu>0\}),\qquad
 \lambda_+:=\min\{-a_\mu/b_\mu:b_\mu<0\},
\end{equation}
where the minimum of the empty set is $+\infty$.
The third test succeeds exactly when $a_\mu\ge0$ for every $b_\mu=0$
and $\lambda_-\le\lambda_+$. In that case define $\lambda:=\lambda_-$.
This specifies every multiplier from the polynomial coefficients.

Start with $[0,1]\times[0,\vartheta_*]$ and apply the three tests in
the displayed order. If none succeeds, split in $z$ when
$d_z<d_\vartheta+3$ and in $\vartheta$ otherwise, where
$(d_z,d_\vartheta)$ are the coordinate depths. Recompute $L$ whenever
the $z$ interval changes. Exact rational evaluation gives $1421$ nodes
and $711$ leaves: $636$ pass the first test, $7$ the second, and $68$
the third. The maximum depths are $14$ in $z$ and $11$ in $\vartheta$.
Each split partitions its parent, and the normalized leaf areas sum
to $\sum 2^{-d_z-d_\vartheta}=1$. This proves
Eq.~\eqref{rank_one_eq_joint_certificate} on the entire feasible domain.
The drift margin follows from Eq.~\eqref{rank_one_eq_parameter_condition};
the final error allowances are treated in
Section~\ref{rank_one_sec_algorithm}.

\subsection{A covariance correction from the local trace}\label{sec_trace_correction}
We now retain a diagonal and both off-diagonal local traces and
control the mixed variation directly. The potential, its regularity,
and endpoint moves are unchanged. In this section and the next,
$\beta,\iota$ denote the polynomially weighted $3/4$-power budget and coupling specified below.
This budget satisfies $\beta(y)\ge1-y^2$ and has positive response
scales, as required by the response and coercivity identities above.

Define $D_i:=\widetilde u_i{\widetilde w}_i+\widetilde v_i{\widetilde z}_i$ and
${{\rho_{\rm rat}}}_i:=(\widetilde u_i{\widetilde z}_i+\widetilde v_i{\widetilde w}_i)/D_i$.
The paired basis gives the exact identities
\begin{equation}\label{eq_trace_identity}
 \langle e_i,\widehat T e_i\rangle=-{{\rho_{\rm rat}}}_i/\Gamma_i,
 \qquad \langle f_i,\widehat T f_i\rangle={{\rho_{\rm rat}}}_i/\Gamma_i.
\end{equation}
Indeed, $\mathscr{U}_{ii}=\iota\widetilde u_i^2$ and
$\mathscr{V}_{ii}=\iota\widetilde v_i^2$. Substitution into the paired
basis proves Eq.~\eqref{eq_trace_identity}. Also
\[
 D_i(1-{{\rho_{\rm rat}}}_i)=(\widetilde u_i-\widetilde v_i)({\widetilde w}_i-{\widetilde z}_i),
 \qquad \partial_i\mathcal P=\tau_i({\widetilde w}_i-{\widetilde z}_i).
\]
Define
\[
 \upsilon_i:=\frac{\widetilde v_i\sqrt{{\widetilde z}_i}-\widetilde u_i\sqrt{{\widetilde w}_i}}
 {\sqrt{\widetilde u_i\widetilde v_i}(\sqrt{{\widetilde w}_i}+\sqrt{{\widetilde z}_i})},
 \qquad L_i:=\iota\sqrt{\widetilde u_i\widetilde v_i}(\widetilde u_i-\widetilde v_i),
 \qquad \delta_i:=\Gamma_i^{-1}.
\]
The additional off-diagonal identity is
\begin{equation}\label{eq_trace_off_diagonal}
 \langle f_i,\widehat T e_i\rangle
 =L_i+\frac{\delta_i(\widetilde u_i-\widetilde v_i)\upsilon_i}{D_i}
          ({\widetilde w}_i-{\widetilde z}_i).
\end{equation}
To verify it, direct substitution gives
\[
 \langle f_i,\widehat T e_i\rangle
 =L_i\frac{(\widetilde u_i+\widetilde v_i)\sqrt{{\widetilde w}_i{\widetilde z}_i}}{D_i},
\]
and the factorization
\[
 (\widetilde u_i+\widetilde v_i)\sqrt{{\widetilde w}_i{\widetilde z}_i}-D_i
 =({\widetilde w}_i-{\widetilde z}_i)\upsilon_i\sqrt{\widetilde u_i\widetilde v_i}
\]
proves Eq.~\eqref{eq_trace_off_diagonal}. Both trace deviations are
explicit gradient terms, including when that gradient coordinate is zero.
The other off-diagonal entry also factors through the gradient:
\begin{equation}\label{eq_trace_upper_off_diagonal}
 \langle e_i,\widehat T f_i\rangle=\Omega_i\partial_i\mathcal P,\qquad
 \Omega_i:=\frac{\iota(\widetilde u_i\widetilde v_i)^{3/2}
       ({\widetilde w}_i+{\widetilde z}_i)}
 {D_i\tau_i\sqrt{{\widetilde w}_i{\widetilde z}_i}}.
\end{equation}
Indeed, substitution into the paired basis gives
$\iota(\widetilde u_i\widetilde v_i)^{3/2}
({\widetilde w}_i^2-{\widetilde z}_i^2)/
(D_i\sqrt{{\widetilde w}_i{\widetilde z}_i})$.
Factoring the difference of squares proves the identity without
any division by a gradient coordinate.

Let $\eta:=1+\kappa_{\rm tr}$, with $\kappa_{\rm tr}$ specified in
Section~\ref{sec_trace_budget}, and define the mixed coefficient and target
\[
 g_i:=\beta_i'/\beta_i,\qquad
 \Lambda_i:=\frac{g_i}{\tau_i\sqrt{\widetilde u_i\widetilde v_i}},\qquad
 k_i:=\frac{\iota(-\beta_i'')}{2\eta p_iq_i}.
\]
We give the local scalar construction before stating the covariance
comparison. Suppress the index, define $\delta:=1/\Gamma$ and ${s_{\rm inv}}:=1/k$,
and define $H_i:=\iota^2\beta_i^2u_iv_i/(1-y_i^2)\in(0,1)$.
Define ${j_{\rm 0}}:={j_{\rm 0}}(|y_i|,H_i)$, $b_0:=b_0(|y_i|,H_i)$, and
$b_1:=b_1(|y_i|,H_i)$ from Section~\ref{sec_trace_budget}. Define
\begin{align*}
 j&:=\frac{\Lambda^2\delta {s_{\rm inv}}}{4},& \ell&:=\Lambda L{s_{\rm inv}},\\
 F&:={j_{\rm 0}}^2+2{j_{\rm 0}}-j-4j{s_{\rm inv}}b_0-4j{s_{\rm inv}}({s_{\rm inv}}-\delta)b_0^2,\\
 B&:=2{j_{\rm 0}}-j-\delta(1+2{j_{\rm 0}})-4j{s_{\rm inv}}b_0-4j{s_{\rm inv}}^2b_0^2,\\
 m_0&:=\frac{1/2+b_0({s_{\rm inv}}-\delta)}{1+{j_{\rm 0}}},& n_0&:=-\frac{b_0}{1+{j_{\rm 0}}},\\
 \mathfrak f&:=F+4j{s_{\rm inv}}n_0^2(F-B)+2F\ell n_0,\\
 \mathfrak u&:=B+\delta F+4j{s_{\rm inv}}m_0n_0(B-F)-F\ell m_0,\\
 \mathcal P_*&:=m_0\mathfrak f+(n_0+b_1)\mathfrak u,& O_*&:=(1+{j_{\rm 0}})b_1,\\
 Z_*&:=2\mathfrak f\mathfrak u-({s_{\rm inv}}-\delta)\mathfrak f^2+4j{s_{\rm inv}}\mathcal P_*^2,\\
 N_*&:=\mathfrak u^2+\delta\mathfrak f^2+4j{s_{\rm inv}}\mathcal P_*^2
       -2\delta\mathfrak f\mathfrak u+2\ell \mathcal P_*\mathfrak u,\\
 M_*&:=\delta(\mathfrak f-\mathfrak u)b_0+\mathcal P_*({j_{\rm 0}}+\delta),\\
 \mathfrak z&:=FZ_*-4j{s_{\rm inv}}O_*^2\mathfrak u^2,\\
 \mathfrak p&:=BZ_*-FN_*-4j{s_{\rm inv}}O_*^2\mathfrak u^2
       +8j{s_{\rm inv}}O_*M_*\mathfrak u+2\ell {j_{\rm 0}}O_*\mathfrak u^2.
\end{align*}
The scalar certificate below proves
\begin{equation}\label{eq_trace_scalar_conditions}
 F>0,\quad B>0,\quad\mathfrak f>0,\quad\mathfrak u>0,
 \quad\mathfrak z>0,\quad\mathfrak p>0.
\end{equation}
Define the resulting weights
\begin{align*}
 a_*&:=\mathfrak f/\mathfrak u,&\gamma_*&:=\mathcal P_*/\mathfrak u,&d_*&:=\Gamma {j_{\rm 0}},\\
 b_*&:=\Lambda {s_{\rm inv}} b_0,&r_*&:=\Lambda {s_{\rm inv}}\gamma_*,&o_*&:=\Lambda {s_{\rm inv}}O_*,\\
 z_*&:=Z_*/\mathfrak u^2,&
 \Delta_*&:=Fz_*-\delta o_*^2=\mathfrak z/\mathfrak u^2,\\
 w_e&:=F/\Delta_*,&w_f&:=\delta z_*/\Delta_*,&
 w_{ef}&:=-\delta o_*/\Delta_*.
\end{align*}
The paired weight block is positive definite, since
\begin{equation}\label{eq_trace_block_weights}
 \begin{pmatrix}w_e&w_{ef}\\w_{ef}&w_f\end{pmatrix}^{-1}
 =\begin{pmatrix}z_*&o_*\\o_*&F/\delta\end{pmatrix},\qquad
 F>0,\quad\Delta_*>0.
\end{equation}
Define its row multipliers by
\begin{align*}
 h_*&:=2(a_*w_e+b_*w_{ef}),&q_*&:=2(a_*w_{ef}+b_*w_f),\\
 p_*&:=2(r_*w_e+d_*w_{ef}),&j_*&:=2(r_*w_{ef}+d_*w_f).
\end{align*}
These helper scalars are distinct from the state variables.
The extra trace information allows both row auxiliaries to use both
paired directions. The weight block also retains an off-diagonal entry;
the corresponding trace corrections are canceled by the drift.

\begin{lemma}\label{lem_trace_covariance}
There is a centered normalized response $\xi$ with covariance ${\mathsf A}$,
positive definite on the response subspace and zero on its orthogonal
complement, such that, with
$\mathsf F_i:=f_i^\top\xi$, $\mathsf Y_i:=e_i^\top(I-\widehat T)\xi$,
$K:=\E[rr^\top]$, and $\pi_i:=\tau_i^2K_{ii}$,
\begin{equation}\label{eq_trace_covariance}
 \E[\mathcal Q(r)]+\sum_i\Lambda_i\E[\mathsf Y_i\mathsf F_i]
 \le\sum_i k_iD_i\pi_i+\nu\cdot\nabla \mathcal P,
\end{equation}
where
\[
 \nu_i:=\frac{\delta_i[h_{*,i}-j_{*,i}+p_{*,i}\upsilon_i]
                 (\widetilde u_i-\widetilde v_i)}{D_i\tau_i}
          +q_{*,i}\Omega_i.
\]
For $R_{\mathsf A}:=(I-\widehat T){\mathsf A}(I-\widehat T)^\top$, there are absolute
$a,C>0$ such that
\begin{equation}\label{eq_trace_movement}
 a k_{\rm act}\le\tr[R_{\mathsf A}]\le C k_{\rm act}.
\end{equation}
\end{lemma}
\begin{proof}
Use the weights above at each coordinate and define
\[
 {\mathsf D}:=\sum_i[w_i^e e_ie_i^\top+w_i^f f_if_i^\top
       +w_i^{ef}(e_if_i^\top+f_ie_i^\top)],
\]
\[
 {\mathsf A}:={\mathsf E}({\mathsf E}^\top{\mathsf D}^{-1}{\mathsf E})^{-1}{\mathsf E}^\top,
 \qquad {\mathsf B}:={\mathsf D}-{\mathsf A},
\]
where ${\mathsf E}$ is an orthonormal basis matrix for the response subspace.
As before, ${\mathsf A},{\mathsf B}\succeq0$,
${\mathsf A}{\mathsf D}^{-1}{\mathsf A}={\mathsf A}$,
${\mathsf A}{\mathsf D}^{-1}{\mathsf B}=0$, and
$\tr[{\mathsf D}^{-1}{\mathsf A}]=k_{\rm act}$. No commutation is required.

Fix one pair and suppress its index. Define
$v_*:=a_*e+b_*f$, $w_*:=r_*e+d_*f$, $\widehat e:=e-\Lambda f/(2k)$,
$t_e:=e^\top\widehat T$, and $t_f:=f^\top\widehat T$.
For $C_*:=\Gamma{\mathsf B}+k{\mathsf A}\succ0$, the projection identities give
\[
 C_*^{-1}=\delta{\mathsf D}^{-1}
 +(1/k-\delta){\mathsf D}^{-1}{\mathsf A}{\mathsf D}^{-1}.
\]
Completing a square in $t_e$ yields
\begin{align*}
 &\Gamma t_e{\mathsf B}t_e^\top
   +k(e^\top-t_e){\mathsf A}(e-t_e^\top)
   -\Lambda(e^\top-t_e){\mathsf A}f+t_e(2{\mathsf D}v_*)\\
 &\qquad\ge2\widehat e^\top{\mathsf A}v_*
   -(1/k-\delta)v_*^\top{\mathsf A}v_*
   -\frac{\Lambda^2}{4k}f^\top{\mathsf A}f
   -\delta v_*^\top{\mathsf D}v_*.
\end{align*}
The response condition gives $t_f{\mathsf A}=f^\top{\mathsf A}$.
A square with the PSD matrix ${\mathsf B}$, which need not be invertible, gives
\[
 \Gamma t_f{\mathsf B}t_f^\top+t_f(2{\mathsf D}w_*)
 \ge2f^\top{\mathsf A}w_*+\delta w_*^\top{\mathsf A}w_*
       -\delta w_*^\top{\mathsf D}w_*.
\]
The two diagonal coefficients of ${\mathsf A}$ in these bounds are
$z_*$ and
$-\Lambda {s_{\rm inv}} b_*-({s_{\rm inv}}-\delta)b_*^2-\Lambda^2{s_{\rm inv}}/4+2d_*+\delta d_*^2
=F/\delta$. The off-diagonal coefficient is
\[
 b_*-\frac{\Lambda {s_{\rm inv}} a_*}{2}-({s_{\rm inv}}-\delta)a_*b_*+r_*+\delta r_*d_*
 =\Lambda {s_{\rm inv}}(1+{j_{\rm 0}})b_1=o_*.
\]
Indeed, the definitions give
$b_0-a_*/2-({s_{\rm inv}}-\delta)a_*b_0+(1+{j_{\rm 0}})\gamma_*=(1+{j_{\rm 0}})b_1$.
Multiplication by $\Lambda {s_{\rm inv}}$ proves the identity, including when $\Lambda=0$.
By Eq.~\eqref{eq_trace_block_weights}, the summed coefficient is
exactly ${\mathsf D}^{-1}$.

For the scalar constant term, define
\[
 \mathcal E:=w_e+w_f+\delta(v_*^\top{\mathsf D}v_*+w_*^\top{\mathsf D}w_*)
                 +(-h_*+j_*)\delta+p_*L.
\]
The exact identity
\begin{equation}\label{eq_trace_row_cost}
 1-\mathcal E=\frac{\mathfrak p}{\mathfrak z}>0
\end{equation}
follows by substituting the three block weights. More explicitly,
$\delta\Lambda^2{s_{\rm inv}}^2=4j{s_{\rm inv}}$ gives
\begin{align*}
 \Delta_*(1-\mathcal E)
 &=Bz_*-F[1+\delta a_*^2+4j{s_{\rm inv}}\gamma_*^2-2\delta a_*+2\ell\gamma_*]\\
 &\quad-4j{s_{\rm inv}}O_*^2+8j{s_{\rm inv}}O_*[\delta(a_*-1)b_0+\gamma_*({j_{\rm 0}}+\delta)]
       +2\ell {j_{\rm 0}}O_*.
\end{align*}
Multiplication by $\mathfrak u^2$ gives $\mathfrak p$, while
$\mathfrak u^2\Delta_*=\mathfrak z$, proving Eq.~\eqref{eq_trace_row_cost}.

Sum the two row inequalities over all coordinates. Coercivity gives
\[
 \tr[\widehat {D_{\rm aux}}{\mathsf A}]
 \le\tr[\widehat {D_{\rm aux}}{\mathsf D}]
       -\tr[\Gamma\widehat T{\mathsf B}\widehat T^\top],
 \qquad \tr[\widehat {D_{\rm aux}}{\mathsf D}]=\sum_i(w_i^e+w_i^f).
\]
The second equality uses the identity local blocks of $\widehat {D_{\rm aux}}$,
including their zero off-diagonal entries.
Using the opposite diagonal traces, the remaining correction to the
weighted movement is at most
\[
 \sum_i[(h_{*,i}-j_{*,i})(\langle e_i,\widehat T e_i\rangle+\delta_i)
       +q_{*,i}\langle e_i,\widehat T f_i\rangle
       +p_{*,i}(\langle f_i,\widehat T e_i\rangle-L_i)].
\]
Here the terms $\mathcal E_i-1$ were dropped using
Eq.~\eqref{eq_trace_row_cost}. Eqs.~\eqref{eq_trace_identity},
\eqref{eq_trace_off_diagonal}, and~\eqref{eq_trace_upper_off_diagonal}
identify this correction as
$\nu\cdot\nabla \mathcal P$. Since
$\E[\mathcal Q(r)]=\tr[\widehat {D_{\rm aux}}{\mathsf A}]$ and
$\langle e_i,R_{\mathsf A}e_i\rangle=D_i\pi_i$,
this proves Eq.~\eqref{eq_trace_covariance}.

For movement, the compact scalar certificate in
Lemma~\ref{lem_trace_surplus} bounds $F,\mathfrak f,\mathfrak u,\mathfrak z$
above and away from zero. The fixed polynomial $b_1$ is bounded, and
$\delta o_*^2=4j{s_{\rm inv}}O_*^2\le C$. The block formula therefore gives
$w_e\le C$, $w_f\le C\delta$, and $|w_{ef}|\le C\sqrt\delta$.
Its inverse has $z_*\le C$, $F/\delta\le C\Gamma$, and
$|o_*|\le C\sqrt\Gamma$. Bounding the cross term by its two
diagonal terms gives
\[
 {\mathsf D}\preceq CI,\qquad
 {\mathsf D}^{-1}\preceq C\Pi_0+C\Gamma\Pi_1,
 \qquad\Pi_1:=I-\Pi_0.
\]
With $C_*:={\mathsf E}^\top{\mathsf D}^{-1}{\mathsf E}$,
the response condition $\Pi_1{\mathsf E}=\Pi_1\widehat T{\mathsf E}$
and coercivity imply
\[
 \tr[C_*]\le Ck_{\rm act}
   +C\|\Gamma^{1/2}\Pi_1\widehat T{\mathsf E}\|_F^2
 \le Ck_{\rm act}+C\tr[\widehat T^\top\Gamma\widehat T]
 \le Ck_{\rm act}.
\]
Here the first step follows from the preceding bound on $\mathsf D^{-1}$, $\mathsf E^\top\mathsf E=I_{k_{\rm act}}$, and the response condition $\Pi_1\mathsf E=\Pi_1\widehat T\mathsf E$. The second step follows from $\mathsf E\mathsf E^\top=\Pi\preceq I$ and $\Gamma\Pi_1=\Pi_1\Gamma$, since orthogonal projections do not increase the Frobenius norm. The third step follows from coercivity, $\widehat T^\top\Gamma\widehat T\preceq\widehat {D_{\rm aux}}$, and $\tr[\widehat {D_{\rm aux}}]=2k_{\rm act}$, after increasing the absolute constant $C$. Weighted Frobenius Cauchy--Schwarz gives
\[
 (k_{\rm act}-\tr[\widehat T\Pi])^2
 \le\|(I-\widehat T){\mathsf E}C_*^{-1/2}\|_F^2
       \|{\mathsf E}C_*^{1/2}\|_F^2
 =\tr[R_{\mathsf A}]\tr[C_*].
\]
Since $\|\widehat T\|_F^2\le2k_{\rm act}/\iota$ and $\iota>2$,
$|\tr[\widehat T\Pi]|\le k_{\rm act}\sqrt{2/\iota}<k_{\rm act}$.
This proves the lower bound independently of the smallest weight.
The upper bound follows from ${\mathsf A}\preceq CI$ and
$\|I-\widehat T\|_F^2\le Ck_{\rm act}$.
\end{proof}

\paragraph{The mixed variation and its correction.}
Retain the exact mixed term in the derivation of
Eq.~\eqref{rank_one_eq_curvature}, before applying Cauchy--Schwarz:
\begin{equation}\label{eq_trace_exact_variation}
 \frac12r^\top\nabla^2\mathcal Pr
 \le\mathcal Q(r)+\sum_i m_i(r)
 -\frac \iota2\sum_i(-\beta_i'')(v_iw_i+u_iz_i)r_i^2,
 \qquad m_i(r):=\iota\beta_i'r_i(w_i\dot v_i+z_i\dot u_i).
\end{equation}
For the normalized response above, write
$\mathsf Y_i=\sqrt{D_i}J_i$ and $\mathsf E_i:=e_i^\top\xi$.
The exact identity
\begin{equation}\label{eq_trace_mixed_decomposition}
 m_i(r)=\Lambda_i\mathsf Y_i\mathsf F_i
 +\partial_i\mathcal P\frac{g_i}{\tau_i^2D_i}
       \mathsf Y_i(\mathsf E_i-\mathsf Y_i+\upsilon_i\mathsf F_i)
\end{equation}
holds even when the gradient coordinate is zero. Indeed, the response
equations give
\[
 m_i(r)=\frac{g_iJ_i}{\tau_i}
 [{\widetilde w}_iP_i+{\widetilde z}_iQ_i+({\widetilde z}_i-{\widetilde w}_i)J_i],
\]
and the paired basis expresses the first two terms as
\[
 {\widetilde w}_iP_i+{\widetilde z}_iQ_i
 =\frac{{\widetilde w}_i-{\widetilde z}_i}{\sqrt{D_i}}\mathsf E_i
 +\frac{(\widetilde u_i+\widetilde v_i)\sqrt{{\widetilde w}_i{\widetilde z}_i}}
 {\sqrt{\widetilde u_i\widetilde v_i}\sqrt{D_i}}\mathsf F_i.
\]
The factorization used in Eq.~\eqref{eq_trace_off_diagonal} proves
Eq.~\eqref{eq_trace_mixed_decomposition} without dividing by
${\widetilde w}_i-{\widetilde z}_i$. Define
\[
 \theta_i:=\frac{g_i}{\tau_i^2D_i}
 \E[\mathsf Y_i(\mathsf E_i-\mathsf Y_i+\upsilon_i\mathsf F_i)].
\]
Combining the mixed identity with Eq.~\eqref{eq_trace_covariance} gives
\begin{equation}\label{eq_trace_augmented_variation}
 \frac12\E[r^\top\nabla^2\mathcal Pr]
 \le\sum_i k_iD_i\pi_i+(\nu+\theta)\cdot\nabla \mathcal P
 -\sum_i\frac{\iota(-\beta_i'')}{2p_iq_i}{{\rho_{\rm rat}}}_iD_i\pi_i.
\end{equation}
Thus the mixed variation is included directly in the covariance bound.

\subsection{The improved constant and its exact certificate}\label{sec_trace_budget}
We use a $3/4$ endpoint power and adjust the curvature by a fixed
polynomial. Define
\[
 \iota_{\rm tr}:=\frac{699}{250},\qquad p:=\frac34,\qquad
 \kappa_{\rm tr}:=\frac1{10^6},\qquad \eta:=1+\kappa_{\rm tr},\qquad
 \beta_{\rm tr}(y):=(1-y^2)^p f_{\rm tr}(|y|),
\]
where the polynomial factor is
\begin{equation}\label{eq:budget:trace:polynomial}
\begin{aligned}
 f_{\rm tr}(z):={}&1+0.1915505z^{2}+0.07716335z^{4}+0.03917979z^{6}\\
 &+0.04234609z^{8}-0.08891557z^{10}+0.40792039z^{12}\\
 &-0.91331458z^{14}+1.30793628z^{16}-1.02145746z^{18}\\
 &+0.35713938z^{20}.
\end{aligned}
\end{equation}
For comparison, the polynomial factor of $\beta_{\rm fast}$ is
\begin{equation}\label{eq:budget:fast:polynomial}
\begin{aligned}
 f_{\rm fast}(z):={}&1+0.0498735123z^{2}-0.005706553z^{3}+0.3937540726z^{4}\\
 &-11.0832914979z^{5}+203.0443576388z^{6}-2445.8186989094z^{7}\\
 &+20520.6541832702z^{8}-124448.8676898058z^{9}+559322.9751101618z^{10}\\
 &-1893873.5450890157z^{11}+4878891.695382722z^{12}-9600597.735909414z^{13}\\
 &+14402676.373907391z^{14}-16328233.13094071z^{15}+13740497.154438846z^{16}\\
 &-8313646.91225735z^{17}+3417391.239657022z^{18}-854236.6692651988z^{19}\\
 &+97990.3280047086z^{20}.
\end{aligned}
\end{equation}
All decimal coefficients in these two expressions are exact.
The notation $f_{\rm fast}$ distinguishes the polynomial $f$ used in
Section~\ref{sec:constant:potential} from $f_{\rm tr}$.
Figure~\ref{fig:budget:comparison} shows all three budgets, the polynomial factors $f_{\rm tr}$ and $f_{\rm fast}$, and the corresponding power functions.

The covariance helpers are the fixed polynomials
\begin{align*}
 {j_{\rm 0}}(z,H)&:=\frac{84047268-52938705z+133367380z^2+49643028z^3}{10^8}\\
 &\quad+H\frac{22863212+66017181z-147568585z^2}{10^8},\\
 b_0(z,H)&:=\frac{-41550398+11091757z-14435171z^2}{10^8}\\
 &\quad+H\frac{10275247-10715575z+21317210z^2}{10^8},\\
 b_1(z,H)&:=\frac{3323951+40587513z-130152459z^2+191253811z^3}{10^8}\\
 &\quad+H\frac{-2355589+49791446z-117532304z^2}{10^8}.
\end{align*}
Here $H_i:=\iota^2\beta_i^2u_iv_i/(1-y_i^2)$ is the normalized local
product. The light inequalities give $0<H_i<1$.
For the remainder of this subsection, define $\iota:=\iota_{\rm tr}$,
$\beta:=\beta_{\rm tr}$, and $f:=f_{\rm tr}$.
For $z\in[0,1)$, define ${r_{\rm aux}}:=1-z^2$ and define the polynomials
\[
 \mathfrak a(z):={r_{\rm aux}}f'(z)-2pzf(z),\qquad
 \mathfrak d(z):=2p[1+(1-2p)z^2]f(z)+4pz{r_{\rm aux}}f'(z)-{r_{\rm aux}}^2f''(z).
\]
Direct differentiation gives
\[
 \beta'(z)={r_{\rm aux}}^{p-1}\mathfrak a(z),\qquad
 -\beta''(z)={r_{\rm aux}}^{p-2}\mathfrak d(z),\qquad
 -\frac{\beta'(z)}{\beta(z)}=-\frac{\mathfrak a(z)}{{r_{\rm aux}}f(z)}.
\]
The exact polynomial certificate gives
\begin{equation}\label{eq_trace_budget_signs}
 \frac{f'(z)}z>0,\quad -\frac{\mathfrak a(z)}z>0,\quad
 \mathfrak d(z)>\frac12,\qquad \frac14<{j_{\rm 0}}(z,H)<3.
\end{equation}
The bounds hold for $z,H\in[0,1]$, with the polynomial continuations
of the quotients at zero. Since
$f(0)=1$, the first inequality implies $f\ge1$. Thus
\begin{equation}\label{eq_trace_budget_basic}
 \beta(0)=1,\quad\beta(\pm1)=0,\quad
 1-y^2\le\beta(y)\le1,\quad
 \beta(y)\ge(1-y^2)^{3/4}.
\end{equation}
The upper bound follows from $\beta'\le0$ and evenness. Also
\[
 0\le-\frac{\beta'(z)}{\beta(z)}
 =\frac{2pz}{1-z^2}-\frac{f'(z)}{f(z)}\le\frac{3}{4(1-z)}.
\]
At a light state with $y_i\ge0$, this gives $p_i\ge1/4$ and
$q_i\ge1$, with the roles exchanged for negative coordinates.
The potential, endpoint, response, and coercivity arguments therefore apply.
The budget is smooth on $(-1,1)$, with $\beta'(0)=0$. On the truncated
interval its fixed polynomial factor gives
$|\beta^{(j)}|\le C\delta_{\rm end}^{3/4-j}$ and
$|\beta^{(j)}|/\beta\le C\delta_{\rm end}^{-j}$ for $1\le j\le3$.

The helpers ${j_{\rm 0}}_i:={j_{\rm 0}}(|y_i|,H_i)$ and $b_{{r_{\rm aux}},i}:=b_{r_{\rm aux}}(|y_i|,H_i)$, ${r_{\rm aux}}=0,1$, are used
pointwise in the covariance construction. No derivatives of them are
needed. The remaining certificate verifies that construction on the
entire light-state domain.

\begin{lemma}\label{lem_trace_surplus}
At every light state, the choice
\begin{equation}\label{eq_trace_surplus}
 k_i:=\frac{\iota(-\beta_i'')}{2\eta p_iq_i}
\end{equation}
satisfies all six conditions in Eq.~\eqref{eq_trace_scalar_conditions}.
The quantities $F,B,\mathfrak f,\mathfrak u,\mathfrak z$ are uniformly
bounded above and away from zero.
\end{lemma}
\begin{proof}
By symmetry take $z=y_i\ge0$ and define
$q_0:=-\beta'(z)/\beta(z)$, ${a_{\rm loc}}:=\iota\beta u_i$, ${b_{\rm loc}}:=\iota\beta v_i$,
and $P:={a_{\rm loc}}{b_{\rm loc}}$.
The light rectangle is $0<{a_{\rm loc}}<1+z$, $0<{b_{\rm loc}}<1-z$.
Writing $\mathcal R:=p_iq_i$, direct substitution gives
\[
 \mathcal R=1+q_0({a_{\rm loc}}-{b_{\rm loc}})-q_0^2P,\qquad
 \delta=\frac P{\iota\beta},\qquad
 \Lambda^2\delta=\frac{\iota(\beta')^2}{\beta\mathcal R},\qquad
 \Lambda L=\frac{1+q_0^2P-\mathcal R}{\mathcal R}.
\]
At fixed $P\in[0,{r_{\rm aux}}]$, the function ${a_{\rm loc}}-P/{a_{\rm loc}}$ is increasing. Its two
boundary values, at ${b_{\rm loc}}=1-z$ and ${a_{\rm loc}}=1+z$, give the complete parametrization
\[
 P={r_{\rm aux}}H,\qquad {a_{\rm loc}}-{b_{\rm loc}}=z(1+H)+(1-H)(2V-1),\qquad H,V\in[0,1].
\]
The boundary values are affine in $P$ and agree at $P={r_{\rm aux}}$.
Define $Q_y:=z(1+H)+(1-H)(2V-1)$ and $\varrho:={r_{\rm aux}}^{1/4}$.
With the budget polynomials evaluated at $z$ and the helpers at $(z,H)$,
the scalar data become
\begin{align*}
 \delta&=\frac{\varrho H}{\iota f},&
 j&=\frac{\eta\mathfrak a^2}{2f\mathfrak d},\\
 {s_{\rm inv}}&=\frac{2\eta\varrho[{r_{\rm aux}}f^2-\mathfrak a fQ_y-H\mathfrak a^2]}
                 {\iota\mathfrak d f^2},&
 \sigma&:=\frac{2\eta\varrho {r_{\rm aux}}}{\iota\mathfrak d},\qquad
 \ell=\sigma+4j\delta-{s_{\rm inv}}.
\end{align*}
These formulas keep the mixed coefficient and off-diagonal reference
correlated. They remain finite at both endpoints.

For an exact certificate without singular fractional powers, define
\[
 z:=1-t^4,\qquad {r_{\rm aux}}=t^4(2-t^4),\qquad
 \varrho=t(2-t^4)^{1/4},\qquad 0\le t\le1.
\]
The six expressions in Eq.~\eqref{eq_trace_scalar_conditions} are now
continuous functions on the closed cube $(t,H,V)\in[0,1]^3$.
Their only nonpolynomial operation is the fourth root of $2-t^4\in[1,2]$.

The first three polynomial inequalities in Eq.~\eqref{eq_trace_budget_signs}
have strictly positive Bernstein coefficients on $[0,1]$; their degrees
are $18,20,22$, respectively. Since ${j_{\rm 0}}$ is affine in $H$, its bounds
reduce to $H=0,1$. Their cubic Bernstein coefficients are positive
on $[0,1]$.
For the full-state inequalities, use intervals with endpoints in
$2^{-100}\mathbb Z$. Define $S:=2^{100}$ and define outward rounding by
\begin{equation}\label{eq:certificate:rounding}
 \lfloor x\rfloor_S:=\frac{\lfloor Sx\rfloor}{S},\qquad
 \lceil x\rceil_S:=\frac{\lceil Sx\rceil}{S}.
\end{equation}
Enclose each rational constant by these two values. Addition and
subtraction use the endpoint formulas; multiplication takes the minimum
and maximum of the four endpoint products and rounds outwards.
For $0\notin[a,b]$, inversion gives
$[\lfloor1/b\rfloor_S,\lceil1/a\rceil_S]$.
Integer powers use their exact range on the interval, including zero
when an even power crosses zero, followed by outward rounding.
For $0\le a\le b$ with integer endpoints, the fourth root of
$[a/S,b/S]$ is enclosed by
\begin{equation}\label{eq:certificate:fourth-root}
 \frac1S[\lfloor(aS^3)^{1/4}\rfloor,
          \lceil(bS^3)^{1/4}\rceil].
\end{equation}
Both integer roots are determined by comparisons of fourth powers.
Define $\widehat f$ by $f(z)=\widehat f(z^2)$. Evaluate
$\widehat f,\widehat f',\widehat f''$ by Horner's rule and use
\[
 f'(z)=2z\widehat f'(z^2),\qquad
 f''(z)=2\widehat f'(z^2)+4z^2\widehat f''(z^2).
\]
Evaluate the helper polynomials by Horner's rule in $z$, retaining
their affine dependence on $H$. Substitute into the displayed scalar
formulas, preserving their factorizations. In these formulas, evaluate
squares of computed quantities as products of two copies of the same
interval. Differentiate by the product, quotient, and chain rules,
using the same outward interval operations to enclose every first
derivative.

Here is the complete subdivision rule. Let $G_1,\ldots,G_6$ be the
expressions in Eq.~\eqref{eq_trace_scalar_conditions}, in that order.
Start with $[0,1]^3$. On a box $Q$ with midpoint $x_0$ and half-widths
$h_1,h_2,h_3$, first compute interval enclosures $I_i(Q)$ of $G_i(Q)$.
Accept $Q$ if every lower endpoint is positive. Otherwise compute
$I_i(x_0)$ and derivative enclosures $D_{ij}(Q)$, and define
\begin{align}
 c_i&:=\inf I_i(x_0),\qquad
 e_{ij}:=\lceil h_j\sup\{|u|:u\in D_{ij}(Q)\}\rceil_S,
 \label{eq:certificate:derivative-error}\\
 L_i&:=\max\{\inf I_i(Q),c_i-\sum_{j=1}^3e_{ij}\}.
 \label{eq:certificate:box-lower}
\end{align}
The mean-value theorem gives $G_i(x)\ge L_i$ throughout $Q$.
Accept $Q$ when all six $L_i$ are positive. If some remain nonpositive,
bisect the coordinate maximizing
\begin{equation}\label{eq:certificate:split-score}
 \max_{i:L_i\le0}\frac{e_{ij}}{\max\{S^{-1},c_i\}},
\end{equation}
breaking ties in the order $t,H,V$. If an interval denominator contains
zero, bisect in $t$ before continuing.

Exact evaluation of these rules gives $636477$ nodes and $318239$
accepted leaves, with maximum coordinate depths $(15,19,6)$.
Every accepted lower bound exceeds $3\cdot10^{-9}$. Each bisection
partitions its parent, so the leaves cover the cube and have disjoint
interiors; their exact volumes sum to one.
This verifies the inequalities on the full closed cube, including all
light-state limits. Compactness gives the stated uniform bounds and
proves the lemma.
\end{proof}

Define the drift
\[
 h_i:=-\frac{k_i}{\tau_i}(\widetilde u_i-\widetilde v_i)\pi_i-\nu_i-\theta_i.
\]
The last two terms cancel the gradient corrections in
Eq.~\eqref{eq_trace_augmented_variation}. The first changes
$k_iD_i\pi_i$ into $k_i{{\rho_{\rm rat}}}_iD_i\pi_i$. Since
${{\rho_{\rm rat}}}_iD_i\pi_i=p_iq_i(v_iw_i+u_iz_i)K_{ii}$,
Eq.~\eqref{eq_trace_surplus} gives
\begin{equation}\label{eq_trace_strict_descent}
 \mathscr{L}\mathcal P\le-\kappa_{\rm tr}\sum_i k_i{{\rho_{\rm rat}}}_iD_i\pi_i<0.
\end{equation}
The weights are positive and Eq.~\eqref{eq_trace_movement} ensures
nonzero movement. No derivative of the covariance or drift is used.
This proves the local assertion in Theorem~\ref{rank_one_thm_main}.
Its polynomial implementation is given in Section~\ref{sec_trace_algorithm}.

%% file: 30_proof.tex
\section{Deterministic construction and shared algorithmic tools}\label{rank_one_sec_algorithm}

Sections~\ref{sec:fast:input-rounding} and~\ref{sec:fast:variance-scaling}
round and rescale the inputs and reduce the number of active coordinates.
Sections~\ref{sec:fast:maintained-model} and~\ref{sec:fast:weighted-norms}
give the maintained model, deferred rounding, and conditioning estimates.
Section~\ref{sec:fast:descent-dichotomy} establishes the shared descent
and directional estimates. Sections~\ref{sec_fast_reset}--\ref{sec_fast_arithmetic}
provide capped initialization, operator actions, regular endpoint bounds,
and arithmetic costs.
Sections~\ref{sec:shared:fast-rounding}--\ref{sec:shared:endpoints-sweeps}
establish fast rounding, smoothed geometry, corrected optimizer
evaluation, and shared endpoint and gradient-sweep proofs.
Section~\ref{sec_trace_algorithm} combines these tools with the
trace-corrected covariance to prove the deterministic algorithm and
running time in Theorem~\ref{rank_one_thm_main}. The randomized construction
and the square-root budget also use the shared tools in this section.

\subsection{Multiplication parameters}\label{sec_multiplication_parameters}

For the operation bounds below, define
\begin{equation}\label{fast_eq_multiplication_parameters}
 \omega_0:=2.371177,\qquad
 {\omega_{\rm rect}}_0:=2.042777,\qquad
 \chi:=3.250036,\qquad
 p_0:=2(\omega_0-{\omega_{\rm rect}}_0)=0.6568.
\end{equation}
Square multiplication has exponent $\omega_0$ by
Dupont et al.~\cite{dekmrszawb26}. The rectangular bounds
$\omega(1,1/2,1)\le2.042776<{\omega_{\rm rect}}_0$ and
$\omega(1,2,1)\le3.250035<\chi$ follow from
Alman et al.~\cite[Table~1]{adwxxz25}. The strict slack permits
fixed recursive algorithms at the stated exponents.

The shared fast-budget estimates use $\beta_{\rm fast}$,
$\iota=14779/2500$, and its parameter $\tau$ from
Section~\ref{sec_rank_one_proof}. Section~\ref{sec_trace_algorithm}
states the specialization to the trace-corrected budget. Constants
$C,a>0$ are absolute, respectively sufficiently large and small.
The covariance and drift prove that a descent direction exists;
the constructions compute directions from the gradient and Hessian.

\subsection{One-time rounding before partial signing}\label{sec:fast:input-rounding}
We first approximate the inputs by PSD rank-one matrices on a common
dyadic grid
and reserve part of the strict slack in the final discrepancy constant
for transferring the resulting signing back to the original inputs.

Absorb the eigenvalue sign of each nonzero rank-one Hermitian input into
the final output sign, writing $B_i:=\operatorname{sign}(\tr[A_i])A_i\succeq0$.
The sign can be read from any nonzero diagonal entry. If all inputs are
zero, stop. Let $a_*$ be the largest diagonal entry among all $B_i$ and
choose a normalization factor ${D_{\rm aux}}$ with
$a_*\le {D_{\rm aux}}<2a_*$. In the arithmetic construction take
${D_{\rm aux}}:=a_*$, using only comparisons and divisions. Work with
$\bar A_i:=B_i/{D_{\rm aux}}$. Then
\[
 \|\bar A_i\|=\tr[\bar A_i]\le n,\qquad
 \zeta_0:=\|\sum_i \bar A_i^2\|\ge1/4.
\]
The second assertion follows from an input with diagonal entry
$a_*/{D_{\rm aux}}>1/2$ and $\sum_i \bar A_i^2\succeq \bar A_j^2$.
Define $\xi:=10^{-8}$ and $\delta:=\xi/(mn)$.
Construct dyadic PSD rank-one matrices $\widetilde{\bar A}_i$ with
\[
 \|\widetilde{\bar A}_i-\bar A_i\|_F\le\delta.
\]
All their entries can use one common denominator $2^{3p}$, where
$p=O(\log(2mn))$.

For a nonzero normalized input, choose a largest diagonal entry
$a:=(\bar A_i)_{jj}$ of each input. Since $\bar A_i$ is PSD and rank one,
$\bar A_i=a ww^*$ with $w:=\bar A_i e_j/a$ and $|w_l|\le1$.
Choose a common dyadic mesh ${h_{\rm aux}}:=2^{-p}$ with
${h_{\rm aux}}\le\delta^2/(10^6n^4)$. Approximate $a$ and the selected column with
absolute error at most ${h_{\rm aux}}$ in each complex entry.
If the approximation $\widehat a\le\delta/(4n)$, replace the input by
zero: $a\le\delta/(2n)$, so $\|\bar A_i\|_F\le na\le\delta/2$.
Otherwise $a\ge\delta/(8n)$ and $\widehat a>0$.
Divide the approximate column by $\widehat a$, round its real and
imaginary parts to the mesh ${h_{\rm aux}}$, and call the result $\widehat w$.
Conservative rounding gives
\[
 |\widehat a-a|\le {h_{\rm aux}},\qquad
 \|\widehat w-w\|_2\le32n^{3/2}{h_{\rm aux}}/\delta.
\]
Define $\widetilde{\bar A}_i:=\widehat a\widehat w\widehat w^*$.
Store the two factors without expanding their outer product.
It is PSD of rank at most one, and
\[
 \|\widetilde{\bar A}_i-\bar A_i\|_F
 \le200n^2{h_{\rm aux}}/\delta\le\delta.
\]
Products of the three dyadic factors have common denominator $2^{3p}$.
Choose the least $p$ satisfying the mesh inequality, so
$p=O(\log(2mn))$. The scalar roundings use bisection on bounded
intervals, with $O(p)$ additions and comparisons per entry. Selecting
the diagonals and forming the normalized columns uses $O(mn)$
arithmetic operations.

Constructing and storing the compact factors therefore takes
$\widetilde O(mn)$ arithmetic operations. Expanding their outer
products costs $\widetilde O(mn^2)$.

Run the algorithm on the $\widetilde{\bar A}_i$ and charge the
partial-signing residual after variance scaling.
For the input-transfer estimate, set $C_*:=4.86276$ and
$\widetilde\zeta:=\|\sum_i\widetilde{\bar A}_i^2\|$. Then
\[
 |\widetilde\zeta-\zeta_0|
 \le\sum_i\delta(2n+\delta)\le3\xi,
 \qquad
 \sqrt{\widetilde\zeta}\le(1+6\xi)\sqrt{\zeta_0}.
\]
Here $\zeta_0\ge1/4$ and $\sqrt{1+12\xi}\le1+6\xi$.
For any signing satisfying
$\|\sum_i\sigma_i\widetilde{\bar A}_i\|\le C_*\sqrt{\widetilde\zeta}$,
we have
\[
 \|\sum_i \sigma_i \bar A_i\|
 \le C_*\sqrt{\widetilde\zeta}+m\delta
 \le [C_*(1+6\xi)+2\xi]\sqrt{\zeta_0}
 <4.8628\sqrt{\zeta_0}.
\]
Multiplying by ${D_{\rm aux}}$ and undoing the eigenvalue signs proves the claimed
guarantee for the original inputs. Thus the rounding does not change
the stated constant. For the rest of the algorithmic proof the inputs
are the dyadic matrices just constructed.

\subsection{Variance scaling and the maintained instance}\label{sec:fast:variance-scaling}

Reduce general rank-one Hermitian matrices to PSD matrices by absorbing the sign of their nonzero eigenvalue into the output sign. Normalize the dyadic inputs so that $\sum_i\tr[\bar A_i]=1$, and assign arbitrary signs to inputs with trace below ${\varepsilon_{\rm discard}}/(m^2n)$. With ${\varepsilon_{\rm scale}}:=10^{-8}$ and ${\varepsilon_{\rm discard}}:=10^{-6}$, choose rational $q$ satisfying $\zeta\le q^2\le(1+{\varepsilon_{\rm scale}})\zeta$ and define $C_i:=\bar A_i/q$. Then

\[
\sum_iC_i^2\preceq I,\qquad \rho:={\varepsilon_{\rm scale}}/n.
\]

The removed inputs have total norm at most ${\varepsilon_{\rm discard}}/(mn)\le{\varepsilon_{\rm discard}}\sqrt\zeta$.
Partial signing leaves $k\le n^2$ fractional coordinates.
Use Lemma~\ref{rank_one_lem_grouped_partial_signing},
with normalized residual allowance $\eta_{\rm ps}:=10^{-7}$ and
arithmetic cost $\widetilde O(mn^2+n^{5.055})$.
If $k=0$, output the signs directly; their residual is at most
$\eta_{\rm ps}$. The two exact alternatives below give comparison bounds.

To compute the variance scale, first form $M_2=\sum_i\bar A_i^2$.
Rank one gives $\bar A_i^2=\tr[\bar A_i]\bar A_i$, so forming this
matrix costs $O(mn^2)$ operations.
Its norm has an inverse-polynomial lower bound, since
$\tr[M_2]=\sum_i\tr[\bar A_i]^2\ge1/m$.
Choose $\ell_{\rm pow}$ as the least power of two at least
$C\varepsilon_{\rm scale}^{-1}\log(2n)$. The trace power satisfies
\[
 \|M_2\|\le\tr[M_2^{\ell_{\rm pow}}]^{1/\ell_{\rm pow}}
 \le n^{1/\ell_{\rm pow}}\|M_2\|.
\]
Choose $C$ so that $n^{1/\ell_{\rm pow}}\le1+\varepsilon_{\rm scale}/4$.
Compute $\tr[M_2^{\ell_{\rm pow}}]$ by exact repeated squaring.
Since $1/(mn)\le\|M_2\|\le1$, bisection of
$q^{2\ell_{\rm pow}}=\tr[M_2^{\ell_{\rm pow}}]$ on $[0,2]$
returns a rational upper endpoint $q$ within
$\varepsilon_{\rm scale}/(16mn)$ of the root using
$O(\log(2mn))$ steps. Each comparison uses exact repeated squaring.
The displayed trace-power bound and this scalar accuracy imply
$\|M_2\|\le q^2\le(1+\varepsilon_{\rm scale})\|M_2\|$. The cost is
$\widetilde O(mn^2+n^3)$ on the short dyadic inputs. A common scalar
normalization does not change linear dependencies among the vectorized inputs.

We reduce the number of fractional coordinates while allowing a controlled
residual. The arithmetic construction uses approximate nullspace directions.

\begin{lemma}[Partial signing with bounded residual]\label{rank_one_lem_regularized_partial_signing}
Let $b_1,\ldots,b_m\in\mathbb R^{d_{\rm vec}}$ have rational entries and
satisfy $\|b_i\|_2\le2$, and let $0<\eta<1$.
Define $\chi:=211/400$. There is a
deterministic algorithm returning $y\in[-1,1]^m$ with at most $d_{\rm vec}$
fractional coordinates and
\[
 \|\sum_i y_i b_i\|_2\le\eta.
\]
Its arithmetic cost is
\[
 \widetilde O(m d_{\rm vec}^{1+\chi}+d_{\rm vec}^{5/2}),
\]
where the suppressed logarithmic factors also include $\log(1/\eta)$.
In particular, for $d_{\rm vec}=n^2$, the arithmetic cost is
$\widetilde O(mn^{3.055}+n^5)$.
\end{lemma}
\begin{proof}
We count scalar additions, subtractions, multiplications, divisions,
and comparisons at unit cost. The ridge matrices and their batched
updates are maintained exactly in this model; the dyadic rounding below
is part of the algorithm.

All unprocessed coefficients start at zero. Retain the fractional
columns in a matrix $A$, insert the next input with coefficient zero,
and, whenever there are ${d_{\rm vec}}+1$ retained columns, perform the following
approximate nullspace move. Choose a positive dyadic $\delta$ with
\[
 \frac12(\frac{\eta}{100m{d_{\rm vec}}})^2
 \le\delta\le(\frac{\eta}{100m{d_{\rm vec}}})^2,
 \qquad G:=\delta I+AA^{\top},
 \qquad {\Pi_{\rm rid}}:=I-A^{\top}G^{-1}A.
\]
The matrix ${\Pi_{\rm rid}}$ is PSD and has trace at least one. Choose a largest
diagonal entry, say ${\Pi_{\rm rid}}_{jj}$, and define $r:={\Pi_{\rm rid}}e_j$. Then
\[
 r_j\ge\frac1{{d_{\rm vec}}+1},\qquad \|Ar\|_2\le\frac{\sqrt\delta}{2}.
\]
The latter inequality follows by applying singular value decomposition
to $A{\Pi_{\rm rid}}=\delta(\delta I+AA^{\top})^{-1}A$: its singular values are
$\delta s/(\delta+s^2)\le\sqrt\delta/2$. Singular value decomposition
is used only in this proof, not by the algorithm.

Define $\tau:=\eta/(10^4m d_{\rm vec}^2)$ and choose a dyadic mesh
$u$ with $\tau/8<u\le\tau/4$. Round the exact diagonal entries to this
grid, choose a largest rounded diagonal, and round the corresponding
exact column to the same grid, obtaining $\widetilde r$.
The estimates below require only that each entrywise error is at most
$\tau$. In particular,
\[
 \|\widetilde r\|_\infty\ge\frac1{d_{\rm vec}+1}-3\tau
 \ge\frac1{2(d_{\rm vec}+1)}.
\]
Move in its positive direction
until the first coefficient reaches a boundary. The step length is at
most $4({d_{\rm vec}}+1)$. The boundary ratios are formed and compared
exactly. Set each hitting coefficient exactly to its boundary value, round the other coefficients
toward zero to the grid of mesh $u$, and remove every coefficient at a
boundary. Each move removes at least one retained column.
Since $\|A\|\le2\sqrt{{d_{\rm vec}}+1}$, the change in the signed sum has norm at
most
\[
 4({d_{\rm vec}}+1)(\frac{\sqrt\delta}{2}+2({d_{\rm vec}}+1)\tau)
       +2({d_{\rm vec}}+1)u.
\]
With $u\le\tau$, the sum of
these bounds over at most $m$ moves is less than $\eta$. Insertions do
not change the signed sum. This proves the output guarantee. We now implement
the exact matrix evaluations and dyadic rounding within the claimed
arithmetic cost.

We use two levels of batching. An epoch ends after ${d_{\rm vec}}$ column
insertions or deletions. At its start let $C$ contain all currently
retained columns and the next at most ${d_{\rm vec}}$ input columns. There are at
most ${d_{\rm vec}}+1$ retained columns, so $C$ has at most $2{d_{\rm vec}}+1$ columns.
Keep this same $C$ throughout the epoch, including columns subsequently
deleted from the active set. Form the current $G$ and
$\mathsf G:=C^{\top}G^{-1}C$ exactly from scratch. To invert $G\succ0$,
recursively invert a leading principal block of half the order and its
Schur complement; both are positive definite. The block inverse formula uses a fixed
number of matrix products at each recursion node. After padding with an identity block to an
order that is a power of two, square multiplication with exponent $5/2$
gives
\[
 T_{\rm inv}(d)\le2T_{\rm inv}(d/2)+O(d^{5/2}),\qquad T_{\rm inv}(1)=O(1).
\]
Thus exact inversion costs $O(d_{\rm vec}^{5/2})$ arithmetic operations.
Forming $G$ and $C^{\top}G^{-1}C$ has the same bound.
During the epoch, inserting or deleting
column $C_i$ changes $G$ by $\varepsilon_{\rm upd} C_iC_i^{\top}$, with
$\varepsilon_{\rm upd}\in\{1,-1\}$, and hence changes ${\mathsf G}$ by
\begin{equation}\label{rank_one_eq_ridge_update}
 {\mathsf G}_{\rm new}={\mathsf G}-
 \frac{\varepsilon_{\rm upd}}{1+\varepsilon_{\rm upd} {\mathsf G}_{ii}}{\mathsf G}_{:,i}{\mathsf G}_{i,:}.
\end{equation}
The insertion denominator is at least one. For deletion, write
$G=G_-+C_iC_i^{\top}$, where $G_-\succeq\delta I$. Then
\[
 1-\mathsf G_{ii}=(1+C_i^{\top}G_-^{-1}C_i)^{-1}
 \ge\frac{\delta}{\delta+4}>0.
\]
Consequently every update is defined in exact arithmetic.

Define ${b_{\rm batch}}:=\lceil {d_{\rm vec}}^\chi\rceil$. Keep these updates as a list of at most
${b_{\rm batch}}$ outer products. Recovering
a requested column costs $O({d_{\rm vec}}{b_{\rm batch}})$; updating all diagonal entries costs
$O({d_{\rm vec}})$. The restriction of $I-{\mathsf G}$ to the retained columns supplies ${\Pi_{\rm rid}}$.
Before the next update would exceed this limit, collect the pending
terms as ${\mathsf A}\operatorname{diag}({\alpha_{\rm aux}}){\mathsf A}^{\top}$, add this product to the explicit base
matrix ${\mathsf G}$, and clear the list. Here ${\mathsf A}$ has $O({d_{\rm vec}})$ rows and at most
${b_{\rm batch}}$ columns. The rectangular multiplication bound
$\omega(1,\chi,1)<2.055$ of Alman et al.~\cite[Table~1]{adwxxz25} gives cost
$O({d_{\rm vec}}^{1+2\chi})$ for this operation. This writes the sum of pending
updates explicitly; it does not recompute an inverse. At an epoch
boundary discard the list and recompute from the actual retained
columns. No product is needed merely to clear the final incomplete
list. If several coordinates hit boundaries, an epoch may end between
their deletions; the new epoch includes every column still retained.

Since $0\preceq\Pi_{\rm rid}\preceq I$, its entries lie in $[-1,1]$.
Binary search on the dyadic grid rounds each requested entry using
$O(\log(1/u))$ arithmetic operations and comparisons. The same bound
applies to rounding each coefficient in $[-1,1]$. Each candidate
boundary ratio for a nonzero direction entry is formed and compared
in $O(1)$ operations. Choosing $\delta$ and $u$
by repeated halving costs $O(\log(2m d_{\rm vec}/\eta))$ operations.
The batch size is the least integer $b$ with $b^{400}\ge d_{\rm vec}^{211}$,
so binary search computes it in $O(\log(2d_{\rm vec}))$ operations.
Thus all scalar choices, rounding, and input inspections together cost
$\widetilde O(m d_{\rm vec})$.

There are at most $2m$ insertions and deletions, at most $m$ nullspace
moves, and $O(m/{b_{\rm batch}})$ products that write pending updates explicitly.
The total cost is therefore
\[ \widetilde O((m/{d_{\rm vec}}+1){d_{\rm vec}}^{5/2} +(m/{b_{\rm batch}}){d_{\rm vec}}^{1+2\chi}+m{d_{\rm vec}}{b_{\rm batch}}) =\widetilde O(m{d_{\rm vec}}^{1+\chi}+{d_{\rm vec}}^{5/2}), \]
because $\chi>1/2$.

\end{proof}

Grouping before applying a small reduction also appears in the
Carath\'eodory construction of Maalouf, Jubran, and Feldman~\cite{mjf20}.
Here a balanced tree of groups combines cube constraints, inherited
coefficients, and controlled numerical error.

\begin{lemma}[Partial signing by grouped coordinates]\label{rank_one_lem_grouped_partial_signing}
Let $b_1,\ldots,b_m\in\mathbb R^{d_{\rm vec}}$ have rational entries and
norm at most two, and let $0<\eta<1$.
A deterministic algorithm returns $y\in[-1,1]^m$ with at most
$d_{\rm vec}$ fractional coordinates and
\[
 \|\sum_i y_ib_i\|_2\le\eta
\]
using $\widetilde O(m d_{\rm vec}+d_{\rm vec}^{2.5275})$ arithmetic
operations. The suppressed factors include $\log(1/\eta)$.
\end{lemma}
\begin{proof}
Write $D:=d_{\rm vec}$ and use the arithmetic model of
Lemma~\ref{rank_one_lem_regularized_partial_signing}. We first give a
small rounding procedure.
Let $A$ have ${r_{\rm aux}}\le2D$ columns of norm at most two and let
${y_{\rm aux}}\in[-1,1]^{r_{\rm aux}}$ be any dyadic coefficient vector. For
$0<\varepsilon<1$, we can leave at most $D$ fractional coordinates
while changing $A{y_{\rm aux}}$ by norm at most $\varepsilon$, at cost
$\widetilde O(D^{2.5275})$ arithmetic operations. Coordinates already at a boundary
are fixed and removed before this procedure.

For each remaining active matrix define
$G:=\delta I+AA^{\top}$ and ${\Pi_{\rm rid}}:=I-A^{\top}G^{-1}A$, with a positive
dyadic $\delta$ satisfying
\[
 \tfrac12(\varepsilon/(100D^2))^2\le\delta
 \le(\varepsilon/(100D^2))^2.
\]
When ${r_{\rm aux}}>D$, $\tr[{\Pi_{\rm rid}}]\ge {r_{\rm aux}}-D\ge1$, so some diagonal entry is at
least $1/(2D)$. As in Lemma~\ref{rank_one_lem_regularized_partial_signing},
the corresponding column $r$ satisfies $\|Ar\|_2\le\sqrt\delta/2$.
Set $\tau:=\varepsilon/(10^4D^3)$ and choose a dyadic mesh $u$ with
$\tau/8<u\le\tau/4$. Round the exact diagonal entries to this grid,
choose a largest rounded diagonal, and round the corresponding exact
column to the same grid. The estimates below require only that the
entrywise errors are at most $\tau$. The resulting direction has maximum
absolute entry at least $1/(2D)-3\tau\ge1/(4D)$. From any ${y_{\rm aux}}$ in the cube, the positive boundary step
therefore has length at most $8D$.
Form and compare the boundary ratios exactly, set all hitting
coordinates exactly to their endpoints, and round the other coordinates
toward zero to the grid of mesh $u$.
Remove every boundary coordinate. At most $D$ moves are needed.
Their total change in $A{y_{\rm aux}}$ has norm at most
\[
 D(8D(\sqrt\delta/2+4D\tau)+4Du)<\varepsilon.
\]
Here the rounding error is at most the sum of the column norms times
the coefficient errors. The argument does not require ${y_{\rm aux}}=0$.

We implement the small procedure by batching its deletion updates.
Keep its original matrix $C$, which has at most $2D$ columns, including
columns subsequently removed from the active set. Define $a:=211/400$
and ${b_{\rm batch}}:=\lceil D^a\rceil$. Initially form
${\mathsf G}:=C^{\top}G^{-1}C$ exactly in $O(D^{5/2})$ arithmetic operations, using square
multiplication and the recursive positive-definite inversion in the
proof of Lemma~\ref{rank_one_lem_regularized_partial_signing}.
Represent subsequent changes to ${\mathsf G}$ by an explicit base
matrix and a list of at most ${b_{\rm batch}}$ rank-one terms. Keep its diagonal
separately, and perform all matrix updates exactly. A requested column
is the corresponding base column plus
the pending terms, and costs $O(D{b_{\rm batch}})$ to recover.

For a deletion at position $j$, recover the full column
$z={\mathsf G}_{:,j}$. Eq.~\eqref{rank_one_eq_ridge_update} adds
$zz^{\top}/(1-{\mathsf G}_{jj})$ to ${\mathsf G}$. Append this term
to the list and update every diagonal entry in $O(D)$ operations.
The deletion-denominator bound in the proof of
Lemma~\ref{rank_one_lem_regularized_partial_signing} gives
$1-\mathsf G_{jj}\ge\delta/(\delta+4)>0$, so every update is defined.
The grid searches and scalar parameter choices in that proof apply
here as well. Rounding each entry costs $O(\log(1/u))$ arithmetic
operations and comparisons, with
$\log(1/u)=O(\log(2D/\varepsilon))$.
Consequently choosing a direction, taking a boundary step, rounding
coefficients, and processing a deletion cost $\widetilde O(D{b_{\rm batch}})$ in total per
deleted coordinate. The inherited vector ${y_{\rm aux}}$ changes only the boundary
ratios; it does not enter $G$ or its update formula. Process simultaneous
boundary hits as consecutive deletions before the next rounding move.
There are at most $2D$ deletions, and no inserted columns in this
small procedure.

After every ${b_{\rm batch}}$ deletions, add the pending sum
${\mathsf U}\operatorname{diag}({\alpha_{\rm aux}}){\mathsf U}^{\top}$ to the explicit base
and clear the list. Here ${\mathsf U}$ has $O(D)$ rows and at most
${b_{\rm batch}}$ columns. The bound
$\omega(1,a,1)\le2.054999<1+2a$ from
Alman et al.~\cite[Table~1]{adwxxz25} gives cost
$O(D^{1+2a})$ per such product. A final incomplete list need not be
written to the base after the stopping condition is met. Therefore
the total arithmetic cost of the small procedure is
\[
 \widetilde O(D^{5/2}+D^2{b_{\rm batch}}+(D/{b_{\rm batch}})D^{1+2a}+D^2)
 =\widetilde O(D^{2+a}),\qquad 2+a=2.5275.
\]

We now organize the full input into a balanced binary tree whose
leaves are the indices $1,\ldots,m$. Each internal node splits its
leaves into two sets with sizes differing by at most one. Precompute
the vector $W_\mathcal J:=\sum_{i\in \mathcal J}b_i$ at every node $\mathcal J$ by additions.
There are fewer than $2m$ nodes, so this costs $O(mD)$ operations.
Initially the root is the only active group and its coefficient is
zero. A group coefficient is shared by all leaves of that group.
A group at coefficient $1$ or $-1$ is frozen permanently.

Let ${L_{\rm tree}}:=\max\{1,\lceil\log_2m\rceil\}$ and choose a power of two
$M_{\rm grp}$ with $m\le M_{\rm grp}<2m$. At each of at most ${L_{\rm tree}}$
rounds, replace each active nonsingleton group by its two children,
with both children inheriting the parent coefficient; retain active
singletons. Since
\[
 {y_{\rm aux}}_\mathcal JW_\mathcal J={y_{\rm aux}}_\mathcal JW_{\mathcal J_1}+{y_{\rm aux}}_\mathcal JW_{\mathcal J_2},
\]
splitting makes no change to the signed sum. There are at most $2D$
groups after splitting, provided there were at most $D$ before it.
Apply the small procedure to the columns $W_\mathcal J/M_{\rm grp}$, whose
norms are at most two, with
$\varepsilon:=\eta/(M_{\rm grp}{L_{\rm tree}})$ and the inherited coefficients.
Freeze every group that reaches a boundary. At most $D$ active
groups remain, and the change in the original signed sum is at most
$M_{\rm grp}\varepsilon=\eta/{L_{\rm tree}}$. If there are already at most $D$
groups, this rounding step is empty.

The root satisfies the required group count. After ${L_{\rm tree}}$ rounds every
remaining active group is a singleton, because its size is at most
$\lceil m/2^{L_{\rm tree}}\rceil=1$. Hence at most $D$ original coordinates are
fractional. Starting from zero and summing the residual allowances
over the rounds gives the asserted bound $\eta$.
Store each coefficient only at its current tree node; a final tree
traversal assigns coefficients to all leaves in $O(m)$ operations.
No update scans the leaves of an active group. The total cost is
\[
 \widetilde O(mD+{L_{\rm tree}}D^{2.5275})=\widetilde O(mD+D^{2.5275}).
\]
The power-of-two scale and the tree depth are found by doubling, within
the same bound. This proves the arithmetic cost.
\end{proof}

For the matrix inputs, use an orthonormal real basis of the Hermitian matrices and
approximate the coordinate vector of each $C_i$ to Euclidean error at most
$\eta_{\rm ps}/(4m)$ on a common dyadic grid. Since
$\|C_i\|_F\le1$, these approximations have norm at most two and
a common dyadic denominator of polynomial size in $mn$. Rank one and the variance computation above
give total cost $\widetilde O(mn^2+n^3)$ for scaling and vectorization.
Apply Lemma~\ref{rank_one_lem_grouped_partial_signing} with
$\eta:=\eta_{\rm ps}/2$ and ${d_{\rm vec}}:=n^2$. Because every output coefficient
has absolute value at most one, the resulting coefficients for the true
matrices satisfy
\[
 e_0:=\sum_i y_iC_i,\qquad \|e_0\|_F\le\eta_{\rm ps}.
\]
The mathematical model below uses the true retained matrices. Only this
single initial residual is charged to the approximate procedure; its
error does not recur in the nonlinear phase. For exact partial signing,
define $e_0:=0$.

\subsection{The maintained model and deferred rounding}\label{sec:fast:maintained-model}
Define

\[
\gamma:=10^{-6}/k,\qquad {\varepsilon_{\rm round}}:=10^{-6}.
\]

Use the budget $\beta(y)=(1-y^2)^{97/200}f(|y|)$ from Section~\ref{sec_rank_one_proof}; $f\ge1$, $0\le\beta\le1$, and $\iota=14779/2500$. At endpoint distance at least $\delta_{\rm end}/2$,

\begin{equation}\label{rank_one_eq_cont_1}
|\beta^{(j)}(y)|\le C\delta_{\rm end}^{97/200-j}\quad(1\le j\le3),
\qquad |\beta'(y)|\le C\delta_{\rm end}^{-1}\beta(y).
\end{equation}

The relative derivative bound follows directly from the logarithmic derivative of $(1-y^2)^{97/200}$ and boundedness of $f'/f$. Third derivatives at zero are interpreted one-sided; $\beta$ is $C^2$ and its second derivative is Lipschitz on every truncated interval.

There are two deliberately separated objects: the exact mathematical model, and rounded numerical data used to evaluate it. All discrepancy and descent inequalities refer to the exact model.

Let $y^0$ be the fractional coefficients from the chosen partial signing. Round them to a common fine dyadic grid, obtaining $\widetilde y^0\in[-1,1]^k$ with each error at most $\delta_0$, where $\delta_0\sqrt{k}\le{\varepsilon_{\rm round}}/2$. Initialize the model with

\[
F:=-\sum_{i=1}^k\widetilde y_i^0C_i,
\qquad M(y):=F+\sum_{i\in I}y_iC_i.
\]

Thus $M(\widetilde y^0)=0$. The actual previously frozen sum is $e_0-\sum_i y_i^0C_i$, so this initialization creates a fixed final discrepancy error of norm at most $\eta_{\rm ps}+{\varepsilon_{\rm round}}/2$. For exact partial signing $e_0=0$ and the $\eta_{\rm ps}$ term is omitted. Indeed

\begin{equation}\label{rank_one_eq_cont_2}
\|\sum_i r_iC_i\|\le\|\sum_i r_iC_i\|_F\le\|r\|_2.
\end{equation}

For rank-one inputs, the adjoint inequality is
$\sum_i\tr[C_i{D_{\rm aux}}]^2\le\tr[{D_{\rm aux}}^2\sum_iC_i^2]\le\|{D_{\rm aux}}\|_F^2$.

Define ${L_{\rm log}}:=\lceil\log_2(4n)\rceil$ and choose a dyadic cutoff
\begin{equation}\label{rank_one_eq_cont_3}
 \frac{{\varepsilon_{\rm round}}^2}{8192{L_{\rm log}}}<\delta_{\rm end}\le\frac{{\varepsilon_{\rm round}}^2}{4096{L_{\rm log}}}.
\end{equation}
When a coordinate $t_i$ has endpoint distance at most $\delta_{\rm end}$, update
$F\leftarrow F+t_iC_i$, remove its coupling slot, and record $t_i$.
Its output sign is chosen jointly with the other recorded coordinates
at the end. This removal preserves the modeled sum and cannot increase
the potential. Process these removals before testing regular endpoint
moves. The continuation paths below remove the coupling while
keeping the modeled sum fixed.
A regular endpoint move fixes its sign immediately.

For the fast-budget estimates, the potential is $\mathcal F:=\mathcal P+\gamma\mathcal B$, with
$\mathcal B(y):=\sum_{i\in I}\beta(y_i)$ and
\[
 \mathcal P(y):=\min\{b+\rho\tr[U+V]:
 U^{-1}+M(y)+\mathcal K_y(V)\preceq bI,\quad
 V^{-1}-M(y)+\mathcal K_y(U)\preceq bI\},
\]
\[
 \mathcal K_y(D):=\iota\sum_{i\in I}\beta(y_i)C_iDC_i.
\]
We work in the original coordinates $y$. Write $A\lesssim B$ when
$A\le P_n B$ for a fixed positive polynomial $P_n$ in $\log(2n)$,
enlarged when needed. Since $\delta_{\rm end}^{-1}=O(\log(2n))$, the explicit
budget satisfies, throughout endpoint distance at least $\delta_{\rm end}/2$,
\begin{equation}\label{fast_eq_scalar}
 -\beta''>1/2,\qquad |\beta^{(j)}|\lesssim-\beta'',\qquad
 \frac{|\beta^{(j)}|^2}{\beta}\lesssim-\beta''\quad(1\le j\le4).
\end{equation}
At zero, the derivatives of orders three and four are interpreted
one-sided. Derivatives through order two are continuous.

Coordinates close to an endpoint can be rounded together within the
reserved discrepancy allowance, completing the partial signing.

\begin{lemma}[Deferred rounding]\label{round_lem_round}
Suppose $C_i\succeq0$ have rank at most one, satisfy $\sum_iC_i^2\preceq I$, and have coefficients $t_i\in[-1,1]$ with $|t_i|\ge1-\delta_{\rm end}$, where Eq.~\eqref{rank_one_eq_cont_3} holds. All the coefficients can be rounded deterministically to signs with
\[
 \lVert \sum_i(\sigma_i-t_i)C_i\rVert<{\varepsilon_{\rm round}}/2.
\]
For at most $k$ input matrices, the cost is
$\widetilde O(kn^3)$ arithmetic operations.
\end{lemma}
\begin{proof}
If there are no inputs, return the empty signing. Otherwise choose a
positive dyadic $u$ with
$\varepsilon_{\rm round}/(256k)<u\le\varepsilon_{\rm round}/(128k)$.
We first construct rational PSD matrices of rank at most one $\widehat C_i$ with
$\|\widehat C_i-C_i\|_F\le u$ using arithmetic operations.
Let $q_i$ be a largest diagonal entry of $C_i$. If
$q_i\le u/(2n)$, set $\widehat C_i=0$; then
$\|C_i\|_F=\tr[C_i]\le nq_i\le u/2$.
Otherwise, for a maximizing index $j$, define
$c_i:=C_i e_j/q_i$, so $C_i=q_i c_i c_i^*$ and each entry of $c_i$
has absolute value at most one. Also $q_i\le1$.
Round $q_i$ and the real and imaginary parts of $c_i$ to a dyadic
mesh of size between $u/(200n)$ and $u/(100n)$, obtaining $\widehat q_i>0$ and
$\widehat c_i$, with $(\widehat c_i)_j=1$.
Set $\widehat C_i:=\widehat q_i\widehat c_i\widehat c_i^*$.
The outer-product perturbation bound gives
$\|\widehat C_i-C_i\|_F\le u$, and every nonzero
$\tr[\widehat C_i]$ is at least $u/(4n)$.
Binary search on bounded intervals performs these scalar roundings
in $O(\log(2n/u))$ arithmetic operations per entry. Constructing all
expanded matrices costs $\widetilde O(kn^2)$ arithmetic operations.
The approximations satisfy
\[
 \sum_i\widehat C_i^2\preceq(1+2ku+ku^2)I\preceq2I,
 \qquad 2ku\le\varepsilon_{\rm round}/64.
\]
Define
\[
 \theta:=8{L_{\rm log}}/{\varepsilon_{\rm round}},\qquad \tau:=\frac1{4\theta}=\frac{{\varepsilon_{\rm round}}}{32{L_{\rm log}}}.
\]
For every matrix with $\widehat{\nu}_i:=\tr[\widehat C_i]>\tau$, choose the nearest sign. Rank one gives $\widehat C_i^2=\widehat{\nu}_i\widehat C_i$, so the norm of the total error from these matrices is at most
\begin{equation}\label{round_eq_large}
 \delta_{\rm end}\lVert \sum_{\widehat{\nu}_i>\tau}\widehat C_i\rVert
 \le 2\delta_{\rm end}/\tau\le{\varepsilon_{\rm round}}/64.
\end{equation}

For $\widehat{\nu}_i\le\tau$, temporarily let $\epsilon_i\in\{-1,1\}$ have mean $t_i$, and define ${\mathsf E}_i:=(\epsilon_i-t_i)\widehat C_i$. Then
\[
 \E[{\mathsf E}_i]=0,\quad
 \E[{\mathsf E}_i^2]=(1-t_i^2)\widehat C_i^2\preceq2\delta_{\rm end}\widehat C_i^2,
 \quad\lVert \theta {\mathsf E}_i\rVert\le1/2.
\]
For $|y|\le1$, $e^y\le1+y+y^2$. Spectral calculus and scalar Jensen therefore give, for either sign,
\begin{equation}\label{round_eq_mgf}
 0\preceq G_i^\pm:=\log\E[e^{\pm\theta {\mathsf E}_i}]
 \preceq2\delta_{\rm end}\theta^2\widehat C_i^2,
 \qquad\lVert \sum_iG_i^\pm\rVert\le4\delta_{\rm end}\theta^2\le{L_{\rm log}}/16.
\end{equation}
For a chosen prefix with error sum $E$, define the estimator
\begin{equation}\label{round_eq_estimator}
 {\mathcal E_{\rm rnd}}(E):=
 \tr[\exp(\theta E+\sum_{\rm remaining}G_i^+)]
 +\tr[\exp(-\theta E+\sum_{\rm remaining}G_i^-)].
\end{equation}
Lieb's concavity theorem, in the form
$\E[\tr[e^{H+U}]]\le\tr[e^{H+\log\E[e^U]}]$, implies that the weighted average of the two next estimators is at most the current one \cite[Corollary~3.3]{t12}. Thus one of the two signs does not increase it. Initially
${\mathcal E_{\rm rnd}}_0\le2n e^{{L_{\rm log}}/16}$.
Choose the smaller estimator with additive evaluation error at most $1/(8\max\{1,k\})$. The total increase is at most $1/4$, hence the final estimator is at most $2{\mathcal E_{\rm rnd}}_0$. At the end, the remaining sums vanish and
\[
 e^{\theta\lVert E\rVert}\le \tr[e^{\theta E}]+\tr[e^{-\theta E}]
 \le4n e^{{L_{\rm log}}/16}.
\]
Consequently
\begin{equation}\label{round_eq_small}
 \lVert E\rVert\le\frac{\log(4n)+{L_{\rm log}}/16}{\theta}
 \le\frac{17{\varepsilon_{\rm round}}}{128}.
\end{equation}
Combining Eqs.~\eqref{round_eq_large} and~\eqref{round_eq_small}
with the matrix approximation error gives
\[
 \lVert \sum_i(\sigma_i-t_i)C_i\rVert
 \le\frac{{\varepsilon_{\rm round}}}{64}+\frac{17{\varepsilon_{\rm round}}}{128}+\frac{{\varepsilon_{\rm round}}}{64}
 =\frac{21{\varepsilon_{\rm round}}}{128}<{\varepsilon_{\rm round}}/2.
\]

It remains to justify the evaluation cost. Rank one gives the explicit expression
\[
 G_i^\pm=\frac{\log m_i^\pm}{\widehat{\nu}_i}\widehat C_i,
\quad m_i^\pm:=\frac{1+t_i}{2}e^{\pm\theta(1-t_i)\widehat{\nu}_i}
             +\frac{1-t_i}{2}e^{\mp\theta(1+t_i)\widehat{\nu}_i},
\]
with zero matrices omitted. Store the signed prefix and the two
remaining sums, updating each by one matrix addition at a decision.
By Eq.~\eqref{round_eq_mgf}, the remaining sums are PSD.
The estimator bound therefore bounds $\|\theta E\|$ by
$\log(4n)+L_{\rm log}/16$ at each accepted prefix. Either candidate
changes this by at most $1/2$. Hence
$M:=4L_{\rm log}+4=O(\log(2n))$ bounds both Hermitian exponential
arguments with more than one unit of spare margin.

For $\|H\|\le M$, the degree-$J$ Taylor polynomial for $e^H$ has
operator error at most $e^M M^{J+1}/(J+1)!$.
Taking $J=O(M+\log(2kn))$ makes the trace error smaller than
$1/(32k)$. Its factorial recurrence uses $J$ ordinary matrix
multiplications. The scalar exponential arguments in $m_i^\pm$
have absolute value at most $1/2$, and $1\le m_i^\pm\le e^{1/2}$.
Taylor series for the exponential and the convergent series
$\log m=2\sum_{j\ge0}z^{2j+1}/(2j+1)$, where $z:=(m-1)/(m+1)$,
therefore give the required scalar accuracies with logarithmically
many arithmetic operations. The lower bound
$\widehat\nu_i\ge u/(4n)$ controls the division in $G_i^\pm$.

Choose each computed $G_i^\pm$ to have operator error at most
$1/(64k^2 n3^{M+1})$. Exact additions of these approximations give
error at most $1/(64kn3^{M+1})$ in each remaining sum. On arguments
of norm at most $M+1$, the trace exponential is
$n e^{M+1}$-Lipschitz in operator norm. Thus these scalar
approximations and the matrix Taylor tail together keep each
candidate evaluation error below $1/(8k)$, as required.
There are at most $k$ decisions, each costing
$\widetilde O(n^3)$ arithmetic operations, and the matrix additions
cost $O(kn^2)$ in total. This proves the claimed arithmetic bound.
\end{proof}
\subsection{KKT conditioning and weighted norms}\label{sec:fast:weighted-norms}

Matrix spaces carry Frobenius norms and pairs carry product norms. Order inequalities for pairs are componentwise. Write ${\mathsf P}:=(U,V)$, ${\mathsf Q}:=(W,Z)$, and $E:=(I,I)$ for the primal and dual pairs at the unique optimizer. The dual and tight-constraint identities give

\[
\tr[W+Z]=1,\quad W\succeq\rho U^2,\quad Z\succeq\rho V^2,
\quad U,V\succeq \kappa_0I,\quad \|U^{-1}\|,\|V^{-1}\|\le C,
\]

\[
\rho\tr[U^2+V^2]\le1.
\]

Let

\[
{\mathscr J}({\mathsf A}_1,{\mathsf A}_2):=(U^{-1}{\mathsf A}_1U^{-1}-\mathcal K_y({\mathsf A}_2),
V^{-1}{\mathsf A}_2V^{-1}-\mathcal K_y({\mathsf A}_1)).
\]

It is self-adjoint and ${\mathscr J}^{-1}$ preserves the PSD order by
Lemma~\ref{rank_one_lem_potential}. Also ${\mathscr J}{\mathsf Q}=\rho E$ and
$\|{\mathscr J}^{-1}\|\le\rho^{-1}$: for $\|B\|_F\le1$, positivity gives
$-{\mathsf Q}/\rho\preceq {\mathscr J}^{-1}B\preceq {\mathsf Q}/\rho$, whose Frobenius
norm is at most $\tr[W+Z]/\rho=\rho^{-1}$. Let ${\mathscr E}$ be the block diagonal Hessian of the Lagrangian in ${\mathsf P}$; its $U$ block is

\[
{\mathscr E}_U({\mathsf A}):=U^{-1}{\mathsf A}U^{-1}WU^{-1}+U^{-1}WU^{-1}{\mathsf A}U^{-1}.
\]

In particular $a\rho^2I\preceq {\mathscr E}\preceq CI$ and $\|{\mathscr J}\|\le C$.
For the lower bound, $W\succeq\rho U^2$ implies
\[
 \tr[WU^{-1}{\mathsf A}U^{-1}{\mathsf A}U^{-1}]
 \ge\rho\tr[{\mathsf A}U^{-1}{\mathsf A}]
 \ge\rho^{3/2}\|{\mathsf A}\|_F^2.
\]
The $V$ block is identical, and $\rho<1$ gives the stated weaker bound. The full KKT Jacobian satisfies

\begin{equation}\label{rank_one_eq_cont_4}
\|{\mathsf R}^{-1}\|\le C\rho^{-2}=Cn^2.
\end{equation}

For completeness, consider the full KKT inverse, including the scalar
normalization. Write $E:=(I,I)$. Its linear system has the form
\[
  {\mathscr J} {\mathsf A}+{d_{\rm b}}E=F_1,\qquad
  {\mathscr J} {\mathsf B}- {\mathscr E} {\mathsf A}=F_2,\qquad \langle E,\mathsf B\rangle=f_3.
\]
Define ${\mathsf f}:= {\mathscr J}^{-1}E$, ${\mathsf g}_0:= {\mathscr J}^{-1}F_1$, and
$D_0:=\langle \mathsf f, {\mathscr E} {\mathsf f}\rangle$. Elimination gives
\[
 {d_{\rm b}}=\frac{\langle \mathsf f, {\mathscr E} {\mathsf g}_0\rangle+\langle \mathsf f,F_2\rangle-f_3}{D_0},\qquad
 {\mathsf A}={\mathsf g}_0-{\mathsf f}{d_{\rm b}},\qquad {\mathsf B}= {\mathscr J}^{-1}( {\mathscr E} {\mathsf A}+F_2).
\]
Use the following projection bounds before estimating these expressions:
\begin{gather*}
 \lVert I-{\mathsf f}{\mathsf f}^* {\mathscr E}/D_0\rVert\le
 \sqrt{\lVert  {\mathscr E}\rVert/\lambda_{\min}( {\mathscr E})}=O(\rho^{-1}),\\
 0\preceq {\mathscr E}- {\mathscr E}{\mathsf f}{\mathsf f}^* {\mathscr E}/D_0\preceq {\mathscr E},\\
 \lVert \mathsf f\rVert^2/D_0=O(\rho^{-2}),\qquad
 \lVert  {\mathscr E} {\mathsf f}\rVert\lVert \mathsf f\rVert/D_0=O(\rho^{-1}),\qquad
 \lVert \mathsf f\rVert\ge\lVert E\rVert/\lVert  {\mathscr J}\rVert\ge a\sqrt n.
\end{gather*}
Applying these separately to the $F_1,F_2,f_3$ terms gives
\[
 \lVert \mathsf A\rVert+\lVert \mathsf B\rVert+|{d_{\rm b}}|\le
 C\rho^{-2}(\lVert F_1\rVert+\lVert F_2\rVert+|f_3|).
\]

For a Hermitian pair $D:=(D_U,D_V)$, write
\[
 \widehat D_U:=U^{-1/2}D_U U^{-1/2},\qquad
 T_U:=U^{-1/2}WU^{-1/2},
\]
and use the analogous notation for the $V$ block. Define
\[
 e(D)^2:=\langle D,{\mathscr E}D\rangle
       =2\tr[T_U\widehat D_U^2]+2\tr[T_V\widehat D_V^2].
\]
Sums over the two blocks below are denoted by $\sum_{A=U,V}$.
We have $\sum_A\tr[T_A]\le C$, $T_U\succeq \kappa_0\rho I$, and
${\mathscr E}\succeq2\rho^{3/2}I$. Consequently
\begin{equation}\label{rank_one_eq_weighted_norms}
 \|D\|_F\le Cn^{3/4}e(D),\qquad
 \max_A\|\widehat D_A\|\le C\sqrt ne(D),\qquad
 e(D)\le C\|D\|_F.
\end{equation}
The second inequality follows by bounding $\tr[\widehat D_A^2]$
using $T_A\succeq \kappa_0\rho I$. We also have
$\|\widehat D_A\|\le C\|D_A\|_F$.

For a Hermitian pair $B$, define the weighted order bound
\[
 q(B):=\inf\{\langle {\mathsf Q},A\rangle:A\succeq0,\ -A\preceq B\preceq A\}.
\]
It is subadditive and homogeneous. Positivity and self-adjointness of
${\mathscr J}^{-1}$ give
\begin{equation}\label{rank_one_eq_weighted_inverse}
 \|{\mathscr J}^{-1}B\|_F\le\rho^{-1}q(B),\qquad
 e({\mathscr J}^{-1}B)\le Cn q(B).
\end{equation}
Indeed, for any admissible $A$, define $C_0:={\mathscr J}^{-1}A\succeq0$.
Then $-C_0\preceq {\mathscr J}^{-1}B\preceq C_0$, and the sum of the nuclear
norms of the two blocks is at most
$\langle E,C_0\rangle=\rho^{-1}\langle {\mathsf Q},A\rangle$.
For this nuclear-norm inequality, pair the two spectral projections of
${\mathscr J}^{-1}B$ with its upper and lower bounds. Taking the infimum proves the
first assertion; ${\mathscr E}\preceq CI$ proves the second.

Define $\mathcal A(D_U,D_V):=(\mathcal K_y(D_V),\mathcal K_y(D_U))$.
For the weighted maps below, $\mathcal A_t^{(j)}$ denotes a derivative
along any path whose coupling coefficients have bounded relative
derivatives. Along a reset path this bound is absolute; along a local
$y$ segment it is at most $P_n$. In the following display, replace $C$
by $P_n$ in the latter case.

Two useful consequences are
\begin{equation}\label{rank_one_eq_weighted_maps}
 q(\mathcal A_t^{(j)}D)\le Ce(D),\qquad q({\mathscr E}D)\le Ce(D).
\end{equation}
For the first, define
$D_U^\#:=U^{1/2}|\widehat D_U|U^{1/2}$, and similarly for $V$.
Then $\pm\mathcal A_t^{(j)}D\preceq C\mathcal A_t(D^\#)$.
The dual equation implies
$\mathcal A_t{\mathsf Q}\preceq {\mathsf S}$, where
${\mathsf S}:=(U^{-1}WU^{-1},V^{-1}ZV^{-1})$. Hence its weighted cost is at most
\[
 C\langle {\mathsf S},D^\#\rangle
 =C\sum_A\tr[T_A|\widehat D_A|]
 \le C\sqrt{\sum_A\tr[T_A]}\sqrt{\sum_A\tr[T_A\widehat D_A^2]}
 \le Ce(D).
\]
For the second, the normalized $U$ block of ${\mathscr E}D$ is
$\widehat D_UT_U+T_U\widehat D_U$.
For any $a>0$, its positive and negative are bounded by
$a\widehat D_U^2+a^{-1}T_U^2$.
The total weighted cost is at most
$a e(D)^2/2+a^{-1}\sum_A\tr[T_A^3]$.
Optimizing $a$ and using $\sum_A\tr[T_A^3]\le C$ proves the claim.
This argument does not require the matrices to commute.

There is also an exact description of the order norm:
\begin{equation}\label{fast_eq_qtrace}
 q(B)=\sum_{A=U,V}\|{\mathsf Q}_A^{1/2}B_A{\mathsf Q}_A^{1/2}\|_*.
\end{equation}
Congruence reduces its defining minimization to the trace of a positive
majorant of a Hermitian matrix and its negative, whose minimum is the
nuclear norm. For an inverse-derivative block
$U^{-1/2}DU^{-1/2}$ this equals $\|T_U^{1/2}DT_U^{1/2}\|_*$, by
polar decomposition. This also justifies the noncommuting products below.
Define ${\mathsf f}:={\mathscr J}^{-1}E={\mathsf Q}/\rho$ and $D_0:=e({\mathsf f})^2$. Then
\begin{equation}\label{fast_eq_normalization}
 \|{\mathsf f}\|_F\le Cn,\qquad
 D_0=2\rho^{-2}\sum_{A=U,V}\tr[T_A^3]
       \ge\sqrt{2/{\varepsilon_{\rm scale}}}.
\end{equation}
Indeed $T_U\succeq\rho U$ implies
$\tr[T_U^2]\ge\rho\tr[T_UU]=\rho\tr[W]$.
Summing both blocks and applying the power-mean inequality on $2n$
eigenvalues gives $\sum_A\tr[T_A^3]\ge\rho^{3/2}/\sqrt{2n}$.

On a fixed-radius neighborhood in the optimizer variables, the KKT map
${\mathcal E_{\rm KKT}}(y,{\mathsf s})$ has $\|{\mathcal E_{\rm KKT}}_{{\mathsf s}{\mathsf s}}\|\le C$: inverse matrices,
$W,Z$, and the coupling have bounded operator norms. Hence the inverse
bound in Eq.~\eqref{rank_one_eq_cont_4} gives a Newton radius
$r_m:=\kappa_mn^{-2}$. Inside this radius,
\begin{equation}\label{rank_one_eq_cont_18}
 \|{\mathsf s}_{j+1}-{\mathsf s}_*\|\le Cn^2\|{\mathsf s}_j-{\mathsf s}_*\|^2.
\end{equation}
Inexact evaluation adds a polynomially controlled error floor.

Write
\[
 \mathcal C(y,{\mathsf P}):=(U^{-1}+M+\mathcal K_y(V),V^{-1}-M+\mathcal K_y(U)),
 \qquad \mathcal L:=b+\rho\langle E,{\mathsf P}\rangle
                   +\langle {\mathsf Q},\mathcal C-bE\rangle.
\]
The optimizer satisfies $\mathcal C=bE$, ${\mathscr J}{\mathsf Q}=\rho E$, and
$\tr[W+Z]=1$. The symbol ${\mathsf s}:=(b,{\mathsf P},{\mathsf Q})$ denotes all optimizer variables.
The matrix ${\mathscr E}$ is the Hessian of $\mathcal L$ in ${\mathsf P}$.

Write $C_i=a_ia_i^*$ and
\[
 u_i:=a_i^*Ua_i,\quad v_i:=a_i^*Va_i,\quad
 w_i:=a_i^*Wa_i,\quad z_i:=a_i^*Za_i,\quad t_i:=v_iw_i+u_iz_i.
\]
There is no numerical choice of phases for these vectors: the expressions can all be computed as matrix traces. Define
\begin{equation}\label{fast_eq_budgetmetric}
 \varpi_i:=(-\beta_i'')(\iota t_i+\gamma),\qquad G:=\operatorname{diag}(\varpi_i).
\end{equation}
At a light state, $u_i,v_i\le2/(\iota\beta_i)$ and $w_i+z_i\le\tr[C_i]\le1$. Consequently
\begin{equation}\label{fast_eq_w_range}
 \gamma/4\le \varpi_i\le P_n.
\end{equation}
For a Euclidean unit direction $r$, define
\begin{equation}\label{fast_eq_theta}
 A_r:=\iota\sum_i(-\beta_i'')t_i r_i^2,
 \qquad \theta:=r^{\top}Gr=A_r+\gamma\sum_i(-\beta_i'')r_i^2.
\end{equation}
In particular $\theta\ge\gamma/4\ge a/n^2$, where $a>0$ is absolute.

\subsection{A budget-weighted gradient/curvature dichotomy}\label{sec:fast:descent-dichotomy}
The following dichotomy selects the next type of local move using
thresholds scaled by the coordinate budgets.

\begin{lemma}\label{fast_lem_dichotomy}
At every light state, either some coordinate has
\begin{equation}\label{fast_eq_gradthreshold}
 |\mathcal F_i|\ge P_n^{-1}\sqrt{\varpi_i/n},
\end{equation}
or there is a Euclidean unit direction $r$ with
\begin{equation}\label{fast_eq_curvthreshold}
 r^{\top}\nabla^2\mathcal Fr\le-P_n^{-1}r^{\top}Gr.
\end{equation}
Both alternatives admit fixed-factor separated numerical thresholds
with inverse-polynomial evaluation margins.
\end{lemma}
\begin{proof}
Use the pointwise drift
$h_i=-C_{2,i}M_iK_{ii}$, where
$M_i:=u_i-v_i+2\iota\beta_i'u_iv_i$ and $C_{2,i}=(1+\tau/2)k_i$.
Lemma~\ref{rank_one_lem_joint_curvature}, the strict surplus in
Eq.~\eqref{eq_fast_strict_surplus}, and the budget descent estimate give
\[
 \mathscr{L}\mathcal P\le-a\sum_ip_iq_it_iK_{ii},\qquad
 \mathscr{L}\mathcal B\le-\frac1{20}\sum_i(-\beta_i'')K_{ii}.
\]
Here $a>0$ is absolute, although small. Since $p_iq_i\ge1/2$
and $-\beta''\le P_n$,
\begin{equation}\label{fast_eq_weighteddrift}
 \mathscr{L}\mathcal F\le-P_n^{-1}\tr[GK].
\end{equation}
On the other hand lightness implies $|M_i|\lesssim u_i+v_i$. Define $\nu_i:=\tr[C_i]$. Rank-one Cauchy--Schwarz and the dual lower bounds give
\[
 w_i\ge\rho u_i^2/\nu_i,\quad z_i\ge\rho v_i^2/\nu_i,
 \quad u_i,v_i\ge \kappa_0\nu_i.
\]
Therefore $t_i\ge \kappa_0\rho(u_i^2+v_i^2)$, and hence
\begin{equation}\label{fast_eq_drift_coordinate}
 (C_{2,i}M_i)^2\lesssim n\varpi_i.
\end{equation}
If every $|\mathcal F_i|$ is smaller than a sufficiently small inverse-polylogarithmic multiple of $\sqrt{\varpi_i/n}$, equations \eqref{fast_eq_weighteddrift}--\eqref{fast_eq_drift_coordinate} imply
\[
 \tr[\nabla^2\mathcal FK]\le-P_n^{-1}\tr[GK].
\]
The covariance is nonzero. Conjugating by $G^{-1/2}$ proves the
second alternative. Otherwise the first holds.

To quantify the evaluation margins, define
$g_i:=\mathcal F_i/\sqrt{\varpi_i/n}$ and
$A:=G^{-1/2}\nabla^2\mathcal FG^{-1/2}$.
The exact argument provides an inverse-polylogarithmic
$\delta_*>0$ such that $\max_i|g_i|<\delta_*$ implies
$\lambda_{\min}(A)\le-\delta_*$.
Suppose $\widehat g$ and a symmetric $\widehat A$ satisfy
$\|\widehat g-g\|_\infty\le\delta_*/8$ and
$\|\widehat A-A\|\le\delta_*/8$.
If some $|\widehat g_i|\ge3\delta_*/4$, then
$|g_i|\ge5\delta_*/8$. Otherwise
$\max_i|g_i|<7\delta_*/8<\delta_*$, so
$\lambda_{\min}(\widehat A)\le-7\delta_*/8$.
There is then a unit vector ${\mathsf h}$ with
${\mathsf h}^\top\widehat A{\mathsf h}\le-3\delta_*/4$, and it satisfies
${\mathsf h}^\top A{\mathsf h}\le-5\delta_*/8$.
For $r:=G^{-1/2}{\mathsf h}/\|G^{-1/2}{\mathsf h}\|_2$, this yields
\[
 r^\top\nabla^2\mathcal F r\le-(5\delta_*/8)r^\top G r.
\]
Enlarge $P_n$ by a fixed factor to obtain the stated thresholds.
By Eq.~\eqref{fast_eq_w_range}, the diagonal scalings, their inverses,
and the normalization are polynomially bounded. The KKT inverse
and derivative bounds make $\|A\|$ polynomially bounded as well.
Thus inverse-polynomial errors in the unscaled quantities suffice
for these tolerances. Reserving a further fixed fraction of each
margin permits approximate diagonal scaling and a direct check of
the final generalized Rayleigh quotient.
\end{proof}

\subsection{Directional estimates retaining the size of the budget}
Derivatives in this section are along the straight original-coordinate line $y+tr$. Work at endpoint distance at least $\delta_{\rm end}/2$. Define ${\mathsf A}:={\mathsf P}'$, ${\mathsf B}:={\mathsf Q}'$, and $F_2:=\mathcal A'{\mathsf Q}$.

The next estimates retain the directional budget scale, which controls
the optimizer response and the length of a local step.

\begin{lemma}\label{fast_lem_directional}
For $1\le j\le4$,
\begin{equation}\label{fast_eq_weightedcoupling}
 q(\mathcal A^{(j)}D)\lesssim\sqrt{A_r}e(D),
 \qquad |\langle F_2,D\rangle|\lesssim\sqrt{A_r}e(D).
\end{equation}
For $j\ge2$, $q(\mathcal C_{t^j})\lesssim A_r$. The second-order envelope identity is
\begin{equation}\label{fast_eq_energyidentity}
 \mathcal P''=-A_r+e(\mathsf A)^2+2\langle F_2,\mathsf A\rangle.
\end{equation}
Consequently, in a region where $\mathcal P''\le C\theta_0$ and $\theta\asymp\theta_0$, one has
\begin{equation}\label{fast_eq_smallE}
 e(\mathsf A)\lesssim\sqrt{\theta_0},\quad
 \lVert \mathsf A\rVert_{\mathrm F}\lesssim n^{3/4}\sqrt{\theta_0},\quad
 \lVert \mathsf B\rVert_{\mathrm F}\lesssim n\sqrt{\theta_0}.
\end{equation}
\end{lemma}
\begin{proof}
For one coupling block, use
\[
 |a_i^*D_Va_i|^2\le v_ia_i^*D_VV^{-1}D_Va_i.
\]
Weighted scalar Cauchy--Schwarz bounds its majorant cost by the product of
\[
 (\iota\sum_i\frac{|\beta_i^{(j)}|^2}{\beta_i}|r_i|^{2j}v_iw_i)^{1/2}
 \quad\text{and}\quad
 (\iota\sum_i\beta_iw_i\frac{|a_i^*D_Va_i|^2}{v_i})^{1/2}.
\]
The first squared factor is at most $P_nA_r$. The second is at most
$\tr[D_VV^{-1}D_V\mathcal K(W)]\le e(D_V)^2/2$, by the dual equation. Combine both blocks. This proves \eqref{fast_eq_weightedcoupling}. For direct partial derivatives use $M^{(j)}=0$ for $j\ge2$ and \eqref{fast_eq_scalar}; their majorant costs are at most $P_nA_r$.
Differentiating the stationary Lagrangian gives \eqref{fast_eq_energyidentity} because $\langle \mathsf Q,\mathcal C_{tt}\rangle=-A_r$. It gives
$\mathcal P''\ge e(\mathsf A)^2-P_n\sqrt{A_r}e(\mathsf A)-A_r$, proving the energy estimate. The remaining estimates follow from Eqs.~\eqref{rank_one_eq_weighted_norms}--\eqref{rank_one_eq_weighted_maps}, ${\mathsf B}={\mathscr J}^{-1}({\mathscr E}{\mathsf A}+F_2)$, and the dual energy bound for $F_2$.
\end{proof}

\subsection{Capped semidefinite initialization}\label{sec_fast_reset}
The following short-step barrier method initializes the optimizer cache
for the square-root budget.
The bounds in this subsection are per initialization; each application
counts its own calls.

The face has a bounded modeled $M$ and a PSD coupling $\mathcal K$. The bounds
\begin{equation}\label{fast_eq_resetbounds}
 \lVert M\rVert\le20,\qquad \mathcal K(I)\preceq32I
\end{equation}
hold for the true model and for its sufficiently accurate short PSD cache. Define
\[
 a=1/1000,\qquad R_c=2\lceil\rho^{-1/2}\rceil,\qquad T_c=200.
\]
For ${\mathsf a}=(b,U,V)$ impose the two affine LMIs
\begin{equation}\label{fast_eq_lmis}
 {D_{\rm sl}}_+({\mathsf a})=\begin{pmatrix}bI-M-\mathcal K(V)&I\\I&U\end{pmatrix}\succeq0,\quad
 {D_{\rm sl}}_-({\mathsf a})=\begin{pmatrix}bI+M-\mathcal K(U)&I\\I&V\end{pmatrix}\succeq0,
\end{equation}
plus $aI\preceq U,V\preceq R_c I$ and $0\le b\le T_c$. Minimize
\[
 \langle {\mathsf c}_{\rm obj},\mathsf a\rangle=b+\rho\tr[U+V].
\]
The independent real variable dimension is $O(n^2)$.

The added caps provide a bounded domain and a strictly feasible
starting point for barrier initialization while preserving the optimum.

\begin{lemma}\label{fast_lem_caps}
These caps do not change the potential optimum. The point
\begin{equation}\label{fast_eq_strictpoint}
 {\mathsf a}_0=(100,I,I)
\end{equation}
is strictly feasible, with every slack matrix bounded below by $I/2$. The domain has diameter $O(n)$.
\end{lemma}
\begin{proof}
At ${\mathsf a}_0$, the top-left block of each LMI is at least $48I$. Subtracting $I/2$ from the complete block matrix and using its Schur complement proves ${D_{\rm sl}}_\pm({\mathsf a}_0)\succeq I/2$. The other slacks have the same lower bound.

The scalar feasible point gives objective at most $100+2{\varepsilon_{\rm scale}}<101$. The unconstrained optimizer therefore has $b_*<101$. Its tight constraints give $U_*^{-1},V_*^{-1}\preceq121I$, so $U_*,V_*\succeq I/121\succ aI$. The dual equations give $\rho\tr[U_*^2+V_*^2]\le1$, hence $U_*,V_*\preceq\rho^{-1/2}I\prec R_c I$. Adding the two tight constraints gives $b_*\ge\sqrt\rho>0$. Thus all added caps have slack at the optimizer. The diameter follows from $R_c=O(\sqrt n)$ and Frobenius dimension $n$ in each matrix block.
\end{proof}

\paragraph{Barrier and local norm.}
Let ${D_{\rm sl}}_1,\ldots,{D_{\rm sl}}_8$ be the two LMIs in Eq.~\eqref{fast_eq_lmis}, the four matrix cap slacks, and the two scalar slacks $b,T_c-b$. Use
\begin{equation}\label{fast_eq_barrier}
 f({\mathsf a})=-\sum_{j=1}^8\log\det {D_{\rm sl}}_j({\mathsf a}),\qquad \nu_b=8n+2.
\end{equation}
Scalars count as order-one matrices. Realification is unnecessary analytically: Hermitian log-determinant derivatives obey the same inequalities. A rational real/imaginary basis has only a constant-factor change in norm.

For $C_j={D_{\rm sl}}_j({\mathsf a})^{-1/2}(D{D_{\rm sl}}_j[\Delta ]){D_{\rm sl}}_j({\mathsf a})^{-1/2}$,
\[
 D^2f[\Delta ,\Delta ]=\sum_j\tr[C_j^2],\quad
 D^3f[\Delta ,\Delta ,\Delta ]=-2\sum_j\tr[C_j^3],
\]
so $|D^3f[\Delta ^3]|\le2(D^2f[\Delta ^2])^{3/2}$ and
$|Df[\Delta ]|\le\sqrt{\nu_b}(D^2f[\Delta ^2])^{1/2}$. Thus $f$ is a self-concordant barrier of parameter at most $\nu_b$. The cap terms also give the global lower bound
\begin{equation}\label{fast_eq_globalbarrierlower}
 \nabla^2 f({\mathsf a})\succeq h_0I,\qquad h_0=(T_c+R_c)^{-2},
\end{equation}
in the product Frobenius norm. For example the $U-aI$ term is at least $R_c^{-2}\lVert D_U\rVert_{\mathrm F}^2$, and $-\log b$ gives at least $T_c^{-2}(nb)^2$.

For the Newton contraction and minimizer-distance estimates we use \cite[Theorems 2.7--2.8]{dkv15}. We give the conditioning and precision estimates for this program below.

\paragraph{Initialization without assuming an analytic center}
For $f({\mathsf a})+\langle b_{\rm obj},\mathsf a\rangle$, denote the Newton decrement by
\[
 \delta_{b_{\rm obj}}({\mathsf a})=\lVert \nabla f({\mathsf a})+{b_{\rm obj}}\rVert_{{\mathsf a},*},\qquad
 \lVert r\rVert_{{\mathsf a},*}^2=\langle r_{\rm aux},\nabla^2f({\mathsf a})^{-1}{r_{\rm aux}}\rangle.
\]
Compute a short rational ${b_{\rm obj}}_0$ approximating $-\nabla f({\mathsf a}_0)$ so that $\delta_{{b_{\rm obj}}_0}({\mathsf a}_0)\le1/64$. The global Hessian bound makes this an inverse-polynomial absolute tolerance.

First track the objectives $f+\mu\langle {b_{\rm obj}}_0,\mathsf a\rangle$ from $\mu=1$ down to a positive rational $\mu_{\min}$ satisfying
\[
 \mu_{\min}\le\frac{\sqrt{h_0}}{128(1+\lVert {b_{\rm obj}}_0\rVert)},
\]
within a factor two of the displayed quantity. Next switch, with a constant number of corrections, to $f+{\alpha_{\rm IPM}}\langle {\mathsf c}_{\rm obj},\mathsf a\rangle$, starting at positive rational
\[
 {\alpha_{\rm IPM}}_0\le\frac{\sqrt{h_0}}{128(1+\lVert {\mathsf c}_{\rm obj}\rVert)}
\]
within a factor two, and increase to ${\alpha_{\rm IPM}}_{\max}\ge4\nu_b/{\varepsilon_{\rm obj}}$. Norms in these choices can be replaced by explicit rational upper bounds. Here ${\varepsilon_{\rm obj}}$ is any specified inverse-polynomial objective accuracy.

In either phase use parameter changes of relative magnitude at most
\[
 \gamma_0=\frac1{16\lceil\sqrt{\nu_b}\rceil}
\]
and at least $\gamma_0/2$, except for the last truncated change. Take one full Newton correction per change. If $\delta\le1/8$, the exact corrected decrement is at most
\[
 (\frac{\delta}{1-\delta})^2\le1/49.
\]
Reserve additive error $1/256$ in decrement for rounding and the linear solve. Define $q_0=1/49+1/256$. Since
\[
 \lVert b_{\rm obj}\rVert_{{\mathsf a},*}\le\sqrt{\nu_b}+\delta_{b_{\rm obj}}({\mathsf a}),\qquad
 q_0+\frac{1+q_0}{16}=\frac{17729}{200704}<\frac18,
\]
the invariant closes with slack. At the phase switch the omitted linear term and the new linear term each have dual local norm at most $1/128$, by Eq.~\eqref{fast_eq_globalbarrierlower}; this also fits the invariant after a constant number of corrections. All parameter endpoints have polynomial magnitude and inverse magnitude. Hence the stage count is
\begin{equation}\label{fast_eq_barrierstages}
 O(\sqrt n\log\frac{2nk}{{\varepsilon_{\rm obj}}})=\widetilde O(\sqrt n).
\end{equation}
Store $\mu$ and ${\alpha_{\rm IPM}}$ on sufficiently fine common dyadic grids. Relative changes between $\gamma_0/2$ and $\gamma_0$ are available because the smallest parameter and smallest increment are inverse-polynomial. Form the linear objective from its current short scalar and fixed short coefficient vector, not by accumulating rounded objective updates.

\paragraph{Polynomial bounds on inverse slacks.}
To bound the required precision, we bound each inverse slack and the condition number of the Newton system.

For any objective vector ${b_{\rm obj}}$ on the two paths, let $\bar {\mathsf a}_{b_{\rm obj}}$ be the exact minimizer of $f+\langle b_{\rm obj},\cdot\rangle$. It exists in the interior because the capped domain is bounded and the barrier diverges at its boundary. Central stationarity and affinity imply the exact identity
\begin{equation}\label{fast_eq_traceidentity}
 \sum_{j=1}^8\tr[{D_{\rm sl}}_j(\bar {\mathsf a}_{b_{\rm obj}})^{-1}{D_{\rm sl}}_j({\mathsf a}_0)]
 =\nu_b+\langle b_{\rm obj},{\mathsf a}_0-\bar {\mathsf a}_{b_{\rm obj}}\rangle.
\end{equation}
Indeed $D{D_{\rm sl}}_j[{\mathsf a}_0-\bar {\mathsf a}_{b_{\rm obj}}]={D_{\rm sl}}_j({\mathsf a}_0)-{D_{\rm sl}}_j(\bar {\mathsf a}_{b_{\rm obj}})$; pair this with each inverse slack and use $\nabla f(\bar {\mathsf a}_{b_{\rm obj}})=-{b_{\rm obj}}$.

Every summand on the left is nonnegative. Since ${D_{\rm sl}}_j({\mathsf a}_0)\succeq I/2$, Lemma \ref{fast_lem_caps} gives
\begin{equation}\label{fast_eq_inverse_slack}
 \sum_j\tr[{D_{\rm sl}}_j(\bar {\mathsf a}_{b_{\rm obj}})^{-1}]
 \le2(\nu_b+\lVert b_{\rm obj}\rVert\operatorname{diam}\mathcal D).
\end{equation}
The right side is polynomial in $n,k,{\varepsilon_{\rm obj}}^{-1}$. Thus \emph{each} slack inverse has polynomial operator norm. A bound on the product of slack eigenvalues would not suffice here.

The self-concordant minimizer-distance bound, at decrement at most $1/8$, gives
\[
 \lVert {\mathsf a}-\bar {\mathsf a}_{b_{\rm obj}}\rVert_{\mathsf a}
 \le\delta+\frac{3\delta^2}{(1-\delta)^3}
 \le\frac18+\frac{24}{343}<\frac14.
\]
Therefore each slack at an iterate is within constant Loewner factors of its slack at the exact center. Eq.~\eqref{fast_eq_inverse_slack} applies to iterates with changed constants. Upper slack norms are polynomial from the caps and bounded affine coefficient maps.

These inverse-slack bounds and Eq.~\eqref{fast_eq_globalbarrierlower}
give polynomial conditioning of the Newton systems.

\paragraph{Objective accuracy and KKT recovery}
At an exact center for ${b_{\rm obj}}={\alpha_{\rm IPM}} {\mathsf c}_{\rm obj}$, and any capped feasible ${\mathsf a}_*$,
\[
 {\alpha_{\rm IPM}}\langle {\mathsf c}_{\rm obj},\bar {\mathsf a}-{\mathsf a}_*\rangle
 =\nu_b-\sum_j\tr[{D_{\rm sl}}_j(\bar {\mathsf a})^{-1}{D_{\rm sl}}_j({\mathsf a}_*)]\le\nu_b.
\]
An iterate of decrement at most $1/8$ is within a fixed local distance of this center. Since
\[
 \lVert {\alpha_{\rm IPM}} {\mathsf c}_{\rm obj}\rVert_{\bar {\mathsf a},*}\le\sqrt{\nu_b},
\]
its additional objective gap is less than $\nu_b/{\alpha_{\rm IPM}}$ after enlarging an absolute constant. Thus ${\alpha_{\rm IPM}}_{\max}\ge4\nu_b/{\varepsilon_{\rm obj}}$ leaves a margin for a true objective gap at most ${\varepsilon_{\rm obj}}$, including the chosen rounding and data errors.

We next recover the primal-dual optimizer from a feasible point of small objective gap. The optimal multipliers satisfy $W_*,Z_*\succeq\rho I/121^2$. On the cap, $U^{-1},V^{-1}\succeq R_c^{-1}I$, so the Lagrangian Hessian in ${\mathsf P}$ is at least $2\rho/(121^2R_c^3)$ times the identity. This is at least a fixed positive multiple of $\rho^{5/2}$. Let $\mathcal L_*$ be the original Lagrangian with the optimal multipliers fixed.
Its $b$ coefficient vanishes because $\tr[W_*+Z_*]=1$, and its gradient
in ${\mathsf P}$ vanishes at ${\mathsf P}_*$. For any capped feasible $\widetilde {\mathsf a}$,
\[
 \langle {\mathsf c}_{\rm obj},\widetilde {\mathsf a}\rangle-\langle {\mathsf c}_{\rm obj},{\mathsf a}_*\rangle
 \ge\mathcal L_*(\widetilde {\mathsf P})-\mathcal L_*({\mathsf P}_*)
 \ge a''\rho^{5/2}\|\widetilde {\mathsf P}-{\mathsf P}_*\|_F^2.
\]
The last inequality follows by integrating the Lagrangian Hessian along
the segment in the matrix caps. A feasible point of gap ${\varepsilon_{\rm obj}}$ therefore satisfies
\begin{equation}\label{fast_eq_primalrecover}
 \lVert \widetilde {\mathsf P}-{\mathsf P}_*\rVert_{\mathrm F}\le C\rho^{-5/4}\sqrt{\varepsilon_{\rm obj}}.
\end{equation}
Recover $\widetilde {\mathsf Q}$ from ${\mathscr J}_{\widetilde {\mathsf P}}\widetilde {\mathsf Q}=\rho E$. Near ${\mathsf P}_*$, the inverse bound and a Neumann perturbation give
\begin{equation}\label{fast_eq_recover}
 \lVert (\widetilde b,\widetilde {\mathsf P},\widetilde {\mathsf Q})-{\mathsf s}_*\rVert
 \le Cn^{9/4}\sqrt{\varepsilon_{\rm obj}},
\end{equation}
plus controlled coefficient and solve errors. The scalar error follows as well from the objective and Eq.~\eqref{fast_eq_primalrecover}. For example ${\varepsilon_{\rm obj}}=a'(2nk)^{-40}$, with sufficiently small fixed $a'>0$, reaches a fixed fraction of the $\kappa_mn^{-2}$ KKT Newton radius. A final KKT correction restores its standard cache floor.

\paragraph{Coefficient errors.}
The solver uses the short PSD rank-one cache and nonnegative dyadic budget coefficients. These preserve the SDP's convexity and strict scalar point. On the capped domain, coefficient errors perturb each Schur-complement constraint by at most a fixed polynomial times the cache tolerance. Increasing $b$ by that amount repairs true feasibility; $b<T_c$ has ample slack near the optimum. The same repair in the opposite direction compares the two optima. Choose cache error so both repairs and objective perturbations are much smaller than ${\varepsilon_{\rm obj}}$. Then Eq.~\eqref{fast_eq_recover} produces an approximation to the \emph{true} optimizer. The fixed contribution is still a symbolic combination of the initial $k$ matrices with bounded coefficients, so its error is bounded afresh, independently of how many resets or local steps occurred.

Each call therefore uses $\widetilde O(\sqrt n)$ barrier stages.

\subsection{Rank-one operator actions}\label{sec_fast_structured_solves}
The rank-one representation reduces coupling evaluations and their
directional derivatives to rectangular matrix products.

\begin{lemma}\label{fast_lem_operator_cost}
For a fixed number of Hermitian arguments, the coupling and its
fixed-order directional derivatives can be applied in
$\widetilde O(n^\chi)$ arithmetic operations, to any prescribed inverse-polynomial accuracy.
The same bound
computes all endpoint scores, gradient coordinates, and weights $\varpi_i$
at a given optimizer.
\end{lemma}
\begin{proof}
For each nonzero input, choose a largest diagonal entry
$\gamma_i:=(C_i)_{jj}>0$ and set $b_i:=C_i e_j/\gamma_i$.
Then $C_i=\gamma_i b_i b_i^*$ and $\|b_i\|_\infty\le1$ by the
PSD minor inequality. These factors require $O(nk)$ arithmetic
operations; omit zero inputs. Put all $k\le n^2$
columns in ${\mathsf C}$ and pad to an $n$-by-$n^2$ matrix. For an argument ${D_{\rm aux}}$,
compute ${D_{\rm aux}}{\mathsf C}$, and then its column inner products with ${\mathsf C}$. This gives
every $b_i^*{D_{\rm aux}}b_i$. A coupling with scalar weights ${\kappa}_i$ is
\[
 {\mathsf C}\operatorname{diag}({\kappa}_i\gamma_i^2(b_i^*{D_{\rm aux}}b_i)){\mathsf C}^*.
\]
The two products have shapes $(n,n,n^2)$ and $(n,n^2,n)$.
Permutation symmetry of the matrix multiplication tensor gives cost
$\widetilde O(n^\chi)$ for each. The column inner products and diagonal
scalings cost $O(nk)\le O(n^3)$.
The same products compute all quadratic forms defining $u_i,v_i,w_i,z_i$
in the potential notation. Directional differentiation changes the
weights to $\iota\beta_i^{(j)}r_i^j$, so a fixed derivative order has the same
cost. These products also form signed sums $\sum_i {\kappa}_iC_i$.
Every fixed-order differentiated KKT right-hand side is a fixed sum of
these expressions and matrix products, so its assembly has the same bound.
The primal inverse-congruence and Lagrangian-Hessian maps require a
fixed number of order-$n$ products, whose cost is smaller.

The fixed-order scalar budget derivatives are evaluated by bisection
and convergent series on the maintained truncated domain. Their
magnitudes and accuracy reciprocals are polynomial in $k,n$, so
this adds only logarithmic factors. Exact matrix arithmetic and these
scalar approximations therefore attain the prescribed accuracy
within the stated arithmetic bound.

\end{proof}

\subsection{Regular endpoint moves}\label{sec_fast_face_changes}
The next lemma bounds regular endpoint moves.

\begin{lemma}\label{fast_lem_regular_resets}
There are $\widetilde O(n)$ regular endpoint moves.
\end{lemma}
\begin{proof}
Process near-endpoint coordinates first. For a regular move to $1$, define
$C_i=a_ia_i^*$, $\nu_i:=\tr[C_i]\le1$, and $d:=1-y_i\ge\delta_{\rm end}$.
At an exact trigger, $\iota\beta_i v_i\ge d$. After changing $M$ by $dC_i$
and deleting this coupling slot, the two slacks at the old optimizer are
$(\iota\beta_i v_i-d)C_i$ and $(\iota\beta_i u_i+d)C_i$.
Define $a_0:=\iota\beta_i u_i+d$ and replace $V$ by
\[
 V_+:=(V^{-1}+a_0C_i)^{-1}.
\]
Keep $b,U$ fixed. The second constraint becomes tight, and the first
remains feasible because $V_+\preceq V$ and the remaining coupling
preserves PSD order. Sherman--Morrison and Cauchy--Schwarz give
\begin{equation}\label{fast_eq_endpoint_gain}
 \mathcal P_{\rm old}-\mathcal P_{\rm new}
 \ge\rho\tr[V-V_+]
 =\rho\frac{a_0a_i^*V^2a_i}{1+a_0v_i}
 \ge\rho\frac{a_0v_i^2}{\nu_i(1+a_0v_i)}.
\end{equation}
The last fraction increases with $a_0,v_i$ and decreases with $\nu_i$.

For an approximate trigger of absolute error at most $\delta$, the exact
score is at least $-2\delta$. Take $\delta\le\delta_{\rm end}/4$.
Then $v_i\ge\delta_{\rm end}/(2\iota)$, since $\beta_i\le1$.
The feasibility repair increases $b$ by at most $2\delta\nu_i$.
Using $a_0\ge\delta_{\rm end}$ and $\delta_{\rm end}\le1$, the gain in
Eq.~\eqref{fast_eq_endpoint_gain}, before that repair, is at least
\[
 \frac{\rho\delta_{\rm end}^3}{4\iota^2+2\iota}.
\]
Choose $\delta$ smaller than a fixed fraction of this quantity. The
repaired move decreases $\mathcal P$ by $a\rho\delta_{\rm end}^3$.
The move to $-1$ is identical with $U,V$ interchanged.
The reservoir also decreases. Since the nonnegative potential starts
at an absolute constant, and $\delta_{\rm end}^{-1}=O(\log(2n))$,
there are $\widetilde O(\rho^{-1})=\widetilde O(n)$ such moves.
All additional comparison margins are inverse polynomial.
\end{proof}

\subsection{Arithmetic cost per stage}\label{sec_fast_arithmetic}
We give the matrix algorithms and count their arithmetic operations to
the inverse-polynomial accuracies required above.

\paragraph{Inversion by Newton--Schulz iteration.}

Here is an explicit substitute for rational elimination with growing intermediate denominators. Suppose a real square matrix $A$ has polynomially bounded $\|A\|$ and $\|A^{-1}\|$. This applies to the KKT Jacobian throughout its Newton neighborhood by Eq.~\eqref{rank_one_eq_cont_4} and the Newton-neighborhood bound above. Choose a polynomial upper bound $L\ge\|A\|$, and initialize

\[
\mathsf Z_0:=A^{\top}/L^2,\qquad \mathsf Z_{{r_{\rm aux}}+1}:=\mathsf Z_{r_{\rm aux}}(2I-A\mathsf Z_{r_{\rm aux}}).
\]

With $E_{r_{\rm aux}}:=I-A\mathsf Z_{r_{\rm aux}}$, exact arithmetic gives $E_{{r_{\rm aux}}+1}=E_{r_{\rm aux}}^2$ and

\[
\|E_0\|\le1-\frac1{L^2\|A^{-1}\|^2}.
\]

Thus an inverse-polynomial residual is reached in $O(\log(2kn))$ matrix multiplications. This works for nonsymmetric and indefinite matrices; it does not rely on positive eigenvalues of $A$.

For order ${d_{\rm sys}}=O(n^2)$, these products and multiplication
by the right-hand sides cost $\widetilde O(n^{2\omega_0})$ arithmetic
operations. The same construction handles the order-$n$ primal inverses.

\paragraph{Fast products and assembly.}

Here assembly also reduces to these products. In real coordinates on the
Hermitian matrices, let $B$ have columns $\operatorname{vec}(C_i)$.
The coupling operator is represented by
$\iota B\operatorname{diag}(\beta_i)B^*$, a product padded to order $O(n^2)$.
The inverse-congruence and Lagrangian-Hessian blocks are sums of a fixed
number of Kronecker products; their $O(n^4)$ entries can be assembled
directly after order-$n$ products. Rank one makes each column of the
derivative right-hand sides a scalar multiple of $C_i$ in each block;
the scalar coefficients require $O(kn^2)$ work in total.
Let ${q_{\rm env}}(y,{\mathsf s})$ be the vector of envelope first derivatives, so that
$\nabla \mathcal F(y)={q_{\rm env}}(y,{\mathsf s}(y))$. The same observation applies to the
$k$-by-$O(n^2)$ matrix ${q_{\rm env}}_{\mathsf s}$, and ${q_{\rm env}}_y$ is diagonal.
Consequently forming ${q_{\rm env}}_y-{q_{\rm env}}_{\mathsf s}{\mathsf R}^{-1}{\mathcal E_{\rm KKT}}_y$ uses a fixed
number of products padded to order $O(n^2)$, together with $O(n^4)$ work.
Assembly, Newton--Schulz inversion, and all Hessian entries therefore cost
$\widetilde O(n^{2\omega_0})$ per curvature event, in the arithmetic model.

For a barrier stage, let $\mathcal A_j$ be the linear part of ${\mathsf a}\mapsto {D_{\rm sl}}_j({\mathsf a})$. Its Hessian is
\[
 \nabla^2 f({\mathsf a})=\sum_j\mathcal A_j^*
 {\mathcal T_{\rm cong}}_{{D_{\rm sl}}_j({\mathsf a})^{-1}}\mathcal A_j,
 \qquad{\mathcal T_{\rm cong}}_{\mathsf E}(D)={\mathsf E}D{\mathsf E}.
\]
After real Hermitian vectorization all maps have order $O(n^2)$. Their coefficient matrices and Kronecker blocks have $O(n^4)$ entries. Fast multiplication forms the initial Hessian and its Newton inverse in $\widetilde O(n^{2\omega_0})$ per face. The ordinary implementation recomputes this inverse at each stage. There are a constant number of blocks. A rational real/imaginary basis has bounded diagonal Gram weights (one or two), so no exact irrational basis or growing-denominator orthogonalization is needed.

\paragraph{A curvature witness by normalized matrix squaring.}
This construction avoids a cubic-time Jacobi sweep and requires no
separation between adjacent eigenvalues. Let $A$ approximate the symmetric scaled Hessian
$G^{-1/2}\nabla^2\mathcal FG^{-1/2}$. Choose a known polynomial bound $L\ge\max\{1,\|A\|\}$,
and let $\delta$ be a sufficiently small inverse-polylogarithmic threshold.
If the gradient test fails, the existing error margins ensure
$\lambda_{\min}(A)\le-2\delta$.
Define
\[
 \mathsf T_0:=I-A/(2L),\qquad {D_{\rm pow}}_0:=\mathsf T_0/\|\mathsf T_0\|_F,
 \qquad {D_{\rm pow}}_{j+1}:={D_{\rm pow}}_j^2/\|{D_{\rm pow}}_j^2\|_F.
\]
In exact arithmetic ${D_{\rm pow}}_{r_{\rm aux}}=\mathsf T_0^q/\|\mathsf T_0^q\|_F$, where $q:=2^{r_{\rm aux}}$.
Choose $q$ as the least power of two satisfying
\[
 q\ge(4L/\delta)\log(8kL/\delta),
 \qquad {r_{\rm aux}}=O(\log(2kn)).
\]
An eigenvalue of $A$ at most $-2\delta$ gives an eigenvalue of $\mathsf T_0$
at least $1+\delta/L$. For eigenvalues of $A$ at least $-\delta$,
the corresponding ratio is at most
\[
 \frac{1+\delta/(2L)}{1+\delta/L}
 \le1-\frac{\delta}{4L}.
\]
The total squared Frobenius weight of these latter eigenvectors in ${D_{\rm pow}}_{r_{\rm aux}}$
is therefore at most
\[
 k\exp(-q\delta/(2L))\le\delta/(8L).
\]
It follows that
\[
 \tr[{D_{\rm pow}}_{r_{\rm aux}}A{D_{\rm pow}}_{r_{\rm aux}}]\le-3\delta/4,
 \qquad \|{D_{\rm pow}}_{r_{\rm aux}}\|_F=1.
\]
Writing this trace as the sum of the column quadratic forms shows that
a column has negative Rayleigh quotient. More quantitatively, discard
columns of squared norm less than $\delta/(16kL)$; their total contribution
in absolute value is at most $\delta/16$.
Some retained column has Rayleigh quotient at most $-\delta/2$.
Compute all column quadratic forms using one product $A{D_{\rm pow}}_{r_{\rm aux}}$, then scan
them. This costs $\widetilde O(k^{\omega_0})\le\widetilde O(n^{2\omega_0})$.

The scalar normalizations can be evaluated to inverse-polynomial
accuracy within the same arithmetic cost. To bound their errors,
define $\mathcal N_s(X):=X^s/\|X^s\|_F$ for nonzero symmetric $X$.
Differentiating the power and its normalization gives
\[
 \|\mathrm D\mathcal N_s(X)[E]\|_F
 \le\frac{s}{\|X\|}\|E\|_F
 \le\frac{s\sqrt k}{\|X\|_F}\|E\|_F.
\]
Thus $\mathcal N_s$ is $2s\sqrt k$-Lipschitz on a segment within
Frobenius distance $1/2$ of a unit-Frobenius-norm symmetric matrix.
Symmetrize each computed iterate. If its initial and local errors are
at most $\eta<1/2$, propagating them through the remaining exact powers
$q,q/2,\ldots,2$ gives
\[
 \|\widehat D_{r_{\rm aux}}-{D_{\rm pow}}_{r_{\rm aux}}\|_F
 \le8q\sqrt k\eta.
\]
Choose $0<\epsilon\le\delta/(100L)$ and
$\eta:=\epsilon/(8q\sqrt k)$. Every normalization is polynomially
conditioned, since a computed iterate of Frobenius norm at least
$1-\eta$ has squared matrix of Frobenius norm at least
$(1-\eta)^2/\sqrt k\ge1/(4\sqrt k)$.
The final quadratic trace changes by at most $L(2+\epsilon)\epsilon$.
The retained-column argument therefore still gives a Rayleigh quotient
at most $-\delta/4$ with conservative comparison margins. Checking the
selected quotient directly certifies the witness, and the Hessian
approximation margin transfers it to the true potential. Each scalar evaluation uses only polylogarithmically many arithmetic
operations, so the arithmetic bound remains $\widetilde O(n^{2\omega_0})$.

\subsection{Fast deferred rounding}\label{sec:shared:fast-rounding}
The rounding rule of Lemma~\ref{round_lem_round} admits a faster
implementation of its trace-exponential comparisons.
\begin{lemma}[Fast evaluation in deferred rounding]\label{sr_lem_round}
Under the hypotheses of Lemma~\ref{round_lem_round}, the deterministic
rounding rule and discrepancy allowance can be obtained using
$\widetilde O(k n^{\omega_0})$ arithmetic operations.
\end{lemma}
\begin{proof}
Retain the heavy/small split, scalar coefficients $G_i^\pm$, and
conditional trace-exponential estimator in that lemma. Its proof gives
known bounds $O(\log(2n))$ for both Hermitian exponential arguments at
every accepted prefix and either next candidate. For example,
$\|\theta E\|\le\log(4n)+\ell/16$ follows from positivity of the
remaining $G_i^\pm$ and the estimator bound. A candidate changes this
by at most $1/2$. A conservative integer $M=4\ell+4$ therefore bounds
all argument norms, including the stipulated evaluation errors.

For $\|A\|\le M$, evaluate $e^A$ by the factorial Taylor recurrence
\[
 T_0=I,\quad T_j=T_{j-1}A/j,\quad E_J=\sum_{j=0}^JT_j.
\]
Its operator tail is at most
$e^M M^{J+1}/(J+1)!$. Since $e<3$ and
$(J+1)!\ge((J+1)/e)^{J+1}$, taking, for example,
\[
 J=\lceil16M+2\log_2(512\max\{1,k\}n)\rceil
\]
makes the trace tail smaller than $1/(256\max\{1,k\})$.
This uses $J=O(\log(2kn))$ order-$n$ multiplications. A recurrence
error inserted at order $i$ is propagated by a product bounded by
$M^{j-i}i!/j!\le e^M$; summing all errors adds only a polynomial
factor in $n,k$. The exact term and partial-sum norms are at most
$e^M$, also polynomial. Exact arithmetic products and inverse-polynomial scalar
approximations therefore enclose each candidate estimator to the
original additive tolerance. The number of products is logarithmic.

The rank-one formula for $G_i^\pm$ uses only scalar exponentials and
logarithms, evaluated once per coefficient. For small traces use the
equivalent divided series from the source. Maintain the two remaining
matrix sums and the signed prefix by rank-one additions, costing
$O(kn^2)$ altogether; their cumulative errors over at most $k$ updates
are bounded by $k$ times the prescribed unit error. Choosing that unit
smaller by this known polynomial keeps the estimator enclosure valid.
There is no full reconstruction for every candidate and no multiplying
of conditioning factors over the sequence.

A constant number of candidate trace evaluations per coordinate now
costs $\widetilde O(n^{\omega_0})$ instead of $\widetilde O(n^3)$.
The scalar work and rank-one updates are smaller. Choose the smaller
candidate using the same conservative error allowance as in
Lemma~\ref{round_lem_round}. Its conditional expectation argument and
all errors then remain unchanged, including
\[
 \|\sum_i(\sigma_i-t_i)C_i\|
 \le21{\varepsilon_{\rm round}}/128<{\varepsilon_{\rm round}}/2.
\]
The matrix algebra uses complex Hermitian inputs directly; real
bilinear products compute their real and imaginary parts at constant
factor overhead. This proves the arithmetic bound.
\end{proof}

\subsection{Smoothed potential and retained geometry}\label{soft_sec_geometry}
We analyze a retained instance of order $n$ with $k\le n^2$ active
coordinates. The same estimates apply after constant-factor padding
whenever the retained count is at most the square of the padded order. The retained matrices are PSD and rank one, and satisfy
\[
 \sum_iV_i^2\preceq I,\qquad
 \lVert \sum_i a_iV_i\rVert_{\mathrm F}\le\lVert a\rVert_2.
\]
The single preprocessing residual has operator norm at most
$\eta_{\rm ps}=10^{-7}$. The deterministic grouped routine in
Lemma~\ref{rank_one_lem_grouped_partial_signing} is available for fallback.
To see the compatibility with the new potential, let $y^0$ be the
partial signing and $e_0=\sum_i y_i^0C_i$. Keep the true retained
matrices $V_i=C_i$ and subtract $e_0$ from the frozen contribution.
Then the modeled sum starts at zero, and its difference from the true
signed sum remains exactly $e_0$. The residual is charged once at the
end. The grouped procedure changes coefficients only; it preserves
rank one, positivity, and the variance bound on the retained inputs.
The Frobenius bound above follows by considering the nonnegative Gram
matrix $J_{ij}=\tr[V_iV_j]$: with $\nu_i=\tr[V_i]>0$,
$J\nu\le\nu$ follows from $\sum_iV_i^2\preceq I$, so $\|J\|\le1$.
The same common-grid construction preserves the residual allowance.
Stop directly if $k=0$. Fix
\[
 \iota=14779/2500,\quad {\varepsilon_{\rm scale}}=10^{-8},\quad\rho={\varepsilon_{\rm scale}}/n,\quad
 \gamma=10^{-6}/k,\quad {\varepsilon_{\rm round}}={\varepsilon_{\rm discard}}=10^{-6}.
\]
Use the degree-$20$ budget $\beta_{\rm fast}$ from Section~\ref{sec_rank_one_proof},
$\beta(y)=(1-y^2)^{97/200}f(|y|)$. These reference estimates take $\gamma$ fixed. The final implementation
uses the phase values in Eq.~\eqref{sp_eq_lambda}, keeping $\gamma$ fixed
within every move and attenuation path. Their additional logarithmic
factors are absorbed in $P_n$. Define ${L_{\rm log}}=\lceil\log_2(4n)\rceil$ and retain a
dyadic ${\delta_{\rm end}}$ with
\[
 \frac{{\varepsilon_{\rm round}}^2}{8192{L_{\rm log}}}<{\delta_{\rm end}}\le\frac{{\varepsilon_{\rm round}}^2}{4096{L_{\rm log}}},
 \qquad \tau_{\rm s}=10^{-8}/{L_{\rm log}}.
\]
The smoothing temperature $\tau_{\rm s}$ is unrelated to the budget parameter
$\tau$ in the scalar certificate. At endpoint distance at least ${\delta_{\rm end}}/2$,
\begin{equation}\label{soft_eq_budget}
 -\beta''>1/2,\quad |\beta^{(j)}|\le P_n(-\beta''),\quad
 |\beta^{(j)}|^2/\beta\le P_n(-\beta'')\quad(1\le j\le3).
\end{equation}
Second derivatives are continuous at zero; higher derivatives below are
one-sided there.

On a face, let $M(y)=F+\sum_i y_iV_i$ and
$\mathcal K_y(D)=\iota\sum_i\beta(y_i)V_iDV_i$. With ${\mathsf P}=(U,V)$, $E=(I,I)$, define
\[
 \mathcal C({\mathsf P},y)=(U^{-1}+M+\mathcal K_y(V),\ V^{-1}-M+\mathcal K_y(U)).
\]
Pairs are identified with block-diagonal matrices, and traces and
Frobenius inner products sum over the two blocks. Define
\begin{equation}\label{soft_eq_potential}
 \begin{split}
 f_{\tau_{\rm s}}(C)&=\tau_{\rm s}\log\tr[\exp(C/\tau_{\rm s})],\\
 \mathcal P_{\rm s}(y)&=\min_{U,V\succ0}\{\rho\tr[U+V]+f_{\tau_{\rm s}}(\mathcal C({\mathsf P},y))\},\\
 \mathcal F_{\rm s}(y)&=\mathcal P_{\rm s}(y)+\gamma\sum_i\beta(y_i).
 \end{split}
\end{equation}
For the original potential $\mathcal P$,
\begin{equation}\label{soft_eq_sandwich}
 \mathcal P\le \mathcal P_{\rm s}\le \mathcal P+\tau_{\rm s}\log(2n)\le \mathcal P+10^{-8},\qquad
 \mathcal P_{\rm s}\ge\lVert M\rVert.
\end{equation}
Indeed these are the eigenvalue/log-sum-exp inequalities before taking
the infimum over ${\mathsf P}$.

The objective is convex: inversion is operator convex, and
$f_{\tau_{\rm s}}$ is convex and PSD-order increasing. Bounded sublevels are
compact inside the positive cone, so a minimizer exists. Its Gibbs state
\[
 {\mathsf Q}=(W,Z)=\exp(\mathcal C/\tau_{\rm s})/\tr[\exp(\mathcal C/\tau_{\rm s})]
\]
satisfies
\begin{equation}\label{soft_eq_dual}
 \tr[{\mathsf Q}]=1,\quad {\mathscr J}{\mathsf Q}=\rho E,\quad
 W=U\mathcal K(Z)U+\rho U^2,\quad Z=V\mathcal K(W)V+\rho V^2,
\end{equation}
where
\[
 {\mathscr J}(D_U,D_V)=(U^{-1}D_UU^{-1}-\mathcal K(D_V),\
                    V^{-1}D_VV^{-1}-\mathcal K(D_U)).
\]
The positive-map argument preceding Eq.~\eqref{rank_one_eq_response}
makes ${\mathscr J}$ invertible with positive inverse.
It uses the strict dual identities, not hard tight constraints.
The objective Hessian in ${\mathsf P}$ is positive definite, as shown below;
thus the optimizer is unique and is $C^2$ on every open face, including
coordinate zeros.

Uniformly at roots on bounded-model sublevels, scalar feasible matrices
and Eq.~\eqref{soft_eq_dual} give
\begin{equation}\label{soft_eq_rootbounds}
 \begin{gathered}
 U,V\succeq a_0I,\quad \lVert U^{-1}\rVert,\lVert V^{-1}\rVert,\lVert \mathcal K(U)\rVert,\lVert \mathcal K(V)\rVert\le C,\\
 W\succeq\rho U^2,\quad Z\succeq\rho V^2,\quad
 \rho\tr[U^2+V^2]\le1,\quad W,Z\succeq a_0^2\rho I.
 \end{gathered}
\end{equation}
Here and below zero inputs may be omitted. The same argument applies
when the modeled sum and coupling are scaled by a common factor in $[0,1]$.

Let ${\mathscr E}$ be the Hessian in ${\mathsf P}$ of $\langle \mathsf Q,\mathcal C\rangle$ with ${\mathsf Q}$ fixed.
Define
\[
 \widehat D_U=U^{-1/2}D_UU^{-1/2},\quad T_U=U^{-1/2}WU^{-1/2},\quad
 e(D)^2=\langle D,{\mathscr E}D\rangle=2\sum_{A=U,V}\tr[T_A\widehat D_A^2].
\]
The dual identities and the bounds above give
\begin{equation}\label{soft_eq_energybounds}
 \begin{gathered}
 {\mathscr E}\succeq a n^{-3/2}I,\quad {\mathscr E}\preceq CI,\quad
 \lVert D\rVert_{\mathrm F}\le C n^{3/4}e(D),\\
 \lVert \widehat D_A\rVert\le C\sqrt ne(D),\quad
 a/n\le\lambda_{\min}(T_A),\quad \sum_A\tr[T_A]\le C.
 \end{gathered}
\end{equation}
The weighted order norm is
\[
 q(B)=\inf\{\langle \mathsf Q,A\rangle:A\succeq0,\ -A\preceq B\preceq A\}
      =\sum_A\lVert {\mathsf Q}_A^{1/2}B_A{\mathsf Q}_A^{1/2}\rVert_*.
\]
The equality follows by congruence and minimizing the trace of a PSD
majorant. In particular $q(\mathcal C_{{\mathsf P}{\mathsf P}}[D,D])=e(D)^2$.

\paragraph{Entropy metric and an inexpensive root inverse}
On the trace-zero space $\mathcal T=\{{\mathsf B}:\tr[{\mathsf B}]=0\}$ define
\[
 B_{\mathsf Q}=\tau_{\rm s}\Pi_{\mathcal T} D\log_{\mathsf Q}|_{\mathcal T},\qquad
 \mathcal E({\mathsf B})^2=\langle \mathsf B,B_{\mathsf Q}{\mathsf B}\rangle.
\]
Its inverse is the Gibbs covariance operator
\begin{equation}\label{soft_eq_q}
 Q_{\mathsf Q}(B)=\frac1{\tau_{\rm s}}(\int_0^1{\mathsf Q}^tB{\mathsf Q}^{1-t}\d t-\tr[{\mathsf Q}B]{\mathsf Q}).
\end{equation}
Adding a scalar multiple of $E$ to its argument has no effect. In a basis
diagonalizing ${\mathsf Q}$, the integral has logarithmic-mean coefficients
$\mathsf L(z_i,z_j)\le(z_i+z_j)/2$. Hence
\begin{equation}\label{soft_eq_variance}
 \langle B,Q_{\mathsf Q}B\rangle\le\tau_{\rm s}^{-1}\tr[{\mathsf Q}B^2],\qquad
 |\langle B,\mathsf B\rangle|\le\tau_{\rm s}^{-1/2}\sqrt{\tr[{\mathsf Q}B^2]}\mathcal E({\mathsf B}).
\end{equation}
The inverse relation also follows by differentiating the normalized
matrix exponential. At roots, $B_{\mathsf Q}$ has eigenvalues between $\tau_{\rm s}$
and $P_n n$.

The following stronger relative bound will be used repeatedly:
\begin{equation}\label{soft_eq_relv}
 \lVert {\mathsf Q}^{-1/2}{\mathsf B}{\mathsf Q}^{-1/2}\rVert
 \le(\tau_{\rm s}\lambda_{\min}({\mathsf Q}))^{-1/2}\mathcal E({\mathsf B})
 \le P_n\sqrt n\mathcal E({\mathsf B}).
\end{equation}
Indeed, if ${\mathsf B}={\mathsf Q}^{1/2}\widehat {\mathsf B}{\mathsf Q}^{1/2}$, the energy coefficient of
$|\widehat {\mathsf B}_{ij}|^2$ is $\tau_{\rm s} z_i z_j/\mathsf L(z_i,z_j)$,
which is at least $\tau_{\rm s}\min(z_i,z_j)$.

For completeness the root preconditioner has a dimension-free bound
apart from $\tau_{\rm s}^{-1}$. Complete positivity applied to
$(\begin{smallmatrix}D_VV^{-1}D_V&D_V\\D_V&V\end{smallmatrix})\succeq0$
gives
$\mathcal K(D_V)^2\preceq\lVert \mathcal K(V)\rVert\mathcal K(D_VV^{-1}D_V)$.
Pair with $W$ and use $\mathcal K(W)\preceq V^{-1}ZV^{-1}$. Together with
$U^{-2}\preceq\lVert U^{-1}\rVert U^{-1}$ and
$(A-B)^2\preceq2A^2+2B^2$, this proves
\begin{equation}\label{soft_eq_cp}
 \tr[{\mathsf Q}({\mathscr J}D)^2]\le K e(D)^2,\qquad
 K=\max\{\lVert U^{-1}\rVert+\lVert \mathcal K(U)\rVert,\
              \lVert V^{-1}\rVert+\lVert \mathcal K(V)\rVert\}\le C.
\end{equation}
Consequently the primal Schur complement obeys
\begin{equation}\label{soft_eq_precond}
 {\mathscr E}\preceq J_{\rm s}:={\mathscr E}+{\mathscr J}Q_{\mathsf Q}{\mathscr J}\preceq(1+K/\tau_{\rm s}){\mathscr E}.
\end{equation}
This also proves the strict convexity and regularity asserted earlier.

To apply ${\mathscr E}^{-1}$, its $U$ equation becomes the Lyapunov equation
\[
 \widehat D_UT_U+T_U\widehat D_U=U^{1/2}F_UU^{1/2}.
\]
Diagonalize the order-$n$ $T_U$ and divide entry $(i,j)$ by $t_i+t_j$;
all denominators are inverse-polynomial. The $V$ block is identical.
A gap-free approximate Hermitian diagonalization, with residual and
orthogonality bounds, costs $\widetilde O(n^3)$. It requires no accurate choice of
individual eigenvectors at a repeated eigenvalue. Evaluating $Q_{\mathsf Q}$ uses
the same kind of decomposition and scalar logarithmic means. For nearby
eigenvalues use the divided exponential series, not cancellation of
nearly equal logarithms.

The rectangular products of Lemma~\ref{fast_lem_operator_cost},
or square products on groups of at most $n$ factors, give cost
\begin{equation}\label{soft_eq_fcost}
 F(n,k):=\widetilde O(\min\{n^\chi,n^3+(k+n)n^{\omega_0-1}\}).
\end{equation}
This covers coupling actions, signed sums and complete scalar score
scans. In particular, an $n$-by-$n^2$ factor matrix reduces these
actions and their adjoints to products of shapes $(n,n,n^2)$ and
$(n,n^2,n)$. Tensor permutation preserves the exponent $\chi$ and
the fixed recursive error bounds. The $\widetilde O(n^3)$ spectral work is
smaller than $\widetilde O(n^\chi)$. It applies to a rational representation $V_i=c_ia_ia_i^*$ without
extracting exact square roots. By Eq.~\eqref{soft_eq_precond}, a Chebyshev solve
for $J_{\rm s}$ uses only polylogarithmically many ${\mathscr E}^{-1}$-preconditioned
actions. Similarity to its symmetric normalization has polynomial
condition number by Eq.~\eqref{soft_eq_energybounds}. Second-kind Chebyshev
bounds control accumulated application errors: an error inserted at
step $i$ is multiplied at step $J$ by a scalar ratio at most one
and a second-kind polynomial of norm at most a polynomial similarity
factor times $J-i+1$. Taking the local action tolerance below the
target divided by this polynomial attains the required arithmetic
accuracy.

\paragraph{Covariance and endpoint moves}
A regular move to $+1$ changes the two $\mathcal C$ blocks, at fixed ${\mathsf P}$,
by
\[
 (1-y_i-\iota\beta_i v_i)V_i,\qquad (-(1-y_i)-\iota\beta_i u_i)V_i.
\]
Both are PSD-nonpositive when the endpoint test holds. Order
monotonicity of $f_{\tau_{\rm s}}$ proves nonincrease. The other sign is symmetric.
A deferred removal preserves $M$ and removes a PSD coupling term.
Conservative score errors are repaired by a uniform $eE$, whose cost is
exactly $e$ because $f_{\tau_{\rm s}}(C+eE)=f_{\tau_{\rm s}}(C)+e$. The same reservoir
margin therefore pays for the endpoint-score errors.

We do not impose $\mathcal C=bE$ on the new root. Instead the inverse
function theorem for ${\mathscr J}$ supplies a local comparison path with
$\mathcal C({\mathsf P}(t),y+tr)=\mathcal C({\mathsf P}(0),y)$ fixed. Its first-order response
satisfies Eq.~\eqref{rank_one_eq_response}. Along it $f_{\tau_{\rm s}}$ is
constant and $\rho\tr[{\mathsf P}(t)]+f_{\tau_{\rm s}}(\mathcal C({\mathsf P}(0),y))$ majorizes
$\mathcal P_{\rm s}$ with first-order equality. Its second variation is exactly
the inversion-energy and mixed-coupling expression preceding
Eq.~\eqref{rank_one_eq_curvature}. Lemma~\ref{rank_one_lem_covariance}
uses that response equation, the dual
identities, rank one and lightness, all valid here. The unchanged exact
scalar certificate therefore yields the same strict generator surplus.

Define $u_i:=\tr[UV_i]$, $v_i:=\tr[VV_i]$,
$w_i:=\tr[WV_i]$, $z_i:=\tr[ZV_i]$, and $t_i:=v_iw_i+u_iz_i$.
Thus $p_i=1+\iota\beta_i'v_i$ and $q_i=1-\iota\beta_i'u_i$
are the two response coefficients. Define
\[
 \varpi_i=(-\beta_i'')(\iota t_i+\gamma),\quad G=\operatorname{diag}(\varpi_i),\quad
 A_r=\iota\sum_i(-\beta_i'')t_ir_i^2,\quad\theta=r^{\top}Gr.
\]
At a light state $\gamma/4\le \varpi_i\le P_n$ and
$\theta\ge1/(P_nk)$. The drift estimate in Lemma~\ref{fast_lem_dichotomy}
$(C_{2,i}M_i)^2\le P_n n \varpi_i$ follows from
$w_i\ge\rho u_i^2/\operatorname{tr}(V_i)$ and its partner, so remains
valid. Dividing the strict generator estimate by its nonzero covariance
proves the numerically separated alternative
\begin{equation}\label{soft_eq_alternative}
 |(\mathcal F_{\rm s})_i|\ge P_n^{-1}\sqrt{\varpi_i/n}\quad\text{for some }i,
 \quad\text{or}\quad r^{\top}\nabla^2\mathcal F_{\rm s}r\le-P_n^{-1}\theta.
\end{equation}
This is the proof of Lemma~\ref{fast_lem_dichotomy} applied to the
fixed-constraint comparison path for $\mathcal P_{\rm s}$.

\paragraph{Reservoir phases without a new discrepancy allowance.}
Let $k_0$ be the initial retained count, $\Lambda=10^{-6}$, and
$L_0=1+\lceil\log_2\max\{2,k_0\}\rceil$. At the start of a phase
record the current active count $K$ and set
\begin{equation}\label{sp_eq_lambda}
 {\gamma}=\frac{\Lambda}{L_0K}.
\end{equation}
Keep this value fixed until the active count is at most $K/2$, then
start the next phase at the new count. Stop if it is zero. Thus at
each light state in a phase, $K/2<k\le K$. There are at most $L_0$
phases, and ${\gamma}\ge\Lambda/(L_0k_0)$ throughout.

A phase change modifies only the scalar reservoir, not $\mathcal P_{\rm s}$, its
optimizer, the signed sum, or any coupling coefficient. Since
$0\le\beta\le1$, each increase in the potential is at most
${\gamma}_{\rm new}k\le\Lambda/L_0$. The initial reservoir is also
at most $\Lambda/L_0$. Therefore
\begin{equation}\label{sp_eq_injections}
 \text{initial reservoir}+\text{all positive phase jumps}\le\Lambda.
\end{equation}
This replaces, rather than adds to, the original initial reservoir
allowance $10^{-6}$. All other moves decrease their current phase
potential. Their total certified decrease is bounded by the same
constant as before, with Eq.~\eqref{sp_eq_injections} accounting for
all jumps. Phase counters and updates cost only linear scalar work.

\subsection{Joint-energy Newton continuation}\label{soft_sec_newton}
We use the joint primal and entropy metrics to control a Newton
correction, and keep the inverse at the base root fixed during each
short continuation step.

Work with $\tr[{\mathsf Q}]=1$ and residual
\begin{equation}\label{soft_eq_residual}
 {\mathcal E_{\rm KKT}}_y({\mathsf P},{\mathsf Q})=(\rho E-{\mathscr J}_{\mathsf P}{\mathsf Q},\ \Pi_{\mathcal T}(\tau_{\rm s}\log {\mathsf Q}-\mathcal C({\mathsf P},y))).
\end{equation}
Its derivative on matrix pairs ${\mathsf A}$ and trace-zero pairs ${\mathsf B}$ is
\begin{equation}\label{soft_eq_jointa}
 \mathbb A=\begin{pmatrix}{\mathscr E}&-{\mathscr J}|_{\mathcal T}\\ \Pi_{\mathcal T} {\mathscr J}&B_{\mathsf Q}\end{pmatrix}.
\end{equation}
The cross terms are skew-adjoint. At a root let
$\mathbb D=\operatorname{diag}({\mathscr E},B_{\mathsf Q})$ and $\|({\mathsf A},{\mathsf B})\|_{\mathbb D}^2=e({\mathsf A})^2+\mathcal E({\mathsf B})^2$.
Then
\begin{equation}\label{soft_eq_coercive}
 \mathbb A+\mathbb A^*=2\mathbb D,\qquad
 \lVert \mathbb D^{1/2}\mathbb A^{-1}\mathbb D^{1/2}\rVert\le1.
\end{equation}
The second assertion follows by pairing $\mathbb A {y_{\rm aux}}={b_{\rm loc}}$ with ${y_{\rm aux}}$ and
Cauchy--Schwarz. Norms of $\mathbb D^{1/2}$ and its inverse are polynomial.

A uniform Newton neighborhood ensures that approximate optimizers can
be corrected during continuation.

Write ${\mathsf s}=({\mathsf P},{\mathsf Q})$ for the optimization variables.

\begin{lemma}[Joint-energy Newton ball]\label{soft_lem_ball}
At every bounded-model root there is a Newton ball
\[
 \|{\mathsf s}-{\mathsf s}_*\|_{\mathbb D_*}\le a/(P_n\sqrt n)
\]
inside the positive cones and the trace-one affine space. On a smaller
such ball, exact Newton iteration satisfies
\begin{equation}\label{soft_eq_energynewton}
 \|{\mathsf s}_{\rm new}-{\mathsf s}_*\|_{\mathbb D_*}
 \le P_n\sqrt n\|{\mathsf s}-{\mathsf s}_*\|_{\mathbb D_*}^2.
\end{equation}
Moreover
\begin{equation}\label{soft_eq_jacvariation}
 \lVert \mathbb D_*^{-1/2}(\mathbb A({\mathsf s})-\mathbb A({\mathsf s}_*))\mathbb D_*^{-1/2}\rVert
 \le P_n\sqrt n\|{\mathsf s}-{\mathsf s}_*\|_{\mathbb D_*}.
\end{equation}
All comparisons and the Newton radius are uniform in the face and in
real or complex Hermitian inputs.
\end{lemma}
\begin{proof}
The primal relative bound in Eq.~\eqref{soft_eq_energybounds} and the dual one
in Eq.~\eqref{soft_eq_relv} keep $U,V,{\mathsf Q}$ within fixed Loewner factors of their
root values in the stated ball. In particular ${\mathsf Q}\succeq a/n$, and
$W\succeq(a/n)U$, $Z\succeq(a/n)V$, throughout it: these last inequalities
follow by comparing $W,U$ separately to their root values. Hence the
bounds $T_A\succeq a/n$ and $\sum\tr[T_A]\le C$ remain valid there.

We detail the differential estimates giving metric comparison, rather
than assuming it. For fixed test $D$, variation in ${\mathsf P}$ gives
$\langle \mathsf Q,\mathcal C_{{\mathsf P}{\mathsf P}{\mathsf P}}[\Delta {\mathsf P},D,D]\rangle$. Its three normalized inverse
words are bounded by
\[
 C\sqrt ne(\Delta {\mathsf P})e(D)^2.
\]
For an exterior $\widehat{\Delta {\mathsf P}}$ use weighted Cauchy--Schwarz and
$\tr[T\widehat D^4]\le\|\widehat D\|^2\tr[T\widehat D^2]$.
For the middle word use $\widehat D T\widehat D\succeq0$.
Variation in ${\mathsf Q}$ contributes
$\langle \Delta {\mathsf Q},\mathcal C_{{\mathsf P}{\mathsf P}}[D,D]\rangle$, of absolute value at most
$\|{\mathsf Q}^{-1/2}\Delta {\mathsf Q} {\mathsf Q}^{-1/2}\|e(D)^2$.
Thus $|\d {\mathscr E}[D,D]|\le P_n\sqrt n\|(\Delta {\mathsf P},\Delta {\mathsf Q})\|_{\mathbb D}e(D)^2$.

For the entropy block, define $R_q=({\mathsf Q}+qI)^{-1}$. The resolvent formula gives
\[
 \langle \mathsf B,D\log_{\mathsf Q}[{\mathsf B}]\rangle=\int_0^\infty\tr[R_q {\mathsf B} R_q {\mathsf B}]\d q,
\]
and differentiation in $\Delta {\mathsf Q}$ bounds the derivative of this form by
\begin{equation}\label{soft_eq_entropyrelative}
 2\|{\mathsf Q}^{-1/2}\Delta {\mathsf Q} {\mathsf Q}^{-1/2}\|
       \langle \mathsf B,D\log_{\mathsf Q}[{\mathsf B}]\rangle.
\end{equation}
Indeed the differentiated integrand is minus twice the real trace of
$(R_q^{1/2}\Delta {\mathsf Q}R_q^{1/2})(R_q^{1/2}{\mathsf B}R_q^{1/2})^2$.
The first factor has operator norm at most the displayed relative norm,
and the squared factor is PSD. No commutation is used.

Finally, the derivative of the cross block ${\mathscr J}$ in $\Delta {\mathsf P}$ pairs with
${\mathsf B}$ as $\langle \mathsf B,\mathcal C_{{\mathsf P}{\mathsf P}}[\Delta {\mathsf P},D]\rangle$. Its $q$-majorant is at most
$Ce(\Delta {\mathsf P})e(D)$, by the matrix Young inequality.
Eq.~\eqref{soft_eq_relv} bounds the pairing by
$P_n\sqrt n\mathcal E({\mathsf B})e(\Delta {\mathsf P})e(D)$.
These prove the normalized differential bound for $\mathbb A$, and the
relative derivative bounds for both diagonal metric blocks.

A first-exit argument along the straight segment from ${\mathsf s}_*$ keeps
$\mathbb D_{\mathsf s}$ within constant factors of $\mathbb D_*$ in the stated ball: until
exit, integrate the preceding differential bounds, and choose $a$ small.
The direct Loewner comparisons already ensure positivity and the
$T_A$ bounds needed along the segment. Integrating the cross estimates
now gives Eq.~\eqref{soft_eq_jacvariation}. By Eq.~\eqref{soft_eq_coercive} and a Neumann
perturbation, the inverse Jacobian in the fixed root metric has norm at
most two. The integral Taylor remainder for ${\mathcal E_{\rm KKT}}_y$ proves
Eq.~\eqref{soft_eq_energynewton}. These arguments use only two derivatives of
$\mathcal C$ in ${\mathsf P}$ and the smooth log map; coordinate zeros do not
enter this fixed-parameter Newton argument.
\end{proof}

Let ${\mathsf A}={\mathsf P}'$ and ${\mathsf B}={\mathsf Q}'$ on a path of exact roots. Differentiating the two
root equations gives
\[
 {\mathscr J}{\mathsf A}+\tau_{\rm s} D\log_{\mathsf Q}[{\mathsf B}]+{d_{\rm norm}}E=F_1,\quad {\mathscr J}{\mathsf B}-{\mathscr E}{\mathsf A}=F_2,\quad\tr[{\mathsf B}]=0,
 \qquad F_1=\mathcal C_t,\quad F_2=\mathcal A'{\mathsf Q}.
\]
Pairing the equations gives
\begin{equation}\label{soft_eq_forced}
 e({\mathsf A})^2+\mathcal E({\mathsf B})^2=\langle F_1,\mathsf B\rangle-\langle F_2,\mathsf A\rangle.
\end{equation}
The scalar ${d_{\rm norm}}$ is only the log-normalization derivative and need not be
stored as an unknown.

Controlling metric and Jacobian variation allows correction systems
along a short root path to reuse a fixed approximate inverse.

\begin{lemma}[Anchored transport and executable correction]\label{soft_lem_transport}
Along a unit-direction local root path, write
$a(t)=e({\mathsf A})+\mathcal E({\mathsf B})$. The metric and joint Jacobian satisfy
\begin{equation}\label{soft_eq_transportrate}
 \|\mathbb D^{-1/2}\mathbb D'\mathbb D^{-1/2}\|,
 \ \|\mathbb D^{-1/2}\mathbb A'\mathbb D^{-1/2}\|
 \le P_n\sqrt n(a(t)+\sqrt{\theta(t)}).
\end{equation}
For a logarithmic data homotopy $M_t=e^tM_0$,
$\mathcal K_t=e^t\mathcal K_0$, replace $\sqrt\theta$ by $1$. If the integral of the right side over a path is
sufficiently small, its base root is a valid Newton start at the end.
A fixed approximate inverse at that base solves every correction system
along the path in $\widetilde O(F(n,k))$ work. It is an implicitly applied inverse,
not a stored dense matrix.
\end{lemma}
\begin{proof}
The metric derivatives and the optimizer-dependent part of $\mathbb A'$
are bounded in Lemma~\ref{soft_lem_ball}. The additional explicit coupling
term in ${\mathscr J}'$ pairs with ${\mathsf B}_0\in\mathcal T$ and $D$ as
$\langle {\mathsf B}_0,\mathcal A'D\rangle$. Lemma~\ref{fast_lem_directional},
specifically Eq.~\eqref{fast_eq_weightedcoupling}, gives
$q(\mathcal A'D)\le P_n\sqrt\theta e(D)$; use Eq.~\eqref{soft_eq_relv} on
${\mathsf B}_0$. This proves Eq.~\eqref{soft_eq_transportrate}. On the homotopy, the
relative coupling bound and the dual identities instead give
$q(\mathcal A'D)\le Ce(D)$.

If ${a_{\rm int}}$ is the accumulated bound, integration gives
$e^{-{a_{\rm int}}}\mathbb D_0\preceq\mathbb D_t\preceq e^{a_{\rm int}}\mathbb D_0$ and
\[
 \|\mathbb D_0^{-1/2}(\mathbb A_t-\mathbb A_0)\mathbb D_0^{-1/2}\|\le e^{a_{\rm int}} {a_{\rm int}}.
\]
Use Eq.~\eqref{soft_eq_coercive} to get constant contraction for refinement
with $\mathbb A_0^{-1}$. A sufficiently short energy arc, measured in the
successive root metrics, also bounds the endpoint distance in its root
metric. Take it below the radius in Lemma~\ref{soft_lem_ball}.
That lemma adds a small perturbation for each off-root Newton iterate,
so the same base inverse still gives contraction at most, say, $1/2$.

At the base, solving $\mathbb A_0({\mathsf A},{\mathsf B})=({r_{\rm aux}},t)$ reduces to
\begin{equation}\label{soft_eq_rootsolve}
 J_{{\rm s},0}{\mathsf A}={r_{\rm aux}}+{\mathscr J}_0Q_0t,\qquad {\mathsf B}=Q_0(t-{\mathscr J}_0{\mathsf A}).
\end{equation}
Eq.~\eqref{soft_eq_precond} and the explicit Lyapunov inverse implement
this in $\widetilde O(F(n,k))$. Build it from a sufficiently accurate base cache.
For a current Jacobian $\widetilde{\mathbb A}$, use
${y_{\rm aux}}_{j+1}={y_{\rm aux}}_j+\widetilde {A_{\rm inv}}_0({b_{\rm loc}}-\widetilde{\mathbb A}{y_{\rm aux}}_j)$, where
$\widetilde {A_{\rm inv}}_0$ means applying Eq.~\eqref{soft_eq_rootsolve} to its prescribed
accuracy. The exact base inverse has contraction bounded away from one;
polynomially small root, coefficient, and application errors preserve it.
All changes of norm have polynomial condition. The error recurrence in
the base norm is at most one half the preceding error plus a polynomial
multiple of the additive evaluation tolerance. Polylogarithmically many
refinements therefore attain the required precision. In particular we do
not assume that the off-root current Lyapunov preconditioner satisfies
Eq.~\eqref{soft_eq_precond} merely from constant Loewner proximity. The use of the
base inverse and the joint transport estimate avoids that inference.
\end{proof}

\subsection{Local moves, curvature tubes, and zeros}\label{soft_sec_local}
The proof of Eq.~\eqref{fast_eq_weightedcoupling} uses only Eq.~\eqref{soft_eq_dual} and
therefore still gives, on a unit direction,
\begin{equation}\label{soft_eq_coupling}
 q(\mathcal A^{(j)}D)\le P_n\sqrt{A_r}e(D),\qquad
 |\langle F_2,D\rangle|\le P_n\sqrt{A_r}e(D),\qquad
 q(\mathcal C_{t^j})\le P_n A_r\ (j\ge2).
\end{equation}
Only $j\le3$ is needed. The last bound uses $M^{(j)}=0$ for $j\ge2$.
The second-order envelope identity now reads
\begin{equation}\label{soft_eq_second}
 \mathcal P_{\rm s}''=-A_r+e({\mathsf A})^2+2\langle F_2,\mathsf A\rangle+\mathcal E({\mathsf B})^2.
\end{equation}
The entropy term is positive, not an uncontrolled remainder.

\paragraph{Coordinates}
For $r=e_i$, $F_1=(p_iV_i,-q_iV_i)$. On expanded light bounds,
$|p_i|,|q_i|\le P_n$ and
\[
 \tr[{\mathsf Q}F_1^2]=\nu_i(p_i^2w_i+q_i^2z_i)
 \le P_n\nu_i(w_i+z_i)\le P_n t_i\le P_n \varpi_i,
 \qquad \nu_i=\tr[V_i].
\]
Here $u_i,v_i\ge a_0\nu_i$. Apply Eq.~\eqref{soft_eq_variance},
Eq.~\eqref{soft_eq_coupling}, and Eq.~\eqref{soft_eq_forced}; then use Eq.~\eqref{soft_eq_second}:
\begin{equation}\label{soft_eq_coord}
 e({\mathsf A})+\mathcal E({\mathsf B})\le P_n\sqrt{\varpi_i},\qquad
 |\partial_i^2\mathcal F_{\rm s}|\le P_n \varpi_i.
\end{equation}
Relative variation satisfies
$|u_i'|/u_i,|v_i'|/v_i\le C\sqrt ne({\mathsf A})$.
For the dual scalars Eq.~\eqref{soft_eq_variance} and
$w_i\ge a_0^2\rho\nu_i$ give
$|w_i'|/w_i,|z_i'|/z_i\le P_n\sqrt n\mathcal E({\mathsf B})$.
Thus
\[
 |\varpi_i'|\le P_n(1+\sqrt{n \varpi_{i,0}})\varpi_i
\]
until a fixed-factor exit. A first-exit argument, also for the expanded
light bounds and endpoint distance, yields the tube
\begin{equation}\label{soft_eq_coordstep}
 |t|\le a_n\min\{{\delta_{\rm end}},(n \varpi_{i,0})^{-1/2}\},\qquad \varpi_i(t)\asymp \varpi_{i,0}.
\end{equation}
At zero the second derivatives are continuous, so this argument integrates
through it. A zero input needs only the scalar reservoir and can also be
rounded directly.

\paragraph{Curvature directions}
Consider a certified base with $\mathcal F_{\rm s}''(0)\le-P_n^{-1}\theta_0$.
In a temporary region $\mathcal P_{\rm s}''\le {D_{\rm aux}}\theta_0$ and
$\theta\asymp\theta_0$, completing the square in Eq.~\eqref{soft_eq_second} gives
\begin{equation}\label{soft_eq_lowenergy}
 e({\mathsf A})+\mathcal E({\mathsf B})\le P_n\sqrt{\theta_0}.
\end{equation}
The needed third derivative is proved from the stationary saddle
$\rho\tr[{\mathsf P}]+\langle \mathsf Q,\mathcal C\rangle-\tau_{\rm s}\tr[{\mathsf Q}\log {\mathsf Q}]$ on $\tr[{\mathsf Q}]=1$.
Let
\[
 B_2=\mathcal C_{tt}+2\mathcal A'{\mathsf A}+\mathcal C_{{\mathsf P}{\mathsf P}}[{\mathsf A},{\mathsf A}],\qquad
 B_3=\mathcal C_{ttt}+3\mathcal A''{\mathsf A}+\mathcal C_{{\mathsf P}{\mathsf P}{\mathsf P}}[{\mathsf A},{\mathsf A},{\mathsf A}].
\]
The envelope identity is
\begin{equation}\label{soft_eq_third}
 \mathcal P_{\rm s}'''=\langle \mathsf Q,B_3\rangle+3\langle \mathsf B,B_2\rangle
                     -\tau_{\rm s} D^3\tr[{\mathsf Q}\log {\mathsf Q}][{\mathsf B},{\mathsf B},{\mathsf B}].
\end{equation}
Terms containing second optimizer derivatives cancel by stationarity.
Eq.~\eqref{soft_eq_coupling} and normalized inverse words give
$q(B_2)\le P_n\theta_0$ and
$|\langle \mathsf Q,B_3\rangle|\le P_n(\theta_0+\sqrt n\theta_0^{3/2})$.
Eq.~\eqref{soft_eq_relv} bounds $|\langle \mathsf B,B_2\rangle|$ by
$P_n\sqrt n\theta_0^{3/2}$.
The resolvent estimate Eq.~\eqref{soft_eq_entropyrelative}, with $\Delta {\mathsf Q}={\mathsf B}$,
bounds the last term in Eq.~\eqref{soft_eq_third} by the same quantity.
Finally differentiating $A_r=-\langle \mathsf Q,\mathcal C_{tt}\rangle$ and using
Eq.~\eqref{soft_eq_coupling} proves
\begin{equation}\label{soft_eq_thirdbounds}
 |\mathcal P_{\rm s}'''|,\ |\mathcal F_{\rm s}'''|,\ |\theta'|
 \le P_n(\theta_0+\sqrt n\theta_0^{3/2}).
\end{equation}
These are one-sided inequalities at zeros.

At the base, $\mathcal P_{\rm s}''\le\theta_0$. A simultaneous first-exit argument
using Eq.~\eqref{soft_eq_thirdbounds} proves Eq.~\eqref{soft_eq_lowenergy} and weight
comparison throughout
\begin{equation}\label{soft_eq_curvstep}
 |t|\le a_n\min\{{\delta_{\rm end}},(n\theta_0)^{-1/2}\}.
\end{equation}
This length is at most ${\delta_{\rm end}}/4$ after fixing its prefactor. It also
keeps the model in a bounded-sum neighborhood. Coordinate-zero crossings
cause no problem: $\mathcal P_{\rm s}''$ and $\theta$ are continuous, and their
one-sided derivatives have the asserted absolute bound. We use a
third-order remainder, not a fourth derivative or a Taylor jet crossing
a jump.

The better of the two signs of Eq.~\eqref{soft_eq_curvstep} has decrease
\begin{equation}\label{soft_eq_curvdrop}
 \Delta_c\ge P_n^{-1}\min\{{\delta_{\rm end}}^2\theta_0,n^{-1}\}.
\end{equation}
Indeed, in the symmetric average the linear terms cancel and the
absolute third-order remainder is at most a small fraction of
$P_n^{-1}\theta_0t^2$. Since $\theta_0\ge a/k$, the accepted count is
\begin{equation}\label{soft_eq_ic}
 I_c\le\widetilde O(k+n).
\end{equation}
Eq.~\eqref{soft_eq_lowenergy} and Eq.~\eqref{soft_eq_curvstep} give an energy arc
small enough for Lemma~\ref{soft_lem_transport} over each whole candidate.
The same base root starts both branches; each needs only a constant
number of corrected stages, including final grid rounding. No
$\widetilde O(n)$ tracking mesh remains.

For later use, Eq.~\eqref{soft_eq_second} also gives a global lower bound
$\nabla^2\mathcal F_{\rm s}\succeq-P_n G$ at a light state, by completing the
square in $e({\mathsf A})$ and retaining $\mathcal E({\mathsf B})^2$. Eq.~\eqref{soft_eq_coord}
gives bounded diagonal entries of the scaled Hessian. Therefore
\begin{equation}\label{soft_eq_scalednorm}
 A:=G^{-1/2}\nabla^2\mathcal F_{\rm s}G^{-1/2}\succeq-P_n I,\qquad
 |A_{ii}|\le P_n,\qquad \lVert A\rVert\le P_n k.
\end{equation}
The last implication uses positivity and the trace bound of $A+P_n I$.

\subsection{Explicit initialization}\label{soft_sec_init}
For a face with modeled $M_0$ and coupling $\mathcal K_0$, use
\[
 M_{r_{\rm hom}}={r_{\rm hom}}M_0,\qquad \mathcal K_{r_{\rm hom}}={r_{\rm hom}}\mathcal K_0,\qquad 0\le {r_{\rm hom}}\le1,
\]
in the smoothed problem, with the same $\rho,\tau_{\rm s}$. This is only an
optimizer-evaluation homotopy; it never changes the signing or its
potential comparisons. At ${r_{\rm hom}}=0$ the exact root is
\begin{equation}\label{soft_eq_explicit}
 U_0=V_0=(2n\rho)^{-1/2}I,\qquad W_0=Z_0=I/(2n).
\end{equation}
The constraints are scalar and the normalized exponential uniform;
substitution in Eq.~\eqref{soft_eq_dual} proves stationarity and thus optimality.
Scalar root evaluation approximates this to any prescribed precision.

The true and cached data satisfy $\|M_0\|\le20$, $\mathcal K_0(I)\preceq32I$.
Taking $U=V=I$ bounds the objective uniformly along the homotopy. Thus
Eq.~\eqref{soft_eq_rootbounds} holds throughout it. In particular the Euclidean
inverse of Eq.~\eqref{soft_eq_jointa} is polynomially bounded, by
Eq.~\eqref{soft_eq_coercive}; a conservative bound on ${\mathsf s}_{r_{\rm hom}}'$ is $P_n n^2$.
The ${r_{\rm hom}}$-forcing has Frobenius norm $O(\sqrt n)$ and the inverse has norm
$O(n^{3/2}+\tau_{\rm s}^{-1})$. An initial positive dyadic
${r_{\rm hom}}_*\asymp a_n n^{-5}$ therefore makes Eq.~\eqref{soft_eq_explicit} a valid
Newton start at ${r_{\rm hom}}_*$. To apply the anchored solve across this first
step, one may also bound the explicit-data cross variation by
$P_n nd{r_{\rm hom}}$: $\|\mathcal A_0D\|_{\mathrm F}\le C\|D\|_{\mathrm F}$,
Eq.~\eqref{soft_eq_energybounds}, and $\|{\mathsf B}\|_{\mathrm F}\le\tau_{\rm s}^{-1/2}\mathcal E({\mathsf B})$
are more than sufficient. The stated ${r_{\rm hom}}_*$ absorbs this bound.

For ${r_{\rm hom}}\ge {r_{\rm hom}}_*$ parameterize by $t=\log {r_{\rm hom}}$. Then
$M_t'=M_t$, $\mathcal K_t'=\mathcal K_t$. At a root,
$F_1=(M_{r_{\rm hom}}+\mathcal K_{r_{\rm hom}}(V),-M_{r_{\rm hom}}+\mathcal K_{r_{\rm hom}}(U))$ has operator norm bounded by a
constant. Eq.~\eqref{soft_eq_variance} gives
$|\langle F_1,\mathsf B\rangle|\le P_n\mathcal E({\mathsf B})$. The dual identities give
$|\langle \mathcal A_{r_{\rm hom}} {\mathsf Q},D\rangle|\le Ce(D)$ by weighted Cauchy--Schwarz.
The energy identity Eq.~\eqref{soft_eq_forced} proves
\begin{equation}\label{soft_eq_initenergy}
 e({\mathsf P}_t')+\mathcal E({\mathsf Q}_t')\le P_n.
\end{equation}
The explicit-data bound for transport is $q(\mathcal A_{r_{\rm hom}}D)\le Ce(D)$.
Choose logarithmic parameter increments between fixed fractions of
$a/(P_n\sqrt n)$. Lemma~\ref{soft_lem_transport} supplies one corrected stage
per increment, at cost Eq.~\eqref{soft_eq_fcost}. The whole face requires
\begin{equation}\label{soft_eq_initcount}
 O(P_n\sqrt n\log(1/{r_{\rm hom}}_*))=\widetilde O(\sqrt n)
\end{equation}
stages. This supplies the first root.

Store ${r_{\rm hom}}$ on a common dyadic grid, with mesh smaller than a fixed
inverse-polynomial fraction of ${r_{\rm hom}}_*/(P_n\sqrt n)$. Round multiplicative
increments so their logarithms stay between the chosen fractions,
except for the final truncated increment. No exact product of all
previous rational increments is retained. The parameter error is below
the same root-transport margin. The initial face starts from Eq.~\eqref{soft_eq_explicit}.
Later face changes use the separately proved continuation paths.

\subsection{Arithmetic evaluation of smoothed states}\label{sec:shared:smoothed-evaluation}
The preceding continuation estimates require matrix actions and value
comparisons at inverse-polynomial accuracy. The following construction
supplies them and preserves the corrected-root invariant.

\begin{lemma}[Matrix functions and corrected states in arithmetic operations]
\label{lem:shared:smoothed-evaluation}
Assume the geometric bounds of Section~\ref{soft_sec_geometry} and a
corrected optimizer in the joint-energy ball of Lemma~\ref{soft_lem_ball}.
The required order-$n$ matrix logarithms, square roots and their inverses,
Lyapunov solves, and entropy covariance actions can be evaluated to
inverse-polynomial error using residual-certified spectral decompositions.
Each decomposition costs $\widetilde O(n^3)$ arithmetic operations.
Together with the anchored inverse of Lemma~\ref{soft_lem_transport},
these actions preserve the corrected-root invariant and provide the
required value, endpoint, and residual enclosures.
\end{lemma}
\begin{proof}
Here is a constructive spectral subroutine for the order-$n$ operations.
A complex Hermitian Jacobi rotation annihilates a largest off-diagonal
entry and reduces squared off-diagonal Frobenius norm by twice its
squared modulus. Thus each rotation reduces this measure by a factor
at most $1-2/(n(n-1))$. After $O(n^2\log({M_{\rm aux}}/\eta))$ rotations it is at
most $\eta^2$, for a known norm bound ${M_{\rm aux}}$ and requested residual $\eta$.
A max-heap for pivot magnitudes and the two changed rows/columns give
$\widetilde O(n)$ work per rotation; accumulating the rotations costs the same.
Only a pivot above the stopping threshold is divided by when finding
its complex phase. Rotations have norm one, so rounded similarity and
orthogonality errors sum polynomially over the rotation count, while
the off-diagonal recurrence has its usual additive-error floor.
For $\eta=(2kn)^{-C}$ this remains $\widetilde O(n^3)$ arithmetic work
with inverse-polynomial local errors. Direct residual and orthogonality checks
supply certified bounds. No small adjacent eigenvalue gap is used.

For a logarithm, the integral resolvent formula bounds perturbations of
$D\log_{\mathsf Q}$ by $C\lambda_{\min}({\mathsf Q})^{-2}$ times the matrix perturbation.
For Lyapunov solves the residual is converted to solution error by
$2\lambda_{\min}(T_U)$. Square-root and inverse-square-root perturbations
have polynomial bounds on the same spectral ranges. These facts turn
the residual-based decompositions into matrix-function actions without
requiring individual eigenvectors to be accurately determined.
Project approximate entropy actions to trace zero on both sides and
symmetrize; on this space their exact smallest eigenvalue is
inverse-polynomial, so the selected accuracy preserves positivity.

Scalar powers with either fixed rational exponent $97/200$ or $3/4$
use certified bisection or series. Matrix logarithms, Lyapunov inverses, and $Q_{\mathsf Q}$
use gap-free order-$n$ spectral decompositions with inverse-polynomial
spectral lower bounds; the requested higher accuracy adds only
polylogarithmic factors. Direct log-trace-exponential value evaluation
uses a common spectral shift. On the bounded model its eigenvalue
magnitudes divided by $\tau_{\rm s}$ are $O(\log n)$ with fixed constants.
No underflow assertion or unit cost for an unbounded exponential is
used. There is also a direct residual-based value evaluation: put
$h=\tr[\mathcal C-\tau_{\rm s}\log {\mathsf Q}]/(2n)$. Since $\tr[{\mathsf Q}]=1$,
$f_{\tau_{\rm s}}(\tau_{\rm s}\log {\mathsf Q}+hE)=h$, so the absolute difference between
$f_{\tau_{\rm s}}(\mathcal C)$ and $h$ is at most
$\|\Pi_{\mathcal T}(\mathcal C-\tau_{\rm s}\log {\mathsf Q})\|$. This supplies an additional value
error enclosure from the joint residual itself. The fixed rational real/imaginary Hermitian basis has polynomial
condition and constant diagonal Gram factors for the full pair space;
projection to trace zero adds only a polynomial factor. All analytical
square roots of metrics used for error bounds need not be computed.

At a corrected Newton stage, approximate the root-anchor inverse and
current actions below a fixed fraction of the contraction margin.
The refinement error recurrence is a contraction plus additive error;
Newton then restores the cache floor. This recurrence does not multiply
per-stage condition numbers over the signing trajectory. Root perturbation,
coordinate-grid, value-comparison, and endpoint-score tolerances are
chosen below fixed fractions of the proved energy radii and true
potential decreases. All minima of weights, endpoint distances, and
spectral values appearing in them have polynomial or inverse-
polylogarithmic bounds. Final Rayleigh and solve residuals include
true-versus-cached errors.
At every accepted state, certify the optimizer cache inside its
joint-energy ball and correct it to the prescribed floor. Use that
base inverse for the next candidate or homotopy increment. The residual
uses logarithms of the current positive $\mathsf Q$. Traceless updates
keep $\tr[\mathsf Q]=1$ exactly, using the rational projection
$\mathsf B\mapsto\mathsf B-\tr[\mathsf B]E/(2n)$. Positivity follows
from the joint ball and errors below a fixed fraction of its radius.
The local inverse bound converts residual enclosures to root-error
enclosures.
\end{proof}

\subsection{Charging face changes to potential decrease}\label{gf_sec_faces}
The explicit homotopy above is used initially and after regular endpoint
moves. Deferred removals are instead followed to their new face. We give
a quantitative bound for these paths because merely accelerating the
curvature search would leave the old initialization cost unchanged.

\begin{lemma}[Regular endpoints and charged attenuation]\label{gf_lem_faces}
There are $\widetilde O(n)$ regular endpoint moves. All deferred removals,
including final corrections, use $\widetilde O(k+\sqrt{nk})$ corrected
stages. All face work, including explicit initializations, therefore uses
$\widetilde O(k+\sqrt{nk}+n^{3/2})$ stages of cost
$\widetilde O(n^\chi)$ arithmetic operations. These bounds hold for
every sequence of certified local moves and for Hermitian inputs.
\end{lemma}
\begin{proof}
For a regular move to $1$, put $d_i=1-y_i\ge{{\delta_{\rm end}}}$,
$\nu_i=\tr[V_i]\le1$, and $h_i={\iota}\beta_i u_i+d_i$.
After changing $M$ by $d_iV_i$ and deleting the coupling slot, test the
new problem at the old $U$ and
\[
 V^+=(V^{-1}+h_iV_i)^{-1}.
\]
The new minus block of $\mathcal C$ equals its old value. At an exact
trigger the plus block does not increase: ${\iota}\beta_i v_i\ge d_i$,
$V^+\preceq V$, and the remaining coupling is positive. Order
monotonicity of $f_{{\tau_{\rm s}}}$ and Sherman--Morrison give
\begin{equation}\label{gf_eq_endpoint}
 \mathcal P_{\rm s,old}-\mathcal P_{\rm s,new}
 \ge{\rho}\tr[V-V^+]
 \ge{\rho}\frac{h_i v_i^2}{\nu_i(1+h_i v_i)}.
\end{equation}
Thus the rank-one adjustment in Lemma~\ref{fast_lem_regular_resets}
also applies to the smoothed maximum; it does not require hard tight
constraints. More explicitly, at a conservative trigger of absolute
score error $\delta_e$, the exact score is at least $-2\delta_e$.
For $\delta_e\le{{\delta_{\rm end}}}/4$, we have $v_i\ge{{\delta_{\rm end}}}/(2{\iota})$, since
$\beta_i\le1$. A uniform $2\delta_e\nu_i E$ repairs the comparison,
at cost exactly $2\delta_e\nu_i$. Before repair the last fraction in
Eq.~\eqref{gf_eq_endpoint} is at least
${\rho}{{\delta_{\rm end}}}^3/(4{\iota}^2+2{\iota})$. Taking the tolerance smaller than a fixed
fraction of this and of the original reservoir allowance leaves decrease
$a{\rho}{{\delta_{\rm end}}}^3$. The other endpoint exchanges $U,V$. All other
accepted moves decrease the same nonnegative, initially bounded
potential, so the regular count is $\widetilde O(n)$. Explicit
initialization thus costs $\widetilde O(n^{3/2})$ stages in total.

For a deferred coordinate, absorb its unchanged fractional coefficient
into $F$, preserving $M$. Write its old slot as
${\mathcal K}_i(D)={\iota}\beta_iV_iDV_i$. Hold the other slots fixed and track
\[
 {\mathcal K}_s={\mathcal K}_{\rm rest}+r{\mathcal K}_i,\qquad r=e^{-s},\qquad 0\le s\le L_*.
\]
The reservoir is constant during this evaluation path and is removed
at its end. All bounded-model root estimates hold, because the coupling
only decreases. Define the positive map on pairs
\[
 K_s(D_U,D_V)=(r{\mathcal K}_i(D_V),r{\mathcal K}_i(D_U)),\qquad
 v(s)=\langle {\mathsf Q},K_s{\mathsf P}\rangle\ge0.
\]
The envelope gives $\mathcal P_{\rm s}'(s)=-v(s)$, and $v(s)\le C$. The forcing is
$F_1=-K_s{\mathsf P}$, $F_2=-K_s{\mathsf Q}$. The pair $K_s{\mathsf P}$ is a PSD subpair of the full
coupling. Hence the entropy variance bound and weighted
Cauchy--Schwarz give
\[
 \tr[{\mathsf Q}F_1^2]\le Cv,
 \quad |\langle F_2,D\rangle|\le C\sqrt v e(D),
 \quad q(K_sD)\le C\sqrt v e(D).
\]
For the last two inequalities, apply the proof of
Eq.~\eqref{soft_eq_coupling} to the single nonnegative coefficient
$r{\iota}\beta_i$: the first squared factor is
$r{\iota}\beta_i(v_iw_i+u_iz_i)=v$, and the second is bounded by the full
dual energy since this slot is included in ${\mathcal K}_s$.
Eq.~\eqref{soft_eq_forced} now yields
\begin{equation}\label{gf_eq_attenuation_energy}
 e({\mathsf P}')+\mathcal E({\mathsf Q}')\le P_n\sqrt v.
\end{equation}
The proof of Lemma~\ref{soft_lem_transport}, with the displayed
$q(K_sD)$ bound as its explicit-data term, gives normalized metric and
Jacobian variation rate at most $P_n\sqrt{nv}$. Lightness is not needed
on this auxiliary path.

There is a relative scalar bound as well. Since $K_s'=-K_s$,
\[
 v'=-v+\langle {\mathsf Q}',K_s{\mathsf P}\rangle+\langle K_s{\mathsf Q},{\mathsf P}'\rangle.
\]
The two pairings are at most
$P_n\sqrt v \mathcal E({\mathsf Q}')$ and $C\sqrt v e({\mathsf P}')$ in absolute value.
Consequently
\begin{equation}\label{gf_eq_attenuation_relative}
 |v'|\le P_nv.
\end{equation}
A zero slot can be omitted directly. At a corrected root compute an upper
approximation $v\le\bar v\le C(v+1/n)$, using absolute error at most
$a/n$, and take a dyadic increment within fixed factors of
\[
 h=\frac{a}{P_n(1+\sqrt{n\bar v})},
\]
except for the final truncation. All other calculations still use the
finer root cache floor. Eq.~\eqref{gf_eq_attenuation_relative}
keeps $v$ comparable over the increment. Its joint energy arc is at most
$a/(P_n\sqrt n)$ and its normalized variation is small. Thus
Lemmas~\ref{soft_lem_ball} and~\ref{soft_lem_transport} justify one
corrected stage, with positive matrices and trace exactly one.

Choose $r_*=e^{-L_*}$ below the root-cache floor divided by a fixed
polynomial in $n,k$. With the prescribed inverse-polynomial root accuracy,
$r_*$ is inverse polynomial and $L_*=O(\log(2kn))$. At $r_*$ drop the remaining slot exactly and
correct once. The direct derivative with respect to $r$ is uniformly
polynomially bounded even at zero: the fixed slot has bounded Frobenius
operator norm, $\|{\mathsf P}\|_F\le C\sqrt n$, and the joint inverse and metric
changes have polynomial norm bounds. Extra fixed guard powers in the
choice of $r_*$ put this last jump inside the joint Newton ball and the
required error floor. The endpoint is precisely the optimizer of the
new face, not a presumed continuation across a large face change.

For $\Delta_i=\int_0^{L_*}v(s)\d s$, the step count is at most
\[
 P_n\int_0^{L_*}(1+\sqrt{nv(s)})\d s+1
 \le P_n(L_*+\sqrt{nL_*\Delta_i})+1.
\]
This follows by integrating over each interval, where $v$ is comparable
to its starting value. The nonnegative $\Delta_i$ sum to at most the
initial potential, since all other moves decrease it and each removal
also drops its reservoir term. Summing over at most $k$ removals gives
$\widetilde O(k+\sqrt{nk})$ stages.

Evaluate $r=e^{-s}$ from the current path parameter. The smallest step
and $r_*$ are inverse polynomial, and the horizon is $O(\log(2kn))$.
Scalar Taylor evaluation and the preceding corrected solves attain
errors below the displayed margins with logarithmic overhead in
arithmetic operations.
The numerical slot stays nonnegative; the true matrices and symbolic
signed contribution are unchanged. Process simultaneous deferred hits
successively, charging each coordinate once. This introduces no extra
rounding or discrepancy allowance.
\end{proof}

\subsection{Shared endpoint continuation and gradient sweeps}\label{sec:shared:endpoints-sweeps}
We work at matrix order $N$, with initial retained count $k_0\le N^2$,
$\rho=\varepsilon_{\rm scale}/N$, and the cutoff and smoothing temperature
specified in Section~\ref{soft_sec_geometry}. The reservoir phases in
Eq.~\eqref{sp_eq_lambda} give $\gamma\ge\Lambda/(L_0k_0)$ and total
positive injections at most $\Lambda$.
Write $P_N$ for a known fixed polynomial in $\log(2N)$, enlarged finitely
many times. Its coefficients follow from the coefficient sums,
truncated budget bounds, and root estimates already proved. The
nonzero input scale retains its established polynomial lower bound.

\paragraph{Positive auxiliary penalties and offsets.}
We enlarge the class of auxiliary evaluation problems, not the signing
potential. Let ${\mathcal K}$ be a positive self-adjoint coupling of the retained
rank-one form, let $M$ be Hermitian, and let the matrix pairs
$\Omega\succeq{\rho} E$ and $D\succeq0$ be fixed.  Define
\begin{equation}\label{am_eq_auxclass}
 \mathcal R_{\Omega,D}
 =\min_{{\mathsf P}=(U,V)\succ0}\{\langle\Omega,{\mathsf P}\rangle+
 f_{{\tau_{\rm s}}}(\mathcal C({\mathsf P})+D)\},\qquad
 f_{{\tau_{\rm s}}}(C)={\tau_{\rm s}}\log\tr[e^{C/{\tau_{\rm s}}}],
\end{equation}
where $C$ denotes the block diagonal matrix associated to its pair and
$\mathcal C({\mathsf P})=(U^{-1}+M+{\mathcal K}(V),V^{-1}-M+{\mathcal K}(U))$.
All instances below have bounded $\|M\|,\|D\|,\|\Omega\|$ and a uniform
constant upper bound on their optimum.

\begin{lemma}[Uniform root machinery with positive additions]
\label{am_lem_auxroots}
The bounded-model root, joint-energy ball, and inexpensive anchored
correction statements hold for Eq.~\eqref{am_eq_auxclass}, with the same
logarithmic losses. A corrected stage costs $\widetilde O(N^\chi)$.
The residual is
\[
 (\Omega-{\mathscr J}_{\mathsf P}{\mathsf Q},\ \Pi_{\mathcal T}({\tau_{\rm s}}\log {\mathsf Q}-\mathcal C({\mathsf P})-D)),\qquad\tr[{\mathsf Q}]=1.
\]
The joint Jacobian and metric are precisely $\mathbb A$ and $\mathbb D$ in
Eq.~\eqref{soft_eq_jointa} and Eq.~\eqref{soft_eq_coercive}.  If only $\Omega,D$
vary, the Jacobian has no additional explicit-data derivative.
\end{lemma}
\begin{proof}
Convexity and attainment follow by the same operator-convex inverse and
positive trace-penalty argument as for $\mathcal P_{\rm s}$.  At a minimizer the Gibbs
multipliers satisfy
\[
 W=U(\Omega_U+{\mathcal K}(Z))U,\qquad
 Z=V(\Omega_V+{\mathcal K}(W))V,\qquad \tr[W+Z]=1.
\]
They are positive definite.  In particular
$W\succeq{\rho} U^2$, $Z\succeq{\rho} V^2$, and
${\rho}\tr[U^2+V^2]\le1$.  The value upper bound, $D\succeq0$, and
bounded $M$ bound the primal inverses and ${\mathcal K}(U),{\mathcal K}(V)$ above.
Thus $U,V\succeq a_0I$, ${\mathsf Q}\succeq a/N$, and all the weighted bounds used
in Lemma~\ref{soft_lem_ball} hold.  Strict convexity of the weighted
inverse terms proves uniqueness.

The additional trace penalty is linear in ${\mathsf P}$ and the offset is
constant in ${\mathsf P}$.  They therefore leave the joint Jacobian unchanged.
Its coercivity identity is algebraic on the trace-zero space and does
not require $\Omega={\rho} E$.  The inverse-word and resolvent proofs
of the joint-energy ball apply without change.  For the inexpensive
root inverse, the complete-positive Schwarz proof uses
${\mathcal K}(W)\preceq V^{-1}ZV^{-1}$ and its other-block counterpart; these
still follow from the displayed dual identities.  Consequently
${\mathscr E}\preceq J_s={\mathscr E}+{\mathscr J}Q_{\mathsf Q}{\mathscr J}\preceq(1+C/{\tau_{\rm s}}){\mathscr E}$ remains valid.  The explicit
Lyapunov inverse and rank-one actions implement the same anchored
correction and its residual certification.  Positive additions have
only a fixed number of rank-one terms in the application below.
\end{proof}

\paragraph{Charging regular endpoints to their true decrease.}
We first treat an exact trigger to $+1$.  At the old root put
$d=1-y_i\ge{{\delta_{\rm end}}}$, $\beta=\beta(y_i)$,
$a=\tr[UV_i]$, $b=\tr[VV_i]$, $p=\tr[WV_i]$, and $q=\tr[ZV_i]$.
Let ${\mathcal K}_i({M_{\rm aux}})={\iota}\beta V_i{M_{\rm aux}}V_i$ and let the target face have
$M_+=M+dV_i$, ${\mathcal K}_+={\mathcal K}-{\mathcal K}_i$.  Define pairs
\begin{equation}\label{am_eq_endpointdata}
 {\mathsf A}=({\mathcal K}_i(Z),{\mathcal K}_i(W)),\qquad
 E_i=(({\iota}\beta b-d)V_i,({\iota}\beta a+d)V_i),\qquad
 v_0=\langle {\mathsf A},{\mathsf P}\rangle={\iota}\beta(bp+aq).
\end{equation}
An exact trigger makes $E_i\succeq0$.  Both ${\mathsf A}$ and $E_i$ have bounded
operator norm.  Also $v_0\le C$, because ${\mathcal K}_i({\mathsf P})$ is a positive
subpair of the full coupling at a bounded-model root.

\begin{lemma}[A charged positive homotopy for a regular endpoint]
\label{am_lem_endpoint}
Let $\Delta=\mathcal P_{s,\rm old}-\mathcal P_{s,\rm new}$ for a regular endpoint,
with the conservative score tolerance tightened as below.  Then
$\Delta\ge a{\rho}{{\delta_{\rm end}}}^3$ and there is an executable continuation
from the old root to the new one with at most
\begin{equation}\label{am_eq_endpointstages}
 \widetilde O(1+\sqrt{N\Delta})
\end{equation}
corrected stages.  All regular endpoint continuations together use
$\widetilde O(N)$ stages, including their initial and final corrections.
No cold initializer is invoked after these endpoints.
\end{lemma}
\begin{proof}
We use a quantitative consequence of Golden--Thompson~\cite{t12}.  If
$0\preceq E'\preceq {M_{\rm aux}}I$, ${M_{\rm aux}}\ge1$, $0<\tau\le1$, and
${\mathsf Q}_C=e^{C/\tau}/\tr[e^{C/\tau}]$, then
\begin{equation}\label{am_eq_softloss}
 f_\tau(C)-f_\tau(C-E')
 \ge\frac{\tau(1-e^{-{M_{\rm aux}}/\tau})}{{M_{\rm aux}}}\tr[{\mathsf Q}_C E']
 \ge\frac{\tau}{2{M_{\rm aux}}}\tr[{\mathsf Q}_C E'].
\end{equation}
Indeed $e^{-E'/\tau}\preceq I-
 (1-e^{-{M_{\rm aux}}/\tau})E'/{M_{\rm aux}}$ by scalar convexity on $[0,{M_{\rm aux}}]$.
Golden--Thompson gives
$\tr[e^{(C-E')/\tau}]\le\tr[e^{C/\tau}e^{-E'/\tau}]$;
then use $-\log(1-r)\ge r$ for $0\le r<1$.
This proof applies directly to the block diagonal pairs and to complex
Hermitian matrices.

At the old ${\mathsf P}$, the new constraint pair is $\mathcal C_{\rm old}-E_i$.
Testing this ${\mathsf P}$ in the new minimization and using
Eq.~\eqref{am_eq_softloss} yields
\[
 \Delta\ge\frac{{\tau_{\rm s}}}{2{M_{\rm aux}}}\langle {\mathsf Q},E_i\rangle.
\]
The self-slot in the old dual equation gives
$q\ge {\iota}\beta b^2p$.  Since ${\iota}\beta b\ge d\ge{{\delta_{\rm end}}}$ and $\beta\le1$,
$b\ge{{\delta_{\rm end}}}/{\iota}$, whence
\begin{equation}\label{am_eq_endpointcharge}
 v_0={\iota}\beta bp+{\iota}\beta aq
 \le (1+{\iota}/{{\delta_{\rm end}}}^2)\langle {\mathsf Q},E_i\rangle
 \le P_N\Delta.
\end{equation}
The separate Sherman--Morrison estimate in Lemma~\ref{gf_lem_faces}
gives $\Delta\ge a{\rho}{{\delta_{\rm end}}}^3$.  We retain this estimate for
counting endpoint events; Eq.~\eqref{am_eq_endpointcharge} supplies the
additional path charge.

For $r=e^{-s}$ consider the auxiliary values
\begin{equation}\label{am_eq_endpointpath}
 F(s)=\min_{\mathsf P}\{{\rho}\tr[{\mathsf P}]+r\langle {\mathsf A},{\mathsf P}\rangle+
 f_{{\tau_{\rm s}}}(\mathcal C_{\rm new}({\mathsf P})+rE_i)\},
\end{equation}
where $\mathcal C_{\rm new}$ denotes the full constraint pair for the
target face.  At $r=1$ the old root is an exact root of this
problem: its constraint pair equals the old one, and
${\mathscr J}_+{\mathsf Q}={\mathscr J}_{\rm old}{\mathsf Q}+({\mathcal K}_i(Z),{\mathcal K}_i(W))={\rho} E+{\mathsf A}$.
The initial value is $\mathcal P_{s,\rm old}+v_0$; the limiting value at $r=0$
is $\mathcal P_{s,\rm new}$.  Both positive additions decrease with $s$, so
\[
 F'(s)=-v(s),\qquad
 v(s)=r\langle {\mathsf A},{\mathsf P}(s)\rangle+r\langle {\mathsf Q}(s),E_i\rangle\ge0,
 \qquad \int_0^\infty v(s)\d s=\Delta+v_0\le P_N\Delta.
\]
All these problems have the uniform bounds in Lemma~\ref{am_lem_auxroots}.

The derivative forcing is $F_1=-rE_i$ and $F_2=-r{\mathsf A}$ in the convention
of Eq.~\eqref{soft_eq_forced}.  The entropy variance bound gives
$|\langle F_1,{\mathsf B}\rangle|\le P_N\sqrt v \mathcal E({\mathsf B})$, since
$\|rE_i\|\le C$ and $\tr[{\mathsf Q}(rE_i)^2]\le C r\langle {\mathsf Q},E_i\rangle$.
For the other forcing use
$r{\mathsf A}\preceq\Omega={\rho} E+r{\mathsf A}\preceq
(U^{-1}WU^{-1},V^{-1}ZV^{-1})$.
For example, after congruence by $U^{1/2}$, write
$J_U=U^{1/2}r{\mathsf A}_UU^{1/2}\preceq T_U$.  Then
\[
 |\tr[r{\mathsf A}_UD_U]|
 \le\sqrt{\tr[J_U]}\sqrt{\tr[J_U\widehat D_U^2]}.
\]
Summing both blocks proves
$|\langle r{\mathsf A},D\rangle|\le\sqrt{v/2} e(D)$.
The joint energy identity therefore gives
\begin{equation}\label{am_eq_endpointenergy}
 e({\mathsf P}')+\mathcal E({\mathsf Q}')\le P_N\sqrt v,\qquad |v'|\le P_Nv.
\end{equation}
For the second inequality differentiate $v$; its two nontrivial pairings
are bounded by the same two estimates.  The normalized metric and
Jacobian variation rate is at most $P_N\sqrt{Nv}$: there is no explicit
coupling derivative on this path.

As in the charged attenuation proof, compute an upper approximation
$v\le\bar v\le C(v+1/N)$ and take increments within fixed factors of
$a/[P_N(1+\sqrt{N\bar v})]$.  Each increment stays in the joint Newton
and anchor-transport neighborhood.  Over a polylogarithmic horizon
$L_*$ its count is at most
$P_N(L_*+\sqrt{NL_*\int v})+1$.
Choose the final $r_*$ below the root cache floor divided by a fixed
polynomial in $N,k_0$; then set $r=0$ and correct once.  Direct sensitivity
in $r$, including at zero, is polynomial because its two fixed forcing
matrices are bounded and the joint inverse has a polynomial norm bound.
This proves Eq.~\eqref{am_eq_endpointstages}.

We detail conservative triggers and numerical coefficients.  A trigger
may have ${\iota}\beta b-d\ge-2\delta_e$.  Replace this coefficient by its
nonnegative part, and round the coefficients of ${\mathsf A},E_i$ to nonnegative
dyadics, at a smaller prescribed accuracy.  The resulting positive
pairs $\widehat {\mathsf A},\widehat E_i$ differ from the exact pairs by
$P_N\delta_e$ in norm.  With $b\ge{{\delta_{\rm end}}}/(2{\iota})$, the same self-slot
argument gives
$v_0\le(1+2{\iota}/{{\delta_{\rm end}}}^2)\langle {\mathsf Q},E_i^+\rangle$.
The uniform repair and Eq.~\eqref{am_eq_softloss} yield
$v_0\le P_N(\Delta+\delta_e)$.  Tighten $\delta_e$ below both the
preceding $a{\rho}{{\delta_{\rm end}}}^3$ margin and a sufficiently small fixed
inverse-polynomial fraction of every root radius.  Then the original
regular decrease remains positive and $v_0\le P_N\Delta$ still holds.

At the old root, the residual for the rounded initial auxiliary
problem is $P_N\delta_e$.  The initial inverse can be built there using
${\mathscr J}_+$, the old ${\mathscr E}$, and the old entropy operator: the inequalities
${\mathcal K}_+(W)\preceq{\mathcal K}(W)$ and ${\mathcal K}_+(Z)\preceq{\mathcal K}(Z)$ prove the
same root Schur preconditioning bound directly.  Thus it is not
assumed to follow from an unproved off-root rule.  The joint ball,
small initial residual and the usual inverse perturbation argument
supply a constant number of initial corrections.  Their value changes
are $P_N\delta_e$.  For the exact rounded auxiliary path its total drop
is at most $P_N\Delta$, after further tightening that accuracy.  This
makes the path executable without knowing any exact old multiplier.

There are $\widetilde O(N)$ regular events.  Their true matrix drops
sum to at most a constant: their reservoir removals are nonnegative,
all other signing moves decrease their phase potential, and the total
positive phase injections are at most $\Lambda$.  Cauchy--Schwarz in
Eq.~\eqref{am_eq_endpointstages} gives
$\widetilde O(I_{\rm reg}+\sqrt{NI_{\rm reg}\sum\Delta})
=\widetilde O(N)$ stages.  The negative endpoint exchanges the two blocks.
\end{proof}

\paragraph{A weighted parameter neighborhood of one light root.}
The coordinate counts can be improved without changing the scalar
budget.  At a light base state $y^0$ with corrected root $z_0=({\mathsf P}_0,{\mathsf Q}_0)$,
fix the current reservoir coefficient and put
\begin{equation}\label{am_eq_frozen}
 \kappa_i^0={\iota}(v_i^0w_i^0+u_i^0z_i^0)+{\gamma},
 \qquad
 \varphi_i(t)=(w_i^0-z_i^0)t+\kappa_i^0\beta(t).
\end{equation}
Each $\varphi_i$ is concave on the truncated interval.  Its derivative
is the true gradient coordinate evaluated at the frozen root, with
only its scalar coordinate argument changed.

\begin{lemma}[Frozen-root weighted neighborhood]\label{am_lem_frozen}
There is a known $P_N$ with the following property.  For $y,y^0$ in the
cube truncated at endpoint distance ${{\delta_{\rm end}}}/2$, let
\[
 d_0(y,y^0)=\sum_i\sqrt{\kappa_i^0}|y_i-y_i^0|.
\]
If $d_0\le a/(P_N^2\sqrt N)$, then the optimizer at $y$ is obtainable from
$z_0$ in $\widetilde O(N^\chi)$ work, using the same base inverse.  Its
root distance and gradient error satisfy
\begin{equation}\label{am_eq_frozenbounds}
 \begin{aligned}
 \|z(y)-z_0\|_{\mathbb D_0}&\le P_Nd_0,\\
 |g_i(y)-\varphi_i'(y_i)|&\le P_N^2\sqrt{\kappa_i^0} d_0,\\
 \tfrac12\kappa_i^0&\le {\iota} t_i(y)+{\gamma}\le2\kappa_i^0.
 \end{aligned}
\end{equation}
It is not required that the whole coordinate segment stay light.
All constants are uniform in the number of coordinates moved and in
real or complex Hermitian inputs.
\end{lemma}
\begin{proof}
At the light base, $u_i^0,v_i^0\le P_N$.  On the entire truncated
scalar interval, $\beta,\beta'$ and $1/\beta_i(y_i^0)$ have
polylogarithmic bounds.  These facts, rather than a unit bound on a
coordinate displacement, suffice below.

For a fixed test scalar $t$ the first forcing is
$F_{1,i}=((1+{\iota}\beta'(t)v_i)V_i,(-1+{\iota}\beta'(t)u_i)V_i)$.
At the base, and throughout a sufficiently small joint-energy ball,
\[
 \tr[{\mathsf Q}F_{1,i}^2]\le P_N\nu_i(w_i+z_i)
 \le P_N(v_iw_i+u_iz_i)\le P_N\kappa_i^0.
\]
Use $u_i,v_i\ge a_0\nu_i$ and constant-factor Loewner comparisons with
the base matrices.  The entropy variance inequality bounds its dual
energy norm by $P_N\sqrt{\kappa_i^0}$.
For $F_{2,i}={\iota}\beta'(t)(z_iV_i,w_iV_i)$, the weighted coupling proof
at the base gives dual energy norm at most $P_N\sqrt{\kappa_i^0}$:
replace its coefficient ratio by
$|\beta'(t)|^2/\beta(y_i^0)\le P_N$.  At nearby positive ${\mathsf P},{\mathsf Q}$ the
scalars $w_i,z_i$ remain within fixed factors of their base values and
the metric remains comparable.  The same estimate follows by absolute
values of the two block pairings.  This argument does not assert an
order comparison of noncommuting products from separate Loewner bounds.

Integration at the fixed $z_0$ therefore gives
\[
 \|\mathcal E_{\rm KKT,y}(z_0)\|_{\mathbb D_0^{-1}}\le P_Nd_0.
\]
The optimizer-dependent Jacobian variation is that of
Lemma~\ref{soft_lem_ball}.  Its explicit coupling change is bounded by
$P_N\sqrt N d_0$ in the normalized joint norm: use the just stated
single-coordinate weighted coupling estimate, pair its value with a
trace-zero dual test through the relative entropy bound, and sum the
absolute coordinate displacements.  In a joint-energy ball this yields
\[
 \|\mathbb D_0^{-1/2}(\mathcal E_{\rm KKT,y}'(z)-\mathbb A_0)\mathbb D_0^{-1/2}\|
 \le P_N\sqrt N(\|z-z_0\|_{\mathbb D_0}+d_0).
\]
The square of $P_N$ in the allowed radius pays for both the residual
constant and the Jacobian Lipschitz constant.  Choose $a$ sufficiently
small.  The chord map
$z\mapsto z-\mathbb A_0^{-1}\mathcal E_{\rm KKT,y}(z)$ is a contraction on a ball of
radius $2P_Nd_0$ about $z_0$, maps the ball to itself, and keeps it in
the positive cones and trace-one affine space.  This proves root
existence there and the first bound in Eq.~\eqref{am_eq_frozenbounds}.
The convex minimization has a unique root, so this is its optimizer.
The inexpensive approximate base inverse and additive-error chord
iteration give the executable correction with the preceding short
precision.  This proof also establishes boundedness of the model;
no large signed-sum change is presumed harmless from its Euclidean
coordinate length alone.

The derivative in root variables of $g_i(z,t)$ is
$\langle F_{2,i},\Delta {\mathsf P}\rangle+\langle F_{1,i},\Delta {\mathsf Q}\rangle$.
The two dual energy bounds just proved hold along the straight segment
inside the base ball.  Integrating gives the gradient error in
Eq.~\eqref{am_eq_frozenbounds}.  Positive scalar trace comparisons give
the stated bounds on ${\iota}t_i+{\gamma}$.  The proof uses only continuous
first and second derivatives at coordinate zero.
\end{proof}

\paragraph{Two separated gradient thresholds.}
Write $\kappa_i={\iota}t_i+{\gamma}$, so $\varpi_i=(-\beta_i'')\kappa_i\ge \kappa_i/2$.
Fix a positive rational inverse-polylogarithmic $h_N$. The high test uses
\begin{equation}\label{am_eq_high}
 |g_i|\le h_N\sqrt{\kappa_i/N}\quad\hbox{for every }i.
\end{equation}
The application chooses $h_N$ small enough for its failed-gradient
condition; the sweep proof below applies to any such threshold. Put $l_N=h_N/64$.  A separated numerical high test either
certifies some $|g_i|\ge(h_N/2)\sqrt{\kappa_i/N}$ or certifies
Eq.~\eqref{am_eq_high}.  For instance compare an approximation with error
at most $(h_N/8)\sqrt{\kappa_i/N}$ against
$(3h_N/4)\sqrt{\kappa_i/N}$, with constant slack for square-root evaluation.

\begin{lemma}[A gradient sweep using one root]\label{am_lem_sweep}
A successful high test at a light root can be handled by a deterministic
gradient sweep.  It ends either saturated, with true potential decrease
at least $1/(P_NN)$, or unsaturated, with all its remaining nondeferred
coordinates satisfying
\begin{equation}\label{am_eq_lowexit}
 |g_i|/\sqrt{{\iota}t_i+{\gamma}}\le h_N/(8\sqrt N).
\end{equation}
The sweep uses one final optimizer correction and a single initial
scan, of total cost $\widetilde O(N^\chi)$, plus $O(N^2)$ work for each
coordinate update.  Over the whole signing trajectory the number of
these coordinate updates is at most $k_0$ plus the number of saturated
sweeps.  Every sweep has a positive inverse-polynomial decrease.
\end{lemma}
\begin{proof}
At its base store Eq.~\eqref{am_eq_frozen} and dyadic weights
$\sqrt{\kappa_i^0}\le r_i\le2\sqrt{\kappa_i^0}$.  Use an arc budget $A_*$, a
positive dyadic within fixed factors of
\begin{equation}\label{am_eq_arcbudget}
 \frac{l_N}{C P_N^2\sqrt N},
\end{equation}
where the fixed $C$ is sufficiently large for
Lemma~\ref{am_lem_frozen} and the margins below.  Track
$Z=\sum_i r_i|y_i-y_i^0|$.
The low test selects coordinates with frozen derivative magnitude at
least $2l_N\sqrt{\kappa_i^0/N}$, evaluated with error at most
$(l_N/4)\sqrt{\kappa_i^0/N}$.  Absence of this test bounds every such
magnitude by $(9l_N/4)\sqrt{\kappa_i^0/N}$.

On selecting $i$, move monotonically downhill for $\varphi_i$ toward
the endpoint target $\pm(1-{{\delta_{\rm end}}}/2)$, stopping early only when the arc
budget is exhausted.  Concavity ensures that the frozen derivative
keeps the same sign and its magnitude does not decrease along this
move.  The gradient error in Eq.~\eqref{am_eq_frozenbounds} is less than
$(l_N/8)\sqrt{\kappa_i^0/N}$, by Eq.~\eqref{am_eq_arcbudget}.  Thus the true
potential decreases throughout, with
\begin{equation}\label{am_eq_sweepdrop}
 \mathcal F_{\rm s}(y^0)-\mathcal F_{\rm s}(y)\ge
 \frac{l_N}{2\sqrt N}\sum_i\sqrt{\kappa_i^0}|y_i-y_i^0|
 \ge\frac{l_NZ}{4\sqrt N}.
\end{equation}
Coordinates that reach their target are marked, but their coupling
slots and reservoir are kept until the end of the sweep.  They remain
inside the same open face at endpoint distance ${{\delta_{\rm end}}}/2$.  No
optimizer is computed during these coordinate updates.  Frozen
functions of untouched coordinates do not change, so one initial
eligible list suffices.

All coordinates lie on one common sufficiently fine dyadic grid.
Take the largest grid displacement in the chosen direction not
exceeding the endpoint or arc limit.  If a last grid step cannot use
the entire arc, declare saturation only after $Z\ge A_*/2$; a mesh
smaller than a fixed inverse-polynomial fraction of $A_*/(k_0\max r_i)$
ensures this.  Unchanged coordinates need no repeated rounding.
Eq.~\eqref{am_eq_sweepdrop} then gives the asserted saturated
decrease.  In the unsaturated case the eligible list is empty and the
same error bound, followed by $\kappa_i(y)\ge \kappa_i^0/2$, gives
Eq.~\eqref{am_eq_lowexit}, with ample slack.

A selected coordinate is processed just once unless it is the final
partial coordinate of a saturated sweep.  Every complete coordinate
update reaches a near-endpoint target and is permanently removed at
the next deferred batch.  Consequently there are at most $k_0$ complete
updates and one partial update per saturated sweep.  Each has only a
constant number of scalar budget evaluations and a rank-one modeled
sum update; charging $\widetilde O(N^2)$ per update is sufficient.
Alternatively keep only the exact coefficient list inside a sweep and
reconstruct its modeled sum at the corrected endpoint by the retained
rank-one action. This avoids accumulation in the numerical sum and
is already included in the stated scan cost.
The value and gradient comparisons at the end use the root correction
from Lemma~\ref{am_lem_frozen}.  A nonempty unsaturated sweep has a
complete coordinate displacement of at least ${{\delta_{\rm end}}}/2$, since its
base has no near-endpoint coordinates.  Its decrease is therefore at
least $l_N{{\delta_{\rm end}}}\sqrt{{\gamma}}/(4\sqrt N)$ up to the fixed evaluation
slack.  This and the saturated bound are inverse polynomial.  The
preceding finer common cache floor certifies the final value decrease.
\end{proof}

\subsection{A polynomial algorithm for the smaller constant}\label{sec_trace_algorithm}
We prove the algorithmic assertion of Theorem~\ref{rank_one_thm_main}.
Use $\beta_{\rm tr},\iota_{\rm tr}$ from Section~\ref{sec_trace_budget},
abbreviated by $\beta,\iota$ here, and the direct mixed covariance of
Lemma~\ref{lem_trace_covariance}.
The two gradient corrections in Eq.~\eqref{eq_trace_augmented_variation} are not proportional
to the covariance diagonal. We first give a direct implementation using
the strict descent of $\mathcal P$. Entropy smoothing then improves its
running time without changing the covariance certificate or constant.

\paragraph{Preprocessing and the maintained model.}
Apply the same one-time input rounding and variance scaling. Use grouped
partial signing with its prescribed residual allowance. The input-rounding transfer will use the internal
constant $C_{\rm tr,*}:=3.34427$ verified below. At most $k\le n^2$
coordinates remain fractional and $\sum_iC_i^2\preceq I$.
Keep ${\varepsilon_{\rm scale}}:=10^{-8}$, $\rho:={\varepsilon_{\rm scale}}/n$, and
${\varepsilon_{\rm round}}={\varepsilon_{\rm discard}}=\chi=e_{\rm end}:=10^{-6}$.
Round the initial coefficients with Euclidean error at most ${\varepsilon_{\rm round}}/2$
and initialize the modeled sum at zero as before.
Small-input rounding is applied in phases, as specified below. It
preserves the modeled sum and uses the same total allowance $\chi$.
Write $t_i:=\tr[C_i]$; throughout, $t_i\le1$.
Choose $\delta_{\rm end}$ by Eq.~\eqref{rank_one_eq_cont_3}.
At endpoint distance at most $\delta_{\rm end}$, record the fractional
coefficient, preserve the current modeled contribution in $F$, and remove
the coupling slot. At the end, apply Lemma~\ref{round_lem_round} jointly
to all recorded coefficients. This lemma uses only the variance bound
and endpoint distances, so it applies to the present budget and pure
matrix potential. It is deterministic and costs less than
${\varepsilon_{\rm round}}/2$ in discrepancy. Together with initial
rounding, the allowance remains ${\varepsilon_{\rm round}}$.
Now $\delta_{\rm end}^{-1}=O(\log(2n))$; in particular,
$\delta_{\rm end}\ge a/n$.
The initial matrix potential is at most $2\sqrt{\iota+2{\varepsilon_{\rm scale}}}$.

Algorithm~\ref{alg:shared:preprocess} collects the common preprocessing for
Algorithm~\ref{alg:trace:signing} and the square-root construction.
Its accuracy inputs are $\eta_{\rm ps}$ and $\varepsilon_{\rm round}$;
the signing algorithms use $10^{-7}$ and $10^{-6}$, respectively.
Fix $\xi:=10^{-8}$, $\varepsilon_{\rm scale}:=10^{-8}$, and
$\varepsilon_{\rm discard}:=10^{-6}$ as in
Sections~\ref{sec:fast:input-rounding} and~\ref{sec:fast:variance-scaling}.
The output $C$ is the active matrix list. The record $\mathcal R$ stores
original indices, eigenvalue signs, common scales, and assigned signs;
callers add each subsequently fixed sign to this record.
The returned state satisfies $|I|\le n^2$, $\sum_{i\in I}C_i^2\preceq I$, and
$F+\sum_{i\in I}y_iC_i=0$. Centering contributes a single error of at most
$\eta_{\rm ps}+\varepsilon_{\rm round}/2$ from
Section~\ref{sec:fast:maintained-model}.
When $I=\varnothing$, $\mathcal R$ already determines a complete signing.

\begin{algorithm}[!htb]
\caption{Shared preprocessing for deterministic signing}
\label{alg:shared:preprocess}
\begin{algorithmic}[1]
\Procedure{Preprocess}{$A_1,\ldots,A_m;\eta_{\rm ps},\varepsilon_{\rm round}$}
    \State Initialize $\mathcal R$ and assign sign $1$ to each zero input; set $J\gets \{i:A_i\ne0\}$.
    \If{$J=\varnothing$}
        \State \Return $(\varnothing,\varnothing,\varnothing,0,\mathcal R)$.
    \EndIf
    \State For $i\in J$, record $s_i\gets \operatorname{sign}(\tr[A_i])$ and set $B_i\gets s_iA_i$.
    \State Apply the common power-of-two scaling and PSD dyadic rounding to $B_i$.
        \Comment{Section~\ref{sec:fast:input-rounding}}
    \State Normalize the rounded matrices to obtain $\bar A_i$ with $\sum_{i\in J}\tr[\bar A_i]=1$; record the scales.
    \State Set $\zeta\gets \|\sum_{i\in J}\bar A_i^2\|$ and compute rational $q>0$ with $\zeta\le q^2\le(1+\varepsilon_{\rm scale})\zeta$.
        \Comment{Section~\ref{sec:fast:variance-scaling}}
    \State Record arbitrary signs and discard indices with $\tr[\bar A_i]<\varepsilon_{\rm discard}/(m^2n)$ from $J$.
    \State Set $C_i\gets \bar A_i/q$ for $i\in J$.
    \State Form common-grid Hermitian coordinates $b_i$ to Euclidean error $\eta_{\rm ps}/(4m)$.
        \Comment{Section~\ref{sec:fast:variance-scaling}}
    \State Apply grouped partial signing to $(b_i)_{i\in J}$ with tolerance $\eta_{\rm ps}/2$, obtaining $y^0$.
        \Comment{Lemma~\ref{rank_one_lem_grouped_partial_signing}}
    \State Set $I\gets \{i\in J:|y_i^0|<1\}$ and $k\gets |I|$; record $y_i^0$ for every $i\in J\setminus I$.
    \If{$k=0$}
        \State \Return $(\varnothing,\varnothing,\varnothing,0,\mathcal R)$.
    \EndIf
    \State Round $y_I^0$ inward to a common dyadic grid of mesh $\delta_0\le\varepsilon_{\rm round}/(2\sqrt{k})$, obtaining $y$.
    \State Set $F\gets -\sum_{i\in I}y_iC_i$.
        \Comment{Section~\ref{sec:fast:maintained-model}}
    \State \Return $((C_i)_{i\in I},y,I,F,\mathcal R)$.
\EndProcedure
\end{algorithmic}
\end{algorithm}

\begin{lemma}\label{trace_lem_small_rounding}
Let $k\le n^2$, let $C_i\succeq0$ have rank at most one, and let
$y_i\in[-1,1]$ be the initial dyadic coefficients. Choose a common dyadic
PSD rank-one cache with $\|\widehat C_i-C_i\|_F\le u$, where
$u\le\chi/(64n^2)$, and define
\[
 \tau_{\rm small}:=\frac{\chi}{4n},\qquad
 S:=\{i:\tr[\widehat C_i]\le\tau_{\rm small}\}.
\]
There is a deterministic choice of signs on $S$ such that
\[
 \|\sum_{i\in S}(\sigma_i-y_i)C_i\|<\chi/2.
\]
Every remaining matrix satisfies $\tr[C_i]\ge\chi/(8n)$.
Given the cached matrices and coefficients, the rounding costs
$O(kn^2)$ arithmetic operations.
\end{lemma}
\begin{proof}
For independent signs with mean $y_i$, centering and rank one give
\[
 \E[\|\sum_{i\in S}(\sigma_i-y_i)\widehat C_i\|_F^2]
 =\sum_{i\in S}(1-y_i^2)\|\widehat C_i\|_F^2
 \le k\tau_{\rm small}^2.
\]
Derandomize this identity by conditional expectation. Starting with
$E=0$, at each coordinate choose the sign minimizing
$\|E+(\sigma_i-y_i)\widehat C_i\|_F^2$, and add its contribution to $E$.
The two possible conditional expectations have weighted average equal
to the current value, so choosing the smaller cannot increase it.
The final cache error is at most
$\sqrt{k}\tau_{\rm small}\le\chi/4$ in Frobenius norm. Consequently
\[
 \|\sum_{i\in S}(\sigma_i-y_i)C_i\|
 \le\chi/4+2ku\le9\chi/32<\chi/2.
\]
If $i\notin S$, then
$\tr[C_i]>\tau_{\rm small}-nu\ge15\chi/(64n)>\chi/(8n)$.
The difference between the squared norms for signs $1$ and $-1$ is
$4(\tr[E\widehat C_i]-y_i\tr[\widehat C_i]^2)$.
Each comparison and update therefore costs $O(n^2)$ operations.
Form the inner products, compare the two candidates, and update
$E$ in exact arithmetic. At most $k$ comparisons and updates give
the stated cost.
\end{proof}

\paragraph{Phases for small-input rounding.}
Set $L:=1+\lceil\log_2(2n^2)\rceil$. At the start of a phase let
$K$ be the current number of active coordinates. Apply
Lemma~\ref{trace_lem_small_rounding} with cutoff
\[
 \tau_K:=\frac{\chi}{4L\sqrt K},\qquad
 u\le\frac{\chi}{64Ln^2}.
\]
The cutoff can be rounded downward by a factor of at most two.
Use one common cache of this accuracy throughout the algorithm.
The same conditional-expectation calculation gives a discrepancy error
of at most $\chi/(4L)+2Ku<\chi/(2L)$ in this phase.
Every retained trace is at least $\chi/(16L\sqrt K)$: indeed the trace
error is at most $nu\le\chi/(64Ln)$ and $K\le n^2$.
Keep the removed fractional contributions in $F$, so this operation
preserves the modeled matrix and only deletes nonnegative coupling terms.

Start another phase whenever the active count becomes at most $K/2$.
If small-input rounding itself causes this, repeat before taking a local
step. There are at most $L$ phases, so the total discrepancy error is
less than $\chi/2$, as before. During local steps of a phase,
$K/2<k\le K$ and consequently
\begin{equation}\label{eq_trace_phase_floor}
 t_i\ge\frac{a}{L\sqrt k},\qquad
 \varpi_i\ge\frac{a}{L^2nk}.
\end{equation}
The second bound follows from $u_i,v_i\ge at_i$,
$w_i,z_i\ge at_i/n$, and $-\beta_i''\ge a$.
In particular the earlier lower bound becomes
$\varpi_i\ge aP_n^{-1}n^{-3}$, which suffices below.
Each phase costs $O(Kn^2)$ operations. The starting counts decrease
geometrically, so all such rounding uses $\widetilde O(n^4)$
arithmetic operations. The signs are chosen by exact dyadic comparisons; these phases
introduce no random bits.

\paragraph{Quantitative drift witnesses.}
In this paragraph, $X\lesssim Y$ means $X\le P_nY$ for a fixed positive
polynomial in $\log(2n)$; $X\gtrsim Y$ has the reverse meaning.
Choose a single sufficiently large polynomial $P_n$ to cover all such
bounds below, enlarging it when finitely many factors are combined.
On a bounded-potential region, the same KKT estimates give
\[
 aI\preceq U,V\preceq C\sqrt n I,\qquad
 an^{-1}I\preceq W,Z\preceq I.
\]
Together with the trace cutoff, the budget bounds, and lightness, these imply
\begin{align*}
 at_i\le u_i,v_i&\le C\sqrt nt_i,&
 an^{-1}t_i\le w_i,z_i&\le t_i,\\
 \beta_i&\gtrsim1,&
 a\le p_i,q_i&\lesssim1,\qquad a\le p_iq_i\lesssim1,\\
 c_i^2,d_i^2&\gtrsim1,&
 c_i^2,d_i^2&\lesssim1,\qquad a\le\tau_i^2\lesssim1.
\end{align*}
Indeed, $\delta_{\rm end}^{-1}=O(\log(2n))$ and
$p_iq_i\le C/(1-|y_i|)$, while $p_i,q_i\ge1/4$.
Lightness also gives the stronger upper bound
\[
 u_i,v_i\le\frac{2}{\iota\beta_i}\lesssim1.
\]
The movement estimate in Eq.~\eqref{eq_trace_movement} gives
\[
 a{k_{\rm act}}\le\tr[R_{\mathsf A}]\le C{k_{\rm act}},
 \qquad R_{\mathsf A}:=(I-\widehat T){\mathsf A}(I-\widehat T)^\top.
\]
Rank-one Cauchy--Schwarz and $W\succeq\rho U^2$, $Z\succeq\rho V^2$ give
\[
 w_i\ge\rho u_i^2/t_i,\qquad z_i\ge\rho v_i^2/t_i.
\]
Consequently, writing $\Delta_i:=D_i\tau_i^2$,
\begin{equation}\label{eq_trace_weighted_denominator}
 \Delta_i=p_i^2u_iw_i+q_i^2v_iz_i
 \ge\frac{a(u_i^3+v_i^3)}{nt_i}\ge an^{-1}t_i^2\ge aP_n^{-1}n^{-3}.
\end{equation}
The budget bounds also give
\[
 a\le k_i=\frac{\iota(-\beta_i'')}{2\eta p_iq_i}
 \le C\delta_{\rm end}^{-5/4}\lesssim1.
\]
Define the diagonal metric and the strict descent by
\begin{align}
 \varpi_i&:=\iota(-\beta_i'')(v_iw_i+u_iz_i),\qquad
 G:=\operatorname{diag}(\varpi_i),
 \label{eq_trace_metric}\\
 D_{\rm tr}&:=\kappa_{\rm tr}\sum_i k_i{{\rho_{\rm rat}}}_iD_i\pi_i
 =\frac{\kappa_{\rm tr}}{2\eta}\tr[GK].
 \label{eq_trace_relative_drift}
\end{align}
The identity preceding Eq.~\eqref{eq_trace_strict_descent} gives
$\varpi_i/\Delta_i=2\eta k_i{{\rho_{\rm rat}}}_i$.
At every light state, $aP_n^{-1}n^{-3}\le\varpi_i\le P_n$.

Suppose that
\begin{equation}\label{eq_trace_small_weighted_gradient}
 |\partial_i\mathcal P|<\kappa_1P_n^{-1}n^{-1/2}\sqrt{\varpi_i}
 \quad\text{for every active }i.
\end{equation}
We first show that $\widetilde w_i/\widetilde z_i\in[1/2,2]$.
The dual lower bounds and $u_i,v_i\ge at_i$ give
\[
 \frac{\varpi_i}{(\widetilde w_i+\widetilde z_i)^2}
 \lesssim\frac{u_i+v_i}{w_i+z_i}
 \lesssim nt_i\frac{u_i+v_i}{u_i^2+v_i^2}\lesssim n.
\]
Since $\partial_i\mathcal P=\tau_i(\widetilde w_i-\widetilde z_i)$
and $\tau_i\ge a$, Eq.~\eqref{eq_trace_small_weighted_gradient} bounds
$|\widetilde w_i-\widetilde z_i|/(\widetilde w_i+\widetilde z_i)$
by $1/3$, after enlarging $P_n$ and choosing $\kappa_1$ sufficiently small.
Consequently
\[
 \frac12\le{{\rho_{\rm rat}}}_i
 =\frac{\widetilde u_i\widetilde z_i+\widetilde v_i\widetilde w_i}
        {\widetilde u_i\widetilde w_i+\widetilde v_i\widetilde z_i}
 \le2,\qquad \varpi_i\asymp\Delta_i
\]
up to the polylogarithmic factors. The movement bound and
$D_i\pi_i=(R_{\mathsf A})_{e_i,e_i}/\tau_i^2$ imply
$D_{\rm tr}\gtrsim k_{\rm act}$.

We bound the drift in the same metric. For its ordinary part, the terms
containing $\beta_i'$ cancel:
$\tau_i(\widetilde u_i+\widetilde v_i)=p_iu_i+q_iv_i=u_i+v_i$.
Thus $|h_i^{(0)}|\lesssim(u_i+v_i)K_{ii}$.
The lower bounds on $w_i,z_i$ imply
\[
 \varpi_i\gtrsim\frac{u_iv_i(u_i+v_i)}{nt_i},\qquad
 \frac{(u_i+v_i)^2}{\varpi_i}
 \lesssim nt_i(1/u_i+1/v_i)\lesssim n.
\]
It follows that
$\sum_i|h_i^{(0)}|\sqrt{\varpi_i}\lesssim\sqrt n\tr[GK]$.

For the trace correction, the compact certificate and block weights give
\[
 |a_{*,i}|+|w_i^e|\le C,\qquad |w_i^{ef}|\le C\sqrt{\delta_i},\qquad
 0<w_i^f\le C\delta_i,
\]
\[
 \delta_i(b_{*,i}^2+r_{*,i}^2)\le C,\qquad
 |\delta_i d_{*,i}|=|{j_{\rm 0}}_i|\le C.
\]
Hence $h_{*,i},j_{*,i},q_{*,i},\delta_i p_{*,i}$ are bounded.
Set $s_i:=\min\{u_i,v_i\}$ and $r_i:=\max\{u_i,v_i\}/s_i\ge1$
within this paragraph. Then $s_i\ge at_i$,
$\max\{u_i,v_i\}\lesssim1$, and $|\upsilon_i|\lesssim\sqrt{r_i}$.
For $r\ge1$,
\[
 \frac{(1+r)^2(1+\sqrt r)^2}{1+r^3}\le16,\qquad
 \frac{r(1+\sqrt r)^2}{1+r^3}\le2.
\]
Since $\delta_i=\iota\beta_i u_iv_i$, Eq.~\eqref{eq_trace_weighted_denominator} gives
\begin{align}
 \frac{(u_i+v_i)(1+|\upsilon_i|)}{\sqrt{\Delta_i}}
 &\lesssim\sqrt{nt_i/s_i}\lesssim\sqrt n,
 \label{eq_trace_drift_ratio}\\
 \frac{\delta_i(1+|\upsilon_i|)^2}{\Delta_i}
 &\lesssim\frac{nt_i}{s_i}\frac{r_i(1+\sqrt{r_i})^2}{1+r_i^3}
 \lesssim n.
 \label{eq_trace_mixed_ratio}
\end{align}
The diagonal and lower off-diagonal correction is
\[
 \nu_i^{(0)}=
 \frac{\delta_i[h_{*,i}-j_{*,i}+p_{*,i}\upsilon_i]
       \tau_i(\widetilde u_i-\widetilde v_i)}{\Delta_i}.
\]
Since $|\tau_i(\widetilde u_i-\widetilde v_i)|\le u_i+v_i$ and
$\varpi_i\lesssim\Delta_i$, Eq.~\eqref{eq_trace_drift_ratio} gives
$\sqrt{\varpi_i}|\nu_i^{(0)}|\lesssim\sqrt n$.
The upper correction is $\nu_i^{(1)}:=q_{*,i}\Omega_i$.
The balance $\widetilde w_i/\widetilde z_i\in[1/2,2]$ and
Eq.~\eqref{eq_trace_upper_off_diagonal} give
$|\Omega_i|\lesssim(u_iv_i)^{3/2}/\Delta_i$.
Since $|q_{*,i}|\lesssim1$ and $\varpi_i\asymp\Delta_i$,
\[
 \sqrt{\varpi_i}|\nu_i^{(1)}|
 \lesssim\frac{(u_iv_i)^{3/2}}{\sqrt{\Delta_i}}
 \lesssim\sqrt{nt_i}\frac{(u_iv_i)^{3/2}}{\sqrt{u_i^3+v_i^3}}
 \lesssim\sqrt n.
\]
The last inequality uses lightness, $u_i,v_i\lesssim1$, and $t_i\le1$.
Thus the weighted sum for $\nu^{(1)}$ is at most
$P_n\sqrt n k_{\rm act}$.
All helper functions are fixed polynomials in $|y_i|$ and
$H_i:=\iota^2\beta_i^2u_iv_i/(1-y_i^2)\in[0,1]$.
No derivatives of these helpers enter the descent argument.

For the mixed correction, retain the coefficient $\Gamma_i$ in the
coercivity inequality. The response condition gives
$\mathsf E_i-\mathsf Y_i=e_i^\top\widehat T\xi$ and
$\mathsf F_i=f_i^\top\widehat T\xi$.
Since $\Gamma$ is scalar on each paired plane and
${\mathsf A}\preceq CI$, Eq.~\eqref{rank_one_eq_energy_dominance} gives
\begin{equation}\label{eq_trace_weighted_response}
 \sum_i\Gamma_i\E[(\mathsf E_i-\mathsf Y_i)^2+\mathsf F_i^2]
 =\tr[\widehat T^\top\Gamma\widehat T{\mathsf A}]
 \le C\tr[\widehat {D_{\rm aux}}]=2Ck_{\rm act}.
\end{equation}
Set $\mathsf Z_i:=\mathsf E_i-\mathsf Y_i+\upsilon_i\mathsf F_i$.
Eq.~\eqref{eq_trace_mixed_ratio} and $\delta_i=1/\Gamma_i$ imply
\[
 \sum_i\frac{\E[\mathsf Z_i^2]}{\Delta_i}
 \le\sum_i\frac{(1+|\upsilon_i|)^2}{\Delta_i}
       \E[(\mathsf E_i-\mathsf Y_i)^2+\mathsf F_i^2]
 \lesssim nk_{\rm act}.
\]
The definition of $\theta_i$, $|g_i|\lesssim1$, and
$\varpi_i\asymp\Delta_i$ now give, by Cauchy--Schwarz,
\[
 \sum_i|\theta_i|\sqrt{\varpi_i}
 \lesssim\sum_i\frac{|\E[\mathsf Y_i\mathsf Z_i]|}{\sqrt{\Delta_i}}
 \le\sqrt{\sum_i\E[\mathsf Y_i^2]}
      \sqrt{\sum_i\frac{\E[\mathsf Z_i^2]}{\Delta_i}}
 \lesssim\sqrt n k_{\rm act}.
\]
Here $\sum_i\E[\mathsf Y_i^2]=\tr[R_{\mathsf A}]\le Ck_{\rm act}$.
Combining the contributions proves, under
Eq.~\eqref{eq_trace_small_weighted_gradient},
\begin{equation}\label{eq_trace_quantitative_drift}
 D_{\rm tr}\gtrsim k_{\rm act},\qquad
 \mathscr L\mathcal P\le-D_{\rm tr},\qquad
 \sum_i|h_i|\sqrt{\varpi_i}\le P_n\sqrt n D_{\rm tr}.
\end{equation}
Thus, for a sufficiently small absolute $\kappa_1>0$,
\begin{equation}\label{eq_trace_alternative}
 \max_i\frac{|\partial_i\mathcal P|}{\sqrt{\varpi_i}}
 \ge\kappa_1P_n^{-1}n^{-1/2}
 \quad\text{or}\quad
 \lambda_{\min}(G^{-1/2}\nabla^2\mathcal P G^{-1/2})\le-\kappa_1.
\end{equation}
Indeed, failure of the first inequality makes the drift term smaller
than $D_{\rm tr}/4$ in absolute value. If the second also fails, then
Eq.~\eqref{eq_trace_relative_drift} gives
$\tr[K\nabla^2\mathcal P]/2>-D_{\rm tr}/4$ after decreasing $\kappa_1$.
Their sum cannot be at most $-D_{\rm tr}$.
The algorithm uses this alternative without computing the covariance,
drift, or helper coefficients.

\paragraph{Multiplicity of negative curvature.}
The same covariance construction can be restricted to a prescribed
subspace of physical directions. Let its codimension be $q\le k/8$.
The response graph is bijective, so the restricted response space has
dimension $\ell:=k-q$. In Lemma~\ref{lem_trace_covariance}, use an
orthonormal basis $E$ of this space and set
\[
 A:=E(E^\top{\mathsf D}^{-1}E)^{-1}E^\top,\qquad B:={\mathsf D}-A.
\]
All row completions and trace identities remain unchanged. The only
rank-dependent equality is now $\tr[{\mathsf D}^{-1}A]=k-q$.
Thus the right side of Eq.~\eqref{eq_trace_covariance} acquires exactly
the additive term $q$.

The movement lower bound remains proportional to $k$. To check this
without using the smallest weight, the block bounds in that lemma give
$A\preceq CI$ and
${\mathsf D}^{-1}\preceq C\Pi_0+C\Gamma\Pi_1$.
The response identity and energy dominance imply
\[
 \tr[E^\top{\mathsf D}^{-1}E]\le Ck,
 \qquad \|\widehat T\|_F^2\le2k/\iota.
\]
Writing $\Pi:=EE^\top$, we have
\[
 \tr[(I-\widehat T)\Pi]
 \ge\ell-\sqrt{2k\ell/\iota}\ge k/16.
\]
For the last inequality, the expression increases for $\ell\ge7k/8$,
and $(13/16)^2>7/(4\iota)$ for $\iota=699/250$.
Weighted Frobenius Cauchy--Schwarz, with
$C_E:=E^\top{\mathsf D}^{-1}E$, gives
\[
 (\tr[(I-\widehat T)\Pi])^2
 \le\tr[(I-\widehat T)A(I-\widehat T)^\top]\tr[C_E].
\]
Hence $ak\le\tr[(I-\widehat T)A(I-\widehat T)^\top]\le Ck$.
The upper bound follows from $A\preceq CI$ and energy dominance.

All the preceding drift estimates hold for this restricted covariance:
they used precisely these movement bounds, $A\preceq CI$, and
Eq.~\eqref{eq_trace_weighted_response}, which still follows from
$\widehat T^\top\Gamma\widehat T\preceq\widehat {D_{\rm aux}}$.
Thus, at a failed gradient test, its physical covariance $K_q$ satisfies
\begin{equation}\label{eq_trace_defect_generator}
 D_q\ge aP_n^{-1}k,\qquad
 D_q=\frac{\kappa_{\rm tr}}{2\eta}\tr[GK_q],\qquad
 \nabla\mathcal P\cdot h_q+\frac12\tr[K_q\nabla^2\mathcal P]
 \le-D_q+q,
\end{equation}
and $\sum_i|(h_q)_i|\sqrt{\varpi_i}\le P_n\sqrt n D_q$.
Choose the common gradient threshold small enough that its drift term
has magnitude at most $D_q/4$, uniformly over the restricted spaces.

There are a fixed $\gamma>0$ and an inverse-polylogarithmic $\pi_n>0$
such that
\begin{equation}\label{eq_trace_inertia}
 \#\{\lambda_j(G^{-1/2}\nabla^2\mathcal P G^{-1/2})\le-8\gamma\}
 \ge\pi_n k.
\end{equation}
Indeed, suppose the number on the left is $q<\pi_n k$, and restrict to
the inverse image of its orthogonal complement. Choose
$\pi_n\le1/8$ and $\pi_n k\le D_q/4$ using the uniform lower bound
in Eq.~\eqref{eq_trace_defect_generator}. On this complement,
\[
 \frac12\tr[K_q\nabla^2\mathcal P]
 \ge-4\gamma\tr[GK_q]\ge-D_q/4
\]
if $\gamma\le\kappa_{\rm tr}/(32\eta)$.
The left side of Eq.~\eqref{eq_trace_defect_generator} is then at
least $-D_q/2$, whereas its right side is at most $-3D_q/4$, a
contradiction. This also covers $\pi_n k<1$, since $q$ is an integer.
These covariances are proof witnesses and are not computed.

\paragraph{Deterministic selection of a spread direction.}
We give the selection procedure explicitly. Suppose a real symmetric
$k\times k$ matrix $H$ has norm at most a polynomial in $n$, and at least
$\pi_n k$ eigenvalues at most $-4\gamma$.
Set $B:=H+2\gamma I$ and define the bounded filter
\[
 F_*:=\frac12(I-B(B^2+\epsilon^2I)^{-1/2}).
\]
Choose a positive inverse-polynomial $\epsilon$ sufficiently small.
The scalar formula shows that $0\preceq F_*\preceq I$,
$\tr[F_*^2]\ge\pi_n k/2$, and
$\|\Pi_{\rm bad}F_*\|\le\epsilon^2/(4\gamma^2)$, where
$\Pi_{\rm bad}$ is the spectral projection of $H$ onto eigenvalues
larger than $-\gamma$.
Compute a symmetric dyadic approximation $F$ such that
\begin{equation}\label{eq_trace_filter_guards}
 \|F\|\le2,\qquad \tr[F^2]\ge\pi_n k/2,\qquad
 \|\Pi_{\rm bad}F\|\le\eta_f,
 \quad \eta_f^2\le\frac{\gamma\pi_n}{16(\|H\|+\gamma)}.
\end{equation}
There is slack for the trace guard: taking $\epsilon\le\gamma/2$
makes $F_*\ge3/4$ on the specified negative eigenspace, and decreasing
$\epsilon$ further makes this eigenvalue arbitrarily close to one.
No spectral projection is needed in the computation.

The inverse square root can be evaluated to inverse-polynomial
accuracy in $\widetilde O(k^{\omega_0})$ arithmetic operations. Define $A:=B^2+\epsilon^2I$, and choose known bounds
$a_0I\preceq A\preceq b_0I$, with $a_0^{-1},b_0$ polynomial in $n$.
For a prescribed inverse-polynomial error $\eta_s$, use
\[
 A^{-1/2}=\frac2\pi\int_0^\infty(A+t^2I)^{-1}\d t.
\]
Choose an integer $J=O(\log n)$ so that
$2^{-J}(a_0^{-1}+1)\le\eta_s/8$. Truncating the integral to
$[2^{-J},2^J]$ has operator error at most $\eta_s/8$ before the
factor $2/\pi$. Split this interval into $[a,2a]$ with $a=2^j$.
Define $R_a:=(A/a^2+(5/2)I)^{-1}$, so $\|R_a\|\le2/5$.
On each interval the geometric expansion gives
\[
 \int_a^{2a}(A+t^2I)^{-1}\d t
 =a^{-1}\sum_{\ell=0}^{s-1}c_\ell R_a^{\ell+1}+E_a,
 \qquad c_\ell:=\int_1^2(5/2-u^2)^\ell \d u.
\]
Since $|5/2-u^2|\le3/2$, the remainder summed over all intervals
has norm at most $(3/5)^s\pi/(2\sqrt{a_0})$. Choose
$s=O(\log n)$ to make this at most $\eta_s/8$. The coefficients are
explicit:
\[
 c_\ell=\sum_{v=0}^{\ell}(-1)^v\binom{\ell}{v}
 (5/2)^{\ell-v}\frac{2^{2v+1}-1}{2v+1}.
\]
The sum of the absolute values of these terms is at most
$2(13/2)^\ell$, which is polynomial for $\ell\le s$.
Evaluate these rational coefficients in exact arithmetic. Form each
positive-definite shifted inverse by recursive block inversion, as
in Lemma~\ref{rank_one_lem_regularized_partial_signing}, using
$O(k^{\omega_0})$ arithmetic operations. Evaluate the finite matrix
sums exactly, and approximate the scalar constant $2/\pi$ by a
convergent series to sufficiently small inverse-polynomial error.
All matrix and coefficient magnitudes are polynomial, so the
truncation bounds and this scalar error give total error at most
$\eta_s$. There are $O(\log n)$ intervals and $O(\log n)$ products
per interval. Thus the total arithmetic cost is
$\widetilde O(k^{\omega_0})$. Choose $\eta_s$ below a
fixed fraction of the filter guards divided by $1+\|B\|$,
symmetrize the output, and use the trace slack above to obtain
Eq.~\eqref{eq_trace_filter_guards}.

To obtain a flat vector, consider independent signs $\xi_j$ only as a
device for conditional expectation, and put
\[
 \Phi(\xi):=\|F\xi\|^2-\lambda_f\sum_i\cosh((F\xi)_i),
 \qquad \lambda_f:=\pi_n/100.
\]
Each row has squared norm at most four. Since
$\prod_j\cosh(F_{ij})\le\exp(\sum_jF_{ij}^2/2)<8$,
Eq.~\eqref{eq_trace_filter_guards} gives
$\E[\Phi]\ge\pi_n k/2-8\lambda_f k>2\pi_n k/5$.
Fix the signs one at a time, choosing the larger conditional
expectation, to additive error at most $\pi_n/100$ per decision.
Then $z:=F\xi$ satisfies
\[
 \Phi(\xi)\ge\pi_n k/4,
 \qquad \|z\|^2\ge\pi_n k/4,
 \qquad \|z\|_\infty\le\log(800k/\pi_n).
\]
For the last inequality use $\|z\|^2\le4k$ and
$\lambda_f\cosh(z_i)\le\|z\|^2$.
Also Eq.~\eqref{eq_trace_filter_guards} gives
\[
 z^\top Hz\le-\gamma\|z\|^2
       +(\|H\|+\gamma)\eta_f^2 k
 \le-3\gamma\|z\|^2/4.
\]

Maintain the prefix sum, the remaining squared column norms, and
the row products in the conditional estimator. Each choice costs
$\widetilde O(k)$ arithmetic operations, hence the total is
$\widetilde O(k^2)$. At an accepted prefix the positive conditional
estimator and the quadratic bound $4k$ give
$\cosh(v_i)\le4k/\lambda_f$ for the prefix sum $v$.
A trial changes an argument by at most two, and each remaining
product is between one and eight. Taylor series therefore evaluate
these scalar candidates to the required absolute tolerance in
polylogarithmically many arithmetic operations per term.

Apply this procedure to a symmetric approximation of
$\widehat G^{-1/2}\nabla^2\mathcal P\widehat G^{-1/2}$ with operator
error at most $\gamma/4$. Eq.~\eqref{eq_trace_inertia} and
$\widehat G\preceq G\preceq2\widehat G$ supply the required
multiplicity with slack. Undo the diagonal scaling and normalize,
$r:=\widehat G^{-1/2}z/\|\widehat G^{-1/2}z\|$.
The computed direction, after sufficiently accurate dyadic rounding
and normalization, has certified bounds
\begin{equation}\label{eq_trace_flat_direction}
 r^\top\nabla^2\mathcal P r\le-a r^\top G r,
 \qquad \|r\|_\infty^2\le P_n n(r^\top G r).
\end{equation}
The curvature margin follows from the preceding quotient and the
operator error. For the second bound, before the harmless rounding,
\[
 \frac{\|r\|_\infty^2}{r^\top G r}
 \le\frac{\max_i(z_i^2/\widehat\varpi_i)}{\|z\|^2}
 \le\frac{P_n}{k\min_i\widehat\varpi_i}\le P_n n
\]
by Eq.~\eqref{eq_trace_phase_floor}. All quantities have polynomial
reciprocals where needed, so the rounding and the final quotient and
spread checks require only inverse-polynomial accuracy. The procedure
is deterministic and its guarantees hold at repeated eigenvalues
of $H$.

\paragraph{Final constant.}
Short-precision partial signing costs at most $\eta_{\rm ps}$; initial rounding and near-endpoint removals cost at most ${\varepsilon_{\rm round}}$;
small-input removals cost at most $\chi$; endpoint repairs cost at most
$e_{\rm end}$. Restoring variance scaling and the earlier discarded inputs,
the internal coefficient is at most
\[
 (2\sqrt{699/250+2{\varepsilon_{\rm scale}}}+{\varepsilon_{\rm round}}+\chi+e_{\rm end}+\eta_{\rm ps})\sqrt{1+{\varepsilon_{\rm scale}}}+{\varepsilon_{\rm discard}}
 < C_{\rm tr,*}=3.34427.
\]
To check the strict inequality, define
$\epsilon:=10^{-8}$ and
$b_*:=3.34427-10^{-6}-(3\cdot10^{-6}+10^{-7})(1+\epsilon)$.
Exact rational arithmetic gives
\begin{equation}\label{eq:trace:final-squared-margin}
 b_*=\frac{3344265899999969}{10^{15}}>0,\qquad
 b_*^2-4(699/250+2\epsilon)(1+\epsilon)>10^{-4}.
\end{equation}
Taking positive square roots and using
$\sqrt{1+\epsilon}\le1+\epsilon$ proves the strict inequality.
The one-time input approximation, with $\xi=10^{-8}$ as above, changes
the coefficient to at most
$C_{\rm tr,*}(1+6\xi)+2\xi<3.3443$.
These allowances give the stated discrepancy constant.

\paragraph{Entropy smoothing for the smaller constant.}\label{trace_sec_entropy}
We now reduce the dimension exponent by smoothing the matrix maximum.
The budget, coupling, preprocessing, and rounding allowances remain
unchanged. Set
\[
 L_{\rm log}:=\lceil\log_2(4n)\rceil,\qquad
 \tau_{\rm s}:=10^{-8}/L_{\rm log},
\]
and define, on matrix pairs,
\[
 \mathcal C({\mathsf P},y):=
 (U^{-1}+M(y)+\mathcal K_y(V),V^{-1}-M(y)+\mathcal K_y(U)),
 \qquad {\mathsf P}:=(U,V).
\]
Use the smoothed potential without a scalar reservoir:
\begin{equation}\label{eq_trace_entropy_potential}
 \mathcal P_{\rm s}(y):=\min_{U,V\succ0}
 \{\rho\tr[U+V]+\tau_{\rm s}\log\tr[\exp(\mathcal C({\mathsf P},y)/\tau_{\rm s})]\}.
\end{equation}
The trace and exponential in this formula act on the block-diagonal
matrix associated with the pair. The log-sum-exp inequalities give
\begin{equation}\label{eq_trace_entropy_sandwich}
 \|M(y)\|\le\mathcal P(y)\le\mathcal P_{\rm s}(y)
 \le\mathcal P(y)+\tau_{\rm s}\log(2n)
 \le\mathcal P(y)+10^{-8}.
\end{equation}
Let ${\mathsf Q}:=(W,Z)$ be its Gibbs state. Stationarity in $U,V$ gives
\[
 \tr[{\mathsf Q}]=1,\qquad {\mathscr J}{\mathsf Q}=\rho E,
 \qquad W=U\mathcal K_y(Z)U+\rho U^2,
 \quad Z=V\mathcal K_y(W)V+\rho V^2.
\]
The bounded-sublevel, positive-inverse, and unique-root arguments in
Section~\ref{soft_sec_geometry} use these identities and nonnegative
coupling coefficients. They therefore apply to $\beta_{\rm tr}$ and
$\iota=699/250$. In particular all the primal and dual bounds used above
remain valid. The budget inequalities needed there follow from the
fixed even polynomial and its $3/4$ endpoint power on the truncated
interval. No reservoir is used in these root estimates.

\paragraph{Transfer of the trace-corrected covariance.}
It is essential to distinguish the comparison response from the
derivative of the smoothed optimizer. At a smoothed root and for a
direction $r$, the inverse function theorem supplies a local comparison
path ${\mathsf P}_{\rm c}(t)$ satisfying
\[
 \mathcal C({\mathsf P}_{\rm c}(t),y+tr)=\mathcal C({\mathsf P},y),
 \qquad {\mathsf P}_{\rm c}(0)={\mathsf P}.
\]
Indeed the derivative with respect to ${\mathsf P}$ is
$-{\mathscr J}$, which is invertible by the strict dual identities.
Its first derivative satisfies Eq.~\eqref{rank_one_eq_response}.
The comparison value
\[
 \varphi_r(t):=\rho\tr[{\mathsf P}_{\rm c}(t)]
       +\tau_{\rm s}\log\tr[\exp(\mathcal C({\mathsf P},y)/\tau_{\rm s})]
\]
majorizes $\mathcal P_{\rm s}(y+tr)$ and agrees with it at zero.
The second term is constant along this path. Pairing the first and
second differentiated constraint equations with ${\mathsf Q}$ gives
\begin{align}
 \partial_i\mathcal P_{\rm s}&=p_iw_i-q_iz_i,
 \label{eq_trace_entropy_gradient}\\
 \frac12r^\top\nabla^2\mathcal P_{\rm s}r
 &\le\mathcal Q(r)+\sum_i m_i(r)-\frac12r^\top Gr.
 \label{eq_trace_entropy_comparison}
\end{align}
Here $G=\operatorname{diag}(\varpi_i)$ is exactly
Eq.~\eqref{eq_trace_metric}, evaluated at the smoothed root;
$\mathcal Q(r)$ and $m_i(r)$ use the comparison derivatives
$\dot U_{\rm c},\dot V_{\rm c}$, as in
Eq.~\eqref{eq_trace_exact_variation}. There is no entropy remainder
in Eq.~\eqref{eq_trace_entropy_comparison}.

The response graph, energy dominance, and local trace identities of
Lemma~\ref{lem_trace_covariance} require the response equation, the
strict dual identities, rank one, and lightness. They do not require
$\mathcal C=bE$. Thus that lemma, the exact mixed correction, and its
scalar certificate apply verbatim to the comparison response, with
$\nabla\mathcal P_{\rm s}$ in place of $\nabla\mathcal P$.
The covariance restriction proof also applies. In particular the drift
estimate in Eq.~\eqref{eq_trace_quantitative_drift}, the separated
gradient test, and the multiplicity in Eq.~\eqref{eq_trace_inertia}
hold for $\mathcal P_{\rm s}$.
The deterministic filter consequently supplies
\begin{equation}\label{eq_trace_entropy_direction}
 r^\top\nabla^2\mathcal P_{\rm s}r\le-a r^\top Gr,
 \qquad \|r\|_\infty^2\le P_n n(r^\top Gr),\qquad \|r\|_2=1.
\end{equation}
The phase trace floor in Eq.~\eqref{eq_trace_phase_floor} is unchanged.

\paragraph{Local steps and optimizer correction.}
We now use derivatives of the actual smoothed optimizer.
On the trace-zero dual space define the entropy energy
\[
 \mathcal E(B)^2:=\tau_{\rm s}\langle B,D\log_{\mathsf Q}[B]\rangle.
\]
For a local line let $A:={\mathsf P}'$, $B:={\mathsf Q}'$,
$F_2:=\mathcal A'{\mathsf Q}$, and $\theta:=r^\top Gr$.
The exact envelope identity and the coupling bound are
\begin{equation}\label{eq_trace_entropy_energy}
 \mathcal P_{\rm s}''=-\theta+e(A)^2+2\langle F_2,A\rangle
                       +\mathcal E(B)^2,
 \qquad |\langle F_2,D\rangle|\le P_n\sqrt\theta e(D).
\end{equation}
The Gibbs variance and relative-dual estimates in
Eqs.~\eqref{soft_eq_variance} and~\eqref{soft_eq_relv} apply here.
Together with the forced energy identity they show, on a coordinate
line with $\theta_0=\varpi_i(y)$,
\[
 e(A)+\mathcal E(B)\le P_n\sqrt{\theta_0},\qquad
 |\mathcal P_{\rm s}''|\le P_n\theta_0,
 \qquad |\varpi_i'|\le P_n(1+\sqrt{n\theta_0})\varpi_i
\]
in its comparison tube. To check the forcing without a reservoir,
write $\nu_i:=\tr[C_i]$. Its signed-matrix term obeys
\[
 \tr[{\mathsf Q}F_1^2]
 =\nu_i(p_i^2w_i+q_i^2z_i)
 \le P_n\nu_i(w_i+z_i)
 \le P_n(v_iw_i+u_iz_i)\le P_n\varpi_i,
\]
using $u_i,v_i\ge a\nu_i$ and the expanded light bounds. This is the
coordinate estimate of Section~\ref{soft_sec_local} with the reservoir
set to zero; its use does not require a floor of order $1/k$.

\paragraph{Face changes for the smoothed implementation.}
We retain the phase cutoff and deterministic small-input rounding.
At a regular endpoint, choose the score accuracy to satisfy
\[
 \delta\le\min\{e_{\rm end}/(16n^2),\delta_{\rm end}/8,
       \rho\delta_{\rm end}^3/(16(4\iota^2+2\iota))\}.
\]
For a move to $1$, write $d_i:=1-y_i\ge\delta_{\rm end}$,
$\nu_i:=\tr[C_i]\le1$, and $h_i:=\iota\beta_i u_i+d_i$.
Test the new face at the old $U$ and
$V^+:=(V^{-1}+h_iC_i)^{-1}$. Its minus constraint block is unchanged;
its plus block is at most the old block plus $2\delta C_i$.
Order monotonicity and translation by a scalar identity give
\[
 \mathcal P_{\rm s,old}-\mathcal P_{\rm s,new}
 \ge\rho\frac{h_iv_i^2}{\nu_i(1+h_iv_i)}-2\delta\nu_i
 \ge a\rho\delta_{\rm end}^3.
\]
For the second bound, the conservative trigger implies
$v_i\ge\delta_{\rm end}/(2\iota)$, while $h_i\ge\delta_{\rm end}$.
The fraction is increasing in $h_i,v_i$ and decreasing in $\nu_i$.
The move to $-1$ exchanges the two blocks. Thus only
$\widetilde O(n)$ regular endpoints occur. Their charged continuation
is verified below using Lemma~\ref{am_lem_endpoint}.

For a deferred or small-input removal, preserve $M$ and attenuate its
coupling slot before deleting it. The argument in
Lemma~\ref{gf_lem_faces} also applies with no reservoir. Explicitly,
write the slot as $e^{-s}\mathcal K_i$, let $K_s$ be its pair map,
and put $v(s):=\langle{\mathsf Q},K_s{\mathsf P}\rangle\ge0$.
The envelope and root forcing give
\[
 \mathcal P_{\rm s}'(s)=-v(s),\quad
 e({\mathsf P}')+\mathcal E({\mathsf Q}')\le P_n\sqrt{v(s)},
 \quad |v'(s)|\le P_nv(s).
\]
These estimates only use positivity and inclusion of the slot in the
full coupling. In particular they apply to small-input removals, not
only coordinates close to an endpoint. A step of order
$1/(P_n(1+\sqrt{nv(s)}))$ admits an anchored correction.
Track to a positive slot weight below the root-cache floor divided by
a fixed polynomial in $n,k$, then delete its remainder and correct once.
Its negative logarithm is polylogarithmic. The root inverse has a
uniform polynomial bound also at zero slot weight, so this last jump
fits the Newton ball. The arithmetic continuation in
Lemma~\ref{gf_lem_faces} supplies the required corrections and residual tests.

For a removal with potential drop $\Delta_i$, integration and
Cauchy--Schwarz bound its corrected stages by
$\widetilde O(1+\sqrt{n\Delta_i})$.
We apply this bound to positive-slot batches below.
The partial-signing residual, small-input signs, and final deferred
rounding have exactly their previous discrepancy allowances.

\paragraph{Hessian assembly.}
For completeness the smoothed Hessian is explicitly constructible.
Let $Lr:=\mathcal C_y[r]$, let
$Tr:=\mathcal A_y'[r]{\mathsf Q}$, and let $Q_{\mathsf Q}$ denote the
Gibbs covariance in Eq.~\eqref{soft_eq_q}. Then
\[
 J_{\rm s}:={\mathscr E}+{\mathscr J}Q_{\mathsf Q}{\mathscr J},\qquad
 B:=T-{\mathscr J}Q_{\mathsf Q}L,
\]
and the ordinary Hessian of the optimized value is
\begin{equation}\label{eq_trace_entropy_hessian}
 \nabla^2\mathcal P_{\rm s}
 =-G+L^\top Q_{\mathsf Q}L-B^\top J_{\rm s}^{-1}B.
\end{equation}
This follows by differentiating the primal objective twice and taking
the Schur complement in its minimizing variables.
In real Hermitian coordinates all maps have order $O(n^2)$ and
$k\le n^2$. The maps $L,T$ have explicit rank-one columns, while
$Q_{\mathsf Q}$ has logarithmic-mean coefficients in a basis diagonalizing
$W,Z$. Forming their coordinate matrices costs $\widetilde O(n^4)$.
Their products, the coupling blocks, and the inverse of $J_{\rm s}$
cost $\widetilde O(n^{2\omega_0})$, including short-precision inversion.
Indeed ${\mathscr E}\preceq J_{\rm s}\preceq P_n{\mathscr E}$ and
${\mathscr E}\succeq an^{-3/2}I$, so all inverse norms are polynomial.
Symmetrize and compute sufficiently accurate entries to enclose the
true scaled Hessian in operator norm.
The deterministic filter already proved above has the same
$\widetilde O(n^{2\omega_0})$ cost.

All root, metric, score, and filter enclosures have polynomial condition
and are evaluated to inverse-polynomial accuracy. The explicit anchored
corrections and scalar logarithmic-mean evaluation from
Section~\ref{sec:shared:smoothed-evaluation} apply unchanged; comparisons reserve
fixed fractions of the certified decreases below.
Finally, smoothing adds at most $10^{-8}$ to the initial potential.
The conservative endpoint comparison and numerical tolerances use
less than $e_{\rm end}/2$ in total, and
$10^{-8}<e_{\rm end}/2$. Their sum therefore fits the same
$e_{\rm end}=10^{-6}$ allowance in the final-constant calculation above.
Its exact rational margin and input-rounding transfer still give
$3.3443$, and hence partition constant $1.67215$.
All sign choices, root solves, and curvature searches in this
implementation are deterministic.

\paragraph{A small reservoir and two diagonal metrics.}
We control the matrix response of a curvature direction as well as
its coordinates.
The coupling and scalar certificate remain unchanged. In a phase
starting with $K$ coordinates, choose a positive dyadic
$\lambda_K$ within fixed factors of $a/(P_n n\sqrt K)$ and set
\begin{equation}\label{eq_trace_matrix_reservoir}
 \mathcal F_{\rm s}:=\mathcal P_{\rm s}+\lambda_K\sum_{i\in I}\beta_i,
 \qquad G_+:=G+\lambda_K\operatorname{diag}(-\beta_i'').
\end{equation}
Here and below $a$ is a sufficiently small fixed constant and $P_n$
is a sufficiently large fixed polylogarithm. Fix these choices once,
including the finite number of inequalities below. The optimizer
does not depend on $\lambda_K$. The total positive reservoir injection,
including initialization and all phase changes, is at most
\[
 \sum_{\rm phases}\lambda_KK\le\frac{Ca}{P_n n}\sum_{\rm phases}\sqrt K
 \le Ca/P_n.
\]
Choose it below $e_{\rm end}/4$. Reservoir removals are nonnegative.
Thus the sum of all accepted potential decreases is still bounded
by an absolute constant.

Gradient tests use $G$, not $G_+$. By
Eq.~\eqref{eq_trace_phase_floor} and the truncated budget bounds,
we can choose $\lambda_K$ so that
\[
 |\lambda_K\beta_i'|\le aP_n^{-1}\sqrt{\varpi_i/n}
\]
with any prescribed fixed slack. A failed separated gradient test
for $\mathcal F_{\rm s}$ therefore implies the hypothesis of
Eq.~\eqref{eq_trace_defect_generator} for $\mathcal P_{\rm s}$.
For every restricted covariance $K_q$, adding the reservoir gives
\[
 \frac12\tr[K_q\nabla^2\mathcal F_{\rm s}]
 \le-\frac34D_q+q
       -\frac{\lambda_K}{2}\tr[K_q\operatorname{diag}(-\beta_i'')].
\]
Here the drift is paired with $\nabla\mathcal P_{\rm s}$, whose
magnitude is at most $D_q/4$. Since $D_q$ is a fixed positive
multiple of $\tr[GK_q]$, the same restriction argument proves
Eq.~\eqref{eq_trace_inertia} with $G_+$ and
$\nabla^2\mathcal F_{\rm s}$, after decreasing its fixed constants.
Moreover, during this phase,
\begin{equation}\label{eq_trace_matrix_floor}
 G_+\succeq\frac{a}{P_n n\sqrt k}I,\qquad K/2<k\le K.
\end{equation}

\paragraph{Deterministic control of the matrix responses.}
Let $\mathcal V$ denote the bounded symmetric filter constructed above,
now for $H:=G_+^{-1/2}\nabla^2\mathcal F_{\rm s}G_+^{-1/2}$.
Its guards give $\|\mathcal V\|\le2$ and
$\tr[\mathcal V^2]\ge\pi_n k/2$. Its leakage outside the negative
spectral subspace can be made any prescribed inverse polynomial.
For a physical direction $r$, let $A(r),B(r)$ be the derivatives
of the actual smoothed primal and dual optimizers. Introduce the
four real-linear Hermitian-valued maps
\begin{align*}
 \mathcal T_1r&:=U^{-1/2}A_U(r)U^{-1/2},&
 \mathcal T_2r&:=V^{-1/2}A_V(r)V^{-1/2},\\
 \mathcal T_3r&:=W^{-1/2}B_W(r)W^{-1/2},&
 \mathcal T_4r&:=Z^{-1/2}B_Z(r)Z^{-1/2}.
\end{align*}
For $r=G_+^{-1/2}\mathcal Vx$, completing the square in
Eq.~\eqref{eq_trace_entropy_energy}, with the reservoir included,
gives
\begin{equation}\label{eq_trace_matrix_energy}
 e(A(r))+\mathcal E(B(r))\le P_n\|x\|_2,\qquad
 \|\mathcal T_\ell G_+^{-1/2}\mathcal V\|_{2\to F}\le P_n\sqrt n.
\end{equation}
Indeed, the direct term is $-r^\top G_+r$, and the coupling term
is bounded by $P_n\sqrt{r^\top Gr}e(A)$.
On the negative subspace the Hessian quadratic form is nonpositive;
the leakage contributes at most $\|x\|^2$ after tightening its error.
The primal relative Frobenius bound follows from
\[
 U^{-1/2}WU^{-1/2}\succeq aI/n,\qquad
 V^{-1/2}ZV^{-1/2}\succeq aI/n.
\]
For the dual bound, let $q_i$ be the eigenvalues of $\mathsf Q$.
The entropy energy of $B$ has coefficient
\[
 \tau_{\rm s}q_iq_j/L(q_i,q_j)\ge a\tau_{\rm s}/n
\]
on the entries of $\mathsf Q^{-1/2}B\mathsf Q^{-1/2}$, where
$L$ is the logarithmic mean and $L(q_i,q_j)\le\max(q_i,q_j)$.
This proves Eq.~\eqref{eq_trace_matrix_energy} for complex Hermitian
matrices as well.

Put $D_{\ell j}:=\mathcal T_\ell G_+^{-1/2}\mathcal Ve_j$.
For each $\ell$,
\begin{equation}\label{eq_trace_matrix_variance}
 \|\sum_jD_{\ell j}^2\|\le P_n n^2.
\end{equation}
To see this, fix a unit vector $v$. The real-linear map $D\mapsto Dv$
from Hermitian matrices with Frobenius norm has norm at most one and
rank at most $2n$. Its composition with the map in
Eq.~\eqref{eq_trace_matrix_energy} therefore has squared Frobenius
norm at most $2n(P_n\sqrt n)^2$. This is precisely
$\sum_j\|D_{\ell j}v\|^2$; enlarging $P_n$ proves the claim.

We now choose signs $\xi_j$ deterministically. The following
conditional-expectation calculation uses independent uniform signs
only as a proof device. Choose $s:=1/(P_n n)$ so that
$s^2\|\sum_jD_{\ell j}^2\|\le1$. Consider
\begin{equation}\label{eq_trace_matrix_objective}
 \|\mathcal V\xi\|^2-\frac{\pi_n}{100}\sum_i\cosh((\mathcal V\xi)_i)
 -\frac{\pi_n k}{1000n}\sum_{\ell=1}^4\sum_{\varepsilon=\pm1}
       \tr[\exp(\varepsilon s\sum_j\xi_jD_{\ell j})].
\end{equation}
Use the exact conditional expectation for the first two terms.
For a partial assignment with remaining set $J$, replace a matrix
penalty by
\[
 \tr[\exp(\varepsilon s S_\ell+(s^2/2)\sum_{j\in J}D_{\ell j}^2)],
 \qquad S_\ell:=\sum_{j\notin J}\xi_jD_{\ell j}.
\]
This is a pessimistic estimator: its two children have average at
most its current value. Apply the trace-exponential concavity
inequality~\cite[Corollary 3.3]{t12} and then
$\log\cosh(sD)\preceq s^2D^2/2$. No commutativity of the different
$D_{\ell j}$ is used. Initially each matrix penalty is at most
$ne^{1/2}$, whereas each scalar penalty has expectation at most $e^2$,
because every row of $\mathcal V$ has squared norm at most four.
The initial lower estimator is consequently at least $\pi_n k/3$.
At each assignment choose the child with the larger lower estimator.
Errors below $\pi_n/100$ per comparison leave a final value at least
$\pi_n k/4$.

Writing $z:=\mathcal V\xi$, positivity of all penalties and
$\|z\|^2\le4k$ give
\[
 \|z\|^2\ge\pi_n k/4,\qquad \|z\|_\infty\le P_n,
 \qquad \|\sum_j\xi_jD_{\ell j}\|\le P_n n.
\]
Both signs in the exponential penalty are needed for the last
operator-norm bound. Normalize $r:=G_+^{-1/2}z/\|G_+^{-1/2}z\|_2$
and put $\theta:=r^\top G_+r$. The filter leakage, chosen smaller
than a fixed inverse-polynomial multiple of $\pi_n$, gives
\begin{equation}\label{eq_trace_matrix_direction}
 r^\top\nabla^2\mathcal F_{\rm s}r\le-a\theta,\qquad
 R:=\|r\|_\infty+\max_\ell\|\mathcal T_\ell r\|
 \le P_n d_k\sqrt\theta,\qquad
 d_k:=\min\{\sqrt n,n/\sqrt k\}.
\end{equation}
For the matrix term, combine the last display with
Eq.~\eqref{eq_trace_matrix_energy}. For the coordinate term,
Eq.~\eqref{eq_trace_matrix_floor} gives
$\|r\|_\infty^2/\theta\le P_n n/\sqrt k$.
This is at most $P_n d_k^2$ because $1\le k\le n^2$.

The sequential sign choices cost $\widetilde O(k^2+kn^3)$ once
the response columns are available. Scalar conditional expectations
are maintained as in the earlier flat-vector construction. For the
matrix terms maintain $S_\ell$ and the sum of the remaining squares,
and evaluate the eight trace exponentials at each child by symmetric
matrix-function evaluation. This costs $\widetilde O(n^3)$ per choice.
Every retained lower estimator is positive, so its matrix penalties
are bounded by $O(n/\pi_n)$. Their variance terms are PSD and bounded
in norm. Including both signs then bounds all their exponent matrices
in norm by $O(\log(2n/\pi_n))$. A trial child changes this norm by
a polynomially bounded amount, in fact $s\|D_{\ell j}\|\le P_n/\sqrt n$.
Thus the scalar and matrix functions, response columns, and all
conditional comparisons can be enclosed at logarithmic working
precision with any prescribed inverse-polynomial error. Choose $P_n$
in $s$ also to bound $s\|D_{\ell j}\|$ by one. The selection has no
randomized fallback.

\paragraph{A finite-step estimate from linear primal and dual paths.}
Matrix response control at one root does not by itself control the
actual optimizer along a longer step. We instead use linear comparison
paths and bound their entropy residual. Write $A:=A(r)$, $B:=B(r)$,
and, for this paragraph only, define
\[
 \mathsf P_t:=\mathsf P+tA,\qquad \mathsf Q_t:=\mathsf Q+tB,
 \qquad y_t:=y+tr.
\]
For $|t|R\le a/P_n$, these matrices remain positive and comparable
to their base values, $\tr[\mathsf Q_t]=1$, and all coordinates remain
at endpoint distance at least $\delta_{\rm end}/2$. Define
\[
 S(t):=\rho\tr[\mathsf P_t]+\langle\mathsf Q_t,
       \mathcal C(\mathsf P_t,y_t)\rangle
       -\tau_{\rm s}\tr[\mathsf Q_t\log\mathsf Q_t]
       +\lambda_K\sum_i\beta(y_i+tr_i).
\]
Stationarity and its first derivative imply
\[
 S(0)=\mathcal F_{\rm s}(y),\qquad S'(0)=\mathcal F_{\rm s}'[r],
 \qquad S''(0)=\mathcal F_{\rm s}''[r,r].
\]
Let $c_0,c_1$ be the scalar multipliers in the Gibbs equation and
its derivative. The Hermitian residual
\[
 E(t):=\mathcal C(\mathsf P_t,y_t)-\tau_{\rm s}\log\mathsf Q_t
                         -(c_0+c_1t)I
\]
satisfies $E(0)=E'(0)=0$. We claim that, on this linear comparison
interval, there is a PSD pair $J(t)$ with
\begin{equation}\label{eq_trace_matrix_residual}
 -J(t)\preceq E(t)\preceq J(t),\qquad
 \|J(t)\|\le P_nR^2t^2,\qquad
 \langle\mathsf Q_t,J(t)\rangle\le P_n\theta t^2,
 \qquad |S'''(t)|\le P_nR\theta.
\end{equation}
Here the bounds use $e(A)^2+\mathcal E(B)^2\le P_n\theta$,
which follows from the envelope identity for the selected negative
direction, and not from the norm of its sign seed.

We give the majorants explicitly enough to verify both different
bounds in Eq.~\eqref{eq_trace_matrix_residual}. Along the linear path,
the second derivative of each constraint block is a sum of an inverse
quadratic, a term $\mathcal K_{yy}[r,r]\mathsf P_t$, and
$2\mathcal K_y'[r]A$. An inverse quadratic is PSD, has operator norm
at most $P_n\max_\ell\|\mathcal T_\ell r\|^2$, and weighted trace at
most $P_ne(A)^2$. The direct coupling term has a PSD absolute majorant
with norm at most $P_n\|r\|_\infty^2$ and weighted trace at most
$P_nr^\top Gr$. Use $|\beta''|/\beta\le P_n$ and the bounded
operator norms of $\mathcal K_y(U),\mathcal K_y(V)$.

For the mixed term, rank one and scalar Young's inequality majorize
each absolute coefficient, up to a fixed factor, by
\[
 (-\beta_i'')r_i^2 C_i P C_i
 +\frac{(\beta_i')^2}{-\beta_i''}C_i A_P P^{-1}A_P C_i,
\]
where $P$ is the relevant primal block. This uses
$|\tr[C_iA_P]|^2\le\tr[C_iP]\tr[C_iA_PP^{-1}A_P]$.
The operator norm of the sum is at most
$P_n(\|r\|_\infty^2+\|P^{-1/2}A_PP^{-1/2}\|^2)$.
Its weighted trace is at most $P_n(\theta+e(A)^2)$: pair the second
sum with the dual block, use
$(\beta_i')^2/[(-\beta_i'')\beta_i]\le P_n$, and then
$\mathcal K_y(Z)\preceq U^{-1}WU^{-1}$ and its exchanged identity.
All these bounds remain valid on the linear paths by fixed-factor
Loewner comparisons and scalar budget comparisons; no order comparison
of noncommuting products is required.

Finally, $-D^2\log_Q[B,B]$ is PSD, has norm at most
$2\|Q^{-1/2}BQ^{-1/2}\|^2$, and satisfies
\[
 -\tr[Q D^2\log_Q[B,B]]=\tr[B D\log_Q[B]].
\]
For the norm assertion use the resolvent representation of $\log$:
if $T:=Q^{-1/2}BQ^{-1/2}$ and $R_u:=(Q+uI)^{-1}$, then
$BR_uB\preceq\|T\|^2Q$, so
$2\int_0^\infty R_uBR_uBR_u \d u\preceq2\|T\|^2I$.
The trace identity follows by differentiating
$\tr[Q D\log_Q[B]]=\tr[B]$ along $Q+tB$.
Multiplication by $\tau_{\rm s}$ gives weighted trace
$\mathcal E(B)^2$. Entropy energies along comparable linear matrices
are comparable by the same resolvent formula.
Integrate these PSD absolute majorants for $E''$ twice. This proves
the first three assertions of Eq.~\eqref{eq_trace_matrix_residual}.

For the last assertion, differentiate $S$ three times. Inverse cubic
words cost at most $P_nR e(A)^2$; differentiating the direct coupling
or its scalar budget costs at most $P_nR\theta$. The terms containing
$B$ paired with the second constraint derivative cost at most its
PSD weighted majorant times $P_n\max_\ell\|\mathcal T_\ell r\|$,
because $-R\mathsf Q_t\preceq B\preceq R\mathsf Q_t$ up to a fixed
factor. The entropy cubic costs at most
$P_nR\mathcal E(B)^2$, by the preceding positive resolvent majorant.
The scalar reservoir cubic is bounded by
$P_n\|r\|_\infty\lambda_K\sum_i(-\beta_i'')r_i^2$.
These are all terms, since both matrix paths are linear, and prove
the last assertion.

For completeness, the entropy residual converts this comparison
into an upper bound on the optimized value. If $Q\succ0$ has trace
one and $\|E\|\le\tau/2$, Golden--Thompson and
$e^x\le1+x+x^2$ for $|x|\le1/2$ give
\begin{equation}\label{eq_trace_matrix_fenchel}
 \tau\log\tr[\exp(\log Q+E/\tau)]-\tr[QE]
 \le\tau^{-1}\tr[QE^2].
\end{equation}
Indeed the exponential trace is at most $\tr[Qe^{E/\tau}]$;
apply the scalar inequality spectrally and then $\log(1+x)\le x$.
If $-J\preceq E\preceq J$, factor
$E=J^{1/2}TJ^{1/2}$ with $\|T\|\le1$. It follows that
$E^2\preceq\|J\|J$. Apply this to
Eq.~\eqref{eq_trace_matrix_residual} and use $\tau_{\rm s}^{-1}\le P_n$.
Evaluating the primal objective at $\mathsf P_t$, rather than its
minimizer, yields the rigorous finite-step bound
\begin{equation}\label{eq_trace_matrix_step_bound}
 \mathcal F_{\rm s}(y_t)\le\mathcal F_{\rm s}(y)
 +t\mathcal F_{\rm s}'[r]+\frac{t^2}{2}\mathcal F_{\rm s}''[r,r]
 +P_nR\theta|t|^3+P_nR^2\theta t^4.
\end{equation}
The smallness $\|E(t)\|\le\tau_{\rm s}/2$ is ensured by the same
choice of $P_n$ and step below. This argument uses no transport
assumption about the optimizer between the two endpoints.

Orient $r$ so that its directional gradient is nonpositive, allowing
error at most $a\theta t$ in this comparison, and take a dyadic step
within fixed factors of
\begin{equation}\label{eq_trace_matrix_steps}
 t:=\frac{a}{P_n^2d_k\sqrt\theta}.
\end{equation}
Eqs.~\eqref{eq_trace_matrix_direction} and~\eqref{eq_trace_matrix_step_bound}
give a true decrease at least $a/(P_n d_k^2)$, after enlarging $P_n$.
The chosen branch and the comparison value are bounded throughout
the segment. We evaluate only this oriented trial. Its new optimizer
is initialized by the explicit data homotopy in
Section~\ref{soft_sec_init}, at cost $\widetilde O(\sqrt n)$ corrected
stages; a single warm correction over the long step is not assumed.
Inward rounding and certified value comparison reserve a fixed
fraction of the decrease. All required thresholds have polynomial
reciprocals by Eq.~\eqref{eq_trace_matrix_floor}.

\paragraph{Gradient sweeps and batched removals for the trace budget.}
To avoid paying for a root after every short coordinate move, use
the deterministic frozen-root sweeps of Lemmas~\ref{am_lem_frozen}
and~\ref{am_lem_sweep}. The necessary adaptation is to keep the
original, smaller weight in the gradient test. At a light base put
\[
 \kappa_i^0:=\iota(v_i^0w_i^0+u_i^0z_i^0),\qquad
 \varphi_i(t):=(w_i^0-z_i^0)t+(\kappa_i^0+\lambda_K)\beta(t),
 \qquad d_0:=\sum_i\sqrt{\kappa_i^0}|y_i-y_i^0|.
\]
The proof of Lemma~\ref{am_lem_frozen} bounds both coordinate forcings
by $P_n\sqrt{\kappa_i^0}$ even without a reservoir. The scalar
reservoir has zero derivative in root variables, so it cancels
exactly from the gradient approximation error. Consequently, for
$d_0\le a/(P_n^2\sqrt n)$, that proof gives
\[
 \|z(y)-z_0\|_{\mathrm{Met}_0}\le P_nd_0,\qquad
 |\partial_i\mathcal F_{\rm s}(y)-\varphi_i'(y_i)|
       \le P_n^2\sqrt{\kappa_i^0}d_0,
 \qquad \kappa_i(y)\asymp\kappa_i^0.
\]
Here $z=(\mathsf P,\mathsf Q)$ and $\mathrm{Met}_0$ is the joint
energy metric. These claims require only a small weighted parameter
neighborhood, not lightness at every intermediate point.
Choose high and low thresholds as in Lemma~\ref{am_lem_sweep}, with
normalization $\sqrt{\kappa_i/n}$. Make the high threshold small
enough for the failed-gradient condition proved above.
Concavity of every $\varphi_i$ then gives a saturated sweep decrease
at least $1/(P_n n)$, or an unsaturated exit with every remaining
normalized gradient below $h_n/(8\sqrt n)$. The same proof gives
at most $k_0$ complete coordinate updates and one partial update
per saturated sweep. Its positive-decrease floor now uses
$\kappa_i\ge a/(P_n nk)$ from the phase cutoff in place of the
reservoir floor; it is still inverse polynomial.

All marked coordinates are removed in one attenuation batch,
preserving the modeled signed sum. Small-input removals are likewise
batched once per phase. The positive-submap calculation for face
changes applies to the sum of any such slots: positivity, rather
than its rank, was the only property used. If its matrix-potential
drop is $\Delta$, its joint energy arc is at most $P_n\sqrt\Delta$
and it costs $\widetilde O(1+\sqrt{n\Delta})$ corrected stages.
Hold $\lambda_K$ fixed during the batch and remove its scalar
reservoir contributions at the end.

For clarity the charging argument also transfers with these smaller
weights. Along a batch, no remaining coefficient changes. For
$\widetilde g_i:=\partial_i\mathcal F_{\rm s}/\sqrt{\kappa_i}$,
the forcing bounds and relative scalar comparisons give
\[
 |\widetilde g_i'|\le P_n a(s)
       +P_n\sqrt n|\widetilde g_i|a(s),\qquad
 a(s):=e(\mathsf P')+\mathcal E(\mathsf Q').
\]
Until $|\widetilde g_i|=h_n/(2\sqrt n)$, the right side is at most
$P_na(s)$. Thus a batch with drop below $1/(P_n' n)$, for a
sufficiently large fixed polylogarithm $P_n'$, cannot turn the low
exit into a successful high test. It also keeps the metric comparable
by the same first-exit argument. The tiny-slot final correction is
chosen below this separation margin. Larger batch drops can occur
only $\widetilde O(n)$ times. A subsequent sweep must otherwise be
preceded by a curvature move, a regular endpoint, or a phase change.

There are $\widetilde O(n)$ curvature moves by
Eq.~\eqref{eq_trace_matrix_steps}, $\widetilde O(n)$ regular endpoints
by their unchanged matrix-potential decrease, and $O(\log n)$ phases.
It follows that there are $\widetilde O(n)$ sweeps and batches, all
deterministically. Summing the stage bounds by Cauchy--Schwarz gives
$\widetilde O(n)$ batch stages. Regular endpoints and the long
curvature trials use at most $\widetilde O(n^{3/2})$ initialization
stages in total. Coordinate updates cost $\widetilde O(n^4)$.
The trace cutoff, recorded coefficients, and final deterministic
rounding have the same discrepancy allowances as before.

Root, Hessian, filter, matrix-exponential, and sign comparisons all
have polynomial condition bounds and inverse-polynomial slack.
Use the anchored enclosures of Section~\ref{sec:shared:smoothed-evaluation},
with sufficiently fine common grids for the new comparisons and
finite-step decrease. The additional response congruences have
polynomial inverse norms by the primal and dual lower bounds.
The extra reservoir injection is below $e_{\rm end}/4$, smoothing
costs at most $10^{-8}$, and all numerical and endpoint errors can
still be kept below $e_{\rm end}/2$. Their sum is below the original
$e_{\rm end}=10^{-6}$ allowance. The discrepancy constant remains
$3.3443$, with partition constant $1.67215$.

\paragraph{A deterministic grouped reduction with dimensional slack.}
The preceding construction can be accelerated without changing its
budget, discrepancy allowances, or curvature step. We first remove the
preprocessing bottleneck. The notation $D$ in the next lemma denotes
a real vector-space dimension.

\begin{lemma}[Deterministic reduction with a factor-two slack]
\label{trace_lem_deterministic_groups}
Under the hypotheses of Lemma~\ref{rank_one_lem_grouped_partial_signing}
with $d_{\rm vec}=D$, a deterministic algorithm returns
$x\in[-1,1]^m$ with at most $2D$ fractional coordinates and
$\|\sum_i x_i b_i\|_2\le\eta$ using
$\widetilde O(mD+D^{\omega_0})$ arithmetic operations.
\end{lemma}
\begin{proof}
Use the balanced tree and exact node sums of
Lemma~\ref{rank_one_lem_grouped_partial_signing}, maintaining at most
$2D$ active groups. A split gives at most $4D$ groups. We first give
a deterministic small reduction for a matrix $C$ with $r_0\le4D$
columns of norm at most two and inherited dyadic coefficients $y$.
Given $0<\varepsilon<1$, it leaves at most $2D$ fractional
coordinates and changes $Cy$ by norm less than $\varepsilon/128$.
If at most $2D$ coordinates are fractional, return $y$.

Choose a dyadic $\sigma_{\rm r}$ with
$\varepsilon/(2^{11}r_0)<\sigma_{\rm r}\le\varepsilon/(2^{10}r_0)$.
Define $\ell:=\lceil\log_2(64r_0)\rceil$, $H:=4\ell$,
$L_{\rm end}:=\lceil\log_2(2/\sigma_{\rm r})\rceil$,
$T_*:=128H^2L_{\rm end}+1$, and
$\nu:=\varepsilon/(2^{14}T_*r_0)$.
Choose a dyadic ridge $\nu^2/2<\delta\le\nu^2$.
If $u_{\rm in}$ is the inherited coefficient mesh, let $u$ be the
largest dyadic number not exceeding
$\min\{\sigma_{\rm r}/(2^{16}H^2),\nu,u_{\rm in}\}$.
All initial coefficients lie on this grid.
At every iteration, freeze coordinates at endpoint distance at
most $\sigma_{\rm r}$ by setting them to their nearest endpoint.
Stop when at most $2D$ coordinates remain active.

For the active matrix $A$ and current active coefficients $x$, define
$s_i:=1-|x_i|$, $\mathsf S:=\operatorname{diag}(s_i)$,
$\mathsf B:=A\mathsf S$, $J_\delta:=\delta I+\mathsf B\mathsf B^\top$,
and $\Pi_\delta:=I-\mathsf B^\top J_\delta^{-1}\mathsf B$.
For $2D<r\le r_0$, the singular-value formulas give
\[
 0\prec\Pi_\delta\preceq I,\qquad
 \tr[\Pi_\delta^2]\ge r-D,\qquad
 \|\mathsf B\Pi_\delta\|\le\sqrt\delta/2.
\]
Indeed the nontrivial eigenvalues are
$\delta/(\delta+\sigma_j(\mathsf B)^2)$, and at least $r-D$
eigenvalues equal one. The residual singular values are
$\delta\sigma_j/(\delta+\sigma_j^2)\le\sqrt\delta/2$.
Form the ridge inverse in exact arithmetic by positive-definite
block inversion, as in Lemma~\ref{rank_one_lem_regularized_partial_signing}.
With multiplication exponent $\omega_0>2$, its recursive cost is
$\widetilde O(D^{\omega_0})$. Round the entries of the symmetric
$\Pi_\delta$ on the largest dyadic mesh not exceeding $\nu/(8r^2)$,
using the same rounding for transposed entries. This gives $V$ with
$\|V-\Pi_\delta\|\le\nu/\sqrt r$. The rounding uses
$O(r^2\log(2r/\nu))$ arithmetic operations.

For independent uniform signs $\xi$, consider the scalar estimator
\begin{equation}\label{trace_eq_group_estimator}
 \Psi(\xi):=\|V\xi\|_2^2-\frac1{100}\sum_{i=1}^r\cosh((V\xi)_i).
\end{equation}
Since $\nu\le1/64$, we have $\|V\|\le65/64$ and
$\tr[V^\top V]\ge r-D-r/16$. Each row of $V$ has squared
norm at most $(65/64)^2$. The elementary inequality
$\cosh t\le e^{t^2/2}$ therefore gives
\[
 \E[\Psi]\ge r-D-r/16-r/50>\frac34(r-D).
\]
Here $r<2(r-D)$, and $e^{(65/64)^2/2}<2$.
Fix the signs successively by conditional expectation. If $u$ is
the current prefix sum, the two conditional expectations of its
quadratic term are
\[
 \|u\pm V_j\|_2^2+\sum_{h>j}\|V_h\|_2^2,
\]
and the conditional expectation of row $i$'s hyperbolic cosine is
\[
 \cosh(u_i\pm V_{ij})\prod_{h>j}\cosh(V_{ih}).
\]
Maintain the prefix, the remaining squared norm, and these row
products. At any accepted prefix the conditional quadratic term is
at most $2r$, while the conditional estimator remains positive.
Hence each conditional hyperbolic cosine is at most $200r$. A trial
changes an argument by at most $65/64$, and each remaining product
is at most $e^{(65/64)^2/2}<2$. The arguments are therefore
$O(\log(2r))$ and the evaluated quantities have polynomial magnitude.
Taylor series with inverse-polynomial error, propagated through the
at most $r$ scalar factors, evaluate each candidate using
$\widetilde O(r)$ arithmetic operations.
Evaluate a candidate to absolute error at most $1/(1000r)$ and
choose the larger approximation. The total comparison loss is at
most $1/500$, so the resulting signs satisfy
$\Psi(\xi)\ge(r-D)/2$.

Put $\widetilde v:=V\xi$, computed exactly on its dyadic grid.
Then $\|\widetilde v\|_2^2\ge(r-D)/2$ and
$\|\widetilde v\|_2^2\le2r$. Positivity of $\Psi$ implies
$\cosh(\widetilde v_i)\le200r$, whence
\[
 \|\widetilde v\|_\infty\le\log(400r)
 <\log_2(64r)\le\ell,\qquad
 \|\widetilde v-\Pi_\delta\xi\|_2\le\nu.
\]
For the strict logarithmic inequality, use $\log400<6$ and
$\log r\le\log_2r$. In particular the direction has squared norm at least $(r-D)/4$
and maximum entry at most $\ell$.

Set $h:=\mathsf S\widetilde v/H$, so $|h_i|\le s_i/4$.
For analysis define $\Phi(x):=\sum_i-\log(1-x_i^2)$ over active
coordinates. Its directional derivative is
\[
 g:=\nabla\Phi(x)^\top h
   =\sum_i\frac{2x_i\widetilde v_i}{H(1+|x_i|)}.
\]
In exact arithmetic form this sum and choose $\tau\in\{-1,1\}$
with $\tau g\ge0$. The estimates also permit a computed sign with
$\tau g\ge-r/(512H^2)$. Round $x+\tau h$ toward zero to the grid
of mesh $u$, then freeze all newly triggered coordinates.

Along the unrounded segment, the distance to the endpoint nearest
$x_i$ is at most $5s_i/4$. Since
$\phi''(z)=(1-z)^{-2}+(1+z)^{-2}$ for $\phi(z):=-\log(1-z^2)$,
Taylor's theorem gives
\[
 \Phi(x+\tau h)-\Phi(x)
 \ge\tau g+\frac8{25H^2}\|\widetilde v\|_2^2
 \ge(\frac2{25}-\frac1{256})\frac{r-D}{H^2}.
\]
Here $r<2(r-D)$. Every endpoint distance before inward rounding
is at least $3\sigma_{\rm r}/4$. On the inward segment
$|\phi'|\le4/\sigma_{\rm r}$, so rounding loses at most
$4ru/\sigma_{\rm r}\le(r-D)/(8192H^2)$.
The net increase exceeds $(r-D)/(16H^2)>D/(16H^2)$.
Assign a frozen coordinate the finite value
$Q_{\rm end}:=\log(2/\sigma_{\rm r})$ in the bookkeeping potential.
A newly frozen coordinate had endpoint distance at least
$3\sigma_{\rm r}/4$, so this assignment cannot reduce that potential.
Starting after the initial freezes, it stays between zero and
$r_0Q_{\rm end}\le4DL_{\rm end}$. Thus fewer than $T_*$ steps
are possible before at most $2D$ coordinates remain.

Each unrounded step has residual
\[
 \|Ah\|_2\le\frac{\sqrt r}{H}(\sqrt\delta/2+2\nu)
 \le5r_0\nu/8.
\]
Rounding contributes at most $2r_0u$ per step. All endpoint freezes
together contribute at most $2r_0\sigma_{\rm r}$, since each
coordinate is frozen once. Hence the total residual is at most
\[
 2r_0\sigma_{\rm r}+T_*(5r_0\nu/8+2r_0u)
 \le\varepsilon(\frac2{2^{10}}+\frac{5/8+2}{2^{14}})
 <\varepsilon/128.
\]
All updates preserve the cube. There are polylogarithmically many
steps, each using $\widetilde O(D^{\omega_0})$ arithmetic operations;
scalar conditional choices cost $\widetilde O(D^2)$ per step.
The derivative sum uses $O(r)$ exact operations, and grid rounding
uses binary search on bounded intervals.

For the full tree, let $R_{\rm tree}:=\max\{1,\lceil\log_2m\rceil\}$
and use the power of two $m\le M_{\rm grp}<2m$ from
Lemma~\ref{rank_one_lem_grouped_partial_signing}. Normalize each
node sum by $M_{\rm grp}$ and use
$\varepsilon:=\eta/(M_{\rm grp}R_{\rm tree})$ at every level.
The root coefficient is zero, on mesh one. Each split inherits
its parent's coefficients; each small reduction leaves at most
$2D$ active groups. All meshes chosen from the displayed bounds
have logarithms $O(\log(2mD/\eta))$: taking the largest admissible
dyadic mesh refines an inherited grid only when a parameter bound
requires it. All scalar parameter and grid searches have logarithmic
cost as well.
After $R_{\rm tree}$ levels, all active groups are singletons.
Their unnormalized residual allowances sum to at most $\eta$.
Exact node-sum preparation costs $O(mD)$, and the final tree traversal
costs $O(m)$. The logarithmically many small reductions therefore
give total arithmetic cost $\widetilde O(mD+D^{\omega_0})$.
\end{proof}

For the matrix problem apply this lemma with $D=n^2$ after the same
input conversion. It leaves at most $2n^2$ fractional slots. Pad all
retained matrices by zeros to order
$N:=2^{\lceil\log_2(2n)\rceil}<4n$, and run the matrix algorithm with
this working order. Then $k\le2n^2\le N^2$. All variances, operator
norms, residual allowances, and returned signs are preserved. In
particular use $\rho=10^{-8}/N$, so $N\rho=10^{-8}$ is unchanged.
The initial potential bound and every discrepancy allowance therefore
remain the same. This deterministic replacement of the grouped step
in Algorithm~\ref{alg:shared:preprocess} costs
$\widetilde O(mn^2+n^{2\omega_0})$. Below we again write $n$ for
the working order; the original and working orders differ by less
than a factor four.

\paragraph{Solving all response columns at one root.}
The primal and dual responses of all coordinate directions share their
optimizer and their linear operator. We compute them together.

\begin{lemma}[Block response construction]\label{trace_lem_block_responses}
At a corrected root with $k\le n^2$ active coordinates, the scaled
Hessian and all four response maps in
Eq.~\eqref{eq_trace_matrix_energy}, composed with the deterministic
filter, can be computed to any prescribed inverse-polynomial
accuracy in
\begin{equation}\label{trace_eq_block_work}
 \widetilde O(n^3+kn^{\omega_0}+n^2k^{\omega_0-1}+k^{\omega_0})
\end{equation}
arithmetic operations. The same bound includes the matrix
conditional selection producing Eq.~\eqref{eq_trace_matrix_direction}.
\end{lemma}
\begin{proof}
Write Hermitian matrices in real orthonormal coordinates and let
$\Phi$ have columns $\operatorname{vec}(C_i)$. Its row dimension is
$O(n^2)$. Rank one gives $C_iXC_i=\tr[C_iX]C_i$, so the coupling
in these coordinates is
\[
 \Phi\operatorname{diag}(\iota\beta_i)\Phi^\top.
\]
An application to $k$ right-hand sides consists of products of
shapes $(k,O(n^2),k)$ and $(O(n^2),k,k)$. Dividing the long dimension
into blocks of $k$ gives cost $\widetilde O(n^2k^{\omega_0-1})$.
Explicitly forming $\Phi$ and all rank-one derivative columns costs
$O(n^2k)$, already covered by this bound.

Prepare approximate spectral decompositions for the two Lyapunov
operators and the Gibbs covariance. For the $U$ block, set
$T_U:=U^{-1/2}WU^{-1/2}$. Applying $\mathscr E^{-1}$ to $F_U$
amounts to solving
\[
 \widehat D_U T_U+T_U\widehat D_U=U^{1/2}F_UU^{1/2}.
\]
In an eigenbasis of $T_U$, divide each entry by the corresponding
$t_i+t_j$. The $V$ block is identical. The primal and dual lower
bounds make these denominators inverse polynomial.
In an eigenbasis of the block-diagonal $\mathsf Q=(W,Z)$, its Gibbs
covariance multiplies entry $(i,j)$ by
$\tau_{\rm s}^{-1}\mathsf L(q_i,q_j)$ and subtracts
$\tau_{\rm s}^{-1}\tr[\mathsf Q F]\mathsf Q$ on an argument $F$,
where $\mathsf L$ is the logarithmic mean.
Use the divided exponential series for nearby eigenvalues and the
quotient formula otherwise.

These decompositions cost $\widetilde O(n^3)$ arithmetic operations.
Indeed, a Hermitian Jacobi rotation at a largest off-diagonal entry
reduces squared off-diagonal Frobenius norm by a factor at most
$1-2/(n(n-1))$. There are $O(n^2\log(2n/\epsilon))$ rotations to
residual $\epsilon$, for the polynomial norm bounds here. Maintain
pivot magnitudes in a heap; updating two rows and columns and the
accumulated basis costs $\widetilde O(n)$ per rotation. Scalar
square roots are approximated by bisection. Orthogonal rotations
accumulate inverse-polynomial approximation errors by a polynomial
factor. Residual and orthogonality checks therefore give the required
spectral accuracy without an eigenvalue-gap assumption. The case
$n=1$ uses scalar evaluation only.
For each of the $k$ right-hand sides, the basis changes, left and
right products, entrywise divisions, and trace centering cost
$\widetilde O(n^{\omega_0})$. Reuse these decompositions for every
column and iteration at this root.

The Gibbs covariance obeys
$\langle F,Q_{\mathsf Q}F\rangle\le\tau_{\rm s}^{-1}\tr[\mathsf QF^2]$,
since $\mathsf L(q_i,q_j)\le(q_i+q_j)/2$ and its centering term is
nonnegative. Complete positivity of $\mathcal K_y$, paired with the
dual identities, gives
$\tr[\mathsf Q(\mathscr JD)^2]\le C\langle D,\mathscr ED\rangle$:
use $\mathcal K_y(D_V)^2\preceq\|\mathcal K_y(V)\|
\mathcal K_y(D_VV^{-1}D_V)$ and the exchanged inequality, together
with $(A-B)^2\preceq2A^2+2B^2$.
Thus
\[
 \mathscr E\preceq J_{\rm s}:=\mathscr E+\mathscr JQ_{\mathsf Q}\mathscr J
 \preceq(1+C/\tau_{\rm s})\mathscr E.
\]
The $\mathscr E$-preconditioned Chebyshev inverse consequently has
polylogarithmic degree for the required inverse-polynomial error.
The polynomial bounds on $\mathscr E$ and $\mathscr E^{-1}$ control
its similarity to a symmetric positive operator. Second-kind
Chebyshev bounds give only polynomial amplification of action
errors through this finite recurrence. Choosing smaller
inverse-polynomial action tolerances therefore attains the target
operator error in the same arithmetic cost.
Apply this one fixed polynomial simultaneously to the columns of
$B=T-\mathscr JQ_{\mathsf Q}L$ in
Eq.~\eqref{eq_trace_entropy_hessian}. Each iteration has the
preceding block cost. Recover the dual response columns from
$Q_{\mathsf Q}$ and these primal responses. The products
$L^\top Q_{\mathsf Q}L$ and $B^\top J_{\rm s}^{-1}B$ have the same
$\widetilde O(n^2k^{\omega_0-1})$ bound. Diagonal scaling then gives
the required Hessian, including its reservoir term.
The filter costs $\widetilde O(k^{\omega_0})$. Composing each
response map with this filter again costs
$\widetilde O(n^2k^{\omega_0-1})$, and all relative congruences
cost $\widetilde O(kn^{\omega_0})$.

Finally form all variance matrices $\sum_jD_{\ell j}^2$ in
$\widetilde O(kn^{\omega_0})$ work. The matrix conditional estimator
used above has order-$n$ Hermitian arguments of norm $O(\log n)$
at every accepted prefix and either next candidate. Indeed its
positive summands bound both signed exponentials, its remaining
variance has norm at most $O(\log n)$, and a single signed increment
has bounded norm by Eq.~\eqref{eq_trace_matrix_variance}.
Use the factorial Taylor evaluation in Lemma~\ref{round_lem_round},
with order-$n$ multiplication at exponent $\omega_0$,
with its error target tightened by the polynomial weights in the
estimator. Each of the $k$ sign choices costs
$\widetilde O(n^{\omega_0})$. Scalar conditional quadratic and
hyperbolic-cosine terms take $\widetilde O(k^2)$ additional work.
All comparisons retain the existing slack, proving
the arithmetic bound in Eq.~\eqref{trace_eq_block_work}.
\end{proof}

\paragraph{Endpoint continuations and the total cost.}
At a regular endpoint, use the charged continuation of
Lemma~\ref{am_lem_endpoint}. We verify its
applicability to the present trace budget. At a positive trigger put
$d=1-y_i$, $\beta=\beta_{\rm tr}(y_i)$ and use $a,b,p,q$ for the
four traces in Eq.~\eqref{am_eq_endpointdata}. Its two additions are
\[
 \mathsf A=(\mathcal K_i(Z),\mathcal K_i(W)),\qquad
 E_i=((\iota\beta b-d)C_i,(\iota\beta a+d)C_i).
\]
They are positive and bounded at the old root. Here
$\mathcal K_i(X)=\iota\beta C_iXC_i$ and $0\le\beta\le1$.
The self-slot inequality $q\ge\iota\beta b^2p$ gives exactly
\[
 v_0:=\langle\mathsf A,\mathsf P\rangle
 \le(1+\iota/\delta_{\rm end}^2)\langle\mathsf Q,E_i\rangle
 \le P_n\Delta,
\]
where $\Delta$ is the true matrix-potential drop. The last inequality
is Eq.~\eqref{am_eq_softloss}. In particular this charge does not
use the fast budget's scalar certificate or any random choice.

The positive auxiliary path in Eq.~\eqref{am_eq_endpointpath}
therefore has initial value $\mathcal P_{{\rm s},{\rm old}}+v_0$
and limiting value $\mathcal P_{{\rm s},{\rm new}}$. Its forcing
energy, root bounds, and anchored corrections use only the positive
coupling and the same lower bound $\rho I$ in the primal equation.
Thus Eq.~\eqref{am_eq_endpointenergy} and its continuation proof
apply with $\iota=699/250$ and $\beta=\beta_{\rm tr}$.
For conservative triggers, clip and round the two additions as in
that lemma, tightening the score tolerance below both the regular
decrease and the root radius. This takes only logarithmic extra
precision. The reservoir is independent of the optimizer and its
removal is nonnegative. Since all other moves decrease their phase
potential and the total positive phase injections are bounded,
$\sum\Delta=O(1)$. There are $\widetilde O(n)$ regular endpoints.
Consequently all regular continuations cost $\widetilde O(n)$
corrected stages. Small and deferred removals still use the charged
batches proved above. The gradient sweeps are unchanged.

For a phase with $K/2<k\le K$, retain the curvature count
$I_K\le\widetilde O(\min\{n,n^2/K\})$ from
Eq.~\eqref{eq_trace_matrix_steps}. Applying
Lemma~\ref{trace_lem_block_responses} gives the following bounds
for its total curvature-search work:
\begin{equation}\label{trace_eq_block_phase_cost}
 \begin{cases}
 \widetilde O(n^4+n^{2+\omega_0}),&K\le n,\\
 \widetilde O(n^5/K+n^{2+\omega_0}
             +n^4K^{\omega_0-2}+n^2K^{\omega_0-1}),&n<K\le n^2.
 \end{cases}
\end{equation}
Both are at most $\widetilde O(n^{2\omega_0})$ for the fixed
$\omega_0=2.371177$. There are only logarithmically many phases.

Curvature endpoints continue to use the explicit cold initializer;
no one-correction claim for those endpoints is needed. Each costs
$\widetilde O(\sqrt n F(n,k))$, with $F$ from
Eq.~\eqref{soft_eq_fcost}. For $K\le n$ this contributes at most
$\widetilde O(n^{3/2}n^3)=\widetilde O(n^{9/2})$ per phase.
For $K>n$ it contributes at most
\[
 \widetilde O(\frac{n^{5/2}}K
                 (n^3+(K+n)n^{\omega_0-1}))
 \le\widetilde O(n^{9/2}+n^{\omega_0+3/2}).
\]
The first initialization is smaller. Regular endpoints, charged
batches, and gradient sweeps together use $\widetilde O(n)$
corrected stages, costing $\widetilde O(n^{1+\chi})$, and all
coordinate updates cost $\widetilde O(n^4)$.
Use Lemma~\ref{sr_lem_round} also for the small-input and final
rounding; together they cost $\widetilde O(n^{2+\omega_0})$.

Adding the deterministic preprocessing gives
\begin{equation}\label{trace_eq_block_runtime}
 T_{\rm arith}\le\widetilde O(mn^2+n^{4.742354}).
\end{equation}
Indeed $2\omega_0=4.742354$ exceeds $9/2$, $1+\chi=4.250036$,
and $2+\omega_0=4.371177$.
All replacements are deterministic. They preserve the input residual,
small-input and deferred-rounding allowances, and the finite-step
proof in Eq.~\eqref{eq_trace_matrix_step_bound}. The more accurate
comparisons fit the same numerical allowances. Hence the discrepancy
and partition constants remain $3.3443$ and $1.67215$, respectively.
This establishes the arithmetic cost in Theorem~\ref{rank_one_thm_main}.

\section{A fast randomized polynomial time algorithm and explicit running time}\label{sec_fast_randomized}
The construction uses the smoothed geometry, optimizer corrections,
rounding, and endpoint and gradient-sweep proofs established in
Section~\ref{rank_one_sec_algorithm}.
Section~\ref{sec:randomized:main-bound} states the randomized running-time bound.
Sections~\ref{soft_sec_filter}--\ref{sp_sec_improvement} construct and certify
spread directions for curvature moves.
Sections~\ref{sr_sec_improvement} and~\ref{oe_sec_improvement} give the grouped
reductions, using Lemma~\ref{sr_lem_round} for final rounding.
Section~\ref{am_sec_improvement} counts the shared sweeps and endpoint
continuations in the small-count finishing routine.
Section~\ref{bs_sec_preprocess} sharpens preprocessing.
Sections~\ref{md_sec_main}--\ref{br_sec_cost} give masked curved moves,
resident batches, and count-sensitive inverse costs.
Section~\ref{sec:randomized:operation-totals} proves the total expected arithmetic cost.
All operation counts in this section are arithmetic operation counts.

\subsection{Randomized running-time bound}\label{sec:randomized:main-bound}

Entropy smoothing permits each local trial to be evaluated with a
constant number of Newton corrections. A verified randomized curved move gives progress depending on the
resident count. Delayed deletion and count-sensitive matrix products
give the expected bound below.

\begin{theorem}[Formal version of Theorem~\ref{thm_intro_randomized}, zero-error randomized polynomial time algorithm]\label{soft_thm_algorithm}
Given rational Hermitian matrices $A_1,\ldots,A_m\in\mathbb C^{n\times n}$
of rank at most one and total binary length $L_{\rm in}$,
a zero-error randomized algorithm finds signs
$\sigma\in\{\pm1\}^m$ such that
\[
 \|\sum_i\sigma_iA_i\|\le4.8628\|\sum_iA_i^2\|^{1/2}.
\]
It terminates for every random-bit sequence. Its expected number of
arithmetic operations satisfies
\[
 \E[T_{\rm arith}]\le\widetilde O(mn^2+n^{3.575374}).
\]
For $A_i=a_ia_i^*$, $\sum_iA_i=I$, and $\|a_i\|^2\le\alpha$,
the two sign classes satisfy
\[
 \|\sum_{i\in I_j}a_ia_i^*-I/2\|\le2.4314\sqrt\alpha,
 \qquad j=1,2.
\]
The partition is computed within the same expected arithmetic bound.
Here $\widetilde O$ suppresses logarithmic factors in $m,n,L_{\rm in}$.
\end{theorem}
\begin{proof}
Lemma~\ref{bs_lem_grouped} gives grouped preprocessing in expected
$\widetilde O(mn^2)$ operations after input conversion. It leaves at
most $2n^2$ original fractional coordinates; constant-factor padding
restores the active-count hypothesis. The single operator-norm residual
uses the existing preprocessing allowance.
Lemmas~\ref{md_lem_step} and~\ref{br_lem_mask} give curved moves with
decrease $\widetilde\Omega(\sqrt k/n^{3/2})$ in the large-count regime.
Lemma~\ref{br_lem_flushes} bounds bulk continuation, and
Lemma~\ref{br_lem_sum} gives the count-sensitive cost of a complete
trial. Section~\ref{am_sec_improvement} supplies the small-count finish.
Section~\ref{sec:randomized:operation-totals} combines these bounds with
final rounding and proves the sharper exponent
$E=1058611/296084<3.575374$. Finite caps and deterministic fallbacks
give correct termination on every random-bit sequence. The unchanged
discrepancy allowances and Eq.~\eqref{eq:intro:partition-from-signing}
give the signing and partition constants.
\end{proof}

Algorithm~\ref{alg:randomized:signing} uses the smoothed potential
$\mathcal F_{\rm s}$ and reservoir phases below. A corrected state
includes its primal-dual optimizer. Resident coordinates still contribute
to the coupling; $D$ records pending near-endpoint coordinates, whereas
$\mathcal D$ records coordinates removed for final rounding.

\begin{algorithm}[!htb]
\caption{Zero-error randomized signing with discrepancy constant $4.8628$}
\label{alg:randomized:signing}
\begin{algorithmic}[1]
\Procedure{RandomizedFastSigning}{$A_1,\ldots,A_m$}
    \Comment{Theorem~\ref{soft_thm_algorithm}}
    \State Convert and scale the input once; apply Lemma~\ref{bs_lem_grouped} if $m>2n^2$.
    \State Retain assigned signs, center the single residual, and pad to $N<4n$ with $k_0\le N^2$.
    \State Initialize the corrected root and reservoir phase; clear the initial near-endpoint batch.
    \State Set $D:=\varnothing$ and count all remaining coordinates as resident.
    \While{the resident count $k>K_c$}
        \State Put newly near-endpoint coordinates in $D$, retaining their coefficients and coupling.
        \If{$|D|\ge\lceil\zeta_N k\rceil$}
            \State Attenuate their combined coupling, delete them into $\mathcal D$, and correct the root.
                \Comment{Lemma~\ref{br_lem_flushes}}
            \State Update the count phase after deletion, set $D:=\varnothing$, and continue.
        \EndIf
        \State Mask $D$ and the conservatively marked non-light coordinates.
            \Comment{Lemma~\ref{br_lem_mask}}
        \State Run capped drift-response trials with the count-sensitive sketch schedule.
            \Comment{Lemma~\ref{br_lem_sum}}
        \If{a direction passes all true curvature, spread, and response tests}
            \State Choose the downhill sign of the quadratic path, predict, and correct its optimizer.
                \Comment{Lemma~\ref{md_lem_step}}
        \Else
            \State Clear $D$ and make one certified old endpoint, gradient, or dense-curvature move.
            \State Update the count phase after any deletion and resume.
        \EndIf
    \EndWhile
    \State Clear the remaining pending batch and keep the current reservoir allocation.
    \While{active coordinates remain}
        \State Process deferred attenuation and count-phase updates.
        \If{no active coordinate remains}
            \State Break.
        \EndIf
        \If{a regular endpoint is certified}
            \State Fix its sign and continue the optimizer by Lemma~\ref{am_lem_endpoint}.
        \ElsIf{the high gradient test passes}
            \State Perform the frozen-root sweep in Lemma~\ref{am_lem_sweep} and correct once.
        \Else
            \State Take a checked spread-curvature move, using the finite deterministic fallback on exhaustion.
                \Comment{Lemmas~\ref{sp_lem_flat} and~\ref{sp_lem_tube}}
        \EndIf
    \EndWhile
    \State Round $\mathcal D$ by verified sampling with deterministic fallback.
        \Comment{Lemma~\ref{oe_lem_round}}
    \State Restore input eigenvalue signs and return the full signing $\sigma$.
\EndProcedure
\end{algorithmic}
\end{algorithm}

The primal matrices are $U,V$ and the dual matrices are $W,Z$.
Define their pairs by ${\mathsf P}:=(U,V)$ and ${\mathsf Q}:=(W,Z)$. All scalar products and traces on pairs sum over both blocks.
Fix the multiplication parameters in Eq.~\eqref{fast_eq_multiplication_parameters}.
The notation $P_n$ denotes a fixed polynomial in $\log(2n)$, enlarged
a finite number of times; constants depend only on the fixed budget
and regularization parameters. All numerical tolerances below are
specified by these uniform bounds.
\subsection{Gibbs-feature preconditioning of curvature search}\label{soft_sec_filter}
The random component still returns only directly verified directions.
The new sketch is not obtained by applying the Hessian to random
vectors. Instead it samples a positive Gram matrix which is comparable
to a positive shift of that Hessian. Its features have rank-two matrix
representations and admit batched construction and implicit application.

At a failed gradient test, work first at the exact root and put
$A=G^{-1/2}\nabla^2\mathcal F_{\rm s}G^{-1/2}$. Later we freeze a sufficiently accurate
structured symmetric surrogate. With fixed slack in the tests, choose
rational bounds
\begin{equation}\label{soft_eq_filterhyp}
 \lambda_{\min}(A)\le-2\delta,\quad -RI\preceq A\preceq {M_{\rm aux}}I,
 \quad 1\le R\le P_n,\quad \delta^{-1}\le P_n,\quad
 0<\delta\le R/4,\quad {M_{\rm aux}}\le P_n k,
\end{equation}
where ${M_{\rm aux}}\ge R$. All bounds may be replaced by rational upper bounds obtained from
the fixed budget coefficient sums, the cutoff, and the stated spectral
intervals; no norm-estimation oracle is used to choose them.
The current face dimension is denoted by $k$ inside a
filter; it never exceeds the initial $k$ used in total counts.

\paragraph{A comparable Gram matrix.}
Retain the exact symmetric Schur formula
\begin{equation}\label{soft_eq_hess}
 \nabla^2\mathcal F_{\rm s}=-G+F_1^*Q_{\mathsf Q}F_1
 -({\mathscr J}Q_{\mathsf Q}F_1-F_2)^*J_s^{-1}({\mathscr J}Q_{\mathsf Q}F_1-F_2).
\end{equation}
One action costs $F(n,k)\le\widetilde O(n^\chi)$, including the root
solves, matrix functions, and adjoints. This identity follows by
substituting the differentiated root equations into the second-order
envelope, so involves the same root and the same potential throughout.

\begin{lemma}[Gibbs Gram comparison]\label{gf_lem_gram}
Set $D=G^{-1/2}F_1^*Q_{\mathsf Q}F_1G^{-1/2}$. There are known rational bounds
$\mathsf T_0\le P_nk$, $\ell_0,b_0\le P_n$ such that
\[
 D\succeq0,\quad\tr[D]\le\mathsf T_0,\quad
 a_{\rm dom}D-(1+2b_0)I\preceq A\preceq D-I,
 \qquad a_{\rm dom}=(1+2\ell_0)^{-1}.
\]
In particular enlarge $R$ so $R\ge1+2b_0$, and put $B=A+2RI$. Then
\begin{equation}\label{gf_eq_gram_comparison}
 \tfrac{a_{\rm dom}}2(D+2RI)\preceq B\preceq D+2RI.
\end{equation}
All these bounds survive a sufficiently small inverse-polynomial
root/coefficient perturbation, with fixed slack in $a_{\rm dom},R,\mathsf T_0$.
\end{lemma}
\begin{proof}
The following square roots are for analysis, not computation:
\[
 {\mathsf A}=Q_{\mathsf Q}^{1/2}F_1G^{-1/2},\qquad
 {\mathsf B}=Q_{\mathsf Q}^{1/2}{\mathscr J}{\mathscr E}^{-1/2},\qquad
 T={\mathscr E}^{-1/2}F_2G^{-1/2}.
\]
Eq.~\eqref{soft_eq_precond} gives $\|{\mathsf B}\|^2\le K/{\tau_{\rm s}}\le\ell_0$.
The weighted coupling estimate gives $\|T\|^2\le b_0$. Coordinate
forcing and the Gibbs variance bound give $D_{ii}\le P_n$, hence the
trace bound, without computing the trace. The Schur formula becomes
\[
 A=-I+{\mathsf A}^*{\mathsf A}-({\mathsf B}^*{\mathsf A}-T)^*(I+{\mathsf B}^*{\mathsf B})^{-1}({\mathsf B}^*{\mathsf A}-T).
\]
For any $y$, its quadratic form plus $\|y\|^2$ is
\[
 \min_{y_{\rm aux}}\{\|{\mathsf A}y-{\mathsf B}{y_{\rm aux}}\|^2+\|{y_{\rm aux}}\|^2+2\langle Ty,{y_{\rm aux}}\rangle\}.
\]
Use $\|{y_{\rm aux}}\|^2+2\langle Ty,{y_{\rm aux}}\rangle\ge\frac12\|{y_{\rm aux}}\|^2-2b_0\|y\|^2$.
The remaining minimum equals
\[\langle {\mathsf A}y,(I+2{\mathsf B}{\mathsf B}^*)^{-1}{\mathsf A}y\rangle\ge a_{\rm dom}\|{\mathsf A}y\|^2.\]
The upper bound follows by dropping the nonnegative Schur term.
This proves all order comparisons. At a root all relevant smallest
eigenvalues are inverse-polynomial; polynomial perturbation bounds
therefore preserve the displayed estimates with their slack. In the
numerical construction keep the exact algebraic Schur form, as specified
below, rather than adding an unrelated Hessian approximation.
\end{proof}

\paragraph{Features whose covariance is $D$.}
For the fixed-depth precursor, first fix an integer $h_{\rm sk}\ge2$
independently of the input dimension. Section~\ref{ue_sec_endpoint}
separately proves the uniform bounds needed to use a growing depth;
it does not substitute a variable depth into a fixed-depth estimate.
The features are real coordinate vectors even when the input is complex.
For one sample choose $t$ uniformly in $[0,1]$. Independently for each
block ${\mathsf Q}_b$ of ${\mathsf Q}$ choose standard circular complex Gaussian vectors
$g_b,h_b$ (each real and imaginary part has variance $1/2$), and put
\begin{equation}\label{gf_eq_feature}
 r_b={\mathsf Q}_b^{t/2}g_b,\quad v_b={\mathsf Q}_b^{(1-t)/2}h_b,\quad
 T_b=(r_bv_b^*+v_br_b^*)/\sqrt2,
 \quad \widetilde T=T-\tr[T]{\mathsf Q},
 \quad z={\tau_{\rm s}}^{-1/2}G^{-1/2}F_1^*\widetilde T.
\end{equation}
These definitions refer to a single fixed root; samples are independent.

\begin{lemma}[Feature identity and a ridge covariance sample]\label{gf_lem_features}
The feature has mean zero and $\E[zz^{\top}]=D$. Let $1\le q\le2n^2$,
$\mu=2R+\mathsf T_0/q$, and let $\ell_{\mathsf Q}$ be an integer upper bound for
$1+\log(\lambda_{\max}({\mathsf Q})/\lambda_{\min}({\mathsf Q}))$. Thus $\ell_{\mathsf Q}=O(\log(2n))$
with constants depending only on the fixed regularization. Put
\[
 p=\lceil128\log_2(2kn\ell_{\mathsf Q}(h_{\rm sk}+1))\rceil,\qquad
 s=\lceil2^{20}q\ell_{\mathsf Q}p^3\rceil,\qquad
 V=[z_1\ \cdots\ z_s],\qquad W=\mu I+s^{-1}VV^{\top}.
\]
With probability at least $0.98$ for ideal samples,
\begin{equation}\label{gf_eq_feature_ridge}
 \tfrac34(D+\mu I)\preceq W\preceq\tfrac54(D+\mu I).
\end{equation}
There is a bounded finite-random-bit construction for which the
comparisons with factors $1/2$ and $2$ hold with probability at least
$9/10$. Its matrix $V$ is
defined by its stored rank-two factors and the displayed centering; it
is not perturbed by an arbitrary final full-rank rounding.
\end{lemma}
\begin{proof}
For Hermitian test pairs $B,C$, the conditional covariance of the
uncentered $T$ is the self-adjoint operator
\[
 \mathcal{T}_t(C)=\tfrac12({\mathsf Q}^tC{\mathsf Q}^{1-t}+{\mathsf Q}^{1-t}C{\mathsf Q}^t).
\]
For one block this follows from
$\tr[BT]=\sqrt2 \operatorname{Re}(v^*Br)$ and independence of the two
circular Gaussians; cross-block covariances are zero. Integrating in $t$
gives $\mathcal{T}(C)=\int_0^1{\mathsf Q}^tC{\mathsf Q}^{1-t}\d t$. For every $t$,
$\mathcal{T}_t(E)={\mathsf Q}$ and $\tr[{\mathsf Q}]=1$. Centering thus has covariance
$\mathcal{T}-{\mathsf Q}\otimes {\mathsf Q}={\tau_{\rm s}}Q_{\mathsf Q}$, proving the feature identity.

In a basis diagonalizing ${\mathsf Q}$, the coefficient of $\mathcal{T}_t$ is no
larger than $(z_i+z_j)/2$. The logarithmic-mean coefficient of
$\mathcal{T}$ obeys
\[
 \frac{(z_i+z_j)/2}{(z_i-z_j)/(\log z_i-\log z_j)}
 =\frac{|\log(z_i/z_j)|}{2}
       \coth (\frac{|\log(z_i/z_j)|}{2})
 \le1+|\log(z_i/z_j)|\le \ell_{\mathsf Q}.
\]
Use the continuous value at equality; $y\coth y\le1+y$ follows from
$e^{2y}-1\ge2y$. Therefore $\mathcal{T}_t\preceq \ell_{\mathsf Q}\mathcal{T}$ as
quadratic forms. Apply the same centering congruence and coordinate
map on both sides to obtain
\begin{equation}\label{gf_eq_conditional_covariance}
 \E[zz^{\top}\mid t]\preceq \ell_{\mathsf Q}D.
\end{equation}

Here is an elementary concentration proof, including the unbounded
Gaussian tail. Define $w=(D+\mu I)^{-1/2}z$ only for analysis. Its mean
covariance is $J=(D+\mu I)^{-1/2}D(D+\mu I)^{-1/2}\preceq I$,
with $\tr[J]\le\mathsf T_0/\mu\le q$. Conditional on $t$, every real
coordinate of $w$ is a bilinear form in two independent real Gaussian
vectors (realify the circular pairs). For such a scalar form $U$,
conditioning and Minkowski's inequality give
\[
 \E[|U|^{2p}]\le((2p-1)!!)^2(\E[U^2])^p.
\]
Indeed condition on the second vector, apply the scalar Gaussian moment,
and diagonalize the nonnegative covariance of its conditional variance;
Minkowski bounds the $p$-norm of that weighted sum of squared Gaussians.
Apply Minkowski once more to the sum of coordinate squares, and use
Eq.~\eqref{gf_eq_conditional_covariance}. This proves
\begin{equation}\label{gf_eq_feature_moment}
 \E[\|w\|^{2p}]\le(4p^2\ell_{\mathsf Q}q)^p.
\end{equation}
Let $L_*=16p^2\ell_{\mathsf Q}q$ and truncate, only in the proof,
$U=ww^{\top}\boldsymbol1_{\{\|w\|^2\le L_*\}}$. The failure probability is
at most $4^{-p}$ and
\[
 \|J-\E[U]\|\le L_*4^{-p},\quad
 \|U-\E[U]\|\le L_*,\quad
 \E[(U-\E[U])^2]\preceq L_*I.
\]
For clarity, the scalar-variance matrix Bernstein bound needed here has
a short proof. If a centered Hermitian ${\mathsf B}$ satisfies $\|{\mathsf B}\|\le L_*$ and
$\E[{\mathsf B}^2]\preceq L_*I$, its exponential series and
$j!\ge2 3^{j-2}$ give
\[
 \E[e^{\theta {\mathsf B}}]
 \preceq\exp (\frac{\theta^2L_*}{2(1-\theta L_*/3)})I
 \quad(0\le\theta L_*<3).
\]
The Golden--Thompson inequality~\cite{golden_1965,thompson_1965} and independence apply this scalar
matrix bound successively inside an expected trace exponential.
Markov's inequality for the largest eigenvalue and then for its negative,
with the optimizing scalar $\theta$, yield
\[
 \Pr[\|s^{-1}\sum_j(U_j-\E[U_j])\|>1/8]
 \le2k\exp(-s/(256L_*)).
\]
This uses a scalar variance bound, not a product of matrix-valued
moment bounds; it is the standard trace-exponential argument,
cf.~\cite{t12}.

Here $s/(256L_*)\ge256p$. Moreover $s4^{-p}$ and $L_*4^{-p}$ are below
$1/100$ and $1/100$, respectively, with ample slack. Use $q\le2n^2$ and the defining lower bound on $p$.
Also $p^5\le2^{p/2}$ for every $p\ge64$. The last inequality follows at $64$ and
then by monotonicity of $p/2-5\log_2p$. A union bound ensures that no
sample was truncated, except with the stated small probability.
The total deviation is less than $1/4$, proving
Eq.~\eqref{gf_eq_feature_ridge}. No truncation test is required by the
algorithm.

For the finite sampler, use midpoints of dyadic intervals for $t$
and the two Box--Muller uniforms for each real Gaussian, and pair
independent real Gaussians for a circular complex Gaussian. Couple
the finite and ideal uniforms by their common intervals. Except on
an event of probability at most $1/100$, all radial uniforms exceed
$(100ns)^{-4}$ and all Gaussian magnitudes are
$O(\sqrt{\log(2ns)})$. On this event the sampling maps have polynomially
bounded derivatives. The spectral bounds for $\mathsf Q$ also give
polynomial divided-difference and parameter-derivative bounds for
$\mathsf Q^{t/2}$, uniformly in $t\in[0,1]$.

Choose the scalar grid and evaluation errors inverse polynomially
small, and evaluate the powers using the spectral construction above.
Scalar logarithms, square roots, and trigonometric functions in the
sampler are approximated by convergent series and bisection to the
same accuracy. Store the resulting dyadic vectors and define each
feature by its rank-two factors and the exact centering formula.
Since all factors have polynomial magnitude, their errors can be
chosen so that the covariance perturbation is at most $\mu/16$.
This changes the ideal factors $3/4,5/4$ to the stated $1/2,2$ with
probability at least $9/10$. A fixed number
$O(\log(2kn(h_{\rm sk}+1)))$ of uniform bits per scalar suffices.
Every radial midpoint is positive; on every sample the Gaussian
output has magnitude $O(\sqrt{\log(2kn(h_{\rm sk}+1))})$.
All sampling and evaluation loops consequently have finite caps.

\end{proof}

\paragraph{Entry bounds and a uniform ridge sampler.}
The trace bound in Lemma~\ref{gf_lem_gram} may be enlarged to
$\mathsf T_0=kd_0$, where $1\le d_0\le P_n$ is a known rational upper
bound on every diagonal entry of $D$. This choice is used below. The
sample matrix is never formed in its entirety.

\begin{lemma}[Bounded entries without materialization]\label{rr_lem_entries}
For the finite features in Lemma~\ref{gf_lem_features}, with
$1\le q\le2n^2$ and the enlarged $p$ specified there, the covariance
comparison with factors $1/2,2$ and the entry bound
\begin{equation}\label{rr_eq_entrybound}
 |V_{ij}|^2\le {\mathscr J}_e:=64p^2\ell_{\mathsf Q}d_0\quad\text{for all }i,j
\end{equation}
hold simultaneously with probability at least $4/5$. This is an
analytical event, not a test requiring all entries of $V$.
\end{lemma}
\begin{proof}
The scalar bilinear Gaussian moment calculation in the feature lemma,
now applied to coordinate $i$ before whitening, gives
$\E[|z_i|^{2p}]\le(4p^2\ell_{\mathsf Q}D_{ii})^p$. Thus the probability that an
ideal feature entry exceeds $16p^2\ell_{\mathsf Q}d_0$ in squared magnitude is at
most $4^{-p}$. Union over $ks$ entries. The proof of the covariance
lemma and its truncation bounds is unchanged for $q\le2n^2$: its
polynomial factors are dominated by $4^{-p}$ with the enlarged $p$.
For example $p^5\le2^{p/2}$ for $p\ge64$, and the definition of $p$
bounds $k,n,\ell_{\mathsf Q},h_{\rm sk}+1$ by fixed small powers of $2^p$.
The same coupling of dyadic Box--Muller samples to ideal uniforms can
make every entry error at most $\sqrt{{\mathscr J}_e}/4$, as well as keeping the
covariance perturbation below $\mu/16$. All required errors are inverse
polynomial. Unioning these failure events leaves probability greater
than $4/5$, with the stated slack. Every finite sample, including the
excluded event, still has the polynomial magnitude bound in that lemma.
\end{proof}

\begin{lemma}[Uniform ridge sampling]\label{rr_lem_sampling}
Let $F\in\mathbb R^{d\times e}$ satisfy $|F_{ij}|^2\le {\mathscr J}_e$, let
${\iota}>0$, ${\gamma}\ge1$, and put ${\mathscr E}_{\gamma}={\gamma} I_d+{\iota}FF^{\top}$.
Sample $r$ column indices independently and uniformly with replacement,
and let $J$ be their list. For
\[
 \mathcal{L}=\max\{1,{\iota}ed{\mathscr J}_e/{\gamma}\},\qquad
 r\ge 2^{16}\mathcal{L}\log(200d/\zeta),
\]
one has, with probability at least $1-\zeta$,
\begin{equation}\label{rr_eq_rowsample}
 \tfrac34{\mathscr E}_{\gamma}\preceq
 \widehat {\mathscr E}_{\gamma}:={\gamma} I_d+{\iota}(e/r)F_{:,J}F_{:,J}^{\top}
 \preceq\tfrac54{\mathscr E}_{\gamma}.
\end{equation}
The statement holds conditionally on any previous choices that fix $F$.
Sampling with replacement, including repeated or zero columns, is allowed.
\end{lemma}
\begin{proof}
Write $f_i$ for a column and define only for analysis
$U=e{\iota}{\mathscr E}_{\gamma}^{-1/2}f_if_i^{\top}{\mathscr E}_{\gamma}^{-1/2}$ for a uniform $i$.
Then $0\preceq U\preceq\mathcal{L} I$ and
$\Gamma=\E[U]={\mathscr E}_{\gamma}^{-1/2}({\mathscr E}_{\gamma}-{\gamma} I){\mathscr E}_{\gamma}^{-1/2}
\preceq I$. For ${\mathsf B}=U-\Gamma$, $\|{\mathsf B}\|\le\mathcal{L}$ and
$\E[{\mathsf B}^2]\preceq\mathcal{L} I$. The exponential-series and
Golden--Thompson argument in Lemma~\ref{gf_lem_features} gives
\[
 \Pr[\|r^{-1}\textstyle\sum_j{\mathsf B}_j\|>1/4]
 \le2d\exp[-r/(256\mathcal{L})]\le\zeta.
\]
Congruence by ${\mathscr E}_{\gamma}^{1/2}$ proves the result. The sampling decisions
need no leverage scores, eigendecomposition, or access to all entries.
\end{proof}

\paragraph{Fixed-depth alternating sketches.}
Fix $h=h_{\rm sk}\ge2$ and rational exponents
\begin{equation}\label{rr_eq_exponents}
 2\ge a_1\ge a_2\ge\cdots\ge a_h\ge1,\qquad
 a_{h-1}\le\tfrac32a_h,\qquad q_j=\lceil n^{a_j}\rceil.
\end{equation}
The depth and all these exponents are fixed independently of $n,m$.
Use $q=q_1$ in the Gibbs feature lemma and
$\mu=2R+kd_0/q_1$. Pad the coordinate rows of $V$ by zeros to
$\bar k=2^{\lceil\log_2 k\rceil}$. For $k=1$ this means $\bar k=1$.
Define the positive system in feature space
\begin{equation}\label{rr_eq_M0}
 K_0=I_s+(s\mu)^{-1}V^{\top}V.
\end{equation}
This is an operator, not a materialized Gram matrix. The exact identity
\begin{equation}\label{rr_eq_outerwood}
 W^{-1}=\mu^{-1}I-(s\mu^2)^{-1}VK_0^{-1}V^{\top}
\end{equation}
will be evaluated by solving $K_0$ accurately; a constant-factor inverse
approximation is not substituted into this subtractive formula.

Choose each $d_j$ to be the least power of two at least
$2^{32}q_j\ell_{\mathsf Q}p^3$. First sample $d_1$ coordinate indices from the padded
$\bar k$ rows, obtaining a list $I_1$, and set
\begin{equation}\label{rr_eq_initial}
 \Phi_1=V_{I_1,:}\in\mathbb R^{d_1\times s},\quad e_1=s,
 \quad {\iota}_1=\frac{\bar k}{d_1s\mu},\quad
 K_1=I_{d_1}+{\iota}_1\Phi_1\Phi_1^{\top}.
\end{equation}
Only the indices and the feature factors are stored. Lemma~\ref{rr_lem_sampling}
applied to $V^{\top}$ shows that
\begin{equation}\label{rr_eq_initial_preconditioner}
 W_0=I_s+{\iota}_1\Phi_1^{\top}\Phi_1,\qquad
 \tfrac34K_0\preceq W_0\preceq\tfrac54K_0,
 \qquad W_0^{-1}=I_s-{\iota}_1\Phi_1^{\top}K_1^{-1}\Phi_1
\end{equation}
on its sampling event.

For $j=1,\ldots,h-1$, suppose $\Phi_j\in\mathbb R^{d_j\times e_j}$ and
$K_j=I+{\iota}_j\Phi_j\Phi_j^{\top}$ have been defined. Put
$\lambda_j=q_j/q_{j+1}\ge1$. Select independently a list $J_{j+1}$ of
$d_{j+1}$ columns from the $e_j$ columns, and define
\begin{equation}\label{rr_eq_recursion}
 \begin{split}
 \Phi_{j+1}&=(\Phi_{j,:,J_{j+1}})^{\top},\\
 e_{j+1}&=d_j,\\
 {\iota}_{j+1}&=\frac{{\iota}_je_j}{d_{j+1}\lambda_j},\\
 K_{j+1}&=I_{d_{j+1}}+{\iota}_{j+1}\Phi_{j+1}\Phi_{j+1}^{\top}.
 \end{split}
\end{equation}
Here $\Phi_{j,:,J}$ means the indicated column submatrix of $\Phi_j$.
The associated preconditioner and its exact inverse are
\begin{equation}\label{rr_eq_woodbury}
 \begin{split}
 W_j&=\lambda_j(I_{d_j}+{\iota}_{j+1}\Phi_{j+1}^{\top}\Phi_{j+1}),\\
 W_j^{-1}&=\lambda_j^{-1}
       (I_{d_j}-{\iota}_{j+1}\Phi_{j+1}^{\top}K_{j+1}^{-1}\Phi_{j+1}).
 \end{split}
\end{equation}
Thus each inverse is reduced to the next, smaller system. At the last
level form $K_h$ and compute a residual-certified short inverse.
No other feature Gram matrix is formed.

An index population of non-power-of-two size is padded by zero columns
to the next power of two before sampling, and its padded size is used
in the scale $e_j/d_{j+1}$. Such padding multiplies the coefficient
invariants below by at most two per level. Alternatively use a fixed
cap of dyadic rejection proposals for exact uniform indices from the
unpadded population; reject the whole trial on exhaustion. To avoid
these factors, use the latter convention for the first feature-column
population $s$; every later population is a power of two. A cap of
$\lceil\log_2(10^6(h+1)\sum_jd_j)\rceil$ proposals per draw fails with
probability at most $10^{-6}$ by a union bound. All conditional accepted
indices are exactly uniform, and all samplers terminate after their caps.

\begin{lemma}[Recursive comparisons and invariants]\label{rr_lem_recursive}
Conditionally on the entry event, with probability at least $99/100$
all the uniform sampling comparisons above and below hold. For every
$j$ the exact coefficient invariant is
\begin{equation}\label{rr_eq_invariant}
 {\iota}_jd_je_j=\frac{\bar k}{\mu}\frac{q_j}{q_1}
 \le\frac{2q_j}{d_0}.
\end{equation}
Moreover
\begin{equation}\label{rr_eq_preconditioned}
 \tfrac34K_j\preceq W_j\preceq\tfrac54\lambda_jK_j
 \quad(1\le j<h).
\end{equation}
All finite-sample $K_j$ and $W_j$ are positive definite, even on failure
of the comparisons, and have polynomially bounded norms and inverse
norms for fixed $h$.
\end{lemma}
\begin{proof}
The invariant follows at $j=1$ from Eq.~\eqref{rr_eq_initial} and is
preserved by Eq.~\eqref{rr_eq_recursion}. Its inequality uses
$\mu\ge kd_0/q_1$ and $\bar k\le2k$. Every $\Phi_j$ is a submatrix of
$V$ or $V^{\top}$, with possible repetitions. Hence its entries still obey
Lemma~\ref{rr_lem_entries} and Eq.~\eqref{rr_eq_entrybound}. The sampling lemma at level $j$ has
\[
 \mathcal{L}\le\max\{1,2q_{j+1}{\mathscr J}_e/d_0\}
 \le129p^2\ell_{\mathsf Q}q_{j+1}.
\]
The initial sample has the same bound with $q_1$. The chosen $d_j$
therefore makes each conditional failure probability at most
$1/(1000(h+1))$. To check the scalar constants, all dimensions are at
most $2^{40}(n+1)^2(\ell_{\mathsf Q}+1)(p+1)^4(h+1)$, whose logarithm is at most
$4p$; the exponential bound in the preceding proof has ample margin.
Union the finitely many conditional failures and the finite-sampling
cap failure. No independence between different levels is asserted or
needed.

The sampling comparison is with
$\lambda_jI+{\iota}_j\Phi_j\Phi_j^{\top}$. This matrix lies between $K_j$ and
$\lambda_jK_j$, proving Eq.~\eqref{rr_eq_preconditioned}.
The identity for $W_j^{-1}$ is Woodbury, with no rank or independence
requirement on the selected columns. It holds for signed feature
entries and repeated columns. Every $K_j\succeq I$ and every
$W_j\succeq\lambda_jI$. The finite feature bounds, prescribed
sample counts, and explicit positive rational scales bound their norms
by a fixed polynomial, also off the good event. This proves the last
assertion without a random conditioning assumption.
\end{proof}

\paragraph{Applying submatrices, not materializing them.}
Define the piecewise affine multiplication upper bound
\begin{equation}\label{rr_eq_theta}
 \vartheta(a)=
 \begin{cases}
 \omega_0+2(a-1)(\xi_0-\omega_0),&1\le a\le3/2,\\
 \xi_0+4(a-3/2)(\xi_1-\xi_0),&3/2\le a\le7/4,\\
 \xi_1+4(a-7/4)(\chi-\xi_1),&7/4\le a\le2,
 \end{cases}
\end{equation}
where $\xi_0=2.794634$ and $\xi_1=3.021592$.
The three pieces follow by tensor interpolation and permutation from
the strict square bound, the bounds for shapes $(1,3/2,1)$ and
$(1,2,1)$ in Table~1 of \cite{adwxxz25}, and
$\omega(1,7/4,1)\le3.021591<\xi_1$ in Table~3 of \cite{lu18}.
The additional shape gives a smaller upper bound between $3/2$ and $2$.
These are upper bounds, not assertions about the exact multiplication exponent.

\begin{lemma}[Implicit restricted feature actions]\label{rr_lem_actions}
A submatrix of $V$, or its transpose, having at most
$\widetilde O(n^a)$ selected coordinate rows and selected feature
columns can be applied, together with its adjoint, in
$\widetilde O(n^{\vartheta(a)})$ arithmetic operations.
The lists may contain repetitions. At one fixed root all initial
sample vectors cost $\widetilde O(n^3+n^{\vartheta(a_1)})$ to prepare.
For the final factor $\Phi_h$, the factor itself, its Gram matrix, and its
inverse approximation cost
\begin{equation}\label{rr_eq_leafsetup}
 \widetilde O(n^{\mathfrak g(a_{h-1},a_h)}+n^{a_h\omega_0}),
 \quad
 \mathfrak g(r,v)=2(\xi_0-\omega_0)r+(3\omega_0-2\xi_0)v.
\end{equation}
\end{lemma}
\begin{proof}
For selected sample columns, a weighted feature sum is first represented
as a pair of weighted rank-two outer-product sums, with the single
centering correction in Eq.~\eqref{gf_eq_feature}. Compute it with
products of shapes $(n,n^a,n)$, and evaluate only the selected original
input quadratic forms with shapes $(n,n,n^a)$. Response coefficients
and diagonal $G^{-1/2}$ are applied before or after these maps as
appropriate. For the adjoint, form the forcing pair from the selected
original coordinates and test it against only the selected samples.
The same products suffice. Repeated indices are treated as repeated
factor columns, or their coefficients can be summed; neither method
requires materializing the feature entries. Padded coordinate indices
denote zero functionals and are skipped; no nonexistent weight is inverted. All these maps are real
coordinate maps even for complex Hermitian data. There is no new
spectral decomposition per action.

The two order-$n$ root spectral decompositions are performed once. To
form the initial $s$ sample vectors, apply the eigenvector matrices to
the arrays of diagonally scaled Gaussians by the same rectangular
products. Scalar powers, sampling, diagonal scalings and lists cost
$\widetilde O(n^{1+a_1})$, which is smaller. Thus the preparation cost
is as asserted, rather than square blocking at exponent
$\omega_0+a_1-1$.

Let $r=a_{h-1}$ and $v=a_h$. The final factor has dimensions
$\widetilde O(n^v)\times\widetilde O(n^r)$; its entries are computed
by a fixed number of products of shape $(n^v,n,n^r)$ and entrywise
operations. Since $v\ge1$, these fit the cost for the terminal Gram,
whose shape is $(n^v,n^r,n^v)$. For $v\le r\le3v/2$, tensor
interpolation of the square and $3/2$ shapes gives
$\mathfrak g(r,v)$. The terminal system has smallest eigenvalue at
least one. The existing normwise-stable Newton--Schulz construction
forms its inverse with $\widetilde O(n^{v\omega_0})$ work, including
residual certification.

The represented factors are fixed by their dyadic rank-two data.
A materialized terminal factor and its implicit actions approximate
the same matrix. Compute the products in exact arithmetic and use
Newton--Schulz iteration until its residual, together with the known
inverse bound, gives the requested error. The polynomial spectral
bounds require only logarithmically many iterations. Real and imaginary
coordinate formulas preserve the adjoint identities at constant cost.
\end{proof}

\paragraph{Executable nested inverse solves.}
Use $W_0$ to solve $K_0$ on a fixed interval $[1/2,2]$, and $W_j$ to
solve $K_j$ on
\begin{equation}\label{rr_eq_intervals}
 [1/(4\lambda_j),4]\quad(1\le j<h).
\end{equation}
The good-event comparisons leave fixed slack in these intervals.
For any of these systems $Kv=b$ and preconditioner $W$, define
$M:=W^{-1}K$, $\iota:=(\alpha+\beta)/2$, $t:=(\beta-\alpha)/2$,
$D:=(\iota I-M)/t$, and $\varrho:=\iota/t$. With $b':=W^{-1}b$, use
\begin{align*}
 a_1&:=1/\varrho,&a_{j+1}&:=(2\varrho-a_j)^{-1},\\
 v_0&:=0,&v_1&:=b'/\iota,\\
 v_{j+1}&:=2a_{j+1}Dv_j-a_ja_{j+1}v_{j-1}+2a_{j+1}b'/t.
\end{align*}
Its exact error identity is
\begin{equation}\label{rr_eq_cheb}
 v_*-v_j=T_j(D)v_*/T_j(\varrho).
\end{equation}
All exact ratios lie in $[0,1]$; clip computed ratios to this interval.
No exponentially large recurrence numerator is stored.
Thus $K_0$ needs polylogarithmically many iterations; $K_j$ needs
$\widetilde O(\sqrt{\lambda_j})$ iterations. Apply $W_j^{-1}$ using
Eq.~\eqref{rr_eq_woodbury} and a recursive solve of $K_{j+1}$.
Use the residual corrections of Lemma~\ref{ue_lem_precision} at each
level, requesting the same inverse-polynomial relative-to-right-hand-side
accuracy from every child. At the leaf use the stored short inverse.
All recursion depths and iteration lengths are prescribed before the
trial. This is an iterative inverse application, not an exact inverse oracle.

For the outer shifted system $B=A+2RI$, apply $W^{-1}$ by
Eq.~\eqref{rr_eq_outerwood}. The feature comparison gives the slack interval
\[
 \frac{a_{\rm dom}R}{4\mu}\le
 {\gamma}(W^{-1}B)\le4,
\]
so one accurate $B$ solve uses
$\widetilde O(\sqrt{1+k/q_1})$ applications.
Use a bounded outer polynomial. Define $T:=RB^{-1}$, so
$0\preceq T\preceq I$, and define
\[
 t_-:=R/(2R-\delta),\qquad t_+:=R/(2R-2\delta),\qquad
 t_0:=(t_-+t_+)/2,\qquad \Delta:=(t_+-t_-)/2.
\]
Here $\Delta\ge\delta/(8R)$. Choose a positive dyadic $\epsilon_f$ with
$\epsilon_f\le1/4$ and
$\epsilon_f^2\le\delta/(64k({M_{\rm aux}}+\delta))$.
Choose an integer
$d_f\ge\max\{2/\Delta,8\Delta^{-2}\log(2/\epsilon_f)\}$ and
define $j_*:=\lceil d_ft_0\rceil$. Since $R/\delta$ is polylogarithmic,
so is $d_f$. The filter is
\begin{equation}\label{rr_eq_outerfilter}
 p_f(t):=\sum_{j=j_*}^{d_f}\binom{d_f}{j}t^j(1-t)^{d_f-j},
 \qquad {y_{\rm aux}}:=p_f(T)\xi,
\end{equation}
where $\xi\in\{\pm1\}^k$ is independent of the sketches.
The Bernoulli logarithmic moment bound, whose second derivative is at
most $1/4$, gives $0\le p_f\le1$ on $[0,1]$,
$p_f\le\epsilon_f$ on $[0,t_-]$, and
$p_f\ge1-\epsilon_f$ on $[t_+,1]$.
Evaluate the polynomial by de Casteljau's recurrence, starting with
$v_j^{(0)}:=\boldsymbol1_{\{j\ge j_*\}}\xi$ and setting
$v_j^{(r)}:=(I-T)v_j^{(r-1)}+Tv_{j+1}^{(r-1)}$.
The $O(d_f^2)$ inverse actions change only logarithmic factors in
the arithmetic cost. In an eigenbasis of $T$, each update is a
convex combination. Jensen's inequality summed over the array proves
contraction in its concatenated Euclidean norm. If each vector
update has error at most $\eta$, summing the errors over $d_f$ levels
gives final error at most $d_f\sqrt{d_f+1}\eta$.
Thus inverse-polynomial action errors attain the prescribed filter
accuracy.

\subsection{Certified states and discrepancy}\label{soft_sec_certified_states}
\paragraph{No hidden oracle at the Newton stages.}
At every face or accepted state an optimizer cache has been certified
inside its joint-energy ball and then corrected to its specified floor.
Use that base inverse for the next candidate or homotopy increment, as
in Lemma~\ref{soft_lem_transport}. The current residual uses logarithms of
positive ${\mathsf Q}$, not a Gibbs state presumed known in advance. Traceless
updates keep $\tr[{\mathsf Q}]=1$ exactly; rational projection subtracts
$\tr[{\mathsf B}]E/(2n)$. Positive definiteness at intermediate points follows
from the proved ball and errors smaller than a fixed fraction of its
radius. Residual enclosures translate to root-error enclosures by the
local inverse bound. No analytic-center, SDP, or exact spectral oracle
is invoked.

The preceding matrix-function constructions and residual inverse
bounds maintain this invariant with inverse-polynomial action errors.
The endpoint and discrepancy allowances are as follows.

\paragraph{Endpoints and error allowances.}
The endpoint-score tolerance is still smaller than a fixed fraction of
${\gamma}{{\delta_{\rm end}}}^{97/200}$. Its uniform PSD repair therefore costs less
than the available reservoir decrease. Deferred removals preserve the
modeled sum. Multiple hits are removed together and count toward at most
$k$ removals; zero matrices may be eliminated directly. No smoothing
penalty is paid at each face: $\mathcal P_{\rm s}$ is one fixed potential for the
maintained model, and Eq.~\eqref{soft_eq_sandwich} is used once at initialization.

Initial-coordinate rounding and deferred rounding cost at most ${{\varepsilon_{\rm round}}}$;
partial signing costs $\eta_{\rm ps}$. Restoring variance scaling and
discarded inputs yields the internal coefficient
\begin{equation}\label{soft_eq_allowance}
 (2\sqrt{{\iota}+2{{\varepsilon_{\rm scale}}}}+10^{-6}+{{\varepsilon_{\rm round}}}+\eta_{\rm ps}+10^{-8})
       \sqrt{1+{{\varepsilon_{\rm scale}}}}+{\varepsilon_{\rm discard}}<4.86276.
\end{equation}
An exact check is obtained by putting $r=121569/25000$ and
$b=r-{\varepsilon_{\rm discard}}-(10^{-6}+{{\varepsilon_{\rm round}}}+\eta_{\rm ps}+10^{-8})(1+{{\varepsilon_{\rm scale}}})$.
Then $b>0$ and
\[
 b^2-4({\iota}+2{{\varepsilon_{\rm scale}}})(1+{{\varepsilon_{\rm scale}}})
   =425477826609165924200044521/10^{32}>0.
\]
The one-time input transfer gives
$4.86276(1+6\cdot10^{-8})+2\cdot10^{-8}=4.8627603117656<4.8628$.
This proves correctness for every sequence of accepted moves, including
every random-bit sequence in the capped randomized algorithm.

\subsection{An endpoint bound with dimension-dependent sketch depth}
\label{ue_sec_endpoint}
We use a dimension-dependent sketch depth $O(\log\log(2n))$.
The following uniform accuracy and arithmetic bounds give a complete
verified curvature search in $\widetilde O(n^\chi)$ operations.

We may pad the matrices with zero rows and columns to order
$N=2^{\lceil\log_2\max\{2,n\}\rceil}$, where $N\le2\max\{2,n\}$.
Run the same normalized algorithm at that order, including its
regularization ${\rho}={{\varepsilon_{\rm scale}}}/N$. Padding preserves the matrix variance,
rank one and the output discrepancy. Its input work remains
$\widetilde O(mn^2)$ arithmetic operations. In the remainder
of this subsection, dimension-dependent bounds refer to $N$. Put
\[
 r=\log_2N,\qquad {\mathscr J}=r+1.
\]
All logarithmic bounds are in ${\mathscr J}$ because $k\le N^2$ in the nonlinear
phase. Original long input coefficients are read and converted only in
the unchanged preprocessing.

\paragraph{A uniform rectangular action bound.}
We will use a deliberately weaker slope than the interpolation bound in
Eq.~\eqref{rr_eq_theta}. Set $\sigma_{\rm mm}=5/6$. For every power of two
$q$ with $N^{4/3}\le q\le N^2$, restricted feature actions, their
adjoints, and the corresponding factor products have uniform cost
\begin{equation}\label{ue_eq_action}
 \widetilde O(N^\chi(q/N^2)^{5/6}).
\end{equation}
Here the constants and logarithmic degree are independent of $q$; this
is important when the shapes are chosen from the input dimension.
Indeed $\chi-\omega_0=0.878859>5/6$. Fix once and for all a common
power-of-two base large enough to implement the square and $(1,2,1)$
bilinear algorithms at strict exponents below $\omega_0$ and $\chi$.
Choose their slack small enough that the difference of those exponents
still exceeds $5/6$. Tensoring $t-r$ square levels and $r$ rectangular
levels gives shape $(b^t,b^t,b^{t+r})$, with cost at most a constant times
\[
 (b^t)^\chi(b^{t+r}/b^{2t})^{5/6}.
\]
Rounding the two dimensions up to adjacent base powers changes them by
fixed factors. Tensor permutation supplies both orientations used in
Lemma~\ref{rr_lem_actions}. Thus one finite collection of fixed bilinear
algorithms implements all the required shapes. Their depth is logarithmic;
fixed coefficients give polynomial normwise error amplification uniformly
in the mix of levels. Lists enlarged by polynomial factors in ${\mathscr J}$ can be
split into that many blocks, contributing only a fixed logarithmic factor.
No new or input-dependent matrix multiplication exponent is used.

\paragraph{Avoiding exponentially large recurrence numerators.}
Consider one of the positive systems $Kv=b$ with the stated
preconditioner $W$, on a slack interval $[\alpha,\beta]$. Retain
$M=W^{-1}K$, ${\iota}=(\alpha+\beta)/2$, $t=(\beta-\alpha)/2$,
$D=({\iota}I-M)/t$, and $\varrho={\iota}/t>1$. Instead of storing the numerators and
$T_j(\varrho)$ in Eq.~\eqref{rr_eq_cheb}, store
\[
 a_j=T_{j-1}(\varrho)/T_j(\varrho).
\]
Use the normalized recurrence
\begin{equation}\label{ue_eq_normalized}
 \begin{split}
 a_1&=1/\varrho,\qquad a_{j+1}=(2\varrho-a_j)^{-1},\\
 v_0&=0,\qquad v_1=b'/{\iota},\qquad b'=W^{-1}b,\\
 v_{j+1}&=2a_{j+1}Dv_j-a_ja_{j+1}v_{j-1}
                    +2a_{j+1}b'/t.
 \end{split}
\end{equation}
It has exactly the same error polynomial:
\begin{equation}\label{ue_eq_error}
 v_*-v_j=T_j(D)v_*/T_j(\varrho).
\end{equation}
In exact arithmetic $0<a_j\le1$. The update of $a_j$ is Lipschitz with
constant at most one on $[0,1]$, since $2\varrho-a_j\ge1$.
The computed ratio can be clipped to $[0,1]$; clipping cannot increase
its distance from the exact ratio. Divisions and the clipping test are
short rational operations.

\begin{lemma}[Uniform arithmetic bounds for nested normalized solves]
\label{ue_lem_precision}
Let $2\le h\le {\mathscr J}^2$ and let $N^{4/3}\le q_h\le\cdots\le q_1\le N^2$
be powers of two. Use the feature and index sample counts in
Lemmas~\ref{gf_lem_features} and~\ref{rr_lem_recursive}, with this $h$.
The entry/covariance and all conditional sampling events have the same
constant joint success lower bound as before. On their intersection,
all completed recursive solves attain
$\|v-K^{-1}b\|\le N^{-B_0}\|b\|$ for a fixed $B_0$ independent of
$h,N$ and the $q_j$. The shifted solves and outer curvature filter
attain their prescribed inverse-polynomial errors.
There is a fixed integer $D\ge1$, independent of these parameters, such
that the full capped arithmetic cost of one trial is at most
\begin{equation}\label{ue_eq_uniform_cost}
 \begin{split}
 &{\mathscr J}^{4D}(N^3+\mathcal {M_{\rm aux}}(N,q_1)+q_h^{\omega_0})
 +{\mathscr J}^{4D}\sqrt{1+k/q_1} N^\chi\\
 &+\sum_{j=1}^{h-1}{\mathscr J}^{D(j+4)}
       \sqrt{(k+q_1)/q_{j+1}} \mathcal {M_{\rm aux}}(N,q_{j-1})\\
 &+{\mathscr J}^{D(h+4)}\sqrt{(k+q_1)/q_h} q_h^2,
 \end{split}
\end{equation}
where $q_0=q_1$ and
$\mathcal {M_{\rm aux}}(N,q)=N^\chi(q/N^2)^{5/6}$.
The setup term presumes $q_{h-1}=q_h$; a fixed factor in their ratio
would give the same conclusion. Each returned direction is verified
against the true model. All samples, including failed sampling events,
have the displayed deterministic cost cap.
\end{lemma}
\begin{proof}
Freeze the root data, feature factors, and index lists during a
trial. The exact invariant in Eq.~\eqref{rr_eq_invariant} gives
$1\le\lambda_{\min}(K_j)$ and
$\|K_j\|\le1+2q_j\max_{a,b}|V_{ab}|^2/d_0$.
The finite sample magnitudes are polynomial uniformly for
$h\le\mathscr J^2$. Hence all system, inverse, Woodbury, and similarity
bounds are $N^{O(1)}$ with exponents independent of depth. The moment
order is $O(\mathscr J)$. Assigning conditional index failure at most
$1/(1000(h+1))$ preserves the preceding joint probability after a
union bound, including the finite sampling caps.

For a normalized Chebyshev solve on the good event, a vector update
error $e_i$ contributes at step $J$ a term
$[T_i(\varrho)/T_J(\varrho)]U_{J-i}(D_0)e_i$.
The scalar ratio is at most one, and the second-kind polynomial has
norm at most a fixed polynomial similarity factor times $J-i+1$.
Thus the accumulated error is at most
$N^C(J+1)^2\max_i\|e_i\|$, for a fixed $C$.
All prescribed iteration counts are polynomial in $N$.

Set $\eta:=N^{-B_0}$. At every level construct a coarse solver $S_j$
using $O(\sqrt{\beta/\alpha}\log N)$ normalized Chebyshev steps and
completed child solves with error at most $\eta$ times their
right-hand-side norm. Normalize each right-hand side within its own
call; a zero right-hand side returns zero. The preceding error bound
makes the effect of child errors at most $N^{C'}\eta\|b\|$, with
$C'$ independent of depth. Choose $B_0$ sufficiently large and local
action tolerances inverse polynomially smaller. The exact truncation
error and these action errors then give
\[
 \|S_j(b)-K_j^{-1}b\|\le\tfrac14\|K_j^{-1}b\|.
\]
Compute fresh residuals and corrections
$v_{\ell+1}:=v_\ell+S_j(b-K_jv_\ell)$ from $v_0:=0$.
With exact residuals the solution error contracts by $1/4$.
Inverse-polynomial residual-action errors add a fixed polynomial
multiple of their local tolerance. Choosing that tolerance smaller
than $\eta N^{-C''}$ and using $O(\log N)$ corrections attains the
same $\eta$ output guarantee at every level. A direct residual check,
using $K_j\succeq I$, certifies it. The terminal inverse starts the
induction; the shifted system is treated in the same way.

The bounded outer array requires only inverse-polynomial action
accuracy by the contraction estimate above. Choose $B_0$ also for
its normalization and quotient margins. Prescribe all iteration
lengths and a polynomial magnitude cap before the trial, and reject
on a failed residual, magnitude, or final witness test. On the good
event these caps and tolerances preserve the successful sign event.
They bound every other trial as well.

Choose a fixed $D$ large enough to absorb all fixed constants, sample
oversampling, matrix-product and spectral overhead, and the accuracy
factor in a Chebyshev length into ${\mathscr J}^D$ at a single level. This $D$
does not depend on the depth. One recursive level therefore has cost
at most ${\mathscr J}^D\sqrt{q_j/q_{j+1}}$ times the sum of its restricted action
cost and one child solve. $K_0$ costs only another ${\mathscr J}^D$ factor, and
the shifted solve costs ${\mathscr J}^D\sqrt{1+k/q_1}$ such applications.
Expanding this recurrence gives Eq.~\eqref{ue_eq_uniform_cost}, with
conservative extra factors ${\mathscr J}^{4D}$. At the leaf, $q_{h-1}=q_h$ allows
the factor, its Gram, and its inverse to be formed by square products
of order $\widetilde O(q_h)$, costing ${\mathscr J}^Dq_h^{\omega_0}$; its inverse
action costs ${\mathscr J}^Dq_h^2$. Preparing all sampled vectors costs
${\mathscr J}^D(N^3+\mathcal {M_{\rm aux}}(N,q_1))$. Index lists and every random bit are
smaller. This proves the arithmetic cost without
suppressing a depth-dependent logarithmic power.
\end{proof}

\paragraph{A sketch schedule absorbing the nesting overhead.}
Fix the integer $D$ from Lemma~\ref{ue_lem_precision}, then fix
\[
 C_*=1000(D+1),\qquad b_{\max}=\lfloor2r/3\rfloor,
 \qquad b_0=\min\{b_{\max},\lceil C_*\log_2 {\mathscr J}\rceil\}.
\]
The first ceiling is computed as the least integer $b$ for which
$2^b\ge {\mathscr J}^{C_*}$; its exponent is a fixed integer. If $b_0=b_{\max}$,
use the two deficits $(b_0,b_0)$. Otherwise form
\begin{equation}\label{ue_eq_deficits}
 b_{t+1}=\min\{b_{\max},\lceil3b_t/2\rceil\}
\end{equation}
until $b_T=b_{\max}$ and use the list
\begin{equation}\label{ue_eq_schedule}
 (\ell_1,\ldots,\ell_h)
  =(b_0,b_1,b_1,b_2,b_2,\ldots,b_T,b_T),
 \qquad q_j=2^{2r-\ell_j},\quad \ell_0=\ell_1.
\end{equation}
This is a completely integer schedule. The sample sizes are nonincreasing,
and
\[
 N^{4/3}\le q_h=q_{h-1}<2N^{4/3},\qquad
 2^{b_0}\le2{\mathscr J}^{C_*},\qquad h=O(\log {\mathscr J}).
\]
More explicitly, $h\le6+4\log_2 {\mathscr J}$, apart from the two-level capped case;
our large fixed $C_*$ makes $h\le {\mathscr J}^2$ in every case. Neither the
matrix algorithms nor their coefficients depend on these variable
exponents; Eq.~\eqref{ue_eq_action} was uniform in the sizes.

\begin{lemma}[Summing all recursive work]\label{ue_lem_sum}
For the schedule in Eq.~\eqref{ue_eq_schedule}, the full capped cost
in Eq.~\eqref{ue_eq_uniform_cost} is
$\widetilde O(N^\chi)$, with a fixed logarithmic degree independent
of $N$. In particular the factor ${\mathscr J}^{Dh}$ is not silently treated as
a polylogarithm.
\end{lemma}
\begin{proof}
The full top-level action term is bounded by
\[
 {\mathscr J}^{4D}\sqrt{1+k/q_1} N^\chi
 \le2{\mathscr J}^{4D+C_*/2}N^\chi.
\]
Setup costs at most a fixed logarithmic factor times
$N^3+N^\chi+N^{4\omega_0/3}$, and $4\omega_0/3<\chi$.

Suppose first $b_0<b_{\max}$. In the $j$th summand in
Eq.~\eqref{ue_eq_uniform_cost}, $k+q_1\le2N^2$ gives, up to a fixed
factor,
\begin{equation}\label{ue_eq_summand}
 N^\chi {\mathscr J}^{D(j+4)}
 2^{-5\ell_{j-1}/6+\ell_{j+1}/2}
 \le2N^\chi {\mathscr J}^{D(j+4)}2^{-\ell_{j-1}/12}.
\end{equation}
The schedule satisfies $\ell_{j+1}\le3\ell_{j-1}/2+1$.
Writing $t=\lfloor(j-1)/2\rfloor$, the deficit $\ell_{j-1}=b_t$
has not yet been capped, so
\[
 \ell_{j-1}\ge(3/2)^tb_0\ge C_*(3/2)^t\log_2{\mathscr J}.
\]
For every integer $j\ge1$,
$j+4\le8(3/2)^t$ and $j\le4(3/2)^t$; induction on the pairs
$(j=2t+1,2t+2)$ proves both inequalities. Our choice of $C_*$ thus gives
\[
 D(j+4)-C_*(3/2)^t/12\le-2j.
\]
Summing Eq.~\eqref{ue_eq_summand} costs at most
$2N^\chi\sum_{j\ge1}{\mathscr J}^{-2j}=O(N^\chi)$.

The remaining leaf inverse-action term is at most a fixed factor times
\[
 {\mathscr J}^{D(h+4)}N^3,
\]
since $q_h<2N^{4/3}$. This term requires its own bound. Before the final
truncation, each increase in $b_t$ is at least $b_0/2$. It follows that
$h+4\le12b_{\max}/b_0$. Therefore
\[
 D(h+4)\log_2{\mathscr J}
 \le(12D/C_*)b_{\max}\le(8D/C_*)r<r/100.
\]
The term is at most a constant times $N^{3.01}$, smaller than $N^\chi$.
This establishes the bound without invoking an unquantified
subpolynomial-overhead argument.

If $b_0=b_{\max}$, the depth is exactly two. The logarithmic factors
are already fixed powers. The single restricted-action exponent has
factor $2^{-b_0/3}\le1$, and the leaf action is $O(N^3)$ up to its
fixed logarithmic factor. The top term obeys the same bound because
$2^{b_0}\le2{\mathscr J}^{C_*}$. All cases are covered.
\end{proof}

\subsection{A negative spectral subspace and larger curvature moves}
\label{sp_sec_improvement}
Lemmas~\ref{ue_lem_precision} and~\ref{ue_lem_sum} make a complete
verified curvature search cost $\widetilde O(n^\chi)$. We retain
the dimension of the negative spectral subspace to bound the number of moves. A bounded
spectral filter, rather than a filter concentrating only on the most
negative eigenvector, produces a direction with a useful coordinate
spread. A logarithmic number of reservoir reallocations and simultaneous
attenuation of deferred coordinates complete the implementation.

We retain the input conversion, variance scaling, batched grouped
partial signing, smoothed potential, root equations, and numerical
root machinery of the preceding subsections. In particular
$\sum_iV_i^2\preceq I$, ${\iota}=14779/2500>5$, ${\rho}=10^{-8}/n$,
${\tau_{\rm s}}=10^{-8}/\lceil\log_2(4n)\rceil$, and the cutoff ${{\delta_{\rm end}}}$ is
unchanged. The initial retained count is $k_0\le n^2$; below $k$ means
the current count. As in the endpoint implementation, zero padding to
a power-of-two matrix order costs only constant factors. All bounds
may be read at that padded order. This does not pad the active
coordinate space in the spectral arguments below.

The symbol $P_n$ continues to denote a fixed positive polynomial in
$\log(2n)$, enlarged finitely many times. All additional constants below
can be replaced by rational upper or lower bounds obtained from the
fixed budget, its derivatives on the truncated interval, and the
root bounds already proved. None requires estimating an unknown
spectrum. The reservoir coefficient will vary between phases; it always
satisfies ${\gamma}\ge 10^{-6}/(L_0k_0)$, where
$L_0=1+\lceil\log_2\max\{2,k_0\}\rceil$. Root geometry is independent
of ${\gamma}$. All weight and numerical bounds used previously remain
valid after this logarithmic change of the lower bound.

\paragraph{A rank-defect version of the covariance construction.}
We first record the linear algebra that supplies multiplicity, not
merely one negative eigenvalue. The normalized response matrices in
Lemma~\ref{rank_one_lem_covariance} act on a real coefficient space of
dimension $2k$. In this
paragraph call them $T,{M_{\rm aux}},\Gamma$, to avoid confusion with the current
scaled Hessian. In the paired orthonormal basis $e_i,f_i$ they satisfy
\[
 {M_{\rm aux}}\succeq T^*\Gamma T,\quad {M_{\rm aux}}\succeq0,\quad
 \langle e_i,{M_{\rm aux}}e_i\rangle=\langle f_i,{M_{\rm aux}}f_i\rangle=1,
 \quad \Gamma e_i=\Gamma_i e_i,\quad\Gamma f_i=\Gamma_i f_i,
\]
with $\Gamma_i\ge {\iota}>5$, and
\[
 \langle e_i,Te_i\rangle+\langle f_i,Tf_i\rangle=0,
 \qquad \langle e_i,Te_i\rangle\le0.
\]
Let $\Pi_0$ project onto the span of the $e_i$.

\begin{lemma}[Projection with a rank defect]\label{sp_lem_rankdefect}
Let $\Pi$ have rank $k-r$ and satisfy
$(I-\Pi_0)(I-T)\Pi=0$. With
\[
 \theta_i=\frac2{\sqrt{1+4\Gamma_i}-1},\qquad
 \kappa_i=\frac{1+\theta_i}{1-\theta_i^3},
\]
there is a PSD matrix $\mathcal {\mathsf A}$, supported on $\Pi$, such that
\begin{equation}\label{sp_eq_defect}
 \tr[{M_{\rm aux}}\mathcal {\mathsf A}]
 \le \sum_i\kappa_i
   \langle e_i,(I-T)\mathcal {\mathsf A}(I-T)^*e_i\rangle+r.
\end{equation}
If $0\le r\le k/4$, the same matrix satisfies
\begin{equation}\label{sp_eq_movement}
 \tr[(I-T)\mathcal {\mathsf A}(I-T)^*]\ge k/175.
\end{equation}
These estimates do not depend on the smallest covariance weight.
\end{lemma}
\begin{proof}
Write $u_i=1+\theta_i^2+\theta_i^3$ and put
\[
 \varpi_i^e=\frac1{(1+\theta_i)u_i},\quad
 \varpi_i^f=\frac{\theta_i}{(1+\theta_i)(2+\theta_i)},\quad
 h_i=\frac2{u_i},\quad j_i=\frac2{2+\theta_i}.
\]
Define $D_*=\sum_i(\varpi_i^e e_ie_i^*+\varpi_i^f f_if_i^*)$. For an
orthonormal basis matrix $E_*$ of $\operatorname{range}\Pi$, set
\[
 C_*=E_*^*D_*^{-1}E_*,\quad
 \mathcal {\mathsf A}=E_*C_*^{-1}E_*^*,\quad \mathcal {\mathsf B}=D_*-\mathcal {\mathsf A}.
\]
Both $\mathcal {\mathsf A},\mathcal {\mathsf B}$ are PSD and
$\mathcal {\mathsf A}D_*^{-1}\mathcal {\mathsf A}=\mathcal {\mathsf A}$,
$\mathcal {\mathsf A}D_*^{-1}\mathcal {\mathsf B}=0$, and
$\tr[D_*^{-1}\mathcal {\mathsf A}]=k-r$.

For completeness, the row completions in
Lemma~\ref{rank_one_lem_weighted} apply to this rank without alteration.
For rows $t_i=e_i^*T$ and $s_i=f_i^*T$, they give
\begin{align*}
 \Gamma_it_i\mathcal {\mathsf B}t_i^*+
 \kappa_i(e_i^*-t_i)\mathcal {\mathsf A}(e_i-t_i^*)+h_it_ie_i
 &\ge \frac{\langle e_i,\mathcal {\mathsf A}e_i\rangle}{\varpi_i^e}
                 -\frac{h_i^2}{4\Gamma_i\varpi_i^e},\\
 \Gamma_is_i\mathcal {\mathsf B}s_i^*+j_is_if_i
 &\ge \frac{\langle f_i,\mathcal {\mathsf A}f_i\rangle}{\varpi_i^f}
                 -\frac{j_i^2}{4\Gamma_i\varpi_i^f}.
\end{align*}
The first is completion with $\Gamma_i\mathcal {\mathsf B}+\kappa_i\mathcal {\mathsf A}$,
whose inverse is
\[
 \Gamma_i^{-1}D_*^{-1}
 +(\kappa_i^{-1}-\Gamma_i^{-1})D_*^{-1}\mathcal {\mathsf A}D_*^{-1}.
\]
The second uses $s_i\mathcal {\mathsf A}=f_i^*\mathcal {\mathsf A}$ and a PSD square
with $\mathcal {\mathsf B}$; its invertibility is not needed. The scalar identity
\[
 \frac{h_i^2}{4\Gamma_i\varpi_i^e}
 +\frac{j_i^2}{4\Gamma_i\varpi_i^f}=1-\varpi_i^e-\varpi_i^f
\]
and $h_i>j_i$ show, on summing, that
\[
 \tr[\Gamma T\mathcal {\mathsf B}T^*]+
 \sum_i\kappa_i\langle e_i,(I-T)\mathcal {\mathsf A}(I-T)^*e_i\rangle
 \ge \tr[{M_{\rm aux}}D_*]-r.
\]
Here the sum of the added diagonal terms is nonpositive.
Since ${M_{\rm aux}}\succeq T^*\Gamma T$,
$\tr[{M_{\rm aux}}\mathcal {\mathsf A}]\le\tr[{M_{\rm aux}}D_*]-\tr[\Gamma T\mathcal {\mathsf B}T^*]$;
this proves Eq.~\eqref{sp_eq_defect}. The loss is exactly the rank
defect $r$, not an unquantified dimension factor.

For movement, $0<\theta_i\le101/200$ implies
$(\varpi_i^e)^{-1}\le3$ and $(\varpi_i^f)^{-1}\le2\Gamma_i$.
Thus $D_*^{-1}\preceq3\Pi_0+2\Gamma\Pi_1$, where
$\Pi_1=I-\Pi_0$. The response identity gives
$\Pi_1E_*=\Pi_1TE_*$. Since $\Gamma$ commutes with $\Pi_1$,
\[
 \tr[C_*]\le3(k-r)+2\|\Gamma^{1/2}\Pi_1TE_*\|_F^2
 \le3k+2\tr[{M_{\rm aux}}]\le7k.
\]
Also $\|T\|_F^2\le2k/{\iota}$. Write $\ell=k-r\ge3k/4$. Then
\[
 \tr[(I-T)\Pi]\ge\ell-\sqrt{2k\ell/{\iota}}\ge k/5.
\]
The last inequality follows from ${\iota}>5$: the expression increases on
$\ell\ge3k/4$, and $(3/4-1/5)^2=121/400>3/10$.
Weighted Frobenius Cauchy--Schwarz now gives
\[
 (\tr[(I-T)\Pi])^2
 \le \|(I-T)E_*C_*^{-1/2}\|_F^2
       \|E_*C_*^{1/2}\|_F^2
 =\tr[(I-T)\mathcal {\mathsf A}(I-T)^*]\tr[C_*].
\]
This proves Eq.~\eqref{sp_eq_movement}.
\end{proof}

We spell out how the rank defect transfers to physical directions.
Retain $u_i,v_i,w_i,z_i,t_i,p_i,q_i$ from the
smoothed section and define
\[
 z_i^{\rm mov}=p_i^2u_iw_i+q_i^{ 2}v_iz_i>0,
 \qquad {M_{\rm aux}}_i=p_i u_i-q_i v_i.
\]
In the response construction put
${\iota}_i=(\beta_ip_i/q_i)^{1/2}$,
$d_i=(\beta_iq_i/p_i)^{1/2}$,
and $\tau_i=(p_iq_i/\beta_i)^{1/2}$.
The two response-coordinate energy diagonals are
$E_{W,i}={\iota}_i^2u_iw_i$ and $E_{Z,i}=d_i^2v_iz_i$; thus
$\tau_i^2(E_{W,i}+E_{Z,i})=z_i^{\rm mov}$.
The response graph in Eq.~\eqref{rank_one_eq_graph} maps bijectively to
physical $h\in\mathbb R^k$. Indeed the two equations
$p-Bq=J$ and $Ap-q=J$ have a unique solution for every $J$, since
$I-BA$ is invertible by the strict dual inequalities; $h_i=J_i/\tau_i$.
Consequently imposing any $r$ independent linear conditions on $h$
reduces its normalized response subspace dimension from $k$ to $k-r$.
All other hypotheses of Lemma~\ref{sp_lem_rankdefect} are unchanged.

\begin{corollary}[Covariance on a prescribed subspace]\label{sp_cor_covariance}
For every subspace $\mathcal {\mathscr E}\subseteq\mathbb R^k$ of codimension
$r\le k/4$ at a light state, there is a finitely supported centered
$h\in\mathcal {\mathscr E}$, with covariance $\Sigma$, such that
\begin{equation}\label{sp_eq_covariance_defect}
 \E[Z(h)]\le\sum_i\kappa_i z_i^{\rm mov}\Sigma_{ii}+r,
 \qquad \sum_i z_i^{\rm mov}\Sigma_{ii}\ge k/175.
\end{equation}
Here $Z(h)$ is the fixed-constraint inversion energy used in the source
covariance proof, not the entropy covariance operator $Q_{\mathsf Q}$.
\end{corollary}
\begin{proof}
Use the restricted response projection in Lemma~\ref{sp_lem_rankdefect}.
The normalized energy matrix has trace $2k$ and the same unit diagonal.
The physical covariance is obtained by applying the fixed response map
to $\mathcal {\mathsf A}^{1/2}\xi$, where $\xi$ consists of independent signs.
The movement matrix is supported on the $e_i$ coordinates and its
$i$th diagonal is $z_i^{\rm mov}\Sigma_{ii}$. The two conclusions
therefore follow directly. This distribution is only an existence
certificate in the proof; the algorithm never forms the response
projection, its covariance, or its square root.
\end{proof}

\paragraph{Many negative eigenvalues after a failed gradient test.}
For any covariance in Corollary~\ref{sp_cor_covariance}, use the preceding
drift $\mu_i=-C_{2,i}{M_{\rm aux}}_i\Sigma_{ii}$, with
$C_{2,i}=(1+\tau/2)\kappa_i$. The same exact cancellation and scalar
certificate as in Eq.~\eqref{eq_fast_strict_surplus} now yield
\begin{equation}\label{sp_eq_generator_defect}
 \mathcal D\mathcal F_{\rm s}\le-\upsilon_n\sum_i\varpi_i\Sigma_{ii}+C_\tau r,
 \qquad C_\tau=1+\tau/2,
\end{equation}
where $\upsilon_n^{-1}\le P_n$ is a known rational bound. For example,
if $\mathfrak b_n\ge\max(-\beta'')$ on the truncated interval, take
\[
 \upsilon_n=\min\{1/40,\kappa_{\rm fast}/(4{\iota}\mathfrak b_n)\}.
\]
In fact the matrix term is at most
$-\kappa_{\rm fast}\sum_iC_{2,i}p_iq_i t_i\Sigma_{ii}
+C_\tau r$, and the reservoir term is at most
$-{\gamma}\sum_i(-\beta_i'')\Sigma_{ii}/20$.
Here $C_{2,i}\ge1$ and $p_iq_i\ge1/2$.
This verifies Eq.~\eqref{sp_eq_generator_defect} with the stated choice.
The fixed-constraint majorization for $\mathcal P_{\rm s}$ is exactly the one already
proved in Section~\ref{soft_sec_geometry}; it does not assume hard
tight equations for the smoothed root.

Choose a known $\mathfrak m_n\le P_n$, enlarged so that
\[
 |C_{2,i}{M_{\rm aux}}_i|\le\mathfrak m_n\sqrt{n\varpi_i},\qquad
 |{M_{\rm aux}}_i|,\ |\beta_i'|,\ p_iq_i\le\mathfrak m_n.
\]
Use separated gradient tests so that failure implies
\begin{equation}\label{sp_eq_gradient_failed}
 |(\mathcal F_{\rm s})_i|\le\frac{\upsilon_n}{8\mathfrak m_n}\sqrt{\varpi_i/n}
 \quad\text{for all }i.
\end{equation}
A successful test certifies a lower bound half this size, which changes
only the logarithmic factors of the retained coordinate-step analysis.
At such a failed state there is a known $\mathfrak z_n\le P_n$ with
\begin{equation}\label{sp_eq_zeta_weight}
 z_i^{\rm mov}\le\mathfrak z_n \varpi_i.
\end{equation}
Indeed the exact identity is
\[
 z_i^{\rm mov}
 =p_iq_i t_i+
 {M_{\rm aux}}_i((\mathcal F_{\rm s})_i-{\gamma}\beta_i').
\]
Since $\varpi_i\ge({\iota}t_i+{\gamma})/2$, the right side is bounded by
$(2\mathfrak m_n/{\iota}+1+2\mathfrak m_n^2)\varpi_i$ using
Eq.~\eqref{sp_eq_gradient_failed}. This value may be used for
$\mathfrak z_n$.

\begin{lemma}[Negative spectral multiplicity]\label{sp_lem_inertia}
Put
\[
 \delta_n=\upsilon_n/16,\qquad
 \pi_n=\min\{1/8,\upsilon_n/(1024C_\tau\mathfrak z_n)\}.
\]
At a light state satisfying Eq.~\eqref{sp_eq_gradient_failed}, the
matrix $A=G^{-1/2}\nabla^2\mathcal F_{\rm s}G^{-1/2}$ has at least $\pi_n k$
eigenvalues smaller than $-4\delta_n$. In particular
$\pi_n^{-1},\delta_n^{-1}\le P_n$. No covariance computation is
part of the algorithmic assertion.
\end{lemma}
\begin{proof}
Let $r$ count eigenvalues below $-4\delta_n$. If $r>k/4$ the result
is immediate. Otherwise take $\mathcal {\mathscr E}$ to be the preimage, under
$G^{1/2}$, of their orthogonal complement, and apply
Corollary~\ref{sp_cor_covariance}. Write ${\mathsf P}_\Sigma=\sum_i\varpi_i\Sigma_{ii}$.
Then ${\mathsf P}_\Sigma\ge k/(175\mathfrak z_n)$ and
\[
 \tfrac12\tr[\nabla^2\mathcal F_{\rm s}\Sigma]\ge-2\delta_n {\mathsf P}_\Sigma
 =-\upsilon_n{\mathsf P}_\Sigma/8.
\]
On the other hand $|\nabla \mathcal F_{\rm s}\cdot\mu|\le\upsilon_n{\mathsf P}_\Sigma/8$,
so Eq.~\eqref{sp_eq_generator_defect} implies
\[
 \tfrac12\tr[\nabla^2\mathcal F_{\rm s}\Sigma]
 \le-7\upsilon_n{\mathsf P}_\Sigma/8+C_\tau r.
\]
It follows that
$r\ge3\upsilon_n k/(700C_\tau\mathfrak z_n)\ge\pi_n k$.
The argument also covers small $k$: the positive lower bound and
integrality force at least one such eigenvalue.
\end{proof}

\paragraph{A bounded spectral filter with a spread guarantee.}
At a failed gradient test, freeze the numerical root and the exact
symmetric Schur surrogate as before. Choose its accuracy so the scaled
surrogate is within $\delta_n/100$ of the true scaled Hessian. Denote
this surrogate by $A$ and write $\delta=\delta_n$, $\pi=\pi_n$.
By Lemma~\ref{sp_lem_inertia} and Weyl's inequality it has at least
$\pi k$ eigenvalues at most $-3\delta$. Enlarge the known rational
spectral bounds to
\[
 -RI\preceq A\preceq {M_{\rm aux}}I,\qquad {M_{\rm aux}}\le P_nk,\qquad
 1\le R\le P_n,\qquad 0<\delta\le R/8.
\]
The PSD shift is $B=A+2RI\succeq RI$. The endpoint construction
of Section~\ref{ue_sec_endpoint} supplies accurate applications of
$B^{-1}$ with complete cost $\widetilde O(n^\chi)$ per invocation,
including sketch preparation, provided the number of outer applications
and their accuracy and magnitude lengths are $O(\log(2kn))$.
We change only its outer polynomial.

\begin{lemma}[A checked flat-filter direction]\label{sp_lem_flat}
Given the preceding spectral bounds, a finite random-sign filter,
using polylogarithmically many applications of $B^{-1}$, has constant
probability of producing a true generalized-curvature witness $r$
with $\|r\|_2=1$ and
\begin{equation}\label{sp_eq_spread}
 r^{\top}\nabla^2\mathcal F_{\rm s}r\le-\tfrac12\delta r^{\top}Gr,
 \qquad \|r\|_\infty^2\le\mathsf s_n r^{\top}Gr,
 \qquad \mathsf s_n\le P_n.
\end{equation}
Both inequalities are verified with enclosures. Combined with the
finite recursive feature and index sampler, a capped trial succeeds
with conditional probability at least $1/64$ and costs
$\widetilde O(n^\chi)$ in arithmetic operations, including
unsuccessful trials.
\end{lemma}
\begin{proof}
Set $T=RB^{-1}$, so $0\preceq T\preceq I$, and define
\[
 t_- =\frac{R}{2R-\delta},\qquad
 t_+ =\frac{R}{2R-3\delta},\qquad
 t_0=(t_-+t_+)/2,\qquad \Delta=(t_+-t_-)/2.
\]
Here $\Delta\ge\delta/(8R)$. Choose a positive dyadic $\varepsilon_f$
with
\[
 \varepsilon_f\le1/16,\qquad
 \varepsilon_f^2\le\frac{\delta\pi}{128({M_{\rm aux}}+\delta)},
\]
and an integer $d_f\ge\max\{2/\Delta, 8\Delta^{-2}\log(2/\varepsilon_f)\}$.
All these quantities have polynomial or polylogarithmic magnitude;
$d_f$ is polylogarithmic. For $j_* =\lceil d_ft_0\rceil$
use the Bernstein polynomial
\begin{equation}\label{sp_eq_flatpoly}
 p_f(t)=\sum_{j=j_*}^{d_f}\binom{d_f}{j}t^j(1-t)^{d_f-j}.
\end{equation}
It lies in $[0,1]$ on $[0,1]$. For completeness, the logarithmic moment
function of a centered Bernoulli variable has value and derivative zero
at zero and second derivative at most $1/4$. Twice integrating gives
$\E[\exp(t(U-\E[U]))]\le\exp(t^2/8)$. Independence and
optimization of the exponential Markov bound therefore give
$\Pr[|\operatorname{Bin}(d_f,t)/d_f-t|\ge a]\le2e^{-2d_fa^2}$;
each one-sided tail has the same bound without the factor two.
It follows that
\[
 p_f(t)\le\varepsilon_f\ (t\le t_-),\qquad
 p_f(t)\ge1-\varepsilon_f\ (t\ge t_+).
\]
For the latter bound, rounding $d_ft_0$ costs at most $1/d_f\le\Delta/2$;
the two binomial tail bounds are at most $\exp(-d_f\Delta^2/2)$.
Thus $F=p_f(T)$ obeys $0\preceq F\preceq I$, is at least
$1-\varepsilon_f$ on eigenvalues of $A$ at most $-3\delta$, and
is at most $\varepsilon_f$ on eigenvalues of $A$ at least $-\delta$.

Draw independent real signs $\xi\in\{\pm1\}^k$ and let ${y_{\rm aux}}=F\xi$.
Writing $M_f=\tr[F^2]$, we have
\[
 M_f\ge\pi k/2,\qquad
 \E[\|{y_{\rm aux}}\|^2]=M_f,\qquad
 \E[\|{y_{\rm aux}}\|^4]\le3M_f^2.
\]
The fourth-moment inequality follows by expanding the quadratic form
$\xi^{\top}F^2\xi$; its second moment is at most
$(\tr[F^2])^2+2\tr[F^4]\le3M_f^2$.
Consequently $\Pr[\|{y_{\rm aux}}\|^2\ge M_f/2]\ge1/12$.
Each coordinate of $F\xi$ has the sign moment-generating-function
bound $\E[e^{t {y_{\rm aux}}_i}]\le e^{t^2/2}$, since its coefficient row has
Euclidean norm at most one. A union bound yields
\[
 \Pr[\|{y_{\rm aux}}\|_\infty^2>2\log(384k)]\le1/192.
\]
With probability greater than $1/16$ both estimates hold, so
\begin{equation}\label{sp_eq_flatmass}
 \|{y_{\rm aux}}\|^2\ge\pi k/4,\qquad
 \|{y_{\rm aux}}\|_\infty^2\le2\log(384k).
\end{equation}
The squared mass on eigenvalues at least $-\delta$ is at most
$\varepsilon_f^2k$. On this event
\[
 \frac{{y_{\rm aux}}^{\top}A{y_{\rm aux}}}{\|{y_{\rm aux}}\|^2}
 \le-\delta+\frac{4({M_{\rm aux}}+\delta)\varepsilon_f^2}{\pi}
 \le-7\delta/8.
\]

Unscale by the diagonal budget and normalize:
$r=G^{-1/2}{y_{\rm aux}}/\|G^{-1/2}{y_{\rm aux}}\|$. In exact arithmetic,
\[
 \frac{\|r\|_\infty^2}{r^{\top}Gr}
 =\frac{\max_i {y_{\rm aux}}_i^2/\varpi_i}{\|{y_{\rm aux}}\|^2}.
\]
The weight lower bound, Eq.~\eqref{sp_eq_lambda}, and $k>K/2$ show that
Eq.~\eqref{sp_eq_flatmass} makes this at most a fixed multiple of
$L_0\log(384(k_0+1))/(\Lambda\pi)$. For example one may take the known
rational upper bound
\[
 \mathsf s_n\ge
 \frac{2^{10}L_0\log(384(k_0+1))}{\Lambda\pi}.
\]
The slack also covers numerical weights within a fixed small relative
factor of the true weights. This proves the ideal spread and negative
quotient with more margin than Eq.~\eqref{sp_eq_spread} requires.

Here is an explicit evaluation and precision argument. Set
$v_j^{(0)}=\boldsymbol1_{\{j\ge j_*\}}\xi$ for $0\le j\le d_f$ and use
\begin{equation}\label{sp_eq_casteljau}
 v_j^{(r)}=(I-T)v_j^{(r-1)}+Tv_{j+1}^{(r-1)},
 \qquad 1\le r\le d_f,
 \quad 0\le j\le d_f-r.
\end{equation}
Its final vector is $F\xi$. This requires $O(d_f^2)$ inverse actions,
which is still a fixed polylogarithm. In an eigenbasis of $T$, each
update is a convex combination. Jensen's inequality, summed over
the shortened array, shows contraction in its concatenated Euclidean
norm. If each vector update has error at most $\eta$, the final error
is at most $d_f\sqrt{d_f+1}\eta$. Choose inverse-polynomial action
errors smaller than the normalization and quotient margins divided
by this factor, using the completed solves of
Lemma~\ref{ue_lem_precision}.

Require a lower norm enclosure at least $\sqrt{\pi k/8}$ and verify
both true inequalities in Eq.~\eqref{sp_eq_spread} before acceptance.
The estimates in Eq.~\eqref{sp_eq_flatmass} and the quotient bound
have slack for these errors, including the root and weight errors.
The feature and conditional-index events have probability at least
$(4/5)(99/100)$, and the independent sign event has probability
greater than $1/16$. Their product, with the reserved numerical
slack, is greater than $1/64$. Prescribed residual, iteration, and
magnitude caps bound every unsuccessful trial. Lemmas~\ref{ue_lem_precision}
and~\ref{ue_lem_sum} give the stated arithmetic cost.

\end{proof}

\paragraph{A curvature tube retaining the maximum coordinate.}
The previous proof bounded every $|r_i|$ by one. Keeping its actual
maximum makes the spread certificate useful.

\begin{lemma}[Spread-sensitive curvature tube]\label{sp_lem_tube}
Let $r$ be a certified Euclidean unit curvature direction at a failed
gradient state, put $\theta_0=r^{\top}Gr$ and $b=\|r\|_\infty$. A sufficiently
small fixed inverse-polylogarithmic $a_n>0$ makes
\begin{equation}\label{sp_eq_step}
 |s|\le a_n\min\{{{\delta_{\rm end}}}/b, (n\theta_0)^{-1/2}\}
\end{equation}
a valid two-sided trial tube. Both candidates can be evaluated from
the base optimizer with a constant number of corrected stages. The
better sign decreases the true current phase potential by at least
\begin{equation}\label{sp_eq_drop}
 P_n^{-1}\min\{{{\delta_{\rm end}}}^2\theta_0/b^2, n^{-1}\}.
\end{equation}
If Eq.~\eqref{sp_eq_spread} holds, this is at least $1/(P_nn)$.
The conclusion is valid across every coordinate-zero crossing.
\end{lemma}
\begin{proof}
Keep the directional maximum in the weighted coupling estimates.
For $1\le j\le3$ and $j\ge2$, respectively, they become
\[
 q(\mathcal A^{(j)}D)\le P_nb^{j-1}\sqrt{A_r} e(D),\qquad
 q(\mathcal C_{s^j})\le P_nb^{j-2}A_r.
\]
Indeed $|r_i|^{2j}\le b^{2j-2}r_i^2$ in weighted Cauchy--Schwarz,
and $|r_i|^j\le b^{j-2}r_i^2$ in the direct term. The bound for
$F_2=\mathcal A'{\mathsf Q}$ is unchanged. In the temporary small-energy region
of the curvature proof, completing the square in the second-order
identity still gives $e({\mathsf P}')+\mathcal E({\mathsf Q}')\le P_n\sqrt{\theta_0}$.
The third-order saddle identity now yields
\begin{equation}\label{sp_eq_third}
 |\mathcal P_{\rm s}'''|,\ |\mathcal F_{\rm s}'''|,\ |\theta'|
 \le P_n(b\theta_0+\sqrt n \theta_0^{3/2}).
\end{equation}
In detail the direct cubic and differentiated-coupling terms acquire
$b\theta_0$; the inverse words, transfer of the dual derivative, and
entropy cubic have the unchanged $\sqrt n\theta_0^{3/2}$ bound.
The reservoir cubic is at most $P_nb\theta_0$. Thus no term without
its required directional factor has been suppressed.

Use a simultaneous first-exit argument for weight comparability,
$\mathcal P_{\rm s}''\le {M_{\rm aux}}_0\theta_0$, expanded light bounds, and a bounded potential
sublevel. The coordinate displacement in Eq.~\eqref{sp_eq_step} is at
most ${{\delta_{\rm end}}}/4$, after fixing $a_n$, so endpoint distance remains at
least ${{\delta_{\rm end}}}/2$. Relative variations of $u_i,v_i,w_i,z_i$ have rate
at most $P_n\sqrt{n\theta_0}$; scalar budget relative variations have
rate at most $P_nb$. The same estimates used in
Section~\ref{soft_sec_local}, now integrated over Eq.~\eqref{sp_eq_step},
therefore close these comparisons and Eq.~\eqref{sp_eq_third}.

One additional point is needed because the Euclidean step can exceed
one: boundedness of the modeled matrix must not be inferred merely
from its Lipschitz bound times $s$. At the base the failed gradient
test gives
\[
 |\mathcal F_{\rm s}'(0)|\le P_n^{-1}\sqrt{k\theta_0/n}.
\]
While the temporary estimates hold, $|\mathcal F_{\rm s}''|\le P_n\theta_0$.
For $|s|\le a_n/\sqrt{n\theta_0}$ this implies
\[
 |\mathcal F_{\rm s}(y+sr)-\mathcal F_{\rm s}(y)|
 \le P_n^{-1}a_n\sqrt{k}/n+P_na_n^2/n<1/2,
\]
using $k\le n^2$ and a small enough prefactor. Hence the potential
cannot exit a fixed larger bounded sublevel; $\|M\|\le \mathcal P_{\rm s}\le \mathcal F_{\rm s}$
then supplies the bounded-model root estimates. This closes the
otherwise missing part of the first-exit argument.

In the symmetric average the linear terms cancel. The cubic remainder
from Eq.~\eqref{sp_eq_third} is at most a fixed small fraction of the
negative quadratic term, because
$P_n|s|(b+\sqrt{n\theta_0})$ is small relative to the certified
curvature threshold. This proves Eq.~\eqref{sp_eq_drop}. If
$b^2\le\mathsf s_n\theta_0$, its first entry is at least
${{\delta_{\rm end}}}^2/\mathsf s_n$, an inverse-polylogarithm, so the minimum
is at least $1/(P_nn)$.

The joint root energy arc is at most $P_n\sqrt{\theta_0}|s|$ and its
explicit-data transport integral at most $P_n\sqrt{n\theta_0}|s|$.
Shrinking the same prefactor makes both meet
Lemmas~\ref{soft_lem_ball} and~\ref{soft_lem_transport} over each
whole candidate. These are the existing executable anchored corrections,
not exact optimizer calls. $\mathcal P_{\rm s}''$ and $\theta$ are continuous at zeros;
one-sided bounds in Eq.~\eqref{sp_eq_third} integrate across any number
of them. No third derivative at a zero or higher Taylor jet is used.
Final common-grid rounding is taken below fixed fractions of the
true decrease, energy radius, and curvature and spread margins.
The longest possible step is polynomial since
$\theta_0\ge\Lambda/(4L_0k_0)$, so this remains a polylogarithmic
precision requirement.
\end{proof}

\paragraph{Batched deferred attenuation.}
At a corrected state, collect all coordinates satisfying the
near-endpoint test into one set $J$. Absorb their unchanged fractional
signed contribution into $F$, preserving $M$, and attenuate their
entire positive coupling
\[
 {\mathcal K}_s={\mathcal K}_{\rm rest}+e^{-s}{\mathcal K}_J,
 \qquad{\mathcal K}_J(D)={\iota}\sum_{i\in J}\beta_iV_iDV_i.
\]
Use one path to the required tiny residual coefficient, then delete
all its remaining slots exactly and correct once. Hold their reservoir
contribution fixed until that deletion. The proof of
Lemma~\ref{gf_lem_faces} applies to this positive submap verbatim:
with $K_s$ the corresponding pair map and $v=\langle {\mathsf Q},K_s{\mathsf P}\rangle$,
\[
 \mathcal P_{\rm s}'=-v,\quad e({\mathsf P}')+\mathcal E({\mathsf Q}')\le P_n\sqrt v,\quad
 |v'|\le P_nv,
 \quad q(K_sD)\le C\sqrt v e(D).
\]
The last inequality is the weighted Cauchy--Schwarz proof summed over
$J$, whose second factor is bounded by the full coupling dual energy;
no factor $|J|$ is lost. Likewise $K_s{\mathsf P}$ is a PSD subpair of the full
coupling, so $\tr[{\mathsf Q}(K_s{\mathsf P})^2]\le Cv$.

If there are $b_{\rm def}$ nonempty batches and their total matrix
potential drop is $\Delta_{\rm def}$, their complete correction cost
in stages is
\begin{equation}\label{sp_eq_batchcount}
 \widetilde O(b_{\rm def}+\sqrt{n b_{\rm def}\Delta_{\rm def}})
 =\widetilde O(b_{\rm def}+\sqrt{n b_{\rm def}}).
\end{equation}
The last equality uses the initial bound and the total phase injections
in Eq.~\eqref{sp_eq_injections}. All final tiny-slot deletions, grid
and scalar-exponential evaluations, positivity and residual enclosures
are included, as in the parent attenuation proof. Computing or applying
${\mathcal K}_J$ uses the same rank-one action bound as a full coupling, so a
batch stage still costs $\widetilde O(n^\chi)$.

Attenuation changes no remaining coefficient, so it cannot create a new
near-endpoint coordinate. A regular endpoint changes only the removed
coefficient, and phase changes alter no coordinate.
No charge proportional to $k_0$ separate optimizer corrections is
inserted for simultaneous removals. Recording their coefficients and
updating active index lists costs $O(k_0)$ over the whole run.

\paragraph{Finite fallback and curvature counts.}
Lemma~\ref{sp_lem_tube} gives
$I_{\rm flat}\le\widetilde O(n)$ for all accepted spread directions,
regardless of intervening fallback moves. Regular endpoints still have
$I_{\rm reg}\le\widetilde O(n)$: their rank-one feasibility adjustment
and net decrease $\widetilde\Omega(1/n)$ are unchanged.

At each failed gradient test, try at most
\[
 r_{\rm tr{y_{\rm aux}}}=\lceil64\{\log(2I_{\max})+20\log(2n)\}\rceil
\]
independent finite feature/index/sign trials. A successful trial must
pass both the true quotient and spread tests. If all fail, form the
full symmetric Hessian and use the existing deterministic witness
routine. This fallback need not provide a spread direction: use the
old length in Eq.~\eqref{soft_eq_curvstep} and its certified decrease.
A fallback costs $\widetilde O(n^{2\omega_0})$, including assembly and
precision. It always succeeds and terminates.

Since ${\gamma}\ge\Lambda/(L_0k_0)$, every curvature move, including a
fallback, has decrease at least $1/(P_n(k_0+n))$. Together with the
bounded initial potential and phase injections this gives the
\emph{deterministic} invocation cap $I_{\max}=P_n(k_0+n)$, after
rounding up its known bounds. At any invocation conditional on its
entire previous history, the fallback probability is at most
\[
 (63/64)^{r_{\rm tr{y_{\rm aux}}}}
 \le[2I_{\max}(2n)^{20}]^{-1}.
\]
Thus
\begin{equation}\label{sp_eq_fallbackexpectation}
 \E[I_{\rm fb}]\le\tfrac12(2n)^{-20}.
\end{equation}
This analysis does not assume independence between different invocations.
There are no unbounded sampling or precision loops on an adverse
random-bit sequence.

The number of successful flat-filter invocations is
$\widetilde O(n)$; each complete capped invocation, including rejected
trials and all implicit inverse setup, costs $\widetilde O(n^\chi)$.
The fallback contribution is charged in Lemma~\ref{am_lem_counts}.

\paragraph{Phase changes and discrepancy accounting.}
The root equations and their numerical residual do not contain
${\gamma}$. Thus phase changes require no optimizer solve; only scalar
weights, gradient reservoir terms and the scaled Hessian definition
change. Known lower bounds use $\Lambda/(L_0k_0)$ uniformly. Phase
coefficients are exact short rationals, and positive phase jumps are
charged analytically by Eq.~\eqref{sp_eq_injections}, not hidden in
value-comparison error.

Choose the root and comparison accuracies below fixed fractions of
the uniform weight lower bound and the margins in
Eq.~\eqref{sp_eq_spread}. The accepted directions then satisfy the
true tests through every phase change, with the same fallback and
input-conversion allowances.

Finally, Eq.~\eqref{sp_eq_injections} accounts for the initial reservoir
and every later reallocation within the \emph{same} $10^{-6}$ allowance.
The smoothing penalty is paid only once, and all other allowances
are unchanged. Consequently the internal coefficient and transfer are
still
\[
 (2\sqrt{{\iota}+2{{\varepsilon_{\rm scale}}}}+10^{-6}+{{\varepsilon_{\rm round}}}+\eta_{\rm ps}+10^{-8})
     \sqrt{1+{{\varepsilon_{\rm scale}}}}+{\varepsilon_{\rm discard}}<4.86276,
\]
\[
 4.86276(1+6\cdot10^{-8})+2\cdot10^{-8}
 =4.8627603117656<4.8628.
\]
Thus spread moves and batched attenuation preserve the discrepancy allowance.

\subsection{Slack-scaled grouped reduction and fast final rounding}
\label{sr_sec_improvement}
We use the spread construction of Section~\ref{sp_sec_improvement}
and replace the small reduction inside the preprocessing group tree. It now leaves at most twice
the vector-space dimension fractional coordinates. Constant-factor
zero padding restores the active-count hypothesis of the subsequent
matrix algorithm. We also accelerate the last deferred-rounding step;
otherwise its $n^5$ term would remain in the total.

\paragraph{A small reduction with dimensional slack.}
In the next lemmas $D$ is the real vector-space dimension, not a matrix
order. The columns need not be rank one or positive; the application to
matrix inputs is given afterward. Fix the same admissible square
multiplication exponent $\omega_0=2.371177$ as in
Eq.~\eqref{fast_eq_multiplication_parameters}. A fixed bilinear
recursion at a strict exponent below $\omega_0$ gives polynomial
normwise rounding amplification. No new multiplication bound is used.

\begin{lemma}[Slack-scaled approximate null directions]\label{sr_lem_direction}
Let $A\in\mathbb R^{D\times r}$ have column norms at most two,
$2D<r\le r_0\le4D$, and let $x\in(-1,1)^r$. Put
\[
 s_i=1-|x_i|,\qquad \mathsf S=\operatorname{diag}(s_i),\qquad
 \mathsf B=A\mathsf S,
\]
and, for $\delta>0$, define
\begin{equation}\label{sr_eq_projection}
 J_\delta=\delta I_D+\mathsf B\mathsf B^{\top},\qquad
 \Pi_\delta=I_r-\mathsf B^{\top}J_\delta^{-1}\mathsf B.
\end{equation}
Then $0\prec\Pi_\delta\preceq I$,
$\tr[\Pi_\delta^2]\ge r-D$, and
$\|\mathsf B\Pi_\delta\|\le\sqrt\delta/2$.
Let $\ell=\lceil\log_2(64r_0)\rceil$, and let $\xi$ have independent
uniform real signs. If a dyadic vector $\widetilde v$ satisfies
\[
 \|\widetilde v-\Pi_\delta\xi\|_2\le\nu\le1/64,
\]
then, with probability at least $1/16$, it passes both exact tests
\begin{equation}\label{sr_eq_tests}
 \|\widetilde v\|_\infty\le\ell,\qquad
 \|\widetilde v\|_2^2\ge(r-D)/4.
\end{equation}
For any vector passing the tests, $h_i=s_i\widetilde v_i/H$ with
$H=4\ell$ satisfies $|h_i|\le s_i/4$.
\end{lemma}
\begin{proof}
The singular-value formula for $\Pi_\delta$ gives eigenvalues
$\delta/(\delta+\sigma_j(\mathsf B)^2)$ and at least $r-D$
eigenvalues equal to one. Its corresponding residual singular values
are $\delta\sigma_j/(\delta+\sigma_j^2)\le\sqrt\delta/2$.
This calculation is only a proof; the implementation uses a ridge
inverse, not a singular-value oracle.

Write $V=\Pi_\delta\xi$ and $a=\tr[\Pi_\delta^2]$. Independent signs
give
\[
 \E[\|V\|^2]=a,\qquad
 \E[\|V\|^4]\le3a^2.
\]
For the latter, expand the square of $\xi^{\top}\Pi_\delta^2\xi$; its
variance is at most $2\tr[\Pi_\delta^4]\le2a^2$.
Cauchy--Schwarz on the event $\|V\|^2\ge a/2$ gives its probability
at least $1/12$. Each row of $\Pi_\delta$ has Euclidean norm at most
one. The elementary estimate $\cosh t\le e^{t^2/2}$ and exponential
Markov inequality therefore imply
\[
 \Pr[\|V\|_\infty>\ell/2]
 \le2r_0e^{-\ell^2/8}
 \le2^{\ell-5}e^{-\ell^2/8}<1/192.
\]
Here $r_0\ge3$ and $\ell\ge8$; the last expression decreases for
$\ell\ge8$, and its value at eight is $8e^{-8}<1/192$.
On the intersection of these two events, the perturbation by at most
$\nu$ preserves the infinity test. Also $\|V\|\le\sqrt r$, so
\[
 |\|\widetilde v\|^2-\|V\|^2|
 \le2\sqrt r\nu+\nu^2\le(r-D)/4
\]
for $r>2D\ge2$. This gives the second test. The intersection has
probability at least $1/12-1/192>1/16$. The last assertion follows
from $H=4\ell$.
\end{proof}

\begin{lemma}[A capped small cube reduction]\label{sr_lem_small}
Let $C\in\mathbb Q^{D\times r_0}$ have $r_0\le4D$ columns of norm
at most two, and let $y\in[-1,1]^{r_0}$ be a dyadic vector.
Given $0<\varepsilon<1$ and $0<\zeta_{\rm f}<1/2$, a finite randomized
procedure returns $x\in[-1,1]^{r_0}$ with at most $2D$ fractional
coordinates and
\begin{equation}\label{sr_eq_smallresidual}
 \|C(x-y)\|_2\le\varepsilon.
\end{equation}
Every random-bit sequence terminates with this guarantee. Its fast
attempt costs $\widetilde O(D^{\omega_0})$ operations, and it calls a
$\widetilde O(D^3)$ deterministic fallback with probability at most
$\zeta_{\rm f}$. Thus its expected cost is
$\widetilde O(D^{\omega_0}+\zeta_{\rm f}D^3)$ arithmetic operations.
The suppressed factors include logarithms of the inverse error and
failure budgets.
\end{lemma}
\begin{proof}
If there are already at most $2D$ fractional coordinates, return $y$.
Otherwise $r_0\ge3$. Choose a dyadic $\sigma_{\rm r}$ satisfying
\[
 \frac{\varepsilon}{2^{11}r_0}<\sigma_{\rm r}
 \le\frac{\varepsilon}{2^{10}r_0}.
\]
Set the following parameters once for the entire small call:
\begin{equation}\label{sr_eq_parameters}
 \begin{split}
 \ell&=\lceil\log_2(64r_0)\rceil,\quad H=4\ell,\quad
 L_{\delta_{\rm end}}=\lceil\log_2(2/\sigma_{\rm r})\rceil,\\
 T_*&=128H^2L_{\delta_{\rm end}}+1,\qquad
 \nu=\frac{\varepsilon}{2^{14}T_*r_0},\qquad
 \nu^2/2<\delta\le\nu^2.
 \end{split}
\end{equation}
The ridge $\delta$ is dyadic. Use a dyadic coefficient grid of mesh
$u$ no larger than
\begin{equation}\label{sr_eq_grid}
 \min\{
 \frac{\sigma_{\rm r}}{2^{16}H^2}, 
 \frac{\varepsilon}{2^{14}T_*r_0}
 \},
\end{equation}
within a factor two of this minimum. Keep the inherited coefficients
exact until the first move. Each inward rounding changes a coordinate
by at most $u$, whether or not its preceding value lies on the grid.

Keep the original $y$ for a possible fallback. Freeze every current
coordinate at distance at most $\sigma_{\rm r}$ from an endpoint by
setting it to that nearest endpoint. This includes the initial scan.
Each coordinate is frozen at most once. If at most $2D$ remain, stop.
Otherwise form Eq.~\eqref{sr_eq_projection} for the current active columns
and current slacks. Draw signs and compute $\widetilde v$ to the
accuracy in Lemma~\ref{sr_lem_direction}. Reject a direction failing
Eq.~\eqref{sr_eq_tests}. For a passing vector put
$h=\mathsf S\widetilde v/H$. Both $x+h$ and $x-h$ are strictly inside
the active cube. For the barrier
\begin{equation}\label{sr_eq_barrier}
 \Phi(x):=\sum_{i\ {\rm active}}\phi(x_i),\qquad
 \phi(t):=-\log(1-t^2),
\end{equation}
use the directional derivative
\begin{equation}\label{sr_eq_derivative_sign}
 g:=\nabla\Phi(x)^\top h
   =\sum_i\frac{2x_ih_i}{1-x_i^2}
   =\sum_i\frac{2x_i\widetilde v_i}{H(1+|x_i|)}.
\end{equation}
Compute $\widehat g$ with $|\widehat g-g|\le r/(512H^2)$ and choose
$\tau:=1$ if $\widehat g\ge0$, and $\tau:=-1$ otherwise.
In the arithmetic model take $\widehat g=g$ by evaluating the displayed
rational sum exactly. Round $x+\tau h$ inward, toward zero, to the coefficient
grid, freeze all newly triggered coordinates, and repeat.

\paragraph{Progress with finite endpoints.}
Put $s_i:=1-|x_i|$. Since $r>2D$, we have $r\le2(r-D)$.
The derivative sign rule gives
\[
 \tau g\ge-|\widehat g-g|\ge-\frac{r-D}{256H^2},
\]
even when the two signs cannot be distinguished at the prescribed
accuracy. For $0\le t\le1$, the distance from $x_i+t\tau h_i$ to
the endpoint nearest $x_i$ is at most $s_i+|h_i|\le5s_i/4$.
One of the two terms in
\begin{equation}\label{sr_eq_barrier_curvature}
 \phi''(z)=\frac1{(1-z)^2}+\frac1{(1+z)^2}
\end{equation}
is therefore at least $16/(25s_i^2)$ throughout this segment.
Taylor's theorem and Eq.~\eqref{sr_eq_tests} imply
\[
 \Phi(x+\tau h)-\Phi(x)
 \ge\tau g+\frac8{25}\sum_i\frac{h_i^2}{s_i^2}
 \ge(\frac2{25}-\frac1{256})\frac{r-D}{H^2}.
\]
Before inward rounding, every active endpoint distance is at least
$3\sigma_{\rm r}/4$. On the inward segment $|\phi'|\le4/\sigma_{\rm r}$,
so Eq.~\eqref{sr_eq_grid} bounds its loss by
\[
 \frac{4ru}{\sigma_{\rm r}}\le\frac{r-D}{8192H^2}.
\]
Since $2/25-1/256-1/8192=15559/204800>1/16$, the net increase is at least
\begin{equation}\label{sr_eq_gain}
 \frac{r-D}{16H^2}>\frac{D}{16H^2}.
\end{equation}

For analysis, assign every frozen coordinate the finite value
$Q_{\delta_{\rm end}}=\log(2/\sigma_{\rm r})$ and leave active values equal to
$\phi(x_i)$. This defines a capped bookkeeping potential
$\mathcal V$. After a move, a newly frozen coordinate was still at
distance at least $3\sigma_{\rm r}/4$; replacing its barrier value by
$Q_{\delta_{\rm end}}$ cannot decrease $\mathcal V$. Initial freezes cause no
problem: start the bookkeeping after that scan. Always
\[
 0\le\mathcal V\le r_0Q_{\delta_{\rm end}}\le4DL_{\delta_{\rm end}}.
\]
Consequently fewer than $64H^2L_{\delta_{\rm end}}+1<T_*$ successful steps are
possible while $r>2D$. This is a deterministic bound for every
sequence of accepted directions, not a drift-only stopping argument.
The active scalar barrier is never evaluated at an exact endpoint.

\paragraph{The residual is a separate budget.}
Lemma~\ref{sr_lem_direction} gives at a successful step
\[
 \|Ah\|_2
 \le\frac{\sqrt r}{H}(\frac{\sqrt\delta}{2}+2\nu)
 \le\frac{5r_0\nu}{8},
\]
where $H\ge4$ and $\sqrt r\le r_0$. Coefficient rounding adds at most
$2r_0u$. All initial and subsequent endpoint freezes together add at
most $2r_0\sigma_{\rm r}$. Thus any fast output obeys
\begin{equation}\label{sr_eq_errorbudget}
 \begin{split}
 \|C(x-y)\|_2
 &\le2r_0\sigma_{\rm r}
       +T_*(5r_0\nu/8+2r_0u)\\
 &\le\varepsilon(\frac2{2^{10}}+
                   \frac{5/8+2}{2^{14}})
 <\varepsilon/128.
 \end{split}
\end{equation}
No error term is charged per original leaf at every group update.
All columns in this small call have norm at most two, so this bound
also covers repeated, zero, or linearly dependent columns.

\paragraph{A finite rational direction oracle.}
At a current state, $\delta I\preceq J_\delta\preceq(\delta+4r_0)I$.
Form $\mathsf B$ and $J_\delta$ exactly. Starting from
$R_0=J_\delta/U^2$, with a rational $U\ge\|J_\delta\|$, apply
$R_{j+1}=2R_j-R_jJ_\delta R_j$. Its residual squares at each iteration,
and its initial spectral gap is at least $\delta^2/U^2$. Thus
logarithmically many exact arithmetic products give an inverse
approximation $R$ to the required accuracy. Require, for example, a true residual enclosure
\[
 \|I-J_\delta R\|_\mathrm F
 \le\frac{\delta\nu}{64r_0^{3/2}}.
\]
The known inverse bound $\|J_\delta^{-1}\|\le\delta^{-1}$ converts
this to an inverse-error bound. Applying
$\xi-\mathsf B^{\top}R\mathsf B\xi$ and rounding to a common dyadic
vector grid, with sufficiently smaller fixed error allocations,
gives $\|\widetilde v-\Pi_\delta\xi\|_2\le\nu$. A square-root
factor in an error bound may be replaced by $r_0$ to use rational
thresholds. The product residual includes construction error in
$J_\delta$; it is not a residual for a succession of different matrices.

The polynomial bounds on $U$, $\delta^{-1}$, and the inverse residual
target prove that the inverse costs
$\widetilde O(D^{\omega_0})$ arithmetic operations, and each subsequent
direction trial costs $\widetilde O(D^2)$.

\paragraph{Arithmetic cost of the sign rule.}
Eq.~\eqref{sr_eq_derivative_sign} cancels the endpoint slack before
evaluation. Each denominator $1+|x_i|$ lies in $[1,2)$, so evaluating
and summing its rational terms costs $O(r)$ arithmetic operations.
The formula is valid also at $x_i=0$. The progress proof above allows
the displayed additive error in this sign test.

\paragraph{Caps, verification, and fallback.}
At any unsuccessful search for a direction, permit at most
\[
 R_* =64\lceil\log_2(2T_*/\zeta_{\rm f})\rceil
\]
independent sign trials. If they all fail, or if any required accuracy
or consistency check fails, abandon the fast attempt and return the
deterministic fallback from the original inherited $y$. Also cap the
number of successful steps at $T_*$. Upon a prospective fast return,
check the cube, the fractional count, and
$\|C(x-y)\|_2^2\le\varepsilon^2$ by exact rational arithmetic. The
matrix-vector product and squared Euclidean norm make this last check
$O(Dr_0)$ arithmetic operations.
Eq.~\eqref{sr_eq_errorbudget} leaves ample slack.

For fallback split the at most $4D$ columns into two sets of at most
$2D$. Apply the deterministic small reducer in the proof of
Lemma~\ref{rank_one_lem_grouped_partial_signing} to each set, with its
inherited coefficients and residual $\varepsilon/2$. It leaves at
most $D$ fractional coordinates per set. Its combined cost is
$\widetilde O(D^3)$ and its residual is at most $\varepsilon$.
The arithmetic construction in that lemma applies to the inherited
dyadic coefficients and preserves its residual estimate.

All matrix and scalar tolerances above guarantee their tests in the
stated caps; only the random-direction event may fail to occur.
Conditionally on every previous history, a sign trial succeeds with
probability at least $1/16$. Thus the probability of a fallback is
at most
\[
 T_*(15/16)^{R_*}\le\zeta_{\rm f}.
\]
Every other fast-attempt cost is bounded deterministically by
$T_*R_*$ times the displayed polynomial work, and $T_*,R_*$ are
polylogarithmic in the parameters. The resulting expected bound is
as stated. Every random-bit sequence terminates, and every returned
vector has the required verified or deterministic guarantee.
\end{proof}

\begin{lemma}[Grouped reduction with a factor-two slack]
\label{sr_lem_grouped}
Under the hypotheses of
Lemma~\ref{rank_one_lem_grouped_partial_signing} with $d_{\rm vec}=D$,
a zero-error randomized algorithm finds $x\in[-1,1]^m$ with at most $2D$ fractional
coordinates and $\|\sum_i x_ib_i\|_2\le\eta$ in expected cost
\begin{equation}\label{sr_eq_groupedcost}
 \widetilde O(mD+D^{\omega_0})
\end{equation}
arithmetic operations.
Every random-bit sequence terminates with the same residual guarantee.
\end{lemma}
\begin{proof}
Retain the balanced tree, the precomputed sums $W_J$, and the power of
two $M_{\rm grp}\in[m,2m)$ from the source. Maintain at most $2D$
active groups, not $D$. Splitting their nonsingleton groups gives at
most $4D$ groups. At each of at most
$R_{\rm tree}=\max\{1,\lceil\log_2m\rceil\}$ levels, apply
Lemma~\ref{sr_lem_small} to their normalized sums $W_J/M_{\rm grp}$
and inherited coefficients, with
\[
 \varepsilon=\eta/(M_{\rm grp}R_{\rm tree}),\qquad
 \zeta_{\rm f}=(2mD)^{-20}/R_{\rm tree}.
\]
All normalized columns have norm at most two. Coefficients are copied
to children before rounding; this changes no signed sum. Each level's
returned residual is at most $\eta/R_{\rm tree}$ in the unnormalized
sum, on every random-bit sequence. Summing gives the same single
allowance $\eta$. After the last level the remaining groups are
singletons, so at most $2D$ original coordinates are fractional.

Precomputing all group sums and writing final leaf coefficients cost
$\widetilde O(mD)$. No accepted small-reduction step scans a group's
leaves. There are logarithmically many fast calls of cost
$\widetilde O(D^{\omega_0})$. The total expected fallback contribution
is at most $(2mD)^{-20}\widetilde O(D^3)$, hence is smaller. This remains
valid conditionally on inherited coefficients and previous random
choices. Each small call uses the mesh prescribed by
Eq.~\eqref{sr_eq_grid}; all grid searches have a uniform logarithmic
arithmetic bound across the tree. The exact matrix residual checks
are included in the small-call cost. This proves the expected
arithmetic bound.
\end{proof}

\paragraph{Restoring the matrix active-count hypothesis.}
Let $n$ be the input matrix order. Put $D=n^2$ and
\[
 N=2^{\lceil\log_2(2n)\rceil},\qquad 2n\le N<4n.
\]
Use the one-time conversion and variance scaling with the conservative
order $N$, and represent the nonzero $n$-by-$n$ block in $D$ real
Hermitian coordinates. The other entries are zero. The isometric
coordinate approximation and residual transfer following
Lemma~\ref{rank_one_lem_grouped_partial_signing} cost
$\widetilde O(mn^2+N^3)$ arithmetic operations. Apply
Lemma~\ref{sr_lem_grouped} with its part of the same
$\eta_{\rm ps}=10^{-7}$ allowance. Its output has
\[
 k_0\le2D=2n^2\le N^2.
\]
Pad each true retained matrix by zeros to order $N$, and run exactly
the spread-direction implementation at that order, including
$\rho={\varepsilon_{\rm scale}}/N$, its cutoff, reservoir phases, and all numerical
margins. Padding preserves positivity, rank one, the variance bound,
and the norm of every signed sum. It is not an additional discrepancy
approximation. The operator inequalities with $I_N$ remain valid.

The centered initial modeled sum is still zero. Every returned group
coefficient has absolute value at most one, so the approximation error
in all vectorized inputs is charged once, as in the source. The
fractional coefficients are on a short dyadic grid; they may subsequently
be rounded more finely as in the inherited initialization, within its
unchanged ${\varepsilon_{\rm round}}/2$ allowance. The initial residual $e_0$ is subtracted
from the fixed contribution and restored once at the end. No retained
matrix is replaced by a group sum in the nonlinear phase.

\paragraph{Nonlinear algorithm, face events, and precision.}
The grouped reduction preserves the centered model and the active-count
hypothesis. It does not change the joint-energy Newton ball or its anchored
inverse. Regular endpoints use Lemma~\ref{am_lem_endpoint}.
All deferred coordinates at a
state are still attenuated as one positive coupling submap, with every
final tiny-slot deletion corrected. Reservoir reallocation is charged
by the same total $10^{-6}$ allowance. Spread and fallback directions
retain their distinct step rules; only successful verified spread
moves use the longer tube. Coordinate-zero crossings use the same
continuous second derivatives and one-sided bounds. The exact ridge
barrier used only in preprocessing is not a replacement for this
matrix potential and imposes no new differentiability requirement on it.

Random signs for the new reduction are independent of future curvature
trials. The latter's conditional success statements remain true for
arbitrary adaptive earlier data. Every small reducer has a finite cap
and a deterministic fallback; so does every curvature invocation.
Their composition therefore terminates correctly on every random-bit
sequence, not merely almost surely. All outputs of preprocessing,
including fallback outputs, meet the same single residual bound.

The nonlinear model continues to use the true retained matrices with
a symbolic fixed contribution whose coefficients are bounded by two.
Caches approximate these objects once at the requested inverse-polynomial
accuracy; no original long rational is reread in a Newton stage, sketch,
or final trace evaluation. Short-grid rounding, positivity, residual,
spread, and Rayleigh enclosures remain those of the spread implementation.
Lemma~\ref{sr_lem_round} only changes how its final estimator is evaluated.

\subsection{Amortized gradient sweeps and regular endpoint continuation}
\label{am_sec_improvement}
We apply the shared endpoint continuation and gradient sweeps from
Section~\ref{sec:shared:endpoints-sweeps} to the randomized construction.
After the slack-scaled grouped reduction of Section~\ref{sr_sec_improvement},
pad to $N=2^{\lceil\log_2(2n)\rceil}<4n$, so the initial retained count
satisfies $k_0\le2n^2\le N^2$. Use the same $\rho$, cutoff, smoothing
temperature, and reservoir phases as in the shared proofs.
For the high test in Eq.~\eqref{am_eq_high}, take
$h_N=\upsilon_N/[64(\mathfrak m_N+1)]$ in the notation of
Lemma~\ref{sp_lem_inertia}. This choice implies its failed-gradient
hypothesis; set $l_N=h_N/64$ as in Lemma~\ref{am_lem_sweep}.
The count below gives expected cost $\widetilde O(N^{\chi+1})$
for all corrected local and face stages.

\paragraph{Why unsaturated sweeps cannot occur too often.}
A short unsaturated sweep need not itself pay $1/(P_NN)$.  It is
important not to charge every such sweep to the larger saturated
progress bound.

\begin{lemma}[Sweep and deferred-batch counts]\label{am_lem_counts}
With the high/low tests above, the total number of gradient sweeps is
$\widetilde O(N+I_{\rm flat}+I_{\rm reg}+I_{\rm fb})$.
All deferred attenuation batches use expected $\widetilde O(N)$
corrected stages.  Together with regular endpoint continuation and
spread-curvature candidates, all corrected local and face stages
number $\widetilde O(N)$ in expectation.
\end{lemma}
\begin{proof}
Saturated sweeps number $\widetilde O(N)$ by
Eq.~\eqref{am_eq_sweepdrop} and the bounded total phase potential decrease.
After an unsaturated sweep, collect all marked coordinates in one
nonempty deferred set and use the retained positive-submap attenuation
path.  No remaining coefficient changes during this path.  Write
$\Delta_J$ for its matrix-potential decrease before the final tiny-slot
correction and let $L_*\le P_N$ be a uniform upper bound on its
logarithmic horizon.  The preceding forcing estimate gives total energy
arc at most
\begin{equation}\label{am_eq_smallatten}
 P_N\int_0^{L_*}\sqrt{v(s)} \d s
 \le P_N\sqrt{L_*\Delta_J}.
\end{equation}
If $\Delta_J\le1/(P_N' N)$ for a sufficiently large known fixed
polylogarithm $P_N'$, this arc and its normalized transport are small
relative to $h_N/\sqrt N$.  The final tiny-slot deletion is below the
same margin by tightening its prescribed cache floor.

Here is the needed gradient comparison along that small path.  For a
remaining coordinate put $\widetilde g_i(s)=g_i(s)/\sqrt{{\iota}t_i(s)+{\gamma}}$.
In a fixed-factor neighborhood of its starting root, the forcing proof
in Lemma~\ref{am_lem_frozen} and the relative scalar bounds give
\[
 |\widetilde g_i'|\le P_N a(s)+P_N\sqrt N|\widetilde g_i|a(s),\qquad
 a(s)=e({\mathsf P}')+\mathcal E({\mathsf Q}').
\]
Until $|\widetilde g_i|$ reaches $h_N/(2\sqrt N)$ the right side is at most
$P_N a(s)$.  Starting from Eq.~\eqref{am_eq_lowexit},
Eq.~\eqref{am_eq_smallatten} prevents that exit and keeps the metric
comparable, by the same first-exit transport argument.  Thus after
small attenuation no high gradient test can pass.  This comparison
uses the fixed remaining coordinates; it does not assume that a
normalized weight has zero derivative.

If a new reservoir phase is opened, charge this event separately;
there are at most $L_0$ phases.  Otherwise a new sweep following an
unsaturated one must be preceded either by attenuation drop at least
$1/(P_N'N)$, or by a regular endpoint or curvature event.  The large
attenuation drops sum to at most the bounded total potential decrease,
so there are only $\widetilde O(N)$ of them.  This proves the stated
sweep count, with no independence assumption between adaptive states.

There is at most one deferred batch per corrected accepted state;
attenuation changes no remaining coefficient.  Therefore the batch
count is bounded by the number of sweeps, curvature events and regular
endpoints, plus one.  Eq.~\eqref{sp_eq_batchcount} gives total batch stages
$\widetilde O(b+\sqrt{Nb})$, because their nonnegative drops sum to a
constant.  Successful spread moves and regular endpoints number
$\widetilde O(N)$.  For the retained capped curvature fallback,
$\E[I_{\rm fb}]\le(2N)^{-20}/2$.  The excess stage count is
bounded by $P_N(1+\sqrt N)I_{\rm fb}$, using
$\sqrt j\le j$ for nonnegative integers.  Taking expectations gives
$\widetilde O(N)$ batch and sweep stages.  Lemma~\ref{am_lem_endpoint}
gives $\widetilde O(N)$ regular stages, and the initial explicit
homotopy uses only $\widetilde O(\sqrt N)$ stages.  Both signs of each
spread-curvature candidate retain their constant number of corrections.
This proves the total claim.
\end{proof}

\paragraph{Algorithmic order and arithmetic cost.}
After the unchanged preprocessing and initial centered homotopy,
process the deferred batch at a corrected state, then update the
reservoir phase if needed.  Recompute conservative regular-endpoint
tests; execute a triggered regular endpoint with
Lemma~\ref{am_lem_endpoint}.  Otherwise the state is light.  The high
gradient test either starts a sweep or satisfies the retained negative
spectral multiplicity hypothesis.  In the latter case use the same
verified spread filter and finite fallback.  All marked sweep
coordinates are deferred together before another high test.  Phases
remain fixed inside a sweep and its deferred attenuation path.

At most a polynomial number of coordinate updates, phases, and slow
fallback moves occur even for an adverse random-bit sequence.  Every
sweep has a certified positive decrease, and the classification in
Lemma~\ref{am_lem_counts} also gives a deterministic polynomial cap.
The curvature invocation cap $I_{\max}=P_N(k_0+N)$ can be enlarged by
fixed factors: every fallback still has its old inverse-polynomial
true decrease.  The same logarithmic trial cap bounds the conditional
fallback probability independently of the preceding sweeps.  No
randomness is used in either new continuation or gradient batching.

The frozen sweep data and coordinate comparisons use errors below
fixed fractions of the proved gradient thresholds and root radii.
The positive continuation ends with a residual-certified correction
at the actual new face.

All paths and products are valid for Hermitian matrices.  Real scalar
coefficient directions and trace-zero projection use the existing
rational real/imaginary coordinates.  The budget is $C^2$; its scalar
concavity and continuous first derivative hold across zero, and the
chord argument does not require third derivatives there.  Neither a
sweep nor the positive regular homotopy changes the recorded input
matrices or the single centered residual.  The fixed signed
contribution has the same bounded symbolic coefficients.  The true
model, short cache and their error enclosures remain separate.
Original ${L_{\rm in}}$-bit data are not reread during any new step.

The entire new local work, including scans and both curvature
candidates, is $\widetilde O(N^{\chi+1})$ in expectation.  The at most
$k_0+\widetilde O(N)$ microsteps may be charged
$\widetilde O((k_0+N)N^2)$ separately.  No regular or deferred face
work is omitted.  The complete total is therefore
\begin{equation}\label{am_eq_refined}
 \E[T_{\rm arith}]\le\widetilde O(
 mn^2+n^{2\omega_0}+N^3+k_0N^{\omega_0}
 +(k_0+N)N^2+N^{\chi+1}).
\end{equation}
Both rare preprocessing and curvature fallbacks keep
their existing expected-cost charges.

Finally, the auxiliary positive penalties are used only while evaluating
an endpoint's optimizer.  Their total path decreases are charged to
actual endpoint drops; they are not additions to the signing potential
or to its final allowance.  Sweeps decrease the same current phase
potential, and reservoir phases still inject at most $10^{-6}$ in
all.  Thus the unchanged transfer is
\[
 4.86276(1+6\cdot10^{-8})+2\cdot10^{-8}
 =4.8627603117656<4.8628.
\]
The isotropic partition coefficient remains $2.4314$.

\subsection{Energy-controlled groups and a sketched preprocessing solve}
\label{oe_sec_improvement}
The preceding implementation is limited by an order-$n^2$ inverse in
preprocessing. We change the preprocessing, not the smoothed potential.
An initial random orientation makes the sum of the squared norms of all
group columns small. A regularized low-rank approximation then accelerates
the ridge solves. Crucially, the repeated actions are applied through the
original rank-one inputs, rather than through their dense group vectors.
This last distinction preserves the input term $mn^2$.

\paragraph{Rational coordinates and the input energy.}
Perform the existing one-time input conversion, discarding and variance
scaling, so the true retained PSD rank-one matrices $C_i$ satisfy
$\sum_i C_i^2\preceq I$. They have compact rational factors and a common
rational scaling, as in Sections~\ref{sec:fast:input-rounding} and
\ref{sr_sec_improvement}. For preprocessing only, use the rational linear
map $\mathcal V$ that lists the diagonal entries and the real and imaginary
parts above the diagonal. Its dimension is $D=n^2$, and
\begin{equation}\label{oe_eq_vectorization}
 \|\mathcal V(H)\|_2\le\|H\|_{\mathrm F}
 \le\sqrt2\|\mathcal V(H)\|_2.
\end{equation}
For a vector $u$, its adjoint test matrix has diagonal entries $u_{ii}$
and off-diagonal real and imaginary parts equal to one half of the
corresponding entries of $u$. Thus
$\langle\mathcal V(H),u\rangle=\tr[H\mathcal V^*(u)]$ exactly.
No irrational orthonormal basis or entrywise approximation that destroys
rank one is needed. Put $b_i=\mathcal V(C_i)$. Then
\begin{equation}\label{oe_eq_inputenergy}
 \sum_i\|b_i\|_2^2\le\sum_i\tr[C_i^2]\le n,
 \qquad \|b_i\|_2\le1.
\end{equation}
These vectors are represented exactly by the converted short inputs and
their common rational scale. The original long input is not read again.
We target operator-norm residual $\eta=\eta_{\rm ps}/2$. This is a
weaker intermediate guarantee than the preceding Frobenius residual, and
is used explicitly below. It suffices: the centered model starts exactly
at zero, and only the operator norm of its fixed residual is charged at
the final discrepancy transfer. The retained matrices, coefficient-grid
rounding, and variance bound do not change. No argument below asserts
that the new small reducer preserves the old Frobenius guarantee.

\begin{lemma}[A certified orientation of the whole group tree]
\label{oe_lem_orientation}
Let a fixed balanced binary tree have leaves $1,\ldots,m$, height at most
$R=\max\{1,\lceil\log_2m\rceil\}$, and columns satisfying
Eq.~\eqref{oe_eq_inputenergy}. There is a capped random-bit procedure costing
$\widetilde O(mD)$ per attempt that either fails or returns signs $e_i$
and all sums $w_J=\sum_{i\in J}e_i b_i$ such that
\begin{equation}\label{oe_eq_treeenergy}
 \sum_{J\text{ a node}}\|w_J\|_2^2\le T:=128(R+1)n.
\end{equation}
One attempt succeeds with probability at least $127/128$. For every
accepted orientation, every subsequently chosen collection of distinct
tree nodes has total squared norm at most $T$, independently of all later
adaptive choices.
\end{lemma}
\begin{proof}
Draw independent uniform signs. For each node,
$\E[\|w_J\|^2]=\sum_{i\in J}\|b_i\|^2$. Each leaf occurs in at
most $R+1$ ancestors, including itself. Therefore the expectation of the
left side of Eq.~\eqref{oe_eq_treeenergy} is at most $(R+1)n$.
Markov's inequality gives the stated probability. Compute every group
sum by additions and test Eq.~\eqref{oe_eq_treeenergy} exactly using their
common denominator. There are fewer than $2m$ nodes, so forming the
sums and their squared norms takes $O(mD)$ arithmetic operations. The final assertion is
deterministic because all summands in the verified total are nonnegative.
To make failure probability at most $\zeta/4$, use
$\lceil\log_2(4/\zeta)\rceil$ independent attempts. On exhaustion use the
existing deterministic grouped reducer on the un-oriented $b_i$.
Run that fallback with Euclidean target $\eta/2$. Its cost is
$\widetilde O(mD+D^3)$, and Eq.~\eqref{oe_eq_vectorization} gives the required
operator residual.
This rule terminates on an all-equal or otherwise adverse bit sequence.
\end{proof}

The group coefficient is now $y_J$, while an original leaf carries
$x_i=e_i y_J$. Splitting still preserves the signed sum, and freezing a
group gives valid leaf signs $e_i y_J$. The true matrix $C_i$ is never
negated in the nonlinear instance. Only its initial coefficient changes.
The group columns themselves may be indefinite matrices; the cube reducer
below is a real-vector procedure and does not require their positivity.

\paragraph{Matrix variance instead of a Frobenius residual.}
The following elementary comparison is the additional gain in the ridge.
The sign guarantee requires a matrix operator norm, not the sum of the
squares of all matrix entries.

\begin{lemma}[Variance of a matrix-valued contraction]\label{oe_lem_variance}
Let $\mathcal L:\mathbb R^r\to\mathbb H_n$ be real linear, with
$\|\mathcal L\|_{2\to\mathrm F}\le a$, and put $R_i=\mathcal L(e_i)$.
Then
\begin{equation}\label{oe_eq_variance}
 \sum_iR_i^2\preceq n a^2I.
\end{equation}
For independent uniform signs $\xi_i$ and $t>0$,
\begin{equation}\label{oe_eq_rademacher}
 \Pr[\|\sum_i\xi_iR_i\|\ge t]
 \le2n\exp(-\frac{t^2}{2na^2}).
\end{equation}
The zero-operator case is interpreted directly.
\end{lemma}
\begin{proof}
Choose a real Frobenius-orthonormal Hermitian basis $H_1,\ldots,H_{n^2}$
only for analysis. The standard diagonal, symmetric and imaginary pairs
give $\sum_jH_j^2=nI$. Write the columns $R_i$ in this basis as a
coefficient matrix $T$. Then $TT^{\top}\preceq a^2I$. The real linear map
\[
 Q\longmapsto\sum_{j,l}Q_{jl}H_jH_l
\]
is positive on real symmetric PSD matrices: for $Q=vv^{\top}$ its value is
$(\sum_jv_jH_j)^2\succeq0$, and decompose a general $Q$ into such terms.
Applying this map proves Eq.~\eqref{oe_eq_variance}. No products are commuted.

For completeness, functional calculus gives
$\E[e^{\theta\xi_iR_i}]=\cosh(\theta R_i)
\preceq e^{\theta^2R_i^2/2}$, so
$\log\E[e^{\theta\xi_iR_i}]\preceq\theta^2R_i^2/2$.
The same Lieb trace inequality used in Lemma~\ref{round_lem_round},
iterated over independent increments, yields
\[
 \E[\tr[e^{\theta\sum_i\xi_iR_i}]]
 \le\tr[\exp(\sum_i\log\E[e^{\theta\xi_iR_i}])]
 \le n e^{\theta^2na^2/2}.
\]
Here the iterative step is conditional expectation with all other terms
held fixed; independence permits their deterministic logarithmic moments.
Exponential Markov bounds the largest eigenvalue, optimization uses
$\theta=t/(na^2)$, and the same calculation for the negative matrix
proves Eq.~\eqref{oe_eq_rademacher}. This is the standard matrix Rademacher
bound with its full matrix variance, cf.~\cite{t12}; replacing it
by the sum of the individual operator-norm bounds would lose the gain.
\end{proof}

Lemma~\ref{oe_lem_variance} gives the following projection tail bound.
In the slack-scaled cube walk let $B=AS$ and
$\Pi_\delta=I-B^{\top}(\delta I+BB^{\top})^{-1}B$. The map
$\mathcal L=\mathcal V^{-1}B\Pi_\delta$ has
$\|\mathcal L\|_{2\to\mathrm F}\le\sqrt{\delta/2}$ by
Eq.~\eqref{oe_eq_vectorization} and the singular-value bound for $B\Pi_\delta$.
Consequently
\begin{equation}\label{oe_eq_projectiontail}
 \Pr[\|\mathcal V^{-1}(B\Pi_\delta\xi)\|\ge t]
 \le2n\exp[-t^2/(n\delta)].
\end{equation}
This removes a factor $n$ from the necessary inverse ridge compared with
the worst-case Euclidean estimate $\|B\Pi_\delta\xi\|\le\sqrt{r\delta}/2$.

\begin{lemma}[A matrix-norm certificate]\label{oe_lem_power}
For rational Hermitian $E$ and positive rational $a$, a test using
$\widetilde O(n^{\omega_0})$ arithmetic operations accepts only if
$\|E\|\le a$ and certainly accepts if $\|E\|\le a/2$.
Its work bound holds for every input.
\end{lemma}
\begin{proof}
Define $Z:=(E/a)^2$ and $Q:=I-Z$. The certificate is a rational
matrix $X$ satisfying
\begin{equation}\label{oe_eq_powercertificate}
 \|X^*QX-I\|_F\le1/2.
\end{equation}
Then $X^*QX\succ0$, so $X$ is nonsingular and $Q\succ0$ by congruence.
This implies $\|E\|<a$.

Set
\[
 c_j:=4^{-j}\binom{2j}{j},\qquad
 X_s:=\sum_{j=0}^{s-1}c_jZ^j,\qquad
 s:=\lceil10\log_2(2n)\rceil.
\]
Compute $X_s$ and $X_s^*QX_s-I$ in exact arithmetic. Accept if its
squared Frobenius norm is at most $1/4$, and reject otherwise.
There are $O(\log(2n))$ order-$n$ products and $O(n^2)$ scalar
operations for the final comparison, giving the stated cost.

If $\|E\|\le a/2$, then $0\preceq Z\preceq I/4$ and
$Q^{-1/2}=\sum_{j\ge0}c_jZ^j$, with $0<c_j\le1$.
The tail has norm at most $4^{1-s}/3$. Since
$\|X_s\|\le4/3$ and $\|Q^{-1/2}\|\le4/3$, this gives
$\|X_sQX_s-I\|_F<1/8$ for the prescribed $s$.
Thus the test accepts in this case. The number of products is fixed
for all other inputs as well.
\end{proof}

\paragraph{A trace-controlled regularized approximation.}
We state the numerical linear-algebra ingredient with its own proof.
Nystr\"om preconditioning is a general method, cf.\
\cite{ftu23}; the precise trace, finite-sampling and
precision bounds needed here are provided below.

\begin{lemma}[Ridge sketch from a trace bound]\label{oe_lem_sketch}
Let $C\succeq0$ have order $D$, let $T\ge\max\{1,\tr[C]\}$, and
$1\le q\le D$. Set $s=256q$, draw a $D$-by-$s$ standard real Gaussian
matrix $\Omega$, and define
\begin{equation}\label{oe_eq_nystrom}
 Y=C\Omega,\quad K=\Omega^{\top}C\Omega+TI_s,\quad
 Z=YK^{-1}Y^{\top}.
\end{equation}
Always $0\preceq Z\preceq C$. With probability greater than $0.98$,
\begin{equation}\label{oe_eq_sketcherror}
 0\preceq C-Z\preceq512Tq^{-1}I.
\end{equation}
For a specified positive rational $\delta$, a bounded finite-random-bit
sampler and arithmetic matrix computations give fixed factors
$\widehat Y,\widehat K=\widehat K^{\top}\succeq(T/2)I$ defining
$\widehat P=\delta I+\widehat Y\widehat K^{-1}\widehat Y^{\top}$.
With probability at least $9/10$, the spectrum of
$\widehat P^{-1}(\delta I+C)$ lies in
\begin{equation}\label{oe_eq_interval}
 [\alpha,\beta],\qquad
 \alpha=1/4,\quad \beta=16(1+2^{12}T/(q\delta)).
\end{equation}
For every finite sample the factors and inverse bounds are polynomial in
$D,T,\delta^{-1}$ and the input magnitudes. The construction attains
any prescribed inverse-polynomial factor accuracy.
\end{lemma}
\begin{proof}
Completing a positive quadratic gives the identity
\begin{equation}\label{oe_eq_variational}
 x^{\top}(C-Z)x=\min_z\{
 \|C^{1/2}x-C^{1/2}\Omega z\|^2+T\|z\|^2\}.
\end{equation}
This proves both order bounds. For analysis only, diagonalize $C$ and
split its first $q$ eigenvalues from the tail $\Lambda_2$. The Gaussian
blocks in this basis are denoted $\Omega_1,\Omega_2$. We verify the two
events
\begin{equation}\label{oe_eq_gaussianevents}
 \sigma_{\min}(\Omega_1)\ge\sqrt{s}/2,\qquad
 \|\Lambda_2^{1/2}\Omega_2\|^2\le8T+32s\|\Lambda_2\|.
\end{equation}
For a fixed unit vector in $\mathbb R^q$, its squared image norm is a
chi-square variable with $s$ degrees of freedom. Its moment-generating
function gives probability at most $2e^{-s/32}$ of leaving
$[9s/16,25s/16]$. For example use parameters $\pm1/8$, and the elementary
bounds $-\log(1-u)\le u+u^2/(2(1-u))$ and
$\log(1+u)\ge u-u^2/2$. A $1/8$-net of at most $17^q$ points then gives
operator norm at most $10\sqrt{s}/7$ and minimum singular value at least
$4\sqrt{s}/7>\sqrt{s}/2$. Its failure probability is at most
$2\exp(q\log17-s/32)<1/64$.

For the tail put $L=\|\Lambda_2\|$. If $L=0$, nothing is needed.
Otherwise for a fixed unit vector in $\mathbb R^s$, the tail image norm
squared is $X=\sum_j\lambda_jg_j^2$ and
$\E[e^{X/(4L)}]\le e^{T/(2L)}$. Thus
$\Pr[X>4T+16sL]\le e^{-4s}$. A $1/4$-net of at most $9^s$ points
and the norm extension factor $4/3$ prove the second event, with failure
at most $e^{-s}<1/256$. The total failure is less than $0.02$.
The elementary net bound follows by comparing volumes of disjoint balls
of radius half the net separation in a maximal separated set.

On these events insert $z=\Omega_1^\dagger x_1$ into
Eq.~\eqref{oe_eq_variational}. The top residual is zero. The tail residual
operator is the concatenation
$[-\Lambda_2^{1/2}\Omega_2\Omega_1^\dagger,\Lambda_2^{1/2}]$,
so its squared norm is at most the sum of the squared norms of these
blocks. The ridge term is at most $4T\|x_1\|^2/s$. Therefore
\[
 \|C-Z\|\le129L+36T/s
 \le130T/q<512T/q,
\]
using $L\le T/(q+1)$; for $q=D$ the tail is zero. No eigendecomposition
or pseudoinverse in this argument is used by the algorithm.

Use midpoint dyadic uniforms and certified Box--Muller evaluation for
the actual sampler. Couple to ideal uniforms. Except on an event of
probability less than $1/100$, all radial uniforms exceed
$(100Ds)^{-4}$ and all normals have a fixed polynomial magnitude bound.
On that event the derivatives of the scalar sampling maps are polynomial,
so a sufficiently fine grid makes $\|\Delta\Omega\|$ as small as any
prescribed inverse polynomial. Every finite output is bounded by
$C\sqrt b$ when $b$ uniform bits are used, even outside that event.
There is no rejection sampler with an unbounded loop.

For $\|\Delta\Omega\|\le1$, expanding Eq.~\eqref{oe_eq_nystrom}, using
$K\succeq TI$ for both samples, bounds its perturbation by
$16T(1+\|\Omega\|)^4\|\Delta\Omega\|$. Make this less than
$\delta/64$. Compute the factors and their Gram to still finer normwise
accuracy, symmetrizing the small Gram. Their polynomial bounds and
$K\succeq TI$ give $\widehat K\succeq(T/2)I$ and permit
\[
 \|\widehat P-(\delta I+Z)\|\le\delta/16.
\]
The same accuracy and positivity guarantees apply to every finite sample.
On the ideal good event,
$\delta I+Z\preceq\delta I+C\preceq
(1+512T/(q\delta))(\delta I+Z)$.
The perturbation leaves ample slack in Eq.~\eqref{oe_eq_interval}. Unioning
the preceding events gives probability at least $9/10$.
All norm and inverse bounds are polynomial. Scalar series and
residual-certified inversion therefore attain the required accuracy
with logarithmic arithmetic overhead.
\end{proof}

\paragraph{Using the original rank-one factors for group actions.}
At a fixed small-reduction state, let $A$ contain the active group vectors,
$x$ their coefficients, $S=\operatorname{diag}(1-|x_i|)$ and $B=AS$. A dense copy of
$B$ is used for the batched sketch setup only. Repeated vector actions
can be faster.

\begin{lemma}[Structured group actions]\label{oe_lem_actions}
For any fixed active antichain of groups and its short slacks, $B$ and
$B^{\top}$, and therefore $\delta I+BB^{\top}$, admit numerical actions at cost
\begin{equation}\label{oe_eq_action}
 \widetilde O((m+n^2)n^{\chi-2})
\end{equation}
arithmetic operations. Each action approximates the same exact rational operator.
Constructing an explicit active group matrix costs $O(n^4)$ after the
one-time group sums. For $q=n=\sqrt D$, all
sketch and preconditioner setup costs are
\begin{equation}\label{oe_eq_setup}
 \widetilde O(D^{\alpha_0}+q^{\omega_0}),\qquad
 \alpha_0=2.042777.
\end{equation}
A subsequent preconditioner application costs
$\widetilde O(Dq+q^2)$.
\end{lemma}
\begin{proof}
Precompute a membership map from each leaf to its current active group,
or to a zero marker, in $O(m)$ work at each successful state. A group
coefficient vector is expanded to leaf coefficients
$e_i(1-|x_{J(i)}|)v_{J(i)}$. Form the weighted sum of the original
rank-one matrices, then apply $\mathcal V$. For the adjoint, apply
$\mathcal V^*$ to the test vector, evaluate its quadratic form on every
original rank-one input, sum over groups with the orientation signs,
and multiply by the slacks. This proves the exact adjoint identity,
including the factor $1/2$ for complex off-diagonal test entries.
No group matrix is treated as rank one.

Store each converted input as $C_i=a_iw_iw_i^*$; put the short scalar
$a_i$ in the coefficient, so a rational square root is unnecessary.
Block the $m$ factor columns into blocks of at most $n^2$.
The products have shapes $(n,n^2,n)$ and their tensor permutations.
The retained exponent $\chi$ gives
$(m/n^2+1)n^\chi$, including both adjoints. Expansion, aggregation,
conversion and all scalar work are smaller. Exact arithmetic products
apply the represented group operator. This proves
Eq.~\eqref{oe_eq_action}, even though the materialized group columns themselves
are dense and generally not rank one.

For setup put $U=B^{\top}\Omega$, $Y=BU$, and $K=U^{\top}U+TI$; then form the
small matrix $K+\delta^{-1}Y^{\top}Y$. These products, for at most $4D$ active
columns and $s=256q$, fit a constant number of products of shape
$(D,D,D^{1/2})$ or its permutations. The bound
$\omega(1,1/2,1)\le2.042776<\alpha_0$ in Table~1 of
\cite{adwxxz25} gives the stated strict exponent. One fixed
recursion and its tensor permutations suffice.
The small inverse costs $\widetilde O(q^{\omega_0})$ using the
residual-certified Newton--Schulz construction. Its minimum
eigenvalue is at least $T/2$ for the represented factors. The exact identity
\begin{equation}\label{oe_eq_woodbury}
 \widehat P^{-1}=\delta^{-1}I-
 \delta^{-2}\widehat Y
 (\widehat K+\delta^{-1}\widehat Y^{\top}\widehat Y)^{-1}\widehat Y^{\top}
\end{equation}
gives the application cost. All additive construction errors are made
small enough for the positivity and comparison margins in the preceding
lemma. The subtractive formula is evaluated to absolute accuracy; no
constant-factor inverse is inserted in it. Neither $BB^{\top}$ nor an
order-$D$ inverse is formed in the fast branch.
\end{proof}

\paragraph{A larger admissible ridge in the cube walk.}
We retain the slack-scaled barrier walk. Its direction accuracy is still
Euclidean, while its signed-sum residual is now certified in matrix
operator norm. These are separate quantities; the latter permits a larger
ridge through Eq.~\eqref{oe_eq_projectiontail}.

\begin{lemma}[Energy-controlled cube reduction]\label{oe_lem_cube}
Suppose $A_0\in\mathbb R^{D\times r_0}$, $D=n^2$, contains the rational
coordinate vectors of $r_0\le4D$ Hermitian group matrices, with
$\|A_0\|_{\mathrm F}^2\le T$ and inherited dyadic coefficients
$y\in[-1,1]^{r_0}$. Put $M$ to be a power of two with
$2\sqrt T\le M<4\sqrt T$. This bounds both vector-operator and individual
matrix norms. Given $0<\varepsilon<1$, a finite randomized procedure
returns at most $2D$ fractional coefficients and
$\|\mathcal V^{-1}(A_0(x-y))\|\le\varepsilon$ for every random-bit sequence.
Its fast branch has a polylogarithmic number of accepted steps and capped
trials. At each trial its arithmetic work, up to logarithms, is
\begin{equation}\label{oe_eq_cubecost}
 D^{\alpha_0}+q^{\omega_0}+D^2+
 (1+\sqrt{T/(q\delta)})
 ((m+n^2)n^{\chi-2}+Dq+q^2),
\end{equation}
where $D=n^2$, $q=n$, and
\begin{equation}\label{oe_eq_ridge_size}
 \delta^{-1}=\widetilde O(n/\varepsilon^2).
\end{equation}
The work statement uses the structured group actions; for arbitrary
Hermitian columns replace their action cost by $O(D r_0)$. The dependence on a variable
$\varepsilon^{-1}$ in Eq.~\eqref{oe_eq_cubecost} is not claimed to be only
logarithmic. In the application $\varepsilon=\eta/R$ with fixed absolute
$\eta$, so all its factors are polylogarithmic in the input parameters.
\end{lemma}
\begin{proof}
If at most $2D$ coefficients are fractional, return immediately. Otherwise
use the same $\ell=\lceil\log_2(64r_0)\rceil$, $H=4\ell$ and choose
\[
 \frac{\varepsilon}{2^{12}r_0M}<\sigma_{\rm r}
 \le\frac{\varepsilon}{2^{11}r_0M}
\]
dyadically. Define
\begin{equation}\label{oe_eq_parameters}
 \begin{aligned}
 L_{\delta_{\rm end}}&=\lceil\log_2(2/\sigma_{\rm r})\rceil,
 &T_*&=128H^2L_{\delta_{\rm end}}+1,\\
 \nu&=\frac{\varepsilon}{2^{16}T_*M},
 &\delta_0&=\frac{\varepsilon^2}{2^{24}nT_*^2L_n},
 \qquad \delta_0/2<\delta\le\delta_0,\\
 L_n&=\lceil\log_2(128n)\rceil.
 \end{aligned}
\end{equation}

Choose the common dyadic coordinate mesh within a factor two of
\begin{equation}\label{oe_eq_grid}
 \min\{\frac{\sigma_{\rm r}}{2^{16}H^2},
                 \frac{\varepsilon}{2^{16}T_*r_0M}\}.
\end{equation}
Keep inherited coefficients exact until the first move. Inward
rounding changes each coordinate by at most the prescribed mesh,
including at that first move. The mesh exponent is $O(\log(2DT/\varepsilon))$.

At a current state form $B=AS$ and set $C=BB^{\top}$ only as an operator.
Its trace is at most $T$. At each independent trial construct a fresh
sketch of Lemma~\ref{oe_lem_sketch}, then draw independent signs $\xi$.
Solve
\[
 (\delta I+BB^{\top})z=B\xi
\]
using Eq.~\eqref{oe_eq_interval} and Eq.~\eqref{oe_eq_woodbury}. Require a certified
true residual at most $\delta\nu/(8M)$. Form
$\widetilde v=\xi-B^{\top}z$ on a fine dyadic grid, including smaller
right-hand-side and multiplication errors so
\[
 \|\widetilde v-\Pi_\delta\xi\|\le\nu,\qquad
 \Pi_\delta=I-B^{\top}(\delta I+BB^{\top})^{-1}B.
\]
The final residual can be evaluated exactly from the explicit short $B$
and the dyadic solution in $\widetilde O(D r_0)$ work, so there is no
unverified norm or inverse oracle in this certificate.

The projection is a contraction, has $r-D$ eigenvalues equal to one,
and satisfies $\|B\Pi_\delta\|\le\sqrt\delta/2$. The sign second- and
fourth-moment argument, and the coordinate sign tail bound, prove the
same acceptance probability $1/16$ as Lemma~\ref{sr_lem_direction} for
\[
 \|\widetilde v\|_\infty\le\ell,
 \qquad\|\widetilde v\|^2\ge(r-D)/4.
\]
These tests are exact on the computed dyadics. Form
$h=S\widetilde v/H$ and the exact Hermitian matrix
$E_h=\mathcal V^{-1}(Ah)$. In addition require the norm certificate
Eq.~\eqref{oe_eq_powercertificate} with target
\begin{equation}\label{oe_eq_steptarget}
 t_{\rm step}=\varepsilon/(2^8T_*).
\end{equation}
This exact check costs $\widetilde O(D r_0+n^{\omega_0})$ and is charged
once per trial, not at each inverse iteration.

The extra test does not remove most of the successful directions.
By Eq.~\eqref{oe_eq_projectiontail}, the ideal residual has norm at most
$t_{\rm step}$ except with probability
\[
 2n\exp[-2^8L_n]<1/1024.
\]
On that event $\|E_h\|\le(t_{\rm step}+M\nu)/H
\le t_{\rm step}/2$, since $H\ge4$. Lemma~\ref{oe_lem_power} therefore
passes. The sketch is chosen before the independent sign vector, so
conditional on a good sketch the ideal sign estimates apply unchanged.
The full success probability is at least
$(9/10)(1/16-1/1024)>1/32$. All inverse and action errors are already
included in $\nu$. Numerical caps below preserve this lower bound.

For a passing direction use this $h$. The two candidates
remain within one quarter of every active slack. Choose $\tau$ by
Eq.~\eqref{sr_eq_derivative_sign}, with the same error bound
$|\widehat g-g|\le r/(512H^2)$, and round $x+\tau h$ inward on
Eq.~\eqref{oe_eq_grid}. The curvature and rounding proof of
Eq.~\eqref{sr_eq_gain} applies verbatim: it uses only the slack and
norm tests on $\widetilde v$. Thus the net gain remains
$(r-D)/(16H^2)>D/(16H^2)$. Freeze all coordinates at endpoint
distance at most $\sigma_{\rm r}$.
Assign each frozen coordinate the finite
bookkeeping value $\log(2/\sigma_{\rm r})$. The total bookkeeping value
never exceeds $4DL_{\delta_{\rm end}}$, and freezing after a step cannot lower it.
The initial near-endpoint scan is handled before starting this accounting.
Thus fewer than $T_*$ successful steps are possible while $r>2D$.
This argument is independent of the signs of the group vectors and of
the initial inherited coefficients.

For the residual, every accepted direction has
$\|\mathcal V^{-1}(Ah)\|\le t_{\rm step}$ by its exact certificate,
regardless of any random event. Its sign choice cannot change this norm.
Grid rounding costs at most $r_0Mu$ per step, and all freezes together
cost at most $r_0M\sigma_{\rm r}$. Hence every fast output obeys
\begin{equation}\label{oe_eq_cube_residual}
 \|\mathcal V^{-1}(A_0(x-y))\|
 \le\varepsilon(2^{-11}+2^{-8}+2^{-16})<\varepsilon/128.
\end{equation}
This is an operator-norm statement. It does not assert a Frobenius bound.
Our choice is
\[
 \delta\asymp\frac{\varepsilon^2}{nT_*^2L_n},
\]
which proves Eq.~\eqref{oe_eq_ridge_size}. The smaller oracle accuracy $\nu$
preserves progress and the residual test's success probability. It does
not determine the ridge.

For completeness use the normalized Chebyshev recurrence
Eq.~\eqref{ue_eq_normalized} for the positive system and preconditioner, on
Eq.~\eqref{oe_eq_interval}. Its exact error is Eq.~\eqref{ue_eq_error}; the
similarity matrix has polynomial condition. Its degree is
$O(\sqrt{\beta/\alpha}\log Q_*)$, where $Q_*$ is a specified polynomial
in $D,T,\varepsilon^{-1}$ and the inverse residual target. This is the
factor in Eq.~\eqref{oe_eq_cubecost}. The $q$-order Woodbury inverse is
formed once per trial, with its residual including its materialization
error. Each solve iteration applies $BB^{\top}$ through original rank-one
factors and applies the stored low-rank preconditioner. No inverse of
order $D$ is formed in this fast branch.

The normalized recurrence error estimate in
Lemma~\ref{ue_lem_precision} bounds accumulated action errors by a
polynomial in the displayed iteration count and the similarity bound.
Choose local errors below the residual target divided by this
polynomial. Exact products and scalar series attain these errors
with logarithmic arithmetic overhead. Evaluate the final residual
against the same fixed group matrix.

Set all loop lengths and a polynomial magnitude cap in advance. Reject
on violation of a cap, the final residual, positivity enclosure, or a
direction test. A bad sketch therefore has bounded cost even when the
claimed spectral interval is false. On the good sketch event the stated
precision and iteration bounds guarantee the residual and magnitude
tests. This establishes the $1/32$ success lower bound.
Permit at most $64\lceil\log_2(2T_*/\zeta)\rceil$ independent trials
per state, and cap successful steps at $T_*$. On exhaustion, restart
from the original inherited $y$ using the deterministic small reducer:
split into two blocks of at most $2D$, scale the columns by $M$, and
use Euclidean target $\varepsilon/(4M)$ in each block. Its combined
Euclidean error in the unscaled vectors is at most $\varepsilon/2$;
Eq.~\eqref{oe_eq_vectorization} gives matrix operator error below $\varepsilon$.
Its fractional count is at most $2D$ and its cost is
$\widetilde O(D^3)$ arithmetic operations. The probability of fallback is
at most $\zeta$, by conditioning at every state and unioning the caps.
Check a prospective fast output's cube and count, and apply the exact
norm certificate to its matrix residual with target $\varepsilon$.
The work is $\widetilde O(D r_0+n^{\omega_0})$. The residual slack in
Eq.~\eqref{oe_eq_cube_residual} guarantees this check. Every random-bit
sequence consequently returns a correct output in finite time.

Evaluate the rational derivative sum in
Eq.~\eqref{sr_eq_derivative_sign} exactly in $O(r)$ arithmetic
operations and choose its sign. The preceding progress estimate
also allows the stated additive error. All successful steps, sign
tests, sketch trials, and comparisons have the announced
polylogarithmic multiplicities.
\end{proof}

\paragraph{The group tree and the exact exponents.}
Apply Lemma~\ref{oe_lem_orientation} once to the entire input tree, with
an orientation cap giving failure at most $(2mn)^{-30}$.
For an accepted tree maintain at most $2D$ active groups. At a level,
split active nonsingletons and copy their coefficients, giving at most
$4D$ active groups. Their total energy is bounded by the same verified
$T$ from Eq.~\eqref{oe_eq_treeenergy}, even though the active groups depend
on all past randomness. Use Lemma~\ref{oe_lem_cube} directly on these
\emph{unscaled} columns with $\varepsilon=\eta/R$ and failure budget
$(2mn)^{-30}/R$. There is no factor $M_{\rm grp}\asymp m$ in this error
target. All group norms are controlled by the certified energy instead.
Each level adds at most $\eta/R$ to the matrix operator residual.
The triangle inequality sums these deterministic residual certificates.
In a leaf return $x_i=e_i y_i$; hence the final true matrix residual has
operator norm at most $\eta<\eta_{\rm ps}$. We do not infer a Frobenius
residual from these certificates.

The fallback within each small call is restarted only for that call,
from its saved inherited coefficients. The top-level orientation fallback
uses the original columns from zero. Each has the required residual on
every outcome. Their total expected additional work is bounded by
$(2mn)^{-30}\widetilde O(mn^2+n^6)$ and is negligible. All preprocessing
randomness precedes and is independent of the future curvature trials;
their conditional guarantees apply to every resulting initial signing.

Choose the sketch size $q=n=\sqrt D$. The small-call parameters give
\[
 T=\widetilde O(n),\quad \delta^{-1}=\widetilde O(n),\quad
 1+\sqrt{T/(q\delta)}=\widetilde O(\sqrt n).
\]
Here $\eta=\eta_{\rm ps}/2$ is fixed absolutely and
$\varepsilon=\eta/R$. Thus the explicit polynomial dependence on
$\varepsilon^{-1}$ in the cube lemma becomes only polylogarithmic in $m$.
The repeated structured-action work is
\[
 \widetilde O((m+n^2)n^{\chi-3/2})
 =\widetilde O(mn^2+n^4),
\]
since $\chi-3/2=1.750036<2$. Preconditioner applications cost
$\widetilde O(n^{7/2})$. Setup costs
\[
 D^{\alpha_0}=n^{2\alpha_0}=n^{4.085554},\qquad
 q^{\omega_0}=n^{\omega_0}.
\]
Materializing active groups, exact residual actions, and all norm
certificates cost $\widetilde O(n^4+n^{\omega_0})$ per capped trial.
Consequently the complete expected preprocessing cost is
\begin{equation}\label{oe_eq_preprocessingtotal}
 \widetilde O(mn^2+n^{4.085554})
\end{equation}
arithmetic operations. This includes orientation, group sums, every successful
step, all failed capped trials, fixed-size sketches and small inverses,
all original-factor actions, random bits, exact residuals and comparison
integers. The complete fallback contribution is negligible as specified
above. The original input is processed once when constructing the
normalized matrices.

\paragraph{A verified randomized final rounding.}
The old deterministic final rounding costs $\widetilde O(k_0N^{\omega_0})$,
which would dominate the improved preprocessing and local stages. We
replace its choice of signs, not its cutoff or reserved error allowance.

\begin{lemma}[Capped final rounding by a norm certificate]\label{oe_lem_round}
For the recorded coefficients on a dyadic mesh of inverse-polynomial
size and the PSD cache in Lemma~\ref{round_lem_round}, a zero-error randomized procedure has error
strictly less than ${\varepsilon_{\rm round}}/2$ and expected arithmetic cost
\[
 \widetilde O(kN^2+N^{\omega_0}).
\]
Every random-bit sequence terminates correctly; the exceptional branch
uses the existing fast deterministic deferred-rounding routine.
\end{lemma}
\begin{proof}
Retain exactly the heavy/small split, $\ell=\lceil\log_2(4N)\rceil$,
$\theta=8\ell/{\varepsilon_{\rm round}}$, and threshold $\tau=1/(4\theta)$ in
Lemma~\ref{round_lem_round}. Heavy matrices use their nearest signs and
have error at most ${\varepsilon_{\rm round}}/64$. For the small ones independently draw
$\epsilon_i\in\{-1,1\}$ with mean equal to the recorded dyadic $t_i$,
and form exactly
\[
 E=\sum_i(\epsilon_i-t_i)\widehat C_i.
\]
The probabilities $(1+t_i)/2$ are dyadic, so a fixed finite number of
uniform bits implements them exactly. These draws are fresh, conditional
on all previous preprocessing and signing choices.

The existing scalar exponential estimate proves, for either sign,
\[
 \sum_i\log\E[e^{\pm\theta(\epsilon_i-t_i)\widehat C_i}]
 \preceq (\ell/16)I.
\]
Applying the same Lieb trace inequality as above and Markov gives
\[
 \Pr[\|E\|>{\varepsilon_{\rm round}}/6]
 \le2N\exp(-61\ell/48)<1/2.
\]
For the last bound use $2N\le2^{\ell-1}$ and $e^{-x}\le2^{-x}$;
the result is at most $2^{-1-13\ell/48}<1/2$.
Accept only when the verified norm certificate
Eq.~\eqref{oe_eq_powercertificate}, at order $N$, passes with threshold
${\varepsilon_{\rm round}}/3$. It always accepts when $\|E\|\le{\varepsilon_{\rm round}}/6$, and every
accepted output has true cached norm at most ${\varepsilon_{\rm round}}/3$.
Including the heavy error and the original true-versus-cache error,
\[
 \|\sum_i(\epsilon_i-t_i)C_i\|
 \le{\varepsilon_{\rm round}}/64+{\varepsilon_{\rm round}}/3+{\varepsilon_{\rm round}}/64
 =35{\varepsilon_{\rm round}}/96<{\varepsilon_{\rm round}}/2.
\]
This bound is deterministic on acceptance; no random event is assumed
for an accepted signing.

Use at most $\lceil30\log_2(2N)\rceil$ independent attempts. On their
exhaustion, invoke Lemma~\ref{sr_lem_round} on the original recorded
coefficients. Its cost is $\widetilde O(kN^{\omega_0})$, but the
conditional failure probability is at most $(2N)^{-30}$, and $k\le N^2$.
Every attempt forms $E$ in $O(kN^2)$ arithmetic operations. The norm
test costs $\widetilde O(N^{\omega_0})$ by
Lemma~\ref{oe_lem_power}, including on failed samples. The coefficient
mesh makes the number of random bits per draw logarithmic.
Thus the expected fallback work is smaller than the displayed costs,
all loops are finite, and the arithmetic bound follows.
\end{proof}

\subsection{A smaller sketch for grouped preprocessing}\label{bs_sec_preprocess}
We refine the implementation in Section~\ref{oe_sec_improvement}.
The group tree, its orientation, the scalar barrier, and the residual
allowance are unchanged. Here $n$ is the original matrix order and
$D:=n^2$. If $m\le2D$, the all-zero fractional vector already has the
required active count and zero residual, so no grouped reduction is needed.

Otherwise retain the certified orientation of
Lemma~\ref{oe_lem_orientation}. Every active antichain has total squared
column norm at most $T:=128(R+1)n$, where
$R:=\max\{1,\lceil\log_2m\rceil\}$. At a small-reduction state let
$A\in\mathbb R^{D\times r}$ contain its $r\le4D$ group columns, and put
$S:=\operatorname{diag}(1-|y_i|)$, $B:=AS$, and $C:=BB^\top$.
Thus $\|B\|_F^2\le T$. By Lemma~\ref{oe_lem_actions}, an action of $B$
or $B^\top$ through the original rank-one inputs costs
\begin{equation}\label{bs_eq_originalaction}
 \mathcal A(m,n):=\widetilde O((m+n^2)n^{\chi-2}).
\end{equation}
The group columns themselves need not have rank one.

\begin{lemma}[Streamed sketch construction]\label{bs_lem_stream}
For any $1\le q\le D$, let $s:=256q$ and use the finite Gaussian sample
$\Omega\in\mathbb R^{D\times s}$ of Lemma~\ref{oe_lem_sketch}. Define
\[
 U:=B^\top\Omega,\quad Y:=BU,\quad K:=U^\top U+TI_s,
 \quad P:=\delta I+YK^{-1}Y^\top.
\]
Its factors and small Woodbury inverse can be constructed to the
prescribed accuracies in
\begin{equation}\label{bs_eq_setup}
 \widetilde O(q\mathcal A(m,n)+Dq^2+q^3+Dq)
\end{equation}
arithmetic operations. An application of the represented $P^{-1}$ costs
$\widetilde O(Dq+q^2)$.
\end{lemma}
\begin{proof}
Compute each column $u_j=B^\top\omega_j$ and then $y_j=Bu_j$ by
Eq.~\eqref{bs_eq_originalaction}. There are $512q$ such actions.
Ordinary dot products form $U^\top U$ and $Y^\top Y$ in $O(Dq^2)$
operations. The matrix $K+\delta^{-1}Y^\top Y$ has order $s$ and
smallest eigenvalue at least $T$ before numerical perturbation.
Residual-certified inversion with ordinary products costs
$\widetilde O(q^3)$. Sampling and writing the factors costs
$\widetilde O(Dq)$. Finally
\begin{equation}\label{bs_eq_woodbury}
 P^{-1}=\delta^{-1}I-
 \delta^{-2}Y(K+\delta^{-1}Y^\top Y)^{-1}Y^\top
\end{equation}
gives the application cost. No order-$D$ inverse is formed. The
construction and solve use one fixed group matrix and one sample.
Let $L_{\rm aux}\ge2$ bound their dimensions, norms, $T$, and
$\delta^{-1}$ polynomially. If each action error is at most $e$,
the product formulas give $\|\widehat Y-Y\|\le2L_{\rm aux}e$ and
$\|\widehat K-K\|\le4L_{\rm aux}^2e$.
Since $K\succeq TI\succeq I$, the inverse perturbation identity
bounds the error in $YK^{-1}Y^\top$ by $32L_{\rm aux}^8e$ for
sufficiently small $e$. Taking this below $\delta/64$ preserves
the ridge-sketch interval and the required action accuracies.
\end{proof}

Choose the least power of two $q$ with $q^3\ge n^2$. Then
$n^{2/3}\le q<2n^{2/3}$, with equality allowed when $n=1$.
All other parameters of Lemma~\ref{oe_lem_cube} are retained. In
particular, at a tree level its allowance is $\varepsilon=\eta/R$, and
\[
 T=\widetilde O(n),\qquad
 \delta^{-1}=\widetilde O(n),\qquad
 J=\widetilde O(1+\sqrt{T/(q\delta)})=\widetilde O(n^{2/3}).
\]
Here $J$ is the normalized Chebyshev iteration count. The ridge-sketch
lemma holds for every $q$ in this range and gives the slack interval
$[1/4,16(1+2^{12}T/(q\delta))]$ with probability at least $9/10$.

\begin{lemma}[Grouped preprocessing within the input term]\label{bs_lem_grouped}
After one-time conversion and scaling, the grouped reduction has expected
arithmetic cost $\widetilde O(mn^2)$. It leaves at most $2n^2$
fractional original coefficients and the same single operator-norm
residual. Every random-bit sequence returns a correct result in finite time.
\end{lemma}
\begin{proof}
One capped trial, including its final residual and norm certificates,
costs at most
\[
 \widetilde O(D^2+n^{\omega_0}+q\mathcal A(m,n)+Dq^2+q^3+Dq
          +J(\mathcal A(m,n)+Dq+q^2)).
\]
The $D^2$ term materializes the short group matrix for final tests only;
it is not incurred at each inverse iteration. Lemma~\ref{oe_lem_power}
supplies the matrix norm certificate. The terms using original inputs are
\[
 (q+J)\mathcal A(m,n)
 =\widetilde O((m+n^2)n^{\chi-4/3})
 \le\widetilde O(mn^2+n^4),
\]
because $\chi-4/3=1.916702\ldots<2$. Also
$Dq^2+JDq=\widetilde O(n^{10/3})$ and
$q^3+Jq^2=\widetilde O(n^2)$.
The same projection norm and spread event has probability at least $1/16$;
its matrix-residual tail removes at most $1/1024$. Hence a trial succeeds
with conditional probability at least
$(9/10)(1/16-1/1024)>1/32$.

The derivative sign rule, inward grid, and endpoint freezing in
Lemma~\ref{oe_lem_cube} are unchanged. There are polylogarithmically many
successful steps and capped trials per tree level. Keep their failure
budgets and deterministic fallback, restarting a failed small call from
its original coefficients. Its cost is $\widetilde O(D^3)$ and its
conditional probability is at most $(2mn)^{-30}/R$. The orientation
fallback has its original charge. When $m>2D$, the separate $n^4$ term
is absorbed by $mn^2$; when $m\le2D$, skip the reduction. Orientation
and group sums also cost $\widetilde O(mn^2)$. The sum of certified
tree-level residuals is the same single allowance. Thus the claimed
expected cost and outcome guarantees follow.
\end{proof}

\subsection{Masked drift and curved moves}\label{md_sec_main}
All dimensions in this subsection refer to the padded matrix order.
We keep the potential, reservoir phases, and notation of
Section~\ref{soft_sec_geometry}. A fixed background of coordinates
will be held at its current coefficients while the other coordinates move.
At a corrected state let
\[
 M=F+\sum_i y_iV_i,\qquad
 \mathcal K(D)=\iota \sum_i\beta_i\tr[V_iD]V_i,\qquad V_i=a_ia_i^*\succeq0,
 \qquad\sum_iV_i^2\preceq I.
\]
Zero inputs can be omitted. The true matrices define the mathematical model, as in Section~\ref{soft_sec_geometry}. Let ${\mathsf P}=(U,V)$ and
${\mathsf Q}=(W,Z)$ be the unique smoothed optimizer and its positive Gibbs state,
with $\tr[{\mathsf Q}]=1$. On bounded-model sublevels,
\begin{gather}
 W=U\mathcal K(Z)U+\rho U^2,\quad
 Z=V\mathcal K(W)V+\rho V^2,                                      \label{md_eq_dual}\\
 U,V\succeq aI,\quad \|U^{-1}\|,\|V^{-1}\|,\|\mathcal K(U)\|,\|\mathcal K(V)\|\le C,
 \quad\rho\tr[U^2+V^2]\le1.                              \label{md_eq_bounds}
\end{gather}
For coordinate $i$ put
\begin{gather*}
 u_i=\tr[UV_i],\quad v_i=\tr[VV_i],\quad w_i=\tr[WV_i],\quad z_i=\tr[ZV_i],
 \quad t_i=v_iw_i+u_iz_i,\\
 p_i=1+\iota \beta_i'v_i,\quad q_i=1-\iota \beta_i'u_i,\quad
 M_i=p_i u_i-q_i v_i,\\
 z_i^{\rm mov}=p_i^2u_iw_i+q_i^{ 2}v_iz_i,\qquad
 \varpi_i=(-\beta_i'')(\iota t_i+\gamma).
\end{gather*}
Here $z_i^{\rm mov}$ is a scalar movement weight, not an optimizer coordinate.
The scalar gradient $g_i=\partial_i \mathcal F_{\rm s}$ satisfies
\begin{equation}
 M_ig_i=z_i^{\rm mov}-p_iq_i t_i+\gamma M_i\beta_i'.
                                                        \label{md_eq_identity}
\end{equation}
On light coordinates, and on their fixed-factor expanded neighborhood,
\[
 p_i,q_i>0,\quad
 p_iq_i\ge1/2,\quad
 |M_i|,|\beta_i'|,p_i,q_i\le P_n,
 \quad |M_i|\le P_n(u_i+v_i),\quad \varpi_i\ge(\iota t_i+\gamma)/2.
\]
All phases use the existing reservoir allocation
$\gamma=\Lambda/(L_0K)$, where $\Lambda=10^{-6}$, $K$ is the count at
phase start and $L_0=1+\lceil\log_2\max(2,k_0)\rceil$.
The initial reservoir plus all positive phase jumps is at most $\Lambda$.
No new allocation is made at the switch to the finishing algorithm.

Let ${\mathscr E}$ be the primal Lagrangian Hessian from Eq.~\eqref{soft_eq_energybounds}. Write
\[
 e(D)^2=\langle D,{\mathscr E}D\rangle,
 \qquad \mathcal E(B)^2=\tau_{\rm s}\langle B,D\log_{\mathsf Q}[B]\rangle,
 \quad\tr[B]=0,
 \qquad \mathbb D=\operatorname{diag}({\mathscr E},B_{\mathsf Q}).
\]
The joint Jacobian $\mathbb A$ has symmetric part $\mathbb D$ and
$\|\mathbb D^{1/2}\mathbb A^{-1}\mathbb D^{1/2}\|\le1$.
\paragraph{Masking non-light coordinates costs only a rank defect.}
\begin{lemma}[A small mask]\label{md_lem_mask}
After removing near-endpoint coordinates, a conservative classification
can mark a set $B$ of at most
\[
 H_n=\lceil16\iota ^2n/(\varepsilon_{\rm scale}\delta_{\rm end}^2)\rceil
\]
coordinates such that every unmarked coordinate is light. The classification costs one full score scan and its true-data enclosures. Marked coordinates are
not changed, and their coupling is not removed.
\end{lemma}
\begin{proof}
At this point $1-|y_i|\ge\delta_{\rm end}$. Mark an index if a certified upper
enclosure of either endpoint score is at least $-\delta_{\rm end}/8$, using width
at most $\delta_{\rm end}/16$. An unmarked score is strictly negative. A marked
index has either $\iota \beta_i u_i\ge\delta_{\rm end}/2$ or
$\iota \beta_i v_i\ge\delta_{\rm end}/2$. Since $\beta_i\le1$, it has
$u_i\ge\delta_{\rm end}/(2\iota )$ or $v_i\ge\delta_{\rm end}/(2\iota )$.
The rank-one variance bound gives
$\sum_i\tr[MV_i]^2\le\|M\|_{\rm F}^2$ for Hermitian $M$.
To see this, the nonnegative Gram matrix $J_{ij}=\tr[V_iV_j]$ obeys
$J\nu\le\nu$, $\nu_i=\tr[V_i]>0$, so its spectral norm is at most one.
Its Gram factor and adjoint are contractions. Therefore the number
marked is at most
$4\iota ^2\delta_{\rm end}^{-2}(\|U\|_{\rm F}^2+\|V\|_{\rm F}^2)
\le4\iota ^2n/(\varepsilon_{\rm scale}\delta_{\rm end}^2)$, leaving the stated slack for enclosures.
\end{proof}

Write $J$ for the unmarked set, $p=|J|$, $q=|B|$, and
$\mathcal K=\mathcal K_J+\mathcal K_B$.
Write $\mathcal A_J(D_U,D_V):=(\mathcal K_J(D_V),\mathcal K_J(D_U))$,
and define $\mathcal A_B$ similarly. All subspaces in the next lemma concern the $p$
unmarked physical coordinates.
\begin{lemma}[Covariance with a fixed background]\label{md_lem_cov}
Let $\mathcal H\subseteq\mathbb R^J$ have codimension $r$. If
$r+2q\le p/4$, there is a finitely supported centered direction in
$\mathcal H$, extended by zero on $B$, whose covariance satisfies
\begin{equation}
 \E[\mathcal Q(h)]\le\sum_{i\in J}\kappa_i z_i^{\rm mov}\Sigma_{ii}+r+2q,
 \qquad\sum_{i\in J}z_i^{\rm mov}\Sigma_{ii}\ge p/175.               \label{md_eq_cov}
\end{equation}
Here $\mathcal Q$ is the original full-coupling fixed-constraint inversion energy.
This covariance is only used in the proof, not computed by the algorithm.
\end{lemma}
\begin{proof}
We verify why a background coupling is not being silently discarded.
The operator ${\mathscr J}_J=\operatorname{diag}(U^{-1}(\cdot)U^{-1},
V^{-1}(\cdot)V^{-1})-\mathcal A_J$ has a positive inverse: its dual
inequalities are strict by Eq.~\eqref{md_eq_dual} and $\mathcal K_J\preceq\mathcal K$ as
positive maps. Let $D_J(h)={\mathscr J}_J^{-1}F_1h$ be its fixed-constraint response.
Impose the additional real linear equations
\[
 \tr[V_b(D_J(h))_U]=\tr[V_b(D_J(h))_V]=0\quad(b\in B).
\]
They add at most $2q$ to the codimension. On this subspace
$\mathcal A_B D_J(h)=0$, whence ${\mathscr J}D_J(h)=F_1h$.
The response is thus the response for the full problem.

For clarity the normalized covariance construction remains valid with
strict dual inequalities, rather than equality for $\mathcal K_J$.
Set $s_i=(\beta_ip_i/q_i)^{1/2}$ and
$t_i^*=(\beta_iq_i/p_i)^{1/2}$, and define response
matrices
\[
 A_{ij}=\iota s_is_j|a_i^*Ua_j|^2,\qquad
 B_{ij}=\iota t_i^*t_j^*|a_i^*Va_j|^2.
\]
The normalized response vectors $\mathbf u,\mathbf v$ obey $\mathbf u-B\mathbf v=\tau h$ and $A\mathbf u-\mathbf v=\tau h$,
where $\tau_i=(p_iq_i/\beta_i)^{1/2}$.
The energy matrices have entries
$s_is_j\operatorname{Re}\tr[WV_iUV_j]$ and
$t_i^*t_j^*\operatorname{Re}\tr[ZV_iVV_j]$.
Normalize their positive diagonals
$E_{U,i}=s_i^2u_iw_i$ and $E_{V,i}=(t_i^*)^2v_iz_i$.
The resulting block energy matrix $M$ has trace $2p$ and unit diagonals.
For the normalized off-diagonal response map $T$, the dual inequalities
imply
\[
 M\succeq T^*\Gamma T,\qquad \Gamma_i=(\iota \beta_i u_i v_i)^{-1}\ge \iota >5.
\]
For example, writing $D=\sum_i s_i\mathbf u_iV_i$, Cauchy--Schwarz gives
\[
 \sum_{j\in J}\frac{\iota \beta_jz_j}{u_j}
       (a_j^*UDUa_j)^2
 \le\tr[\mathcal K_J(Z)UDUDU]\le\tr[WDUD].
\]
This is the required block inequality; the other block is identical.
The light inequalities and $\beta_i\ge1-y_i^2$ give $\Gamma_i\ge \iota $.
In each pair use the unit vector proportional to
$(\sqrt{E_{U,i}},-\sqrt{E_{V,i}})$ and its orthogonal complement.
Their energy diagonals remain one, the two response diagonals sum to
zero, and the first is nonpositive. The response graph is mapped by
$I-T$ onto the span of the first paired vectors. Its dimension is $p$,
since the strict dual inequalities make the response system invertible.

Lemma~\ref{sp_lem_rankdefect} therefore applies
to its subspace of dimension at least $p-r-2q$. Its loss is its codimension,
at most $r+2q$, and its movement is at least $p/175$ when this codimension
is at most $p/4$. The physical movement identity is
$\tau_i^2(E_{U,i}+E_{V,i})=z_i^{\rm mov}$. On the constrained response space the
energy is the full problem's energy, as proved above. This gives
Eq.~\eqref{md_eq_cov}. The second fixed-constraint derivative gives the same
energy expression even though the background can affect the second
response: pairing its defining equation with the full ${\mathsf Q}$ uses
${\mathscr J}{\mathsf Q}=\rho E$ and leaves exactly the original energy and mixed terms.
No second-response vanishing is assumed.
\end{proof}

\paragraph{A drift-adjusted Hessian at every masked light state.}
Let $C_{2,i}:=(1+\tau/2)\kappa_i$ and $C_\tau:=1+\tau/2$, as in Eq.~\eqref{sp_eq_generator_defect}. Its exact scalar certificate and Eq.~\eqref{md_eq_cov} give
\begin{equation}
 \tfrac12\tr[\nabla^2\mathcal F_{\rm s}\Sigma]
 -\sum_{i\in J}C_{2,i}M_i g_i\Sigma_{ii}
 \le-\upsilon_n\sum_{i\in J}\varpi_i\Sigma_{ii}+C_\tau(r+2q),   \label{md_eq_generator}
\end{equation}
where $\upsilon_n^{-1}\le P_n$. This is the retained generator calculation;
there is no failed-gradient hypothesis in Eq.~\eqref{md_eq_generator}.

Choose known $C_w\le P_n$ so that on light coordinates
\[
 p_iq_i t_i+\gamma|M_i\beta_i'|\le C_w\varpi_i.
\]
For example the displayed light bounds and $\varpi_i\ge(\iota t_i+\gamma)/2$
give such a coefficient by sums of fixed budget bounds. Set
\begin{equation}\label{md_eq_modified}
 \begin{gathered}
 a_*:=\frac{\upsilon_n}{4(1+C_w)},\qquad
 d_i:=(C_{2,i}+a_*)M_i,\qquad
 \varpi_i^\sharp:=\varpi_i+z_i^{\rm mov},\\
 G_\sharp:=\operatorname{diag}_{J}(\varpi_i^\sharp),\qquad
 \mathcal H_\sharp:=(\nabla^2\mathcal F_{\rm s})_{JJ}
                  -2\operatorname{diag}_{J}(d_i g_i).
 \end{gathered}
\end{equation}
Using Eq.~\eqref{md_eq_identity} in Eq.~\eqref{md_eq_generator} proves
\begin{equation}
 \tfrac12\tr[\mathcal H_\sharp\Sigma]
 \le-a_*\tr[G_\sharp\Sigma]+C_\tau(r+2q).                  \label{md_eq_newgen}
\end{equation}
All coefficients in Eq.~\eqref{md_eq_modified} are evaluated at the base root
and frozen during the following path.

\begin{lemma}[Negative multiplicity without a gradient test]\label{md_lem_inertia}
Put
\[
 K_*(n)=\lceil\max\{16H_n,4000C_\tau H_n/a_*\}\rceil,
 \qquad\delta_n=a_*/4,\qquad\pi_n=a_*/(700C_\tau).
\]
For $k>K_*(n)$ the matrix
$A_\sharp=G_\sharp^{-1/2}\mathcal H_\sharp G_\sharp^{-1/2}$
has at least $\pi_np$ eigenvalues below $-4\delta_n$. Its spectrum is
contained in $[-P_n,P_np]$, and its diagonal is bounded in absolute value
by $P_n$. Moreover $K_*(n)=nP_n$.
\end{lemma}
\begin{proof}
We have $q\le H_n$, $p\ge k/2$, and
$2q\le a_*p/(700C_\tau)$. Let $r$ count eigenvalues below $-a_*$.
If $r\ge p/8$ the conclusion follows. Otherwise $r+2q\le p/4$, so use
Eq.~\eqref{md_eq_cov} on the preimage of the complementary spectral space.
Write $D_\Sigma=\tr[G_\sharp\Sigma]\ge p/175$. On that space the left
side of Eq.~\eqref{md_eq_newgen} is at least $-a_*D_\Sigma/2$.
Thus
\[
 r\ge \frac{a_*p}{350C_\tau}-2q\ge\pi_np.
\]
This also deals with nonintegral thresholds by the integer eigenvalue count.

Eq.~\eqref{soft_eq_second} gives $\nabla^2\mathcal F_{\rm s}\succeq-P_nG$ and
$|\partial_i^2\mathcal F_{\rm s}|\le P_n\varpi_i$ on the light subset. Eq.~\eqref{md_eq_identity} gives $|M_ig_i|\le z_i^{\rm mov}+C_w\varpi_i$.
The added diagonal in Eq.~\eqref{md_eq_modified} is therefore bounded by
$P_nG_\sharp$. The lower bound and diagonal bound follow, and the trace
of $A_\sharp+P_nI$ gives the upper norm bound. Finally all quantities in
$K_*$ except its explicit $n$ factor are fixed polylogarithms.
\end{proof}

\paragraph{Computing a direction and its true relative response.}
The Schur formula in Eq.~\eqref{soft_eq_hess} is restricted to $J$.
Under the $G_\sharp$ normalization its diagonal term changes, by
Eq.~\eqref{md_eq_modified}, to a diagonal of norm at most $P_n$. Set
\[
 D_\sharp=G_\sharp^{-1/2}F_1^*Q_{\mathsf Q}F_1G_\sharp^{-1/2}.
\]
The same completing-square argument as Lemma~\ref{gf_lem_gram} gives
\[
 a_{\rm dom}D_\sharp-P_nI\preceq A_\sharp
          \preceq D_\sharp+P_nI,
 \quad\tr[D_\sharp]\le P_np,\quad a_{\rm dom}^{-1}\le P_n.
\]
Indeed the $F_2$ bound uses $\varpi_i\le \varpi_i^\sharp$, and
$\tr[{\mathsf Q}F_{1,i}^2]\le P_nz_i^{\rm mov}$ since $u_i,v_i\ge a_0\tr[V_i]$.
The fixed-root Schur preconditioner for ${\mathscr J}$ is unchanged; it includes
$\mathcal K_B$. A sufficiently large known positive shift is therefore
comparable to $D_\sharp+RI$. The preceding Gibbs feature identity,
finite sampler, alternating row/column sketches, and inverse recurrences
apply with $G_\sharp$ and this diagonal shift. They require no zero-gradient
condition, only these order bounds. The weights have lower bound
$\Lambda/(2L_0K)$, and all surrogate perturbations have the same
polynomial conditioning margins.

We record the additional response test and its proof to identify the
curved-path gain. For $D=(D_U,D_V)$ put
$\widehat D_A=A^{-1/2}D_AA^{-1/2}$ and
$T_U=U^{-1/2}WU^{-1/2}$, $T_V=V^{-1/2}ZV^{-1/2}$.
Define
\begin{equation}
 \mathfrak r(D,B)=\max_{A=U,V}
 \{\|\widehat D_A\|,
 \|T_A^{1/2}\widehat D_AT_A^{-1/2}\|,
 \|{\mathsf Q}_A^{-1/2}B_A{\mathsf Q}_A^{-1/2}\|\}.                     \label{md_eq_relative}
\end{equation}
The middle norm is a singular-value norm of a generally non-Hermitian
matrix; it must not be replaced by its spectral radius or omitted.
Each of the six maps in Eq.~\eqref{md_eq_relative} has norm at most $P_n\sqrt n$
from joint energy to Frobenius norm, by Eq.~\eqref{soft_eq_energybounds}.

\begin{lemma}[A verified drift-response direction]\label{md_lem_direction}
At $k>K_*(n)$ a finite randomized trial, with an absolute conditional
success probability, finds a real unit direction $r$ supported on $J$
satisfying, for $\theta=r^{\top}G_\sharp r$ and $z'_r=(D,B)$,
\begin{gather}
 r^{\top}\mathcal H_\sharp r\le-\delta_n\theta/2,
 \qquad \|r\|_\infty\le P_n\sqrt\theta,                  \label{md_eq_dir}\\
 \mathfrak r(z'_r)\le
 P_n\min\{\sqrt n,n/\sqrt p\}\sqrt\theta.               \label{md_eq_response}
\end{gather}
All inequalities and the derivative solve are checked against the
true model. The trial uses matrix actions and residual tests; it
constructs neither a covariance nor a negative-eigenspace basis.
\end{lemma}
\begin{proof}
Use Lemma~\ref{md_lem_inertia} and the bounded flat polynomial of the shifted inverse, equal to within a
small leakage of one on eigenvalues below $-3\delta_n$, and of zero on
those above $-\delta_n$. Call it $F$; $0\preceq F\preceq I$.
Choose leakage at most $1/(P_n\sqrt p)$ as well as the quotient error requirement in Lemma~\ref{sp_lem_flat}. The Bernstein-polynomial construction and finite
inverse evaluation are unchanged.

The response map $\mathcal R$ from $G_\sharp^{-1/2}y$ to $z'_r$ obeys
$\|\mathcal RF\|_{2\to\mathbb D}\le P_n$.
To verify this point, negative curvature of $A_\sharp$ bounds the
\emph{true} Hessian above by $P_n\|y\|^2$ on that band, because the
removed drift diagonal has norm at most $P_n$. In
\[
 r^{\top}\nabla^2\mathcal F_{\rm s}r=-r^{\top}Gr+e(D)^2+\mathcal E(B)^2+2\langle F_2r,D\rangle
\]
complete the square, using $r^{\top}Gr\le\theta$ and the retained $F_2$
bound. This bounds the response energy on the band. The whole-space
response norm is at most $P_n\sqrt p$; the chosen leakage handles its
complement.

For a real-linear map $\mathcal M:\mathbb R^p\to\mathbb C^{n\times n}$
with $\|\mathcal M\|_{2\to F}\le a$, its coefficient matrices obey
$\sum_iM_iM_i^*,\sum_iM_i^*M_i\preceq2na^2I$.
This follows by applying the positive coefficient-covariance map to
$a^2I$ in the real matrix-unit and imaginary-unit basis. The Hermitian
case has $n$ instead of $2n$. The matrix Rademacher bound hence controls
all six relative maps composed with $\mathcal RF$ by $P_nn$, with a
fixed small failure probability. Meanwhile the flat-filter sign moments
give $\|F\xi\|^2\ge\pi_np/4$ and
$\|F\xi\|_\infty^2\le2\log(384p)$ on an event of fixed probability.
A union bound, not an independence assumption between these tests, gives
their simultaneous event. Divide by $\|F\xi\|$ and then normalize
$G_\sharp^{-1/2}F\xi$ physically. Since $p\ge k/2>K/4$ in its reservoir
phase and $\varpi_i^\sharp\ge\Lambda/(2L_0K)$, this proves the spread bound.
The relative response bound is $P_nn/\sqrt p$ times $\sqrt\theta$;
the deterministic energy-to-relative comparison also gives its
$\sqrt n$ alternative. The modified Rayleigh inequality follows from the
same bounded filter.

Compute $z'_r$ through the actual differentiated joint equations and a
true residual enclosure. Order-$n$ matrix-function and singular-norm
enclosures check Eq.~\eqref{md_eq_response}; no individual eigenvector gap
is required. These costs are included in the invocation below. A fixed
smaller choice of all thresholds leaves numerical slack. On any trial,
only a direction passing the true enclosures is accepted.
\end{proof}

\paragraph{The curved trial and its predictor.}
Freeze $d_i$ from Eq.~\eqref{md_eq_modified} and set it to zero on $B$.
For either sign of $s$ use the quadratic coordinate path
\begin{equation}
 y_i(s)=y_i+s r_i-s^2d_ir_i^2\quad(i\in J),\qquad
 y_i(s)=y_i\quad(i\in B).                                 \label{md_eq_path}
\end{equation}
Its second potential derivative at zero is exactly
$r^{\top}\mathcal H_\sharp r$. In particular the drift is retained rather than
estimated as a small gradient error.

\begin{lemma}[Curved prediction and progress]\label{md_lem_step}
A direction passing Eq.~\eqref{md_eq_dir}--\eqref{md_eq_response} permits
\begin{equation}
 s_*=\frac{a_n\Gamma(n,p)}{\sqrt{n\theta}},\qquad
 \Gamma(n,p)=\max\{1,(p/n)^{1/4}\},                        \label{md_eq_length}
\end{equation}
where $a_n>0$ is a fixed sufficiently small inverse-polylogarithm.
Choose a sign with certified nearly nonpositive initial linear derivative.
A first-order optimizer predictor followed by a true root correction
then gives a decrease at least
\begin{equation}
 P_n^{-1}n^{-1}\max\{1,\sqrt{p/n}\}.                       \label{md_eq_progress}
\end{equation}
The complete chosen candidate costs a constant number of root solves and
true value/derivative evaluations, up to logarithms. Coordinates in the
mask stay fixed. All coordinate-zero crossings are allowed.
\end{lemma}
\begin{proof}
Write $b=\|r\|_\infty$, $z'_0=z'_r$, and
$\ell=\mathfrak r(z'_0)+b$. The proof's reference norms are at the base.
For the straight predictor $z_p(s)=z_0+sz'_0$, we establish
\begin{align}
 \|\mathcal E_{{\rm KKT},y(s)}(z_p(s))\|_{\mathbb D_0^{-1}}
 &\le P_ns^2\ell\sqrt\theta,                              \label{md_eq_predictor_res}\\
 \|z(s)-z_p(s)\|_{\mathbb D_0}
 &\le P_ns^2\ell\sqrt\theta,                              \label{md_eq_predictor_err}\\
 |(\mathcal F_{\rm s}\circ y)''(s)-(\mathcal F_{\rm s}\circ y)''(0)|
 &\le P_n\theta(|s|\ell+\sqrt n s^2\ell\sqrt\theta),
                                                               \label{md_eq_curved_hess}
\end{align}
provided $P_n(|s|\ell+\sqrt n s^2\ell\sqrt\theta)$ is a sufficiently
small fraction of $\delta_n$ and $|s|b$ is a sufficiently small fraction
of $\delta_{\rm end}$. We give the extra estimates for the curved path explicitly.

First $|d_i|\le P_n(u_i+v_i)$ and $|d_i|\le P_n$.
A positive majorant for $-2\sum_i d_ir_i^2V_i=M''$ has norm at most
$P_nb^2$: use $\beta_i^{-1}\le P_n$ and
$\mathcal K(U)+\mathcal K(V)\preceq CI$. Its weighted trace is at most $P_n\theta$,
since
\[
 |d_i|(w_i+z_i)\le P_n(u_i+v_i)(w_i+z_i)
 \le P_n(t_i+z_i^{\rm mov})\le P_n \varpi_i^\sharp.
\]
The middle inequality uses the positive lower bounds for $p_i$ and
$q_i$ on the expanded light set. The remaining terms of
$\mathcal C_{ss}$ are weighted by $\beta_i''r_i^2+
\beta_i'(-2d_ir_i^2)$ and have the same operator-majorant and trace
bounds. If $-B_0\preceq B\preceq B_0$, then
$B^2\preceq\|B_0\|B_0$, by writing
$B=B_0^{1/2}CB_0^{1/2}$ with $\|C\|\le1$.
The entropy variance inequality therefore bounds their dual
energy norm by $P_nb\sqrt\theta$.
The primal component $\mathcal A''{\mathsf Q}$ has the same bound by weighted
Cauchy--Schwarz with $|r_i|^4\le b^2r_i^2$.
Along a short segment $y_i'(s)=r_i(1-2sd_ir_i)$ remains within a fixed
relative factor, and the same assertions hold with base norms and weights.
There is no cubic signed-sum term, since $M'''=0$.
The direct budget/coupling cubic terms are bounded by $P_nb\theta$;
terms containing $\beta_i''y_i'y_i''$ obey this same bound.

For optimizer variation, let $R$ be a normalized Hermitian test matrix. The inverse-Hessian words are
$\tr[T(\widehat D R^2+R\widehat D R+R^2\widehat D)]$.
The middle word is bounded by $\|\widehat D\|\tr[TR^2]$;
the exterior words are bounded by
$\|T^{1/2}\widehat D T^{-1/2}\|\tr[TR^2]$.
Dual variation uses $\|{\mathsf Q}^{-1/2}\Delta {\mathsf Q} {\mathsf Q}^{-1/2}\|$.
For the logarithm block, the resolvent formula gives the relative bound
\[
 |D\langle B,D\log_{\mathsf Q}[B]\rangle[\Delta {\mathsf Q}]|
 \le2\|{\mathsf Q}^{-1/2}\Delta {\mathsf Q} {\mathsf Q}^{-1/2}\|
                  \langle B,D\log_{\mathsf Q}[B]\rangle.
\]
Signed coefficient variations bounded by $t$ times the original positive
coupling have weighted output norm at most $Ct e(D)$, by splitting the
two sign sets and applying complete-positive Schwarz. The explicit
coefficient changes here have relative size $P_n|s|b$.
These identities prove normalized Jacobian variation at the predictor
at most $P_n|s|\ell$; they retain matrix order and apply to Hermitian data.
The first residual derivative at zero vanishes. The pure second parameter
forcing just bounded and the mixed and optimizer-only terms are at most
$P_n\ell\sqrt\theta$. Integration proves
Eq.~\eqref{md_eq_predictor_res}.

Set $E_s$ to twice its right side. On the energy ball of radius $E_s$
around the predictor, additional relative variation is at most
$P_n\sqrt n E_s$. The chord map using the exact base joint inverse is
contractive and maps this ball to itself when the stated bound is small.
All four matrix blocks remain positive, with $\tr[\mathsf Q]=1$. This proves
Eq.~\eqref{md_eq_predictor_err}; primal convexity identifies the root with the
original optimizer. The auxiliary normalized square roots are not
computed by the algorithm.

For Eq.~\eqref{md_eq_curved_hess}, write the parameter forcing in the joint
residual as $f_s=\partial_s\mathcal E_{{\rm KKT},y(s)}$.
The root response is $\dot z(s)=-\mathbb A_s^{-1}f_s$.
The force variation is bounded by $P_n|s|b\sqrt\theta$ plus
$P_nb\|z(s)-z_0\|_{\mathbb D_0}$; the acceleration terms satisfy the
same bound by the displayed $M''$ majorant. Thus
\[
 \|\dot z(s)-z'_0\|_{\mathbb D_0}\le
 P_n(|s|\ell+\sqrt n E_s)\sqrt\theta.
\]
The direct second-parameter term includes the reservoir acceleration.
Its change is bounded by
\[
 P_n\theta(|s|\ell+\sqrt n E_s).
\]
Indeed its root differential is paired with the positive majorants above,
while its explicit differential has the stated $b\theta$ bound.
The energy, metric, and mixed forcing terms obey the same estimate by
Cauchy--Schwarz. Substituting in the second-order envelope identity proves
Eq.~\eqref{md_eq_curved_hess}. This argument includes the gradient times
coordinate-acceleration term; it has not been discarded.

By Eq.~\eqref{md_eq_response},
$\ell\le P_n\sqrt\theta\min\{\sqrt n,n/\sqrt p\}$.
Substitution of Eq.~\eqref{md_eq_length} shows
$|s_*|\ell\le P_na_n$ and
$\sqrt n s_*^2\ell\sqrt\theta\le P_na_n^2$ in both ranges of $p$.
Also $s_*b\le P_na_n n^{-1/4}$ since $p\le n^2$.
The quadratic displacement in Eq.~\eqref{md_eq_path} is at most
$P_n(s_*b)^2$. Choosing $a_n$ smaller guarantees endpoint distance at
least $\delta_{\rm end}/2$ and all required expanded light bounds. These are
comparisons for the originally unmasked set, not a claim that the whole
state becomes light.

Compute $d_0=\sum_i g_i r_i$ with an enclosure of width at most
$\delta_n s_*\theta/100$. Choose the sign so its initial derivative is
at most that width. On the chosen side, Eq.~\eqref{md_eq_curved_hess}
keeps the second derivative at most $-\delta_n\theta/4$.
Integration yields decrease $\delta_n s_*^2\theta/16$, after reserving
fixed fractions for all errors. This proves Eq.~\eqref{md_eq_progress}.
Only the chosen side is needed. Its potential is bounded by the old value
plus the reserved arbitrarily small initial slack; boundedness is not
inferred from its Euclidean length.

The base inverse is applied using the true Schur solve and fixed-anchor
refinement. Predictor, derivative, matrix-function, coordinate-grid, and
value errors are below fixed fractions of the preceding radii and gains.
All reciprocal margins and step lengths are polynomial, so
inverse-polynomial evaluation errors preserve the stated gains. At zeros the budget is $C^2$ and its
one-sided third derivatives are bounded. The force estimates integrate
through those points; no third-order jet crosses a discontinuity.
\end{proof}

\subsection{Resident coordinates and delayed deletion}\label{br_sec_resident}
A resident coordinate still contributes its signed coefficient, coupling
slot, and scalar reservoir to the potential. Let $k$ count these
coordinates, and let $D$ be the pending subset whose coefficients have
reached the near-endpoint region. Keep each pending coefficient fixed at
its recorded value. Entering $D$ changes neither the potential nor its
optimizer. The set $\mathcal D$ of coordinates already removed for final
rounding is separate from $D$.

With the constants of Section~\ref{md_sec_main}, define
\begin{equation}\label{br_eq_flushparams}
 \zeta_n:=\min\{1/100,a_*/(10000C_\tau)\},\qquad
 K_c:=\lceil\max\{32,n/\zeta_n,H_n/\zeta_n\}\rceil.
\end{equation}
Thus $\zeta_n^{-1}\le P_n$ and $K_c=nP_n$, with a known fixed
polylogarithmic factor. Reservoir phases count all resident coordinates.

\begin{lemma}[A resident mask]\label{br_lem_mask}
Suppose $k>K_c$ and $|D|<\lceil\zeta_n k\rceil$. Place every newly
near-endpoint coordinate into $D$. If this trigger is still false, let
$B$ contain $D$ and the non-light coordinates marked by
Lemma~\ref{md_lem_mask}, and let $J$ be its complement. Write
$q:=|B|$ and $p:=|J|$. Then
\begin{equation}\label{br_eq_maskbounds}
 q\le2\zeta_n k,\qquad p\ge(1-2\zeta_n)k,\qquad
 2q\le a_*p/(700C_\tau).
\end{equation}
The masked direction and curved-move lemmas apply with this background,
giving decrease at least $P_n^{-1}\sqrt p/n^{3/2}$.
\end{lemma}
\begin{proof}
Integrality gives $|D|<\zeta_n k$ when the trigger fails. Every movable
coordinate has endpoint distance greater than $\delta_{\rm end}$, so
Lemma~\ref{md_lem_mask} marks at most $H_n\le\zeta_n k$ others.
This proves the first two bounds. The third follows from
$4\zeta_n k\le a_*k/(2500C_\tau)$ and $p\ge0.98k$.

Lemma~\ref{md_lem_cov} permits any fixed background with rank loss $2q$.
If the negative eigenvalue count is at least $p/8$, the required
multiplicity already holds. Otherwise its proof and
Eq.~\eqref{md_eq_newgen} give
\[
 r\ge a_*p/(350C_\tau)-2q\ge a_*p/(700C_\tau).
\]
The movement weight is at least $p/175$ and $r+2q\le p/4$.
Thus Lemmas~\ref{md_lem_direction} and~\ref{md_lem_step} apply, with
smaller fixed numerical margins if necessary. Their derivative bounds
are used only on $J$. Pending coordinates have zero parameter
derivatives, even if their endpoint distance is smaller than the cutoff.
The full background coupling remains in every root and response solve.
\end{proof}

After the initial corrected root, process its near-endpoint coordinates
once by positive-submap attenuation. Subsequently, collect pending
coordinates until $|D|\ge\lceil\zeta_n k\rceil$, then attenuate their
combined coupling and delete them. Keep the old reservoir phase fixed
through final deletion and root correction. Update the count phase only
afterward. When $k\le K_c$, clear the pending set once and switch
permanently to the finishing algorithm of Section~\ref{am_sec_improvement},
without allocating a new reservoir. Every newly pending coordinate after
a curved trial has endpoint distance at least $\delta_{\rm end}/2$;
the initial scan separately handles exact endpoints.

\begin{lemma}[Bulk continuation and move counts]\label{br_lem_flushes}
Apart from failure-induced work, the large-count regime has at most
\[
 b\le3+\zeta_n^{-1}\log\max\{2,k_0\}
\]
attenuation calls, including initial and final cleanup. All these calls
and the initial data homotopy use $\widetilde O(\sqrt n)$ corrected
stages. In a phase $K/2<k\le K$, successful curved moves number
$\widetilde O(n^{3/2}/\sqrt K)$.
\end{lemma}
\begin{proof}
Each triggered deletion removes at least a $\zeta_n$ fraction of the
resident count. Count these multiplicative drops and add the two
untriggered cleanups. For a pending batch use
$\mathcal K_s=\mathcal K_{\rm rest}+e^{-s}\mathcal K_D$.
The proof of Lemma~\ref{gf_lem_faces} and the batched argument in
Section~\ref{sp_sec_improvement} use positivity of this submap and
complete-positive Schwarz. They incur no factor $|D|$. A batch with
matrix-potential decrease $\Delta_j$ uses
$\widetilde O(1+\sqrt{n\Delta_j})$ stages, including its final tiny
coefficient, exact deletion, and root correction. All decreases sum to
a constant by Eq.~\eqref{sp_eq_injections}. Therefore
\[
 \sum_j(1+\sqrt{n\Delta_j})
 \le b+\sqrt{nb\sum_j\Delta_j}=\widetilde O(\sqrt n).
\]
The initial homotopy has the same stage bound. In a count phase
$p\ge0.98k>K/3$; summing the decrease from Lemma~\ref{br_lem_mask}
proves the move count. Pending status itself adds no potential increase.
\end{proof}

\subsection{Matrix functions without diagonalization}\label{br_sec_functions}
We replace the cubic spectral evaluations in the preceding implementation.
All matrices in this subsection have known polynomial norm and inverse
norm bounds, and the requested absolute Frobenius error is inverse
polynomial. The number of products below is polylogarithmic.

\begin{lemma}[Logarithm and derivative actions]\label{br_lem_log}
For $\ell I\preceq A=A^*\preceq UI$, with $\ell>0$, both $\log A$ and
$D\log_A[E]$ can be approximated to the prescribed error in
$\widetilde O(n^{\omega_0})$ arithmetic operations.
\end{lemma}
\begin{proof}
Use the ordered resolvent identities
\begin{align*}
 \log A&=\int_0^\infty((1+t)^{-1}I-(A+tI)^{-1}) \d t,\\
 D\log_A[E]&=\int_0^\infty(A+tI)^{-1}E(A+tI)^{-1} \d t.
\end{align*}
The first follows from scalar functional calculus; differentiation under
the integral gives the second. For $M\ge\max\{1,\|E\|_F\}$, the omitted
tails outside $[a_0,b_0]$ are bounded respectively by
\[
 \sqrt n(a_0(1+\ell^{-1})+(U+1)/b_0),\qquad
 a_0M/\ell^2+M/b_0.
\]
Choose positive dyadic endpoints to make these smaller than one eighth
of the target. Split the interval into dyadic bands $[a,b]$, $b\le2a$.
Put $c:=(a+b)/2$, $h:=(b-a)/2$, $R:=(A+cI)^{-1}$, and $W:=hR$.
Then $\|W\|\le1/3$, and for $t=c+hv$, $|v|\le1$,
\[
 (A+tI)^{-1}=\sum_{j=0}^d(-v)^jW^jR+\mathcal R_d(v),
 \qquad \|\mathcal R_d(v)\|\le\tfrac32\|R\|3^{-d-1}.
\]
Integrate monomials using $\mu_j:=\int_{-1}^1v^j \d v$, equal to zero
for odd $j$ and $2/(j+1)$ for even $j$. The logarithm uses
$h\sum_j(-1)^j\mu_jW^jR$ and the analogous scalar series at $1+c$.
The derivative uses
\[
 h\sum_{i,j=0}^d(-1)^{i+j}\mu_{i+j}W^iRERW^j.
\]
No factor crosses $E$. The band errors are bounded by a constant times
$h(\|R\|+(1+c)^{-1})3^{-d}$ and $hM\|R\|^2 3^{-d}$, respectively,
with the Frobenius conversion for the first. The known bounds determine
a logarithmic $d$ and a logarithmic number of bands. Compute each shifted
inverse by residual-certified Newton--Schulz iteration, and symmetrize
the outputs. All operations are order-$n$ products or inverses.
\end{proof}

\begin{lemma}[Exponentials, powers, and Gibbs actions]\label{br_lem_exp}
For Hermitian $L$ with $\|L\|=O(\log n)$, $e^L$ has an
$\widetilde O(n^{\omega_0})$ arithmetic implementation. This also
implements $A^t$, $|t|\le1$, and the operators $Q_{\mathsf Q}$ and
$B_{\mathsf Q}$ at a root with polynomial spectral bounds. For $s$
vectors and possibly different $t_j\in[0,1]$, all $A^{t_j}g_j$ can
be computed by polylogarithmically many products of shape $(n,n,s)$
after one matrix logarithm.
\end{lemma}
\begin{proof}
The exponential series truncated at degree $d$ has norm error at most
$e^M M^{d+1}/(d+1)!$ for $M\ge\|L\|$. A sufficiently large constant
times $M+\log(n/\varepsilon)+1$ suffices for Frobenius error
$\varepsilon$. Evaluate the normalized terms $L^j/j!$. For powers
use $A^t=\exp(t\log A)$ and Lemma~\ref{br_lem_log}.
For the Gibbs covariance put $L:=\log\mathsf Q$ on the two blocks.
The ordered expansion is
\[
 \int_0^1e^{tL}Ee^{(1-t)L} \d t
 \simeq\sum_{i,j=0}^d\frac{L^iEL^j}{(i+j+1)!}.
\]
Indeed the beta integral cancels the two factorials. Evaluate with
normalized powers and bounded beta-integral coefficients. The two
exponential tails control the error. Subtract
$\tr[\mathsf Q E]\mathsf Q$ and divide by $\tau_{\rm s}$ for
$Q_{\mathsf Q}$; use $\tau_{\rm s}\Pi_{\mathcal T}D\log_{\mathsf Q}$
for $B_{\mathsf Q}$.

For the batched assertion set $G:=[g_1\ \cdots\ g_s]$ and use
\[
 Y_0:=G,\qquad Y_j:=L Y_{j-1}\operatorname{diag}(t_1/j,\ldots,t_s/j).
\]
The sum of these arrays gives the different exponential actions.
The columnwise tail bound, summed in Frobenius norm, fixes their common
accuracy. For Gibbs features this prepares
$\mathsf Q^{t_j/2}g_j$ and $\mathsf Q^{(1-t_j)/2}h_j$ without an
eigenvector matrix. Scalar work and writing vectors cost $O(ns)$ per term.
\end{proof}

\begin{lemma}[Dyadic-shift Lyapunov solves]\label{br_lem_lyapunov}
If $\ell I\preceq T\preceq UI$, the equation $TX+XT=B$ can be solved
to the prescribed Frobenius accuracy in $\widetilde O(n^{\omega_0})$
arithmetic operations, without an eigenvalue-separation assumption.
\end{lemma}
\begin{proof}
For a positive dyadic shift $p$, put
$R_p:=(T+pI)^{-1}$ and $A_p:=(T-pI)R_p$. Starting from zero, update
\begin{equation}\label{br_eq_adi}
 X^+:=A_pXA_p+2pR_pBR_p.
\end{equation}
The exact solution $X_*$ satisfies
\[
 (T+pI)X_*(T+pI)-(T-pI)X_*(T-pI)=2pB,
\]
so $X^+-X_*=A_p(X-X_*)A_p$. Cycle through shifts from
$2^{\lfloor\log_2\ell\rfloor}$ to $2^{\lceil\log_2U\rceil}$.
Every eigenvalue $t$ has one shift $p\le t<2p$, giving
$|(t-p)/(t+p)|\le1/3$; all other factors are at most one.
The $A_p$ commute because they are functions of $T$. Thus one cycle
contracts error by at most $1/9$, and
$\|X_*\|_F\le\|B\|_F/(2\ell)$ supplies the initial error bound.
A logarithmic number of cycles suffices. A final residual evaluation,
together with the operator lower bound $2\ell$, certifies the solution.
\end{proof}

\begin{lemma}[Count-sensitive root and feature actions]\label{br_lem_actions}
For $n\le k\le n^2$, put $u:=\log_n k$ and use $\vartheta$ from
Eq.~\eqref{rr_eq_theta}. Every true residual, corrected root stage,
response solve, full resident score scan, and relative-response norm
certificate costs
\begin{equation}\label{br_eq_action}
 \widetilde O(n^{\vartheta(u)}).
\end{equation}
For $k\le n$, use $u:=1$. Implicit feature actions with at most
$\widetilde O(n^u)$ selected coordinates and features have the same
bound, including initial sample preparation.
\end{lemma}
\begin{proof}
The coupling and its adjoint, signed sums, and score scans use products
of shapes $(n,n,k)$ and $(n,k,n)$ and $O(nk)$ scalar work. Fix a common
power-of-two block base, with exponent divisible by four, for the strict
square, $(1,3/2,1)$, $(1,7/4,1)$ and $(1,2,1)$ bounds. At $O(\log n)$ recursive levels,
mix the two relevant endpoint algorithms; choose the least number of
wider levels covering $k$. Padding costs a constant factor. The product
costs interpolate to $\vartheta(u)$, uniformly in the integer sizes.
Tensor permutations give the adjoints.

The root Schur system in Eq.~\eqref{soft_eq_precond} is
polylogarithmically conditioned relative to $\mathscr E$.
Its inverse blocks reduce to
\[
 \widehat D_UT_U+T_U\widehat D_U=U^{1/2}F_UU^{1/2},
 \qquad T_U=U^{-1/2}WU^{-1/2},
\]
and the analogous $V$ equation. Lemmas~\ref{br_lem_log}--\ref{br_lem_lyapunov}
implement these and the Gibbs actions in $\widetilde O(n^{\omega_0})$.
The root bounds in Eq.~\eqref{soft_eq_rootbounds} give all spectral
floors and ceilings. Anchored correction in Lemma~\ref{soft_lem_transport}
uses the actual residual and the actual $D\log_{\mathsf Q}$ block off
the root. It does not presume a current Gibbs identity at an approximate
iterate. The same correction count therefore has the cost in
Eq.~\eqref{br_eq_action}.

Lemma~\ref{br_lem_exp} prepares the sampled power vectors. The product
shapes are those of Lemma~\ref{rr_lem_actions}. Its rank-two feature
and quadratic-form formulas then give restricted actions and adjoints,
including repeated indices. A terminal factor of order
$\widetilde O(q)$ and its Gram use square products when the last two
sketch sizes equal $q\ge n$. The explicit ridge supplies its inverse
floor. Every action refers to one frozen covariance, Hessian surrogate,
and set of feature factors. Relative-response certificates, including
the non-Hermitian similarities in Eq.~\eqref{md_eq_relative}, use the
certificate of Lemma~\ref{oe_lem_power} with $Z:=M^*M/a^2$
for a response matrix $M$. The proof applies because $Z\succeq0$;
when $\|M\|\le a/2$, the same series certifies $I-Z\succ0$.
Reserve the factor-two gap in the response tests. The cost is
$\widetilde O(n^{\omega_0})$, within the displayed bound.
\end{proof}

\subsection{Count-sensitive recursive inverse costs}\label{br_sec_cost}
We use the alternating ridge systems in Eqs.~\eqref{rr_eq_M0}--\eqref{rr_eq_woodbury},
restricted to the movable coordinates and normalized by $G_\sharp$.
All coupling actions still include the full resident background.
Write the padded order as $n=2^r$, put $L:=2+r$, and set
\[
 \kappa:=2^{\lceil\log_2k\rceil}=2^t,\qquad
 u:=t/r,\qquad r\le t\le2r.
\]
The case $n=1$ is handled directly. The large-count regime has $k>n$.
For $n\le q\le\kappa$ define
\begin{equation}\label{br_eq_restricted}
 \mathcal M(n,q;\kappa):=n^{\vartheta(u)}(q/\kappa)^{5/6}.
\end{equation}
All three slopes of $\vartheta$ exceed $5/6$, so Lemma~\ref{br_lem_actions}
bounds a restricted action by this quantity, up to a fixed logarithmic
factor. Enlarged lists can be split into polylogarithmically many blocks.

\begin{lemma}[Expanded recursive cost]\label{br_lem_trial}
Let $2\le h\le L^2$ and let $n\le q_h\le\cdots\le q_1\le\kappa$
be powers of two with $q_{h-1}=q_h$. There is a fixed integer $D\ge1$
such that a complete capped drift-response trial uses at most the following
number of arithmetic operations:
\begin{equation}\label{br_eq_uniformcost}
 \begin{split}
 &L^{4D}(n^{\omega_0}+\mathcal M(n,q_1;\kappa)+q_h^{\omega_0}
                +\sqrt{1+\kappa/q_1} n^{\vartheta(u)})\\
 &+\sum_{j=1}^{h-1}L^{D(j+4)}
       \sqrt{(\kappa+q_1)/q_{j+1}} \mathcal M(n,q_{j-1};\kappa)\\
 &+L^{D(h+4)}\sqrt{(\kappa+q_1)/q_h} q_h^2,
 \qquad q_0:=q_1.
 \end{split}
\end{equation}
The constant $D$ is independent of all sizes and depth. The trial
retains an absolute conditional success probability and verifies the
true direction and response inequalities on every accepted sample.
\end{lemma}
\begin{proof}
The invariant in Eq.~\eqref{rr_eq_invariant}, now with the padded
movable count in place of $\bar k$, remains at most $2q_j/d_0$.
All ridges, factors, and inverse bounds are polynomial uniformly in
depth. The sampling proofs in Lemmas~\ref{rr_lem_entries}
and~\ref{rr_lem_recursive} require polynomial sizes and positive
ridges; they do not require the earlier terminal floor $n^{4/3}$.
Take moment order
$\lceil128\log_2(2kn\ell_{\mathsf Q}(h+1))\rceil=O(L)$ and allocate
each conditional index failure at most $1/(1000(h+1))$.
The same union bound gives constant joint success probability.
Lemma~\ref{md_lem_direction} adds its six norm tests with their reserved
slack. Finite caps bound work on all other samples.

Use the normalized recurrence in Eq.~\eqref{ue_eq_normalized}, with
the residual corrections in Lemma~\ref{ue_lem_precision}.
That proof uses the telescoping coefficient invariant, positive
ridges, and polynomial sizes. It therefore applies also to terminal
size $q_h\ge n$: its norm, similarity, and error exponents remain
independent of depth. Completed child solves use the same
$\eta=n^{-B_0}$ accuracy relative to their own right-hand sides,
and fresh residual corrections attain that accuracy at every parent. Choose $D$ large enough to include the fixed per-level
logarithmic overhead, including residual correction. One internal level
costs at most $L^D\sqrt{q_j/q_{j+1}}$ times its restricted action plus
one child solve. The initial $K_0$ solve has constant condition number;
the shifted physical solve contributes $L^D\sqrt{1+\kappa/q_1}$.
Expanding gives the sum and terminal inverse-action term in
Eq.~\eqref{br_eq_uniformcost}. The last two equal sizes make terminal
construction, Gram formation, and inversion cost $L^Dq_h^{\omega_0}$;
the inverse application costs $L^Dq_h^2$.

Initial sample preparation costs
$L^D(n^{\omega_0}+\mathcal M(n,q_1;\kappa))$ by
Lemma~\ref{br_lem_exp}. Full resident actions cost
$\widetilde O(n^{\vartheta(u)})$. Index lists, random bits, all response
and quotient checks, and the chosen curved candidate and correction
fit the displayed terms. No child inverse or terminal setup is omitted.
\end{proof}

Fix this $D$, set $C_*:=2000(D+1)$, and define
\[
 b_{\max}:=t-r,\qquad
 b_0:=\min\{b_{\max},C_*\lceil\log_2L\rceil\}.
\]
If $b_0=b_{\max}$, use the two deficits $(b_0,b_0)$. Otherwise set
$b_{i+1}:=\min\{b_{\max},\lceil3b_i/2\rceil\}$ until $b_T=b_{\max}$,
and use
\begin{equation}\label{br_eq_deficits}
 (\ell_1,\ldots,\ell_h):=(b_0,b_1,b_1,\ldots,b_T,b_T),
 \quad q_j:=2^{t-\ell_j},\quad \ell_0:=\ell_1.
\end{equation}
This integer schedule has $h=O(\log L)$, $h\le L^2$,
$q_{h-1}=q_h=n$, and $2^{b_0}\le L^{2C_*}$.

\begin{lemma}[Uniform count-sensitive trial bound]\label{br_lem_sum}
The schedule in Eq.~\eqref{br_eq_deficits} makes the complete cost in
Eq.~\eqref{br_eq_uniformcost} equal to
$\widetilde O(n^{\vartheta(u)})$, with a fixed logarithmic degree.
\end{lemma}
\begin{proof}
The top term is at most $2L^{4D+C_*}n^{\vartheta(u)}$; setup obeys the
same bound because $q_h=n$. An internal summand is at most
\[
 2n^{\vartheta(u)}L^{D(j+4)}2^{-5\ell_{j-1}/6+\ell_{j+1}/2}
 \le4n^{\vartheta(u)}L^{D(j+4)}2^{-\ell_{j-1}/12},
\]
using $\ell_{j+1}\le3\ell_{j-1}/2+1$.
When $b_0<b_{\max}$, put $v:=\lfloor(j-1)/2\rfloor$.
Then $\ell_{j-1}\ge C_*(3/2)^v\log_2L$ before the terminal truncation,
while $j+4\le8(3/2)^v$ and $j\le4(3/2)^v$. Consequently
\[
 L^{D(j+4)}2^{-\ell_{j-1}/12}\le L^{-2j}.
\]
All internal terms sum to $O(n^{\vartheta(u)})$.

The leaf overhead needs a separate estimate. Each preterminal increase
is at least $b_0/2$, so $h+4\le12b_{\max}/b_0$ and
\[
 D(h+4)\log_2L\le(12D/C_*)b_{\max}<r/100.
\]
The leaf inverse applications cost at most
$2\sqrt\kappa n^{3/2+1/100}$. They fit because
$\vartheta(u)-u/2\ge\omega_0-1/2=1.871177>1.51$.
If $b_0=b_{\max}$, the depth is two, all logarithmic powers are fixed,
and the sole internal term has the additional factor $2^{-b_0/3}$.
The same comparison handles the leaf. Thus no depth-dependent
logarithmic power is treated as a fixed polylogarithm without a bound.
\end{proof}

\subsection{Operation totals}\label{sec:randomized:operation-totals}
Lemma~\ref{bs_lem_grouped} leaves $k_0\le\min\{m,2n^2\}$ original
fractional coordinates. Pad to $N=2^{\lceil\log_2(2n)\rceil}<4n$.
The nonlinear instance contains the original retained PSD rank-one
matrices, not the generally indefinite group sums. Center its single
residual and use the original coordinate-rounding allowance.

\paragraph{Curvature phases.}
Temporarily write $n$ for the padded order. In a phase $K/2<k\le K$,
Lemma~\ref{br_lem_flushes} bounds the number of successful moves by
$\widetilde O(n^{3/2}/\sqrt K)$. Lemma~\ref{br_lem_sum} bounds each
trial cost by $\widetilde O(n^{\vartheta(\log_nK)})$.
Constant-factor padding of $K$ is harmless.
Put
\[
 \phi(u):=3/2-u/2+\vartheta(u),\qquad
 s_2:=4(\xi_1-\xi_0)=0.907832,
\]
with $\xi_0,\xi_1$ as in Eq.~\eqref{rr_eq_theta}. The three slopes of
$\vartheta$ are $0.846914$, $0.907832$, and $0.913776$.
Hence every piece of $\phi$ increases, and every piece of
$\phi(u)-(u+2)$ decreases. Their crossing lies in $(3/2,7/4)$:
\begin{equation}\label{br_eq_exponent}
 u_*:=\frac{\xi_0-3s_2/2-1/2}{3/2-s_2}
       =\frac{466443}{296084},\qquad
 E:=2+u_* =\frac{1058611}{296084}<3.575374.
\end{equation}
Thus
\[
 n^{\phi(u)}\le
 \begin{cases}
 n^E,&1\le u\le u_*,\\
 Kn^2,&u_*\le u\le2,\quad K=n^u.
 \end{cases}
\]
Every dyadically padded phase ceiling satisfies $K\le2m$.
Summing the logarithmically many phases therefore costs
$\widetilde O(mn^2+n^E)$, including sketch construction, response
certificates, curved predictors, and true root corrections.

\paragraph{Continuation and finishing.}
Lemma~\ref{br_lem_flushes} bounds initial and bulk continuation by
$\widetilde O(\sqrt n)$ stages. If $k_0\ge n$, put $u_0:=\log_n k_0$.
The slopes of $\vartheta$ are less than one, so
\[
 \vartheta(u_0)-u_0\le\omega_0-1<3/2,
 \qquad \sqrt n n^{\vartheta(u_0)}\le k_0n^2\le mn^2.
\]
For $k_0<n$, their cost is $\widetilde O(n^{\omega_0+1/2})$, smaller
than $n^3$. At the permanent switch $k\le K_c=nP_n$, use the finishing
algorithm of Section~\ref{am_sec_improvement}. Its
$\widetilde O(n)$ expected corrected and search stages now cost
$\widetilde O(n^{\omega_0})$ each. In this range a full sampled
feature Gram has order $\widetilde O(n)$, and its construction and
inverse fit the same bound. Thus finishing costs
$\widetilde O(n^{1+\omega_0})$. Its coordinate updates cost
$\widetilde O(n^3)$.

\paragraph{Failure branches.}
Every trial has fixed sample, iteration, precision, and magnitude caps.
Every accepted curved move has an inverse-polynomial true decrease.
On cap exhaustion, clear the pending set with the certified attenuation
routine and perform one old guaranteed move: a regular endpoint if
available, otherwise a coordinate-gradient step or the deterministic
dense curvature witness with its original step length. Then resume.
These moves retain their decrease or removal guarantees. Bounded
potential injections and at most $k_0$ removals give a deterministic
polynomial invocation bound $I_{\max}$. If $n^C$ bounds the polynomial
emergency work, choose the logarithmic trial cap to make its conditional
exhaustion probability at most $(2I_{\max}n^{2C+100})^{-1}$.
The expected total emergency work, including extra batches, is then
negligible. Preprocessing, finishing, and final rounding retain their
own capped verified fallbacks. Correctness and finite termination hold
for every random-bit sequence; the faster bound is in expectation.

\paragraph{The complete arithmetic bound.}
Return to original dimension $n$ and padded order $N<4n$.
One-time conversion and scaling cost $\widetilde O(mn^2+n^3)$;
Lemma~\ref{bs_lem_grouped} adds $\widetilde O(mn^2)$.
Each removed coefficient updates the fixed symbolic sum once.
Lemma~\ref{oe_lem_round} costs
$\widetilde O(k_0N^2+N^{\omega_0})$ for final rounding.
Since $k_0\le m$, these are input costs, even when the weaker estimate
$k_0\le N^2$ would give a separate quartic term. Combining the bounds gives
\begin{equation}\label{soft_eq_total}
 \E[T_{\rm arith}]
 \le\widetilde O(mn^2+n^3+n^{1+\omega_0}+n^E)
 =\widetilde O(mn^2+n^E).
\end{equation}
Here $1+\omega_0=3.371177<E$.

All mathematical paths use the full true coupling. Pending coefficients
are fixed, then attenuated as a positive submap before exact deletion.
Reservoir phases stay fixed on each path and their total injection is
unchanged. Coordinate-zero crossings use the same continuous second
derivatives and one-sided bounds. Consequently Eq.~\eqref{soft_eq_allowance}
and the one-time input transfer still give
\[
 4.86276(1+6\cdot10^{-8})+2\cdot10^{-8}
 =4.8627603117656<4.8628.
\]
Padding preserves the signed norm, and the partition error is half
the signed bound. This proves Theorem~\ref{soft_thm_algorithm}.

%% file: 40_bit_complexity.tex
\section{Finite-precision implementation and bit complexity}\label{sec_bit_complexity}

Sections~\ref{rank_one_sec_algorithm} and~\ref{sec_fast_randomized}
describe the algorithms, prove their progress, and count arithmetic
operations to prescribed accuracies. Theorem~\ref{bit_thm_totals} gives
bit-complexity versions of Theorems~\ref{rank_one_thm_main}
and~\ref{soft_thm_algorithm} with the same discrepancy and partition
guarantees. This section proves that finite precision attains the
prescribed accuracies and establishes those bit bounds. We retain
the notation of each algorithm and refer to its conditioning and
discrepancy estimates.

\subsection{Computational model and accuracy requirements}\label{bit_sec_model}
Let $L_{\rm in}$ be the total binary input length. We charge bit
operations for reading the original rational inputs and for all
subsequent integer arithmetic. After the one-time input conversion,
the preprocessing grids use $O(\log(2mn))$ bits per numerical entry,
with fixed multipliers. The nonlinear phase has $k=O(n^2)$ active
coordinates and uses $O(\log(2kn))=O(\log(2n))$ accuracy and magnitude
bits per numerical entry. The constants in these bounds also cover
the fixed discrepancy and residual allowances.

These bounds also cover temporary scalar operands after input
conversion. Section~\ref{bit_sec_comparisons} chooses the cube-barrier
sign by a short approximation to its directional derivative; it does
not form a product of all coordinate denominators. A common dyadic
accumulator for polynomially many short summands needs only an
additional logarithmic number of bits. Storage for an array of
$O(n^2)$ entries is separate from the precision of one entry.
For unrestricted $m$, the preprocessing bound remains $O(\log(2mn))$;
it is $O(\log(2n))$ when $m$ is polynomial in $n$. Original rational
inputs may be longer and are charged separately in $L_{\rm in}$.

The arithmetic analyses specify residual, quotient, endpoint, and
value tolerances before an operation is accepted. We implement those
same tests using enclosures smaller than their reserved margins.
The potential, covariance inequalities, step sizes, and progress
counts are unchanged. Conditioning bounds used in those analyses
remain there; the arguments below convert them to working precision.

\subsection{One-time input conversion}\label{bit_sec_input}
We first implement the input approximation used in Section~\ref{sec:fast:input-rounding}. Preserving positive semidefiniteness and rank one allows the subsequent algorithms to use the rounded matrices directly.

\begin{lemma}[Rank-preserving input rounding]\label{bit_lem_input}
Let $\bar A_1,\ldots,\bar A_m$ be the normalized rational PSD rank-one inputs of Section~\ref{sec:fast:input-rounding}, and put $\delta:=10^{-8}/(mn)$. There are dyadic matrices $\widetilde{\bar A}_i$ of rank at most one such that
\[
 \widetilde{\bar A}_i\succeq0,\qquad
 \|\widetilde{\bar A}_i-\bar A_i\|_F\le\delta.
\]
Each matrix has a factorization $\widehat a_i\widehat w_i\widehat w_i^*$ on a common mesh of exponent $p=O(\log(2mn))$, and the expanded matrices have common denominator $2^{3p}$. The factors can be constructed in $\widetilde O(L_{\rm in}+mn)$ bit operations. Constructing all expanded matrices costs $\widetilde O(L_{\rm in}+mn^2)$ bit operations.
\end{lemma}
\begin{proof}
Implement the construction in Section~\ref{sec:fast:input-rounding}.
Choose the normalization factor $D_{\rm aux}$ to be a power of two
satisfying $a_*\le D_{\rm aux}<2a_*$, within the allowed range there.
Its exponent is obtained from binary lengths and one rational
comparison and is stored separately.
Balanced comparison tournaments select the largest diagonal entries
in $\widetilde O(L_{\rm in})$ bit operations: at each level the total
length of the rational numbers compared is $O(L_{\rm in})$.

Use the common mesh $h_{\rm aux}=2^{-p}$ from that construction.
To approximate a bounded rational number after multiplication by a
power of two, retain the leading $p+O(1)$ bits needed for a certified
quotient interval. Numbers below the absolute error threshold are
replaced by zero. Reading the original numerators and denominators
and performing these truncated divisions costs their total input
length plus a polynomial in $p$ per selected entry. Large shifts are
handled by exponent arithmetic. The original inputs are charged once.

After this scan, each division and rounding in the factor construction
has $O(\log(2mn))$-bit operands. In particular, the branch that forms
$\widehat w$ has $\widehat a>\delta/(4n)$, so its division is
polynomially conditioned. The error bound, positive semidefiniteness,
and rank bound follow from the construction already proved in
Section~\ref{sec:fast:input-rounding}. The three dyadic factors give
common denominator $2^{3p}$ for the expanded matrices. Constructing
and storing all compact factors costs
$\widetilde O(L_{\rm in}+mn)$ bit operations. Their outer products
have logarithmic-length operands and require $O(mn^2)$ scalar
products, giving total bit cost $\widetilde O(L_{\rm in}+mn^2)$.
\end{proof}

\subsection{Stable partial signing and grouped reduction}\label{bit_sec_reduction}
The following lemmas implement the corresponding arithmetic reductions. Each uses the local notation of the reduction it implements.

The ridge reduction makes many updates between recomputations. The required stability estimate controls an entire consecutive sequence of updates.

\begin{lemma}[Stability of a ridge-update epoch]\label{bit_lem_ridge}
Under the hypotheses of Lemma~\ref{rank_one_lem_regularized_partial_signing},
assume additionally that each rational input entry has
$O(\log(2m d_{\rm vec}/\eta))$ bits. Consider an epoch of its algorithm, with its fixed column matrix $C$, ridge $\delta$, dimension $d_{\rm vec}$, and entry tolerance $\tau$. Put $\mathfrak b:=1+\|C\|^2/\delta$. Suppose the initial matrix error is at most $\epsilon_0$, and each update or product that writes pending updates introduces matrix error at most $\rho$. For a sufficiently small absolute constant $c>0$, if $\epsilon_0+d_{\rm vec}\rho\le c\mathfrak b^{-3}$, the error at every epoch prefix is
\[
 O(\mathfrak b^2(\epsilon_0+d_{\rm vec}\rho)).
\]
The prescribed entry tolerance and residual allowance are attained with $O(\log(2md_{\rm vec}/\eta))$ working bits, with a fixed multiplier. The resulting bit cost is the arithmetic bound in Lemma~\ref{rank_one_lem_regularized_partial_signing}, up to logarithmic factors.
\end{lemma}
\begin{proof}
Throughout an epoch,
$\delta I\preceq G\preceq(\delta+4({d_{\rm vec}}+1))I$ and
$\|{\mathsf G}\|\le\|C\|^2/\delta=O({d_{\rm vec}}/\delta)$. The insertion denominator in
Eq.~\eqref{rank_one_eq_ridge_update} is at least one. For deletion,
write $G=G_-+C_iC_i^{\top}$, with $G_-\succeq\delta I$. Then
\[
 1-{\mathsf G}_{ii}=(1+C_i^{\top}G_-^{-1}C_i)^{-1}
          \ge\frac{\delta}{\delta+4}.
\]
All exact quantities and all divisions therefore have polynomial
magnitude bounds in $m,{d_{\rm vec}},1/\eta$.

For a consecutive sequence of updates let $D$ be the diagonal matrix
of its net column changes. Every input is inserted and removed at
most once, so $\|D\|\le1$. The exact composition is
\[
 F_D({\mathsf G}):=(I+{\mathsf G}D)^{-1}{\mathsf G}.
\]
If $G'$ is the Gram matrix at the end of this sequence, the identity
\[
 (I+{\mathsf G}D)^{-1}=I-C^{\top}(G')^{-1}CD
\]
shows that the inverse on the left has norm at most
$\mathfrak b:=1+\|C\|^2/\delta$. This bound holds for every consecutive
subsequence, including one containing both insertions and deletions.
For a symmetric perturbation $E$ with $\|E\|\le1/(2\mathfrak b)$,
\begin{equation}\label{rank_one_eq_ridge_stability}
 \begin{aligned}
 F_D({\mathsf G}+E)-F_D({\mathsf G})
 &=(I+{\mathsf G}D)^{-1}E(I+D({\mathsf G}+E))^{-1},\\
 \|F_D({\mathsf G}+E)-F_D({\mathsf G})\|&\le2\mathfrak b^2\|E\|.
 \end{aligned}
\end{equation}
The second inverse is bounded by $2\mathfrak b$ by the Neumann series.

Here is the resulting error accounting over a whole epoch. Suppose each update, evaluated
at the currently represented matrix, introduces an additive matrix
error of norm at most $\rho$, and each product that writes the
pending updates explicitly has the same error bound. Such a product
leaves the represented matrix unchanged in exact arithmetic, so it
is an identity map with an additive error. Compare a computed prefix with
the exact prefix by inserting one error at a time and composing all
remaining exact updates. Eq.~\eqref{rank_one_eq_ridge_stability}, applied
to every such suffix, bounds the prefix error by
$O(\mathfrak b^2(\epsilon_0+{d_{\rm vec}}\rho))$, where $\epsilon_0$ is the
initial error. A bootstrap with this bound smaller than $1/(4\mathfrak b)$
justifies the same bound for the perturbed suffixes. In particular no
factor is raised to the epoch length.

The lazy representation satisfies this additive-error model. Summing
at most ${b_{\rm batch}}$ stored outer products recovers a column with error at most
a fixed polynomial in ${d_{\rm vec}},{b_{\rm batch}},\delta^{-1}$ times its scalar rounding
unit. The lower bounds on the denominators then give the same type of
bound for the error of an update. The separately stored diagonal has
at most ${b_{\rm batch}}$ such rounding errors. Coefficient updates also use only
polynomially bounded quantities. Direction entries and retained
coefficients lie on a common dyadic grid. The boundary ratios are
formed and compared exactly using a constant multiple of their word
length. Setting the hitting coefficients exactly to their boundary
values and rounding the others toward zero restores the grid and
preserves the cube. Taking
$O(\log(2m{d_{\rm vec}}/\eta))$ bits, with a sufficiently large absolute multiplier,
makes all matrix evaluation errors smaller than $\tau/4$ entrywise.
Since $u\le\tau/4$, the subsequent grid rounding keeps the total
entry errors below $\tau$. The
explicit matrix is symmetrized after each product, and its diagonal
replaces the separately stored diagonal. The factors in each rectangular
product have polynomially bounded entries. A fixed bilinear recursive
algorithm implements the strict rectangular exponent bound; its depth
is $O(\log {d_{\rm vec}})$ and all coefficients at a level are fixed. Hence its
intermediate magnitudes and normwise error amplification are polynomial,
and only a constant multiple of the same logarithmic precision is
needed. The strict slack $2.054999<2.055$ permits this fixed algorithm.
Each new epoch is recomputed from its retained input columns; errors
are not passed through an unbounded sequence of epochs.

The epoch inverses are well conditioned up to a fixed polynomial, so
the short-precision Newton--Schulz construction and stable fast products
of Section~\ref{bit_sec_linear_algebra} apply, with logarithmic factors in $m{d_{\rm vec}}/\eta$.
They give the stated bit cost as well as the arithmetic cost.
\end{proof}

We next apply the epoch stability bound to the small reduction used at each level of the group tree.

\begin{lemma}[Finite-precision grouped reduction]\label{bit_lem_grouped}
Under the hypotheses of Lemma~\ref{rank_one_lem_grouped_partial_signing},
put $D:=d_{\rm vec}$ and assume that the input entries have a common
rational denominator, with numerators and denominator of
$O(\log(2mD/\eta))$ bits.
At a level of its group tree, set
$\varepsilon:=\eta/(M_{\rm grp}L_{\rm tree})$ and consider the small
procedure on at most $2D$ columns, with inherited coefficient word
length $p_y$. The procedure preserves the cube,
leaves at most $D$ fractional coordinates, and changes the represented
sum by norm at most $\varepsilon$, using
$O(p_y+\log(2D/\varepsilon))$ working bits, including temporary operands,
with a fixed multiplier. Its bit cost is $\widetilde O(D^{2.5275})$,
with the input word length and $\log(1/\varepsilon)$ included in the
suppressed factors.
The full grouped algorithm returns $y\in[-1,1]^m$ with at most $D$
fractional coordinates and $\|\sum_i y_i b_i\|_2\le\eta$, using
$O(\log(2mD/\eta))$ working bits and
$\widetilde O(mD+D^{2.5275})$ bit operations.
\end{lemma}
\begin{proof}
Throughout this procedure,
$\delta I\preceq G\preceq(\delta+8D)I$, and every deletion
denominator is at least $\delta/(\delta+4)$. For any consecutive
sequence of deletions, Lemma~\ref{bit_lem_ridge} applies
with $\mathfrak b\le1+8D/\delta$. It bounds the amplification of an
additive matrix error through the entire sequence by
$O(\mathfrak b^2)$. Writing a pending sum to the base is an identity
map with an additive product error. Recovering a column from at most
${b_{\rm batch}}$ terms and then computing its deletion update introduce errors
bounded by a fixed polynomial in $D,\delta^{-1}$ times the scalar
rounding unit. The separately stored diagonal has the same bound;
refresh it from the base whenever a pending sum is written out.
The suffix-error accounting in
Lemma~\ref{bit_lem_ridge} therefore applies
to all $O(D)$ deletions and products, including simultaneous hits.

Each rectangular product uses fixed bilinear recursion. Its factors
have polynomially bounded entries, and its logarithmic recursion
depth gives polynomial error amplification. The strict exponent slack
permits a fixed algorithm at exponent $1+2a$. Together with the initial
inverse, these bounds make the matrix evaluation errors smaller than
$\tau/4$ entrywise, using a sufficiently large constant times
$\log(2D/\varepsilon)$ guard bits. Since $u\le\tau/4$, rounding each
queried entry to the direction grid keeps its total error below $\tau$.
Boundary ratios are formed and compared exactly from the dyadic
coefficients and directions, using a constant multiple of their word
length. The hitting coefficients are set exactly to their endpoints,
and rounding the others toward zero restores the common grid.
The initial word length $p_y$ is included in the working precision
and bit cost. Thus batching changes neither the residual allowance nor the
cube and stopping guarantees of Lemma~\ref{rank_one_lem_grouped_partial_signing}.

For the full group tree, write the input entries over their common
denominator. Each node sum contains at most $m$ inputs, so its numerator
length increases by at most $O(\log(2m))$. All node sums are therefore
computed exactly in $\widetilde O(mD)$ bit operations with
$O(\log(2mD/\eta))$ bits per entry. Dividing by the power of two
$M_{\rm grp}$ adds only $O(\log(2m))$ denominator bits. With
$\varepsilon=\eta/(M_{\rm grp}L_{\rm tree})$, one has
\[
 \log(2D/\varepsilon)=O(\log(2mD/\eta)).
\]
Choose the same coefficient grid at every level. The root coefficient
is zero, and each child inherits its parent's coefficient. Every small
reduction restores this grid, so all inherited coefficients have
$O(\log(2mD/\eta))$ bits. The preceding bounds therefore apply uniformly
to all $L_{\rm tree}$ small reductions, including their temporary
operands. Their total bit cost is
$\widetilde O(L_{\rm tree}D^{2.5275})=\widetilde O(D^{2.5275})$.
The final tree traversal copies the coefficients to the leaves in
$\widetilde O(m)$ bit operations. Adding the group-sum cost proves the
claimed total. The residual accounting and fractional-coordinate bound
in Lemma~\ref{rank_one_lem_grouped_partial_signing} apply at every level,
since each finite reduction satisfies its prescribed error allowance.
\end{proof}

The randomized small reducer also needs certified directions from a fixed ridge inverse. Its numerical evaluations and derivative sign rule are both implemented on short data.

\begin{lemma}[Finite ridge-direction evaluations]\label{bit_lem_ridge_direction}
Use the parameters of Lemma~\ref{sr_lem_small}.
Assume the rational columns share a common
denominator and have word length $p_C$, and the inherited dyadic
coefficients have word length $p_y$. Its ridge inverse and direction trials attain their prescribed residual, norm, and maximum-entry tolerances using
\[
 O(p_C+p_y+\log(2D/\varepsilon)+\log(2T_*))
\]
working bits, with a fixed multiplier. Forming the inverse costs $\widetilde O(D^{\omega_0})$ bit operations, and each direction trial after fixing the inverse costs $\widetilde O(D^2)$ bit operations.
If $p_C+p_y=O(\log(2D/(\varepsilon\zeta_{\rm f})))$, its full
small-reduction algorithm has expected bit cost
$\widetilde O(D^{\omega_0}+\zeta_{\rm f}D^3)$, including the
$\widetilde O(D^3)$ fallback, whose probability is at most
$\zeta_{\rm f}$. Every random-bit sequence returns at most $2D$
fractional coordinates and residual at most $\varepsilon$.

For Lemma~\ref{sr_lem_grouped}, assume a common rational input
denominator and numerator and denominator lengths
$O(\log(2mD/\eta))$. Its full grouped algorithm has expected bit
cost $\widetilde O(mD+D^{\omega_0})$, with at most $2D$ fractional
coordinates and residual at most $\eta$ on every outcome. Its
working precision, including temporary operands, is
$O(\log(2mD/\eta))$.
\end{lemma}
\begin{proof}
All magnitudes and inverse bounds here are polynomial in
$D,\varepsilon^{-1}$ and the displayed logarithmic factors.
Newton--Schulz reaches the stated accuracy in logarithmically many
products: from $R_0=J_\delta/U^2$ with $U\ge\|J_\delta\|$, its exact
residual squares at each iteration, and its initial spectral gap is
at least $\delta^2/U^2$. Rounded residuals satisfy
$e_{t+1}\le e_t^2+\epsilon_{\rm unit}$. Taking the product-error floor
below a fixed fraction of the squared initial gap and final target
proves the same count, with polynomial iterate norms.
The fixed stable square recursion gives the required accuracy with
$O(p_C+p_y+\log(2D/\varepsilon)+\log(2T_*))$ bits, up to a fixed
constant multiplier. Forming or inverting an order-$D$ matrix costs
$\widetilde O(D^{\omega_0})$ in both models. A direction trial after
this inverse is fixed costs $\widetilde O(D^2)$, including its exact
short-dyadic norm and maximum-entry tests. No singular values,
nullspace basis, or exact rational inverse is computed.

For each derivative sign, evaluate the summands
$2x_i\widetilde v_i/[H(1+|x_i|)]$ to error $1/(512H^2)$ on a
common dyadic grid and add them exactly. Their denominators lie in
$[1,2)$ and their magnitudes are polynomial. The sum has error at
most $r/(512H^2)$ and uses only an additional $O(\log(2r))$
accumulator bits. This is the permitted sign error in
Lemma~\ref{sr_lem_small}. Compute each rational step from the short
data, round inward to its prescribed grid, and freeze endpoints by
exact comparisons. The common column denominator and dyadic
coefficient grid make the final matrix residual and squared-norm
comparison exact with the same logarithmic working bound.

Lemma~\ref{bit_lem_grouped} implements the two deterministic
fallback blocks with the required inherited coefficients and
residuals. The deterministic progress, residual, trial-cap, and
conditional failure proofs of Lemma~\ref{sr_lem_small} therefore
apply unchanged. Its polylogarithmic number of inverse constructions
and sign trials gives the stated expected bit bound.

For the group tree, exact node sums add only $O(\log(2m))$ bits
to the common input denominator and numerators. Division by the
power of two $M_{\rm grp}$ has the same overhead. Each small call
uses $\varepsilon=\eta/(M_{\rm grp}R_{\rm tree})$ and
$\zeta_{\rm f}=(2mD)^{-20}/R_{\rm tree}$. After its first inward
update its coefficients lie on the grid in Eq.~\eqref{sr_eq_grid};
inherited coefficients are retained exactly until then. The grid
exponents have a uniform $O(\log(2mD/\eta))$ bound, so coefficient
lengths do not multiply through the tree. Node sums and the leaf
traversal cost $\widetilde O(mD)$ bit operations. Summing the
logarithmically many small calls and their fallback probabilities
gives the expected grouped bound. Lemma~\ref{bit_lem_input} charges
the original binary input length once when these routines follow
the matrix input conversion.
\end{proof}

For energy-controlled groups, the sampled preconditioner is held fixed throughout a solve. This permits one error estimate for the normalized recurrence.

\begin{lemma}[Structured reductions and norm certificates]\label{bit_lem_group_solve}
Consider a solve in Lemma~\ref{oe_lem_cube}, with failure parameter $0<\zeta<1$ and with the group map, dyadic sample, and represented factors fixed. On the sampling event giving its stated preconditioned spectral interval, the true residual and direction tolerances are attained with $O(\log(2mDT/(\varepsilon\zeta)))$ working bits, including the input word length and a fixed multiplier. The structured actions and the solve have bit cost equal to their arithmetic bound up to logarithmic factors.
This includes the finite ridge sketch of Lemma~\ref{oe_lem_sketch},
the setup and actions of Lemma~\ref{oe_lem_actions}, and the full
capped reduction of Lemma~\ref{oe_lem_cube}, with the same residual,
fractional-count, and fallback-probability guarantees, provided the
input entries and inherited coefficients have logarithmic word
length in these parameters and a common rational input denominator.
The orientation of Lemma~\ref{oe_lem_orientation} costs
$\widetilde O(mD)$ bit operations per attempt on such input.

For the norm certificate of Lemma~\ref{oe_lem_power}, if the input
entries and threshold have $b$ bits and the entries share a common
rational denominator, the working precision is $O(b+\log(2n))$.
For logarithmic input length the bit cost is
$\widetilde O(n^{\omega_0})$, on both accepted and rejected inputs.
The test accepts only if $\|E\|\le a$ and always accepts if
$\|E\|\le a/2$.
\end{lemma}
\begin{proof}
Freeze $B$, the dyadic sample,
and the represented factors throughout the solve. They define one
positive system and one positive preconditioner. The small Woodbury
matrix has smallest eigenvalue at least $T/2$ and polynomial norm, so
Newton--Schulz and fixed recursive products attain any required
inverse-polynomial absolute accuracy in logarithmic working precision.
The factors and $\delta^{-2}$ in Lemma~\ref{oe_lem_actions} and Eq.~\eqref{oe_eq_woodbury} have polynomial
magnitude. Cancellation is covered by a finer absolute error, not by
assuming a relative inverse approximation suffices.

In the normalized Chebyshev recurrence, the scalar ratios lie in $[0,1]$
and their update is Lipschitz with constant at most one. An error inserted
at step $i$ propagates through
\[
 [T_i(\varrho)/T_J(\varrho)]U_{J-i}(D_0),
\]
where $D_0$ is the scaled preconditioned operator. On the good interval its norm is at most a
polynomial similarity factor times $J-i+1$. Vector and scalar errors
therefore accumulate with a polynomial bound in the prescribed iteration
count, not an exponential product of condition factors. All iterates
and intermediate right-hand sides have polynomial magnitudes on that
event. A sufficiently large constant times
$\log(2mDT/(\varepsilon\zeta))$ precision, including the input word
length, meets the true residual and direction targets. The structured
matrix actions and the explicit residual use the same rational
vectorization and exact group sums, so no cache mismatch is omitted.

\textbf{Sampling, products, and exact comparisons.}
The dyadic Box--Muller construction in Lemma~\ref{oe_lem_sketch}
has polynomial scalar derivative bounds on its coupling event and
polynomial output magnitudes for every finite sample. A fixed
logarithmic precision therefore preserves its covariance interval
and its success probability. Form and symmetrize the small Gram
matrix below the stated $\delta/64$ error allocation. The inverse
perturbation identity preserves its positive lower bound and the
represented preconditioner. Fixed recursive matrix products have
logarithmic depth and fixed coefficients, hence polynomial error
amplification and intermediate magnitudes. A larger fixed guard
precision gives the setup and structured-action bit bounds at their
arithmetic exponents.

All orientation sums and squared norms are exact sums of at most
polynomially many common-denominator products. Their word lengths
remain logarithmic, giving $\widetilde O(mD)$ bit operations for
Lemma~\ref{oe_lem_orientation}. The same argument covers explicit
group residuals and direction norm tests. For the barrier sign,
approximate each bounded rational summand to error $1/(512H^2)$
on one dyadic grid, then add exactly, as in
Lemma~\ref{bit_lem_ridge_direction}. The accumulated sign error is
the one allowed in Lemma~\ref{oe_lem_cube}. It retains inherited coefficients exactly until the first inward update;
the grid in Eq.~\eqref{oe_eq_grid} then has a uniform logarithmic mesh exponent.
Thus every accepted step preserves the arithmetic progress and
residual estimates. Prescribed iteration and magnitude caps bound
failed sketches; on the good sketch event the preceding error
bounds preserve all tests. The fallback is implemented by
Lemma~\ref{bit_lem_grouped}, after the stated scaling and splitting.
This proves the full capped reduction and its fallback probability.

\textbf{Matrix-norm certificates.}
Use $Z=(E/a)^2$, $Q=I-Z$, and the degree
$s=\lceil10\log_2(2n)\rceil$ binomial sum from
Lemma~\ref{oe_lem_power}. Compute on a common dyadic grid and reject
if a fixed polynomial magnitude cap is exceeded. In the good case
$0\preceq Z\preceq I/4$, the powers are contractive, the coefficients
have $O(s)$ bits, and the exact congruence residual is below $1/8$.
Stable products approximate the sum and the final congruence with
Frobenius error below $1/16$ in each required enclosure. Scaling
and input errors are included against the true rational $E,a$.
Using $O(b+\log(2n))$ working bits provides these enclosures. The
good case reaches no magnitude cap and is accepted. Every accepted
case has a certified upper bound at most $1/2$ for
$\|X^*QX-I\|_F$, which proves $Q\succ0$. There are
$O(\log(2n))$ matrix products on every outcome. The same argument
applies to a non-Hermitian matrix by taking $Z=M^*M/a^2$.

Summing the verified orientation, sketch, solve, residual, and
fallback costs over the group tree gives the bit bound in
Eq.~\eqref{oe_eq_preprocessingtotal}. Original input length is
charged once by Lemma~\ref{bit_lem_input}.
\end{proof}

\subsection{Numerical caches and acceptance margins}\label{sec_fast_precision}
The exact retained instance and its numerical cache have different roles. The next lemma controls cache error against the true model at each accepted state.

\begin{lemma}[Short caches and certified acceptance]\label{bit_lem_cache}
Consider a retained instance with $k\le n^2$ active matrices, the symbolic fixed contribution, and the conditioning and acceptance margins specified in Section~\ref{rank_one_sec_algorithm}. A one-time PSD rank-one cache with entry accuracy $u=a(kn)^{-C_0}$, for a sufficiently large fixed $C_0$, permits the KKT residuals, predictors, gradient and curvature tests, endpoint scores, and value comparisons to be evaluated within their reserved errors using $O(\log(2kn))$ accuracy and magnitude bits. Each accepted correction restores the prescribed root-error floor. The initial coordinate rounding uses only the reserved $\varepsilon_{\rm round}/2$ allowance, and subsequent solves use the short cache.
For the recorded coefficients in Lemma~\ref{round_lem_round}, given
on a dyadic grid with $O(\log(2kn))$ bits and with the corresponding
short PSD rank-one cache already available, deterministic deferred
rounding uses $\widetilde O(kn^3)$ bit operations and
$O(\log(2kn))$ working bits per entry, including temporary operands.
Its discrepancy is less than $\varepsilon_{\rm round}/2$, including
cache and comparison errors.
For the small-trace rounding of Lemma~\ref{trace_lem_small_rounding},
cached entries and initial coefficients of $O(\log(2kn))$ bits
give $\widetilde O(kn^2)$ bit cost, including exact comparisons.
The discrepancy and surviving-trace guarantees in that lemma
are preserved.
\end{lemma}
\begin{proof}

\textbf{One-time data approximation.} 

After partial signing, the algorithm needs only its $k$ active matrices. Store a short rational PSD rank-one approximation to each $C_i$, with Frobenius error at most $u$, where $u:=a(kn)^{-C_0}$ and $C_0$ is a sufficiently large fixed constant.

Such approximations can be obtained without assuming a lower bound on a nonzero eigenvalue. Discard a matrix whose largest diagonal entry is below $u/(Cn)$; its Frobenius norm equals its trace and is $O(u)$. Otherwise choose a largest diagonal entry $a:=(C_i)_{jj}\ge u/(Cn)$ and write $C_i=a {\mathsf c}{\mathsf c}^*$, where ${\mathsf c}:=C_i e_j/a$. Here every $|{\mathsf c}_i|\le1$ by the PSD minor inequality and maximality of $a$. Approximate $a$ by a positive rational and ${\mathsf c}$ by dyadics to enough additional $O(\log(2kn))$ bits. The product remains PSD and rank at most one and has the required error. This is a numerical cache; the mathematical model continues to use the true $C_i$.

Maintain $F$ symbolically as a combination of the initial $k$ columns. Each coefficient has absolute value at most two: it is $-\widetilde y_i^0$ plus the cached removal coefficient, if any. All coefficients use the same fine dyadic grid. Errors in $F$, ${D_{\rm pow}}$, the coupling, and its first two derivatives can consequently be bounded afresh by a fixed polynomial in $k,n,\delta_{\rm end}^{-1}$ times $u$. They do not accumulate with the number of local steps. No long original rational coefficient is used in subsequent solves.

For initialization, the cap and feasibility-repair argument in
Section~\ref{sec_fast_reset} controls coefficient errors before
recovering the true optimizer. For local tracking, short-cache evaluations
are inexact Newton steps for the true KKT system.
The forcing for a predictor is evaluated and solved to a sufficiently
small inverse-polynomial error, so its product with the mesh length is
below a fixed fraction of the Newton margin. The finitely many derivative
formulas involved have polynomial perturbation bounds.

The initial coordinate rounding costs only the ${\varepsilon_{\rm round}}/2$ allowance already reserved in the initialization argument. It is what permits the nonlinear phase to forget the original partial-signing denominators.

\textbf{Accuracy margins and tracking error.} 
The weights satisfy $\varpi_i\ge\gamma/4$, and all derivative bounds and
matrix condition numbers are polynomial in $k,n$ for the fixed cubic predictor.
The gradient and curvature tests in Lemma~\ref{fast_lem_dichotomy}
and the Newton radius therefore have inverse-polynomial margins.
The accepted decreases are bounded separately in each application. Choose the cache, derivative,
jet, and coordinate-grid accuracies below a fixed fraction of these margins.
The potential and KKT perturbation bounds validate comparisons for the
exact model. Inexact Newton corrections contract errors to their specified
floor at every stage, including initialization.

For endpoint scores use error at most $\delta$, and trigger if an
approximation to $\iota\beta_i v_i-(1-y_i)$ or $\iota\beta_i u_i-(1+y_i)$ is at
least $-\delta$. If neither triggers, the exact state is light.
A triggered move needs a feasibility repair at most $2\delta\|C_i\|$;
choose $\delta$ smaller than a fixed fraction of
$\min\{\gamma\delta_{\rm end}^{97/200},\rho\delta_{\rm end}^3\}$.
The reservoir pays for the repair, and the regular endpoint decrease
in Lemma~\ref{fast_lem_regular_resets} retains a fixed fraction of its margin.
Local trial coordinates are evaluated as $y+tr$ from the original dyadic
base, without accumulating intermediate coordinate errors. At acceptance,
round once and correct the optimizer again. Fixed-order powers and budget
derivatives have polylogarithmic evaluation cost: the fixed algebraic
power $(1-y^2)^{97/200}$ is computed by rational bisection.

\textbf{Deferred rounding.}
For the recorded matrices, choose the fixed cache exponent large
enough that $2ku\le\varepsilon_{\rm round}/64$ and
$2ku+ku^2\le1$. The cache then satisfies all approximation
hypotheses used in the proof of Lemma~\ref{round_lem_round}.
Its heavy/small split and nearest-sign decisions are exact
comparisons of short rational numbers. The recorded dyadic
coefficients remain exact throughout this final rounding.

For the small matrices, the exact signed prefix and the two
remaining logarithmic moment sums give Hermitian exponential
arguments of norm at most $M:=4L_{\rm log}+4$ in every candidate.
Nonzero cache traces are bounded below by $2^{-O(\log(2kn))}$.
The scalar exponentials and logarithms in
Lemma~\ref{round_lem_round} therefore need only
$O(\log(2kn))$ accuracy bits, including the division by a nonzero
trace. Their convergent series provide certified truncation tails.
For a matrix argument $H$, evaluate its exponential by the factorial
recurrence $T_0:=I$, $T_j:=T_{j-1}H/j$, summing through
$J=O(\log(2kn))$. The exact term and partial-sum norms are at most
$e^M$, which is polynomial in $n$. An additive recurrence error at
order $i$ is propagated by factors of norm at most
$M^{j-i}i!/j!\le e^M$. Thus ordinary rounded multiplication at
inverse-polynomial entry accuracy, together with the Taylor tail,
certifies the trace to additive error $1/(64\max\{1,k\})$.

Maintain the prefix and remaining sums on a common dyadic grid.
There are at most $k$ additions or subtractions to each sum, so their
errors are at most $k$ times the local entry error, with the matrix
dimension factor included. The trace-exponential Lipschitz bound
$n e^{M+1}$ converts these argument errors into a polynomially
bounded evaluation error. Taking a sufficiently large fixed
multiple of $\log(2kn)$ working bits keeps the full two-trace
estimator error below $1/(8\max\{1,k\})$ at every decision.
Choose the smaller computed estimator. Lemma~\ref{round_lem_round}
then bounds the accumulated comparison increase by $1/4$ and the
discrepancy, including cache error, by
$21\varepsilon_{\rm round}/128<\varepsilon_{\rm round}/2$.
Each of the at most $k$ decisions uses a logarithmic number of
order-$n$ ordinary matrix products. Short scalar work and the
$O(kn^2)$ matrix-sum updates are smaller, proving the
$\widetilde O(kn^3)$ bit bound.

\textbf{Small-trace rounding.}
Use the dyadic PSD cache with the accuracy required in
Lemma~\ref{trace_lem_small_rounding}. Its conditional choice compares
$\tr[E\widehat C_i]-y_i\tr[\widehat C_i]^2$ with zero.
The coefficients and cache entries share dyadic denominators of
$O(\log(2kn))$ bits. Each prefix entry is a sum of at most $k$
products of such numbers, and each comparison is a sum of at most
$n^2$ further products. Their exact numerators and denominators
therefore have $O(\log(2kn))$ bits, with a fixed multiplier.
Each comparison and prefix update uses $O(n^2)$ short arithmetic
operations, and there are at most $k$ of them. This proves the
$\widetilde O(kn^2)$ bit bound. Since every comparison is exact,
the conditional-expectation proof and the cache-error transfer of
Lemma~\ref{trace_lem_small_rounding} apply without an additional loss.
Summing over the geometrically decreasing phase counts gives
$\widetilde O(n^4)$ bit operations for all small-trace rounding.
\end{proof}

\subsection{Stable matrix products and inverse solves}\label{bit_sec_linear_algebra}

The arithmetic constructions use products and inverses whose norms and condition numbers have already been bounded. These bounds determine the required guard precision.

\begin{lemma}[Stable products and residual-certified inverses]\label{bit_lem_linear_algebra}
Under the polynomial magnitude and inverse bounds in Sections~\ref{sec_fast_reset} and~\ref{sec_fast_arithmetic}, the required matrix products and inverse solves attain their prescribed inverse-polynomial errors with $O(\log(2kn))$ working bits. An order-$d_{\rm sys}$ product or inverse costs $\widetilde O(d_{\rm sys}^{\omega_0})$ bit operations; in particular, $d_{\rm sys}=O(n^2)$ gives cost $\widetilde O(n^{2\omega_0})$. For the structured rank-one actions of Lemma~\ref{fast_lem_operator_cost},
the coupling, fixed-order directional derivatives, endpoint scores,
gradient coordinates, and weights $\varpi_i$ attain their prescribed
errors in $\widetilde O(n^\chi)$ bit operations with the same
logarithmic guard precision.
Given the dyadic coefficient and PSD-cache data of
Lemma~\ref{bit_lem_cache}, the fast deterministic deferred rounding
of Lemma~\ref{sr_lem_round} costs
$\widetilde O(kn^{\omega_0})$ bit operations. The capped randomized
rounding of Lemma~\ref{oe_lem_round} has expected bit cost
$\widetilde O(kN^2+N^{\omega_0})$. Both retain discrepancy less than
$\varepsilon_{\rm round}/2$, and the randomized routine terminates
correctly on every random-bit sequence.
\end{lemma}
\begin{proof}
For the inverse, use $A,\mathsf Z_r,E_r,L$ and the exact residual recurrence from Section~\ref{sec_fast_arithmetic}.
Round each matrix product and iterate to the common grid. If its induced residual error is at most $\delta$, then

\[
\|E_{{r_{\rm aux}}+1}\|\le\|E_{r_{\rm aux}}\|^2+\delta.
\]

Take $\delta$ smaller than a fixed fraction of both the square of the initial spectral gap and the target residual. The same logarithmic iteration bound holds: while the residual is near one, its gap from one grows by a constant factor; below a fixed constant, the usual squaring convergence applies down to the error floor. Throughout, $\|\mathsf Z_{r_{\rm aux}}\|\le\|A^{-1}\|(1+\|E_{r_{\rm aux}}\|)$ is polynomial. Hence $O(\log(2kn))$ bits per entry suffice for all products. A direct residual computation bounds the final inverse error by $\|A^{-1}\|\|E_{r_{\rm aux}}\|$.

For order ${d_{\rm sys}}=O(n^2)$, implement the products with the fast multiplication routine below. Inversion and multiplication by the right-hand sides then cost $\widetilde O(n^{2\omega_0})$ arithmetic and bit operations. The same procedure handles the order-$n$ primal inverses. The residual and precision arguments above are unchanged. Approximate linear solves and KKT evaluations give the additive error asserted after Eq.~\eqref{rank_one_eq_cont_18}.

\textbf{Fast products.} 

For the exponent $\omega_0$ in Eq.~\eqref{fast_eq_multiplication_parameters},
stable recursive multiplication with exponent $\omega_0$
is available by Demmel, Dumitriu, Holtz and
Kleinberg~\cite{ddhk07}.
We may choose $\omega_0:=2.371177$ using the strict bound of
Dupont et al.~\cite{dekmrszawb26}.
The coarser choice $\omega_0:=5/2$ already follows from
Coppersmith and Winograd~\cite{cw90}.
For polynomially bounded matrices and $p=O(\log(2kn))$ requested
accuracy bits, their polynomial normwise error bounds require only
$O(p+\log(2kn))$ working bits. Thus an order-${d_{\rm sys}}$ product costs
$\widetilde O({d_{\rm sys}}^{\omega_0})$ bit operations as well as arithmetic operations.

\textbf{Rank-one actions.} 
All factors have polynomial magnitude, including the fixed-order jets.
Choose fixed bilinear recursive algorithms at the strict exponent
bounds in Eq.~\eqref{fast_eq_multiplication_parameters}. Each has
logarithmic recursion depth and fixed coefficients at each level, so
intermediate magnitudes and normwise error amplification are bounded
by a fixed polynomial. Tensoring or permuting these fixed algorithms
preserves that property. A fixed multiple of the logarithmic working precision therefore attains the prescribed error.
This proves the bit bound.
\textbf{Fast deferred rounding.}
In the proof of Lemma~\ref{bit_lem_cache}, replace ordinary matrix
multiplication in the factorial exponential recurrence by the fixed
fast products just proved. Each argument has norm $O(\log(2n))$;
the recurrence terms, intermediate norms, and error amplification
are polynomial. The same logarithmic working precision and
comparison margins therefore suffice for
Lemma~\ref{sr_lem_round}. There are $O(\log(2kn))$ products per
candidate and $O(k)$ candidates. Prefix and remaining-sum updates
cost $\widetilde O(kn^2)$ bit operations and preserve the same
conditional-estimator errors. This proves the fast deterministic
bit bound with its original discrepancy allowance.

For Lemma~\ref{oe_lem_round}, the recorded coefficients have
$O(\log(2kN))$ dyadic bits, so the Bernoulli draws with probabilities
$(1+t_i)/2$ are exact with that many uniform bits. Each attempt
forms its cached residual by $O(kN^2)$ common-denominator operations
on logarithmic-length entries. Lemma~\ref{bit_lem_group_solve}
implements the norm certificate in $\widetilde O(N^{\omega_0})$
bit operations, including on a failed sample. Its acceptance and
completeness thresholds are precisely those used in the arithmetic
probability proof. The failure cap has probability at most
$(2N)^{-30}$; the fast deterministic fallback costs
$\widetilde O(kN^{\omega_0})$. The same expectation calculation
therefore proves the randomized bit bound, finite termination,
and the strict discrepancy allowance.
\end{proof}

The initializer must preserve feasibility and the Newton decrement at every accepted stage. We apply the same residual-certified linear algebra to its capped barrier path.

\begin{lemma}[Finite-precision capped initialization]\label{bit_lem_initialization}
Assume the bounds on inverse slacks, Hessian conditioning, and centers in Section~\ref{sec_fast_reset}. Its initializer attains the requested objective accuracy with $O(\log(2nk/\varepsilon_{\rm obj}))$ working bits. It preserves the feasibility and decrement margins within the same stage count. Its bit cost is the corresponding arithmetic cost up to logarithmic factors.
\end{lemma}
\begin{proof}
Use the inverse-slack bounds and barrier from Section~\ref{sec_fast_reset}, and Lemma~\ref{bit_lem_linear_algebra} for the numerical solves.
The Hessian is a sum of products of these inverses and coefficient maps, so its norm is polynomial. Its inverse norm is polynomial by Eq.~\eqref{fast_eq_globalbarrierlower}. All coordinates, gradients, Newton directions, and objective parameters have polynomial magnitude. Consequently $O(\log(2nk/{\varepsilon_{\rm obj}}))$ accuracy and magnitude bits suffice to preserve feasibility and the decrement slack at every stage. Rounded Newton corrections need not accumulate errors: the decrement invariant is re-established after each step.

This also gives a constructive implementation: compute inverse slacks and the initial Newton-system inverse by short-precision Newton--Schulz iteration, with residual-based error bounds and safe polynomial norm/condition bounds. Recompute the Newton-system inverse at each subsequent stage. Stable fast multiplication supplies the required inverse-polynomial product accuracy with logarithmic guard precision. Determinants need not be evaluated by the algorithm; the barrier gradient and Hessian use only inverses and traces. Approximate ${b_{\rm obj}}_0$ and every oracle to a sufficiently small inverse-polynomial floor. The exact-center and Dikin arguments in Section~\ref{sec_fast_reset} ensure that the prescribed floor works at every stage, without assuming a priori precision at unvisited centers.
\end{proof}

\subsection{Smoothed root evaluations and matrix functions}\label{soft_sec_precision}
At a corrected smoothed root, residual bounds suffice to evaluate the matrix functions used by the algorithm. The spectral implementation below uses their established spectral ranges.

\begin{lemma}[Matrix functions and corrected smoothed states]\label{bit_lem_smoothed}
Assume the geometric bounds of Section~\ref{soft_sec_geometry} and the state invariant of Section~\ref{soft_sec_certified_states}. The required order-$n$ matrix logarithms, square roots and their inverses, Lyapunov solves, and entropy covariance actions attain the prescribed errors with $O(\log(2kn))$ working bits. An order-$n$ spectral decomposition, certified by its residual, costs $\widetilde O(n^3)$ bit operations. If inverse and structured actions meet their error bounds, rounded Newton corrections preserve the state invariant and provide the required value, endpoint, and residual enclosures.
\end{lemma}
\begin{proof}
The input conversion and grouped partial signing use their proved
accuracy and additive ${L_{\rm in}}$ accounting. After that phase, the
true model uses its retained matrices; numerical computations use PSD
rank-one approximations. The fixed contribution is stored symbolically
as a combination of the initial $k$ matrices with coefficients of
absolute value at most two. Its matrix error, all coupling errors, and
their derivative errors are bounded afresh by a fixed polynomial times
the chosen cache error, independently of the number of steps or resets.

Use $p=C\log(2kn)$ accuracy bits in the nonlinear cache for a
sufficiently large fixed constant $C$. Choose $C$ to cover the prescribed inverse-polynomial action errors and the scalar and matrix-product guard bits. The input
conversion is not rerun on the original ${L_{\rm in}}$-bit data at each
stage. Retained normalized rational inputs can be approximated to the
new accuracy in the same one-time cache construction; magnitudes and
nonzero scale lower bounds are polynomial in the input parameters.
A common trace/variance scalar is also approximated only once.

Lemma~\ref{lem:shared:smoothed-evaluation} supplies the residual-based
spectral construction, matrix-function perturbation bounds, and
corrected-state invariant in the arithmetic model. Its Jacobi routine
uses $\widetilde O(n^2)$ rotations and $\widetilde O(n)$ work per rotation.
All magnitudes and error amplification factors are polynomial in $n,k$.
The norm-one rotations accumulate similarity and orthogonality errors
only polynomially. Thus the chosen $O(\log(2kn))$ accuracy and guard bits
preserve the spectral residual tests and give
$\widetilde O(n^3)$ bit operations per decomposition.
The same accuracy covers scalar powers, logarithmic means, shifted
exponentials, and the rational real/imaginary Hermitian coordinates.

Rounded anchored refinement satisfies the contraction-plus-additive-error
recurrence proved in Lemma~\ref{lem:shared:smoothed-evaluation}.
At a curvature filter, use its explicit second-kind recurrence bound
instead. Neither estimate multiplies per-stage condition numbers over
the signing trajectory. Choosing the cache floor below fixed fractions
of the root, endpoint, and potential-decrease margins preserves all
comparisons, including the true-versus-cached residual errors. The
one-time input conversion is the only step charged to $L_{\rm in}$.
\end{proof}

\subsection{Deterministic conditional choices and response blocks}\label{bit_sec_deterministic}

The deterministic choices use conditional estimators with polynomially bounded values and derivatives. This supplies a common finite-precision implementation for the scalar and grouped choices.

\begin{lemma}[Certified deterministic conditional choices]\label{bit_lem_conditional}
The scalar conditional choices in Section~\ref{sec_trace_algorithm} take $\widetilde O(k^2)$ bit operations. Certified evaluation within their comparison margins uses $O(\log(2kn))$ working bits. For the grouped algorithm of Lemma~\ref{trace_lem_deterministic_groups},
assume the input entries share a common rational denominator and
their numerators and denominator have $O(\log(2mD/\eta))$ bits.
It returns at most $2D$ fractional coordinates and residual at most
$\eta$ using $\widetilde O(mD+D^{\omega_0})$ bit operations
and $O(\log(2mD/\eta))$ working bits per entry, including temporary
operands. In both cases the accepted choices preserve the conditional-estimator guarantee, including when the two candidate values nearly tie.
\end{lemma}
\begin{proof}
Use the filter $F$, coefficients $\lambda_f,\pi_n$, and conditional estimator from Section~\ref{sec_trace_algorithm}.
For the scalar choices,
if $v$ is the sum of the fixed columns with their chosen signs and
$J$ indexes the remaining columns, then
\[
 \E[\|F\xi\|^2\mid v]=\|v\|^2+\sum_{j\in J}\|F_{:j}\|^2,
 \quad
 \E[\cosh((F\xi)_i)\mid v]
   =\cosh(v_i)\prod_{j\in J}\cosh(F_{ij}).
\]
Maintain $v$ and these products; each candidate decision requires
$O(k)$ scalar operations. There is no large-exponent precision cost:
the maintained conditional value of $\Phi$ is positive, while the
conditional quadratic expectation is at most $4k$. Thus
$\cosh(v_i)\le4k/\lambda_f$. A trial adds at most two to $|v_i|$,
and the remaining products are between one and eight.
All numbers and derivatives used in a comparison are therefore
polynomially bounded. Certified scalar Taylor approximations with
logarithmic precision suffice, including when the two choices nearly tie.

\textbf{Grouped deterministic directions.} 
In Lemma~\ref{trace_lem_deterministic_groups}, the conditional estimator uses the short matrix $V$.
At every chosen
prefix their value stays positive, whereas the conditional quadratic
term is at most $\|V\|^2r<2r$. Hence each conditional hyperbolic
cosine is at most $200r$. At either next candidate it increases by
at most a fixed factor, since $|V_{ij}|\le65/64$. The remaining
products are bounded by $e^{(65/64)^2/2}<2$. All arguments are
$O(\log r)$ and all intermediate magnitudes are polynomial.
Taylor evaluation and outward error bounds consequently need only
$O(\log(2D/\varepsilon))$ bits, with the inherited short-data factors.
The divisions removing a factor $\cosh(V_{ij})\ge1$ are stable.

The small reduction in Lemma~\ref{trace_lem_deterministic_groups}
has $\delta I\preceq J_\delta\preceq(\delta+4r_0)I$ and uses the
same ridge and direction tolerances as Lemma~\ref{bit_lem_ridge_direction}.
That lemma computes a residual-certified inverse with
$O(p_C+p_y+\log(2D/\varepsilon)+\log(2T_*))$ working bits.
Form and symmetrize $I-\mathsf B^\top R\mathsf B$, then round it
to the direction grid. Polynomially smaller intermediate errors
ensure $\|V-\Pi_\delta\|\le\nu/\sqrt r$, including this final
rounding, in $\widetilde O(D^{\omega_0})$ bit operations.
The conditional comparisons proved here cost only
$\widetilde O(D^2)$ per direction. The vector $V\xi$ and its norm
and maximum-entry tests are computed exactly from its dyadic entries.

For the derivative sign, each summand
$2x_i\widetilde v_i/[H(1+|x_i|)]$ has bounded magnitude and
denominator $1+|x_i|\in[1,2)$. Approximate each summand to error
$1/(512H^2)$ on one common dyadic grid and sum exactly. This gives
$|\widehat g-g|\le r/(512H^2)$ in $\widetilde O(r)$ bit operations.
Choosing its sign is the permitted inexact rule in
Lemma~\ref{trace_lem_deterministic_groups}. Compute the rational
step from the short dyadic data, round toward zero, and freeze
endpoints by exact comparisons. These operations use a fixed
multiple of the same working precision. Thus the deterministic
progress, cube, and residual estimates of that lemma hold in fewer
than $T_*$ steps, and a complete small call costs
$\widetilde O(D^{\omega_0})$ bit operations.

For the full tree, exact node sums retain the common denominator;
adding at most $m$ numerators adds only $O(\log(2m))$ bits.
Division by $M_{\rm grp}$ adds the same order of denominator bits.
The common-grid argument in Lemma~\ref{trace_lem_deterministic_groups}
and $\varepsilon=\eta/(M_{\rm grp}R_{\rm tree})$ give
$p_C+p_y+\log(2D/\varepsilon)+\log(2T_*)=O(\log(2mD/\eta))$
uniformly at every level. Node-sum construction and the final leaf
traversal cost $\widetilde O(mD)$ bit operations. The logarithmically
many small calls give the claimed total, with exactly the residual
and fractional-count guarantees of the arithmetic algorithm.
For the matrix application, Lemma~\ref{bit_lem_input} charges the
original binary input length once in the input conversion.
\end{proof}

All response columns at one corrected root use the same spectral operators. Their simultaneous approximation can therefore be controlled in Frobenius norm.

\begin{lemma}[Precision for block responses]\label{bit_lem_responses}
Under the hypotheses of Lemma~\ref{trace_lem_block_responses}, use
the logarithmic-length cache and corrected-root data of
Lemmas~\ref{bit_lem_cache} and~\ref{bit_lem_smoothed}.
The Hessian and response blocks and their filter compositions attain
the prescribed inverse-polynomial operator tolerances with
$O(\log(2kn))$ working bits per entry, including temporary operands.
The matrix conditional comparisons attain their reserved additive
accuracies and return a direction satisfying
Eq.~\eqref{eq_trace_matrix_direction}.
The total bit cost is the arithmetic bound in
Lemma~\ref{trace_lem_block_responses}, up to logarithmic factors.
\end{lemma}
\begin{proof}
The spectral constructions of Lemma~\ref{bit_lem_smoothed} apply
to the polynomial spectral bounds in
Lemma~\ref{trace_lem_block_responses}. They compute all order-$n$
decompositions in $\widetilde O(n^3)$ bit operations.
Use the same approximate spectral operators for all columns. The energy preconditioning bounds give polynomial
similarity condition numbers. The normalized Chebyshev recurrence has polynomial error amplification: the scalar propagation ratios are at most one, and a second-kind polynomial of degree $j$ on the preconditioned spectral interval has norm at most a polynomial similarity factor times $j+1$. The same bound applies to a matrix of right-hand sides in Frobenius norm. Demand a polynomially smaller
per-column error, so the sum of squared column errors is below the
prescribed squared operator tolerance. This changes only logarithmic
precision and iteration factors. No eigengap is needed, and no
order-$n^2$ dense inverse is formed.

All block products and relative congruences use
Lemma~\ref{bit_lem_linear_algebra}; splitting the long dimension
into $k$-sized blocks preserves the arithmetic counts in
Lemma~\ref{trace_lem_block_responses}. Summing their errors over
polynomially many blocks changes only the fixed guard factor.

For the deterministic filter, use its finite resolvent expansion
from Section~\ref{sec_trace_algorithm}. Each shifted inverse is
polynomially conditioned, every resolvent power is contractive,
and the sums of absolute scalar coefficient terms are polynomial.
Lemma~\ref{bit_lem_linear_algebra} and common-grid scalar evaluation
therefore give inverse-polynomial filter error with
$O(\log(2kn))$ working bits in $\widetilde O(k^{\omega_0})$ bit
operations. Tighten this error by the polynomial norms of the
response maps before composition; the stated operator tolerances
and filter guards are then preserved.

Forming the variance matrices costs
$\widetilde O(kn^{\omega_0})$ bit operations. At an accepted
conditional prefix and either next candidate, all matrix
exponential arguments have norm $O(\log n)$.
The factorial recurrence and additive-error accounting in
Lemma~\ref{bit_lem_cache} apply, with fast products from
Lemma~\ref{bit_lem_linear_algebra} in place of ordinary products.
The polynomial estimator weights require only a fixed increase of
the logarithmic precision. Each candidate is enclosed to a small
fixed fraction of its per-choice allowance, so at most $k$
comparisons consume the reserved total slack, even in a near tie.
The scalar terms use Lemma~\ref{bit_lem_conditional}.
The matrix choices cost $\widetilde O(kn^{\omega_0})$ bit operations
and their scalar terms cost $\widetilde O(k^2)$.
Together with the spectral setup and response solves, these costs
give the bound in Lemma~\ref{trace_lem_block_responses} in the bit model.
\end{proof}

\subsection{Finite sampling}\label{bit_sec_sampling}
We realize the feature distribution using finitely many fair bits. A coupling to the ideal samples quantifies the effect on the feature covariance.

\begin{lemma}[Finite feature sampling]\label{bit_lem_sampling}
Under the hypotheses and sample count $s$ of Lemma~\ref{gf_lem_features}, its feature sampler has a finite implementation using dyadic uniforms with $b=O(\log(2kn))$ bits, with a sufficiently large fixed multiplier. It can be coupled to the ideal sampler so that, except on an event of probability at most $1/100$, the feature covariance changes by operator norm at most $\mu/16$. On the intersection with the ideal feature event, the represented features satisfy the two-sided bound with slack $1/2,2$. Every finite sample has polynomially bounded magnitude, including samples outside that event. The sampler and its matrix-function evaluations use logarithmic working precision.
The restricted feature actions, sample preparation, and terminal
factor, Gram, and inverse construction in Lemma~\ref{rr_lem_actions}
have their stated arithmetic costs also in bit operations, up to
logarithmic factors, with the same fixed represented feature matrix.
\end{lemma}
\begin{proof}
For the finite sampler use midpoints of dyadic uniform intervals for
$t$ and the Box--Muller uniforms. Pair independent real normals into
circular complex normals. Couple these finite samples to ideal samples
using the same uniform intervals. Except on an event of probability
$1/100$, every radial uniform is at least $(100ns)^{-4}$ and every
normal is at most $C\sqrt{\log(2ns)}$. On this event all sampling errors
are a fixed polynomial times the scalar grid error. The eigenvalues of
${\mathsf Q}$ lie in an inverse-polynomial interval, and $t\in[0,1]$;
the Frobenius divided-difference bounds for ${\mathsf Q}^{t/2}$, as well as
its $t$ derivative, are polynomial uniformly in $t$. A gap-free order-$n$
diagonalization and certified scalar powers therefore approximate all
$r_b,v_b$ to any fixed inverse-polynomial accuracy, without identifying
individual eigenvectors at repeated eigenvalues. The normalization
constants can also be replaced by short dyadics at this accuracy.

Store these vectors and constants, and define the represented features
by the exact rank-two and centering formulas using the same short
${\mathsf Q},F_1,G$. The feature perturbation and hence the perturbation of
$VV^{\top}/s$ are as small as prescribed, since all factors have polynomial
magnitude. Choose them to give operator error at most $\mu/16$.
The finite version of the two-sided bound, with slack $1/2,2$, follows.
Every finite Box--Muller output is bounded by $C\sqrt b$ if $b$ bits are
used, even on the excluded event. Thus all represented features have
known polynomial magnitude for every sample. Taking $b$ a sufficiently
large fixed multiple of $\log(2kn)$, with the polylogarithmic factors
already displayed, suffices. There is no exact-real or unbounded
rejection-sampling oracle.

For the actions of Lemma~\ref{rr_lem_actions}, use the stable products
of Lemma~\ref{bit_lem_linear_algebra} with the same rectangular
shapes and their tensor permutations. The order-$n$ spectral data
are prepared once by Lemma~\ref{bit_lem_smoothed}. Fixed recursive
products have polynomial intermediate magnitudes and normwise error
amplification. Taking the scalar roundoff inverse polynomially
smaller attains every action tolerance at logarithmic precision.
The terminal factor and the implicit products use the same stored
rank-two factors and exact centering definition, including repeated
indices. Their errors are additive errors relative to that one
matrix. Its ridge supplies the positive lower bound for the
residual-certified terminal inverse. These operations preserve the
real/complex adjoint formulas and give all the bit costs stated
for restricted actions and terminal setup.
\end{proof}

\subsection{Uniform precision for nested solves}\label{bit_sec_nested}
The recursion in Lemma~\ref{ue_lem_precision} increases the number of calls. Residual correction allows every level to use the same accuracy target.

\begin{lemma}[Depth-independent precision for nested solves]\label{bit_lem_nested}
Use the parameters and frozen feature and index data of Lemma~\ref{ue_lem_precision}, with $2\le h\le\mathscr J^2$ and $\mathscr J=1+\log_2 N$. There are fixed constants $B_0,D_0$, independent of the depth and sketch sizes, with the following properties. On the simultaneous good sampling event, each completed recursive solve, with matrix $K$ and right-hand side $b$, returns $v$ satisfying
\[
 \|v-K^{-1}b\|\le\eta\|b\|,\qquad \eta:=N^{-B_0},
\]
and the shifted solves attain their prescribed errors. All numerical entries use at most $D_0\mathscr J$ accuracy and magnitude bits. The finite conditional index events have joint probability at least $99/100$ given the feature event, and the original verified curvature trial retains conditional success probability at least $1/24$. All samples obey prescribed iteration and magnitude caps; the arithmetic call recurrence in Lemma~\ref{ue_lem_precision} therefore also gives its bit bound with the stated logarithmic overhead.
\end{lemma}
\begin{proof}
Freeze one
short root, its consistently defined symmetric Hessian surrogate, all
finite feature factors, and all index lists. No action changes these
objects. The invariant in Eq.~\eqref{rr_eq_invariant} is algebraic and
holds for any depth. Store each coefficient from its closed formula
${\iota}_j=\bar k q_j/(\mu q_1d_je_j)$ rather than by accumulating rounded
products of previous coefficients. All sizes are integers and $\mu$ is
a fixed short rational, so these positive coefficients have uniformly
$O(\log N)$ encoding lengths. On every finite sample the feature entries have
magnitude at most $N^{C_0}$, for a fixed $C_0$, since $h\le {\mathscr J}^2$ makes
the moment order and the finite-sampling word lengths $O({\mathscr J})$.
Consequently, at every level,
\[
 1\le\lambda_{\min}(K_j),\qquad
 \|K_j\|\le1+{\iota}_jd_je_j\max_{a,b}|V_{ab}|^2\le N^{C_1}.
\]
Similarly $\lambda_{\min}(W_j)\ge\lambda_j\ge1$ and
$\|W_j\|\le N^{C_1}$. The factors in their Woodbury formulas, their
positive scales and inverse scales, and all individual operator norms
are bounded by $N^{C_2}$. These exponents do not grow with depth:
${\iota}_jd_je_j\le2q_j/d_0$ telescopes exactly, rather than accumulating a
new condition factor at every level. Sample-list sizes are at most
$N^2{\mathscr J}^{C_3}$, with a fixed exponent; indices may be composed once into
lists of original coordinates and sample columns.

The proofs of the simultaneous entry and covariance event and of uniform
ridge sampling now use $p=\lceil128\log_2(2kN\ell_{\mathsf Q}(h+1))\rceil$.
For $h\le {\mathscr J}^2$ this is still $O({\mathscr J})$. The same moment and finite-grid
bounds dominate the polynomial factors in all dimensions. Each
conditional index-sampling failure is at most $1/(1000(h+1))$;
unioning them and the finite rejection-cap failure leaves probability
at least $99/100$, conditionally on the feature event. Thus these
probability bounds are uniform in the present range of depths.

On the simultaneous good event the preconditioned systems are similar
to symmetric positive matrices in their prescribed slack intervals,
through transformations whose condition numbers are at most $N^{C_4}$.
In Eq.~\eqref{ue_eq_error}, $|T_j|\le1$ on $[-1,1]$ and
$T_j(\varrho)\ge1$. Therefore each exact normalized iterate has norm at
most $N^{C_5}\|b\|$, independently of the requested accuracy and the
iteration count. Normalize every right-hand side by a dyadic scalar
before a recursive call, then restore that scale. This avoids propagating
large right-hand sides as unbounded intermediate integers.

Here is an explicit error bound for the normalized recurrence. A vector
update error $e_i$ contributes at a later index $J$ the term
\[
 \frac{T_i(\varrho)}{T_J(\varrho)}{\mathsf A}_{J-i}(D)e_i
\]
(up to the first-step convention). The scalar ratio is at most one and
$\|{\mathsf A}_{J-i}(D)\|\le N^{C_4}(J-i+1)$. Thus the total vector
error is at most $N^{C_4}(J+1)^2\max_i\|e_i\|$.
Errors in the clipped scalar ratios accumulate at most linearly,
since their update is nonexpansive on $[0,1]$. Their contribution,
and the contributions of action and coefficient errors, are bounded
by a fixed polynomial in $N,J$ times the scalar unit and the
right-hand-side norm.

We use residual correction to keep the accuracy request independent
of depth. Fix $\eta:=N^{-B_0}$, with a sufficiently large constant
$B_0$ chosen below. Every completed recursive solve is required to
return $v$ with $\|v-K^{-1}b\|\le\eta\|b\|$.
Normalize each right-hand side by a dyadic scalar within its own call;
restore that scalar when returning, without multiplying scales along
the recursion tree. A zero right-hand side returns zero.

First construct a coarse solve $S_j(b)$ at each level. Run the normalized
Chebyshev recurrence for $O(\sqrt{\beta/\alpha}\log N)$ steps, using
completed child solves to accuracy $\eta$ relative to their right-hand
sides. On the good event, all condition, similarity, Woodbury, and
recurrence factors are bounded by $N^{C_8}$ for a fixed $C_8$ independent
of $h$ and the level. The preceding error convolution therefore bounds
the effect of child errors by $N^{C_9}\eta\|b\|$, for one fixed $C_9$.
Local arithmetic errors have the same bound with the scalar unit $u$
in place of $\eta$. The exact recurrence error can be made smaller
than $N^{-C_1-3}\|b\|$ by its fixed logarithmic multiplier.
Choose $B_0>C_9+C_1+10$ and $u$ sufficiently smaller than $\eta$.
Since $\|K^{-1}b\|\ge N^{-C_1}\|b\|$, these choices give
\[
 \|S_j(b)-K^{-1}b\|\le\tfrac14\|K^{-1}b\|.
\]
This assertion needs only the same completed-child accuracy $\eta$,
not a more accurate child when the parent requests a smaller residual.

Starting at $v_0:=0$, compute fresh residuals and corrections,
\[
 r_\ell:=b-Kv_\ell,\qquad
 v_{\ell+1}:=v_\ell+S_j(r_\ell),
\]
rounding each operation. Exact residuals would contract solution error
by a factor at most $1/4$. With residual and arithmetic errors, all
iterates stay polynomially bounded and
\[
 \|v_{\ell+1}-K^{-1}b\|
 \le\tfrac14\|v_\ell-K^{-1}b\|+N^{C_{10}}u\|b\|,
\]
where $C_{10}$ is uniform in depth. The same estimates follow
inductively for rounded iterates. Choose
$u\le\eta N^{-C_{10}-C_1-10}$ and a fixed $O(\log N)$ correction cap.
Then the solution error is below $\eta\|b\|/(4N^{C_1})$;
a direct residual enclosure certifies the requested
$\eta\|b\|$ solution error, using $K\succeq I$.
Stopping and scaling tests use the same polynomial margins.
The terminal inverse supplies the induction base. Induction from the
leaf now proves the same output accuracy at every level, with
$O(\log N)$ accuracy and magnitude bits throughout. The number of
levels changes the number of calls but not their accuracy exponents.
The shifted system is handled identically using $B\succeq RI$.

The terminal Gram and all materialized actions have inverse-polynomial
error, and the explicit ridge and Gram forms supply the positivity
bounds. No intermediate spectral decomposition is charged implicitly.
Root and feature matrix functions approximate the same frozen data;
their logarithmic precision is included in $u$. The bounded outer
filter requires only inverse-polynomial action errors by its array
contraction bound. Choose $B_0$ also large enough for these final norm
and quotient margins. Structured rational real/imaginary coordinates
and their fixed Gram weights add only polynomial condition factors.

Prescribe the common logarithmic precision, all iteration lengths,
and a polynomial magnitude cap $N^{C_{11}}$ in advance. Normalize
recursive right-hand sides as above and reject on any violated cap,
required residual or positivity enclosure, or failed final quotient.
Even off the sampling event, every operation and loop has these caps.
On the good event the induction proves that no accuracy test removes
the successful independent sign-overlap event. The conditional success
probability remains at least $1/24$.
\end{proof}

\subsection{Certified spectral filters}\label{bit_sec_filters}

The normalized-squaring witness is evaluated on symmetric matrices at every step. Its error can be bounded by propagating each local perturbation through the remaining exact powers.

\begin{lemma}[Stable normalized squaring]\label{bit_lem_squaring}
Use the normalized-squaring sequence of Section~\ref{sec_fast_arithmetic}, with $q=2^r$ and matrices of order $k$; write $D_j$ and $\widehat D_j$ for the exact and computed iterates. If the initial error and every local normalized-squaring error are at most $\eta<1/2$, with symmetrization after each product, then
\[
 \|\widehat D_r-D_r\|_F\le8q\sqrt{k}\eta.
\]
For the prescribed polynomially bounded $q$, inverse-polynomial local errors preserve a computed Rayleigh quotient at most $-\delta/4$ and certify negative curvature for the true potential. The complete witness computation uses $O(\log(2kn))$ working bits and $\widetilde O(n^{2\omega_0})$ bit operations.
\end{lemma}
\begin{proof}
For a nonzero symmetric
matrix $X$ and an integer $s\ge1$, define
$\mathcal N_s(X):=X^s/\|X^s\|_F$. The product rule gives
$\|\mathrm D(X^s)[E]\|_F\le s\|X\|^{s-1}\|E\|_F$ for symmetric $E$.
Since $\|X^s\|_F\ge\|X\|^s$, differentiating the normalization gives
\[
 \|\mathrm D\mathcal N_s(X)[E]\|_F
 \le\frac{s}{\|X\|}\|E\|_F
 \le\frac{s\sqrt{k}}{\|X\|_F}\|E\|_F.
\]
Consequently, on a segment within Frobenius distance $1/2$ of a
symmetric matrix of unit Frobenius norm, $\mathcal N_s$ is
$2s\sqrt{k}$-Lipschitz. Also,
$\mathcal N_2^{\circ j}=\mathcal N_{2^j}$ for $j\ge1$.

Symmetrize after each product. Let $\widehat D_j$ be the computed iterates,
with initial error and local errors bounded by $\eta<1/2$:
\[
 \|\widehat D_0-{D_{\rm pow}}_0\|_F\le\eta,
 \qquad
 \|\widehat D_{j+1}-\mathcal N_2(\widehat D_j)\|_F\le\eta.
\]
Each computed iterate has Frobenius norm between $1-\eta$ and $1+\eta$.
Insert these errors one at a time and propagate each through the remaining
exact normalized squarings. The remaining powers are $q,q/2,\ldots,2$;
the final local error is unchanged. The preceding Lipschitz bound and
this geometric sum give
\[
 \|\widehat D_{r_{\rm aux}}-{D_{\rm pow}}_{r_{\rm aux}}\|_F
 \le8q\sqrt{k}\eta.
\]
Choose an inverse-polynomial accuracy $0<\epsilon\le\delta/(100L)$ and
define $\eta:=\epsilon/(8q\sqrt{k})$. Since $q$ is polynomially bounded
in $k,n$, this local accuracy is inverse polynomial as well. Moreover,
$\|\widehat D_j^2\|_F\ge(1-\eta)^2/\sqrt{k}\ge1/(4\sqrt{k})$.
Thus all normalizations are polynomially conditioned. The stable fast
products above, Frobenius norms, and scalar square roots achieve the
required local errors with $O(\log(2kn))$ working bits per entry and
polylogarithmic overhead. The final trace changes by at most
$L(2+\epsilon)\epsilon$, leaving a fixed fraction of its negative margin.
Compute column norms and quadratic forms with conservative margins,
discarding only columns below comparable inverse-polynomial thresholds.
The trace and retained-column bounds then still supply a witness with
Rayleigh quotient at most $-\delta/4$. Finally normalize that column
and check its Rayleigh quotient against the computed Hessian directly.
The Hessian approximation margin transfers this negative curvature to
the true potential. The complete filter has arithmetic and bit cost
$\widetilde O(n^{2\omega_0})$.
\end{proof}

\paragraph{Variance scaling.}
The trace-power enclosure in Section~\ref{sec:fast:variance-scaling}
also has a short-precision implementation. Write
$\ell_{\rm pow}=2^r$ and $M_2=\sum_i\bar A_i^2$ after trace
normalization, so $1/(mn)\le\|M_2\|\le1$. Form $M_2$ using the
rank-one formula and a common dyadic accumulator. For the powers,
use the exact reference quantities
\[
 \begin{aligned}
 D_0&:=M_2/\|M_2\|_F,& S_0&:=\|M_2\|_F,\\
 D_{j+1}&:=D_j^2/\|D_j^2\|_F,&
 S_{j+1}&:=S_j\|D_j^2\|_F^{1/2^{j+1}}.
 \end{aligned}
\]
Then $D_j=M_2^{2^j}/\|M_2^{2^j}\|_F$ and
$S_j=\|M_2^{2^j}\|_F^{1/2^j}$, whence
\[
 \tr[M_2^{\ell_{\rm pow}}]^{1/\ell_{\rm pow}}
 =S_r(\tr[D_r])^{1/\ell_{\rm pow}}.
\]
All reference matrices are PSD. Hence
$\|D_j^2\|_F\ge1/\sqrt n$, $1\le\tr[D_r]\le\sqrt n$, and
$1/(mn)\le S_j\le1$. The normalized-power error estimate just proved,
which also holds for Hermitian matrices, applies with power at most
$\ell_{\rm pow}$ and bounds matrix error amplification
by a polynomial in $m,n,\ell_{\rm pow}$. The scalar roots and the
$S_j$ updates have polynomial perturbation bounds on these ranges.
Bisection with outward error bounds evaluates these roots; its
power comparisons use rounded repeated squaring. There are only
$O(\log\ell_{\rm pow})$ updates. Thus a sufficiently large fixed
multiple of $\log(2mn)$ accuracy and guard bits encloses the positive
square root of the displayed trace-power expression to absolute
error at most $\varepsilon_{\rm scale}/(32mn)$.
Choose its upper endpoint as $q$. The slack in the trace-power bound
then gives $\|M_2\|\le q^2\le(1+\varepsilon_{\rm scale})\|M_2\|$.
Stable products, the accumulators, and these scalar evaluations use
$O(\log(2mn))$-bit operands and
$\widetilde O(mn^2+n^3)$ bit operations.

The outer filters use a fixed self-adjoint contraction throughout their evaluation. The following array estimate controls additive action and rounding errors.

\begin{lemma}[Stability of the de Casteljau array]\label{bit_lem_array}
Let $0\preceq T\preceq I$ be fixed, and evaluate a degree-$d_f$ de Casteljau array by the updates $(a,b)\mapsto(I-T)a+Tb$. Each level is a contraction in the Euclidean norm of the concatenated array. If every vector update has error at most $\eta$, the final output error is at most $d_f\sqrt{d_f+1}\eta$. Consequently the bounded polynomial arrays used in Sections~\ref{soft_sec_filter} and~\ref{sp_sec_improvement} attain their prescribed inverse-polynomial errors with $O(\log(2kn))$ working bits.
\end{lemma}
\begin{proof}
For any vectors $a,b$,
\[
 \|(I-T)a+Tb\|^2
 \le\langle a,(I-T)a\rangle+\langle b,Tb\rangle.
\]
Summing over adjacent pairs proves that each level is a contraction
in the Euclidean norm of the concatenated vector array. An error of
norm at most $\eta$ in each vector update therefore causes final error
at most $d_f\sqrt{d_f+1}\eta$. Every exact intermediate vector has norm
at most $\sqrt{k}$. All inverse actions approximate the same frozen
$T$, so their errors enter as these additive update errors.
Inverse-polynomial action and rounding errors give inverse-polynomial
output error. The $O(d_f^2)$ actions change only logarithmic
factors in the cost, and all vector entries use $O(\log(2kn))$ bits.
\end{proof}

We combine the finite sampler and nested solves with the array estimate to implement the spread-filter certificate.

\begin{lemma}[Certified finite flat-filter trials]\label{bit_lem_flat_filter}
Under the hypotheses of Lemma~\ref{sp_lem_flat}, one capped finite trial uses $O(\log(2kn))$ working bits and $\widetilde O(n^\chi)$ bit operations, including unsuccessful trials. Its conditional success probability is at least $1/64$. Every accepted direction is normalized and satisfies both true curvature and spread inequalities in that lemma; all other samples obey the same finite work and magnitude caps.
\end{lemma}
\begin{proof}
For Lemma~\ref{sp_lem_flat}, use the de Casteljau array in Eq.~\eqref{sp_eq_casteljau}, with degree $d_f$ and fixed operator $T$.
Each exact intermediate is a
polynomial between zero and one on $[0,1]$, applied to $\xi$, so its
norm is at most $\sqrt k$. Lemma~\ref{bit_lem_array} applies: if each vector update
has error at most $\eta$, the final error is at most
$d_f\sqrt{d_f+1}\eta$. Choose inverse-polynomial action and scalar
errors below the normalization and quotient margins divided by this
factor. Thus $O(\log(2kn))$ working bits suffice, with a polynomial
magnitude cap.

Freeze one symmetric surrogate, its exact matrix-function definitions,
all feature factors and index lists during this computation. Every
inexact application approximates the same $T$. Lemma~\ref{bit_lem_nested} supplies these inverse-action errors with the same logarithmic precision. Its residual corrections add only fixed logarithmic
factors at each level; choose its fixed overhead integer $D$ before
the integer sketch schedule. No recursion depth is hidden as a constant.
The feature and conditional index events retain probability at least
$(4/5)(99/100)$; the independent sign event has probability greater
than $1/16$. After numerical slack their product is still greater
than $1/64$.

Reject a candidate unless a lower enclosure for its norm is at least
$\sqrt{\pi k/8}$ and both inequalities in Eq.~\eqref{sp_eq_spread}
are certified for the true model. The norm threshold has an
inverse-polylogarithmic lower bound, so normalization is conditioned.
The true weight and Hessian enclosures absorb the surrogate discrepancy
and the final grid errors. On the good events none of these tests
rejects the ideal estimates with their slack. On every other sample,
all lists, loops, precisions and magnitude checks retain their fixed
caps. Thus neither failed sketches nor failed sign trials introduce
unbounded work. No negative eigenspace or individual eigenvector is
computed by this routine.
\end{proof}

\subsection{Phases, sweeps, and endpoint continuation}\label{bit_sec_sweeps}
The remaining implementation issue is compatibility across accepted moves and changes of face. The phase parameters and corrected-state margins provide a common precision for these transitions.

\begin{lemma}[Certified sweeps and face continuation]\label{bit_lem_continuation}
Assume the corrected-state invariant and the phase, sweep, and endpoint hypotheses in Sections~\ref{sp_sec_improvement} and~\ref{am_sec_improvement}. The reservoir phases, frozen-root sweeps, deferred attenuation, and regular endpoint homotopies admit finite-precision implementations using $O(\log(2kn))$ working bits. Their accepted moves preserve the true feasibility, decrease, and discrepancy allowances, and each terminal correction represents the actual resulting face. Finite trial caps and certified filter success margins are retained. These routines have their arithmetic costs up to logarithmic factors in the bit model, with the original input length charged only in preprocessing.
In particular, the regular endpoints and deferred removals in
Lemma~\ref{gf_lem_faces} retain their respective
$\widetilde O(n)$ and $\widetilde O(k+\sqrt{nk})$ counts. Including
initializations, their bit cost is
$\widetilde O((k+\sqrt{nk}+n^{3/2})n^\chi)$.
\end{lemma}
\begin{proof}

\textbf{Endpoint and attenuation stages.}
For Lemma~\ref{gf_lem_faces}, store the attenuation parameter $s$
on a common dyadic grid and evaluate $r=e^{-s}$ from that parameter.
The smallest step and terminal coefficient are inverse polynomial,
and the horizon is $O(\log(2kn))$. A fixed logarithmic guard
precision puts scalar, coefficient, and solve errors below its
trigger, energy-arc, and root-residual margins. Nonnegative numerical
slots are retained throughout, and the last correction refers to
the exact new face. Lemmas~\ref{bit_lem_linear_algebra},
\ref{bit_lem_initialization}, and~\ref{bit_lem_smoothed} give the
per-stage bit costs. The arithmetic charging proof consequently
retains its stage counts and total bit bound.

\textbf{Reservoir phases.} 
Use the phase weights and spread filter of Section~\ref{sp_sec_improvement}.
The exact numerical Schur surrogate, matrix-function definitions,
root data, and weights are frozen within each filter. True-versus-cache
errors include the new phase weight and flat-polynomial accuracy.
Lemma~\ref{bit_lem_nested} remains applicable because the new
outer degree, bit target and list counts are fixed polylogarithms.
By Lemma~\ref{bit_lem_flat_filter}, every successful direction satisfies true, enclosed quotient and spread bounds; each chosen candidate has true certified decrease.
Conservative root, endpoint, value, grid, phase-weight, and action
accuracies are below fixed fractions of the same inverse-polynomial
or inverse-polylogarithmic margins. A root correction restores its
cache floor rather than multiplying past error factors.

All real/complex matrix operations use the same rational Hermitian
coordinates and exact trace-zero projection. No phase choices of
input vectors or accurate individual eigenvectors at a repeated
eigenvalue are needed. The rank-defect covariance is solely a proof
of spectral multiplicity; it creates no covariance or SDP oracle.
The polynomial filter needs no negative-eigenspace basis. The existing
finite fallback certifies arbitrary curvature when necessary. Its
slower progress has been included in both deterministic caps and
expected costs, rather than presuming a spread fallback.

All frozen signed coefficients still refer to the original retained
matrices with absolute symbolic coefficients at most two. Batched
attenuation deletes only positive coupling slots and never changes
the recorded fractional signs. The final joint rounding is identical.
The one-time input conversion is not repeated, and original long
rational inputs do not enter nonlinear arithmetic. Polylogarithmic
working precision therefore retains additive ${L_{\rm in}}$ in the bit
bound and the $mn^2$ input term in both models.

\textbf{Frozen sweeps and positive continuation.} 
We implement the sweeps and regular endpoint homotopies of Section~\ref{am_sec_improvement}.
The implementation never evaluates an exact optimizer during a sweep.
Its frozen scalar data come from a corrected short root, and their
certified errors are below fixed fractions of $l_N\sqrt{\kappa_i^0/N}$.
Positive $\kappa_i^0\ge{\gamma}$ have known inverse-polynomial lower bounds.
Square roots, arc limits, and all coordinate endpoints have dyadic
upper/lower enclosures.  The common grid is chosen once below the
smallest of the arc, cutoff, value-decrease and root-radius margins.
All these minima are inverse polynomial up to fixed logarithms.
Coefficient and scalar errors are included in the chord contraction,
not multiplied through all sweep updates.  A final root and value
correction certifies the true sweep decrease.  No eigenvector or
curvature oracle is inserted into the deterministic sweep.

For regular continuation, store the nonnegative scalar coefficients of
$\widehat {\mathsf A},\widehat E_i$ multiplying the true retained rank-one matrix;
the numerical cache uses the corresponding PSD approximations.  Their
errors, including the conservative trigger repair, are chosen below
the initial auxiliary residual and endpoint-drop margins.  Coefficient
rounding preserves the positive additions.  Store $s$ on a common
dyadic grid and evaluate $r=e^{-s}$ from it; do not accumulate rounded
multiplicative updates.  A stage uses the joint residual and a
certified anchored inverse, all using $O(\log(2kn))$ working bits.
The extra linear penalty introduces no new determinant or SDP solve.
The final $r=0$ correction is an optimizer of the actual new face.
\end{proof}

\subsection{Short-precision barrier signs}\label{bit_sec_comparisons}
The cube reductions need the progress in Eq.~\eqref{sr_eq_gain}, which
allows an approximate derivative sign. Exact comparison of the two
barrier values is unnecessary.

\begin{lemma}[Short-precision derivative sign]\label{bit_lem_barrier_sign}
Use $r,H,x,\widetilde v$ and $g$ from
Eq.~\eqref{sr_eq_derivative_sign}. If $x$ and $\widetilde v$ have at most
$b$ bits per entry, a deterministic procedure computes $\widehat g$ with
\[
 |\widehat g-g|\le\frac{r}{512H^2}
\]
using $O(b+\log(2rH))$ bits for every scalar operand, including
temporaries. Its cost is $\widetilde O(r(b+\log(2rH)))$ bit operations
and $O(r)$ arithmetic operations. The sign rule preserves the progress,
step caps, and residual bounds of Lemmas~\ref{sr_lem_small},
\ref{trace_lem_deterministic_groups}, and~\ref{oe_lem_cube}.
\end{lemma}
\begin{proof}
Take a common dyadic denominator $Q=2^{b_0}$ for the represented
$x_i=X_i/Q$ and $\widetilde v_i=V_i/Q$, with $b_0=O(b)$.
The integer $H=4\ell$ has $O(\log H)$ bits. Each summand is
\[
 g_i=\frac{2X_iV_i}{HQ(Q+|X_i|)}.
\]
Choose $p:=\lceil\log_2(512H^2)\rceil$ and compute by integer division
\[
 a_i:=\lfloor\frac{2^{p+1}X_iV_i}{HQ(Q+|X_i|)}\rfloor,
 \qquad \widehat g:=2^{-p}\sum_i a_i.
\]
Each numerator and denominator contains only a fixed number of short
factors and has $O(b+p+\log H)$ bits. Also
$|2x_i/(1+|x_i|)|\le1$ and $|\widetilde v_i|/H\le1/4$, so
$|a_i|\le2^p/4+1$. The sum is accumulated exactly as an integer with
$O(p+\log(2r))$ bits on this common grid. No product over coordinates
is formed. Since $|2^{-p}a_i-g_i|\le2^{-p}$, the stated error bound
and costs follow. The displayed floor means ordinary integer division.

Choose $\tau$ from the sign of $\widehat g$. Then
$\tau g\ge-r/(512H^2)\ge-(r-D)/(256H^2)$.
The curvature calculation in the proof of Lemma~\ref{sr_lem_small}
and its inward-rounding bound give
\[
 (\frac2{25}-\frac1{256}-\frac1{8192})\frac{r-D}{H^2}
 >\frac{r-D}{16H^2}.
\]
Thus its original bookkeeping cap and residual budget remain valid.
The sign of a direction does not change its matrix residual or spread
tests. The deterministic grouped procedure and the energy-controlled
cube reducer use these same tests and grids, so they have the same
conclusion. A near tie causes only the displayed derivative loss and
never requests more precision.
\end{proof}

\subsection{Resident batches and count-sensitive solves}\label{bit_sec_resident}
The final randomized implementation changes the preprocessing sketch,
the matrix-function evaluations, and the sketch sizes. We verify that
these changes preserve logarithmic numerical precision and the arithmetic
counts in Section~\ref{sec:randomized:operation-totals}.

\begin{lemma}[Precision for streamed factors and rational matrix functions]
\label{bit_lem_resident_functions}
The construction in Lemma~\ref{bs_lem_stream} has its stated arithmetic
cost also in bit operations, up to logarithmic factors, with
$O(\log(2mn))$ numerical bits. In the nonlinear phase,
Lemmas~\ref{br_lem_log}--\ref{br_lem_actions} have their stated
arithmetic bounds also as bit costs with $O(\log n)$ accuracy and magnitude bits per entry.
\end{lemma}
\begin{proof}
Freeze the group matrix and finite sample. Let $L_{\rm aux}\ge2$ be a
known polynomial bound for
$1+D+r+s+T+\delta^{-1}+\nu^{-1}+\|\Omega\|+\|B\|$.
If the streamed $U$ error is at most $e$ and the subsequent $Y$ action
error is at most $e$, then
\[
 \|\widehat Y-Y\|\le2L_{\rm aux}e,\qquad
 \|\widehat K-K\|\le4L_{\rm aux}^2e.
\]
These estimates include the Gram error and symmetrization. Since
$K\succeq TI\succeq I$, for $e\le(2L_{\rm aux})^{-16}$ the inverse
perturbation identity gives
\[
 \widehat K\succeq(T/2)I,\qquad
 \|\widehat Y\widehat K^{-1}\widehat Y^\top-YK^{-1}Y^\top\|
 \le32L_{\rm aux}^8e<\delta/64.
\]
Allocate each column error below $e/s$ and the small inverse error
below its explicit polynomial amplification bound. This preserves the
ridge-sketch interval. Lemma~\ref{bit_lem_group_solve} applies to the
same frozen factors with the new polynomial iteration count
$J=\widetilde O(n^{2/3})$. Its additive recurrence-error estimate
requires only a larger fixed logarithmic multiplier. Lemma~\ref{bit_lem_barrier_sign} implements the scalar derivative
sign rule with the same short precision, including its temporary
operands.

For nonlinear matrix functions, the known spectral endpoints and
requested inverse-polynomial error have $O(\log n)$ bits.
The shifted inverses in Lemma~\ref{br_lem_log} use
Lemma~\ref{bit_lem_linear_algebra}. Their normalized powers have norm
at most $1/3$, so truncation and arithmetic errors add with polynomial
amplification over logarithmically many bands and terms. In
Lemma~\ref{br_lem_exp}, normalized Taylor terms have norm at most
$e^{\|L\|}=n^{O(1)}$. A term error is amplified by at most this
factor in later terms. The ordered covariance expansion and batched
different-power recurrence obey the same bound. In
Eq.~\eqref{br_eq_adi}, exact updates are nonexpansive in the current
error; errors in their shifted inverses and products add along the
prescribed finite sequence. The iterate norms remain polynomial.
Reserve half the target for truncation or contraction, and half for
these explicit arithmetic errors. Stable recursive products therefore
give the claimed bit cost at a fixed $O(\log n)$ precision.

For a possibly non-Hermitian response matrix $M$, the certificate in
Lemma~\ref{oe_lem_power} extends by setting
$Z:=M^*M/a^2$ and $Q:=I-Z$. If $\|M\|\le a/2$, then
$0\preceq Z\preceq I/4$. Its logarithmic binomial series constructs
$X$ with a verified enclosure $\|X^*QX-I\|_F<1/2$ using the same
short products. Conversely this certificate implies $Q\succ0$ by
congruence, hence $\|M\|<a$. Reserve this factor-two gap in the six
relative-response tests. Thus the non-Hermitian similarities require
no long exact trace powers or spectral decomposition. Every finite
evaluation has prescribed iteration and magnitude caps and rejects
on failure of its required enclosure.
\end{proof}

\begin{lemma}[Uniform precision for the resident trials]
\label{bit_lem_resident_trials}
The trial in Lemma~\ref{br_lem_trial}, the resident masks, and the
curved and bulk continuations can be implemented with $O(\log n)$
numerical bits per entry. Their bit costs equal their arithmetic bounds
up to the logarithmic factors explicitly allowed there.
\end{lemma}
\begin{proof}
The four fixed multiplication algorithms in Eq.~\eqref{rr_eq_theta}
have logarithmic recursion depth and fixed coefficients. Their mixed
recursions therefore have polynomial intermediate magnitudes and
normwise error amplification, by the proof of
Lemma~\ref{bit_lem_linear_algebra}. Rounding each scalar operation to
a sufficiently fine common dyadic grid gives the stated inverse-polynomial
errors with $O(\log n)$ bits for every scalar, including temporary operands.

The proof of Lemma~\ref{bit_lem_nested} extends from terminal size
$n^{4/3}$ to $n$. Its norm, inverse, and coefficient estimates use the
invariant in Eq.~\eqref{rr_eq_invariant}, positive ridges, and polynomial
sizes. None uses the stronger terminal floor. With $h\le L^2$,
the moment order and sample words remain $O(\log n)$. The last two
equal sizes allow the leaf factor, Gram, and inverse to be materialized
at their charged square-product cost.

Freeze every feature factor and index list. At every level require the
same completed-solve guarantee
$\|v-K^{-1}b\|\le\eta\|b\|$, $\eta:=n^{-B_0}$, with fixed $B_0$.
Normalize each right-hand side within its own call. The normalized
Chebyshev error convolution gives a coarse solve $S$ with
$\|S(b)-K^{-1}b\|\le\|K^{-1}b\|/4$ using this same child accuracy.
Fresh residual correction $v^+:=v+S(b-Kv)$ then obeys
\[
 \|v^+-K^{-1}b\|
 \le\tfrac14\|v-K^{-1}b\|+n^{C}u\|b\|,
\]
for a fixed $C$ independent of depth and common rounding unit $u$.
The residual is a direct action of the same frozen operator; it does
not call a recursive inverse. Choose $u=n^{-D_0}$ with fixed sufficiently
large $D_0$, and perform the fixed $O(\log n)$ correction cap from
Lemma~\ref{bit_lem_nested}. Induction from the leaf proves the common
$\eta$ interface at all levels. Child accuracy does not grow with depth.
The bounded filter array in Lemma~\ref{bit_lem_array} then requires
only inverse-polynomial inverse-action errors.

The per-level correction factor is included in the fixed $D$ of
Eq.~\eqref{br_eq_uniformcost}. Choose $C_*=2000(D+1)$ after $D$.
Lemma~\ref{br_lem_sum} explicitly absorbs the resulting depth overhead;
it is not treated as one global logarithmic factor. Lemma~\ref{bit_lem_resident_functions}
supplies the root primitives and norm certificates used in these calls.

Mask scores, drift coefficients, predictor radii, and certified gains
have inverse-polynomial margins by Lemmas~\ref{md_lem_mask}
and~\ref{md_lem_step}. Choose one finer common grid for these finitely
many margins. Pending coefficients are stored without alteration and
have zero derivatives. Their true coupling and reservoir remain until
bulk attenuation; symbolic coefficients in the fixed contribution keep
their original bounded lengths. Lemma~\ref{bit_lem_cache} bounds cache
errors afresh against this true model. The charged attenuation uses
the same positive-submap estimates, root corrections, and final
tiny-slot accuracy as Lemma~\ref{bit_lem_continuation}. No derivative
bound on a pending endpoint is used. Each correction restores the same
root-error floor, and no long original input is read again. Bad trials
obey the prescribed caps; successful sampling events have slack for
all the stated enclosures. This proves the precision and cost claims.
\end{proof}

\subsection{Total bit complexity}\label{bit_sec_totals}
The arithmetic analyses determine the number of accepted stages, trials, and fallback calls. We now charge their finite implementations using the preceding lemmas.

\begin{theorem}[Bit complexity of the signing algorithms]\label{bit_thm_totals}
Let $A_1,\ldots,A_m\in\mathbb C^{n\times n}$ be rational Hermitian matrices of rank at most one and total binary length $L_{\rm in}$.
The algorithms of Theorems~\ref{rank_one_thm_main} and~\ref{soft_thm_algorithm} admit finite-precision implementations with the same discrepancy and partition guarantees.
The deterministic implementation has bit cost
\begin{equation}\label{bit_eq_deterministic_total}
 T_{\rm bit}\le\widetilde O(L_{\rm in}+mn^2+n^{4.742354}).
\end{equation}
The zero-error randomized implementation has expected bit cost
\begin{equation}\label{bit_eq_randomized_total}
 \E[T_{\rm bit}]\le\widetilde O(L_{\rm in}+mn^2+n^{3.575374}).
\end{equation}
The randomized implementation terminates and returns a correct signing for every random-bit sequence. Here $\widetilde O$ suppresses logarithmic factors in $m,n,L_{\rm in}$. After input conversion, preprocessing numerical entries use $O(\log(2mn))$ bits and nonlinear numerical entries use $O(\log(2n))$ bits, with fixed multipliers. Lemma~\ref{bit_lem_barrier_sign} gives these same bounds for all temporary operands in the barrier sign rule; no multiple-word product over coordinates is required. The preprocessing precision is $O(\log n)$ when $m$ is polynomial in $n$.
\end{theorem}
\begin{proof}

\textbf{Input conversion and working precision.} 
Lemma~\ref{bit_lem_input} charges the original input length once and gives total input-conversion cost $\widetilde O(L_{\rm in}+mn^2)$, including materialization. Lemmas~\ref{bit_lem_ridge}, \ref{bit_lem_grouped}, \ref{bit_lem_ridge_direction}, and~\ref{bit_lem_group_solve} implement the preprocessing solves on short data. Lemma~\ref{bit_lem_barrier_sign} implements their derivative sign decisions without long products.

The nonlinear phase has $k=O(n^2)$. Lemmas~\ref{bit_lem_cache}, \ref{bit_lem_linear_algebra}, \ref{bit_lem_initialization}, and~\ref{bit_lem_smoothed} give logarithmic working precision for its caches, initialization, and corrected states. Lemma~\ref{bit_lem_nested} makes this bound uniform in the permitted recursion depth. Taking the largest of the finitely many fixed precision multipliers supplies all the prescribed tolerances. The input conversion is not repeated in these stages.

\textbf{Deterministic algorithm.} 
Lemmas~\ref{bit_lem_conditional} and~\ref{bit_lem_responses} implement the deterministic grouped reduction, conditional choices, and block-response bounds. Lemmas~\ref{bit_lem_squaring}, \ref{bit_lem_array}, and~\ref{bit_lem_continuation} implement the required filters, sweeps, and face changes within their reserved errors. These bounds preserve the arithmetic counts in Section~\ref{sec_trace_algorithm} up to logarithmic factors. Adding the one-time input cost proves Eq.~\eqref{bit_eq_deterministic_total}.

\textbf{Randomized algorithm.} 
Lemmas~\ref{bit_lem_sampling}, \ref{bit_lem_nested}, and~\ref{bit_lem_flat_filter} give finite feature samples, certified solves, and capped trials with the stated conditional success margins. Lemma~\ref{bit_lem_continuation} implements the sweeps and endpoint continuations. Lemmas~\ref{bit_lem_resident_functions} and~\ref{bit_lem_resident_trials} implement the smaller preprocessing sketch, resident masks, curved trials, and count-sensitive solves with logarithmic numerical precision. Lemma~\ref{bit_lem_group_solve} supplies the matrix-norm certificates, and Lemma~\ref{bit_lem_linear_algebra} gives the fast deterministic and randomized final-rounding bit bounds. Every trial has bounded bit cost, so the conditional fallback charges and expected stage counts in Section~\ref{sec:randomized:operation-totals} apply to the finite implementation. Together with the input cost, they prove Eq.~\eqref{bit_eq_randomized_total}. The verified tests and deterministic fallbacks preserve correctness and termination for every random-bit sequence.

All error allowances are those reserved in the arithmetic analysis, so the discrepancy and partition constants are unchanged.
\end{proof}

%% file: _3_app.tex
\input{49_llm_disclaimer} 
\input{60_one_half}

%% file: 49_llm_disclaimer.tex
\section*{Acknowledgments}
The authors would like to thank Victor Reis and Lichen Zhang for very helpful discussions about the lower bound of discrepancy problems such as the Kadison-Singer problem and Matrix Spencer problem. The authors would also like to thank Lichen Zhang for very insightful discussions about the bit-complexity.

\paragraph{Independent and Concurrent Works}
After completing this draft, we became aware of two independent and concurrent works: \cite{ej26} gives a polynomial-time algorithm with constant 13 for Eq.~(1), and \cite{k26} gives a polynomial-time algorithm with constant 35 for Eq.~(1). Due to time constraints, we will provide a more detailed discussion of these works in a future version.

\paragraph{A Follow-up Paper}
There is a follow-up paper \cite{sy26_high_rank} by the same authors as this paper. It gives constructive results for the high-rank Kadison-Singer problem by generalizing the potential function used in this paper to the Schatten norm.

\section*{AI Usage Disclosure}
We used Claude Code Fable 5, ChatGPT Pro 5.6, and Codex Sol 5.6 in preparing this draft. We used these tools to generate major proof ideas. To improve the running time, we instructed the tools to draw on ideas from the SDP approach in \cite{hjstz22} and the LP approach in \cite{cls21}. The authors have carefully rewritten and verified all proofs. The authors take full responsibility for their correctness.

%% file: 60_one_half.tex
\section{The square-root budget}
\label{sec_one_half}

This section records the bound obtained from the pure square-root budget.
In this section only, define
\[
 \beta_{\rm sqrt}(t):=(1-t^2)^{1/2},\qquad
 \iota_{\rm tr}:=\frac{303}{100},\qquad
 \kappa_{\rm tr}:=10^{-6},\qquad \eta:=1+\kappa_{\rm tr}.
\]
There is no polynomial factor in this budget. The covariance below uses
the constant helper ${j_{\rm 0}}=43/50$ and retains one off-diagonal local trace.
We reuse the potential regularity and response identities from
Section~\ref{sec_rank_one_proof}, but give the covariance comparison,
scalar certificate, and quantitative descent bounds for this budget here.

\begin{theorem}\label{one_half_rank_one_thm_main}
For Hermitian matrices $A_1,\ldots,A_m\in\C^{n\times n}$ of rank at most one,
there are signs $\sigma\in\{\pm1\}^m$ such that
\[
 \|\sum_i \sigma_iA_i\|\le3.4814\|\sum_iA_i^2\|^{1/2}.
\]
For rational inputs, a deterministic
algorithm finds such signs using
\[
 T_{\rm arith}\le\widetilde O(mn^2+n^{56})
\]
arithmetic operations.
If $A_i=a_ia_i^*$, $\sum_iA_i=I$, and $\|a_i\|^2\le\alpha$, the two
sign classes satisfy
\[
 \|\sum_{i\in I_j}a_ia_i^*-I/2\|\le1.7407\sqrt\alpha,
 \qquad j=1,2.
\]
For rational matrices $a_ia_i^*$,
the partition has the same arithmetic bound. Here $\widetilde O$ suppresses
logarithmic factors in $m,n$.
\end{theorem}
\begin{proof}
Use the budget $\beta_{\rm sqrt}$ and $\iota_{\rm tr}=303/100$ from
Section~\ref{one_half_sec_trace_budget}. For nonzero PSD inputs, write
$\zeta:=\|\sum_i\bar A_i^2\|$ and fix $\rho>0$. The continuous matrix
potential $\mathcal P$ attains a minimum on the compact cube. Among its minimizers,
choose one with the fewest active coordinates. An available endpoint move
does not increase $\mathcal P$ and removes a coordinate, contradicting this choice.
Otherwise all active coordinates are light. The active gradient vanishes
and the active Hessian is PSD, contradicting the strict generator bound
in Eq.~\eqref{one_half_eq_trace_strict_descent}. Thus some minimizer is a vertex,
and the scalar feasible point in Eq.~\eqref{rank_one_eq_initial} gives
\[
 \|\sum_i \sigma_i\bar A_i\|\le2\sqrt{\iota_{\rm tr}\zeta+2n\rho}.
\]
Let $\rho$ tend to zero and take a vertex that occurs infinitely often.
Since $4\iota_{\rm tr}<(3.4814)^2$, this proves the claimed bound.
For general Hermitian rank-one inputs, apply the result to $|A_i|$ and
absorb the eigenvalue signs; $|A_i|^2=A_i^2$ preserves the variance.
The deterministic construction and its arithmetic cost are proved in
Section~\ref{one_half_sec_trace_algorithm}, including the required
approximation-error allowances. Finally $A_i^2\preceq\alpha A_i$ in the isotropic case,
and each partition error is half the signed sum.
\end{proof}

Algorithm~\ref{alg:square-root:signing} implements the square-root construction
of Section~\ref{one_half_sec_trace_algorithm}. It uses the shared preprocessing
of Algorithm~\ref{alg:shared:preprocess} and maintains the unsmoothed potential
$\mathcal P$ with modeled sum $M(y):=F+\sum_{i\in I}y_iC_i$.
The record $\mathcal R$ stores assigned signs and the input conversion data.
Fix $\varepsilon_{\rm scale}:=10^{-8}$ and
$\varepsilon_{\rm round}=\chi=e_{\rm end}:=10^{-6}$, with the internal
input-rounding constant $C_{\rm tr,*}:=3.4813833$.
The constants $\kappa_1,\kappa_2,a>0$ are sufficiently small as in that section.
All optimizer solves, derivative estimates, and potential comparisons use its
ordinary matrix routines at the prescribed accuracies. Trial optimizers are corrected from
the current cache; after a removal, the capped initializer is used afresh.

\begin{algorithm}[!htb]
\caption{Deterministic square-root signing with discrepancy constant $3.4814$}
\label{alg:square-root:signing}
\begin{algorithmic}[1]
\Procedure{SquareRootSigning}{$A_1,\ldots,A_m$}
    \State Set $\beta(t)\gets\sqrt{1-t^2}$, $\iota\gets303/100$, and $\rho\gets\varepsilon_{\rm scale}/n$.
    \State $(C,y,I,F,\mathcal R)\gets\Call{Preprocess}{A_1,\ldots,A_m;10^{-7},\varepsilon_{\rm round}}$.
    \If{$I=\varnothing$}
        \State Recover and return the original-input signing from $\mathcal R$.
    \EndIf
    \State Set $k_0\gets|I|$ and choose dyadic $\varepsilon_{\rm round}/(4\sqrt{k_0})<\delta_{\rm end}\le\varepsilon_{\rm round}/(2\sqrt{k_0})$.
    \State Set $\varepsilon_{\rm grad}\gets\kappa_1n^{-12}/4$, $t\gets\kappa_2n^{-19}\le\delta_{\rm end}/4$, and $0<\delta\le e_{\rm end}/(4k_0)$.
    \State Set $S\gets\{i\in I:\tr[C_i]<\chi/(2n^2)\}$ and record sign $1$ for each $i\in S$.
    \State Set $F\gets F+\sum_{i\in S}y_iC_i$ and $I\gets I\setminus S$.
    \If{$I\ne\varnothing$}
        \State Initialize the optimizer cache for $\mathcal P(y)$ to the prescribed accuracy by the capped solver.
    \EndIf
    \While{$I\ne\varnothing$}
        \State Set $S\gets\{i\in I:1-|y_i|\le\delta_{\rm end}\}$.
        \If{$S\ne\varnothing$}
            \State Record $\operatorname{sign}(y_i)$ for $i\in S$; set $F\gets F+\sum_{i\in S}y_iC_i$ and $I\gets I\setminus S$.
        \Else
            \State Compute $u_i\gets\tr[C_iU]$, $v_i\gets\tr[C_iV]$, and $\beta_i\gets\beta(y_i)$.
            \State Enclose the scores $\iota\beta_iv_i-(1-y_i)$ and $\iota\beta_iu_i-(1+y_i)$ to error $\delta$.
            \If{an approximate score is at least $-\delta$}
                \State Let $s\in\{1,-1\}$ be its endpoint; record $s$, set $F\gets F+sC_i$, and remove $i$ from $I$.
            \Else
                \State Compute $\widehat g$ with $\|\widehat g-\nabla\mathcal P(y)\|_\infty\le\varepsilon_{\rm grad}/4$; choose $j\in\arg\max_i|\widehat g_i|$.
                \If{$|\widehat g_j|\ge2\varepsilon_{\rm grad}$}
                    \State Set $r\gets-\operatorname{sign}(\widehat g_j)e_j$.
                \Else
                    \State Form a symmetric approximation to the ordinary Hessian.
                    \State Find $r$ with $\|r\|_2\le1$ and certified $r^\top\nabla^2\mathcal P(y)r\le-\varepsilon_{\rm grad}/4$.
                \EndIf
                \State Round $y+tr$ and $y-tr$ inward to the prescribed grid; correct both trial optimizers.
                \State Accept a trial with certified decrease at least $an^{-50}$; update $y$ and its cache.
                \State \textbf{continue}.
            \EndIf
        \EndIf
        \If{$I\ne\varnothing$}
            \State Reinitialize the optimizer cache for the reduced face to the prescribed accuracy.
        \EndIf
    \EndWhile
    \State Recover the original-input signs from $\mathcal R$ and \Return $\sigma$.
\EndProcedure
\end{algorithmic}
\end{algorithm}

\subsection{Potential and response notation}
\label{sec_one_half_setup}
Discard zero matrices and absorb the eigenvalue sign of each input.
After a common scaling, write the PSD inputs as $\bar A_i=a_ia_i^*$,
with $\sum_i\tr[\bar A_i]=1$, and define
$\zeta:=\|\sum_i\bar A_i^2\|$. Throughout the proof define
$\beta:=\beta_{\rm sqrt}$ and $\iota:=\iota_{\rm tr}$.
For $\rho>0$, define the potential
\begin{align*}
 \mathcal P(y):=\min_{b,\ U,V\succ0}\ &b+\rho\tr[U+V]\\
 \text{subject to}\quad
 &U^{-1}+M(y)+\mathcal K_y(V)\preceq bI,\\
 &V^{-1}-M(y)+\mathcal K_y(U)\preceq bI,
\end{align*}
where $M(y):=\sum_i y_i\bar A_i$ and
$\mathcal K_y(D):=\iota\sum_i\beta(y_i)\bar A_iD\bar A_i$.
The scalar $\eta=1+\kappa_{\rm tr}$ is distinct from the coupling map $\mathcal K_y$.
The budget is smooth in the open interval, nonnegative on the closed
interval, and vanishes at both endpoints. Thus the proof of
Lemma~\ref{rank_one_lem_potential} applies to this potential, as do its
KKT identities, endpoint moves, and initial bound
\[
 \|M(y)\|\le \mathcal P(y),\qquad \mathcal P(0)\le2\sqrt{\iota\zeta+2n\rho}.
\]

At the optimizer let $W,Z$ be the dual matrices and write
$u_i:=a_i^*Ua_i$, $v_i:=a_i^*Va_i$,
$w_i:=a_i^*Wa_i$, and $z_i:=a_i^*Za_i$.
A state is light when both inequalities in
Eq.~\eqref{rank_one_eq_light} hold at every active coordinate.
Define the response scales
\[
 p_i:=1+\iota\beta_i'v_i,\quad q_i:=1-\iota\beta_i'u_i,\quad
 c_i:=\sqrt{\beta_ip_i/q_i},\quad
 d_i:=\sqrt{\beta_iq_i/p_i},\quad
 \tau_i:=\sqrt{p_iq_i/\beta_i},
\]
and define
\[
 \widetilde u_i:=c_iu_i,\quad\widetilde v_i:=d_iv_i,\quad
 \widetilde w_i:=c_iw_i,\quad\widetilde z_i:=d_iz_i,\qquad
 \Gamma_i:=\frac1{\iota\beta_iu_iv_i}.
\]
For $y_i\ge0$, lightness and $-\beta_i'/\beta_i=y_i/(1-y_i^2)$
give $p_i\ge1/(1+y_i)\ge1/2$ and $q_i\ge1$;
for $y_i<0$ the roles are exchanged. Also $\beta_i\ge1-y_i^2$,
so $\Gamma_i>\iota>2$. Consequently the response-space construction in the
proof of Lemma~\ref{rank_one_lem_covariance} applies unchanged.
We use its matrices $\widehat {D_{\rm aux}},\widehat T$, paired orthonormal vectors
$e_i,f_i$, response projection $\Pi$, and energy $\mathcal Q(r)$.
In particular,
\[
 \widehat {D_{\rm aux}}\succeq\widehat T^\top\Gamma\widehat T,
 \qquad (I-\Pi_0)(I-\widehat T)\Pi=0,
\]
where $\Pi_0$ projects onto the span of the $e_i$.
These are the response and coercivity identities; the numerical curvature
bounds for either budget in the main text are not used.

\subsection{A covariance correction from the local trace}\label{one_half_sec_trace_correction}
We now retain both a diagonal and an off-diagonal local trace and
control the mixed variation directly. We use the potential and response notation from
Section~\ref{sec_one_half_setup}.

Define $D_i:=\widetilde u_i{\widetilde w}_i+\widetilde v_i{\widetilde z}_i$ and
${{\rho_{\rm rat}}}_i:=(\widetilde u_i{\widetilde z}_i+\widetilde v_i{\widetilde w}_i)/D_i$.
The paired basis gives the exact identities
\begin{equation}\label{one_half_eq_trace_identity}
 \langle e_i,\widehat T e_i\rangle=-{{\rho_{\rm rat}}}_i/\Gamma_i,
 \qquad \langle f_i,\widehat T f_i\rangle={{\rho_{\rm rat}}}_i/\Gamma_i.
\end{equation}
Indeed, $\mathscr{U}_{ii}=\iota\widetilde u_i^2$ and
$\mathscr{V}_{ii}=\iota\widetilde v_i^2$. Substitution into the paired
basis proves Eq.~\eqref{one_half_eq_trace_identity}. Also
\[
 D_i(1-{{\rho_{\rm rat}}}_i)=(\widetilde u_i-\widetilde v_i)({\widetilde w}_i-{\widetilde z}_i),
 \qquad \partial_i\mathcal P=\tau_i({\widetilde w}_i-{\widetilde z}_i).
\]
Define
\[
 \upsilon_i:=\frac{\widetilde v_i\sqrt{{\widetilde z}_i}-\widetilde u_i\sqrt{{\widetilde w}_i}}
 {\sqrt{\widetilde u_i\widetilde v_i}(\sqrt{{\widetilde w}_i}+\sqrt{{\widetilde z}_i})},
 \qquad L_i:=\iota\sqrt{\widetilde u_i\widetilde v_i}(\widetilde u_i-\widetilde v_i),
 \qquad \delta_i:=\Gamma_i^{-1}.
\]
The additional off-diagonal identity is
\begin{equation}\label{one_half_eq_trace_off_diagonal}
 \langle f_i,\widehat T e_i\rangle
 =L_i+\frac{\delta_i(\widetilde u_i-\widetilde v_i)\upsilon_i}{D_i}
          ({\widetilde w}_i-{\widetilde z}_i).
\end{equation}
To verify it, direct substitution gives
\[
 \langle f_i,\widehat T e_i\rangle
 =L_i\frac{(\widetilde u_i+\widetilde v_i)\sqrt{{\widetilde w}_i{\widetilde z}_i}}{D_i},
\]
and the factorization
\[
 (\widetilde u_i+\widetilde v_i)\sqrt{{\widetilde w}_i{\widetilde z}_i}-D_i
 =({\widetilde w}_i-{\widetilde z}_i)\upsilon_i\sqrt{\widetilde u_i\widetilde v_i}
\]
proves Eq.~\eqref{one_half_eq_trace_off_diagonal}. Both trace deviations are
explicit gradient terms, including when that gradient coordinate is zero.

Let $\eta:=1+\kappa_{\rm tr}$, with $\kappa_{\rm tr}$ specified in
Section~\ref{one_half_sec_trace_budget}, and define the mixed coefficient and target
\[
 g_i:=\beta_i'/\beta_i,\qquad
 \Lambda_i:=\frac{g_i}{\tau_i\sqrt{\widetilde u_i\widetilde v_i}},\qquad
 k_i:=\frac{\iota(-\beta_i'')}{2\eta p_iq_i}.
\]
We give the local scalar construction before stating the covariance
comparison. Suppress the index, define $\delta:=1/\Gamma$, and define
${j_{\rm 0}}:=43/50$. Define
\begin{align*}
 j&:=\frac{\Lambda^2\delta}{4k},&
 F&:={j_{\rm 0}}^2+2{j_{\rm 0}}-j,& B&:=2{j_{\rm 0}}-j-\delta(1+2{j_{\rm 0}}),\\
 \mathfrak u&:=B+\delta F-\frac{F\Lambda L}{2k(1+{j_{\rm 0}})},\\
 \mathfrak q&:=\delta({j_{\rm 0}}^2+2\delta {j_{\rm 0}}+\delta)
       +\frac1k[B+\frac{(F-B)j}{(1+{j_{\rm 0}})^2}],\\
 \mathfrak z&:=2\mathfrak u+\delta F
       -\frac Fk(1-\frac j{(1+{j_{\rm 0}})^2}).
\end{align*}
The scalar certificate below proves
\begin{equation}\label{one_half_eq_trace_scalar_conditions}
 F>0,\quad B>0,\quad\mathfrak u>0,\quad\mathfrak z>0,
 \quad\mathfrak u^2-F\mathfrak q>0.
\end{equation}
Define the resulting positive weights
\begin{align*}
 a_*&:=F/\mathfrak u,& d_*&:=\Gamma {j_{\rm 0}},&
 r_*&:=\frac{\Lambda a_*}{2k(1+{j_{\rm 0}})},\\
 z_*&:=2a_*-a_*^2(1/k-\delta)+\delta r_*^2
        =\frac{F\mathfrak z}{\mathfrak u^2},&
 w_e&:=z_*^{-1},& w_f&:=\delta/F,\\
 h_*&:=2a_*w_e,& p_*&:=2r_*w_e,& j_*&:=2d_*w_f.
\end{align*}
These helper scalars are distinct from the state variables. In
particular, $j_*$ is the row multiplier and $j$ is the scalar above.

The extra trace information allows the covariance to include the
mixed response itself, with a gradient correction canceled by the drift.

\begin{lemma}\label{one_half_lem_trace_covariance}
There is a centered normalized response $\xi$ with covariance ${\mathsf A}$,
positive definite on the response subspace and zero on its orthogonal
complement, such that, with
$\mathsf F_i:=f_i^\top\xi$, $\mathsf Y_i:=e_i^\top(I-\widehat T)\xi$,
$K:=\E[rr^\top]$, and $\pi_i:=\tau_i^2K_{ii}$,
\begin{equation}\label{one_half_eq_trace_covariance}
 \E[\mathcal Q(r)]+\sum_i\Lambda_i\E[\mathsf Y_i\mathsf F_i]
 \le\sum_i k_iD_i\pi_i+\nu\cdot\nabla \mathcal P,
\end{equation}
where
\[
 \nu_i:=\frac{\delta_i[h_{*,i}-j_{*,i}+p_{*,i}\upsilon_i]
                 (\widetilde u_i-\widetilde v_i)}{D_i\tau_i}.
\]
For $R_{\mathsf A}:=(I-\widehat T){\mathsf A}(I-\widehat T)^\top$, there are absolute
$a,C>0$ such that
\begin{equation}\label{one_half_eq_trace_movement}
 a k_{\rm act}\le\tr[R_{\mathsf A}]\le C k_{\rm act}.
\end{equation}
\end{lemma}
\begin{proof}
Use the weights above at each coordinate and define
\[
 {\mathsf D}:=\sum_i(w_i^e e_ie_i^\top+w_i^f f_if_i^\top),\qquad
 {\mathsf A}:={\mathsf E}({\mathsf E}^\top{\mathsf D}^{-1}{\mathsf E})^{-1}{\mathsf E}^\top,
 \qquad {\mathsf B}:={\mathsf D}-{\mathsf A},
\]
where ${\mathsf E}$ is an orthonormal basis matrix for the response subspace.
The weighted projection identities give
${\mathsf A},{\mathsf B}\succeq0$,
${\mathsf A}{\mathsf D}^{-1}{\mathsf A}={\mathsf A}$,
${\mathsf A}{\mathsf D}^{-1}{\mathsf B}=0$, and
$\tr[{\mathsf D}^{-1}{\mathsf A}]=k_{\rm act}$. No commutation is required.

Fix one pair and suppress its index. Define
$v_*:=a_*e$, $w_*:=r_*e+d_*f$, $\widehat e:=e-\Lambda f/(2k)$,
$t_e:=e^\top\widehat T$, and $t_f:=f^\top\widehat T$.
For $C_*:=\Gamma{\mathsf B}+k{\mathsf A}\succ0$, the projection identities give
\[
 C_*^{-1}=\delta{\mathsf D}^{-1}
 +(1/k-\delta){\mathsf D}^{-1}{\mathsf A}{\mathsf D}^{-1}.
\]
Completing a square in $t_e$ yields
\begin{align*}
 &\Gamma t_e{\mathsf B}t_e^\top
   +k(e^\top-t_e){\mathsf A}(e-t_e^\top)
   -\Lambda(e^\top-t_e){\mathsf A}f+t_e(2{\mathsf D}v_*)\\
 &\qquad\ge2\widehat e^\top{\mathsf A}v_*
   -(1/k-\delta)v_*^\top{\mathsf A}v_*
   -\frac{\Lambda^2}{4k}f^\top{\mathsf A}f
   -\delta v_*^\top{\mathsf D}v_*.
\end{align*}
The response condition gives $t_f{\mathsf A}=f^\top{\mathsf A}$.
A square with the PSD matrix ${\mathsf B}$, which need not be invertible, gives
\[
 \Gamma t_f{\mathsf B}t_f^\top+t_f(2{\mathsf D}w_*)
 \ge2f^\top{\mathsf A}w_*+\delta w_*^\top{\mathsf A}w_*
       -\delta w_*^\top{\mathsf D}w_*.
\]
The two diagonal coefficients of ${\mathsf A}$ in these bounds are
$z_*=w_e^{-1}$ and
$-\Lambda^2/(4k)+2d_*+\delta d_*^2=F/\delta=w_f^{-1}$.
The off-diagonal coefficient vanishes because
\[
 -\frac{\Lambda a_*}{2k}+r_*+\delta r_*d_*=0.
\]
Thus the summed coefficient is exactly ${\mathsf D}^{-1}$.

For the scalar constant term, define
\[
 \mathcal E:=w_e+w_f+\delta[(a_*^2+r_*^2)w_e+d_*^2w_f]
                 +(-h_*+j_*)\delta+p_*L.
\]
The exact identity
\begin{equation}\label{one_half_eq_trace_row_cost}
 1-\mathcal E=\frac{\mathfrak u^2-F\mathfrak q}{z_*\mathfrak u^2}\ge0
\end{equation}
follows by putting
$N_*:=1+\delta(a_*^2+r_*^2-2a_*)+2r_*L$ and expanding
\[
 z_*F(1-\mathcal E)=z_*B-FN_*
 =-F+2a_*\mathfrak u-a_*^2\mathfrak q.
\]
Substitute $a_*=F/\mathfrak u$ to obtain Eq.~\eqref{one_half_eq_trace_row_cost}.

Sum the two row inequalities over all coordinates. Coercivity gives
\[
 \tr[\widehat {D_{\rm aux}}{\mathsf A}]
 \le\tr[\widehat {D_{\rm aux}}{\mathsf D}]
       -\tr[\Gamma\widehat T{\mathsf B}\widehat T^\top],
 \qquad \tr[\widehat {D_{\rm aux}}{\mathsf D}]=\sum_i(w_i^e+w_i^f).
\]
Using the opposite diagonal traces, the remaining correction to the
weighted movement is at most
\[
 \sum_i[(h_{*,i}-j_{*,i})(\langle e_i,\widehat T e_i\rangle+\delta_i)
       +p_{*,i}(\langle f_i,\widehat T e_i\rangle-L_i)].
\]
Here the terms $\mathcal E_i-1$ were dropped using
Eq.~\eqref{one_half_eq_trace_row_cost}. Eqs.~\eqref{one_half_eq_trace_identity}
and~\eqref{one_half_eq_trace_off_diagonal} identify this correction as
$\nu\cdot\nabla \mathcal P$. Since
$\E[\mathcal Q(r)]=\tr[\widehat {D_{\rm aux}}{\mathsf A}]$ and
$\langle e_i,R_{\mathsf A}e_i\rangle=D_i\pi_i$,
this proves Eq.~\eqref{one_half_eq_trace_covariance}.

For movement, the compact scalar certificate in
Lemma~\ref{one_half_lem_trace_surplus} bounds $F,\mathfrak u,\mathfrak z$
above and away from zero. Therefore $w_e$ and $w_e^{-1}$ are bounded,
while $w_f=1/(\Gamma F)$. In particular
\[
 {\mathsf D}\preceq CI,\qquad
 {\mathsf D}^{-1}\preceq C\Pi_0+C\Gamma\Pi_1,
 \qquad\Pi_1:=I-\Pi_0.
\]
With $C_*:={\mathsf E}^\top{\mathsf D}^{-1}{\mathsf E}$,
the response condition $\Pi_1{\mathsf E}=\Pi_1\widehat T{\mathsf E}$
and coercivity imply
\[
 \tr[C_*]\le Ck_{\rm act}
   +C\|\Gamma^{1/2}\Pi_1\widehat T{\mathsf E}\|_F^2
 \le Ck_{\rm act}+C\tr[\widehat T^\top\Gamma\widehat T]
 \le Ck_{\rm act}.
\]
Weighted Frobenius Cauchy--Schwarz gives
\[
 (k_{\rm act}-\tr[\widehat T\Pi])^2
 \le\|(I-\widehat T){\mathsf E}C_*^{-1/2}\|_F^2
       \|{\mathsf E}C_*^{1/2}\|_F^2
 =\tr[R_{\mathsf A}]\tr[C_*].
\]
Since $\|\widehat T\|_F^2\le2k_{\rm act}/\iota$ and $\iota>2$,
$|\tr[\widehat T\Pi]|\le k_{\rm act}\sqrt{2/\iota}<k_{\rm act}$.
This proves the lower bound independently of the smallest weight.
The upper bound follows from ${\mathsf A}\preceq CI$ and
$\|I-\widehat T\|_F^2\le Ck_{\rm act}$.
\end{proof}

\paragraph{The mixed variation and its correction.}
Retain the exact mixed term in the derivation of
Eq.~\eqref{rank_one_eq_curvature}, before applying Cauchy--Schwarz:
\begin{equation}\label{one_half_eq_trace_exact_variation}
 \frac12r^\top\nabla^2\mathcal Pr
 \le\mathcal Q(r)+\sum_i m_i(r)
 -\frac \iota2\sum_i(-\beta_i'')(v_iw_i+u_iz_i)r_i^2,
 \qquad m_i(r):=\iota\beta_i'r_i(w_i\dot v_i+z_i\dot u_i).
\end{equation}
For the normalized response above, write
$\mathsf Y_i=\sqrt{D_i}J_i$ and $\mathsf E_i:=e_i^\top\xi$.
The exact identity
\begin{equation}\label{one_half_eq_trace_mixed_decomposition}
 m_i(r)=\Lambda_i\mathsf Y_i\mathsf F_i
 +\partial_i\mathcal P\frac{g_i}{\tau_i^2D_i}
       \mathsf Y_i(\mathsf E_i-\mathsf Y_i+\upsilon_i\mathsf F_i)
\end{equation}
holds even when the gradient coordinate is zero. Indeed, the response
equations give
\[
 m_i(r)=\frac{g_iJ_i}{\tau_i}
 [{\widetilde w}_iP_i+{\widetilde z}_iQ_i+({\widetilde z}_i-{\widetilde w}_i)J_i],
\]
and the paired basis expresses the first two terms as
\[
 {\widetilde w}_iP_i+{\widetilde z}_iQ_i
 =\frac{{\widetilde w}_i-{\widetilde z}_i}{\sqrt{D_i}}\mathsf E_i
 +\frac{(\widetilde u_i+\widetilde v_i)\sqrt{{\widetilde w}_i{\widetilde z}_i}}
 {\sqrt{\widetilde u_i\widetilde v_i}\sqrt{D_i}}\mathsf F_i.
\]
The factorization used in Eq.~\eqref{one_half_eq_trace_off_diagonal} proves
Eq.~\eqref{one_half_eq_trace_mixed_decomposition} without dividing by
${\widetilde w}_i-{\widetilde z}_i$. Define
\[
 \theta_i:=\frac{g_i}{\tau_i^2D_i}
 \E[\mathsf Y_i(\mathsf E_i-\mathsf Y_i+\upsilon_i\mathsf F_i)].
\]
Combining the mixed identity with Eq.~\eqref{one_half_eq_trace_covariance} gives
\begin{equation}\label{one_half_eq_trace_augmented_variation}
 \frac12\E[r^\top\nabla^2\mathcal Pr]
 \le\sum_i k_iD_i\pi_i+(\nu+\theta)\cdot\nabla \mathcal P
 -\sum_i\frac{\iota(-\beta_i'')}{2p_iq_i}{{\rho_{\rm rat}}}_iD_i\pi_i.
\end{equation}
Thus the mixed variation is included directly in the covariance bound.

\subsection{The square-root certificate}\label{one_half_sec_trace_budget}
Define
\[
 \iota_{\rm tr}:=\frac{303}{100},\qquad
 \kappa_{\rm tr}:=\frac1{10^6},\qquad
 \eta:=1+\kappa_{\rm tr},\qquad {j_{\rm 0}}:=\frac{43}{50},
 \qquad\beta_{\rm sqrt}(y):=\sqrt{1-y^2}.
\]
Abbreviate these by $\iota,\beta$ in this section. At $z:=|y|<1$,
\begin{gather}
 \beta(0)=1,\quad\beta(\pm1)=0,\quad1-y^2\le\beta(y)\le1,
 \quad\beta(y)=\sqrt{1-y^2},\label{one_half_eq_trace_budget_basic}\\
 \beta'(z)=-\frac z{\sqrt{1-z^2}},\qquad
 -\beta''(z)=\frac1{(1-z^2)^{3/2}}\ge1,\qquad
 -\frac{\beta'(z)}{\beta(z)}=\frac z{1-z^2}.
 \label{one_half_eq_trace_budget_signs}
\end{gather}
The budget is smooth on $(-1,1)$. On the truncated interval,
$|\beta^{(j)}|\le C\delta_{\rm end}^{1/2-j}$ and
$|\beta^{(j)}|/\beta\le C\delta_{\rm end}^{-j}$ for $1\le j\le3$.
Lightness gives $p_i\ge1/(1+z)\ge1/2$ and $q_i\ge1$ for
$y_i\ge0$, with the roles exchanged for negative coordinates.
The potential, endpoint, response, and coercivity arguments therefore apply.

The remaining certificate verifies the direct mixed covariance on the
entire light-state domain.

\begin{lemma}\label{one_half_lem_trace_surplus}
At every light state, the choice
\begin{equation}\label{one_half_eq_trace_surplus}
 k_i:=\frac{\iota(-\beta_i'')}{2\eta p_iq_i}
\end{equation}
satisfies all five conditions in Eq.~\eqref{one_half_eq_trace_scalar_conditions}.
The quantities $F,B,\mathfrak u,\mathfrak z$ are uniformly bounded
above and away from zero.
\end{lemma}
\begin{proof}
By symmetry take $z=y_i\ge0$ and define
$w:=\sqrt{1-z^2}$, $q_0:=z/w^2$,
${a_{\rm loc}}:=\iota\beta u_i$, ${b_{\rm loc}}:=\iota\beta v_i$, and $P:={a_{\rm loc}}{b_{\rm loc}}$.
The light rectangle is $0<{a_{\rm loc}}<1+z$, $0<{b_{\rm loc}}<1-z$.
Writing $\mathcal R:=p_iq_i$, direct substitution gives
\[
 \mathcal R=1+q_0({a_{\rm loc}}-{b_{\rm loc}})-q_0^2P,\qquad
 \delta=\frac P{\iota w},\qquad
 \Lambda^2\delta=\frac{\iota(\beta')^2}{\beta\mathcal R},\qquad
 \Lambda L=\frac{1+q_0^2P-\mathcal R}{\mathcal R}.
\]
In particular $j=\Lambda^2\delta/(4k)=\eta z^2/2$.
Let ${s_{\rm inv}}:=1/k=2\eta\mathcal Rw^3/\iota$ and
$\sigma:=2\eta w^3/\iota$. The scalar expressions become
\begin{align*}
 F&={j_{\rm 0}}^2+2{j_{\rm 0}}-j,& B&=2{j_{\rm 0}}-j-\delta(1+2{j_{\rm 0}}),\\
 \mathfrak u&=B+\delta F+\frac{F({s_{\rm inv}}-\sigma-4j\delta)}{2(1+{j_{\rm 0}})},\\
 \mathfrak q&=\delta({j_{\rm 0}}^2+2\delta {j_{\rm 0}}+\delta)
       +{s_{\rm inv}}[B+\frac{(F-B)j}{(1+{j_{\rm 0}})^2}],\\
 \mathfrak z&=2\mathfrak u+\delta F-{s_{\rm inv}}F(1-\frac j{(1+{j_{\rm 0}})^2}).
\end{align*}
Here ${s_{\rm inv}}$ is an inverse target coefficient.

At fixed $P\in[0,w^2]$, the function ${a_{\rm loc}}-P/{a_{\rm loc}}$ is increasing.
The two boundary values, at ${b_{\rm loc}}=1-z$ and ${a_{\rm loc}}=1+z$, show that
$(\delta,\mathcal R)$ lies in the triangle with vertices
\[
 (0,\frac1{1+z}),\qquad(0,\frac1{1-z}),\qquad
 (\frac w\iota,\frac1{w^2}).
\]
Both bounds are affine in $P$ and agree at $P=w^2$.
Thus the complete domain is covered by
\[
 \delta=\frac w\iota H,\qquad
 {s_{\rm inv}}=\frac{2\eta}\iota w[1+z(1-H)(2V-1)],\qquad H,V\in[0,1].
\]
At $z=0$ the triangle degenerates harmlessly. The limiting boundary
$z=1$ is used only for certification and uniform bounds.

Define $z:=2t/(1+t^2)$ and $w:=(1-t^2)/(1+t^2)$, with $0\le t\le1$.
Define the following rational polynomials in $(t,H,V)$:
\begin{align*}
 g&:=1+t^2,& w_n&:=1-t^2,& z_n&:=2t,\\
 j_n&:=\eta z_n^2/2,& h_n&:=w_nH/\iota,\\
 r_n&:=\frac{2\eta}\iota w_n[g+z_n(1-H)(2V-1)],&
 {s_{\rm inv}}_n&:=\frac{2\eta}\iota w_n^3,\\
 F_n&:=({j_{\rm 0}}^2+2{j_{\rm 0}})g^2-j_n,&
 B_n&:=2{j_{\rm 0}}g^2-j_n-(1+2{j_{\rm 0}})h_ng,\\
 T_n&:=r_ng-{s_{\rm inv}}_n-4j_nh_n,\\
 U_n&:=B_ng^3+h_nF_ng^2+\frac{F_nT_n}{2(1+{j_{\rm 0}})},\\
 Q_{{b_{\rm loc}},n}&:={j_{\rm 0}}^2h_ng+(2{j_{\rm 0}}+1)h_n^2,&
 Q_{1,n}&:=B_ng^2+\frac{(F_n-B_n)j_n}{(1+{j_{\rm 0}})^2},\\
 Q_n&:=Q_{{b_{\rm loc}},n}g^4+r_nQ_{1,n},&
 P_n&:=U_n^2-F_nQ_ng^2,\\
 Z_n&:=2U_ng+h_nF_ng^3-r_nF_ng^2+\frac{r_nF_nj_n}{(1+{j_{\rm 0}})^2}.
\end{align*}
Substitution gives
\[
 F=F_n/g^2,\quad B=B_n/g^2,\quad\mathfrak u=U_n/g^5,
 \quad\mathfrak q=Q_n/g^6,\quad\mathfrak z=Z_n/g^6,
 \quad\mathfrak u^2-F\mathfrak q=P_n/g^{10}.
\]
All five numerator polynomials are strictly positive on $[0,1]^3$.
The exact tensor-product Bernstein certificate is
\begin{center}
\begin{tabular}{c|c|c}
Polynomial & Degree in $(t,H,V)$ & Certified $t$ intervals\\\hline
$F_n$ & $(4,0,0)$ & $[0,1]$\\
$B_n$ & $(4,1,0)$ & $[0,1]$\\
$U_n$ & $(10,1,1)$ & $[0,1]$\\
$Z_n$ & $(12,1,1)$ & $[0,1]$\\
$P_n$ & $(20,2,2)$ & $[0,1/4],\ [1/4,1/2],\ [1/2,1]$\\
\end{tabular}
\end{center}
The other two coordinates retain their full unit intervals.
Apply Eq.~\eqref{eq:certificate:bernstein} after the affine change of
variables from each listed box to $[0,1]^3$. The first four polynomials
have $5,10,44,52$ coefficients, respectively; $P_n$ has $189$ on each
of its three boxes. Exact rational substitution shows that all $678$
coefficients exceed $1/400$. The Bernstein basis is nonnegative and
sums to one, so each numerator is positive throughout its listed boxes.
These boxes cover the full domain for every polynomial.
Since $1\le g\le2$, the first four polynomials also give the stated
uniform bounds. This proves the lemma.
\end{proof}

\paragraph{Strict descent.}
For a drift $h$ and centered direction $r$ with covariance $K$, define
$\mathscr{L}\mathcal P:=h\cdot\nabla \mathcal P+\frac12\tr[K\nabla^2\mathcal P]$.
Define the drift
\[
 h_i:=-\frac{k_i}{\tau_i}(\widetilde u_i-\widetilde v_i)\pi_i-\nu_i-\theta_i.
\]
The last two terms cancel the gradient corrections in
Eq.~\eqref{one_half_eq_trace_augmented_variation}. The first changes
$k_iD_i\pi_i$ into $k_i{{\rho_{\rm rat}}}_iD_i\pi_i$. Since
${{\rho_{\rm rat}}}_iD_i\pi_i=p_iq_i(v_iw_i+u_iz_i)K_{ii}$,
Eq.~\eqref{one_half_eq_trace_surplus} gives
\begin{equation}\label{one_half_eq_trace_strict_descent}
 \mathscr{L}\mathcal P\le-\kappa_{\rm tr}\sum_i k_i{{\rho_{\rm rat}}}_iD_i\pi_i<0.
\end{equation}
The weights are positive and Eq.~\eqref{one_half_eq_trace_movement} ensures
nonzero movement. No derivative of the covariance or drift is used.
This proves the local assertion in Theorem~\ref{one_half_rank_one_thm_main}.
Its polynomial implementation is given in Section~\ref{one_half_sec_trace_algorithm}.

\subsection{Deterministic arithmetic algorithm}\label{one_half_sec_trace_algorithm}
We prove the arithmetic assertion of Theorem~\ref{one_half_rank_one_thm_main}.
Use $\beta_{\rm sqrt},\iota_{\rm tr}$ from Section~\ref{one_half_sec_trace_budget},
abbreviated by $\beta,\iota$ here, and the direct mixed covariance of
Lemma~\ref{one_half_lem_trace_covariance}.
The two gradient corrections in Eq.~\eqref{one_half_eq_trace_augmented_variation} are not proportional
to the covariance diagonal. We therefore use the strict descent of $\mathcal P$
itself and give a separate polynomial bound.

\paragraph{Preprocessing and the maintained model.}
Apply the one-time input rounding and variance scaling from
Section~\ref{rank_one_sec_algorithm}, and the grouped reduction in
Lemma~\ref{rank_one_lem_grouped_partial_signing}, with residual allowance
$\eta_{\rm ps}:=10^{-7}$. The input-rounding transfer will use the internal
constant $C_{\rm tr,*}:=3.4813833$ verified below. At most $k\le n^2$
coordinates remain fractional and $\sum_iC_i^2\preceq I$.
Keep ${\varepsilon_{\rm scale}}:=10^{-8}$, $\rho:={\varepsilon_{\rm scale}}/n$, and
${\varepsilon_{\rm round}}={\varepsilon_{\rm discard}}=\chi=e_{\rm end}:=10^{-6}$.
Round the initial coefficients with Euclidean error at most ${\varepsilon_{\rm round}}/2$
and initialize the modeled sum at zero as before.
Remove each active matrix with $\tr[C_i]<\chi/(2n^2)$, retain its
current fractional contribution in $F$, and choose any output sign.
The total discrepancy error of these removals is at most $\chi$.
The remaining traces $t_i:=\tr[C_i]$ satisfy
\[
 an^{-2}\le t_i\le1.
\]
For this algorithm choose ${\varepsilon_{\rm round}}/(4\sqrt k)<\delta_{\rm end}\le{\varepsilon_{\rm round}}/(2\sqrt k)$.
At endpoint distance at most $\delta_{\rm end}$, record the nearest sign, preserve
the current modeled contribution, and remove the coupling slot.
Eq.~\eqref{rank_one_eq_cont_2} bounds the sum of these errors by
${\varepsilon_{\rm round}}/2$. Thus $\delta_{\rm end}\ge a/n$ and deferred rounding is unnecessary.
The initial matrix potential is at most $2\sqrt{\iota+2{\varepsilon_{\rm scale}}}$.

The preparatory approximations have explicit arithmetic constructions.
Discard zero inputs. For a nonzero normalized PSD rank-one input $B$, choose a largest diagonal
entry $a=B_{jj}$ and put $w:=Be_j/a$. Then $B=aww^*$ and $|w_l|\le1$.
For target error $\delta_{\rm in}:=10^{-8}/(mn)$, choose a common dyadic
mesh $h\le\delta_{\rm in}^2/(10^6n^4)$. Approximate $a$ and the selected
column to error $h$ per real and imaginary part. If
$\widehat a\le\delta_{\rm in}/(4n)$, output zero; its Frobenius error is
at most $\delta_{\rm in}/2$. Otherwise $a\ge\delta_{\rm in}/(8n)$.
Divide the approximate column by $\widehat a$ and round to mesh $h$,
obtaining $\widehat w$ with
$\|\widehat w-w\|_2\le32n^{3/2}h/\delta_{\rm in}$. Expanding the outer
product gives
\[
 \|\widehat a\widehat w\widehat w^*-B\|_F
 \le200n^2h/\delta_{\rm in}\le\delta_{\rm in}.
\]
Bounded-interval binary searches perform the scalar roundings in
$O(\log(2mn))$ arithmetic operations per entry. The resulting PSD
rank-one factors cost $\widetilde O(mn)$ operations, and their outer
products cost $\widetilde O(mn^2)$.

After trace normalization, the variance matrix $M_2:=\sum_i\bar A_i^2$
satisfies $1/(mn)\le\|M_2\|\le1$. Choose a power of two
$\ell_{\rm pow}\ge8\varepsilon_{\rm scale}^{-1}\log(2n)$ within a
factor two of this bound. Exact repeated squaring computes
$\tr[M_2^{\ell_{\rm pow}}]$, and
\[
 \|M_2\|\le\tr[M_2^{\ell_{\rm pow}}]^{1/\ell_{\rm pow}}
 \le n^{1/\ell_{\rm pow}}\|M_2\|.
\]
Bisection with rational comparisons between $q^{2\ell_{\rm pow}}$ and
this trace selects an upper approximation to its positive
$2\ell_{\rm pow}$-th root within relative error
$\varepsilon_{\rm scale}/8$. It therefore gives
$\|M_2\|\le q^2\le(1+\varepsilon_{\rm scale})\|M_2\|$.
The lower norm bound makes the number of bisection steps logarithmic
in $mn$. Forming $M_2$ and performing these operations costs
$\widetilde O(mn^2+n^3)$. Including Lemma~\ref{rank_one_lem_grouped_partial_signing},
the arithmetic preprocessing cost is
$\widetilde O(mn^2+n^{5.055})$.

\paragraph{Quantitative drift witnesses.}
On a bounded-potential region, the same KKT estimates give
\[
 aI\preceq U,V\preceq C\sqrt n I,\qquad
 an^{-1}I\preceq W,Z\preceq I.
\]
Together with the trace cutoff, the budget bounds, and lightness, these imply
\begin{align*}
 at_i\le u_i,v_i&\le C\sqrt nt_i,&
 an^{-1}t_i\le w_i,z_i&\le t_i,\\
 an^{-1/2}\le\beta_i&\le1,&
 a\le p_i,q_i&\le Cn,\qquad a\le p_iq_i\le Cn,\\
 an^{-3/2}\le c_i^2,d_i^2&\le Cn,&
 a\le \tau_i^2&\le Cn^{3/2},\\
 D_i&\ge an^{-13/2},&
 \iota<\Gamma_i&\le Cn^{9/2},\qquad {{\rho_{\rm rat}}}_i\ge an^{-3/2}.
\end{align*}
For example $D_i=c_i^2u_iw_i+d_i^2v_iz_i$.
The ratio ${{\rho_{\rm rat}}}_i$ lies between $\widetilde u_i/\widetilde v_i$ and its
reciprocal, and that ratio lies between $an^{-3/2}$ and $Cn^{3/2}$.
The new movement estimate in Eq.~\eqref{one_half_eq_trace_movement} gives
\[
 a{k_{\rm act}}\le\tr[R_{\mathsf A}]\le C{k_{\rm act}},
 \qquad R_{\mathsf A}:=(I-\widehat T){\mathsf A}(I-\widehat T)^\top.
\]
Eq.~\eqref{one_half_eq_trace_strict_descent} therefore implies
\begin{equation}\label{one_half_eq_trace_quantitative_drift}
 \mathscr{L}\mathcal P\le-a{k_{\rm act}}n^{-3/2}\le-a{k_{\rm act}}n^{-4}.
\end{equation}
The identity
\[
 D_i\tau_i^2=p_i^2u_iw_i+q_i^2v_iz_i\ge an^{-5}
\]
gives $K_{ii}=(R_{\mathsf A})_{e_i,e_i}/(D_i\tau_i^2)$ and hence
$\tr[K]\le C{k_{\rm act}}n^5$.

The scalar certificate gives $k_i\ge a>0$ and
\[
 k_i=\frac{\iota}{2\eta\beta_i^3p_iq_i}
 \le C\delta_{\rm end}^{-3/2}\le Cn^{3/2}.
\]
For the ordinary part of the drift,
$\tau_i(\widetilde u_i+\widetilde v_i)=u_i+v_i\le C\sqrt n$.
Thus $|h_i^{(0)}|\le k_i(u_i+v_i)K_{ii}\le Cn^2K_{ii}$, giving
$\|h^{(0)}\|_1\le Ck_{\rm act}n^7$.

For the trace correction, the helper scalars $a_{*,i},w_i^e,h_{*,i}$
and $j_{*,i}=2{j_{\rm 0}}/F_i$ are bounded. Moreover
\[
 \delta_i r_{*,i}^2
 =\frac{a_{*,i}^2j_i}{k_i(1+{j_{\rm 0}})^2}\le C,\qquad
 |\delta_i p_{*,i}|\le C,
\]
where the second inequality uses $\delta_i\le1/\iota$ and bounded $w_i^e$.
Rewrite the correction as
\[
 \nu_i=
 \frac{\delta_i[h_{*,i}-j_{*,i}+p_{*,i}\upsilon_i]
       \tau_i(\widetilde u_i-\widetilde v_i)}{D_i\tau_i^2}.
\]
The ratio bounds above give $|\upsilon_i|\le Cn^{3/4}$.
Using $|\tau_i(\widetilde u_i-\widetilde v_i)|\le u_i+v_i\le C\sqrt n$
and $D_i\tau_i^2\ge an^{-5}$ yields
$|\nu_i|\le Cn^{25/4}$ and $\|\nu\|_1\le Ck_{\rm act}n^{25/4}$.
For the mixed correction, $|g_i|/(\tau_i^2D_i)\le Cn^6$ and
\[
 |\upsilon_i|\le\max\{\sqrt{\widetilde u_i/\widetilde v_i},
                    \sqrt{\widetilde v_i/\widetilde u_i}\}\le Cn^{3/4}.
\]
Since $\mathsf E_i-\mathsf Y_i=e_i^\top\widehat T\xi$,
${\mathsf A}\preceq CI$, the movement upper bound, and Cauchy--Schwarz give
\[
 \sum_i|\E[\mathsf Y_i(\mathsf E_i-\mathsf Y_i)]|\le Ck_{\rm act},
 \qquad\sum_i|\E[\mathsf Y_i\mathsf F_i]|\le Ck_{\rm act}.
\]
Thus $\|\theta\|_1\le C{k_{\rm act}}n^{27/4}$.
Combining the contributions gives the convenient weaker bounds
\[
 \|h\|_1\le C{k_{\rm act}}n^8,\qquad
 \tr[K]\le C{k_{\rm act}}n^7.
\]
Taking a sufficiently small absolute $\kappa_1>0$, these bounds and
Eq.~\eqref{one_half_eq_trace_quantitative_drift} force
\begin{equation}\label{one_half_eq_trace_alternative}
 \|\nabla \mathcal P\|_\infty\ge \kappa_1n^{-12}
 \quad\text{or}\quad
 \lambda_{\min}(\nabla^2\mathcal P)\le-\kappa_1n^{-12}.
\end{equation}
Otherwise both terms of the generator would be too small in magnitude.
The algorithm uses this alternative without computing its witnesses.
In particular, it need not compute the covariance or its helper coefficients.

\paragraph{Derivative bounds and local steps.}
On the truncated interval, the square-root budget satisfies
$|\beta^{(j)}|\le C\delta_{\rm end}^{1/2-j}$ and
$|\beta^{(j)}|/\beta\le C\delta_{\rm end}^{-j}$.
Write ${\mathsf P}:=(U,V)$ and ${\mathsf Q}:=(W,Z)$.
For a unit $y$ direction, the KKT elimination leading to
Eq.~\eqref{rank_one_eq_cont_18} gives
\[
 \|{\mathsf P}'\|_F\le Cn\delta_{\rm end}^{-1},\qquad
 \|{\mathsf Q}'\|_F\le Cn^2\delta_{\rm end}^{-1},\qquad |b'|\le C\delta_{\rm end}^{-1}.
\]
The second-order envelope identity then gives
$|\mathcal P''|\le Cn^{5/2}\delta_{\rm end}^{1/2-2}\le Cn^5$.
For clarity, the third-order identity contains only the fixed-tangent
third derivatives of the Lagrangian. Its inverse terms have either three
primal factors paired with ${\mathsf Q}$, or two primal factors paired with ${\mathsf Q}'$.
Their norms are bounded by $Cn^3\delta_{\rm end}^{-3}$ and $Cn^4\delta_{\rm end}^{-3}$,
respectively. The differentiated coupling terms have no larger bounds,
by $|\beta^{(j)}|/\beta\le C\delta_{\rm end}^{-j}$ and the dual identities.
Thus $|\mathcal P'''|\le Cn^4\delta_{\rm end}^{-3}\le Cn^7$ on the truncated domain,
giving the same Lipschitz bound for $\mathcal P''$.

Take a step of length $t:=\kappa_2n^{-19}\le\delta_{\rm end}/4$, with $\kappa_2>0$
sufficiently small. A gradient step from Eq.~\eqref{one_half_eq_trace_alternative}
decreases $\mathcal P$ by at least $an^{-31}$. Averaging the two signs of a curvature
step and using the third-derivative bound gives decrease at least $an^{-50}$
for the better sign. The latter is a common lower bound.

\paragraph{Endpoint tests.}
Compute endpoint scores to error at most $\delta\le e_{\rm end}/(4k)$,
and trigger at an approximate score of at least $-\delta$.
A true score is then at least $-2\delta$; increasing $b$ by
$2\delta\|C_i\|\le2\delta$ repairs feasibility. There are at most $k$
such events, so their total possible increase of $\mathcal P$ is less than
$e_{\rm end}$. If no event triggers, the exact state is light.
Near-endpoint and small-input removals preserve the modeled sum and
remove a PSD coupling term, so they cannot increase $\mathcal P$.
The potential therefore remains bounded by its initial value plus
$e_{\rm end}$. Since it is nonnegative, there are $O(n^{50})$ local steps.

\paragraph{Arithmetic implementation.}
The KKT inverse and Newton radius remain $Cn^2$ and $an^{-2}$.
A local step moves the optimizer by at most
$Cn^2\delta_{\rm end}^{-1}t\le Cn^{-16}$, inside the Newton neighborhood.
All direction thresholds, score tolerances, step lengths, and accepted
decreases have polynomial reciprocals. On the truncated interval,
$1-y_i^2\ge a/n$, so rational bisection evaluates $\beta(y_i)$ and its
fixed-order derivatives to inverse-polynomial error with logarithmic
arithmetic work. Choose the coordinate, value, and derivative errors
below fixed fractions of the corresponding acceptance margins.

The KKT Jacobian has order $O(n^2)$ and polynomial norm and inverse
norm. For such a matrix $J$, choose a polynomial bound $L\ge\|J\|$
and start Newton--Schulz iteration at $R_0:=J^*/L^2$. The exact
update $R_{j+1}:=R_j(2I-JR_j)$ satisfies
\[
 I-JR_{j+1}=(I-JR_j)^2.
\]
Its initial residual norm is at most
$1-(L\|J^{-1}\|)^{-2}$. Logarithmically many ordinary matrix products
therefore attain the inverse-polynomial residual required by each
Newton correction, at cost $\widetilde O(n^6)$. The inverse bound converts
this residual to a true optimizer-error bound. The Newton contraction
restores the same error floor after each local move.

The rank-one coupling and its derivatives, the KKT blocks, and the
ordinary Hessian can be assembled with ordinary matrix products of
order at most $O(n^2)$, at the same cost. For a symmetric Hessian
approximation $A$, choose $L\ge\max\{1,\|A\|\}$ and a threshold
$\delta_{\rm curv}:=\kappa_1n^{-12}/4$. If the gradient test fails, the reserved Hessian
error leaves $\lambda_{\min}(A)\le-2\delta_{\rm curv}$. Take $q$ to be
a power of two with
\[
 q\ge(4L/\delta_{\rm curv})\log(8kL/\delta_{\rm curv}).
\]
Form $(I-A/(2L))^q$ by exact repeated squaring and compute the
Rayleigh quotients of its nonzero columns. The spectral-weight
argument of Section~\ref{sec_fast_arithmetic} gives a column with
quotient at most $-\delta_{\rm curv}/2$. Squared norms and quadratic
forms are rational computations. Divide a retained column by its
largest absolute entry, so its norm lies in $[1,\sqrt k]$, and use
bisection to normalize it to the prescribed accuracy. The Hessian and normalization
margins give a true negative-curvature direction. Since $q$ is
polynomially bounded, this construction uses $O(\log(2n))$ products
and costs $\widetilde O(n^6)$ arithmetic operations. Both trial
optimizers lie in the Newton neighborhood, so their corrections and
potential comparisons have the same bound.

The capped initialization in Section~\ref{sec_fast_reset} applies:
$\beta\le1$, the coupling is $\iota_{\rm tr}$, and the coefficients
remain fixed during each solve. There are at most $k+1=O(n^2)$
initializations, each with $\widetilde O(\sqrt n)$ barrier stages.
Ordinary multiplication therefore gives total initialization cost
$\widetilde O(n^{17/2})$. Together with the $O(n^{50})$ local steps,
grouped preprocessing, and coordinate updates, this proves
\[
 T_{\rm arith}\le\widetilde O(mn^2+n^{56}).
\]

\paragraph{Final constant.}
Partial signing contributes at most $\eta_{\rm ps}$; initial rounding and near-endpoint removals cost at most ${\varepsilon_{\rm round}}$;
small-input removals cost at most $\chi$; endpoint repairs cost at most
$e_{\rm end}$. Restoring variance scaling and the earlier discarded inputs,
the internal coefficient is at most
\[
 (2\sqrt{303/100+2{\varepsilon_{\rm scale}}}+{\varepsilon_{\rm round}}+\chi+e_{\rm end}+\eta_{\rm ps})\sqrt{1+{\varepsilon_{\rm scale}}}+{\varepsilon_{\rm discard}}
 < C_{\rm tr,*}=3.4813833.
\]
For the strict inequality, define
$\epsilon:=10^{-8}$ and
$b_*:=3.4813833-10^{-6}-(3\cdot10^{-6}+10^{-7})(1+\epsilon)$.
Exact rational arithmetic gives
\begin{equation}\label{eq:square-root:final-squared-margin}
 b_*=\frac{3481379199999969}{10^{15}}>0,\qquad
 b_*^2-4(303/100+2\epsilon)(1+\epsilon)>9\cdot10^{-7}.
\end{equation}
Taking positive square roots and using
$\sqrt{1+\epsilon}\le1+\epsilon$ proves the strict inequality.
The one-time input approximation, with $\xi=10^{-8}$ as above, changes
the coefficient to at most
$C_{\rm tr,*}(1+6\xi)+2\xi<3.4814$.
This proves the algorithmic assertion of Theorem~\ref{one_half_rank_one_thm_main}.

\subsection{Bit complexity of the square-root-budget algorithm}
\label{sec:square-root:bit-complexity}
The polynomial conditioning and acceptance margins in
Section~\ref{one_half_sec_trace_algorithm} permit a finite-precision
implementation using the shared numerical routines.

\begin{theorem}[Bit complexity]\label{thm:square-root:bit-complexity}
Let $A_1,\ldots,A_m\in\mathbb C^{n\times n}$ be rational Hermitian
matrices of rank at most one and total binary length $L_{\rm in}$.
A deterministic finite-precision implementation of the algorithm in
Theorem~\ref{one_half_rank_one_thm_main} returns signs satisfying
\[
 \|\sum_i\sigma_iA_i\|\le3.4814\|\sum_iA_i^2\|^{1/2}
\]
with bit cost
\begin{equation}\label{eq:square-root:bit-cost}
 T_{\rm bit}\le\widetilde O(L_{\rm in}+mn^2+n^{56}).
\end{equation}
For $A_i=a_ia_i^*$, $\sum_iA_i=I$, and $\|a_i\|^2\le\alpha$, its sign
classes satisfy
\[
 \|\sum_{i\in I_j}a_ia_i^*-I/2\|\le1.7407\sqrt\alpha,
 \qquad j=1,2,
\]
within the same bit bound. After input conversion, preprocessing uses
$O(\log(2mn))$ bits per numerical entry and the nonlinear phase uses
$O(\log(2n))$ bits per numerical entry, including temporary operands,
with fixed multipliers. The suppressed factors in
Eq.~\eqref{eq:square-root:bit-cost} are logarithmic in $m,n,L_{\rm in}$.
\end{theorem}
\begin{proof}
Lemma~\ref{bit_lem_input} performs the one-time input conversion in
$\widetilde O(L_{\rm in}+mn^2)$ bit operations. For variance scaling,
use the normalized-power estimate of Lemma~\ref{bit_lem_squaring}
and the stable products of Lemma~\ref{bit_lem_linear_algebra}.
The inverse-polynomial norm lower bound in
Section~\ref{one_half_sec_trace_algorithm} permits scalar enclosures
with $O(\log(2mn))$ bits. Lemma~\ref{bit_lem_grouped} implements grouped
partial signing on the converted data. Thus preprocessing has bit cost
$\widetilde O(L_{\rm in}+mn^2+n^{5.055})$. Its residual and the initial
coordinate-rounding error remain within the prescribed allowances.
The original inputs are charged once.

For the nonlinear phase, $k\le n^2$ and $\delta_{\rm end}\ge a/n$.
Hence $1-y_i^2\ge a/n$ on every retained trial interval, and the
square-root budget and its fixed-order derivatives have polynomial
magnitudes and perturbation bounds. Bisection evaluates them using
$O(\log(2n))$ bits. The KKT inverse norm is at most $Cn^2$, its
Newton radius is at least $an^{-2}$, and all gradient, curvature,
endpoint, and decrease margins have inverse-polynomial size.
Choose a common numerical unit $(2n)^{-C}$, with a sufficiently large
fixed exponent, below these margins after all polynomial error
amplification and the $O(n^{50})$ step count.

The cache argument of Lemma~\ref{bit_lem_cache} applies with these
square-root-budget bounds. The fixed matrix contribution is stored
symbolically using bounded coefficients, so its error is estimated
afresh from the cache. Every accepted optimizer correction restores
the prescribed root-error floor. Lemma~\ref{bit_lem_linear_algebra}
implements the inverse solves and products, and
Lemma~\ref{bit_lem_initialization} preserves the capped initializer's
feasibility and decrement margins. For curvature selection,
Lemma~\ref{bit_lem_squaring} evaluates the normalized powers and
certifies a column's Rayleigh quotient. The power exponent is
polynomially bounded for $\delta_{\rm curv}=\kappa_1n^{-12}/4$,
so this lemma requires only logarithmic working precision.
All temporary magnitudes are polynomial. Taking the largest of these
fixed precision multipliers supplies a common $O(\log(2n))$ bound.

The endpoint comparisons use the same conservative trigger and
feasibility repair as in Section~\ref{one_half_sec_trace_algorithm}.
The coordinate and value errors are chosen below fixed fractions
of the accepted decrease and the discrepancy allowances. These
enclosures preserve feasibility, the gradient/curvature alternative,
and a decrease of order $n^{-50}$ at each local step. The potential
argument therefore gives $O(n^{50})$ steps also for the finite
implementation. Its rounding, input residual, and endpoint repairs
fit the same final-constant calculation, giving the stated signing
bound. The partition bound follows by halving the signed sum and
using $A_i^2\preceq\alpha A_i$.

With ordinary multiplication, a corrected local step has bit cost
$\widetilde O(n^6)$, including its derivative, curvature, and value
tests. There are at most $O(n^2)$ capped initializations, each with
$\widetilde O(\sqrt n)$ stages, so all initializations cost
$\widetilde O(n^{17/2})$ bit operations. Adding preprocessing and the
local steps gives
\[
 \begin{aligned}[b]
 T_{\rm bit}&\le\widetilde O(L_{\rm in}+mn^2+n^{5.055}
                         +n^{17/2}+n^{50}n^6)\\
 &\le\widetilde O(L_{\rm in}+mn^2+n^{56}),
 \end{aligned}
\]
which proves Eq.~\eqref{eq:square-root:bit-cost}.
\end{proof}

%% file: main.bbl
\newcommand{\etalchar}[1]{$^{#1}$}
\begin{thebibliography}{ADW{\etalchar{+}}25}

\bibitem[ADW{\etalchar{+}}25]{adwxxz25}
Josh Alman, Ran Duan, Virginia~Vassilevska Williams, Yinzhan Xu, Zixuan Xu, and
  Renfei Zhou.
\newblock More asymmetry yields faster matrix multiplication.
\newblock In {\em Proceedings of the 2025 Annual ACM-SIAM Symposium on Discrete
  Algorithms (SODA)}, pages 2005--2039. SIAM, 2025.

\bibitem[AOSS18]{aoss18}
Nima Anari, Shayan {Oveis Gharan}, Amin Saberi, and Nikhil Srivastava.
\newblock Approximating the largest root and applications to interlacing
  families.
\newblock In {\em Proceedings of the Twenty-Ninth Annual ACM-SIAM Symposium on
  Discrete Algorithms (SODA)}, pages 1015--1028. SIAM, 2018.

\bibitem[CLS21]{cls21}
Michael~B Cohen, Yin~Tat Lee, and Zhao Song.
\newblock Solving linear programs in the current matrix multiplication time.
\newblock {\em Journal of the ACM (JACM)}, 68(1):1--39, 2021.

\bibitem[CW90]{cw90}
Don Coppersmith and Shmuel Winograd.
\newblock Matrix multiplication via arithmetic progressions.
\newblock {\em Journal of Symbolic Computation}, 9(3):251--280, 1990.

\bibitem[DDHK07]{ddhk07}
James Demmel, Ioana Dumitriu, Olga Holtz, and Robert Kleinberg.
\newblock Fast matrix multiplication is stable.
\newblock {\em Numerische Mathematik}, 106(2):199--224, 2007.

\bibitem[DEK{\etalchar{+}}26]{dekmrszawb26}
Emilien Dupont, Marvin Eisenberger, Borislav Kozlovskii, Abbas Mehrabian,
  Francisco J.~R. Ruiz, Abigail See, Renfei Zhou, Josh Alman,
  Virginia~Vassilevska Williams, and Matej Balog.
\newblock Improving the matrix multiplication exponent with modern optimization
  and {AlphaEvolve}.
\newblock arXiv preprint arXiv:2608.16884, 2026.

\bibitem[dKV15]{dkv15}
Etienne de~Klerk and Frank Vallentin.
\newblock On the {Turing} model complexity of interior point methods for
  semidefinite programming.
\newblock arXiv preprint arXiv:1507.03549, 2015.

\bibitem[EJ26]{ej26}
Ekene Ezeunala and Haotian Jiang.
\newblock Rank-one matrix discrepancy and algorithmic {Kadison--Singer}.
\newblock {\em arXiv preprint arXiv:2609.17266}, 2026.

\bibitem[FTU23]{ftu23}
Zachary Frangella, Joel~A. Tropp, and Madeleine Udell.
\newblock Randomized {Nystr\"om} preconditioning.
\newblock {\em SIAM Journal on Matrix Analysis and Applications},
  44(2):718--752, 2023.

\bibitem[Gol65]{golden_1965}
Sidney Golden.
\newblock Lower bounds for the {Helmholtz} function.
\newblock {\em Physical Review}, 137(4B):B1127--B1128, 1965.

\bibitem[HJS{\etalchar{+}}22]{hjstz22}
Baihe Huang, Shunhua Jiang, Zhao Song, Runzhou Tao, and Ruizhe Zhang.
\newblock Solving {SDP} faster: A robust {IPM} framework and efficient
  implementation.
\newblock In {\em 2022 IEEE 63rd Annual Symposium on Foundations of Computer
  Science (FOCS)}, pages 233--244. IEEE, 2022.

\bibitem[JMS22]{jms22}
Ben Jourdan, Peter Macgregor, and He~Sun.
\newblock Is the algorithmic {Kadison--Singer} problem hard?
\newblock arXiv preprint arXiv:2205.02161, 2022.

\bibitem[Kat26]{k26}
Tarun Kathuria.
\newblock A walk from free probability to matrix discrepancy {II}: {Weaver}'s
  problem and the {Kadison--Singer} conjecture.
\newblock {\em arXiv preprint arXiv:2609.18913}, 2026.

\bibitem[KLS20]{kls20}
Rasmus Kyng, Kyle Luh, and Zhao Song.
\newblock Four deviations suffice for rank 1 matrices.
\newblock {\em Advances in Mathematics}, 375:107366, 2020.

\bibitem[KS59]{ks59}
Richard~V. Kadison and I.~M. Singer.
\newblock Extensions of pure states.
\newblock {\em American Journal of Mathematics}, 81(2):383--400, 1959.

\bibitem[Leh99]{l99}
Franz Lehner.
\newblock Computing norms of free operators with matrix coefficients.
\newblock {\em American Journal of Mathematics}, 121(3):453--486, 1999.

\bibitem[LU18]{lu18}
Fran{\c c}ois {Le Gall} and Florent Urrutia.
\newblock Improved rectangular matrix multiplication using powers of the
  {Coppersmith--Winograd} tensor.
\newblock In {\em Proceedings of the Twenty-Ninth Annual ACM-SIAM Symposium on
  Discrete Algorithms (SODA)}, pages 1029--1046. SIAM, 2018.

\bibitem[MJF20]{mjf20}
Alaa Maalouf, Ibrahim Jubran, and Dan Feldman.
\newblock Fast and accurate least-mean-squares solvers.
\newblock arXiv preprint arXiv:1906.04705v2, 2020.

\bibitem[MSS15]{mss15}
Adam~W. Marcus, Daniel~A. Spielman, and Nikhil Srivastava.
\newblock Interlacing families {II}: Mixed characteristic polynomials and the
  {Kadison--Singer} problem.
\newblock {\em Annals of Mathematics}, 182(1):327--350, 2015.

\bibitem[SY26]{sy26_high_rank}
Zhao Song and Song Yue.
\newblock Square-root log-rank standard deviations for the higher-rank
  {Kadison--Singer} problem: Polynomial-time algorithms via {Schatten}-norm
  potentials, 2026.
\newblock Preprint.

\bibitem[Tho65]{thompson_1965}
Colin~J. Thompson.
\newblock Inequality with applications in statistical mechanics.
\newblock {\em Journal of Mathematical Physics}, 6(11):1812--1813, 1965.

\bibitem[Tro12]{t12}
Joel~A. Tropp.
\newblock User-friendly tail bounds for sums of random matrices.
\newblock {\em Foundations of Computational Mathematics}, 12:389--434, 2012.

\bibitem[Wea04]{w04}
Nik Weaver.
\newblock The {Kadison--Singer} problem in discrepancy theory.
\newblock {\em Discrete Mathematics}, 278(1--3):227--239, 2004.

\bibitem[XXZ22]{xxz22}
Jiaxin Xie, Zhiqiang Xu, and Ziheng Zhu.
\newblock Upper and lower bounds for matrix discrepancy.
\newblock {\em Journal of Fourier Analysis and Applications}, 28(6):81, 2022.

\end{thebibliography}
